%% file: thesis_fb.tex
\documentclass[12pt,italian,english]{book}
\usepackage[T1]{fontenc}
\usepackage[latin9]{inputenc}
\usepackage[a4paper]{geometry}
\usepackage{color}
\usepackage[english]{babel}
\usepackage{units}
\usepackage{mathrsfs}
\usepackage{amsthm}
\usepackage{amsmath}
\usepackage{amssymb}
\usepackage{makeidx}
\makeindex
\usepackage{graphicx}
\usepackage{setspace}
\usepackage[numbers]{natbib}
\usepackage[unicode=true,pdfusetitle,
 bookmarks=true,bookmarksnumbered=false,bookmarksopen=false,
 breaklinks=true,pdfborder={0 0 0},backref=false,colorlinks=true]
 {hyperref}

\makeatletter

\newcommand{\noun}[1]{\textsc{#1}}

\numberwithin{equation}{section}
\numberwithin{figure}{section}
\theoremstyle{plain}
\newtheorem{thm}{\protect\theoremname}[section]
  \theoremstyle{definition}
  \newtheorem{defn}[thm]{\protect\definitionname}
  \theoremstyle{remark}
  \newtheorem{rem}[thm]{\protect\remarkname}
  \theoremstyle{definition}
  \newtheorem{example}[thm]{\protect\examplename}
  \theoremstyle{plain}
  \newtheorem{prop}[thm]{\protect\propositionname}
  \theoremstyle{plain}
  \newtheorem{lem}[thm]{\protect\lemmaname}
  \theoremstyle{plain}
  \newtheorem{cor}[thm]{\protect\corollaryname}
  \theoremstyle{plain}
  \newtheorem{assumption}[thm]{\protect\assumptionname}

\usepackage{ae,aecompl} 

\usepackage{hyperref}
\usepackage{color}
\usepackage{graphicx}
\usepackage{multicol}
\usepackage{url}
\usepackage{fancyhdr}

\RequirePackage[dvipsnames]{xcolor} 
\definecolor{halfgray}{gray}{0.55} 
\definecolor{webgreen}{rgb}{0,.5,0}
\definecolor{webbrown}{rgb}{.6,0,0}
\definecolor{Maroon}{cmyk}{0, 0.87, 0.68, 0.32}
\definecolor{RoyalBlue}{cmyk}{1, 0.50, 0, 0}
\definecolor{Black}{cmyk}{0, 0, 0, 0}

\hypersetup{
 colorlinks=true, linktocpage=true, pdfstartpage=3, pdfstartview=FitV,%
breaklinks=true, pdfpagemode=UseNone, pageanchor=true, pdfpagemode=UseOutlines,%
plainpages=false, bookmarksnumbered, bookmarksopen=true, bookmarksopenlevel=1,%
hypertexnames=true, pdfhighlight=/O,
urlcolor=webbrown, linkcolor=RoyalBlue, citecolor=webgreen, 
pdftitle = {Relative Cauchy Evolution for Spin 1 Fields},
pdfsubject = {Tesi di Laurea Magistrale in Scienze Fisiche},
pdfkeywords = {locally, covariant, quantum, field, theory, QFT, LCQFT,  generally, locality, principle, GCLP, relative, Cauchy, evolution, RCE},
pdfauthor = {Marco Benini},
pdfcreator = {\LaTeX\ with package \flqq hyperref\frqq},
pdfproducer = {pdfeTeX-0.\the\pdftexversion\pdftexrevision},
plainpages=false,
pdfpagelabels
}

\makeatother

  \addto\captionsenglish{\renewcommand{\assumptionname}{Assumption}}
  \addto\captionsenglish{\renewcommand{\corollaryname}{Corollary}}
  \addto\captionsenglish{\renewcommand{\definitionname}{Definition}}
  \addto\captionsenglish{\renewcommand{\examplename}{Example}}
  \addto\captionsenglish{\renewcommand{\lemmaname}{Lemma}}
  \addto\captionsenglish{\renewcommand{\propositionname}{Proposition}}
  \addto\captionsenglish{\renewcommand{\remarkname}{Remark}}
  \addto\captionsenglish{\renewcommand{\theoremname}{Theorem}}
  \addto\captionsitalian{\renewcommand{\assumptionname}{Assunzione}}
  \addto\captionsitalian{\renewcommand{\corollaryname}{Corollario}}
  \addto\captionsitalian{\renewcommand{\definitionname}{Definizione}}
  \addto\captionsitalian{\renewcommand{\examplename}{Esempio}}
  \addto\captionsitalian{\renewcommand{\lemmaname}{Lemma}}
  \addto\captionsitalian{\renewcommand{\propositionname}{Proposizione}}
  \addto\captionsitalian{\renewcommand{\remarkname}{Osservazione}}
  \addto\captionsitalian{\renewcommand{\theoremname}{Teorema}}
  \providecommand{\assumptionname}{Assumption}
  \providecommand{\corollaryname}{Corollary}
  \providecommand{\definitionname}{Definition}
  \providecommand{\examplename}{Example}
  \providecommand{\lemmaname}{Lemma}
  \providecommand{\propositionname}{Proposition}
  \providecommand{\remarkname}{Remark}
\providecommand{\theoremname}{Theorem}

\begin{document}
\begin{frontmatter}

\input{1_frontpage.tex}

\cleardoublepage{}

\input{3_abstract.tex}

\input{4_ToC.tex}

\end{frontmatter} 

\begin{mainmatter}

\input{5_introduction.tex}

\input{6_chapter1_MathPrel.tex}

\input{7_chapter2_GCLP.tex}

\input{8_chapter3_RCE.tex}

\input{9_conclusions.tex}

\end{mainmatter}

\begin{backmatter}

\input{12_bibliografy.tex}

\input{13_subject_index.tex}

\end{backmatter}

\end{document}

%% file: 1_frontpage.tex
\begin{spacing}{0.90}
\begin{center}
\thispagestyle{empty}
\par\end{center}
\end{spacing}

\vspace{-1cm}

\noindent \begin{center}
\includegraphics[width=3.5cm]{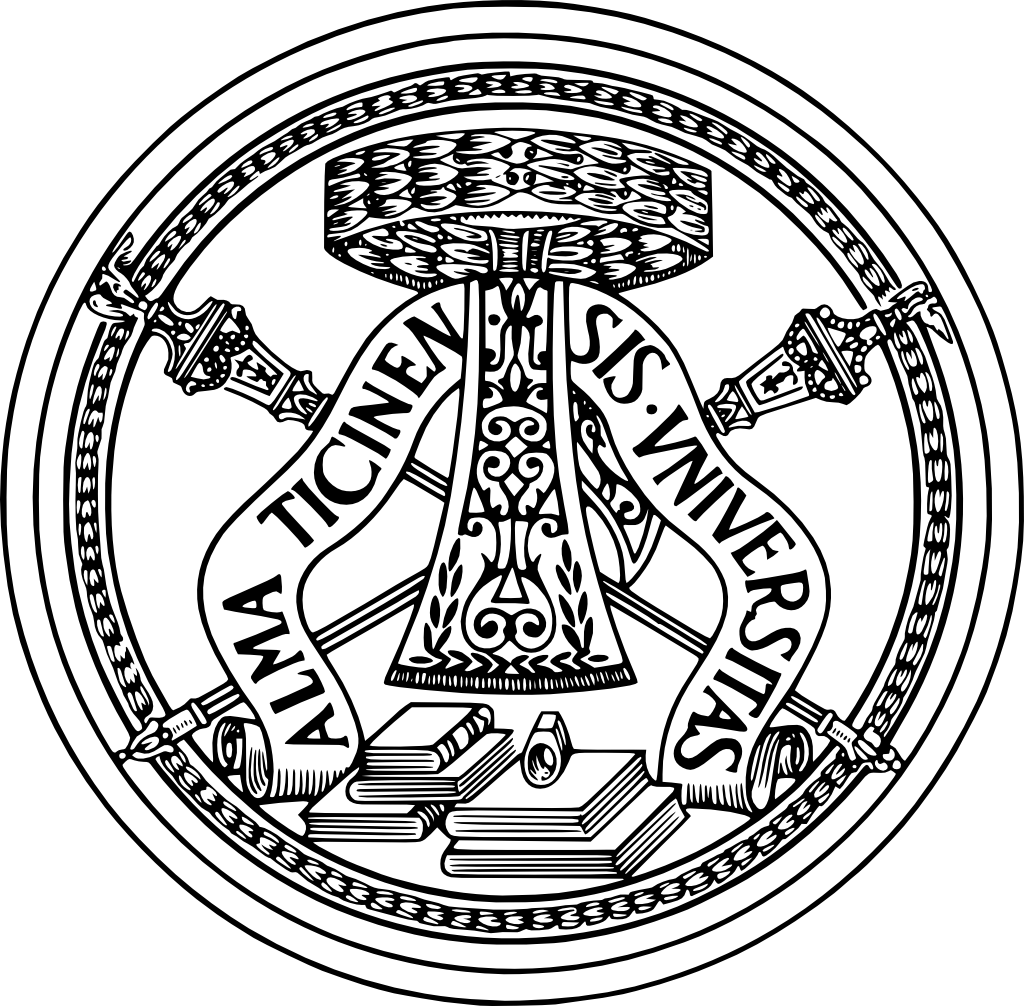}
\par\end{center}

\noindent \begin{center}
\noun{\LARGE Università degli Studi di Pavia}
\par\end{center}{\LARGE \par}

\noindent \begin{center}
{\Large \vspace{-0.5cm}
}\noun{\Large Facoltà di Scienze MM. FF. NN.}
\par\end{center}{\Large \par}

\noindent \begin{center}
{\Large \vspace{-0.5cm}
}\noun{\large Corso di Laurea Magistrale in Scienze Fisiche}
\par\end{center}{\large \par}

\vspace{3.5cm}

\begin{center}
\textsc{\LARGE Relative Cauchy Evolution}\\
\textsc{\LARGE for Spin 1 Fields}
\par\end{center}{\LARGE \par}

\begin{spacing}{0.90}
\vspace{3.5cm}
{\large Relatore:}{\large \par}
\end{spacing}

{\large Dott. Claudio Dappiaggi}{\large \par}

\begin{spacing}{0.90}
\bigskip{}
{\large \hfill{}Tesi di Laurea}{\large \par}
\end{spacing}

{\large \hfill{}di Marco Benini}{\large \par}

\begin{spacing}{0.90}
{\large \hfill{}Matr. N. 386572}{\large \par}
\end{spacing}

\vspace{3.5cm}

\begin{center}
{\large Anno Accademico 2010/2011}
\par\end{center}{\large \par}

\begin{center}
{\Large }
\par\end{center}

%% file: 3_abstract.tex
\selectlanguage{italian}%

\chapter*{Abstract (Italiano)}

Nel corso degli anni '60 del secolo scorso è iniziata la ricerca di
una formulazione matematicamente rigorosa della teoria quantistica
dei campi. Uno dei primi rilevanti successi in questo ambito si deve
all'approccio algebrico e assiomatico proposto da Haag e Kastler (si
veda \cite{HK64}). Tale formalismo consente di definire la teoria
quantistica dei campi sullo spaziotempo di Minkowski in un ben preciso
contesto matematico, quello algebrico, e di implementare in maniera
naturale all'interno di questa teoria i concetti di causalità e di
covarianza di Lorentz.

Precisiamo che questo tipo di approccio non genera una teoria nuova
rispetto alla teoria quantistica dei campi sullo spaziotempo di Minkowski
nella sua formulazione originaria. Al contrario riproduce i medesimi
risultati, presentando tuttavia due vantaggi significativi: in primo
luogo la formulazione della teoria avviene in un contesto matematico
ben precisato, che consente di motivare in maniera rigorosa i risultati
ottenuti, e in secondo luogo l'approccio si rivela adatto a notevoli
estensioni. Infatti nel corso degli anni le idee originali di Haag
e Kastler si sono sviluppate e hanno visto ampliare il proprio dominio
di applicazione, pur conservando in buona parte la loro identità,
sino a giungere alla formulazione della teoria quantistica dei campi
su spazitempi curvi.

A quasi 40 anni di distanza dal lavoro di Haag e Kastler, Brunetti,
Fredenhagen e Verch (\cite{BFV03}) hanno proposto un approccio alla
teoria dei campi su spazitempi curvi che va sotto il nome di \textsl{principio
di località generalmente covariante}. Questo approccio è da considerarsi
come complementare a quello originale in quanto non introduce nuovi
assiomi nella teoria e consente di recuperare in maniera naturale
l'approccio algebrico e assiomatico di Haag e Kastler. D'altra parte
ha il merito di porre l'accento sugli aspetti che accomunano le procedure
di quantizzazione su spazitempi distinti (ovvero la struttura funtoriale
soggiacente) e sulle caratteristiche che invece le contraddistinguono
(ovvero gli spazi di stati). Inoltre la struttura funtoriale di questo
approccio implementa naturalmente la proprietà di covarianza nella
teoria quantistica di campo, così come è previsto dalla relatività
generale per ogni teoria fisica.

Nella tesi è presentato in un contesto generale il \textsl{principio
di località generalmente covariante}\textit{.} Questo postula che
ogni teoria quantistica di campo sia formulata come una \textsl{teoria
quantistica di campo localmente covariante} (nel seguito talvolta
abbreviata dall'acronimo \textsl{LCQFT}). Senza la pretesa di essere
esaustivi, possiamo dire che una \textsl{LCQFT} consiste in un funtore
covariante che a ogni spaziotempo globalmente iperbolico associa un'algebra
e a ogni embedding isometrico tra spazitempi globalmente iperbolici
fa corrispondere un omomorfismo iniettivo tra le algebre associate
a tali spazitempi. Due ulteriori proprietà possono essere richieste
a una \textsl{LCQFT}: la \textsl{causalità}, ovvero, semplificando,
il fatto che commutino tra loro gli elementi di due algebre associate
a spazitempi che ammettono embedding isometrici con immagini causalmente
separate in uno spaziotempo comune, e il \textsl{time slice axiom},
ossia la richiesta che sia suriettivo ogni omomorfismo associato a
un embedding isometrico la cui immagine contiene una superficie di
Cauchy del suo codominio.

Ribadiamo che la covarianza generale è implementata all'interno della
teoria grazie alla proprietà di covarianza del funtore che realizza
certa \textsl{teoria quantistica di campo localmente covariante}.
Sulla scia di quanto provato da Brunetti, Fredenhagen e Verch, riproponiamo
la dimostrazione del fatto che da ogni \textsl{LCQFT} \textsl{causale}
verificante il \textsl{time slice axiom} è possibile recuperare lo
schema assiomatico di Haag e Kastler, il quale coinvolge reti di algebre
locali e automorfismi covarianti associati alle isometrie dello spaziotempo
soggiacente. Questo fatto consente di interpretare una opportuna sottoalgebra
dell'algebra associata da una \textsl{LCQFT} ad un dato spaziotempo
come l'algebra delle osservabili fisiche associate a tale spaziotempo.

L'approccio alla teoria quantistica di campo suggerito dal \textsl{principio
di località generalmente covariante} è completamente indipendente
dal particolare modello fisico che di volta in volta può essere preso
in considerazione, tuttavia, affinché il principio si dimostri fisicamente
rilevante, occorre verificare la possibilità di realizzare una \textsl{teoria
quantistica di campo localmente covariante} che soddisfi sia la \textsl{causalità}
che il \textsl{time slice axiom} in tutte le situazioni di interesse
fisico. Nella tesi si riprendono i risultati ottenuti in \cite{BFV03}
per il campo di Klein-Gordon e si discutono i casi del campo di Proca
e del campo elettromagnetico. Cogliamo l'occasione per ricordare che
il caso del campo di Dirac è stato affrontato in \cite{San10a}.

Come si vedrà, di fatto la realizzazione di una \textsl{teoria quantistica
di campo localmente covariante} per un campo bosonico riposa soltanto
sulla possibilità di costruire uno spazio simplettico di soluzioni
per le equazioni di campo classiche per ogni spaziotempo globalmente
iperbolico e sulla individuazione di una mappa simplettica in corrispondenza
di ogni embedding isometrico tra spazitempi globalmente iperbolici,
mappa simplettica che ha come dominio e codominio gli spazi simplettici
associati agli spazitempi che fanno da dominio e da codominio per
l'embedding assegnato. Per quanto riguarda il soddisfacimento della
\textsl{causalità} e del \textsl{time slice axiom} di una \textsl{LCQFT}
ottenuta in questo modo, di nuovo il problema si riduce a livello
classico a questioni di supporto delle soluzioni di problemi di Cauchy
per le equazioni di campo e alla suriettività della mappe simplettiche.

Obiettivo principale di questa tesi è lo studio di un particolare
tipo di dinamica introdotto in \cite{BFV03} che va sotto il nome
di \textsl{evoluzione relativa di Cauchy} (\textsl{RCE}). La caratteristica
peculiare della \textsl{RCE} risiede nella sua capacità di evidenziare
la sensibilità di una \textsl{teoria quantistica di campo localmente
covariante} alle fluttuazioni della metrica dello spaziotempo sottostante.
Precisamente ci si pone lo scopo di studiare la relazione che intercorre
tra la \textsl{RCE} e il tensore energia-impulso nel caso delle \textsl{LCQFT}
costruite per il campo di Klein-Gordon, per il campo di Proca e per
il campo elettromagnetico. L'interesse nei confronti di tale relazione
nasce dall'intento di incorporare il valore di aspettazione del tensore
energia-impulso di un campo quantistico assegnato nel membro di destra
dell'equazione di Einstein (per maggiori dettagli sull'equazione di
Einstein semiclassica rimandiamo a \cite{Wal94}).

Seguendo la definizione proposta recentemente da Fewster e Verch in
\cite{FV11}, limitatamente a quelle \textsl{teorie quantistiche di
campo localmente covarianti} che soddisfano il \textsl{time slice
axiom}, definiamo l'\textsl{evoluzione relativa di Cauchy} come un
automorfismo sull'algebra associata a un dato spaziotempo globalmente
iperbolico indotto da una perturbazione locale della metrica spaziotemporale.
La definizione stessa della \textsl{RCE} consente di interpretarla
come una sorta di reazione dinamica della teoria quantistica di campo
a una fluttuazione della metrica dello spaziotempo sottostante. Riesamineremo
alcune proprietà della \textsl{RCE} ponendo l'accento sulla sua insensibilità
a perturbazioni della metrica indotte da diffeomorfismi e sul fatto
che, di conseguenza, la derivata funzionale della \textsl{RCE} rispetto
alla metrica abbia divergenza nulla.

In \cite{BFV03} è sviluppato nel dettaglio lo studio dell'\textsl{evoluzione
relativa di Cauchy} per il campo di Klein-Gordon. In particolare Brunetti,
Fredenhagen e Verch giungono a dimostrare una particolare relazione
che in questa situazione intercorre tra \textsl{RCE} e tensore energia-impulso.
Qui questo caso è riesaminato a scopo esemplificativo e ci si pone
l'obiettivo di estendere la relazione tra \textsl{RCE} e tensore energia-impulso
dimostrata in \cite{BFV03} per il campo di Klein-Gordon anche ai
casi del campo di Proca e del campo elettromagnetico (per l'analogo
problema nel caso del campo di Dirac si rimanda di nuovo a \cite{San10a}).
In questo modo il significato dell'\textsl{evoluzione relativa di
Cauchy} in relazione al tensore energia-impulso risulta esteso dal
caso del campo di Klein-Gordon ai casi del campo di Proca e del campo
elettromagnetico. In particolare questo fatto motiva l'introduzione
del valore di aspettazione del tensore energia-impulso nel membro
di destra dell'equazione di Einstein anche per i casi del campo di
Proca e del campo elettromagnetico.

\selectlanguage{english}%

\chapter*{Abstract (English)}

During the Sixties of the last century the search for a mathematically
rigorous formulation of quantum field theory has begun. One of the
first and most prominent successes in this area is due to the algebraic
and axiomatic approach proposed by Haag and Kastler (refer to \cite{HK64}).
This formalism allows the definition of quantum field theory over
Minkowski spacetime in a precisely specified mathematical context,
namely the algebraic one, and the natural implementation of the notions
of causality and Lorentz covariance in such theory.

We specify that this approach does not produce a new theory with respect
to the original formulation of quantum field theory on Minkowski spacetime.
On the contrary it gives rise to equivalent results, yet presenting
two significant advantages: in first place the theory is formulated
in a precise mathematical context, that allows to motivate rigorously
the results one obtains, and in second place the approach proves suitable
to remarkable extensions. As a matter of fact over the years the original
ideas of Haag and Kastler were significantly developed and went through
an enlargement of their range of applicability, while largely preserving
their original identity, until the formulation of quantum field theory
on curved spacetimes.

Almost 40 years after the work made by Haag and Kastler, Brunetti,
Fredenhagen and Verch (\cite{BFV03}) proposed a new approach to quantum
field theories on curved spacetimes named \textsl{generally covariant
locality principle}. On one hand this approach is to be considered
as complementary to the original one since it does not add new axioms
to the theory and allows the natural recovering of the algebraic and
axiomatic approach by Haag and Kastler. On the other hand it has the
merit of highlighting the common aspects of quantization procedures
on different spacetimes (namely the underlying functorial structure)
and the distinguishing features (namely state spaces). Furthermore
the functorial structure of this approach naturally implements covariance
in quantum field theories, as it is expected by each physical theory
according to general relativity.

In this thesis the \textsl{generally covariant locality principle}
is presented in a general setting. It postulates that each quantum
field theory be formulated as a \textsl{locally covariant quantum
field theory} (sometimes denoted by the acronym \textsl{LCQFT}). Without
pretending to be exhaustive, we may say that a \textsl{LCQFT} consists
of a covariant functor mapping each globally hyperbolic spacetime
to an algebra and each isometric embedding between two globally hyperbolic
spacetimes to an injective homomorphism between the algebras associated
to such spacetimes. Other two properties can be required to a \textsl{LCQFT}:
\textsl{causality}, which, simplifying, means that elements coming
from two algebras associated to spacetimes isometrically embedded
in causally separated subregions of a common spacetime commute, and
the \textsl{time slice axiom}, which requires that each homomorphism
associated to an isometric embedding whose image includes a Cauchy
surface of its codomain be surjective.

We repeat that general covariance is implemented in the theory as
a consequence of the covariance property of the functor giving rise
to a \textsl{locally covariant quantum field theory}. Following what
was shown by Brunetti, Fredenhagen and Verch, we present the proof
of the fact that, starting from a \textsl{LCQFT} fulfilling both \textsl{causality}
and the \textsl{time slice axiom}, it is possible to recover the Haag-Kastler
scheme, involving nets of local algebras and covariant automorphisms
associated to isometries of the underlying spacetime. This fact makes
it possible to interpret a proper subalgebra of the algebra provided
by a \textsl{LCQFT} on a given spacetime as the algebra of physical
observables associated to that spacetime.

The approach to quantum field theory suggested by the \textsl{generally
covariant locality principle} is completely independent of the specific
physical model considered from time to time, yet we must check the
possibility of realizing a \textsl{locally covariant quantum field
theory} fulfilling both \textsl{causality} and the \textsl{time slice
axiom} in each situation of physical interest in order to have a physically
relevant principle. In this thesis the results obtained in \cite{BFV03}
for the Klein-Gordon field are recovered and the cases of the Proca
and the electromagnetic fields are discussed. We take the chance to
remind that the case of the Dirac field was handled in \cite{San10a}.

As we will see, the construction of a \textsl{locally quantum field
theory} for a bosonic field actually relies only on the possibility
of building a symplectic space of solutions for the classical field
equations for each globally hyperbolic spacetime and on the specification
of a symplectic map for each isometric embedding between two globally
hyperbolic spacetimes, the domain and codomain of the symplectic map
being the symplectic spaces associated to the domain and codomain
of the given embedding. As for the \textsl{causality property} and
the \textsl{time slice axiom} of a \textsl{LCQFT} built in this way,
again the problem is reduced at a classical level to a matter of support
for solutions of Cauchy problems for the field equations and to the
surjectivity of the symplectic maps.

The main purpose of this thesis is to study a particular type of dynamics
proposed by \cite{BFV03} named \textsl{relative Cauchy evolution}
(briefly \textsl{RCE}). The distinctive feature of the \textsl{RCE}
relies in its ability of highlighting the sensitivity of a \textsl{locally
covariant quantum field theory} to fluctuations of the metric of the
underlying spacetime. In particular our aim is to study a relation
between the \textsl{RCE} and the stress-energy tensor for the \textsl{LCQFTs}
built for the Klein-Gordon field, the Proca field and the electromagnetic
field. The interest in such relation arises from the intention of
including the expectation value of the stress-energy tensor of a given
quantum field in the right hand side of the Einstein's equation (for
further details on the semiclassical Einstein's equation we refer
to \cite{Wal94}).

Following the definition recently proposed by Fewster and Verch in
\cite{FV11}, only for those \textsl{locally covariant quantum field
theories} fulfilling the \textsl{time slice axiom}, we define the
\textsl{relative Cauchy evolution} as an automorphism on the algebra
associated to a given globally hyperbolic spacetime induced by a local
perturbation of the spacetime metric. The definition of the \textsl{RCE}
suggests its interpretation as a dynamical reaction of the quantum
field theory to a fluctuation of the metric of the underlying spacetime.
We will re-examine some properties of the RCE with particular attention
to its insensitivity to perturbations of the metric induced by diffeomorphisms
and to the fact that, consequently, the functional derivative of the
RCE with respect to the spacetime metric has null divergence.

In \cite{BFV03} the relative Cauchy evolution for the Klein-Gordon
field is thoroughly analyzed. In particular Brunetti, Fredenhagen
and Verch were successful in showing that in this case a particular
relation between the RCE and the stress-energy tensor holds. Here
we re-examine this case as an example and we have as our goal to extend
to the cases of the Proca and the electromagnetic fields the relation
between the RCE and the stress-energy tensor proved in \cite{BFV03}
for the Klein-Gordon field (for the similar problem in the case of
the Dirac field we refer again to \cite{San10a}). In this way the
meaning of the RCE in relation to the stress-energy tensor is extended
from the case of the Klein-Gordon field to the Proca and the electromagnetic
fields. In particular this fact motivates the insertion of the expectation
value of the stress-energy tensor on the right hand side of the Einstein
equation for the Proca and the electromagnetic fields too.

%% file: 4_ToC.tex
\begin{flushleft}
\newpage
\par\end{flushleft}

\tableofcontents{}

%% file: 5_introduction.tex
\chapter*{Introduction}

\addcontentsline{toc}{chapter}{Introduction}

In the mid Sixties Haag and Kastler proposed an algebraic approach
to quantum field theory on Minkowski spacetime (\cite{HK64}). Although
it is equivalent to the original formulation of quantum field theory
arising from the Wightman axioms (\cite{SW64}), this approach proved
to be very successful since it provided a mathematically precise framework
for quantum field theories which could be easily applied on curved
spacetimes.

In this context a further milestone ahead was unveiled by Brunetti,
Fredenhagen and Verch in \cite{BFV03}. To wit they formulated the
\textsl{generally covariant locality principle} (in the following
denoted by \textsl{GCLP}), postulating that each quantum field theory
on an arbitrary globally hyperbolic spacetime must be provided by
a \textsl{locally covariant quantum field theory} (\textsl{LCQFT}),
i.e. a covariant functor from the category of globally hyperbolic
spacetimes to the category of algebras. The result is a formulation
of quantum field theory that naturally exhibits the covariance property
required by general relativity, this being a direct consequence of
the functorial structure of each LCQFT.

As suggested in \cite{BFV03}, one can require two additional properties
to a LCQFT:
\begin{itemize}
\item \textsl{causality}, which, roughly speaking, means that we require
that elements of the algebras, which are associated via a fixed LCQFT
to globally hyperbolic spacetimes embedded in causally separated subregions
of another globally hyperbolic spacetime, must commute;
\item the \textsl{time slice axiom}, which requires that each morphism of
the category of algebras must be surjective if it is obtained applying
a given LCQFT to a morphism of the category of globally hyperbolic
spacetimes, whose image includes a Cauchy surface of the target spacetime.
\end{itemize}
Causality forces the absence of causal relations between observables
localized in causally separated subregions of a globally hyperbolic
spacetime. This simply means that we do not admit causal effects between
events not connected by causal curves. As for the time slice axiom,
we can interpret it as a sort of causal determinacy, in analogy with
the classical case. As much as we know everything about a classical
dynamical system once suitable initial data on a Cauchy surface of
a globally hyperbolic spacetime are assigned, likewise the whole algebra
of observables associated to a quantum field on a globally hyperbolic
spacetime is contained in the algebra of observables associated to
a suitable neighbourhood of a Cauchy surface.

In \cite{BFV03} it was shown that, on each globally hyperbolic spacetime,
an arbitrary causal LCQFT automatically gives rise to a quantum field
theory satisfying the Haag-Kastler axioms (\cite{HK64}). Hence we
may regard the GCLP as a natural criterion to realize on curved spacetimes
the approach to quantum field theory originally proposed by Haag and
Kastler. Moreover we may borrow the interpretation of the Haag-Kastler
axioms saying that a proper subalgebra of the algebra assigned by
a fixed LCQFT applied to any but fixed globally hyperbolic spacetime
is the algebra of the quantum observables admitted by the physics
on the given spacetime.

The GCLP proved to be very successful. A number of results in various
topics about quantum field theories on curved spacetimes were proved
in this framework. For example LCQFTs fulfilling both causality and
the time slice axiom were built for free field models of physical
interest (namely Klein-Gordon, Dirac, Proca and electromagnetic fields)
and questions about what it is meant for a theory to produce the same
physics in all spacetimes arose. A few references are \cite{BFV03,BGP07,San10a,Dap11,FV11}.

Indeed this is not the whole story for quantum field theories on curved
spacetimes. In fact at this point we are not able to get physical
predictions from the algebra of observables. What we need is a notion
of state to be evaluated on the observables in order to get predictions
exactly as we do in quantum mechanics. This issue is not touched by
the GCLP, nor we discuss it in this thesis. Yet we feel worth to say
that relevant results were obtained also in this sector. For example
it is known that there exist states for quantum field theories on
globally hyperbolic spacetimes which satisfy properties that are known
to hold for the vacuum states of quantum field theories on Minkowski
spacetime (e.g. the Hadamard condition and the Reeh-Schlieder property).
Some references for these topics are \cite{KW91,Rad96,SV01,SVW02,FV03,FP03,San10b,Dap11}.

Another interesting application of the GCLP consists in the realization
of a particular form of dynamics known as \textsl{relative Cauchy
evolution} (\textsl{RCE}). The RCE is an algebraic automorphism that
can be defined on each globally hyperbolic spacetime and for each
LCQFT fulfilling the time slice axiom. Its relevance relies in the
fact that it accounts for the effects that a fluctuation of the spacetime
metric produces on the algebra provided by the LCQFT on a given globally
hyperbolic spacetime.

The study of the RCE is interesting in first place because indeed
we want to deal with a stable theory, which is to say that it would
be unlikely to have a quantum field theory on a globally hyperbolic
spacetime with observables that are so much sensitive to small changes
in the spacetime metric that they disappear (or maybe appear) only
because of a small change in the spacetime geometry. In the second
place the interest in the analysis of the reaction of a quantum field
theory to fluctuations of the spacetime metric comes from the attempt
to solve the semiclassical Einstein's equation (we only give a sketch
of the problem). Up to now our quantum field theories (and this is
the case of the GCLP too) are settled on spacetimes which are given
once and for all. Yet, as far as we know, the spacetime where we live
is a solution of the Einstein's equation. To simplify the situation
assume that in the whole universe there is nothing but a quantum field.
Then one should insert the expectation value of the stress-energy
tensor associated to such field on the RHS of the Einstein's equation
(the equation that arises is the above mentioned semiclassical Einstein's
equation, see \cite{Wal94} for further reference). When one tries
to solve the semiclassical Einstein's equation, serious difficulties
emerge: As the solution develops, the quantum field given at the beginning
is affected by the new geometry of the spacetime where it lives. Hence
we have a back-reaction effect, namely the quantum field, whose stress-energy
tensor appears on the RHS of the semiclassical Einstein's equation,
is affected by the solution of such equation. If the quantum field
theory is too much sensitive to a change in the spacetime structure
(essentially a change in the metric), it may happen that the stress-energy
tensor appearing on the RHS of the semiclassical Einstein's equation
loses its meaning while we solve the equation (as a matter of fact
it happens that we no longer have any equation to solve). Being able
to properly control the RCE means that the algebra of observables
provided by a given LCQFT on some globally hyperbolic spacetime is
not severely distorted by a small change in the spacetime metric,
hence we can expect that the stress-energy tensor associated to the
quantum field preserves its meaning while we solve the semiclassical
Einstein's equation, i.e. it still describes the stress-energy tensor
associated to the quantum field taken into account even when the spacetime
geometry has changed due to the fact that we are solving the Einstein's
equation.

Now that we have given a sketch of the topics we are going to deal
with and we have presented the motivation that pushed us to their
study, we would like to briefly summarize the content of the thesis.

In Chapter \ref{chapMathematicalPreliminaries} we present almost
all the mathematical tools that will be needed for the next chapters.
We devote Section \ref{secDifferentialGeometry} to introduce some
notions in differential geometry, namely manifolds and vector bundles.
Particular attention is devoted to differential forms and integration
over manifolds. In Section \ref{secLorentzianGeometry} we specialize
to the case of Lorentzian manifolds, being interested in the notion
of global hyperbolicity. With these concepts at hand, in Section \ref{secWaveEquations}
we turn our attention to the discussion of wave equations on globally
hyperbolic spacetime. In first place we define what we mean by wave
equation (or normally hyperbolic equation to be more precise) and
in second place we present a theorem about the existence and uniqueness
of solutions for Cauchy problems associated to normally hyperbolic
equations, we introduce Green operators and we study some of their
properties. In Section \ref{secAlgebrasAndStates} we completely change
the subject in order to deal with algebras and states. We are mainly
interested in unital C{*}-algebras (in particular Weyl systems and
CCR representations, which are special C{*}-algebras that bestly fit
the canonical commutation relations) and states defined on them. We
conclude the first chapter with Section \ref{secCategoryTheory},
where we recall some basic concepts from category theory.

The main discussion begins with Chapter \ref{chapGCLP}. In Section
\ref{secLCQFT} the generally covariant locality principle (GCLP)
is formulated defining the notion of locally covariant quantum field
theory (LCQFT) and a physical interpretation of the principle is provided,
interpretation that is essentially borrowed from that of the Haag-Kastler
axioms (refer to \cite{HK64}). We conclude this section showing that
it is possible to rigorously recover the Haag-Kastler axioms (hence
their interpretation) once that a LCQFT fulfilling the causality condition
and the time slice axiom is given. We devote Section \ref{secBuildingALCQFT}
to show a procedure to build a LCQFT starting from the assignment
of a proper normally hyperbolic equation involving sections in a general
vector bundle over a globally hyperbolic spacetime. Such procedure
essentially consists in the construction of a covariant functor describing
the theory of the classical field and in the quantization of this
theory via composition with a properly defined covariant functor which
embodies the quantization scheme. Section \ref{secExamples} concludes
the second chapter presenting the realization of LCQFTs for three
models of physical interest, namely the Klein-Gordon field, the Proca
field and the electromagnetic field. While the Klein-Gordon field
is a mere specialization of the general procedure presented in Section
\ref{secBuildingALCQFT}, the other two require significant modifications
due to the fact that their classical dynamics is not ruled by a normally
hyperbolic equation.

We conclude the thesis with Chapter \ref{chapRCE} discussing the
relative Cauchy evolution (RCE). In Section \ref{secRCEDefinitionAndProperties}
we define the RCE for a LCQFT fulfilling the time slice axiom and
we study its insensitivity to fluctuations of the perturbed spacetime
metric produced by diffeomorphisms. After that we introduce the functional
derivative of the RCE with respect to the spacetime metric as a section
in the symmetrized tensor product of two copies of the tangent bundle
and we show that its divergence (with respect to the Levi-Civita connection)
is null. These properties, namely symmetry and null divergence, are
hints for a strict relation between the functional derivative of the
RCE and the stress-energy tensor associated to some quantum field.
The study of this relation for the specific cases of the Klein-Gordon,
the Proca and the electromagnetic fields concludes the thesis. Specifically
in Section \ref{secRCEForConcreteFields}, after a brief summary of
some of the properties satisfied by quasifree Hadamard states, we
present the calculation originally performed in \cite{BFV03} to prove
that a strict relation between the RCE and the quantized stress-energy
tensor holds for the Klein-Gordon field and we show that an identical
relation holds for the Proca and the electromagnetic fields too.

%% file: 6_chapter1_MathPrel.tex
\chapter{\label{chapMathematicalPreliminaries}Mathematical preliminaries}

We devote the present chapter to the introduction of the main mathematical
tools which will be indispensable for the discussion in the following
chapters. All the topics presented here are discussed very briefly
and the interested reader is invited to refer to the specific literature
of each sector. For this scope at the beginning of all sections we
provide some reference for the subject discussed.

The first section is devoted to the definition of manifolds, vector
bundles and connections, differential forms and integration. In the
second section we present few arguments concerning Lorentzian geometry.
Then the third section is devoted to some basic topics about wave
equations on globally hyperbolic spacetimes: we present a theorem
about existence and uniqueness of solutions to such equations with
proper initial data and then we will introduce the advanced and retarded
Green operators together with their properties. In the fourth section
of this chapter we turn our attention to the mathematical ingredients
that will be essential in the construction of the algebraic approach
to quantum field theory, specifically C{*}-algebras and states. Finally
the last section presents some very useful concepts of category theory
that will be widely applied in the next chapters.

\section{\label{secDifferentialGeometry}Differential geometry}

This section is a very concise (and far from complete) recollection
of the notions in differential geometry that are unavoidable for our
discussion. Besides the efforts spent in making this section self
sufficient, almost all topics are presented in a manner that is too
brief to be clear for a reader that approaches to them for the first
time. For this reason the author strongly encourages the reader to
refer to any book concerning differential geometry (for example \cite{Jos95}
or \cite{Boo86}) to clarify the omissions to which we are forced.

\subsection{Manifolds and tensor bundles}

We begin defining manifolds. These objects will provide the playground
for the entire thesis. The notion of manifold that we present is not
the more general one. To be precise we define smooth connected Hausdorff
manifolds with a countable basis of open subsets. This is a sufficiently
wide class of manifolds and at the same time it incorporates a number
of properties we are interested in.
\begin{defn}
\label{defManifold}\index{manifold}\index{coordinate neighborhood}A
\textsl{$d$-dimensional manifold} $M$ is a connected Hausdorff topological
space with a countable basis of open subsets such that for each point
$p\in M$ there exists a triple $\left(U,\Omega,\phi\right)$, called
\textsl{coordinate neighborhood} (or \textsl{local chart}), where
$U$ is an open neighborhood of $p$ in $M$, $\Omega$ is an open
neighborhood of 0 in $\mathbb{R}^{d}$ and $\phi:U\rightarrow\Omega$
is a homeomorphism. There are two other requirements:
\begin{itemize}
\item \index{atlas}\index{transition chart}there exists a \textsl{(smooth)
atlas}, which is a collection $\left\{ \left(U_{\alpha},\Omega_{\alpha},\phi_{\alpha}\right)\right\} _{\alpha\in I}$
of coordinate neighborhoods in $M$, where $I$ is an index set, such
that $\left\{ U_{\alpha}\right\} _{\alpha\in I}$ is an open covering
of $M$ and the map, called \textsl{transition chart}, 
\begin{eqnarray*}
T_{\phi_{\alpha}}^{\phi_{\beta}}:\Omega_{\alpha}\cap\Omega_{\beta} & \rightarrow & \Omega_{\alpha}\cap\Omega_{\beta}\\
x & \mapsto & \left(\phi_{\beta}\circ\phi_{\alpha}^{-1}\right)\left(x\right)
\end{eqnarray*}
is a diffeomorphism for each $\alpha$, $\beta\in I$ such that $U_{\alpha}\cap U_{\beta}\neq\emptyset$;
\item \index{maximal atlas}there exists a \textsl{maximal atlas}, i.e.
an atlas that contains each coordinate neighborhood $\left(U,\Omega,\phi\right)$
such that the transition maps $T_{\phi_{\alpha}}^{\phi}$ and $T_{\phi}^{\phi_{\alpha}}$
are diffeomorphisms for each $\alpha\in I$ with $U_{\alpha}\cap U\neq\emptyset$.
\end{itemize}
\end{defn}
We would like to make some remarks concerning this definition. In
the first place each atlas of a manifold is contained in a maximal
one, so that it is sufficient to find an atlas and then the maximal
atlas is automatically obtained. This implies that a connected Hausdorff
space with a countable basis becomes a manifold if it possesses an
atlas, even if not maximal. Secondly we observe that the topology
of each manifold defined here is such that it is also a paracompact
space and this implies that for each of our manifolds there exists
a partition of unity (cfr. \cite[Chap. V, Sect. 4, p. 193]{Boo86}).

Now that we have a notion of manifold, we would like to define {}``regular''
functions between manifolds (continuous functions are already defined
since manifolds are topological spaces).
\begin{defn}
\label{defRegularFunction}\index{function of class mathrm{C}^{k}@function of class $\mathrm{C}^{k}$}\index{smooth function}\index{diffeomorphism}Let
$M$ and $N$ be two manifolds and let $f$ be a continuous function
from $M$ to $N$. We say that $f$ is \textsl{a $\mathrm{C}^{k}$-function}
if for each $p\in M$, each coordinate neighborhood $\left(U,\Omega,\phi\right)$
of $p$ in $M$ and each coordinate neighborhood $\left(V,\Theta,\psi\right)$
of $f\left(p\right)$ in $N$, the function
\begin{eqnarray*}
f_{U,V}:\phi\left(U\cap f^{-1}\left(V\right)\right) & \rightarrow & \psi\left(f\left(U\right)\cap V\right)\\
x & \mapsto & \left(\psi\circ f\circ\phi^{-1}\right)\left(x\right)
\end{eqnarray*}
is of class $\mathrm{C}^{k}$ (in the sense of functions between open
subsets of Euclidean spaces).

Moreover $f$ is a \textsl{smooth function} if it is a $\mathrm{C}^{k}$-function
for each $k\in\mathbb{N}$ and we say that $f$ is a \textsl{diffeomorphism}
if it is a homeomorphism which is smooth together with its inverse.
\end{defn}
Given a manifold $M$ and a notion of smooth function, for each $p\in M$
it is possible to introduce a vector space $\mathrm{T}_{p}M$, called
tangent space that proves very useful when one wants to speak of {}``derivatives''
at the point $p$ of real valued functions defined on $M$.
\begin{defn}
\label{defTangentSpace}\index{tangent space}Let $M$ be a $d$-dimensional
manifold and let $p\in M$. Consider the set $\mathscr{C}_{p}$ of
smooth curves $c:I\rightarrow M$, where $I$ is an open interval
of $\mathbb{R}$ containing 0, such that $c\left(0\right)=p$ . We
say that two curves $c_{1}$, $c_{2}\in\mathscr{C}_{p}$ are equivalent
(and we write $c_{1}\sim c_{2}$) if there exists a coordinate neighborhood
$\left(U,\Omega,\phi\right)$ of $p$ such that $\left(\phi\circ c_{1}\right)^{\prime}\left(0\right)=\left(\phi\circ c_{2}\right)^{\prime}\left(0\right)$%
\footnote{Here the composition $\circ$ is to be intended in a proper sense:
$\phi\circ c_{1}$ denotes the composition of $\phi$ with a function
$d_{1}$ from an open interval $J$ of $\mathbb{R}$ containing 0
(eventually smaller than the domain $I$ of $c_{1}$) to $U$ defined
by $d_{1}\left(t\right)=c_{1}\left(t\right)$ for each $t\in I$.%
}, where $^{\prime}$ denotes the usual derivative of a function from
an open interval of $\mathbb{R}$ containing 0 to an open subset of
a Euclidean space. Then we define the \textsl{tangent space $\mathrm{T}_{p}M$}
as the quotient of $\mathscr{C}_{p}$ with respect of the equivalence
relation $\sim$.
\end{defn}
It is possible to show that $\mathrm{T}_{p}M$ is actually a $d$-dimensional
$\mathbb{R}$-vector space and that its elements act as {}``derivatives''
on real valued functions defined on neighborhoods of $p$. To be precise
by {}``derivative'' we mean the following: let $f$ be a smooth
real valued function defined at least on a neighborhood of $p\in M$
and let $v$ be an element of the tangent space $\mathrm{T}_{p}M$;
we define the application of $v$ to $f$ as the real number $\left(f\circ c\right)^{\prime}\left(0\right)$
where $c$ is any of the curves in the equivalence class $v$. To
see how this works refer to \cite[Chap. 2]{Ish99}: there the tangent
space is seen both as a {}``set of derivatives'' and as a set of
equivalence classes of curves (as in the above definition) and the
equivalence of this two approaches is thoroughly analyzed.
\begin{rem}
\label{remTensorBundles}\index{cotangent space}\index{tensor space of type left(i,jright)@tensor space of type $\left(i,j\right)$}\index{tensor space}Thanks
to the $\mathbb{R}$-vector structure of $\mathrm{T}_{p}M$, it is
possible to introduce the \textsl{cotangent space} $\mathrm{T}_{p}^{*}M$
as its dual: We define the elements of $\mathrm{T}_{p}^{*}M$ as linear
maps from $\mathrm{T}_{p}M$ to $\mathbb{R}$. We obtain again a $d$-dimensional
$\mathbb{R}$-vector space and then we can build via tensor products
a new $d^{i+j}$-dimensional $\mathbb{R}$-vector space called \textsl{tensor
space of type $\left(i,j\right)$}:
\[
\mathrm{T}_{p}^{\left(i,j\right)}M=\left(\mathrm{T}_{p}M\right)^{\otimes i}\otimes\left(\mathrm{T}_{p}^{*}M\right)^{\otimes j}\mbox{.}
\]
By convention we set $\mathrm{T}_{p}^{\left(0,0\right)}M=\mathbb{R}$.
Finally we build the \textsl{tensor space} via direct sum:
\[
\mathscr{T}_{p}M=\bigoplus_{\left(i,j\right)\in\mathbb{N}\times\mathbb{N}}\mathrm{T}_{p}^{\left(i,j\right)}M\mbox{.}
\]
This is again a real vector space (this time $\dim\mathscr{T}_{p}M=\infty$)
and it can be even shown that $\left(\mathscr{T}_{p}M,\otimes\right)$
is an associative algebra generated by $\mathbb{R}$, $\mathrm{T}_{p}M$
and $\mathrm{T}_{p}^{*}M$.

\index{tangent bundle}\index{cotangent bundle}\index{tensor bundle of type left(i,jright)@tensor bundle of type $\left(i,j\right)$}\index{tensor bundle}Once
that we have the notion of tangent space, we can define the \textsl{tangent
bundle $\mathrm{T}M$} of a manifold $M$ as the disjoint union on
the manifold of the tangent spaces at each point:
\[
\mathrm{T}M=\bigsqcup_{p\in M}\mathrm{T}_{p}M\mbox{.}
\]
Similarly we define the \textsl{cotangent bundle} $\mathrm{T}^{*}M$,
the \textsl{tensor bundle of type $\left(i,j\right)$} $\mathrm{T}^{\left(i,j\right)}M$
and the \textsl{tensor bundle $\mathscr{T}M$}. Notice that $\mathrm{T}^{\left(0,0\right)}M$
is simply $M\times\mathbb{R}$.

At this point $\mathrm{T}^{\left(i,j\right)}M$ are merely sets. Hereafter
we will endow them with a far richer structure.

Our knowledge about tangent spaces allows us to introduce a notion
of differential at a point that can be patched on the entire manifold
giving rise to the so called pushforward. This new differential at
a fixed point indeed reduces to the usual differential when the manifolds
involved are open subsets of Euclidean spaces endowed with the trivial
atlas (the canonical identification of each tangent space at a point
of an open subset of a Euclidean space with the same Euclidean space
is understood).\end{rem}
\begin{defn}
\label{defPushforward}\index{differential at a point}\index{push-forward}Let
$M$ and $N$ be two manifolds. Consider a smooth map $f:M\rightarrow N$
and a point $p\in M$. We define the \textsl{differential of $f$
at $p$} as the map
\begin{eqnarray*}
\mathrm{d}_{p}f:\mathrm{T}_{p}M & \rightarrow & \mathrm{T}_{f\left(p\right)}N\mbox{,}\\
\left[c\right] & \mapsto & \left[f\circ c\right]\mbox{,}
\end{eqnarray*}
where $\left[\cdot\right]$ denotes the equivalence class in the appropriate
tangent space that has $\cdot$ as representative.

We define the \textsl{push-forward through $f$} as the map $f_{*}:\mathrm{T}M\rightarrow\mathrm{T}N$
such that $\left.f_{*}\right|_{\mathrm{T}_{p}M}=\left(p,\mathrm{d}_{p}f\right)$
for each $p\in M$.

\index{immersion}\index{submersion}\index{embedding}Moreover we
say that $f$ is:
\begin{itemize}
\item an \textsl{immersion} if $\dim M\leq\dim N$ and $\mathrm{d}_{p}f$
is injective for each $p\in M$;
\item a \textsl{submersion} if $\dim M\geq\dim N$ and $\mathrm{d}_{p}f$
is surjective for each $p\in M$;
\item an \textsl{embedding} if it is an immersion and $f$ maps $M$ homeomorphically
onto its image $f\left(M\right)$ (endowed with the topology induced
by that of $N$), i.e. the map
\begin{eqnarray*}
f^{\prime}:M & \rightarrow & f\left(M\right)\\
p & \mapsto & f\left(p\right)
\end{eqnarray*}
is a homeomorphism.
\end{itemize}
\end{defn}
\index{differential}It is possible to show that the definition of
differential at a point is well posed and it is easy to see that it
reduces to the usual notion of differential when $M$ and $N$ are
open subsets of Euclidean spaces, as anticipated. For this reason
often the push-forward through $f$ is also called \textsl{differential}
and is denoted with $\mathrm{d}f$. Instead the name {}``push-forward''
is due to the fact that in some sense $f_{*}$ {}``pushes'' through
$f$ each element $v\in\mathrm{T}M$ to an element $f_{*}v\in\mathrm{T}N$
in such a way that if $v\in\mathrm{T}_{p}M$ then $f_{*}v\in\mathrm{T}_{f\left(p\right)}N$.

Embeddings allow us to recognize submanifolds.
\begin{defn}
\index{submanifold}\index{inclusion map}Let $M$ be a manifold and
let $S$ be a a manifold whose underlying set is included in $M$.
We say that a manifold $S$ is a \textsl{submanifold of $M$} if the
inclusion map $\iota_{S}^{M}:S\rightarrow M$, $p\mapsto p$ is an
embedding from $S$ to $M$.\end{defn}
\begin{rem}
\label{remSubmanifold}An important example of submanifold of a given
$d$-dimensional manifold $M$ is the following. Suppose that $S$
is a connected open subset of $M$. We can endow $S$ with the topology
induced by the topology of $M$ and we immediately recognize that
$S$ is a connected Hausdorff topological space with a countable basis
of open subsets. We can define a coordinate neighborhood for $S$
taking a coordinate neighborhood $\left(U,\Omega,\phi\right)$ for
$M$: We take $U\cap S$ as open subset of $S$ (notice that this
is also an open subset of $M)$ and we use the fact that $\phi$ is
a homeomorphism from $U$ to $\Omega$ to deduce that we can take
$\phi\left(U\cap S\right)\subseteq\Omega$ as open subset of $\mathbb{R}^{d}$.
Then we define $\phi^{\prime}:U\cap S\rightarrow\phi\left(U\cap S\right)$,
$p\mapsto\phi\left(p\right)$ and we observe that $\phi^{\prime}$
is a homeomorphism (it is bijective by construction and it is continuous
with its inverse as a consequence of the same property for $\phi$).
Hence $\left(U\cap S,\phi\left(U\cap S\right),\phi^{\prime}\right)$
is a coordinate neighborhood for $S$ (if it happens that $\phi\left(U\cap S\right)$
is not a neighborhood of 0, a translation in $\mathbb{R}^{d}$ is
sufficient to satisfy also this requirement). Applying this construction
to all the elements of the maximal atlas of $M$, we obtain the maximal
atlas of $S$ end we recognize that $S$ is actually a $d$-dimensional
manifold. The inclusion map $\iota_{S}^{M}$ is smooth because the
coordinate neighborhoods for $S$ are the restrictions (in the sense
of the construction above) of the coordinate neighborhoods for $M$
and the transition charts of $M$ are smooth by definition of manifold.
For each $p\in S$, $\mathrm{d}_{p}\iota_{S}^{M}$ is injective because
each curve contained in a neighborhood of $p$ in $S$ is mapped through
$\iota_{S}^{M}$ to the same curve in the same neighborhood of $p$,
regarded now as a neighborhood with respect to the topology of $M$.
This shows that $\iota_{S}^{M}$ is an immersion. Consider now the
map $\iota_{S}^{M\prime}:S\rightarrow\iota_{S}^{M}\left(S\right)=S$,
$p\mapsto\iota_{S}^{M}\left(p\right)=p$. If on the image $\iota_{S}^{M\prime}\left(S\right)$
we consider the topology that is induced by the topology of $M$,
we realize that the topological space $\iota_{S}^{M\prime}\left(S\right)$
coincides with the topological space $S$, hence it is trivial to
check that $\iota_{S}^{M\prime}$ is a homeomorphism because it is
nothing but the identity map of $S$. Then we realize that the $d$-dimensional
manifold $S$ constructed above is also a submanifold of $M$. Moreover
$\iota_{S}^{M\prime}$ is a diffeomorphism as a consequence of the
fact that all the transition charts for $S$ are diffeomorphisms (this
being a consequence of the existence of a maximal atlas for $S$).
Moreover notice that $\iota_{S}^{M}$ is an open map because $S$
is an open subset of $M$: Take an open subset $\Omega$ of $S$ and
note that trivially $\iota_{S}^{M}\left(\Omega\right)=\Omega$; since
the topology on $S$ is induced by that of $M$, we find an open subset
$\Omega^{\prime}$ of $M$ such that $\Omega=\Omega^{\prime}\cap S$;
we deduce that $\Omega$ is also an open subset of $M$ and we conclude
that $\iota_{S}^{M}$ maps open subsets of $S$ to open subsets of
$M$, i.e. it is an open map.

A special case of this situation is the following. Let $M$ and $N$
be $d$-dimensional manifolds and suppose that $f:M\rightarrow N$
is an embedding whose image $f\left(M\right)$ is an open subset of
$N$. Notice that $f\left(M\right)$ is also connected: We can find
a curve contained in $f\left(M\right)$ connecting two arbitrary points
$p$ and $q$ of $f\left(M\right)$ composing $f$ with a curve $\gamma$
in $M$ that connects the preimages of $p$ and $q$ (the existence
of $\gamma$ follows from the hypothesis of connectedness of $M$).
Applying the construction given above to the connected open subset
$f\left(M\right)$ of $N$, we realize that $f\left(M\right)$ becomes
a $d$-dimensional manifold that is a submanifold of $N$. Since $f$
is an embedding, we have that $f^{\prime}$ is a homeomorphism. Now
also $f\left(M\right)$ is a manifold so that we can ask whether $f^{\prime}$
has some more regularity beyond the continuity of itself and its inverse.
To this end consider a point $p\in M$. We take a coordinate neighborhood
$\left(U,\Omega,\phi\right)$ of $p$ in $M$ and a coordinate neighborhood
$\left(V,\Omega,\psi\right)$ of $f^{\prime}\left(p\right)$ in $f\left(M\right)$.
We recognize immediately that $\left(V,\Omega,\psi\right)$ is also
a coordinate neighborhood of $f\left(p\right)$ in $N$ because $V$,
being an open neighborhood of $f^{\prime}\left(p\right)=f\left(p\right)$
in the topology of $f\left(M\right)$, is also an open neighborhood
of $f\left(p\right)$ in the topology of $N$. Recalling Definition
\ref{defRegularFunction}, we have that $f_{U,V}$ is smooth by hypothesis
and that $f_{U,V}^{\prime}=f_{U,V}$ because $f^{\prime}$ and $f$
coincide on $U\cap f^{-1}\left(V\right)$. Hence $f_{U,V}^{\prime}$
is smooth too and the arbitrariness in the choice of the point $p\in M$
and of the coordinate neighborhoods implies that $f^{\prime}$ is
smooth. On the one hand $\mathrm{d}_{p}f^{\prime}$ is injective for
each $p\in M$ because $f^{\prime}$ is injective. On the other hand
$\mathrm{d}_{p}f^{\prime}$ must be also surjective otherwise $\dim f\left(M\right)>\dim M$.
Then the inverse function theorem implies that $f^{\prime}$ is a
diffeomorphism. Using the inclusion map $\iota_{\psi\left(M\right)}^{N}:\psi\left(M\right)\rightarrow N$
(that is actually an embedding, as we saw above), we can decompose
$\psi$ in $\psi=\iota_{\psi\left(M\right)}^{N}\circ\psi^{\prime}$.
In particular this implies that $\psi$ is an open map because $\psi^{\prime}$
is a homeomorphism and $\iota_{\psi\left(M\right)}^{N}$ is an open
map as seen above.
\end{rem}
Exploiting the definition of the cotangent space as dual of the tangent
space, we can introduce a {}``dual'' of the notion of push-forward.
\begin{defn}
\index{pull-back}Let $M$ and $N$ be two manifolds and let $f:M\rightarrow N$
be a smooth function. We call \textsl{pull-back through $f$} the
map $f^{*}:\bigsqcup_{q\in f\left(M\right)}\mathrm{T}_{q}^{*}N\rightarrow\mathrm{T}^{*}M$
defined as the pointwise dual of the push-forward $f_{*}:\mathrm{T}M\rightarrow\mathrm{T}N$,
i.e. for each $p\in M$, each $\omega\in\mathrm{T}_{f\left(p\right)}^{*}N$
and each $v\in\mathrm{T}_{p}M$ we require that
\[
\left(\left.f^{*}\right|_{\mathrm{T}_{f\left(p\right)}^{*}N}\omega\right)v_{p}=\omega\left(\left.f_{*}\right|_{\mathrm{T}_{p}M}v_{f\left(p\right)}\right)\mbox{,}
\]
where the dual pairings between the vector spaces $\mathrm{T}_{p}^{*}M$
and $\mathrm{T}_{p}M$ and between the vector spaces $\mathrm{T}_{f\left(p\right)}^{*}N$
and $\mathrm{T}_{f\left(p\right)}N$ are taken into account.
\end{defn}
The reader should bear in mind that the dual pairing between $\mathrm{T}_{p}^{*}M$
and $\mathrm{T}_{p}M$ is actually part of the definition of $\mathrm{T}_{p}^{*}M$
as the vector space dual to $\mathrm{T}_{p}M$ (recall the definition
of cotangent space in Remark \ref{remTensorBundles}).
\begin{rem}
\label{remPushforwardPullback}Let $M$ and $N$ be two manifolds
and let $f:M\rightarrow N$ be a smooth function. An extension of
the notions of push-forward and pull-back is possible using the tensor
structure of $\mathrm{T}_{p}^{\left(i,j\right)}M$:
\begin{itemize}
\item the push-forward $f_{*}:\mathrm{T}^{\left(i,0\right)}M\rightarrow\mathrm{T}^{\left(i,0\right)}N$
through $f$ is defined by
\[
\left.f_{*}\right|_{\mathrm{T}_{p}^{\left(i,0\right)}M}\left(v_{1}\otimes\dots\otimes v_{i}\right)=\left.f_{*}\right|_{\mathrm{T}_{p}M}v_{1}\otimes\dots\otimes\left.f_{*}\right|_{\mathrm{T}_{p}M}v_{i}\mbox{,}
\]
for each $p\in M$ and each $v_{1}$, $\dots$, $v_{i}\in\mathrm{T}_{p}M$,
where $f_{*}$ on the RHS%
\footnote{Here, and in the rest of this thesis, the acronym {}``LHS'' stands
for {}``left hand side'', while the acronym {}``RHS'' stands for
{}``right hand side''.%
} is the push forward through $f$ from $\mathrm{T}M$ to $\mathrm{T}N$;
\item the pull-back $f^{*}:\bigsqcup_{q\in f\left(M\right)}\mathrm{T}_{q}^{\left(0,j\right)}N\rightarrow\mathrm{T}^{\left(0,j\right)}M$
through $f$, defined by
\[
\left.f^{*}\right|_{\mathrm{T}_{f\left(p\right)}^{\left(0,j\right)}N}\left(\omega_{1}\otimes\dots\otimes\omega_{j}\right)=\left.f^{*}\right|_{\mathrm{T}_{f\left(p\right)}^{*}N}\omega_{1}\otimes\dots\otimes\left.f^{*}\right|_{\mathrm{T}_{f\left(p\right)}^{*}N}\omega_{j}\mbox{,}
\]
for each $p\in M$ and each $\omega_{1}$, $\dots$, $\omega_{j}\in\mathrm{T}_{f\left(p\right)}^{*}N$,
where $f^{*}$ on the RHS is the pull-back through $f$ from $\bigsqcup_{q\in f\left(M\right)}\mathrm{T}_{q}^{*}N$
to $\mathrm{T}^{*}M$.
\end{itemize}
We can enlarge the notion of push-forward and pull-back much more
if we suppose that $f:M\rightarrow N$ is a diffeomorphism: in such
case the smooth map $f$ is bijective and we have at our disposal
also the smooth bijective map $f^{-1}:N\rightarrow M$, hence we can
push forward through $f^{-1}$ all the elements of $\mathrm{T}^{\left(i,0\right)}N$
to $\mathrm{T}^{\left(i,0\right)}M$ and we can pull back through
$f^{-1}$ all the elements of $\mathrm{T}^{\left(0,j\right)}M$ to
$\mathrm{T}^{\left(0,j\right)}N$. This allows us to define a new
push-forward and a new pull-back through $f$:
\begin{itemize}
\item the push-forward $f_{*}:\mathrm{T}^{\left(i,j\right)}M\rightarrow\mathrm{T}^{\left(i,j\right)}N$
through $f$ is defined by
\[
\left.f_{*}\right|_{\mathrm{T}_{p}^{\left(i,j\right)}M}\left(v\otimes\omega\right)=\left.f_{*}\right|_{\mathrm{T}_{p}^{\left(i,0\right)}M}v\otimes\left.\left(f^{-1}\right)^{*}\right|_{\mathrm{T}_{p}^{\left(0,j\right)}M}\omega\mbox{,}
\]
for each $p\in M$, each $v\in\mathrm{T}_{p}^{\left(i,0\right)}M$
and each $\omega\in\mathrm{T}_{p}^{\left(0,j\right)}M$, where on
the RHS $f_{*}$ denotes the push-forward through $f$ from $\mathrm{T}^{\left(i,0\right)}M$
to $\mathrm{T}^{\left(i,0\right)}N$, while $\left(f^{-1}\right)^{*}$denotes
the pull-back through $f^{-1}$ from $\mathrm{T}^{\left(0,j\right)}M$
to $\mathrm{T}^{\left(0,j\right)}N$;
\item the pull-back $f^{*}:\mathrm{T}^{\left(i,j\right)}N\rightarrow\mathrm{T}^{\left(i,j\right)}M$
through $f$ is defined by
\[
\left.f^{*}\right|_{\mathrm{T}_{q}^{\left(i,j\right)}N}\left(v\otimes\omega\right)=\left.\left(f^{-1}\right)_{*}\right|_{\mathrm{T}_{q}^{\left(i,0\right)}N}v\otimes\left.f^{*}\right|_{\mathrm{T}_{q}^{\left(0,j\right)}N}\omega\mbox{,}
\]
for each $q\in N$, each $v\in\mathrm{T}_{q}^{\left(i,0\right)}N$
and each $\omega\in\mathrm{T}_{q}^{\left(0,j\right)}N$, where on
the RHS $\left(f^{-1}\right)_{*}$ denotes the push-forward through
$f^{-1}$ from $\mathrm{T}^{\left(i,0\right)}N$ to $\mathrm{T}^{\left(i,0\right)}M$,
while $f^{*}$denotes the pull-back through $f$ from $\mathrm{T}^{\left(0,j\right)}N$
to $\mathrm{T}^{\left(0,j\right)}M$.
\end{itemize}
In this way both $f_{*}$ and $f^{*}$ are extended to the whole tensor
bundles over the appropriate manifolds. It turns out that this new
$f_{*}:\mathscr{T}M\rightarrow\mathscr{T}N$ and $f^{*}:\mathscr{T}N\rightarrow\mathscr{T}M$
are inverses of each other.

\end{rem}

\begin{rem}
Suppose that $M$ is a $d$-dimensional manifold. Then our knowledge
about push-forwards and pull-backs through smooth functions between
manifolds allows us to recognize a manifold structure in $\mathrm{T}^{\left(i,j\right)}M$.
We give a sketch of how this is done for the case of $\mathrm{T}M$
(all other cases are similar). First of all we need a topology on
tangent spaces. The fact that $\mathrm{T}_{p}M$ is a $d$-dimensional
$\mathbb{R}$-vector space allows us to naturally identify it with
$\mathbb{R}^{d}$. Using this identification we can also induce on
each $\mathrm{T}_{p}M$ the usual topology of $\mathbb{R}^{d}$. $\mathrm{T}M$
becomes a topological space when endowed with the topology naturally
induced by the disjoint union. Then we notice that this topology is
Hausdorff and it admits a countable basis of open subsets as a consequence
of the topologies on $M$ and on each of the tangent spaces $\mathrm{T}_{p}M$.
Moreover $\mathrm{T}M$ is connected because $M$ and all its tangent
spaces are connected. Now we choose $v\in\mathrm{T}M$. Since $\mathrm{T}M$
is the disjoint union over $M$ of the tangent spaces $\mathrm{T}_{p}M$,
$v$ is of the form $\left(p,u\right)$ for some $p\in M$ and some
$u\in\mathrm{T}_{p}M$. We consider a coordinate neighborhood $\left(U,\Omega,\phi\right)$
and we keep in mind that $\phi$ is a diffeomorphism (this follows
from the maximality of the atlas of $M$). Then we take $V=\bigsqcup_{q\in U}\mathrm{T}_{q}M$
and we realize that this is indeed a neighborhood of $v$ in the topology
of $\mathrm{T}M$. Furthermore $\mathrm{T}_{q}U=\mathrm{T}_{q}M$
for each $q\in U$ since $U$ is an open neighborhood of each $q\in U$
with respect to the topology of $M$, hence $\mathrm{T}U=V$. With
the identification of each $\mathrm{T}_{\phi\left(q\right)}\Omega$
with $\mathbb{R}^{d}$, we have that
\[
\phi_{*}\left(q,w\right)=\left(\phi\left(q\right),\left(\mathrm{d}_{q}\phi\right)w\right)\in\left\{ \phi\left(q\right)\right\} \times\mathbb{R}^{d}
\]
for each $\left(q,w\right)\in V$ by definition of $\phi_{*}$. We
take $\Theta=\bigsqcup_{q\in U}\mathrm{T}_{\phi\left(q\right)}\Omega$
and the above identification implies $\Theta=\Omega\times\mathbb{R}^{d}$
(notice that on $\Omega\times\mathbb{R}^{d}$ we consider the topology
induced by the disjoint union otherwise the identification is not
a homeomorphism). Considering $\phi_{*}$ as a map from $\mathrm{T}U=V$
to $\Theta=\Omega\times\mathbb{R}^{d}$, we easily conclude that $\phi_{*}$
is a homeomorphism. Therefore $\left(V,\Theta,\phi_{*}\right)$ is
a coordinate neighborhood of $v$ in $\mathrm{T}M$. All transition
maps are immediately diffeomorphisms (in the sense of functions between
Euclidean spaces) and the maximal atlas of $\mathrm{T}M$ is easily
built starting from the maximal atlas of $M$.

From this observation we can deduce that $f_{*}:\mathrm{T}^{\left(i,j\right)}M\rightarrow\mathrm{T}^{\left(i,j\right)}N$
and $f^{*}:\mathrm{T}^{\left(i,j\right)}N\rightarrow\mathrm{T}^{\left(i,j\right)}M$
are diffeomorphisms between the manifolds $\mathrm{T}^{\left(i,j\right)}M$
and $\mathrm{T}^{\left(i,j\right)}N$ if $f:M\rightarrow N$ is a
diffeomorphism. We show this fact in the case of $\mathrm{T}M$, but
the same proof works for any other tensor bundle of type $\left(i,j\right)$.
As a matter of fact it suffices to show that both $f_{*}:\mathrm{T}M\rightarrow\mathrm{T}N$
and $f^{*}:\mathrm{T}N\rightarrow\mathrm{T}M$ are smooth functions
between the manifolds $\mathrm{T}M$ and $\mathrm{T}N$ since, as
we had already observed in Remark \ref{remPushforwardPullback}, $f_{*}$
and $f^{*}$ are inverses of each other. We focus on $f_{*}$. Suppose
that $O$ is an open subset of $\mathrm{T}N$. Then $O$ is of the
form
\[
O=\bigsqcup_{q\in\Omega}\Omega_{q}=\left\{ \left(q,w\right):\, q\in\Omega,w\in\Omega_{q}\right\} \mbox{,}
\]
where $\Omega$ is an open subset of $N$ and $\Omega_{q}$ is an
open subset of $\mathrm{T}_{q}N$ for each $q\in N$. Then we have
that
\begin{eqnarray*}
\left(f_{*}\right)^{-1}\left(\Omega\right) & = & f^{*}\left(\left\{ \left(q,w\right):\, q\in\Omega,w\in\Omega_{q}\right\} \right)\\
 & = & \left\{ \left(f^{-1}\left(q\right),\left.f^{*}\right|_{\mathrm{T}_{q}N}w\right):\, q\in\Omega,w\in\Omega_{q}\right\} \\
 & = & \left\{ \left(p,v\right):\, p\in f^{-1}\left(\Omega\right),v\in\left.f^{*}\right|_{\mathrm{T}_{f\left(p\right)}N}\left(\Omega_{f\left(p\right)}\right)\right\} \\
 & = & \bigsqcup_{p\in f^{-1}\left(\Omega\right)}\left.f^{*}\right|_{\mathrm{T}_{f\left(p\right)}N}\left(\Omega_{f\left(p\right)}\right)\mbox{.}
\end{eqnarray*}
$f^{-1}\left(\Omega\right)$ is an open subset of $M$ because $f$
is continuous. Since for each $p\in M$ the map $\left.f_{*}\right|_{\mathrm{T}_{p}M}$
is linear between the finite dimensional topological vector spaces
$\mathrm{T}_{p}M$ and $\mathrm{T}_{f\left(p\right)}N$, it must be
continuous too. Then it follows that
\[
\left.f^{*}\right|_{\mathrm{T}_{f\left(p\right)}N}\left(\Omega_{f\left(p\right)}\right)=\left(\left.f_{*}\right|_{\mathrm{T}_{p}M}\right)^{-1}\left(\Omega_{f\left(p\right)}\right)
\]
is an open subset of $\mathrm{T}_{p}M$ for each $p\in\Omega$. We
conclude that $\left(f_{*}\right)^{-1}\left(\Omega\right)$ has exactly
the shape of an open subset of $\mathrm{T}M$ and this implies that
$f_{*}$ is continuous. Similarly we see that $f^{*}$ is continuous
and hence both $f_{*}$ and $f^{*}$ are homeomorphisms. Finally the
smoothness of these maps easily follows from the smoothness of $f$
and $f^{-1}$.
\end{rem}

\subsection{\label{subVectorBundlesConnectionsInnerProducts}Vector bundles,
connections and inner products}

Till this point we have spoken of $\mathrm{T}^{\left(i,j\right)}M$
as a manifold. However it is possible to recognize a richer structure
on it. This structure is a special case of that presented in the next
definition.
\begin{defn}
\index{vector bundle}\index{total space}\index{base (of a vector bundle)}\index{projection (of a vector bundle)}\index{fiber}\index{local trivialization}A
\textsl{vector bundle} of rank $n$ over a manifold of dimension $d$
is a triple $\left(E,M,\pi\right)$, where $E$, called \textsl{total
space}, and $M$, called \textsl{base}, are manifolds of dimension
respectively $n+d$ and $d$ and $\pi:E\rightarrow M$, called \textsl{projection},
is a smooth surjective map such that the following conditions hold:
\begin{itemize}
\item for each $p\in M$ the set $E_{p}=\pi^{-1}\left(p\right)$, called
\textsl{fiber}, carries the structure of an $n$-dimensional $\mathbb{R}$-vector
space;
\item for each $p\in M$ there exists a pair $\left(U,\Phi\right)$, called
\textsl{local trivialization at }$p$ of $\left(E,M,\pi\right)$,
where $U$ is an open neighborhood of $p$ in $M$ and $\Phi:\pi^{-1}\left(U\right)\rightarrow U\times\mathbb{R}^{n}$
is a diffeomorphism such that

\begin{itemize}
\item $\mathrm{pr}_{1}\left(\Phi\left(\mu\right)\right)=\pi\left(\mu\right)$
for each $\mu\in\pi^{-1}\left(U\right)$, where $\mathrm{pr}_{1}$
denotes the projection on the first factor of the Cartesian product,
\item for each $q\in U$ the map $\Phi_{q}:E_{q}\rightarrow\left\{ q\right\} \times\mathbb{R}^{n}$,
defined by $\Phi_{q}\mu=\Phi\left(\mu\right)$ for each $\mu\in E_{q}$,
is linear and bijective.
\end{itemize}
\end{itemize}
\end{defn}
Note that the projection $\pi$ is an open map, i.e. it maps open
sets to open sets. This property is a consequence of the fact that
the projection on an argument of a Cartesian product is always an
open map.

Usually we will denote vector bundles only with their total space.
However the choice of a base space and a projection is always understood.

\index{Whitney sum}The vectorial structure of each fiber allows us
to construct other vector bundles using the vectorial operations (for
example duality, tensor product, direct sum) fiberwise, provided that
the vector bundles involved share the same base manifold. For example
we can define the dual vector bundle $E^{*}$ of the vector bundle
$E$ simply taking the dual spaces (in the usual sense of vector spaces)
of the original fibers. Notice that the direct sum $E\oplus F$ of
the vector bundles $E$ and $F$ is called \textsl{Whitney sum}.
\begin{rem}
bear in mind that each tensor bundle of type $\left(i,j\right)$ can
be endowed with a vector bundle structure. For example, in the case
of the tangent bundle $\mathrm{T}M$ this is done considering $\mathrm{T}M$
as total space, $M$ as base, the projection $\pi:\mathrm{T}M\rightarrow M$
naturally induced by the disjoint union of tangent spaces as the projection,
and $\left\{ \left(U_{\alpha},\phi_{\alpha*}\right)\right\} $ as
local trivializations (identification of $\mathrm{T}\Omega$ with
$\Omega\times\mathbb{R}^{d}$ is understood), where $\left\{ \left(U_{\alpha},\Omega_{\alpha},\phi_{\alpha}\right)\right\} $
is the maximal atlas of $M$ and $d=\dim M$. Then from now on, when
we speak of $\mathrm{T}^{\left(i,j\right)}M$, we refer to it as endowed
with their natural vector bundle structure.

Notice that each tensor bundle of type $\left(j,i\right)$ is the
dual of the tensor bundle of type $\left(i,j\right)$ and also the
Whitney sum of $j$ copies of $\mathrm{T}M$ and $i$ copies of $\mathrm{T}^{*}M$.
\end{rem}
We can define maps between vector bundles that respect the vector
bundle structures.
\begin{defn}
\index{vector bundle homomorphism}Let $\left(E,M,\pi\right)$ and
$\left(F,N,\sigma\right)$ be two vector bundles. We call \textsl{vector
bundle homomorphism} the pair $\left(\psi,\Psi\right)$ where $\psi$
is a smooth function from the base manifold $M$ to the base manifold
$N$ and $\Psi$ is a smooth function from the total space $E$ to
the total space $F$ such that the following conditions hold:
\begin{itemize}
\item \textsl{compatibility with projections}: $\psi\circ\pi=\sigma\circ\Psi$;
\item \textsl{fiberwise linearity}: $\Psi$ is fiberwise a vector space
homomorphism, i.e. the map
\begin{eqnarray*}
\Psi_{p}:E_{p} & \rightarrow & F_{\psi\left(p\right)}\\
\mu & \rightarrow & \Psi\left(\mu\right)
\end{eqnarray*}
is linear for each $p\in M$.
\end{itemize}
\index{vector bundle isomorphism}Then we say that $\left(\psi,\Psi\right)$
is a \textsl{vector bundle isomorphism} if it is a bijective vector
bundle homomorphism whose inverse is still a vector bundle homomorphism
such that $\Psi$.
\end{defn}
In the next remark we show a construction that allows to build a vector
bundle whose total space and base space are submanifolds of the total
space and the base space of a given vector bundle. We didn't include
such construction immediately after the definition of vector bundles
because we wanted to show also that the inclusion maps of the base
space and of the total space as submanifolds give rise to a vector
bundle homomorphism.
\begin{rem}
\label{remRestrictionOfVectorBundles}Suppose that a vector bundle
$E$ of rank $n$ over a $d$-dimensional manifold $M$ is given and
assume that $S$ is a connected open subset of $M$. In Remark \ref{remSubmanifold}
we saw that it is possible to use the manifold structure of $M$ to
to make $S$ a $d$-dimensional manifold itself. We also recognized
that the new manifold $S$ is a submanifold of $M$ and that the inclusion
map $\iota_{S}^{M}$ is an embedding of $S$ into $M$. Now we consider
the subset $\pi^{-1}\left(S\right)$ of the $\left(n+d\right)$-dimensional
manifold $E$. Since $S$ is an open subset of $M$ and $\pi$ is
continuous, $\pi^{-1}\left(S\right)$ is an open subset of $E$. One
can check by contradiction that $\pi^{-1}\left(S\right)$ is connected
exploiting the following properties: $\pi$ is an open map, $E$ is
locally trivial and $S$ is connected. Since $S$ is connected Then
it is possible to apply Remark \ref{remSubmanifold} to the connected
open subset $\pi^{-1}\left(S\right)$ of the manifold $E$. In this
way we obtain a new $\left(n+d\right)$-dimensional manifold (which
is actually a submanifold of $E$) that we denote with $\left.E\right|_{S}$.
We define the map $\left.\pi\right|_{S}:\left.E\right|_{S}\rightarrow S$,
$\mu\mapsto\pi\left(\mu\right)$ and we note that its image is
\[
\left.\pi\right|_{S}\left(\left.E\right|_{S}\right)=\pi\left(\pi^{-1}\left(S\right)\right)=S\mbox{,}
\]
hence $\left.\pi\right|_{S}$ is surjective. Since the topologies
and the atlases of the manifolds $S$ and $\left.E\right|_{S}$ are
inherited via restriction from the topologies and the atlases of $M$
and respectively $E$, it follows that $\left.\pi\right|_{S}$ is
continuous and also smooth. Then $\left(\left.E\right|_{S},S,\left.\pi\right|_{S}\right)$
is our candidate to become a new vector bundle of rank $n$. The first
thing to be checked is that $\left.\pi\right|_{S}^{-1}\left(p\right)$
is an $n$-dimensional vector space for each $p\in S$: this fact
is trivial because $\left.\pi\right|_{S}^{-1}\left(p\right)=\pi^{-1}\left(p\right)=E_{p}$
and $E_{p}$ is of course an $n$-dimensional vector space. It remains
only the problem of the existence of local trivializations in neighborhoods
of arbitrary points of $S$, but this difficulty is easily overcome
in the following manner. Consider a point $p\in S$ and take a local
trivialization $\left(U,\Phi\right)$ of $E$ at $p$. We note that
$U\cap S$ is an open neighborhood of $p$ in the topology of $S$
and that we can define the map
\begin{eqnarray*}
\Phi^{\prime}:\left.\pi\right|_{S}^{-1}\left(U\cap S\right)=\pi^{-1}\left(U\cap S\right) & \rightarrow & \left(U\cap S\right)\times\mathbb{R}^{n}\\
\mu & \mapsto & \Phi\left(\mu\right)
\end{eqnarray*}
which satisfies
\[
\mathrm{pr}_{1}\left(\Phi^{\prime}\left(\mu\right)\right)=\mathrm{pr}_{1}\left(\Phi\left(\mu\right)\right)=\pi\left(\mu\right)=\left.\pi\right|_{S}\left(\mu\right)
\]
for each $\mu\in\left.\pi\right|_{S}^{-1}\left(U\cap S\right)$ and
is such that the map
\begin{eqnarray*}
\Phi_{p}^{\prime}:\left.\pi\right|_{S}^{-1}\left(p\right)=E_{p} & \rightarrow & \left\{ p\right\} \times\mathbb{R}^{n}\\
\mu & \mapsto & \Phi^{\prime}\left(\mu\right)=\Phi\left(\mu\right)=\Phi_{p}\mu
\end{eqnarray*}
is linear for each $p\in U\cap S$. These properties follow from the
properties of $\Phi$. Hence we have proved that for each point of
$S$ there exists a local trivialization. This implies that $\left(\left.E\right|_{S},S,\left.\pi\right|_{S}\right)$
is a vector bundle in its own right. We will usually denote it simply
with its total space $\left.E\right|_{S}$ as it is customary for
vector bundles.

Side by side with this construction, we can also introduce the inclusion
maps $\iota_{\left.E\right|_{S}}^{E}:\left.E\right|_{S}\rightarrow E$,
$\mu\mapsto\mu$ and $\iota_{S}^{M}:S\rightarrow M$, $p\rightarrow p$.
At this point we think $E$ and $\left.E\right|_{S}$ as $\left(n+d\right)$-dimensional
manifolds and we keep in mind that$\left.E\right|_{S}$ is a submanifold
of $E$. By definition of submanifold $\iota_{\left.E\right|_{S}}^{E}:\left.E\right|_{S}\rightarrow E$
is an embedding, hence, in particular, a smooth map. The same is true
for $\iota_{S}^{M}$. We note that $\pi_{E}\circ\iota_{\left.E\right|_{S}}^{E}=\iota_{S}^{M}\circ\pi_{\left.E\right|_{S}}$
and for each $p\in S$ we realize that the map
\begin{eqnarray*}
\iota_{\left.E\right|_{S}\, p}^{E}:\left.E\right|_{S\, p}=E_{p} & \rightarrow & E_{\iota_{S}^{M}\left(p\right)}=E_{p}\\
\mu & \mapsto & \iota_{\left.E\right|_{S}}^{E}\left(\mu\right)=\mu
\end{eqnarray*}
is a vector space isomorphism. These facts imply that $\left(\iota_{S}^{M},\iota_{\left.E\right|_{S}}^{E}\right)$
is a vector bundle homomorphism which is fiberwise a vector space
isomorphism.
\end{rem}
The following remark focuses the attention on vector bundle homomorphisms.
It provides a procedure to restrict certain vector bundle homomorphisms
to vector bundle isomorphisms.
\begin{rem}
\label{remRestrictionOfVectorBundleHomomorphisms}Let $E$ and $F$
be vector bundles of rank $n$ over $d$-dimensional manifolds $M$
and $N$ and consider a vector bundle homomorphism $\left(\psi,\Psi\right)$
from $E$ to $F$. Suppose that $\psi$ is an embedding of $M$ into
$N$ whose image $\psi\left(M\right)$ is open in $N$ and that $\Psi$
is fiberwise a vector space isomorphism. The first step is the application
of the last part of Remark \ref{remSubmanifold} from which we deduce
that $\psi\left(M\right)$ is a $d$-dimensional submanifold of $N$
and that the map $\psi^{\prime}:M\rightarrow\psi\left(M\right)$,
$p\mapsto\psi\left(p\right)$ is a diffeomorphism such that $\psi=\iota_{\psi\left(M\right)}^{N}\circ\psi^{\prime}$.
In particular we note that $\psi\left(M\right)$ is a connected open
subset of $N$, hence it is possible to apply Remark \ref{remRestrictionOfVectorBundles}
obtaining the new vector bundle $\left.F\right|_{\psi\left(M\right)}$.
Defining the map $\Psi^{\prime}:E\rightarrow\left.F\right|_{\psi\left(M\right)}$,
$\mu\mapsto\Psi\left(\mu\right)$, we can check that it is continuous
with respect to the topologies of $E$ and $\left.F\right|_{\psi\left(M\right)}$
because $\Psi$ is continuous with respect to the topologies of $E$
and $F$ and the topology on $\left.F\right|_{\psi\left(M\right)}$,
whose underlying set is an open subset of $F$, is induced by that
of $F$. Moreover $\Psi^{\prime}$ is smooth because $\Psi$ is smooth
and the atlas of $\left.F\right|_{\psi\left(M\right)}$ is nothing
but the restriction of the atlas of $F$. We can even draw more accurate
conclusions noting that
\[
\pi_{\left.F\right|_{\psi\left(M\right)}}\left(\Psi^{\prime}\left(\mu\right)\right)=\pi_{F}\left(\Psi\left(\mu\right)\right)=\psi\left(\pi_{E}\left(\mu\right)\right)
\]
for each $\mu\in\left.F\right|_{\psi\left(M\right)}$ and that for
each $p\in M$ the map
\begin{eqnarray*}
\Psi_{p}^{\prime}:E_{p} & \rightarrow & \left.F\right|_{\psi\left(M\right)\,\psi\left(p\right)}=F_{\psi\left(p\right)}\\
\mu & \mapsto & \Psi^{\prime}\left(\mu\right)=\Psi\left(\mu\right)=\Psi_{p}\mu
\end{eqnarray*}
is linear. This shows that $\left(\psi^{\prime},\Psi^{\prime}\right)$
is a vector bundle homomorphism. We assumed that $\Psi_{p}=\Psi_{p}^{\prime}$
is a vector space isomorphism for each $p\in M$, hence $\Psi^{\prime}$
is exactly defined as the restriction of $\Psi$ to its image (for
this reason from now on we will denote the vector bundle $\left.F\right|_{\psi\left(M\right)}$
with $\Psi\left(E\right)$). This shows that $\Psi^{\prime}$ is a
bijective smooth function and that $\left(\psi^{\prime},\Psi^{\prime}\right)$
is a bijective vector bundle homomorphism. Some work with local trivializations
and coordinate neighborhoods shows that for each $p\in M$ there exists
an open neighborhood $U$ of $p$ in $M$ such that the Jacobian determinant
of $\Psi^{\prime}$ (locally trivialized and written in local coordinate)
at $p$ is not null. Hence also $\Psi^{\prime-1}$ is a smooth function
between the manifolds $\Psi\left(E\right)$ and $E$ as a consequence
of the inverse function theorem. It is easy to check that
\[
\pi_{E}\circ\Psi^{\prime-1}=\psi^{\prime-1}\circ\pi_{\Psi\left(E\right)}
\]
and that for each $q\in\psi\left(M\right)$ the map
\begin{eqnarray*}
\Psi_{q}^{\prime-1}:\Psi\left(E\right)_{q} & \rightarrow & E_{\psi^{\prime-1}\left(q\right)}\\
\nu & \mapsto & \Psi^{\prime-1}\left(\nu\right)
\end{eqnarray*}
coincides with the inverse of $\Psi_{\psi^{\prime-1}\left(q\right)}^{\prime}$
(hence, in particular, it is a vector space homomorphism). Then we
conclude that $\left(\psi^{\prime-1},\Psi^{\prime-1}\right)$ is a
vector bundle homomorphism from the vector bundle $\Psi\left(E\right)$
to the vector bundle $E$ and that it is the inverse of $\left(\psi^{\prime},\Psi^{\prime}\right)$
so that $\left(\psi^{\prime},\Psi^{\prime}\right)$ is a vector bundle
isomorphism. Remark \ref{remRestrictionOfVectorBundles} provides
a vector bundle homomorphism $\left(\iota_{\psi\left(M\right)}^{N},\iota_{\Psi\left(E\right)}^{F}\right)$
from $\Psi\left(E\right)$ to $F$ which is fiberwise a vector space
isomorphism. This gives us the opportunity to decompose the original
vector bundle homomorphism $\left(\psi,\Psi\right)$: We already know
that $\psi=\iota_{\psi\left(M\right)}^{N}\circ\psi^{\prime}$ and
it can be directly checked that $\Psi=\iota_{\Psi\left(E\right)}^{F}\circ\Psi^{\prime}$,
hence we deduce that 
\[
\left(\psi,\Psi\right)=\left(\iota_{\psi\left(M\right)}^{N}\circ\psi^{\prime},\iota_{\Psi\left(E\right)}^{F}\circ\Psi^{\prime}\right)=\left(\iota_{\psi\left(M\right)}^{N},\iota_{\Psi\left(E\right)}^{F}\right)\circ\left(\psi^{\prime},\Psi^{\prime}\right)\mbox{.}
\]

\end{rem}
Now we want to introduce a particular class of smooth functions from
a manifold to the total space of a vector bundle whose base is such
manifold. The peculiarity of such maps resides in their compatibility
with the projection of the vector bundle.
\begin{defn}
\index{section of class mathrm{C}^{k}@section of class $\mathrm{C}^{k}$}\index{section}Let
$E$ be a vector bundle over a manifold $M$. A \textsl{$\mathrm{C}^{k}$-section
in $E$} is a $\mathrm{C}^{k}$-function $s$ from the base manifold
$M$ to the total space manifold $E$ such that $\pi\circ s=\mathrm{id}_{M}$.

A \textsl{(smooth) section in $E$} is a $\mathrm{C}^{k}$-section
in $E$ for each $k$ or, equivalently, is a smooth function $s$
from the base manifold $M$ to the total space manifold $E$ such
that $\pi\circ s=\mathrm{id}_{M}$.

\index{space of sections}The \textsl{space of $\mathrm{C}^{k}$-sections
}$\mathrm{C}^{k}\left(M,E\right)$ is the set comprised by all $\mathrm{C}^{k}$-sections
in $E$, the \textsl{space of (smooth) sections }$\mathrm{C}^{\infty}\left(M,E\right)$
is the set comprised by all the smooth sections in $E$ and finally
the \textsl{space of smooth sections with compact support }$\mathscr{D}\left(M,E\right)$
(or $\mathrm{C}_{0}^{\infty}\left(M,E\right)$) is the set comprised
by all the smooth sections in $E$ with compact support.
\end{defn}
Note that if $s$ is a smooth section on a vector bundle, we will
often simply say that $s$ is a section. On the contrary for $\mathrm{C}^{k}$-sections
we will never omit the prefix $\mathrm{C}^{k}$. We observe that the
fiberwise vector structure of each vector bundle induces a vector
structure on the set of sections in such vector bundle. This fact
motivates the word {}``space'' (in the sense of vector space) used
to denote the set of $\mathrm{C}^{k}$-sections, the set of sections
and the set of compactly sections defined above.
\begin{rem}
\label{remPushforwardPullbackSections}Let $E$ and $F$ be vector
bundles over the manifolds $M$ and respectively $N$ and let $s$
be a section in $E$. Consider a vector bundle homomorphism $\left(\psi,\Psi\right)$
from $E$ to $F$, where $\psi$ is an embedding of the manifold $M$
into the manifold $N$ whose image $\psi\left(M\right)$ is an open
subset of $N$ and $\Psi$ is fiberwise a vector space isomorphism.
Then we can apply Remark \ref{remRestrictionOfVectorBundleHomomorphisms}
and use the vector bundle isomorphism $\left(\psi^{\prime},\Psi^{\prime}\right)$
from $E$ to $\Psi\left(E\right)$ to define the function $\Psi^{\prime}\circ s\circ\psi^{\prime-1}$
from $\psi\left(M\right)$ to $\Psi\left(E\right)$. This is undoubtedly
a smooth map because it is a composition of smooth maps and we can
ask whether it is a section in the vector bundle $\Psi\left(E\right)$.
The answer is positive because
\[
\pi_{\Psi\left(E\right)}\circ\Psi^{\prime}\circ s\circ\psi^{\prime-1}=\psi^{\prime}\circ\pi_{E}\circ s\circ\psi^{\prime-1}=\psi^{\prime}\circ\mathrm{id}_{M}\circ\psi^{\prime-1}=\mathrm{id}_{\psi\left(M\right)}\mbox{.}
\]
Note that, when $\left(\psi,\Psi\right)$ is a vector bundle isomorphism
form $E$ to $F$, we can directly use it to obtain the section $\Psi\circ s\circ\psi^{-1}$
in $F$ from a section $s$ in $E$ and its inverse $\left(\psi^{-1},\Psi^{-1}\right)$
to obtain the section $\Psi^{-1}\circ t\circ\psi$ from a section
$t$ in $F$. With an abuse of language we say that $\Psi\circ s\circ\psi^{-1}$
and $\Psi^{-1}\circ t\circ\psi$ are respectively the push-forward
of $s$ and the pull-back of $t$ through $\left(\psi,\Psi\right)$.

From Remark \ref{remPushforwardPullback} we can deduce that, given
a diffeomorphism $f$ between the manifolds $M$ and $N$, $\left(f,f_{*}\right)$
can be recognized as a vector bundle homomorphism between the tangent
bundles $\mathrm{T}M$ and $\mathrm{T}N$ (intended as vector bundles)
and similarly $\left(f^{-1},f^{*}\right)$ can be recognized as a
vector bundle homomorphism between the tangent bundles $\mathrm{T}^{*}N$
and $\mathrm{T}^{*}M$. Moreover we can extend them to tensor bundles
of arbitrary type respectively over $M$ and $N$ and realize that
$\left(f,f_{*}\right):\mathrm{T}^{\left(i,j\right)}M\rightarrow\mathrm{T}^{\left(i,j\right)}N$
and $\left(f^{-1},f^{*}\right):\mathrm{T}^{\left(i,j\right)}N\rightarrow\mathrm{T}^{\left(i,j\right)}M$
are inverses of each other so that are both vector bundle isomorphisms.
These observations allow us to push forward and pull back sections
in tensor bundles of any type through diffeomorphisms of the base
manifolds exactly as we do with vector bundle isomorphisms.\end{rem}
\begin{example}
\label{exaCinftyM}A simple example of a space of sections is provided
by the set $\mathrm{C}^{\infty}\left(M\right)$ of smooth real valued
functions over the manifold $M$. As a matter of fact in such case
we can identify each $f\in\mathrm{C}^{\infty}\left(M\right)$ with
the map
\begin{eqnarray*}
M & \rightarrow & M\times\mathbb{R}\\
p & \mapsto & \left(p,f\left(p\right)\right)
\end{eqnarray*}
(still called $f$) which is immediately recognized as a section in
the trivial tensor bundle $\mathrm{T}^{\left(0,0\right)}M=M\times\mathbb{R}$.
\end{example}
\index{vector field}\index{1-form}\index{tensor fields}We take
the chance to introduce some nomenclature: Sections in the tangent
bundle of a manifold are usually called \textsl{vector fields}, while
sections in the cotangent bundle are known as \textsl{1-forms}. Moreover
sections in each tensor bundle of type $\left(i,j\right)$ are generally
called \textsl{tensor fields}.

In a vector bundle there is no natural notion of differentiation,
so that we must provide such notion together with the vector bundle
in order to be able to do calculus.
\begin{defn}
\index{connection}Let $M$ be a manifold and let $E$ be a vector
bundle over $M$. A \textsl{(linear) connection on $E$} is a map
\begin{eqnarray*}
\nabla:\mathrm{C}^{\infty}\left(M,\mathrm{T}M\right)\times\mathrm{C}^{\infty}\left(M,E\right) & \rightarrow & \mathrm{C}^{\infty}\left(M,E\right)\\
\left(X,s\right) & \mapsto & \nabla_{X}s
\end{eqnarray*}
that satisfies the following properties:
\begin{itemize}
\item \textsl{$\mathrm{C}^{\infty}\left(M,\mathbb{R}\right)$-linearity
in the first argument}: for each $f$, $h\in\mathrm{C}^{\infty}\left(M\right)$,
each $X$, $Y\in\mathrm{C}^{\infty}\left(M,\mathrm{T}M\right)$ and
each $s\in\mathrm{C}^{\infty}\left(M,E\right)$ it holds
\[
\nabla_{\left(fX+gY\right)}s=f\nabla_{X}s+g\nabla_{Y}s\mbox{;}
\]

\item \textsl{$\mathbb{R}$-linearity in the second argument}: for each
$a$, $b\in\mathbb{R}$, each $X\in\mathrm{C}^{\infty}\left(M,\mathrm{T}M\right)$
and each $s$, $t\in\mathrm{C}^{\infty}\left(M,E\right)$ it holds
\[
\nabla_{X}\left(as+bt\right)=a\nabla_{X}s+b\nabla_{Y}t\mbox{;}
\]

\item \textsl{Leibniz rule in the second argument}: for each $s\in\mathrm{C}^{\infty}\left(M,E\right)$,
each $f\in\mathrm{C}^{\infty}\left(M\right)$ and each $X\in\mathrm{C}^{\infty}\left(M,\mathrm{T}M\right)$
it holds that
\[
\nabla_{X}\left(fs\right)=\left(\partial_{X}f\right)s+f\nabla_{X}s\mbox{,}
\]
where $\partial_{X}f$ is the section in $\mathrm{T}M$ defined by
$\left(\partial_{X}f\right)\left(p\right)=\left(\mathrm{d}_{p}f\right)\left(X\left(p\right)\right)$
for each $p\in M$.
\end{itemize}
\end{defn}
The properties required allow us to think a connection $\nabla$ as
a map 
\[
\mathrm{C}^{\infty}\left(M,\mathrm{T}M\right)\otimes\mathrm{C}^{\infty}\left(M,E\right)=\mathrm{C}^{\infty}\left(M,\mathrm{T}M\otimes E\right)\rightarrow\mathrm{C}^{\infty}\left(M,E\right)
\]
or also as a map
\[
\mathrm{C}^{\infty}\left(M,E\right)\rightarrow\mathrm{C}^{\infty}\left(M,\mathrm{T}^{*}M\right)\otimes\mathrm{C}^{\infty}\left(M,E\right)=\mathrm{C}^{\infty}\left(M,\mathrm{T}^{*}M\otimes E\right)\mbox{.}
\]

We want to stress that on a given vector bundle there may be several
possible inequivalent connections. This is indeed the case also for
tensor bundles of each type. A concrete example of a connection on
the trivial tensor bundle $\mathrm{T}^{\left(0,0\right)}M=M\times\mathbb{R}$
is provided by the map
\[
\partial:\mathrm{C}^{\infty}\left(M,\mathrm{T}M\right)\times\mathrm{C}^{\infty}\left(M,M\times\mathbb{R}\right)\rightarrow\mathrm{C}^{\infty}\left(M,M\times\mathbb{R}\right)
\]
defined in the statement of the Leibniz rule for a connection (the
identification of $\mathrm{C}^{\infty}\left(M\right)$ with the space
of sections $\mathrm{C}^{\infty}\left(M,M\times\mathbb{R}\right)$
presented in Example \ref{exaCinftyM} is understood).

Notice that there is a natural way to induce a connection on a vector
bundle built through fiberwise vectorial operations (e.g. duality,
tensor product and Whitney sum) starting from the connections on the
original vector bundles. Examples are provided by the following formulas
(we put superscripts on $\nabla$ to indicate the vector bundle on
which the connection is defined):
\begin{eqnarray*}
\left(\nabla_{X}^{E^{*}}\nu\right)\left(\mu\right) & = & \partial_{X}\left(\nu\left(\mu\right)\right)-\nu\left(\nabla_{X}^{E}\mu\right)\mbox{,}\\
\nabla_{X}^{E\otimes F}\left(\mu\otimes\rho\right) & = & \left(\nabla_{X}^{E}\mu\right)\otimes\rho+\mu\otimes\left(\nabla_{X}^{F}\rho\right)\mbox{,}\\
\nabla_{X}^{E\oplus F}\left(\mu\oplus\rho\right) & = & \left(\nabla_{X}^{E}\mu\right)\oplus\left(\nabla_{X}^{F}\rho\right)\mbox{,}
\end{eqnarray*}
where $E$ and $F$ are vector bundles over a manifold $M$ endowed
with connections $\nabla^{E}$ and respectively $\nabla^{F}$, $X$
is an arbitrary vector field over $M$ and $\mu$, $\nu$ and $\rho$
are arbitrary sections respectively in $E$, $E^{*}$ and $F$.

Now we want to define an object that characterizes the behavior of
each connection on a given vector bundle. To do this we need the following
construction. Suppose that $\nabla$ is a connection over the vector
bundle $\left(E,M,\pi\right)$ and fix a point $p\in M$. We denote
with $d$ the dimension of $M$ and with $n$ the rank of $E$. There
exists a neighborhood $U$ of $p$ such that $\left(U,\Omega,\phi\right)$
is a coordinate neighborhood of $p$ in $M$ and $\left(U,\Phi\right)$
is a local trivialization at $p$ of $E$. On the one hand, using
the coordinate neighborhood, we can obtain a set of local vector fields
(i.e. sections in $\mathrm{T}U$) $\left\{ \partial_{1},\dots\partial_{d}\right\} $
that are pointwise linearly independent: This is done pushing forward
through the diffeomorphism $\phi^{-1}:\Omega\rightarrow U$ the vector
fields $\left\{ v_{1},\dots,v_{d}\right\} $ on $\mathrm{T}\Omega$
(identified with $\Omega\times\mathbb{R}^{d}$) that are defined by
$v_{i}\left(x\right)=e_{i}$ for each $x\in\Omega$ and each $i\in\left\{ 1,\dots,d\right\} $,
where $\left\{ e_{1},\dots,e_{d}\right\} $ is the standard orthonormal
base of $\mathbb{R}^{d}$. On the other hand, once chosen an orthonormal
base $\left\{ f_{1},\dots,f_{n}\right\} $ of $\mathbb{R}^{n}$, we
obtain a set of sections $\left\{ \mu_{1},\dots,\mu_{n}\right\} $
in $\left.E\right|_{U}=\pi^{-1}\left(U\right)$ that are pointwise
linearly independent setting $\mu_{j}\left(q\right)=\Phi^{-1}\left(q,f_{j}\right)$
for each $q\in U$ and each $j\in\left\{ 1,\dots,n\right\} $. That
done, we can define the Christoffel symbols.
\begin{defn}
\label{defChristoffelSymbols}\index{Christoffel symbols}Let $\nabla$
be a connection over the vector bundle $\left(E,M,\pi\right)$. With
the construction given above, we can define the\textsl{ Christoffel
symbols} $\Gamma_{ij}^{k}$ of the connection $\nabla$ in a neighborhood
$U$ of a point $p$ in $M$ imposing $\Gamma_{ij}^{k}\mu_{k}=\nabla_{\partial_{i}}\mu_{j}$
(summation over $k$ is implied).
\end{defn}
Consider a vector bundle $E$ over a manifold $M$ endowed with a
connection $\nabla$ and fix a smooth curve $c:\left[a,b\right]\rightarrow M$
and $s_{0}\in E_{c\left(a\right)}$. We can consider the following
problem: Determine $s$ from $\left[a,b\right]$ to $E$ satisfying
\begin{eqnarray*}
\nabla_{X_{t}}s\left(t\right) & = & 0\quad\mbox{for }t\in\left[a,b\right]\mbox{,}\\
s\left(a\right) & = & s_{0},
\end{eqnarray*}
where $X_{t}\in\mathrm{T}_{c\left(t\right)}M$ is the vector tangent
to $c$ in $c\left(t\right)$. Written in local coordinates such problem
reduces to a system of linear first order ordinary differential equations,
hence the solution exists and is unique once that $s_{0}\in E_{c\left(a\right)}$
is given. In particular we obtain $s\left(b\right)$. This allows
us to give the next definition.
\begin{defn}
\label{defParallelTransport}\index{parallel transport}Let $\left(E,M,\pi\right)$
be a vector bundle and let $\nabla$ be a connection on it. For each
smooth curve $c:\left[a,b\right]\rightarrow M$ we define the \textsl{parallel
transport along $c$} as the linear function $Y_{c}:E_{c\left(a\right)}\rightarrow E_{c\left(b\right)}$
that maps each $s_{0}\in E_{c\left(a\right)}$ to $s\left(b\right)$
as above.
\end{defn}
We underline that in general the parallel transport depends upon the
choice of the curve connecting its endpoints, but, once that a curve
is chosen, the connection gives us a way to {}``connect'' different
fibers of the vector bundle through parallel transport.

We want to present another object that characterizes a connection
on a vector bundle. However its definition requires a new tool.
\begin{defn}
\index{Lie bracket}Let $M$ be a manifold. We call \textsl{Lie bracket}
the map
\[
\left[\cdot,\cdot\right]:\mathrm{C}^{\infty}\left(M,\mathrm{T}M\right)\times\mathrm{C}^{\infty}\left(M,\mathrm{T}M\right)\rightarrow\mathrm{C}^{\infty}\left(M,\mathrm{T}M\right)
\]
uniquely determined by the following condition:
\[
\partial_{\left[X,Y\right]}f=\partial_{X}\partial_{Y}f-\partial_{Y}\partial_{X}f\quad\forall X,Y\in\mathrm{C}^{\infty}\left(M,\mathrm{T}M\right),\,\forall f\in\mathrm{C}^{\infty}\left(M,M\times\mathbb{R}\right)\mbox{.}
\]

\end{defn}
\index{Jacobi identity}We take the chance to state the properties
of the Lie bracket: it is $\mathbb{R}$-bilinear, antisymmetric and
satisfies the \textsl{Jacobi identity}, i.e. for each $X$, $Y$,
$Z\in\mathrm{C}^{\infty}\left(M,\mathrm{T}M\right)$ it holds that
\[
\left[\left[X,Y\right],Z\right]+\left[\left[Y,Z\right],X\right]+\left[\left[Z,X\right],Y\right]=0\mbox{.}
\]

Now we are in position to properly define the curvature of a connection
on a vector bundle.
\begin{defn}
\index{curvature}Let $\left(E,M,\pi\right)$ be a vector bundle endowed
with a connection $\nabla$. We call \textsl{curvature of the connection
$\nabla$} the map
\[
C:\mathrm{C}^{\infty}\left(M,\mathrm{T}M\right)\times\mathrm{C}^{\infty}\left(M,\mathrm{T}M\right)\times\mathrm{C}^{\infty}\left(M,E\right)\rightarrow\mathrm{C}^{\infty}\left(M,E\right)
\]
defined by
\[
C\left(X,Y\right)s=\nabla_{X}\nabla_{Y}s-\nabla_{Y}\nabla_{X}s-\nabla_{\left[X,Y\right]}s\mbox{,}
\]
where $X$, $Y\in\mathrm{C}^{\infty}\left(M,\mathrm{T}M\right)$ and
$s\in\mathrm{C}^{\infty}\left(M,E\right)$.\end{defn}
\begin{rem}
From its definition, we deduce that $C$ is $\mathbb{R}$-bilinear
and antisymmetric in the first two arguments and $\mathbb{R}$-linear
in the last argument. Therefore, denoting with $\otimes_{a}$ the
antisymmetrized tensor product, we can interpret $C$ as a map from
$\mathrm{C}^{\infty}\left(M,\mathrm{T}M\right)\otimes_{a}\mathrm{C}^{\infty}\left(M,\mathrm{T}M\right)\otimes\mathrm{C}^{\infty}\left(M,E\right)$
to $\mathrm{C}^{\infty}\left(M,E\right)$. Moreover its value at each
point $p\in M$ depends only on the values of $X$, $Y$ and $s$
in an arbitrary neighborhood of $p$ so that we are allowed to think
$C$ as a section in the vector bundle $\mathrm{T}^{*}M\otimes_{a}\mathrm{T}^{*}M\otimes E^{*}\otimes E$.
\end{rem}
$C$ can be locally written in components following a procedure analogous
to that used to define Christoffel symbols (see before Definition
\ref{defChristoffelSymbols}): For a fixed point $p\in M$, we can
find an open neighborhood $U$ of $p$ in $M$, a set of pointwise
linearly independent local vector fields $\left\{ v_{1},\dots,v_{d}\right\} $
over $U$ and a set of pointwise linearly independent local sections
$\left\{ \mu_{1},\dots\,\mu_{n}\right\} $ in $\left.E\right|_{U}$,
where we set $d=\dim M$ and $h=\mathrm{rank}E$, and we can define
$C_{ijk}^{\hphantom{ijk}l}$ imposing $C_{ijk}^{\hphantom{ijk}l}\mu_{l}=C\left(v_{i},v_{j}\right)\mu_{k}$.
Using this definition it is possible to obtain the expression of $C_{ijk}^{\hphantom{ijk}l}$
in terms of the Christoffel symbols and their derivatives $\partial$
along the local vector fields.

Now we define inner products on vector bundles. This additional structure
allows us to to pick out a specific connection on the tangent bundle
of a manifold that has particular importance for General Relativity.
\begin{defn}
\index{inner product}Consider a vector bundle $E$ over the manifold
$M$. We call \textsl{inner product on $E$} a section $g$ in $E^{*}\otimes E^{*}$
that fulfils the following requirements:
\begin{itemize}
\item \textsl{(fiberwise) symmetry}: for each $p\in M$ and each $u$, $v\in E_{p}$
it holds that
\[
g\left(p\right)\left(u\otimes v\right)=g\left(p\right)\left(v\otimes u\right)\mbox{;}
\]

\item \textsl{(fiberwise) non degeneracy}: for each $p\in M$ we have the
implication
\[
u\in E_{p}:\, g\left(p\right)\left(u\otimes v\right)=0\,\forall v\in E_{p}\;\implies\; u=0\mbox{.}
\]

\end{itemize}
\index{metric}\index{Riemannian metric}\index{Lorentzian metric}Inner
products on $\mathrm{T}M$ are called \textsl{metrics on $M$}. \textsl{Riemannian
metrics} are those whose signature is of type $\left(+,\dots,+\right)$
at any point, while Lorentzian metrics have signature of type $\left(-,+,\dots,+\right)$.
\end{defn}
In some situations it is customary to define Lorentzian metrics with
the requirement that their signature is of type $\left(+,-,\dots,-\right)$.
We can pass from our definition to this one simply taking $-g$ in
place of $g$.

Usually we will denote $g\left(p\right)\left(u\otimes v\right)$ with
$u\cdot_{g,p}v$ and, if there is no risk of misunderstanding, we
will also omit $g$ in our notation so that $u\cdot_{g,p}v$ becomes
$u\cdot_{p}v$. In the case of a metric we will write $g_{p}\left(u,v\right)$
in place of $g\left(p\right)\left(u\otimes v\right)$.
\begin{rem}
\label{remInnerProductsAsSections}Notice that each inner product
on $E$ is automatically a (fiberwise) non degenerate, symmetric,
bilinear form from $E\times E$ to the trivial vector bundle $M\times\mathbb{R}$.
From another point of view, we could define inner products on a vector
bundle $E$ as (fiberwise) non degenerate sections in the vector bundle
$E^{*}\otimes_{s}E^{*}$, where $\otimes_{s}$ denotes the symmetrized
tensor product. In this way the set of inner products on $E$ becomes
a subset of the vector space $\mathrm{C}^{\infty}\left(M,E^{*}\otimes_{s}E^{*}\right)$,
which becomes a Fréchet space when endowed with the usual topology
of $\mathrm{C}^{\infty}$ sections.
\end{rem}
With the usual procedure we can locally rewrite in components an inner
product $g$ on a vector bundle $E$. We must only fix $p\in M$,
consider a local trivialization of $E$ in an open neighborhood $U$
of $p$, find a set $\left\{ \mu_{1},\dots,\mu_{n}\right\} $ of pointwise
linearly independent sections in $\left.E\right|_{U}$ and set $g_{ij}\left(q\right)=\mu_{i}\cdot_{q}\mu_{j}$
for each $q\in U$, where $n=\mathrm{rank}E$. The property of fiberwise
symmetry implies that $g_{ij}\left(q\right)=g_{ji}\left(q\right)$,
while non degeneracy implies that $\left(g_{ij}\left(q\right)\right)$
is an invertible $n\times n$ matrix. This holds for each $q\in U$.
We denote by $\left(g^{ij}\left(q\right)\right)$ the inverse of $\left(g_{ij}\left(q\right)\right)$
for each $q\in U$.

Using an inner product on a vector bundle $E$ we can define the so
called musical isomorphisms between $E$ and its dual $E^{*}$.
\begin{defn}
\label{defMusicalIsomorphisms}\index{musical isomorphisms}\index{lowering isomorphism}\index{rising isomorphism}Let
$E$ be a vector bundle endowed with an inner product $g$. We define
\begin{itemize}
\item the \textsl{lowering isomorphism}:
\begin{eqnarray*}
\flat:E & \rightarrow & E^{*}\mbox{,}\\
\mu & \mapsto & \left.g\right|_{\pi_{E}\left(\mu\right)}\left(\mu\otimes\cdot\right)\mbox{;}
\end{eqnarray*}

\item the \textsl{rising isomorphism}:
\[
\sharp=\flat^{-1}:E^{*}\rightarrow E\mbox{.}
\]

\end{itemize}
The raising isomorphism and the lowering isomorphism are collectively
called \textsl{musical isomorphisms}.
\end{defn}
As suggested by their names, $\flat$ and $\sharp$ are both vector
bundle isomorphisms.

We anticipated that we can uniquely determine a specific connection
on the tangent bundle of a manifold endowed with a metric.
\begin{thm}
\index{Levi-Civita connection}Consider a manifold $M$ endowed with
a metric $g$. Then there exists a unique connection $\nabla$ on
$\mathrm{T}M$ that satisfies the following requirements:
\begin{itemize}
\item \index{metric connection}$\nabla$ is metric, i.e. for each $X$,
$Y$, $Z\in\mathrm{T}M$ it holds that
\[
\partial_{X}\left(g\left(Y,Z\right)\right)=g\left(\nabla_{X}Y,Z\right)+g\left(Y,\nabla_{X}Z\right)\mbox{;}
\]

\item \index{torsion free connection}$\nabla$ is torsion free, i.e. for
each $X$, $Y\in\mathrm{T}M$ it holds that
\[
\nabla_{X}Y-\nabla_{Y}X=\left[X,Y\right]\mbox{.}
\]

\end{itemize}
This connection $\nabla$ on $\mathrm{T}M$ is called \textsl{Levi-Civita
connection}.
\end{thm}
From the requirements singling out the Levi-Civita connection among
all possible connections on $\mathrm{T}M$, we can determine the Christoffel
symbols of the Levi-Civita connection (this fact actually guarantees
uniqueness of the Levi-Civita connection). This is done by fixing
a point $p\in M$ and choosing a coordinate neighborhood of $p$ and
a set of pointwise orthonormal (with respect to the metric on $M$)
local vector fields $\left\{ v_{1},\dots,v_{d}\right\} $ ($d=\dim M$):
we easily find
\begin{equation}
\Gamma_{ij}^{k}=g^{kl}\frac{1}{2}\left(\partial_{i}g_{lj}+\partial_{j}g_{il}-\partial_{l}g_{ij}\right)\mbox{,}\label{eqChristoffelSymbolsOfTheLeviCivitaConnection}
\end{equation}
where $\partial_{i}$ denotes $\partial_{v_{i}}$. Note that the symmetry
$g_{ij}=g_{ji}$ implies that the Christoffel symbols of the Levi-Civita
connection satisfy
\begin{equation}
\Gamma_{ij}^{k}=\Gamma_{ji}^{k}\mbox{.}\label{eqChristoffelSymbolsOfTheLeviCivitaConnectionSymmetry}
\end{equation}

We stress that each time that we will encounter a connection on the
tangent bundle of a manifold endowed with a metric, such connection
will be the Levi-Civita one.

\index{Ricci tensor}\index{scalar curvature}Till now we have dealt
with the curvature of a connection on an arbitrary vector bundle.
In the special case of the tangent bundle $\mathrm{T}M$ of a manifold
$M$ endowed with a metric $g$ we can define other associated objects,
namely the \textsl{Ricci tensor} $R_{ij}$ and the \textsl{scalar
curvature} $S$. We define them locally starting from the local definitions
of $C_{ijk}^{\hphantom{ijk}l}$ and $g^{ij}$: $R_{ik}=C_{ijk}^{\hphantom{ijk}j}$
and $S=g^{ij}R_{ij}$.

We present here the expressions of the curvature and of the Ricci
tensor for the Levi-Civita connection on a manifold endowed with a
metric:
\begin{eqnarray}
C_{ijk}^{\hphantom{ijk}l} & = & \partial_{j}\Gamma_{ik}^{l}-\partial_{i}\Gamma_{jk}^{l}+\Gamma_{ik}^{m}\Gamma_{jm}^{l}-\Gamma_{jk}^{m}\Gamma_{im}^{l}\mbox{;}\nonumber \\
R_{ij} & = & \partial_{k}\Gamma_{ij}^{k}-\partial_{i}\Gamma_{kj}^{k}+\Gamma_{ij}^{l}\Gamma_{kl}^{k}-\Gamma_{kj}^{l}\Gamma_{il}^{k}\mbox{.}\label{eqRicciTensorForTheLeviCivitaConnection}
\end{eqnarray}

\subsection{Differential forms on a manifold}

In this subsection we discuss a specific class of tensor fields over
an arbitrary manifold $M$, called differential forms. A much more
detailed discussion in this topic can be found in \cite[Chap. V]{Boo86}.

Let $M$ be a $d$-dimensional manifold and fix $p\in M$. For $k\in\mathbb{N}$,
we consider $\mathrm{T}_{p}^{\left(0,k\right)}M$. We would like to
pick out a subspace $\mathrm{T}_{p}^{\left(0,k\right)}M$, specifically
the one consisting of such elements that are skew-symmetric when intended
as $k$-linear maps from $\mathrm{T}_{p}M\times\cdots\times\mathrm{T}_{p}M$
($k$ times) to $\mathbb{R}$. To recognize these elements we need
a new tool, the alternating map.
\begin{defn}
\index{alternating map}Let $M$ be a manifold and consider $p\in M$
and $k\in\mathbb{N}$. We define the \textsl{alternating map at $p$}
\[
\mathfrak{a}:\mathrm{T}_{p}^{\left(0,k\right)}M\rightarrow\mathrm{T}_{p}^{\left(0,k\right)}M
\]
setting for each $\omega\in\mathrm{T}_{p}^{\left(0,k\right)}M$, each
$v_{1}$, $\dots$, $v_{k}\in\mathrm{T}_{p}M$
\[
\left(\mathfrak{a}\omega\right)\left(v_{1},\dots,v_{k}\right)=\frac{1}{k!}\sum_{\sigma}\left(\mathrm{sgn}\sigma\right)\omega\left(v_{\sigma\left(1\right)},\dots,v_{\sigma\left(k\right)}\right)\mbox{,}
\]
where $\sigma$ is a permutation of $1$,$\dots$, $k$ and $\mathrm{sgn}\sigma$
is its sign.
\end{defn}
Using the alternating map $\mathfrak{a}$ defined just above, we can
introduce alternating tensor bundles of type $k$ over a manifold
$M$.
\begin{defn}
\label{defAlternatingTensor}\index{alternating tensor}\index{alternating tensor space of type $k$}\index{alternating tensor bundle of type $k$}Let
$M$ be a $d$-dimensional manifold and consider $p\in M$ and $k\in\mathbb{N}$.
An \textsl{alternating tensor of type $k$ over a manifold $M$ at
$p$} is an element $\omega\in\mathrm{T}_{p}^{\left(0,k\right)}M$
such that $\mathfrak{a}\omega=\omega$. The \textsl{alternating tensor
space of type $k$}, denoted by $\mathrm{\Lambda}_{p}^{k}M$, is the
set of all alternating tensors of type $k$ over a manifold $M$ at
$p$ and the \textsl{alternating tensor bundle of type $k$}, denoted
by $\mathrm{\Lambda}^{k}M$, is the disjoint union over $p\in M$
of $\mathrm{\Lambda}_{p}^{k}M$.
\end{defn}
By convention we set $\mathrm{\Lambda}_{p}^{0}M=\mathbb{R}$ for each
$p\in M$ and $\mathrm{\Lambda}^{0}M=M\times\mathbb{R}$.

It turns out that $\mathrm{\Lambda}_{p}^{k}M$ is a real vector space
for each $p\in M$ and each $k\in\mathbb{N}$ and that $\mathrm{\Lambda}^{k}M$
is a vector bundle for each $k\in\mathbb{N}$. It can be shown that
$\Lambda_{p}^{k}M=\left\{ 0\right\} $ for each $k>d$ and that $\dim\left(\mathrm{\Lambda}_{p}^{k}M\right)=\binom{d}{k}$. 

Once that a point $p\in M$ is fixed, it is possible to define a new
algebra in a way similar to that followed for the definition of the
tensor algebra $\left(\mathscr{T}_{p}M,\otimes\right)$. The underlying
set of such algebra is
\[
\mathrm{\Lambda}_{p}M=\bigoplus_{k=0}^{d}\mathrm{\Lambda}_{p}^{k}M\mbox{.}
\]
The vector structure on $\mathrm{\Lambda}_{p}M$ is naturally induced
by the direct sum $\oplus$, while the algebraic structure requires
the introduction of a new operation, the so called wedge product.
\begin{defn}
\index{wedge product}Let $M$ be a $d$-dimensional manifold and
consider $k$, $k^{\prime}\in\mathbb{N}$ and $p\in M$. The wedge
product $\wedge$ is the map from $\mathrm{\Lambda}_{p}^{k}M\times\mathrm{\Lambda}_{p}^{k^{\prime}}M\rightarrow\mathrm{\Lambda}_{p}^{k+k^{\prime}}M$
defined by the formula
\[
\eta\wedge\xi=\binom{k+k^{\prime}}{k}\mathfrak{a}\left(\eta\otimes\xi\right)
\]
for each $\eta\in\mathrm{\Lambda}{}_{p}^{k}M$ and each $\xi\in\mathrm{\Lambda}_{p}^{k^{\prime}}M$.
\end{defn}
It can be shown that $\wedge$ can be naturally extended to an operation
on $\Lambda_{p}M$ (still called wedge product and denoted by $\wedge$)
that is binary, internal, bilinear and associative. This allows us
to conclude that $\left(\mathrm{\Lambda}_{p}M,\wedge\right)$ is an
associative algebra. The fact that $\mathrm{\Lambda}_{p}^{k}M=\mathrm{T}_{p}^{*}M\wedge\cdots\wedge\mathrm{T}_{p}^{*}M$
($k$ times) implies that $\left(\mathrm{\Lambda}_{p}M,\wedge\right)$
is generated by $\mathbb{R}$ and $\mathrm{T}_{p}^{*}M$.

\index{alternating tensor bundle}In addition to such pointwise algebraic
structure, it is possible to define a new vector bundle $\mathrm{\Lambda}M$
through the disjoint union of $\mathrm{\Lambda}_{p}M$ over $p\in M$.
$\mathrm{\Lambda}M$ is called \textsl{alternating tensor bundle}.
As a by product of this construction we obtain an extension of the
wedge product to an operation on the alternating tensor bundle. In
particular we have that
\[
\mathrm{\Lambda}^{k}M=\bigwedge^{k}\mathrm{T}^{*}M\;\forall k\qquad\mbox{and}\qquad\mathrm{\Lambda}M=\bigoplus_{k=0}^{d}\mathrm{\Lambda}^{k}M\mbox{.}
\]

The above preparation allows us to define $k$-forms.
\begin{defn}
\index{differential form}We say that a \textsl{$k$-form over $M$}
(also called \textsl{differential form of order $k$ over $M$}) is
a section in the alternating tensor bundle of order $k$ $\mathrm{\Lambda}^{k}M$.
The \textsl{space of $k$-forms over $M$} is denoted by $\mathrm{\Omega}^{k}M$.
We define the \textsl{space of differential forms over $M$} $\mathrm{\Omega}M$
as the direct sum of all the non trivial spaces of $k$-forms.
\end{defn}
Notice that $\mathrm{\Omega}^{0}M=\mathrm{C}^{\infty}\left(M\right)$
and that $\mathrm{\Omega}^{k}M=\left\{ 0\right\} $ for each $k>\dim M$.
We easily recognize that $\mathrm{\Omega}^{k}M$ is a vector spaces
for each $k$. This fact motivates the word {}``space'' (intended
in the sense of vector space) used in the last definition.

\index{exterior algebra}Previously we defined the wedge product pointwisely.
It is possible to extend this operation from the alternating tensor
spaces at each point to the space of differential forms $\mathrm{\Omega}M$
simply imposing $\left(\Xi\wedge\Theta\right)\left(p\right)=\Xi\left(p\right)\wedge\Theta\left(p\right)$
for each $p\in M$ and each $\Xi$, $\Theta\in\mathrm{\Omega}M$.
It turns out that $\left(\mathrm{\Omega}M,\wedge\right)$ is an associative
algebra, known as \textsl{exterior algebra of $M$}.

As a consequence of its definition, $\mathrm{\Omega}^{k}M$ is a subspace
of $\mathrm{C}^{\infty}\left(M,\mathrm{T}^{\left(0,k\right)}M\right)$
for each $k\in\mathbb{N}$. This fact guarantees that the observations
made about push-forwards and pull-backs through a diffeomorphism of
sections in tensor bundles of any type (see Remark \ref{remPushforwardPullback})
applies also in this case, hence we can push forward and pull back
any $k$-form using a diffeomorphism.

We take the chance to state some useful properties of the wedge product.
\begin{prop}
Let $M$ and $N$ be manifolds. Then for each $k$, $k^{\prime}\in\mathbb{N}$
the wedge product $\wedge$ fulfils the following properties:
\begin{itemize}
\item $\Xi\wedge\Theta=\left(-1\right)^{kk^{\prime}}\Xi\wedge\Omega$ for
each $\Xi\in\mathrm{\Omega}^{k}M$ and each $\Theta\in\mathrm{\Omega}^{k^{\prime}}M$;
\item $\left(f\Xi\right)\wedge\Theta=$f$\left(\Xi\wedge\Theta\right)$
for each $\Omega\in\mathrm{\Omega}^{k}M$, each $\Theta\in\mathrm{\Omega}^{k^{\prime}}M$
and each $f\in\mathrm{C}^{\infty}\left(M\right)$;
\item if $f:M\rightarrow N$ is a smooth function, for each $\Xi\in\mathrm{\Omega}^{k}N$
and each $\Theta\in\mathrm{\Omega}^{k^{\prime}}N$
\[
f^{*}\left(\Xi\wedge\Theta\right)=\left(f^{*}\Xi\right)\wedge\left(f^{*}\Theta\right)\mbox{,}
\]
where the wedge on the LHS is defined on $\mathrm{\Omega}N$, while
the wedge on the RHS is defined on $\mathrm{\Omega}M$.
\end{itemize}
\end{prop}
As a consequence of the last theorem we can conclude that, for each
diffeomorphism $f:M\rightarrow N$, the vector bundle isomorphism
$\left(f^{-1},f^{*}\right):\mathrm{\Lambda}N\rightarrow\mathrm{\Lambda}M$
induces an algebraic isomorphism between the exterior algebras $\mathrm{\Omega}N$
and $\mathrm{\Omega}M$. Notice that one can similarly consider the
vector bundle isomorphism $\left(f,f_{*}\right):\mathrm{\Lambda}M\rightarrow\mathrm{\Lambda}N$
and conclude that this induces an algebraic isomorphism between the
exterior algebras $\mathrm{\Omega}M$ and $\mathrm{\Omega}N$. Moreover
these algebraic isomorphisms are inverses of each other.

Thanks to the following theorem it is possible to define a new operation
on the exterior algebra of $M$ that is a sort of special case of
the push-forward of a real valued smooth function (also called differential,
see Definition \ref{defPushforward}).
\begin{prop}
\label{propExteriorDerivative}\index{exterior derivative}For each
manifold $M$ there exists a unique $\mathbb{R}$-linear map $\mathrm{d}_{M}:\mathrm{\Omega}M\rightarrow\mathrm{\Omega}M$,
called \textsl{exterior derivative}, fulfilling the following properties:
\begin{itemize}
\item the exterior derivative coincides with the differential on $\mathrm{\Omega}^{0}M=\mathrm{C}^{\infty}\left(M\right)$,
i.e. $\mathrm{d}_{M}f=\mathrm{d}f$ for each $f\in\mathrm{\Omega}^{0}M=\mathrm{C}^{\infty}\left(M\right)$;
\item for each $\Xi\in\mathrm{\Omega}^{k}M$ and each $\Theta\in\mathrm{\Omega}^{k^{\prime}}M$
it holds that
\[
\mathrm{d}_{M}\left(\Xi\wedge\Theta\right)=\mathrm{d}_{M}\Xi\wedge\Theta+\left(-1\right)^{k}\Xi\wedge\mathrm{d}_{M}\Theta\mbox{;}
\]

\item $\mathrm{d}_{M}^{2}=\mathrm{d}_{M}\circ\mathrm{d}_{M}=0$.
\end{itemize}
Moreover the exterior derivative satisfies another property: For each
smooth map $f$ from a manifold $M$ to a manifold $N$ we have
\[
f^{*}\circ\mathrm{d}_{N}=\mathrm{d}_{M}\circ f^{*}\mbox{.}
\]

\end{prop}
Notice that the last property of the exterior derivative may also
be read in this way if $f:M\rightarrow N$ is a diffeomorphism:
\[
f_{*}\circ\mathrm{d}_{M}=\mathrm{d}_{N}\circ f_{*}\mbox{.}
\]

In the following we will denote the exterior derivative simply with
$\mathrm{d}$, omitting the subscript referred to the manifold. Notice
that there is no risk of confusion between exterior derivative and
differential because they coincide in the only situation in which
they may be confused, that is $\mathrm{\Omega}^{0}M=\mathrm{C}^{\infty}\left(M\right).$

Using the exterior derivative, we can introduce a classification of
$k$-forms and then define the de Rham cohomology groups that will
be used to introduce an hypothesis when we will discuss the electromagnetic
field.
\begin{defn}
\label{defdeRhamCohomologyGroup}\index{closed $k$-form}\index{exact $k$-form}\index{de Rham cohomology groups}Let
$M$ be a $d$-dimensional manifold and consider $k\in\left\{ 1,\dots,d\right\} $.
We say that $\Theta\in\mathrm{\Omega}^{k}M$ is \textsl{closed} if
$\mathrm{d}\Theta=0$ while we say that it is \textsl{exact} if there
exists $\Xi\in\Omega^{k-1}M$ such that $\mathrm{d}\Xi=\Theta$. We
also denote with $Cl^{k}\left(M\right)$ the \textsl{space of closed
$k$-forms over $M$} and with $Ex^{k}\left(M\right)$ the \textsl{space
of exact $k$-forms over $M$}.

We call \textsl{$k$-th de Rham cohomology group of $M$} the quotient
space $H^{k}\left(M\right)=\nicefrac{Cl^{k}\left(M\right)}{Ex^{k}\left(M\right)}\mbox{.}$
\end{defn}
Notice that $Cl^{k}\left(M\right)$ and $Ex^{k}\left(M\right)$ are
actually vector spaces because $\mathrm{d}$ is linear on $\mathrm{\Omega}M$
and $Cl^{k}\left(M\right)$ is the kernel of $\mathrm{d}$ when restricted
to $\mathrm{\Omega}^{k}M$, while $Ex^{k}\left(M\right)$ is the image
of $\mathrm{\Omega}^{k-1}M$ through $\mathrm{d}$. Moreover since
$\mathrm{d}^{2}=0$, $Ex^{k}\left(M\right)\subseteq Cl^{k}\left(M\right)$.
Hence $H^{k}\left(M\right)$ is a well defined vector space.

In a $d$-dimensional manifold $M$, $d$-forms are of particular
importance because we can use them to define the orientability and
the orientation of a manifold. These notions will become relevant
in the next subsection.
\begin{defn}
\index{orientable manifold}\index{orientation}\index{oriented manifold}Let
$M$ be a $d$-dimensional manifold. We say that $M$ is \textsl{orientable}
if there exists a $d$-form $\Theta$ over $M$ which is nowhere null.
If $M$ is orientable and $\Theta$ is a choice of a nowhere null
$d$-form over $M$, we say that $\Theta$ fixes an \textsl{orientation
on $M$} and we call $M$ an \textsl{oriented manifold}.

\index{orientation preserving embedding}Let $M$ and $N$ be two
orientable manifolds and let $f:M\rightarrow N$ be an embedding.
Choose a nowhere null $d$-form $\Theta$ over $M$ and a nowhere
null $d$-form $\Xi$ over $N$ so that $M$ and $N$ are oriented.
We say that $f$ is \textsl{orientation preserving} if there exists
a strictly positive real valued smooth function $\lambda$ on $M$
such that $f^{*}\Xi=\lambda\Theta$.
\end{defn}
Notice that on a given orientable manifold $M$ there are different
possible choices of nowhere null $d$-forms that induce the same orientation.
It turns out that there are exactly two classes of such forms, each
one comprised by all the nowhere null $d$-forms that differ for a
strictly positive factor $\lambda\in\mathrm{C}^{\infty}\left(M\right)$,
such that each element of a class induce the same orientation on $M$.
Usually an oriented manifold $M$ is denoted by $\left(M,\mathfrak{o}\right)$,
where $\mathfrak{o}$ is one of the above mentioned classes of nowhere
null $d$-forms.

\index{oriented base}Once that an orientation $\mathfrak{o}$ on
$M$ is chosen, for each point $p\in M$ it is possible to find a
base $\left\{ v_{1},\dots,v_{d}\right\} $ of $\mathrm{T}_{p}M$ such
that $\Omega\left(p\right)\left(v_{1},\dots,v_{d}\right)>0$ for each
$\Omega\in\mathfrak{o}$. We say that such base is \textsl{oriented}.
If $M$ is also endowed with a metric $g$, $g_{p}$ defines an inner
product on the vector space $\mathrm{T}_{p}M$ for each $p\in M$.
This allows us to choose $g_{p}$-orthonormal bases of $\mathrm{T}_{p}M$
for any point $p$ in $M$. The next theorem puts together the choice
of an orientation $\mathfrak{o}$ and the presence of a metric to
provide a univocal way to choose a $d$-form in $\mathfrak{o}$.
\begin{thm}
\label{thmVolumeForm}\index{volume form}Let $\left(M,\mathfrak{o}\right)$
be an oriented $d$-dimensional manifold endowed with a metric $g$.
Then there exists a unique nowhere null $d$-form $d\mu_{g}\in\mathfrak{o}$,
called \textsl{volume form over $\left(M,\mathfrak{o}\right)$ induced
by $g$}, such that for each $p\in M$ $\mathrm{d}\mu_{g}$ takes
the value $+1$ on each oriented orthonormal base of $\mathrm{T}_{p}M$.
\end{thm}
It turns out that for each local coordinate neighborhood $\left(U,\Omega,\phi\right)$
of $M$ the following equation holds on every point of $\Omega$:
\begin{equation}
\phi_{*}\left(\mathrm{d}\mu_{g}\right)=\sqrt{\left|\det g\right|}\mathrm{d}x^{1}\wedge\cdots\wedge\mathrm{d}x^{d}=\sqrt{\left|\det g\right|}\mathrm{d}V\mbox{,}\label{eqVolumeFormInACoordinateNeighborhood}
\end{equation}
where $\left\{ \mathrm{d}x^{1},\dots,\mathrm{d}x^{d}\right\} $ is
the base of $\mathrm{T}^{*}\Omega$ (identified with $\Omega\times\mathbb{R}^{d}$)
defined by $\mathrm{d}x^{i}\left(x\right)=e_{i}$ for each $i\in\left\{ 1,\dots,d\right\} $
and each $x\in\Omega$, where $\left\{ e_{1},\dots,e_{d}\right\} $
is an oriented orthonormal base of $\mathbb{R}^{d}$ endowed with
a (non necessarily positive definite) inner product with the same
signature of $g$.

The volume form $\mathrm{d}\mu_{g}$ provided by the last theorem
will become very useful in the next subsection when we will introduce
a notion of integral on a manifold. An example of volume form is the
standard measure $\mathrm{d}V=\mathrm{d}x^{1}\wedge\dots\wedge\mathrm{d}x^{d}$
of $\mathbb{R}^{d}$ that appears in eq. \eqref{eqVolumeFormInACoordinateNeighborhood}
above.

Before we proceed with the next subsection, we want to introduce two
new operators. The first one is the Hodge dual. The detailed procedure
used to define it can be found in \cite[Sect. 2.1, pp. 87-90]{Jos95}.

Consider an oriented $d$-dimensional manifold $\left(M,\mathfrak{o}\right)$
endowed with a metric $g$. Let $\mathrm{d}\mu_{g}\in\mathfrak{o}$
be the volume form over $\left(M,\mathfrak{o}\right)$ induced by
$g$. Since $g$ defines a non degenerate inner product on each cotangent
space $\mathrm{T}_{p}^{*}M$, we can use it, together with the volume
form, to choose an orthonormal base $\left\{ \omega_{1}^{\left(p\right)},\dots,\omega_{d}^{\left(p\right)}\right\} $
of $\mathrm{T}_{p}^{*}M$ for each $p\in M$ such that $\mathrm{d}\mu_{g}\left(p\right)\left(\omega_{1}^{\left(p\right)},\dots,\omega_{d}^{\left(p\right)}\right)=+1$.
Notice that a base of $\mathrm{\Lambda}_{p}^{k}M$ is provided by
\[
B_{p,k}=\left\{ \omega_{i_{1}}^{\left(p\right)}\wedge\cdots\wedge\omega_{i_{k}}^{\left(p\right)}:1\leq i_{1}<\dots<i_{k}\leq d\right\} \mbox{.}
\]
We are ready to define the Hodge dual.
\begin{defn}
\index{Hodge dual}Let $\left(M,\mathfrak{o}\right)$ be a $d$-dimensional
oriented manifold endowed with a metric $g$ and let $\mathrm{d}\mu_{g}\in\mathfrak{o}$
be the volume form over $\left(M,\mathfrak{o}\right)$ induced by
$g$. For each $p\in M$ and each $k\in\left\{ 1,\dots,d\right\} $,
we define the \textsl{Hodge dual $*$} as the unique linear map from
$\mathrm{\Lambda}_{p}^{k}M$ to $\mathrm{\Lambda}_{p}^{d-k}M$ satisfying
the following condition for each element of $B_{p,k}$: 
\[
*\left(\omega_{i_{1}}^{\left(p\right)}\wedge\cdots\wedge\omega_{i_{k}}^{\left(p\right)}\right)=\omega_{j_{1}}^{\left(p\right)}\wedge\cdots\wedge\omega_{j_{d-k}}^{\left(p\right)}\mbox{,}
\]
where $j_{1}$, $\dots,$ $j_{d-k}\in\left\{ 1,\dots,d\right\} $
are chosen in such a way that
\[
\left\{ \omega_{i_{1}}^{\left(p\right)},\dots,\omega_{i_{k}}^{\left(p\right)},\omega_{j_{1}}^{\left(p\right)},\dots,\omega_{j_{d-k}}^{\left(p\right)}\right\} 
\]
is an oriented base of $\mathrm{T}_{p}^{*}M$.
\end{defn}
It can be shown that this definition is well posed so that for each
$p\in M$ and each $k\in\left\{ 1,\dots,d\right\} $ we have at our
disposal the operator $*$. If we consider $\Theta\in\mathrm{\Omega}^{k}M$,
we can take $*\left(\Theta\left(p\right)\right)$ for each $p\in M$.
It turns out that the map
\begin{eqnarray*}
M & \rightarrow & \mathrm{\Lambda}^{d-k}M\\
p & \mapsto & *\left(\Theta\left(p\right)\right)
\end{eqnarray*}
is a smooth section in $\mathrm{\Lambda}^{d-k}M$ that we denote with
$*\Theta$. Then the Hodge dual naturally defines an operator $*$
from $\mathrm{\Omega}^{k}M$ to $\Omega^{d-k}M$. This can be done
for each $k\in\left\{ 1,\dots,d\right\} $ so that the Hodge dual
is defined as an operator on $\mathrm{\Omega}M$.
\begin{rem}
From the last definition it is possible to deduce a formula for the
components of the Hodge dual of a $k$-form. Let $\omega\in\mathrm{\Omega}^{k}M$
and consider a point $p\in M$ and a coordinate neighborhood $\left(U,V,\phi\right)$
of $p$ in $M$. On $\mathrm{T}_{p}^{*}M=\mathrm{\Lambda}_{p}^{1}M$
we choose the oriented orthonormal basis $\left\{ \mathrm{d}x^{1},\dots,\mathrm{d}x^{d}\right\} $
and we denote the totally antisymmetric symbol with $\varepsilon_{i_{1}\dots i_{d}}$.
Then the components of $*\omega$ in $p$ in the basis of $\mathrm{\Lambda}_{p}^{d-k}M$
are given by the formula
\[
\left(*\omega\right)_{i_{1}\dots i_{d-k}}=\frac{1}{k!}\omega_{j_{1}\dots j_{k}}g^{j_{1}j_{1}^{\prime}}\cdots g^{j_{k}j_{k}^{\prime}}\varepsilon_{j_{1}^{\prime}\dots j_{k}^{\prime},i_{1}\dots i_{d-k}}\sqrt{\left|\det g\right|},
\]
where $\omega_{j_{1}\dots j_{k}}$ are the components of $\omega$
at $p$ in the basis of $\mathrm{\Lambda}_{p}^{k}M$, i.e.
\[
\omega\left(p\right)=\frac{1}{k!}\omega_{j_{1}\dots j_{k}}\mathrm{d}x^{j_{1}}\wedge\cdots\wedge\mathrm{d}x^{j_{k}}\mbox{,}
\]
and $\left(g_{\vphantom{ij}}^{ij}\right)$ is the inverse of the matrix
$\left(g_{ij}\right)$, whose coefficients are given by
\[
g_{ij}=g_{p}\left(\left(\mathrm{d}x^{i}\right)^{\sharp},\left(\mathrm{d}x^{j}\right)^{\sharp}\right)\mbox{.}
\]

\end{rem}
There are some other very important properties of the Hodge dual.
We recollect them in the following theorem.
\begin{prop}
\label{propHodgeDualProperties}Let $\left(M,\mathfrak{o}\right)$
be a $d$-dimensional oriented manifold endowed with a metric $g$
with signature $s=\pm1$. The Hodge dual $*:\mathrm{\Lambda}M\rightarrow\mathrm{\Lambda}M$
satisfies the following properties:
\begin{itemize}
\item for each $p\in M$, each $k\in\left\{ 1,\dots,d\right\} $ and each
$\omega\in\mathrm{\Lambda}_{p}^{k}M$ it holds that
\[
**\omega=s\left(-1\right)^{k\left(d-k\right)}\omega\mbox{;}
\]

\item for each $p\in M$, each $k\in\left\{ 1,\dots,d\right\} $ and each
$\omega$, $\theta\in\mathrm{\Lambda}_{p}^{k}M$ it holds that
\[
*\left(\omega\wedge*\theta\right)=s\left\langle \omega,\theta\right\rangle _{g,k}\mbox{,}
\]
where $\left\langle \cdot,\cdot\right\rangle _{g,k}$ denotes the
inner product of the vector bundle $\mathrm{\Lambda}^{k}M$ induced
by the metric $g$.
\end{itemize}
Moreover, if $\left(N,\mathfrak{p}\right)$ is an oriented manifold
endowed with a metric $h$ with signature $s^{\prime}=s$ and $f$
is an orientation preserving embedding such that $g=f^{*}h$, we have
that
\[
f^{*}\circ\overset{N}{*}=\overset{M}{*}\circ f^{*}\mbox{.}
\]

\end{prop}
The last theorem has three very important consequences:
\begin{itemize}
\item For each $p\in M$ and each $k\in\left\{ 1,\dots,d\right\} $ $*:\mathrm{\Lambda}_{p}^{k}M\rightarrow\mathrm{\Lambda}_{p}^{d-k}M$
is a vector space isomorphism whose inverse $*^{-1}=s\left(-1\right)^{k\left(d-k\right)}*$
can be extended to $\mathrm{\Lambda}^{d-k}M$ because it coincides
with $*:\Lambda_{p}^{d-k}M\rightarrow\mathrm{\Lambda}_{p}^{k}M$ (up
to a $\pm1$ factor). Hence $*:\mathrm{\Lambda}^{k}M\rightarrow\mathrm{\Lambda}^{d-k}M$
is a vector bundle isomorphism.
\item It is easy to show that $*1=\mathrm{d}\mu_{g}\left(p\right)$, where
$1$ is in $\mathrm{\Lambda}_{p}^{0}M=\mathbb{R}$.
\item We can use the wedge product and the Hodge dual to completely characterize
the inner product $\left\langle \cdot,\cdot\right\rangle _{g,k}$
induced by the metric $g$ on $\mathrm{\Lambda}^{k}M$ and we note
that the section $\left\langle \omega,\theta\right\rangle _{g,k}\in\mathrm{C}^{\infty}\left(M\right)$
coincides with the section $s*\left(\omega\wedge*\theta\right)$ for
each $\omega$, $\theta\in\mathrm{\Omega}^{k}M$.
\end{itemize}
As anticipated, we conclude this subsection with the introduction
of the codifferential.
\begin{defn}
\label{defCodifferential}\index{codifferential}\index{coclosed form}Let
$\left(M,\mathfrak{o}\right)$ be a $d$-dimensional oriented manifold
endowed with a metric $g$ with signature $s=\pm1$. For each $k\in\left\{ 1,\dots,d\right\} $
we call \textsl{codifferential} the map $\mathrm{\delta}:\mathrm{\Omega}^{k}M\rightarrow\mathrm{\Omega}^{k-1}M$
defined by 
\[
\mathrm{\delta}=\left(-1\right)^{k}*^{-1}\circ\mathrm{d}\circ*=s\left(-1\right)^{dk+d+1}*\circ\mathrm{d}\circ*\mbox{.}
\]

We say that a $k$-form $\Theta$ is \textsl{coclosed} when $\mathrm{\delta}\Theta=0$.
\end{defn}
Notice that, as a consequence of the property $\mathrm{d}^{2}=0$,
it follows also that $\mathrm{\delta}^{2}=0$. Moreover, if $f$ is
an orientation preserving embedding from the oriented manifold $\left(M,\mathfrak{o}\right)$
to the oriented manifold $\left(N,\mathfrak{p}\right)$ and if $M$
is endowed with a metric $g$ of signature $s$, while $N$ is endowed
with a metric $h$ with signature $s^{\prime}=s$ such that $g=f^{*}h$,
then it holds that
\[
f^{*}\circ\mathrm{\delta}_{N}=\mathrm{\delta}_{M}\circ f^{*}\mbox{.}
\]

\subsection{Integration on a manifold}

In the last subsection we discussed some questions about the calculus
of differential forms. In particular, considering a $d$-dimensional
manifold $M$, we used the space $\Omega^{d}M$ of $d$-forms over
$M$ to introduce the orientability of a manifold. This concept allows
us to define a notion of integral on a manifold. The precise procedure
to define the integral on a manifold is shown in detail, for example,
in \cite[Chap. VI]{Boo86}. Here we briefly present such construction
restricting to smooth functions.

\index{integrable form}\index{integral}Suppose that $M$ is an orientable
$d$-dimensional manifold and that we have chosen a nowhere null $d$-form
$\Theta$ that defines an orientation $\mathfrak{o}$ on $M$ so that
$\left(M,\mathfrak{o}\right)$ becomes an oriented manifold. It is
possible to express any other $d$-form $\Xi$ over $M$ as a product
$f\Theta$, where $f\in\mathrm{C}^{\infty}\left(M\right)$. We say
that a function of $\mathrm{C}^{\infty}\left(M\right)$ is integrable
if it has compact support, i.e. if it belongs to $\mathscr{D}\left(M\right)$,
and moreover a $d$-form over $M$ is said to be \textsl{integrable}
if it can be expressed as a product $f\Theta$, with an integrable
function $f$. This definition of integrable $d$-form does not depend
on the choice of the particular $d$-form $\Theta$ used to define
the orientation $\mathfrak{o}$ on $M$. Notice that in our simplified
treatment the set of integrable $d$-forms coincides exactly with
the space of $d$-forms with compact support, denoted by $\mathrm{\Omega}_{0}^{d}M$.

The \textsl{integral of an integrable $d$-form} is defined in first
place on a particular subset of $\mathrm{\Omega}_{0}^{d}M$ constituted
by those $d$-forms $\Xi$ whose support is contained in some coordinate
neighborhood $\left(U,V,\phi\right)$: using the local coordinates,
we write
\[
\phi_{*}\left.\Xi\right|_{U}=h\left(x\right)\mathrm{d}x^{1}\wedge\cdots\wedge\mathrm{d}x^{d}\quad\forall x\in V\mbox{,}
\]
where $h\in\mathscr{D}\left(V\right)$, and then we set
\[
\int\limits _{M}\Xi=\int\limits _{V}h\left(x\right)\mathrm{d}V\mbox{,}
\]
where $\mathrm{d}V$ is the standard measure on $\mathbb{R}^{d}$.
It can be shown that this definition is independent of the choice
of $\left(U,V,\phi\right)$ (provided that only coordinate neighborhoods
having transition charts with positive Jacobian determinant are considered).
In second place such definition is extended to any integrable $d$-form
$\Xi$ with the help of a particular partition of unity that reduces
$\Xi$ to a finite sum of $d$-forms of the type considered in first
place. Again it is possible to prove that this definition does not
depend on the particular choices made.

The next theorem recollects some properties of the integral.
\begin{thm}
Let $M$ be an orientable $d$-dimensional manifold. Let $\Theta$
be a $d$-form over $M$ defining an orientation $\mathfrak{o}$ on
$M$. The construction above defines the integral of integrable $d$-forms
over $M$, specifically the map $\Xi\in\mathrm{\Omega}_{0}^{d}M\mapsto\int_{M}\Xi\in\mathbb{R}$.
Such map fulfils the following properties:
\begin{itemize}
\item $\mathbb{R}$-linearity: for each $a$, $b\in\mathbb{R}$ and each
$\Xi$, $\Xi^{\prime}\in\mathrm{\Omega}_{0}^{d}M$ it holds that
\[
\int\limits _{M}\left(a\Xi+b\Xi^{\prime}\right)=a\int\limits _{M}\Xi+b\int\limits _{M}\Xi^{\prime}\mbox{;}
\]

\item if $\Xi\in\mathrm{\Omega}_{0}^{d}M$ can be expressed as $h\Theta$
with some non negative real valued smooth function $h$, we have $\int_{M}\Xi\geq0$
and $\int_{M}\Xi=0$ if and only if $h=0$;
\item if $f:M\rightarrow N$ is an orientation preserving embedding between
the oriented $d$-dimensional manifolds $\left(M,\mathfrak{o}\right)$
and $\left(N,\mathfrak{p}\right)$, the following equation holds for
each $\Xi\in\mathrm{\Omega}_{0}^{d}N$:
\[
\int\limits _{M}f^{*}\Xi=\int\limits _{N}\Xi\mbox{.}
\]

\end{itemize}
\end{thm}
Till now we considered only the integration on an orientable $d$-dimensional
manifold $M$ of $d$-forms. However we would like to integrate also
functions of $\mathscr{D}\left(M\right)$ as in the case of ordinary
integrals on Euclidean spaces. In the general case this cannot be
done because a measure on an arbitrary orientable manifold is missing.
As a matter of fact, once that an orientation $\mathfrak{o}$ on $M$
is chosen, each $\Omega\in\mathfrak{o}$ provides a possible measure
on $M$ and it is not possible for us to make a particular choice
that reduces to the standard measure $\mathrm{d}V$ when $M=\mathbb{R}^{d}$.
Nevertheless, when $M$ is endowed with a metric $g$ and an orientation
$\mathfrak{o}$ has been chosen, we are able to pick out the volume
form $\mathrm{d}\mu_{g}\in\mathfrak{o}$ exploiting Theorem \ref{thmVolumeForm}.
Using $\mathrm{d}\mu_{g}$ we are able to evaluate in an unambiguous
way the integrals of functions in $\mathscr{D}\left(M\right)$. Moreover
it can be shown that, when $M$ is an open subset of the vector space
$\mathbb{R}^{d}$ endowed with the usual inner product of Euclidean
spaces as metric, $\mathrm{d}\mu_{g}$ reduces to the ordinary measure
$\mathrm{d}V$.

In the development of the thesis we will make extensive use of Stokes'
theorem on manifolds. Before we are ready to present its statement,
we must introduce a slight extension of the notion of manifold. This
extension requires the introduction of the half plane, i.e. the following
subset of $\mathbb{R}^{d}$:
\[
H^{d}=\left\{ x=\left(x_{1},\dots x_{d}\right)\in\mathbb{R}^{d}:x_{d}\geq0\right\} \mbox{.}
\]
We also take the chance to define the boundary of $H^{d}$ as
\[
\partial H^{d}=\left\{ x=\left(x_{1},\dots x_{d}\right)\in\mathbb{R}^{d}:x_{d}=0\right\} \mbox{.}
\]

\begin{defn}
\index{manifold (with boundary)}A \textsl{$d$-dimensional manifold
(with boundary)} $M$ is a connected Hausdorff topological space with
a countable basis of open subsets such that for each point $p\in M$
there exists a triple $\left(U,\Omega,\phi\right)$, called \textsl{coordinate
neighborhood} (or \textsl{local chart}), where $U$ is an open neighborhood
of $p$ in $M$, $\Omega$ is an open subset of $H^{d}$ and $\phi:U\rightarrow\Omega$
is a homeomorphism. Moreover there are other two requirements:
\begin{itemize}
\item there exists a \textsl{(smooth) atlas}, which is a collection $\left\{ \left(U_{\alpha},\Omega_{\alpha},\phi_{\alpha}\right)\right\} _{\alpha\in I}$
of coordinate neighborhoods in $M$, where $I$ is an index set, such
that $\left\{ U_{\alpha}\right\} _{\alpha\in I}$ is an open covering
of $M$ and the map, called \textsl{transition chart}, 
\begin{eqnarray*}
T_{\phi_{\alpha}}^{\phi_{\beta}}:\Omega_{\alpha}\cap\Omega_{\beta} & \rightarrow & \Omega_{\alpha}\cap\Omega_{\beta}\\
x & \mapsto & \left(\phi_{\beta}\circ\phi_{\alpha}^{-1}\right)\left(x\right)
\end{eqnarray*}
is a diffeomorphism for each $\alpha$, $\beta\in I$ such that $U_{\alpha}\cap U_{\beta}\neq\emptyset$;
\item there exists a \textsl{maximal atlas}, i.e. an atlas that contains
each coordinate neighborhood $\left(U,\Omega,\phi\right)$ such that
the transition maps $T_{\phi_{\alpha}}^{\phi}$ and $T_{\phi}^{\phi_{\alpha}}$
are diffeomorphisms for each $\alpha\in I$ with $U_{\alpha}\cap U\neq\emptyset$.
\end{itemize}
\end{defn}
For a detailed discussion about manifolds with boundary the reader
is referred to \cite[Chap. VI, Sect. 4]{Boo86}

\index{boundary}For a $d$-dimensional manifold with boundary $M$,
it makes sense to define a subset $\partial M$, called \textsl{boundary
of $M$}. Such subset consists of the points of $M$ that are preimages
of points of $\partial H^{d}$ through some coordinate neighborhood.
It turns out that $\partial M$ is a $\left(d-1\right)$-dimensional
manifold with topology and differentiable structure induced by those
of $M$ and that the inclusion map $\iota:\partial M\rightarrow M$
is an embedding. Notice that $M\setminus\partial M$ is a manifold
in the ordinary sense and that $M$ itself is actually a manifold
in the ordinary sense if $\partial M$ is empty.

Everything we said till this point about manifolds can be extended
to manifolds with boundary in an almost straightforward way. The only
situation in which it is possible to face some troubles is the definition
of the tangent space at a point of the boundary. A possible approach
to such problem is presented in \cite[p. 254]{Boo86}. Once that such
problem is overcome we can indeed define differential forms on these
new type of manifolds.

Suppose we are dealing with an oriented manifold with boundary and
we want to make an integral on its boundary submanifold. In order
to give sense to integrals on the boundary we need a notion of orientability
of the boundary and the choice of a specific orientation. The following
theorems answers to our question.
\begin{thm}
Let $\left(M,\mathfrak{o}\right)$ be an oriented manifold with non
empty boundary $\partial M$. Then $\partial M$ is itself an orientable
manifold and the orientation $\mathfrak{o}$ of $M$ determines uniquely
an orientation $\mathfrak{o}^{\prime}$ on $\partial M$.
\end{thm}
\index{inward pointing vector}\index{outward pointing vector}\index{tangent to the boundary vector}Consider
an oriented $d$-dimensional manifold $\left(M,\mathfrak{o}\right)$,
a point $p\in\partial M$ and a coordinate neighborhood $\left(U,\Omega,\phi\right)$
of $p$. We have that $\phi\left(p\right)\in\partial H^{d}$ and that
each $v\in\mathrm{T}_{p}M$ can be classified as \textsl{inward pointing},
\textsl{outward pointing} or \textsl{tangent to $\partial M$} if
its last component in the basis induced by the chosen coordinate neighborhood
is respectively positive, negative or null. Such classification turns
out to be independent of the particular coordinate neighborhood and
of the orientation of $M$. The orientation $\mathfrak{o}^{\prime}$
provided by the last theorem can be characterized in the following
way: if $p$ is a point of the boundary $\partial M$ and $v\in\mathrm{T}_{p}M$
is outward pointing, a base $\left\{ v_{1},\dots,v_{d-1}\right\} $
of $\mathrm{T}_{p}\partial M$ is oriented if and only if $\left\{ v_{1},\dots,v_{d-1},v\right\} $
is an oriented base of $\mathrm{T}_{p}M$.

If $\left(M,\mathfrak{o}\right)$ is an oriented $d$-dimensional
manifold with non empty boundary $\partial M$ and $M$ is endowed
with a metric $g$, we immediately have an orientation $\mathfrak{o}^{\prime}$
on $\partial M$ provided by the last theorem and a metric $g^{\prime}$
obtained via pull-back of $g$ through the inclusion map $\iota$
of $\partial M$ into $M$ (remember that $\iota_{\partial M}^{M}$
is actually an embedding). Then we have a volume form on the boundary
$\partial M$, that we denote by $\mathrm{d}S_{g}$, that provides
a precise notion of integral on the boundary.

We are now able to state Stokes' theorem on an arbitrary manifold
with boundary. A thorough discussion about this topic can be found,
for example, in \cite[Chap. VI, Sect. 5]{Boo86}.
\begin{thm}
Let $\left(M,\mathfrak{o}\right)$ be an oriented $d$-dimensional
manifold with (eventually empty) boundary $\partial M$, let $\mathfrak{o}^{\prime}$
denote the orientation of $\partial M$ determined by $\mathfrak{o}$
and let $\iota:\partial M\rightarrow M$ be the inclusion map (actually
an embedding). For each $\Xi\in\mathrm{\Omega}_{0}^{d-1}M$ we have
that
\[
\int\limits _{M}\mathrm{d}\Xi=\int\limits _{\partial M}\iota^{*}\Xi\mbox{.}
\]

\end{thm}
Notice that, if $\partial M$ is empty, then the RHS is always null.
For us this will always be the case since we will always consider
manifolds as defined in Definition \ref{defManifold}, which is to
say manifolds with empty boundary.

With the help of the Hodge dual, defined in the previous subsection,
we can introduce an inner product between $k$-forms with compact
support on an oriented manifold endowed with a metric. With such notion
we can prove that the codifferential $\mathrm{\delta}$ is formally
adjoint to the exterior derivative $\mathrm{d}$. 
\begin{prop}
\label{propInnerProductOnkForms}Let $\left(M,\mathfrak{o}\right)$
be an oriented $d$-dimensional manifold endowed with a metric $g$
and let $*$ be the Hodge dual. For each $k\in\left\{ 1,\dots,d\right\} $
consider the set
\[
S^{k}=\left\{ \left(\Xi,\Xi^{\prime}\right)\in\mathrm{\Omega}^{k}M\times\mathrm{\Omega}^{k}M:\,\mathrm{supp}\left(\omega\right)\cap\mathrm{supp}\left(\theta\right)\mbox{ is a compact subset of }M\right\} \mbox{.}
\]
We have that the map
\begin{alignat*}{2}
\left(\cdot,\cdot\right)_{g,k} & : & S^{k} & \rightarrow\mathbb{R}\\
 &  & \left(\Xi,\Xi^{\prime}\right) & \mapsto\int\limits _{M}\left(\Xi\wedge*\Xi^{\prime}\right)
\end{alignat*}
defines a non degenerate inner product on the vector space $\mathrm{\Omega}_{0}^{k}M$.
\end{prop}
The integrand $\Xi\wedge*\Xi^{\prime}$ in the definition above may
be rewritten as
\[
\Xi\wedge*\Xi^{\prime}=*^{-1}*\left(\Xi\wedge*\Xi^{\prime}\right)=\left\langle \Xi,\Xi^{\prime}\right\rangle _{g,k}\mathrm{d}\mu_{g}\mbox{.}
\]
Then we can express $\left(\Xi,\Xi^{\prime}\right)_{g,k}$ for $\Xi$,
$\Xi^{\prime}\in\mathrm{\Omega}^{k}M$ as the integral of the section
$\left\langle \Xi,\Xi^{\prime}\right\rangle _{g,k}\in\mathrm{\Omega}^{0}M$:
\[
\int\limits _{M}\left(\Xi\wedge*\Xi^{\prime}\right)=\int\limits _{M}\left\langle \Xi,\Xi^{\prime}\right\rangle _{g,k}\mathrm{d}\mu_{g}\mbox{.}
\]

Notice that, if $g$ is a Riemannian metric, then $\left(\cdot,\cdot\right)_{g,k}$
is even a scalar product and so $\left(\mathrm{\Omega}_{0}^{k}M,\left(\cdot,\cdot\right)_{g}\right)$
is a pre-Hilbert space.

As announced $\left(\cdot,\cdot\right)_{g,k}$ allows us to establish
a particular relation between the exterior derivative and the codifferential.
Such a relation is a direct consequence of Stokes' theorem.
\begin{prop}
\label{propCodifferentialIsFormallyAdjointToExteriorDerivative}Let
$\left(M,\mathfrak{o}\right)$ be an oriented $d$-dimensional manifold
with empty boundary endowed with a metric $g$ with signature $s=\pm1$.
Then the codifferential $\mathrm{\delta}$ is formally adjoint to
the exterior derivative $\mathrm{d}$, i.e. for each $k\in\left\{ 1,\dots,d\right\} $,
each $\Xi\in\mathrm{\Omega}_{0}^{k-1}M$ and each $\Xi^{\prime}\in\mathrm{\Omega}_{0}^{k}M$
the following equation holds:
\[
\left(\Xi,\mathrm{\delta}\Xi^{\prime}\right)_{g,k-1}=\left(\mathrm{d}\Xi,\Xi^{\prime}\right)_{g,k}\mbox{.}
\]
\end{prop}
\begin{proof}
Fix $k\in\left\{ 1,\dots,d\right\} $, $\Xi\in\mathrm{\Omega}_{0}^{k-1}M$
and $\Xi^{\prime}\in\mathrm{\Omega}_{0}^{k}M$. Since $M$ has empty
boundary, Stokes' theorem implies that
\[
\int\limits _{M}\mathrm{d}\left(\Xi\wedge*\Xi^{\prime}\right)=0\mbox{.}
\]
On the other hand we have:
\[
\mathrm{d}\left(\Xi\wedge*\Xi^{\prime}\right)=\mathrm{d}\Xi\wedge*\Xi^{\prime}+\left(-1\right)^{k-1}\Xi\wedge\mathrm{d}*\Xi^{\prime}\mbox{.}
\]
Recalling the definition of the codifferential (cfr. Definition \ref{defCodifferential}),
we see that
\[
\mathrm{d}*\Xi^{\prime}=**^{-1}\mathrm{d}*\Xi^{\prime}=\left(-1\right)^{k}*\mathrm{\delta}\Xi^{\prime}\mbox{.}
\]
Hence we deduce that
\[
\mathrm{d}\left(\Xi\wedge*\Xi^{\prime}\right)=\mathrm{d}\Xi\wedge*\Xi^{\prime}-\Xi\wedge*\mathrm{\delta}\Xi^{\prime}\mbox{,}
\]
from which it follows
\[
0=\int\limits _{M}\mathrm{d}\left(\Xi\wedge*\Xi^{\prime}\right)=\int\limits _{M}\left(\mathrm{d}\Xi\wedge*\Xi^{\prime}\right)-\int\limits _{M}\left(\Xi\wedge*\mathrm{\delta}\Xi^{\prime}\right)\mbox{.}
\]
Then the definition of $\left(\cdot,\cdot\right)_{g,k}$ allows us
to conclude the proof.
\end{proof}

\section{\label{secLorentzianGeometry}Lorentzian geometry}

This section is devoted to the presentation of some notions concerning
Lorentzian geometry and in particular global hyperbolicity. The interested
reader should refer to \cite{O'N83} for a deeper insight in this
subject.

\subsection{Lorentzian manifolds}
\begin{defn}
\index{Lorentzian manifold}We call \textsl{Lorentzian manifold} a
pair $\left(M,g\right)$ where $M$ is an orientable $d$-dimensional
manifold and $g$ is a Lorentzian metric on $M$.
\end{defn}
Notice that we have included the requirement of orientability in the
definition of Lorentzian manifold. This is indeed not necessary if
one wants to study Lorentzian manifolds in general, however in the
development of this thesis we will often make use of Stokes' theorem,
which requires the orientability of the manifold to hold.

We take the chance to introduce some notions which prove to be very
helpful in the discussion of the causal structure of a Lorentzian
manifold.
\begin{defn}
Consider a Lorentzian manifold $\left(M,g\right)$. For each point
$p\in M$ and each tangent vector $v\in\mathrm{T}_{p}M$ we say that
$v$ is
\begin{itemize}
\item \index{timelike vector field}\textsl{$g$-timelike} if $g_{p}\left(v,v\right)<0$,
\item \index{lightlike vector field}\textsl{$g$-lightlike} if $g_{p}\left(v,v\right)=0$,
\item \index{causal vector field}\textsl{$g$-causal} if $g_{p}\left(v,v\right)\leq0$,
\item \index{spacelike vector field}\textsl{$g$-spacelike} if $g_{p}\left(v,v\right)>0$.
\end{itemize}
\end{defn}
Note that, if there is no risk of misunderstanding (e.g. when we consider
only one metric on a specified manifold), we often do not make explicit
the metric so that, for example, we simply speak of timelike tangent
vectors, instead of $g$-timelike tangent vectors. Anyway in this
section such omission is not adopted in order to underline the dependence
on the metric of the objects that we define.

With the definitions given above we can define a new property of Lorentzian
manifolds, called time orientability. This concept is associated to
the idea of finding some {}``preferred direction'' on our manifold
that can be interpreted as a direction of {}``time progress'' in
accordance with the given metric.
\begin{defn}
\index{time orientable Lorentzian manifold}\index{time orientation}We
say that a Lorentzian manifold $\left(M,g\right)$ is \textsl{time
orientable} if there exists a vector field $T\in\mathrm{C}^{\infty}\left(M,\mathrm{T}M\right)$
over $M$ such that $T\left(p\right)$ is $g$-timelike for each $p\in M$.
We call \textsl{time orientation} of the time orientable Lorentzian
manifold $\left(M,g\right)$ each one of the connected components
of the set of everywhere $g$-timelike vector fields over $M$.

\index{oriented and time oriented Lorentzian
manifold@oriented and time oriented Lorentzian\\
manifold}Then we call \textsl{oriented and time oriented Lorentzian manifold}
a quadruple $\mathscr{M}=\left(M,g,\mathfrak{o},\mathfrak{t}\right)$
where
\begin{itemize}
\item $\left(M,g\right)$ is a time orientable Lorentzian manifold,
\item $\mathfrak{o}$ is an orientation on $M$,
\item $\mathfrak{t}$ is a time orientation on $\left(M,g\right)$.
\end{itemize}
\end{defn}
With the last two definition we are able to introduce a classification
of the curves in an oriented and time oriented Lorentzian manifold
$\mathscr{M}$.
\begin{defn}
\index{vector tangent to a curve}\index{timelike curve}\index{lightlike curve}\index{causal curve}\index{spacelike curve}Consider
a Lorentzian manifold $\left(M,g\right)$ and a $\mathrm{C}^{1}$
curve $\gamma:I\rightarrow M$, where $I\subseteq\mathbb{R}$ is an
interval.

We define the \textsl{vector $\dot{\gamma}\left(p\right)$ tangent
to the curve $\gamma$ in the point $p$ along the curve} in the following
way: If $t\in I$ such that $\gamma\left(t\right)=p$, we consider
the curve $\gamma_{t}\left(s\right)=\gamma\left(s+t\right)$ defined
for $s$ in a sufficiently small interval containing 0 and we set
$\dot{\gamma}\left(p\right)=\left[\gamma_{t}\left(s\right)\right]$
(for the meaning of $\left[\cdot\right]$ see Definition \ref{defTangentSpace}).

We say that $\gamma$ is \textsl{$g$-timelike}, \textsl{$g$-lightlike},
\textsl{$g$-causal} or \textsl{$g$-spacelike} if $\dot{\gamma}\left(p\right)$
is such for each $p$ along $\gamma$.

\index{future directed causal curve}\index{past directed causal curve}If
$\left(M,g\right)$ is time orientable, $\mathfrak{o}$ is an orientation
of $M$ and $\mathfrak{t}$ is a time orientation of $\left(M,g\right)$
(so that $\mathscr{M}=\left(M,g,\mathfrak{o},\mathfrak{t}\right)$
is an oriented and time oriented Lorentzian manifold) and if the curve
$\gamma$ is $g$-causal we say that it is:
\begin{itemize}
\item \textsl{$\mathfrak{t}$-future directed} if $g_{p}\left(\dot{\gamma}\left(p\right),\mathfrak{t}\left(p\right)\right)<0$
for each $p$ along $\gamma$,
\item \textsl{$\mathfrak{t}$-past directed} if $g_{p}\left(\dot{\gamma}\left(p\right),\mathfrak{t}\left(p\right)\right)>0$
for each $p$ along $\gamma$.
\end{itemize}
This definition extends to piecewise $\mathrm{C}^{1}$ curves considering
separately each $\mathrm{C}^{1}$ piece.
\end{defn}
It may happen that we omit the explicit indication of the metric and
the time orientation chosen on the manifold when we deal with curves.
Clearly such omission will be done only if there is no possibility
of misunderstanding. For example, when we deal with an oriented and
time oriented Lorentzian manifold $\mathscr{M}=\left(M,g,\mathfrak{o},\mathfrak{t}\right)$
and there is no other oriented and time oriented Lorentzian manifold
with the same underlying manifold $M$ but with different metric or
time orientation, you may find the expression {}``future directed
timelike curve'', instead of {}``$\mathfrak{t}$-future directed
$g$-timelike curve''.

Now we define some particular subsets of $M$. These subsets, as we
will see, are very helpful in the characterization of the causal structure
of $\mathscr{M}=\left(M,g,\mathfrak{o},\mathfrak{t}\right)$.
\begin{defn}
\index{chronological future}\index{causal future}\index{chronological past}\index{causal past}\index{Cauchy development}Consider
an oriented and time oriented Lorentzian manifold $\mathscr{M}=\left(M,g,\mathfrak{o},\mathfrak{t}\right)$,
a subset $S\subseteq M$ and a point $p\in S$. We define:
\begin{itemize}
\item the \textsl{$\mathscr{M}$-chronological future of the point $p$
in $S$}, denoted by $I_{+}^{\mathscr{M},S}\left(p\right)$, as the
subset of $S$ constituted by the points $q\in S\setminus\left\{ p\right\} $
such that there exists a $\mathfrak{t}$-future directed $g$-timelike
curve starting from $p$ and ending in $q$ which is entirely contained
in $S$;
\item the \textsl{$\mathscr{M}$-causal future of the point $p$ in $S$},
denoted by $J_{+}^{\mathscr{M},S}\left(p\right)$, as the subset of
$S$ constituted by $p$ and the points $q\in S$ such that there
exists a $\mathfrak{t}$-future directed $g$-causal curve starting
from $p$ and ending in $q$ which is entirely contained in $S$.
\end{itemize}
We also define the \textsl{$\mathscr{M}$-chronological past of the
point $p$ in $S$}, denoted by $I_{-}^{\mathscr{M},S}\left(p\right)$,
and the \textsl{$\mathscr{M}$-causal past of the point $p$ in $S$},
denoted by $J_{-}^{\mathscr{M},S}\left(p\right)$, with the substitution
of the word {}``future'' with the word {}``past'' in the definitions
of $I_{+}^{\mathscr{M},S}\left(p\right)$ and $J_{+}^{\mathscr{M},S}\left(p\right)$.

We extend the definitions of these subsets from arbitrary points $p\in M$
to arbitrary subsets $\Omega\subseteq S$ taking the union over the
points in $\Omega$, e.g. we define the \textsl{$\mathscr{M}$-chronological
future of the subset $\Omega$ in $S$} as $I_{+}^{\mathscr{M},S}\left(\Omega\right)=\bigcup_{p\in\Omega}I_{+}^{\mathscr{M},S}\left(p\right)$,
and we denote the unions $I_{+}^{\mathscr{M},S}\left(\Omega\right)\cup I_{-}^{\mathscr{M},S}\left(p\right)$
and $J_{+}^{\mathscr{M},S}\left(p\right)\cup J_{-}^{\mathscr{M},S}\left(p\right)$
with $I^{\mathscr{M},S}\left(p\right)$ and respectively with $J^{\mathscr{M},S}\left(p\right)$.

Finally we define the \textsl{Cauchy development of $S$ in $\mathscr{M}$}
as the subset $D^{\mathscr{M}}\left(S\right)$ comprised by the points
$q\in M$ such that every inextensible $\mathfrak{t}$-future directed
(or equivalently $\mathfrak{t}$-past directed) $g$-causal curve
in $M$ passing through $q$ meets $S$.
\end{defn}
We invite the reader to bear in mind that, when there is no risk of
ambiguity, we may write $I_{+}^{S}\left(p\right)$ in place of $I_{+}^{\mathscr{M},S}\left(p\right)$.
Moreover in our notation we always omit the subset $S$ in when $S$
is the entire manifold so that we write $I_{+}^{\mathscr{M}}\left(p\right)$
in place of $I_{+}^{\mathscr{M},M}\left(p\right)$.

The notions of causal future and causal past allow us to define future
compact and past compact subsets.
\begin{defn}
\index{past compactness}\index{future compactness}Let $\mathscr{M}=\left(M,g,\mathfrak{o},\mathfrak{t}\right)$
be an oriented and time oriented Lorentzian manifold and let $S$
be a subset of $M$. We say that $S$ is
\begin{itemize}
\item \textsl{$\mathscr{M}$-past compact} if $S\cap J_{-}^{\mathscr{M}}\left(p\right)$
is compact for each $p\in M$;
\item \textsl{$\mathscr{M}$-future compact} if $S\cap J_{+}^{\mathscr{M}}\left(p\right)$
is compact for each $p\in M$.
\end{itemize}
\end{defn}
Past compact and future compact subsets will play an important role
in the next section, when we will study the properties of Green operators.

When dealing with Lorentzian manifolds, we can establish the notion
of causal separation. We will use such notion when we will introduce
the generally covariant locality principle in Chapter \ref{chapGCLP}.
\begin{defn}
\index{causally separated subsets}Let $\left(M,g\right)$ be a Lorentzian
manifold. We say that two subsets $S_{1}$ and $S_{2}$ of $M$ are
\textsl{$\left(M,g\right)$-causally separated} (or simply \textsl{causally
separated}, when there is no risk of misunderstanding) if there is
no $g$-causal curve on $M$ that connects a point of $S_{1}$ and
a point of $S_{2}$.
\end{defn}
Sometimes we may say that $S_{1}$ is $\left(M,g\right)$-causally
separated from $S_{2}$, meaning that there is no point of $S_{1}$
that can be connected through some causal curve to a point of $S_{2}$.
Obviously this is equivalent to saying that $S_{1}$ and $S_{2}$
are $\left(M,g\right)$-causally separated.
\begin{rem}
\label{remCausalSeparation}We can give a condition that is equivalent
to causal separation on an oriented and time oriented Lorentzian manifold
$\mathscr{M}$. We can show that $S_{1}$ and $S_{2}$ are $\mathscr{M}$-causally
separated if and only if $J^{\mathscr{M}}\left(S_{1}\right)\cap S_{2}=\emptyset$
(or equivalently $J^{\mathscr{M}}\left(S_{2}\right)\cap S_{1}=\emptyset$).
This follows from the fact that the points of $J^{\mathscr{M}}\left(S_{1}\right)$
are by definition connected to points of $S_{1}$ through some $g$-causal
curve in $M$. Hence the intersection $J^{\mathscr{M}}\left(S_{1}\right)\cap S_{2}$
consists exactly of those points of $S_{2}$ that are connected to
points of $S_{1}$ through some $g$-causal curve in $M$. Then $J^{\mathscr{M}}\left(S_{1}\right)\cap S_{2}=\emptyset$
means that there are not points of $S_{2}$ that are connected to
points of $S_{1}$ through some $g$-causal curve in $M$, i.e. $S_{2}$
is $\mathscr{M}$-causally separated from $S_{1}$ or, equivalently,
$S_{1}$ and $S_{2}$ are $\mathscr{M}$-causally separated.
\end{rem}
Now we introduce the notion of causal compatibility and the notion
of causal convexity. Loosely speaking causal compatibility means that
the causal future (or past) of a point in a subset $S$ of an oriented
and time oriented Lorentzian manifold coincides with the intersection
with $S$ of the causal future (or past) of such point taken in the
whole manifold. Instead causal convexity requires that each pair of
points in a subset can be connected by a causal curve contained in
such subset.
\begin{defn}
\index{causal compatibility}\index{causal convexity}Let $\mathscr{M}=\left(M,g,\mathfrak{o},\mathfrak{t}\right)$
be an oriented and time oriented Lorentzian manifold and let $S$
be a subset of $M$. We say that $S$ is
\begin{itemize}
\item \textsl{$\mathscr{M}$-causally compatible} if $J_{\pm}^{\mathscr{M},S}\left(p\right)=J_{\pm}^{\mathscr{M}}\left(p\right)\cap S$
for each $p\in S$;
\item \textsl{$\mathscr{M}$-causally convex} if each $\mathfrak{t}$-future
(or equivalently $\mathfrak{t}$-past) directed $g$-causal curve
in $M$ that starts and ends in $S$ is entirely contained in $S$.
\end{itemize}
\end{defn}
We observe that, since each $\mathfrak{t}$-future/past directed $g$-causal
curve that is contained in $S\subseteq M$ can be directly seen also
as a $\mathfrak{t}$-future/past directed $g$-causal curve contained
in $M$, it always holds the inclusion $J_{\pm}^{\mathscr{M},S}\left(p\right)\subseteq J_{\pm}^{\mathscr{M}}\left(p\right)\cap S$
for each $p\in S$. Hence the real condition of causal compatibility
is the other inclusion.
\begin{rem}
\label{remCausalConvexityImpliesCausalCompatibility}It is easily
seen that causal convexity implies causal compatibility. Suppose that
$S$ is an $\mathscr{M}$-causally convex subset of $M$. Once that
a point $p\in S$ is fixed, we can consider a point $q$ in $J_{+}^{\mathscr{M}}\left(p\right)\cap S$
(or in $J_{-}^{\mathscr{M}}\left(p\right)\cap S$). Because of causal
convexity each $\mathfrak{t}$-future (or respectively $\mathfrak{t}$-past
directed) $g$-causal curve in $M$ from $p$ to $q$ is entirely
contained in $S$. From the definition of causal future (respectively
causal past) at least one such curve exists and this implies that
$q$ falls in $J_{\pm}^{\mathscr{M},S}\left(p\right)$ as required
by causal compatibility.
\end{rem}

\begin{rem}
\label{remRestrictionOfOrientedAndTimeOrientedLorentzianManifolds}Each
causally compatible connected open subset of an oriented and time
oriented Lorentzian manifold can be interpreted as an oriented and
time oriented Lorentzian manifold in its own right. For example take
the $d$-dimensional oriented and time oriented Lorentzian manifold
$\mathscr{M}=\left(M,g,\mathfrak{o},\mathfrak{t}\right)$ and let
$\Omega$ be a causally compatible connected open subset of $M$.
Then undoubtedly $\Omega$ can be seen as a $d$-dimensional submanifold
of $M$ (cfr. Remark \ref{remSubmanifold}) and hence a $d$-dimensional
manifold in its own right. Moreover it becomes an oriented and time
oriented Lorentzian manifold when endowed with $\left.g\right|_{\Omega}$,
$\left.\mathfrak{o}\right|_{\Omega}$ and $\left.\mathfrak{t}\right|_{\Omega}$,
where for $\left.\mathfrak{o}\right|_{\Omega}$ we mean the class
of nowhere null $d$-forms over $\Omega$ that includes the restrictions
to $\Omega$ of the nowhere null $d$-forms over $M$ contained in
$\mathfrak{o}$. We denote the oriented and time oriented Lorentzian
manifold $\left(\Omega,\left.g\right|_{\Omega},\left.\mathfrak{o}\right|_{\Omega},\left.\mathfrak{t}\right|_{\Omega}\right)$
with $\left.\mathscr{M}\right|_{\Omega}$.

Notice that $J_{\pm}^{\left.\mathscr{M}\right|_{\Omega}}\left(p\right)=J_{\pm}^{\mathscr{M},\Omega}\left(p\right)$
for each $p\in\Omega$ as a direct consequence of the definition of
causal future/past.

Since causal convexity implies causal compatibility, the same conclusions
hold also for causally convex subsets of oriented and time oriented
Lorentzian manifolds.
\end{rem}

\subsection{Globally hyperbolic spacetimes}

Now we present the notion of global hyperbolicity. Such concept is
the key hypothesis for a theorem that states existence and uniqueness
of global solutions for a wave equation with proper initial data on
an oriented and time oriented Lorentzian manifold. Hence global hyperbolicity
will be an unavoidable request throughout the rest of the thesis.
\begin{defn}
\index{globally hyperbolic subset}Let $\mathscr{M}=\left(M,g,\mathfrak{o},\mathfrak{t}\right)$
be an oriented and time oriented Lorentzi\-an manifold. A subset
$S\subseteq M$ is said to be \textsl{$\mathscr{M}$-globally hyperbolic}
if the following conditions hold:
\begin{itemize}
\item \index{causality condition}$S$ fulfils the \textsl{$g$-causality
condition}, i.e. there are no closed $g$-causal curves in $S$;
\item $J_{+}^{\mathscr{M},S}\left(p\right)\cap J_{-}^{\mathscr{M},S}\left(q\right)$
is a compact subset of $S$ for each $p$, $q\in S$ with respect
to the topology naturally induced on $S$ by the topology of $M$.
\end{itemize}
\index{globally hyperbolic spacetime}We call \textsl{globally hyperbolic
spacetime} each oriented and time oriented Lorentzian manifold $\mathscr{M}=\left(M,g,\mathfrak{o},\mathfrak{t}\right)$
such that $M$ is a $\mathscr{M}$-globally hyperbolic subset.
\end{defn}
\index{strong causality condition}Originally global hyperbolicity
of (oriented and) time oriented Lorentzian manifolds required a stricter
condition then that of causality, which is called \textsl{strong causality
condition}. Such condition requires that there are no {}``almost
closed'' $g$-causal curves. A precise statement of the strong causality
condition on an (oriented and) time oriented Lorentzian manifold $\mathscr{M}=\left(M,g,\mathfrak{o},\mathfrak{t}\right)$
is the following: For each point $p\in M$ and for each open neighborhood
$U$ of $p$ in $M$ there exists an open neighborhood $V\subseteq U$
of $p$ in $M$ such that each $\mathfrak{t}$-future directed (or
equivalently $\mathfrak{t}$-past directed) $g$-causal curve which
starts and ends in $V$ must be entirely contained in $U$. However
this stricter requirement is equivalent to the causality condition
in the present context as was shown by Bernal and Sanchez in \cite{BS07}.
\begin{rem}
\label{remRestrictionToGloballyHyperbolicSubsets}Globally hyperbolic
connected open subsets of oriented and time oriented Lorentzian manifolds
can be considered as globally hyperbolic spacetimes in their own right.
To see this, consider an oriented and time oriented Lorentzian manifold
$\mathscr{M}=\left(M,g,\mathfrak{o},\mathfrak{t}\right)$ and a $\mathscr{M}$-globally
hyperbolic open subset $\Omega$ of $M$. From Remark \ref{remRestrictionOfOrientedAndTimeOrientedLorentzianManifolds}
we have that $\left.\mathscr{M}\right|_{\Omega}$ is itself an oriented
and time oriented Lorentzian manifold and that $J_{\pm}^{\left.\mathscr{M}\right|_{\Omega}}\left(p\right)=J_{\pm}^{\mathscr{M},\Omega}\left(p\right)$
for each $p\in\Omega$. Since per hypothesis $\Omega$ is $\mathscr{M}$-globally
hyperbolic, it follows also that $\Omega$ is $\left.\mathscr{M}\right|_{\Omega}$-globally
hyperbolic too. Therefore $\left.\mathscr{M}\right|_{\Omega}$ is
itself a globally hyperbolic spacetime.\end{rem}
\begin{defn}
\index{achronality}\index{acausality}Let $\mathscr{M}=\left(M,g,\mathfrak{o},\mathfrak{t}\right)$
be an oriented and time oriented Lorentzi\-an manifold and let $S$
be a subset of $M$. $S$ is \textsl{achronal in $\mathscr{M}$} if
it is met at most once by each $\mathfrak{t}$-future directed (or
equivalently $\mathfrak{t}$-past directed) $g$-timelike curve in
$M$. $S$ is \textsl{acausal in $\mathscr{M}$} if it is met at most
once by each $\mathfrak{t}$-future directed (or equivalently $\mathfrak{t}$-past
directed) $g$-causal curve in $M$.

\index{Cauchy surface}We say that $\Sigma$ is a \textsl{Cauchy surface}
\textsl{of} $\mathscr{M}$ if each inextensible $\mathfrak{t}$-future
directed (or equivalently $\mathfrak{t}$-past directed) $g$-timelike
curve in $M$ passing through $p$ meets $\Sigma$ exactly once.
\end{defn}
Obviously each acausal subset is also achronal and each Cauchy surface
is achron\-al. It can be proved that a Cauchy surface $\Sigma$ is
a closed achronal topological hypersurface met by each inextensible
causal curve at least once \cite[Chap. 14, Lem. 29, p. 415]{O'N83},
hence its Cauchy development $D^{\mathscr{M}}\left(\Sigma\right)$
coincides with $M$.

With the last definition we have at our disposal all the material
needed to state a very important theorem that provides two handy conditions
that are equivalent to global hyperbolicity.
\begin{thm}
\label{thmGlobalHyperbolicity}Let $\mathscr{M}=\left(M,g,\mathfrak{o},\mathfrak{t}\right)$
be an oriented and time oriented Lorentzian manifold. Then the following
conditions are equivalent:
\begin{itemize}
\item $\mathscr{M}$ is globally hyperbolic;
\item there exists a Cauchy surface $\Sigma$ of $\mathscr{M}$;
\item there exists a diffeomorphism from the manifold $M$ to the manifold
$\mathbb{R}\times\Sigma$, where $\Sigma$ is a $\left(d-1\right)$-dimensional
manifold, such that the push-forward of the metric $g$ through such
diffeomorphism takes the form $-\beta\mathrm{d}t^{2}+g_{t}$, where
$\beta$ is a smooth strictly positive function of $t\in\mathbb{R}$,
$g_{t}$ is a Riemannian metric on $\left\{ t\right\} \times\Sigma$
for each $t\in\mathbb{R}$ and the family of Riemannian metrics $\left\{ g_{t},t\in\mathbb{R}\right\} $
varies smoothly with $t$. Moreover we have that for each $t\in\mathbb{R}$
$\left\{ t\right\} \times\Sigma$ is the image through the diffeomorphism
of a smooth spacelike Cauchy surface of $\mathscr{M}$.
\end{itemize}
\end{thm}
We do not include the proof of this theorem here, however we give
some references. That the second condition implies the first is proved
in \cite[Chap. 14, Cor. 39, p.422]{O'N83}. Moreover in \cite{BS05}
Bernal and Sanchez showed that the third condition follows from the
first one. With these facts the proof is completed since the implication
from the third condition to the second one is trivial.

The next proposition shows that causal convexity entails global hyperbolicity
for open subsets of an arbitrary globally hyperbolic spacetime.
\begin{prop}
\label{propCausalConvexityImpliesGlobalHyperbolicity}Let $\mathscr{M}=\left(M,g,\mathfrak{o},\mathfrak{t}\right)$
be a globally hyperbolic spacetime and let $\Omega$ be a subset of
$M$. Then if $\Omega$ is $\mathscr{M}$-causally convex, it is also
$\mathscr{M}$-globally hyperbolic.\end{prop}
\begin{proof}
We suppose that $\Omega$ is $\mathscr{M}$-causally convex and we
try to show that $\Omega$ is also $\mathscr{M}$-globally hyperbolic.
Since $\mathscr{M}$ is a globally hyperbolic spacetime, the $g$-causality
condition is fulfilled by the entire set underlying $\mathscr{M}$,
hence it is fulfilled also by $\Omega$. Then we must only check the
other condition for global hyperbolicity. To this end we fix $p$,
$q\in\Omega$. Since causal convexity implies causal compatibility
(cfr. Remark \ref{remCausalConvexityImpliesCausalCompatibility}),
we have that $J_{\pm}^{\mathscr{M},\Omega}\left(r\right)=J_{\pm}^{\mathscr{M}}\left(r\right)\cap\Omega$
for each $r\in\Omega$. It follows that
\[
J_{+}^{\mathscr{M},\Omega}\left(p\right)\cap J_{-}^{\mathscr{M},\Omega}\left(q\right)=J_{+}^{\mathscr{M}}\left(p\right)\cap J_{-}^{\mathscr{M}}\left(q\right)\cap\Omega\mbox{.}
\]
Since $\mathscr{M}$ is globally hyperbolic, we deduce that $J_{+}^{\mathscr{M}}\left(p\right)\cap J_{-}^{\mathscr{M}}\left(q\right)$
is compact with respect to the topology of $M$. If we can show that
it is also contained in $\Omega$, then it is compact also with respect
to the topology induced on $\Omega$ by the topology of $M$ and we
also have
\[
J_{+}^{\mathscr{M}}\left(p\right)\cap J_{-}^{\mathscr{M}}\left(q\right)\cap\Omega=J_{+}^{\mathscr{M}}\left(p\right)\cap J_{-}^{\mathscr{M}}\left(q\right)\mbox{.}
\]
This would complete the proof. Consider then an arbitrary point $r$
in $J_{+}^{\mathscr{M}}\left(p\right)\cap J_{-}^{\mathscr{M}}\left(q\right)$.
Recalling the definitions of causal future and causal past, we find
a $\mathfrak{t}$-future directed $g$-causal curve $\gamma_{1}$
in $M$ from $p$ to $r$ and a $\mathfrak{t}$-past directed $g$-causal
curve $\gamma_{2}$ in $M$ from $q$ to $r$. Reversing $\gamma_{2}$
and pasting the result with $\gamma_{1}$, we obtain a $\mathfrak{t}$-future
directed $g$-causal curve $\gamma$ in $M$ from $p$ to $q$. Since
$p$ and $q$ are points of $\Omega$ and $\Omega$ is causally convex,
we deduce that $\gamma$ is entirely contained in $\Omega$. Since
$r$ is in the image of $\gamma$, it turns out that $r\in\Omega$
and so the inclusion $J_{+}^{\mathscr{M}}\left(p\right)\cap J_{-}^{\mathscr{M}}\left(q\right)\subseteq\Omega$
actually holds.\end{proof}
\begin{rem}
\label{remUsefulSubsetsOfGloballyHyperbolicSpacetimes}Once that a
globally hyperbolic spacetime $\mathscr{M}=\left(M,g,\mathfrak{o},\mathfrak{t}\right)$
is provided, we can build a wide class of $\mathscr{M}$-causally
convex connected open subsets of $M$ that include a Cauchy surface
of $\mathscr{M}$. Applying Theorem \ref{thmGlobalHyperbolicity}
to $\mathscr{M}$, we obtain a diffeomorphism $f$ that factorizes
$M$ into $\mathbb{R}\times\Sigma$ such that $f^{-1}\left(\left\{ t\right\} \times\mathbb{R}\right)$
is a smooth spacelike Cauchy surface of $\mathscr{M}$ for each $t\in\mathbb{R}$.
Then we can consider $\left(-\varepsilon,\varepsilon\right)\times\Sigma$
for an arbitrary $\varepsilon>0$ and define $\Omega_{\varepsilon}=f^{-1}\left(\left(-\varepsilon,\varepsilon\right)\times\Sigma\right)$.
We immediately deduce that $\Omega_{\varepsilon}$ is a connected
open subset of $M$ that includes $f^{-1}\left(\left\{ t\right\} \times\Sigma\right)$
for each $t\in\left(-\varepsilon,\varepsilon\right)$, which are all
smooth spacelike Cauchy surfaces for $\mathscr{M}$. It remains only
to check that $\Omega_{\varepsilon}$ is $\mathscr{M}$-causally convex.
Consider a $\mathfrak{t}$-future directed $g$-causal curve $\gamma:\left[a,b\right]\rightarrow M$
which starts and ends in $\Omega_{\varepsilon}$. Using the factorization
of $\mathscr{M}$ in $\mathbb{R}\times\Sigma$ and noting that the
projection $\pi_{1}:\mathbb{R}\times\Sigma\rightarrow\mathbb{R}$
on the first argument of the Cartesian product $\mathbb{R}\times\Sigma$
is continuous, we deduce that $\pi_{1}\circ f\circ\gamma:\left[a,b\right]\rightarrow\mathbb{R}$
is continuous. If, by contradiction, along $\gamma$ there is a point
$r$ that is outside $\Omega_{\varepsilon}$, then we find $c\in\left(a,b\right)$
such that one of the following inequalities holds:
\begin{alignat*}{4}
\left(\pi_{1}\circ f\circ\gamma\right)\left(c\right) & > & \varepsilon & > & \left(\pi_{1}\circ f\circ\gamma\right)\left(b\right) & > & \left(\pi_{1}\circ f\circ\gamma\right)\left(a\right)\mbox{;}\\
\left(\pi_{1}\circ f\circ\gamma\right)\left(c\right) & < & -\varepsilon & < & \left(\pi_{1}\circ f\circ\gamma\right)\left(a\right) & < & \left(\pi_{1}\circ f\circ\gamma\right)\left(b\right)\mbox{.}
\end{alignat*}
Consider for example the first case (the other one is similar). As
a consequence of the intermediate value theorem, we find $d\in\left[a,c\right]$
such that
\[
\left(\pi_{1}\circ f\circ\gamma\right)\left(d\right)=\left(\pi_{1}\circ f\circ\gamma\right)\left(b\right)\mbox{,}
\]
which is to say that $\gamma$ meets twice the smooth spacelike Cauchy
surface for $\mathscr{M}$ of the form $\Sigma^{\prime}=f^{-1}\left(\left\{ \left(\pi_{1}\circ f\circ\gamma\right)\left(b\right)\right\} \times\Sigma\right)$.
Exploiting \cite[Chap. 14, Lem. 42, p. 425]{O'N83}, we find that
$\Sigma^{\prime}$ is acausal because it is a spacelike Cauchy surface.
Then we have found a contradiction, hence $\gamma$ is contained in
$\Omega_{\varepsilon}$ and so $\Omega_{\varepsilon}$ is actually
$\mathscr{M}$-causally convex.
\end{rem}
Indeed there are more powerful constructions that allow us to obtain
subsets with good topological and causal properties starting from
a globally hyperbolic spacetime. The next proposition is devoted to
the recollection of some results that go in this direction.
\begin{prop}
\label{propUsefulSubsetsOfGloballyHyperbolicSpacetimes}Let $\mathscr{M}=\left(M,g,\mathfrak{o},\mathfrak{t}\right)$
be a globally hyperbolic spacetime.
\begin{itemize}
\item If $\Sigma$ is a Cauchy surface for $\mathscr{M}$ and $K$ is a
compact subset of $M$, then both $\Sigma\cap J_{\pm}^{\mathscr{M}}\left(K\right)$
and $J_{\pm}^{\mathscr{M}}\left(\Sigma\right)\cap J_{\mp}^{\mathscr{M}}\left(K\right)$
are compact subsets of $M$.
\item If $K$ and $K^{\prime}$ are compact subsets of $M$, then $J_{\pm}^{\mathscr{M}}\left(K\right)\cap J_{\mp}^{\mathscr{M}}\left(K^{\prime}\right)$
is a compact subset of $M$ too.
\item If $A$ and $B$ are two non empty subsets of $M$, then $\Omega=I_{+}^{\mathscr{M}}\left(A\right)\cap I_{-}^{\mathscr{M}}\left(B\right)$
is a $\mathscr{M}$-causally convex open subset of $M$. Furthermore
if $A$ and $B$ are relatively compact in $M$, $\Omega$ is relatively
compact in $M$ too.
\item If $K$ is a compact subset of $M$, then there exists a $\mathscr{M}$-causally
convex relatively compact connected open subset $\Omega$ of $M$
including $K$.
\end{itemize}
\end{prop}
\begin{proof}
The proof of the first three points can be found in \cite[Cor. A.5.4, p. 175]{BGP07},
\cite[Lem. A.5.7, p. 176]{BGP07} and \cite[Lem. A.5.12, p. 178]{BGP07}.
However for the third point the thesis of Bär, Ginoux and Pfäffle
is that $\Omega$ is $\mathscr{M}$-globally hyperbolic and $\mathscr{M}$-causally
compatible in place of $\mathscr{M}$-causally convex. Anyway we can
directly check that $\Omega$ is $\mathscr{M}$-causally convex in
the following manner. Suppose that $\gamma$ is a $\mathfrak{t}$-future
directed $g$-causal curve in $M$ starting from $p\in\Omega$ and
ending in $q\in\Omega$. Then each point $r$ along $\gamma$ is contained
in $J_{+}^{\mathscr{M}}\left(p\right)\cap J_{-}^{\mathscr{M}}\left(q\right)$.
Since $p\in I_{+}^{\mathscr{M}}\left(A\right)$, we find $p^{\prime}\in A$
such that $p\in I_{+}^{\mathscr{M}}\left(p^{\prime}\right)$ and,
since $q\in I_{-}^{\mathscr{M}}\left(B\right)$, we find $q^{\prime}\in B$
such that $q\in I_{-}^{\mathscr{M}}\left(q^{\prime}\right)$. From
\cite[Chap. 14, Cor. 1, p. 402]{O'N83} we deduce that the following
implication holds: if $t\in I_{\pm}^{\mathscr{M}}\left(s\right)$
and $u\in J_{\pm}^{\mathscr{M}}\left(t\right)$, then $u\in I_{\pm}^{\mathscr{M}}\left(s\right)$.
Hence $J_{+}^{\mathscr{M}}\left(p\right)\subseteq I_{+}^{\mathscr{M}}\left(p^{\prime}\right)$
and $J_{-}^{\mathscr{M}}\left(q\right)\subseteq I_{-}^{\mathscr{M}}\left(q^{\prime}\right)$
so that
\[
J_{+}^{\mathscr{M}}\left(p\right)\cap J_{-}^{\mathscr{M}}\left(q\right)\subseteq I_{+}^{\mathscr{M}}\left(p^{\prime}\right)\cap I_{-}^{\mathscr{M}}\left(q^{\prime}\right)\subseteq I_{+}^{\mathscr{M}}\left(A\right)\cap I_{-}^{\mathscr{M}}\left(B\right)=\Omega\mbox{.}
\]
This inclusion implies that $\gamma$ is completely included in $\Omega$.

The proof of the fourth point is obtained modifying in a proper way
\cite[Lem. A.5.13, p. 178]{BGP07}. As a first step we apply Theorem
\ref{thmGlobalHyperbolicity} to $\mathscr{M}$ and we find a diffeomorphism
$f$ that factorizes $M$ in $\mathbb{R}\times\Sigma$ so that $\Sigma_{t}=f^{-1}\left(\left\{ t\right\} \times\Sigma\right)$
is a smooth spacelike Cauchy surface for $\mathscr{M}$ for each $t\in\mathbb{R}$.
The projection $\pi_{1}:\mathbb{R}\times\Sigma\rightarrow\mathbb{R}$
on the first factor of the Cartesian product is a continuous map so
that $\pi_{1}\circ f:M\rightarrow\mathbb{R}$ is continuous. Then
the image of the compact subset $K$ of $M$ through $\pi_{1}\circ f$
is compact in $\mathbb{R}$, so that we easily find $t_{-}$ and $t_{+}$
in $\mathbb{R}$ such that $t_{-}<t<t_{+}$ for each $t\in\left(\pi_{1}\circ f\right)\left(K\right)$.
Now we take $C=J_{+}^{\mathscr{M}}\left(K\right)\cap\Sigma_{t_{+}}$
and, applying the first point, we conclude that it is a compact subset
of $M$. Hence we can easily find a relatively compact connected open
subset $A$ of $M$ including $C$. Since $\Sigma_{t_{+}}$ is a smooth
spacelike Cauchy surface for $\mathscr{M}$, it is easy to check that
$K\subseteq J_{-}^{\mathscr{M}}\left(C\right)$. But $C$ is closed
since the topology of $M$ is Hausdorff, while $A$ is open by construction
and $C\subseteq A$. Then it follows that $J_{-}^{\mathscr{M}}\left(C\right)\subseteq I_{-}^{\mathscr{M}}\left(A\right)$:
A point $p$ in $J_{-}^{\mathscr{M}}\left(C\right)$ is connected
to a point $q$ of $C$ via a $\mathfrak{t}$-past directed $g$-causal
curve $\gamma$ in $M$ starting at $q$ and ending at $p$; we can
find a neighborhood $O$ of $p$ included in $A$ so that we can deform
$\gamma$ in a way that it becomes timelike in $O$; hence we obtain
a new $\mathfrak{t}$-past directed $g$-causal curve $\gamma^{\prime}$
that starts in a point $r$ of $O\subseteq A$ and ends in $p$ and
we notice that it cannot be a null curve, i.e. causal, but nowhere
timelike, so that we can make a fixed endpoint deformation of $\gamma^{\prime}$
(cfr. \cite[Chap. 10, Prop. 46, p. 294]{O'N83}) to obtain a $\mathfrak{t}$-past
directed $g$-timelike curve $\gamma^{\prime\prime}$ that starts
in $r\in A$ and ends in $p$. Returning to our main proof, we conclude
that $K\subseteq I_{-}^{\mathscr{M}}\left(A\right)$. We immediately
deduce also that $I_{-}^{\mathscr{M}}\left(K\right)\subseteq I_{-}^{\mathscr{M}}\left(A\right)$.
Take now $D=J_{-}^{\mathscr{M}}\left(\overline{A}\right)\cap\Sigma_{t_{-}}$
and applying again the first point, we find that it is a compact subset
of $M$. Hence we can easily find a relatively compact open subset
$B$ of $M$ including $D$. Keeping in mind that
\[
D=J_{-}^{\mathscr{M}}\left(\overline{A}\right)\cap\Sigma_{t_{-}}\supseteq I_{-}^{\mathscr{M}}\left(A\right)\cap\Sigma_{t_{-}}\supseteq I_{-}^{\mathscr{M}}\left(K\right)\cap\Sigma_{t_{-}}
\]
and that $\Sigma_{t_{-}}$ is a Cauchy surface for $\mathscr{M}$,
we easily check that $\overline{A}\subseteq J_{+}^{\mathscr{M}}\left(D\right)$
and that $K\subseteq I_{+}^{\mathscr{M}}\left(D\right)\subseteq I_{+}^{\mathscr{M}}\left(B\right)$.
With a procedure similar to that applied above, we conclude that $\overline{A}\subseteq I_{+}^{\mathscr{M}}\left(B\right)$,
hence in particular $A\subseteq I_{+}^{\mathscr{M}}\left(B\right)$.
We can apply the second point to the relatively compact open subsets
$A$ and $B$ of $M$ and conclude that $\Omega=I_{-}^{\mathscr{M}}\left(A\right)\cap I_{+}^{\mathscr{M}}\left(B\right)$
is a $\mathscr{M}$-causally convex relatively compact open subset
of $M$. Since by the way we noticed that $K$ is included in both
$I_{-}^{\mathscr{M}}\left(A\right)$ and $I_{+}^{\mathscr{M}}\left(B\right)$,
we deduce that $K\subseteq\Omega$. The proof is completed if we can
show that $\Omega$ is connected. To this end take two arbitrary points
$p$ and $q$ in $\Omega$. Then we can find a $\mathfrak{t}$-future
directed $g$-causal curve $\gamma_{1}$ in $M$ that goes from $p$
to some point $r$ in $A$ and a $\mathfrak{t}$-past directed $g$-causal
curve $\gamma_{3}$ in $M$ that goes from some point $s$ in $A$
to $q$. Since $A$ is open, $A\subseteq I_{-}^{\mathscr{M}}\left(A\right)$.
This fact, together with the inclusion $A\subseteq I_{-}^{\mathscr{M}}\left(B\right)$
shown above, implies that $A\subseteq\Omega$. Hence both $\gamma_{1}$
and $\gamma_{3}$ start and end in $\Omega$. By construction $\Omega$
is causally convex and so $\gamma_{1}$ and $\gamma_{3}$ are completely
included in $\Omega$. In our construction we choose $A$ to be connected,
hence we can find a curve $\gamma_{2}$ from $r$ to $s$ which is
included in $A$ (and therefore in $\Omega$ too). Pasting $\gamma_{1}$,
$\gamma_{2}$and $\gamma_{3}$, we obtain a curve that goes from $p$
to $q$ and the proof of the fourth point is complete.
\end{proof}

\section{\label{secWaveEquations}Wave equations}

In this section we face the problem of the existence and uniqueness
of global solutions to a given wave equation on a globally hyperbolic
spacetime $\mathscr{M}=\left(M,g\right)$ with compactly supported
smooth initial data on a Cauchy surface $\Sigma$ of $\mathscr{M}$.
The discussion here involves smooth sections in an arbitrary $\mathbb{R}$-vector
bundle $E$ over $M$. The results that we recall without proof can
be found in \cite[Chap. 3]{BGP07}. For a complete discussion on the
existence and uniqueness of (local) solutions to wave equations on
time oriented Lorentzian manifolds the reader is referred to \cite{BGP07}.

\subsection{Linear differential operators}

Since we are going to speak of wave equations in vector bundles over
manifolds, we must previously introduce some notions about linear
differential operator that will allow us to recognize which differential
equations are wave equations in which are not.
\begin{defn}
\label{defLinearDifferentialOperator}\index{linear differential operator of order at
most k@linear differential operator of order at\\
most $k$}Let $M$ be a $d$-dimensional manifold and let $E$ and $F$ be two
vector bundles over $M$ respectively of rank $n$ and $m$. A \textsl{linear
differential operator} $L$ \textsl{of order at most} $k$ from $E$
to $F$ is a $\mathbb{R}$-linear map
\[
L:\mathrm{C}^{\infty}\left(M,E\right)\rightarrow\mathrm{C}^{\infty}\left(M,F\right)
\]
that can be locally written in the following way: For each $p\in M$
there exists an open coordinate neighborhood $\left(U,\Omega,\phi\right)$
of $p$ in $M$ on which both $E$ and $F$ are locally trivialized
by the maps $\Phi:\pi_{E}^{-1}\left(U\right)\rightarrow U\times\mathbb{R}^{n}$
and $\Psi:\pi_{F}^{-1}\left(U\right)\rightarrow U\times\mathbb{R}^{m}$
and there exists a family of local sections $\left\{ A_{\alpha}\in\mathrm{C}^{\infty}\left(\Omega,\Omega\times\mathrm{Hom}\left(\mathbb{R}^{n},\mathbb{R}^{m}\right)\right)\right\} _{\left|\alpha\right|\leq k}$
such that on $\Omega$ we can write
\begin{equation}
\Psi\circ\left.Lu\right|_{U}\circ\phi^{-1}=\sum_{\left|\alpha\right|\leq k}A_{\alpha}D_{\alpha}\left(\Phi\circ\left.u\right|_{U}\circ\phi^{-1}\right)\label{eqLinearDifferentialOperator}
\end{equation}
for each section $u\in\mathrm{C}^{\infty}\left(M,E\right)$, where
$\alpha\in\mathbb{N}^{d}$ is a multi-index, $\left|\alpha\right|=\sum_{i=1}^{d}\alpha_{i}$,
$x^{1},\dots,x^{d}$ are the local coordinates on $U$ and 
\[
D_{\alpha}=\frac{\partial^{\left|\alpha\right|}}{\partial^{\alpha_{1}}x^{1}\cdots\partial^{\alpha_{d}}x^{d}}\mbox{.}
\]

\index{linear differential operator of order $k$}A \textsl{linear
differential operator} $L$ \textsl{of order} $k$ is a linear differential
operator of order at most $k$, but not of order at most $k-1$.
\end{defn}
At this point our aim is to identify a specific class of linear differential
operators of order 2, but to do this we need to introduce another
tool.
\begin{defn}
\index{principal symbol}Let $M$ be a $d$-dimensional manifold and
let $E$ and $F$ be two vector bundles over $M$ respectively of
rank $n$ and $m$. Consider a linear differential operator $L$ of
order $k$ from $E$ to $F$. We say that the \textsl{principal symbol}
$\sigma_{L}$ of the linear differential operator $L$ is the map
\[
\sigma_{L}\in\mathrm{T}^{*}M\rightarrow\mathrm{Hom}\left(E,F\right)
\]
locally defined in a way that is based on Definition \ref{defLinearDifferentialOperator}:
For each $p\in M$ there exists a coordinate neighborhood $\left(U,\Omega,\phi\right)$
of $p$ in $M$ on which both $E$ and $F$ are locally trivialized
by the maps $\Phi$ and $\Psi$ and there exists a family of local
smooth sections $\left\{ A_{\alpha}\right\} _{\left|\alpha\right|\leq k}$
such that on $\Omega$ eq. \eqref{eqLinearDifferentialOperator} holds
for each section $u\in\mathrm{C}^{\infty}\left(M,E\right)$; hence
for each $q\in U$, each $\omega\in\mathrm{T}_{q}^{*}M$ and each
$\mu\in E_{q}$ we set
\[
\Psi_{q}\left(\left(\sigma_{L}\left(\omega\right)\right)\mu\right)=\sum_{\left|\alpha\right|=k}\omega_{1}^{\alpha_{1}}\cdots\omega_{d}^{\alpha_{d}}\left(A_{\alpha}\left(\phi^{-1}\left(q\right)\right)\right)\left(\Phi_{q}\mu\right)\mbox{,}
\]
where $\left\{ \omega_{1},\dots,\omega_{d}\right\} $ are the components
of $\omega=\omega_{i}\mathrm{d}x^{i}$ in the basis $\left\{ \mathrm{d}x^{1},\dots,\mathrm{d}x^{d}\right\} $
of $\mathrm{T}_{p}^{*}M=\mathrm{T}_{p}^{*}U$ obtained via pull back
through $\phi$ from the orthonormal basis $\left\{ e_{1},\dots,e_{d}\right\} $
of $\mathbb{R}^{d}=\mathrm{T}_{\phi\left(p\right)}^{*}\Omega$.\end{defn}
\begin{example}
\label{exaLinearDifferentialOperator}The formulation of the last
definitions may appear very abstract (at least this was the impression
of the author when he saw them for the first time), but they are much
more concrete and close to the usual idea of partial derivative than
it seems. However to realize this fact we must restrict ourselves
to a more customary situation. Consider for example $M=\mathbb{R}^{d}$
and $E=M\times\mathbb{R}^{n}$ and $F=M\times\mathbb{R}^{m}$. In
this case there are a global coordinate neighborhood for $M$ and
global trivializations for $E$ and $F$, while $\mathrm{T}M$ reduces
to $M\times\mathbb{R}^{d}$ so that we can identify it with $\mathrm{T}^{*}M$.
Moreover a section $u$ in $E$ is nothing but an $\mathbb{R}^{n}$-valued
smooth function defined on $M=\mathbb{R}^{d}$. In this situation
one recognizes that partial derivatives of order at most $k$ and
their linear combinations with $\mathrm{Hom}\left(\mathbb{R}^{n},\mathbb{R}^{m}\right)$-valued
smooth functions defined on $M$ as coefficients are undoubtedly linear
differential operators from $E$ to $F$ of order at most $k$. If
there is a partial derivative of order $k$ with non null coefficient,
the operator is exactly of order $k$. The local sections $\left\{ A_{\alpha}\right\} $
in this case are actually global and coincide with the $\mathrm{Hom}\left(\mathbb{R}^{n},\mathbb{R}^{m}\right)$-valued
smooth functions defined on $M$ that we used as coefficients. The
principal symbol is simply a function from $\mathrm{T}^{*}M=M\times\mathbb{R}^{d}$
to $\mathrm{Hom}\left(E,F\right)=M\times\mathrm{Hom}\left(\mathbb{R}^{n},\mathbb{R}^{m}\right)$
that maps each $\left(x,\omega\right)=\left(\left(x^{1},\dots,x^{d}\right),\left(\omega_{1},\dots,\omega_{d}\right)\right)\in\mathrm{T}^{*}M$
to a linear combination of the coefficients $A_{\alpha}$ corresponding
to the derivatives of highest order weighted with products of the
components of $\omega$ with powers that equal the order of the partial
derivatives along each direction. For example,
\begin{multline*}
\left(\begin{array}{ccc}
\mathrm{e}^{x^{1}x^{4}} & 0 & \cos x^{4}\\
4 & \tanh x^{3} & 7
\end{array}\right)\frac{\partial^{4}}{\partial x^{1}\partial\left(x^{3}\right)^{3}}+\left(\begin{array}{ccc}
\left(x^{2}\right)^{2} & 1 & 0\\
3 & x^{3} & -1
\end{array}\right)\frac{\partial^{4}}{\partial\left(x^{2}\right)^{2}\partial\left(x^{4}\right)^{2}}\\
+\left(\begin{array}{ccc}
5 & \sinh\left(x^{2}x^{3}\right) & 0\\
0 & x^{1}+x^{3} & \frac{x^{4}}{2}
\end{array}\right)\frac{\partial^{3}}{\partial x^{1}\partial x^{2}\partial x^{3}}+\left(\begin{array}{ccc}
x^{2} & 1 & \sin x^{1}\\
3 & 0 & x^{4}-x^{2}
\end{array}\right)\frac{\partial^{2}}{\partial\left(x^{4}\right)^{2}}
\end{multline*}
is a linear differential operator from $E=M\times\mathbb{R}^{2}$
to $F=M\times\mathbb{R}^{3}$ of order 4, where $M=\mathbb{R}^{4}$,
whose principal symbol is the map
\[
\sigma_{L}:\mathrm{T}^{*}M=M\times\mathbb{R}^{4}\rightarrow\mathrm{Hom}\left(E,F\right)=M\times\mathrm{Hom}\left(\mathbb{R}^{2},\mathbb{R}^{3}\right)
\]
defined by
\[
\sigma_{L}\left(x,\omega\right)=\omega_{1}\omega_{3}^{3}\left(\begin{array}{ccc}
\mathrm{e}^{x^{1}x^{4}} & 0 & \cos x^{4}\\
4 & \tanh x^{3} & 7
\end{array}\right)+\omega_{2}^{2}\omega_{4}^{2}\left(\begin{array}{ccc}
\left(x^{2}\right)^{2} & 1 & 0\\
3 & x^{3} & -1
\end{array}\right)\mbox{.}
\]

\end{example}
We conclude this subsection with the notion of formal selfadjointness.
\begin{defn}
\label{defFormally(Anti)SelfadjointLinearOperator}\index{formally selfadjoint linear operator}\index{formally antiselfadjoint linear operator}Let
$\left(M,\mathfrak{o}\right)$ be a $d$-dimensional oriented manifold
endowed with a metric $g$ and let $E$ be a vector bundle over $M$
of rank $n$ endowed with an inner product that we denote with $\overset{E}{\cdot}$.
A linear operator $L:\mathrm{C}^{\infty}\left(M,E\right)\rightarrow\mathrm{C}^{\infty}\left(M,E\right)$
is said to be \textsl{formally selfadjoint} if for each $u$, $v\in\mathscr{D}\left(M,E\right)$
we have
\[
\int\limits _{M}\left(Lu\right)\overset{E}{\cdot}v\mathrm{d}\mu_{g}=\int\limits _{M}u\overset{E}{\cdot}\left(Lv\right)\mathrm{d}\mu_{g}\mbox{,}
\]
where $\mathrm{d}\mu_{g}$ is volume form over $\left(M,\mathfrak{o}\right)$
induced by $g$. If the equation above holds with a minus sign at
the RHS then $L$ is \textsl{formally antiselfadjoint}.
\end{defn}
Indeed we will apply the last definition to linear differential operators,
but more in general it can be applied to operators acting linearly
on smooth sections in a vector bundle.

\subsection{Normally hyperbolic equations and Cauchy problems}

We are ready to pick out a particular class of linear differential
operators of order 2 which are at the core of the theory of wave equations
on globally hyperbolic spacetimes. Probably the reader has some notion
of what it is generally meant as a wave equation. However, in the
present context, for wave equation we intend a class of linear differential
equations of second order that may be a little bit larger then what
it is usually intended. In order to avoid misunderstanding, we take
the chance to define our notion of wave equation (to be more precise,
of normally hyperbolic equation).
\begin{defn}
\index{normally hyperbolic operator}\index{metric type principal symbol}Let
$\left(M,g\right)$ be a Lorentzian manifold and let $E$ be a vector
bundle over $M$ of rank $n$. A \textsl{normally hyperbolic operator
$P$ on $E$ over $\left(M,g\right)$} is a linear differential operator
of order 2 from $E$ to $E$ whose principle symbol $\sigma_{P}$
is of \textsl{metric type}, i.e. for each $p\in M$ and each $\omega\in\mathrm{T}_{p}^{*}M$
\[
\sigma_{P}\left(p,\omega\right)=-g_{p}\left(\omega^{\sharp},\omega^{\sharp}\right)\mathrm{id}_{E_{p}}\mbox{,}
\]
where $^{\sharp}:\mathrm{T}^{*}M\rightarrow\mathrm{T}M$ is the raising
isomorphism induced by the metric $g$ (see Definition \ref{defMusicalIsomorphisms}).

\index{normally hyperbolic equation}\index{wave equation}A \textsl{normally
hyperbolic equation }(or \textsl{wave equation}) on a vector bundle
$E$ over a Lorentzian manifold $\left(M,g\right)$ is a linear differential
equation of the form
\[
Pu=v\mbox{,}
\]
where $P$ is a normally hyperbolic operator on $E$ over $\left(M,g\right)$
and $u$ is a smooth section in $E$ over $M$ to be determined, while
$v\in\mathrm{C}^{\infty}\left(M,E\right)$ is given.\end{defn}
\begin{example}
Consider the Minkowski spacetime, i.e. the manifold $M=\mathbb{R}^{4}$
endowed with a metric $g$ that is everywhere represented by the matrix
$\left(g_{ij}\right)=\mathrm{diag}\left(-1,+1,+1,+1\right)$, and
the vector bundle $E=M\times\mathbb{R}^{4}$. Recalling our Example
\ref{exaLinearDifferentialOperator}, we see that the Klein-Gordon
operator on Minkowski spacetime
\[
-g^{ij}\frac{\partial^{2}}{\partial x^{i}\partial x^{j}}+m^{2}\mathrm{id}_{\mathrm{C}^{\infty}\left(M,E\right)}:\mathrm{C}^{\infty}\left(M,E\right)\rightarrow\mathrm{C}^{\infty}\left(M,E\right)\mbox{,}
\]
where $\left(g_{\vphantom{ij}}^{ij}\right)$ is the inverse of the
matrix $\left(g_{ij}\right)$ and $m\geq0$ is a parameter (the mass
of the Klein-Gordon field), is a linear differential operator from
$E$ to $E$ of order 2. Its principal symbol is provided by the function
that maps each $\left(x,\omega\right)\in\mathrm{T}^{*}M$ to
\begin{alignat*}{1}
-\omega_{i}\omega_{j}g^{ij}\mathrm{id}_{E} & =-\left(\omega^{\sharp}\right)^{k}g_{ki}\left(\omega^{\sharp}\right)^{h}g_{hj}g^{ij}\mathrm{id}_{E}=-\left(\omega^{\sharp}\right)^{k}\left(\omega^{\sharp}\right)^{h}g_{kh}\mathrm{id}_{E}\\
 & =-g_{x}\left(\omega^{\sharp},\omega^{\sharp}\right)\mathrm{id}_{E}\mbox{,}
\end{alignat*}
hence we recognize that the Klein-Gordon operator in Minkowski spacetime
is a normally hyperbolic operator.
\end{example}
Maybe the most common prototype of wave equation is the d'Alembert
equation. The d'Alembert operator
\begin{eqnarray*}
\Box^{\nabla}:\mathrm{C}^{\infty}\left(M,E\right) & \rightarrow & \mathrm{C}^{\infty}\left(M,\mathrm{T}^{*}M\otimes\mathrm{T}^{*}M\otimes E\right)\\
u & \mapsto & \left(-\left(\mathrm{tr}_{\mathrm{T}^{*}M}\otimes\mathrm{id}_{E}\right)\circ\nabla\circ\nabla\right)u
\end{eqnarray*}
induced by a connection $\nabla$ on a vector bundle $E$ over a Lorentzian
manifold $\left(M,g\right)$ is indeed a normally hyperbolic operator
on $E$ over $\left(M,g\right)$ (for a proof of this fact refer to
\cite[Ex. 1.5.2, p. 34]{BGP07}) and hence the d'Alembert equation
is a normally hyperbolic equation on $E$ over $\left(M,g\right)$.

However there exist many other normally hyperbolic equations. For
example notice that each equation involving the d'Alembert operator
defined above together with other linear differential terms of order
at most 1 is still a normally hyperbolic equation.

\index{connection compatible with a normally hyperbolic operator}It
can be even shown that each normally hyperbolic operator $P$ on a
vector bundle $E$ over a Lorentzian manifold $\left(M,g\right)$
can be written as the sum of the d'Alembert operator $\Box^{\nabla}$
associated to some connection $\nabla$ on $E$ with a section $B$
in $\mathrm{End}\left(E,E\right)$ (cfr. \cite[Lem. 1.5.5, p. 35]{BGP07}).
In this case the connection $\nabla$ is called \textsl{$P$-compatible}.

These observations are made to underline that the typical wave equations
are indeed included in our class of normally hyperbolic equation,
but there are also other (although quite similar) partial differential
equations that fall in our class.

We have defined all the ingredients needed to state a theorem about
the existence and uniqueness of solutions for a non homogeneous normally
hyperbolic equation.
\begin{thm}
\index{Cauchy problem}\label{thmCauchyProblem}Consider a globally
hyperbolic spacetime $\mathscr{M}=\left(M,g,\mathfrak{o},\mathfrak{t}\right)$
and a vector bundle $E$ over $M$. Let $\Sigma$ be a spacelike smooth
Cauchy surface of $\mathscr{M}$, let $\mathfrak{n}\in\mathrm{C}^{\infty}\left(\Sigma,\mathrm{T}M\right)$
be a unit $\mathfrak{t}$-future directed $g$-timelike vector field
over $\Sigma$ normal to $\Sigma$ and let $P$ be a normally hyperbolic
operator on $\left(E,\mathscr{M}\right)$. Denote the $P$-compatible
connection with $\nabla$. Then for each $f\in\mathscr{D}\left(M,E\right)$
and each $u_{0}$, $u_{1}\in\mathscr{D}\left(\Sigma,\pi_{E}^{-1}\left(\Sigma\right)\right)$
there exists a unique solution $u\in\mathrm{C}^{\infty}\left(M,E\right)$
to the \textsl{Cauchy problem}
\[
\left\{ \begin{array}{rcl}
Pu & = & f\mbox{,}\\
\left.u\right|_{\Sigma} & = & u_{0}\mbox{,}\\
\left.\nabla_{\mathfrak{n}}u\right|_{\Sigma} & = & u_{1}\mbox{.}
\end{array}\right.
\]
Moreover we have $\mathrm{supp}\left(u\right)\subseteq J\left(K\right)$,
where $K=\mathrm{supp}\left(u_{0}\right)\cup\mathrm{supp}\left(u_{1}\right)\cup\mathrm{supp}\left(f\right)$.
\end{thm}
The proof of the last theorem is based on the determination of the
so called fundamental solutions. Even if we do not discuss here such
proof, it is useful for us to introduce fundamental solutions in view
of the construction of Green operators. We face these problems after
having introduced the necessary material, specifically distributions
on manifolds. To such topic we devote the next subsection.

\subsection{Distributions on manifolds}

To introduce the notion of fundamental solution we cannot restrict
to sections over vector bundles. We need to introduce {}``sections''
in a broader sense. Distributions on manifolds are the right tools
for our aims. Before we define such objects we need to provide a notion
of convergence in $\mathscr{D}\left(M,E\right).$ This requires some
preparation.

Let $M$ be a manifold with a Riemannian metric $g\in\mathrm{C}^{\infty}\left(M,\mathrm{T}^{*}M\otimes_{s}\mathrm{T}^{*}M\right)$
and let $E$ be a vector bundle over $M$ endowed with a connection
\[
\nabla:\mathrm{C}^{\infty}\left(M,E\right)\rightarrow\mathrm{C}^{\infty}\left(M,\mathrm{T}^{*}M\otimes E\right)
\]
and a positive definite inner product $h\in\mathrm{C}^{\infty}\left(M,E^{*}\otimes_{s}E^{*}\right)$.
Use again $\nabla$ to denote the Levi-Civita connection on $\mathrm{T}M$:
\[
\nabla:\mathrm{C}^{\infty}\left(M,\mathrm{T}M\right)\rightarrow\mathrm{C}^{\infty}\left(M,\mathrm{T}^{*}M\otimes\mathrm{T}M\right)\mbox{.}
\]
Notice that $\nabla$ and $g$ induce (via duality and tensor product)
a connection
\[
\nabla:\mathrm{C}^{\infty}\left(M,\mathrm{T}^{\left(i,j\right)}M\right)\rightarrow\mathrm{C}^{\infty}\left(M,\mathrm{T}^{*}M\otimes\mathrm{T}^{\left(i,j\right)}M\right)
\]
and respectively a positive definite inner product $g\in\mathrm{C}^{\infty}\left(M,\mathrm{T}^{\left(j,i\right)}M\otimes_{s}\mathrm{T}^{\left(j,i\right)}M\right)$
on each $\mathrm{T}^{\left(i,j\right)}M$. Putting together the connections
and the inner products on $\mathrm{T}^{\left(i,j\right)}M$ and $E$
we can obtain (via tensor product) a connection
\[
\nabla:\mathrm{C}^{\infty}\left(M,\mathrm{T}^{\left(i,j\right)}M\otimes E\right)\rightarrow\mathrm{C}^{\infty}\left(M,\mathrm{T}^{*}M\otimes\mathrm{T}^{\left(i,j\right)}M\otimes E\right)
\]
and respectively an inner product
\[
k\in\mathrm{C}^{\infty}\left(M,\left(\mathrm{T}^{\left(j,i\right)}M\otimes E^{*}\right)\otimes\left(\mathrm{T}^{\left(j,i\right)}M\otimes E^{*}\right)\right)
\]
on each vector bundle $\mathrm{T}^{\left(i,j\right)}M\otimes E$.
For each $p\in M$, $k$ induces a norm $\left|\cdot\right|_{p}$
on the fiber $E_{p}$ defined by
\[
\left|\mu\right|_{p}^{2}=\mu\cdot_{k,p}\mu
\]
for each $\mu\in\mathrm{T}_{p}^{\left(i,j\right)}M\otimes E_{p}$.
Then we can use the collection $\left\{ \left|\cdot\right|_{p}:\, p\in M\right\} $
of fiberwise norms and the connection in $\mathrm{T}^{\left(i,j\right)}M\otimes E$
to define a family of seminorms on the space $\mathrm{C}^{\infty}\left(M,E\right)$:
For each compact subset $K$ of $M$ we set
\[
\left|u\right|_{K}=\sup_{i\in\mathbb{N}}\left(\sup_{p\in K}\left|\nabla^{i}u\right|_{p}\right)\mbox{,}
\]
where $\nabla^{i}u$ means the application of the connection
\[
\nabla:\mathrm{C}^{\infty}\left(M,\mathrm{T}^{\left(0,i-1\right)}M\otimes E\right)\rightarrow\mathrm{C}^{\infty}\left(M,T^{\left(0,i\right)}M\otimes E\right)
\]
to the $\mathrm{C}^{\infty}$-section $\nabla^{i-1}u$.
\begin{defn}
Let $M$ be a manifold endowed with a metric $g$ and let $E$ be
a vector bundle over $M$ endowed with a connection $\nabla$ and
an inner product $h$. Endow $\mathrm{T}M$ with the Levi-Civita connection
still denoted by $\nabla$. Following the construction above we define
a notion of convergence in $\mathscr{D}\left(M,E\right)$: We say
that a sequence $\left\{ u_{i}\right\} \subseteq\mathscr{D}\left(M,E\right)$
converges to $u\in\mathscr{D}\left(M,E\right)$ if there exists a
compact subset $K$ of $M$ such that $\mathrm{supp}\left(u_{n}\right)$
and $\mathrm{supp}\left(u\right)$ are contained in $K$ for each
$i\in\mathbb{N}$ and the sequence $\left\{ \left|u_{i}-u\right|_{K}\right\} $
converges to zero.
\end{defn}
Notice that, since we always consider compact subsets, it can be proved
that different choices of inner products (provided that they are positive
definite) and connections yield equivalent seminorms, hence the notion
of convergence the we defined on $\mathscr{D}\left(M,E\right)$ does
not depend on the choices made in the preparatory construction.

Now we can speak of distributions on manifolds.
\begin{defn}
\textsl{\index{distribution on a manifold}}\index{space of distributions}Consider
a manifold $M$, a vector bundle $E$ over $M$ and a finite dimensional
$\mathbb{R}$-vector space $V$. A \textsl{$V$-valued distribution
in $E$} is a linear map $U:\mathscr{D}\left(M,E^{*}\right)\rightarrow V$
that is continuous with respect to the convergence in $\mathscr{D}\left(M,E^{*}\right)$.

$\mathscr{D}^{\prime}\left(M,E,V\right)$ denotes the vector \textsl{space
of $V$-valued distributions in $E$}.
\end{defn}
In the definition given above the choice of a norm on the vector space
$V$ is implied. However $\dim V<\infty$, hence all norms are equivalent
and hence the definition does not depend on the choice of the norm
on $V$.
\begin{rem}
\label{remLinearDifferentialOperatorInDistributionalSense}\index{linear differential operator in distributional sense@linear differential operator in distributio\-nal sense}\index{formal adjoint of a linear differential operator}Consider
an oriented manifold $\left(M,\mathfrak{o}\right)$ endowed with a
metric $g$, two vector bundles $E$ and $F$ over $M$ and a finite
dimensional $\mathbb{R}$-vector space $V$. There is a procedure
to extend any linear differential operator $L:\mathrm{C}^{\infty}\left(M,E\right)\rightarrow\mathrm{C}^{\infty}\left(M,F\right)$
to a \textsl{linear differential operator in distributional sense},
that is a linear map from $\mathscr{D}^{\prime}\left(M,E,V\right)$
to $\mathscr{D}^{\prime}\left(M,F,V\right)$ which we still denote
with $L$.

The first thing to be done is to define the \textsl{formal adjoint}
of $L$, denoted by $L^{*}$. Precisely, there exists a unique linear
differential operator $L^{*}:\mathrm{C}^{\infty}\left(M,F^{*}\right)\rightarrow\mathrm{C}^{\infty}\left(M,E^{*}\right)$
such that
\[
\int\limits _{M}\left(L^{*}u\right)\left(v\right)\mathrm{d}\mu_{g}=\int\limits _{M}u\left(Lv\right)\mathrm{d}\mu_{g}
\]
for each $u\in\mathscr{D}\left(M,F^{*}\right)$ and each $v\in\mathscr{D}\left(M,E\right)$,
where the dual pairing between the proper vector bundles is taken
into account and $\mathrm{d}\mu_{g}$ is the volume form induced by
$g$ on $\left(M,\mathfrak{o}\right)$. Notice that the canonical
identification $\left(E^{*}\right)^{*}=E$ implies $\left(L^{*}\right)^{*}=L$,
where for $\left(L^{*}\right)^{*}$ we mean the formal adjoint of
$L^{*}$ defined repeating the procedure just shown.

At this point we are ready to extend $L^{*}$ to a linear operator
from $\mathscr{D}^{\prime}\left(M,E,V\right)$ to $\mathscr{D}^{\prime}\left(M,F,V\right)$,
that we denote again with $L$. This is the linear differential operator
in distributional sense that extends the {}``original'' $L$. Such
extension is obtained imposing
\[
\left(LU\right)\left[v\right]=U\left[L^{*}v\right]
\]
for each $U\in\mathscr{D}^{\prime}\left(M,E,V\right)$ and each $v\in\mathscr{D}\left(M,F^{*}\right)$.

Note that, in the case $V=\mathbb{R}$, the {}``new'' $L$ acts
exactly as the {}``original'' $L$ on sections of $\mathrm{C}^{\infty}\left(M,E\right)$
(to be precise, we should say that, for each $u\in\mathrm{C}^{\infty}\left(M,E\right)$,
there exists a unique section $v\in\mathrm{C}^{\infty}\left(M,F\right)$
that generates the image through the {}``new'' $L$ of the distribution
generated by $u$ and that such $v$ coincides with the image through
the {}``original'' $L$ of $u$). This fact is a consequence of
the identity $\left(L^{*}\right)^{*}=L$ shown above.
\end{rem}
Before proceeding with the next subsection, we want to make some remarks
about formally selfadjoint linear differential operators and their
extensions in distributional sense.
\begin{rem}
\label{remFormalSelfadjointness}Assume that $E$ is a vector bundle
over an oriented manifold $\left(M,\mathfrak{o}\right)$ endowed with
a metric $g$ and consider an inner product on $E$. Suppose that
$L:\mathrm{C}^{\infty}\left(M,E\right)\rightarrow\mathrm{C}^{\infty}\left(M,E\right)$
is a formally selfadjoint linear differential operator. Considering
the musical isomorphisms defined using the inner product on $E$ (cfr.
Definition \ref{defMusicalIsomorphisms}), we realize that the condition
of formal selfadjointness (cfr. Definition \ref{defFormally(Anti)SelfadjointLinearOperator})
can be rewritten in the following form:
\[
\int\limits _{M}\left(Lu\right)^{\flat}\left(v\right)\mathrm{d}\mu_{g}=\int\limits _{M}u^{\flat}\left(Lv\right)\mathrm{d}\mu_{g}\quad\forall u,v\in\mathscr{D}\left(M,E\right)\mbox{,}
\]
where $\mathrm{d}\mu_{g}$ is the volume form induced by $g$ on $\left(M,\mathfrak{o}\right)$.
In the present situation the formal adjoint of $L$ is given by $\flat\circ L\circ\sharp$.
We can check this fact verifying that, because of the formal selfadjointness
of $L$, $\flat\circ L\circ\sharp$ satisfies the formula, given in
our last remark, that defines uniquely the formal adjoint of a linear
differential operator: for each $u\in\mathscr{D}\left(M,E^{*}\right)$
and each $v\in\mathscr{D}\left(M,E\right)$ we have
\[
\int\limits _{M}\left(\left(\flat\circ L\circ\sharp\right)u\right)\left(v\right)\mathrm{d}\mu=\int\limits _{M}\left(L\left(u^{\sharp}\right)\right)^{\flat}\left(v\right)\mathrm{d}\mu=\int\limits _{M}\left(u^{\sharp}\right)^{\flat}\left(Lv\right)\mathrm{d}\mu=\int\limits _{M}u\left(Lv\right)\mathrm{d}\mu\mbox{.}
\]
At this point we have $L^{*}=\flat\circ L\circ\sharp$. Now identify
$E$ and $E^{*}$ through the musical isomorphisms and we deduce $L^{*}=L$.
Then, after the identification of $E^{*}$ with $E$ as done before,
formal selfadjointness of $L$ means that the formal adjoint of $L$
coincides with $L$. This fact trivially leads also to the coincidence
of the extensions of $L$ and $L^{*}$ as linear differential operators
in distributional sense.
\end{rem}

\subsection{\textmd{\normalsize \label{subGreenOperators}}Fundamental solutions
and Green operators}

Once that we are given a globally hyperbolic spacetime $\mathscr{M}=\left(M,g,\mathfrak{o},\mathfrak{t}\right)$,
a vector bundle $E$ over $M$, a normally hyperbolic operator $P:\mathrm{C}^{\infty}\left(M,E\right)\rightarrow\mathrm{C}^{\infty}\left(M,E\right)$
on $\left(E,\mathscr{M}\right)$ and a vector space $V$, applying
Remark \ref{remLinearDifferentialOperatorInDistributionalSense},
the distributional extension $P:\mathscr{D}^{\prime}\left(M,E,V\right)\rightarrow\mathscr{D}^{\prime}\left(M,E,V\right)$
of the {}``original'' $P$. For our current scope, that is the determination
of global fundamental solutions for each point of $M$, we need to
consider a different vector space each time and hence we have to define
a {}``new'' $P$ for each $p\in M$. The reason that induces us
to do this will become clear in the next definition.
\begin{defn}
\label{defFundamentalSolution}\index{fundamental solutions}\index{delta distribution}Consider
a globally hyperbolic spacetime $\mathscr{M}=\left(M,g,\mathfrak{o},\mathfrak{t}\right)$,
a vector bundle $E$ over $M$ and a linear differential operator
$P:\mathrm{C}^{\infty}\left(M,E\right)\rightarrow\mathrm{C}^{\infty}\left(M,E\right)$
on $E$ over $\mathscr{M}$. Then for each $p\in M$ we consider the
linear differential operator in distributional sense $P:\mathscr{D}^{\prime}\left(M,E,E_{p}^{*}\right)\rightarrow\mathscr{D}^{\prime}\left(M,E,E_{p}^{*}\right)$
(obtained from the given $P$ exploiting Remark \ref{remLinearDifferentialOperatorInDistributionalSense})
and we call \textsl{fundamental solution for $P$ at the point $p$}
each of the distributions of $\mathscr{D}^{\prime}\left(M,E,E_{p}^{*}\right)$
that solve the equation $PU=\delta_{p}$ in distributional sense,
where $\delta_{p}:\mathscr{D}\left(M,E^{*}\right)\rightarrow E_{p}^{*}$
is the \textsl{$E_{p}^{*}$-valued delta distribution at $p$ on $E$
over $M$} defined by $\delta_{p}\left[w\right]=w\left(p\right)$
for each $w\in\mathscr{D}\left(M,E^{*}\right)$.
\end{defn}
Now the reason for which we consider a different {}``new'' $P$
for each $p\in M$ becomes clear: once that $p\in M$ is fixed, we
have to consider $E_{p}^{*}$ as target vector space for the space
of distributions in which we search fundamental solutions in order
to get compatibility between the RHS and the LHS of the distributional
equation $PU=\delta_{p}$.

We have defined fundamental solutions at a given point. Now we have
the problem of their existence and, in case, their uniqueness. The
next theorem provides us a tool that ensures uniqueness under certain
hypotheses.
\begin{lem}
\label{lem PU=00003D0 implies U=00003D0}Consider a globally hyperbolic
spacetime $\mathscr{M}=\left(M,g,\mathfrak{o},\mathfrak{t}\right)$,
a vector bundle $E$ over $M$, a vector space $V$ and a normally
hyperbolic operator $P$ on $E$ over $\mathscr{M}$. Then each solution
$u\in\mathscr{D}^{\prime}\left(M,E,V\right)$ of the equation $Pu=0$
(in distributional sense) with past compact or future compact support
must identically vanish.
\end{lem}
We stress that this lemma guarantees uniqueness only for fundamental
solutions with past compact or future compact support. Nothing is
implied for fundamental solutions with different supports.

The hypothesis of global hyperbolicity in this lemma can be weakened
without modifying the thesis. We keep such hypothesis also here since
anyway throughout our discussion it will always be assumed being indispensable
for many essential results, e.g. the next theorem, in which a relaxation
of the hypothesis of global hyperbolicity leads to the loss of the
result of existence for global fundamental solutions with past compact
or future compact support.

In the statement of the next theorem, besides existence for global
fundamental solutions with past compact or future compact support,
we have included uniqueness too, which is a direct consequence of
Lemma \ref{lem PU=00003D0 implies U=00003D0}.
\begin{thm}
\label{thmFundamentalSolutions}Consider a globally hyperbolic spacetime
$\mathscr{M}=\left(M,g,\mathfrak{o},\mathfrak{t}\right)$, a vector
bundle $E$ over $M$ and a normally hyperbolic operator $P$ on $E$
over $\mathscr{M}$. Then for each $p\in M$ there exists exactly
one fundamental solution for $P$ at $p$ with past compact support
(we denote it with $U^{a}\left(p\right)$) and exactly one fundamental
solution for $P$ at $p$ with future compact support (we denote it
with $U^{r}\left(p\right)$). Such fundamental solutions satisfy the
following properties:
\begin{itemize}
\item $\mathrm{supp}\left(U^{a}\left(p\right)\right)\subseteq J_{+}\left(p\right)$
and $\mathrm{supp}\left(U^{r}\left(p\right)\right)\subseteq J_{-}\left(p\right)$;
\item for each $u\in\mathscr{D}\left(M,E^{*}\right)$ the maps $p\mapsto U^{a/r}\left(p\right)\left[u\right]$,
denoted by $U^{a/r}\left(\cdot\right)\left[u\right]$, are (smooth)
sections in $E^{*}$ over $M$ and satisfy the differential equation
\[
P^{*}\left(U^{a/r}\left(\cdot\right)\left[u\right]\right)=u\mbox{.}
\]

\end{itemize}
\end{thm}
Beyond Lemma \ref{lem PU=00003D0 implies U=00003D0}, the proof of
this last result requires Theorem \ref{thmCauchyProblem} and another
theorem (not included here) that guarantees the linearity and continuity
of the map that associates to each proper initial data the corresponding
solution of the Cauchy problem presented in Theorem \ref{thmCauchyProblem}.
Such theorem can be found in \cite[Thm. 3.2.12, p. 86]{BGP07}. Both
Theorem \ref{thmCauchyProblem} and the omitted theorem are applied
to $P^{*}$ in place of $P$. The hypotheses of these theorems require
that $P^{*}$ is normally hyperbolic. Indeed this follows from the
hypothesis of normal hyperbolicity of $P$.

Now we can use fundamental solutions and their properties to define
a pair of operators that will allow us to obtain the full set of solutions
of the homogeneous Cauchy problems with compactly supported initial
data starting from the space of compactly supported sections. We begin
with a definition.
\begin{defn}
\label{defGreenOperators}\index{advanced Green operator}\index{retarded Green operator}Consider
a globally hyperbolic spacetime $\mathscr{M}=\left(M,g,\mathfrak{o},\mathfrak{t}\right)$,
a vector bundles $E$ over $M$ and a linear differential operator
$P:\mathrm{C}^{\infty}\left(M,E\right)\rightarrow\mathrm{C}^{\infty}\left(M,E\right)$.
We call \textsl{advanced Green operator for $P$} each map
\[
e^{a}:\mathscr{D}\left(M,E\right)\rightarrow\mathrm{C}^{\infty}\left(M,E\right)
\]
that is linear and fulfils the following requirements for each $f\in\mathscr{D}\left(M,E\right)$:
\begin{itemize}
\item $Pe^{a}f=f$;
\item $e^{a}Pf=f$;
\item $\mathrm{supp}\left(e^{a}f\right)\subseteq J_{+}\left(\mathrm{supp}\left(f\right)\right)$.
\end{itemize}
Similarly, we call \textsl{retarded Green operator for $P$} each
map
\[
e^{r}:\mathscr{D}\left(M,E\right)\rightarrow\mathrm{C}^{\infty}\left(M,E\right)
\]
that is linear and fulfils the same requirements with $J_{+}$ replaced
by $J_{-}$.
\end{defn}
Theorem \ref{thmFundamentalSolutions} implies existence and uniqueness
of both an advanced Green operator and a retarded Green operator for
a normally hyperbolic operator (we call them respectively the advanced
Green operator and the retarded Green operator since they are unique).
We present such result in the following corollary.
\begin{cor}
\label{corGreenOperators}Consider a globally hyperbolic spacetime
$\mathscr{M}=\left(M,g,\mathfrak{o},\mathfrak{t}\right)$, a vector
bundle $E$ over $M$ and a normally hyperbolic operator $P$ on $E$
over $\mathscr{M}$. Then two families $\left\{ U^{a}\left(x\right)\right\} $
and $\left\{ U^{r}\left(x\right)\right\} $ of fundamental solutions
for $P^{*}$ with past compact and, respectively, future compact support
define advanced and retarded Green operators $e^{a}$ and $e^{r}$
for $P$ in the following way: $e^{a}f=U^{r}\left(\cdot\right)\left[f\right]$
and $e^{r}f=U^{a}\left(\cdot\right)\left[f\right]$ for each $f\in\mathscr{D}\left(M,E\right)$.

Also the converse is true, i.e. advanced and retarded Green operators
$e^{a}$ and $e^{r}$ for $P$ define two families $\left\{ U^{a}\left(x\right)\right\} $
and $\left\{ U^{r}\left(x\right)\right\} $ of fundamental solutions
for $P^{*}$ with past compact and, respectively, future compact support
through the formulas given above applied in reverse sense.

In particular it follows that uniqueness for fundamental solutions
with past/future compact support implies uniqueness for Green operators.
\end{cor}
The existence of two families of fundamental solutions with the proper
support properties is assured by Theorem \ref{thmFundamentalSolutions}
applied to $P^{*}$, which is normally hyperbolic because we supposed
that $P$ is normally hyperbolic. As for uniqueness of Green operators,
if we assume that there exist two pairs of Green operators, we can
obtain two pairs of families of fundamental solutions with the right
support properties. Then Lemma \ref{lem PU=00003D0 implies U=00003D0}
implies the coincidence of the new families with the original ones
and this fact in turn implies the coincidence of the Green operators
used to build such families of fundamental solutions.

We devote the last part of this section to the presentation of some
properties related to the Green operators. The first one is an extension
of the second property in Definition \ref{defGreenOperators}. Its
validity is essentially based on Lemma \ref{lem PU=00003D0 implies U=00003D0}.
\begin{lem}
\label{lemExtensionOf ea(Pu)=00003Du}Consider a globally hyperbolic
spacetime $\mathscr{M}=\left(M,g,\mathfrak{o},\mathfrak{t}\right)$,
a vector bundle $E$ over $M$, a normally hyperbolic operator $P$
on $E$ over $\mathscr{M}$ and denote with $e^{a}$ and $e^{r}$
its advanced and retarded Green operators. Then for all $u\in\mathrm{C}^{\infty}\left(M,E\right)$
such that $Pu\in\mathscr{D}\left(M,E\right)$ it holds that:
\begin{itemize}
\item if $u$ has past compact support, $e^{a}Pu=u$;
\item if $u$ has future compact support, $e^{r}Pu=u$.
\end{itemize}
\end{lem}
\begin{proof}
Fix $u\in\mathrm{C}^{\infty}\left(M,E\right)$ with past compact support
such that $Pu\in\mathscr{D}\left(M,E\right)$. Then $Pu$ is in the
domain of $e^{a}$ and hence we can consider $e^{a}Pu$. From the
first property in Definition \ref{defGreenOperators} we deduce $Pe^{a}Pu=Pu$
and this identity can be rewritten in the form $P\left(e^{a}Pu-u\right)=0$.
We observe that $\mathrm{supp}\left(e^{a}Pu\right)\subseteq J_{+}\left(\mathrm{supp}\left(Pu\right)\right)$
and that $J_{+}\left(\mathrm{supp}\left(Pu\right)\right)$ is past
compact (this follows from Proposition \ref{propUsefulSubsetsOfGloballyHyperbolicSpacetimes}).
Since $u$ has past compact support by hypothesis, we deduce that
$e^{a}Pu-u$ has past compact support. Consider the distribution $F\in\mathscr{D}^{\prime}\left(M,E,\mathbb{R}\right)$
generated by the section $e^{a}Pu-u$:
\begin{eqnarray*}
F:\mathscr{D}\left(M,E^{*}\right) & \rightarrow & \mathbb{R}\\
f & \mapsto & \int\limits _{M}f\left(e^{a}Pu-u\right)\mathrm{d}\mu_{g}\mbox{,}
\end{eqnarray*}
where $\mathrm{d}\mu_{g}$ denotes the volume form on $\mathscr{M}$
and the dual pairing between $E^{*}$ and $E$ is taken into account.
We obtain $PF=0$ in distributional sense: for each $f\in\mathscr{D}\left(M,E^{*}\right)$
\[
\left(PF\right)\left[f\right]=F\left[P^{*}f\right]=\int\limits _{M}\left(P^{*}f\right)\left(e^{a}Pu-u\right)\mathrm{d}\mu_{g}=\int\limits _{M}f\left(P\left(e^{a}Pu-u\right)\right)\mathrm{d}\mu_{g}=0\mbox{.}
\]
Therefore Lemma \ref{lem PU=00003D0 implies U=00003D0} entails that
$F$ is the null distribution. Since the only section that generates
the null distribution is the null section, we conclude that $e^{a}Pu-u$
vanishes everywhere, which is to say $e^{a}Pu=u$. The proof of $e^{r}Pu=u$
for $u$ with future compact support is similar.
\end{proof}
Before the definition of Green operators, we have anticipated that
they allow us to build the full space of solutions of the homogeneous
Cauchy problems for a normally hyperbolic equation with compactly
supported initial data. Now we see how this is obtained.
\begin{defn}
\index{causal propagator}Consider a globally hyperbolic spacetime
$\mathscr{M}=\left(M,g,\mathfrak{o},\mathfrak{t}\right)$, a vector
bundle $E$ over $M$ and a linear differential operator $P:\mathrm{C}^{\infty}\left(M,E\right)\rightarrow\mathrm{C}^{\infty}\left(M,E\right)$
admitting advanced and retarded Green operators $e^{a}$ and $e^{r}$.
We call \textsl{causal propagator for $P$} the operator $e=e^{a}-e^{r}$.
\end{defn}
The support properties of the advanced and retarded Green operators
explain the reason why the operator $e=e^{a}-e^{r}$ is called causal
propagator for $P$: one may say that $e$ {}``propagates'' each
compactly supported section $f$ to the causal future and past of
its support providing a section $ef$ whose support is contained in
$J\left(\mathrm{supp}\left(f\right)\right)$.
\begin{cor}
\label{corSpaceOfSolutions}Consider a globally hyperbolic spacetime
$\mathscr{M}=\left(M,g,\mathfrak{o},\mathfrak{t}\right)$, a vector
bundle $E$ over $M$ and a normally hyperbolic operator $P$ on $E$
over $\mathscr{M}$. Let $e^{a}$ and $e^{r}$ be the advanced and
retarded Green operators for $P$. Then the space of solutions $\mathcal{S}$
of the homogeneous Cauchy problems associated to $P$ with compactly
supported initial data coincides with the image through the causal
propagator $e$ of $\mathscr{D}\left(M,E\right)$.\end{cor}
\begin{proof}
Before starting with the main part of the proof, we notice that we
can find a spacelike smooth Cauchy surface $\Sigma$ of $\mathscr{M}$
since $\mathscr{M}$ is a globally hyperbolic spacetime (see Theorem
\ref{thmGlobalHyperbolicity}). We set a unit future directed timelike
vector field $\mathfrak{n}$ over $\Sigma$ which is normal to $\Sigma$.

We begin from the inclusion $e\left(\mathscr{D}\left(M,E\right)\right)\subseteq\mathcal{S}$.
Fix $f\in\mathscr{D}\left(M,E\right)$ and define $u_{0}=\left.ef\right|_{\Sigma}$
and $u_{1}=\left.\nabla_{\mathfrak{n}}\left(ef\right)\right|_{\Sigma}$.
$u_{0}$, $u_{1}\in\mathscr{D}\left(\Sigma,\pi^{-1}\left(\Sigma\right)\right)$:
$\mathrm{supp}\left(ef\right)\cap\Sigma$ is compact because it is
closed and contained in $J\left(\mathrm{supp}\left(f\right)\right)\cap\Sigma$
which is compact too (cfr. Proposition \ref{propUsefulSubsetsOfGloballyHyperbolicSpacetimes}).
The first defining property of Green operators (see Definition \ref{defGreenOperators})
implies trivially that $P\left(ef\right)=0$. Moreover $\left.ef\right|_{\Sigma}=u_{0}$
and $\left.\nabla_{\mathfrak{n}}\left(ef\right)\right|_{\Sigma}=u_{1}$
by construction, where $u_{0}$ and $u_{1}$ are proper initial data
for a homogeneous Cauchy problem associated to $P$. Hence $ef\in\mathcal{S}$.

Now we turn our attention to the converse inclusion, i.e. $\mathcal{S}\subseteq e\left(\mathscr{D}\left(M,E\right)\right)$.
Fix $u\in\mathcal{S}$. Since $u$ is a solution of a homogeneous
Cauchy problem associated to $P$ with compactly supported initial
data, we find proper initial data that generate such solution simply
imposing $u_{0}=\left.u\right|_{\Sigma}$ and $u_{1}=\left.\nabla_{\mathfrak{n}}u\right|_{\Sigma}$.
As above $u_{0}$, $u_{1}\in\mathscr{D}\left(\Sigma,\pi^{-1}\left(\Sigma\right)\right)$
because from \ref{thmCauchyProblem} it follows that there exists
a compact subset $K$ of $M$ such that $\mathrm{supp}\left(u\right)\subseteq J\left(K\right)$.
Therefore we have that $u$ is the unique solution of the following
homogeneous Cauchy problem:
\[
\left\{ \begin{array}{rcl}
Pu & = & 0\mbox{,}\\
\left.u\right|_{\Sigma} & = & u_{0}\mbox{,}\\
\left.\nabla_{\mathfrak{n}}u\right|_{\Sigma} & = & u_{1}\mbox{.}
\end{array}\right.
\]
It is easy to find a compact subset $K^{\prime}$ of $M$ that includes
the supports of $u_{0}$ and $u_{1}$ and a relatively compact open
subset $\Omega$ of $M$ that includes $K$. We define the open subsets
$\Omega_{\pm}=J_{\pm}\left(\Omega\right)$ and $\Omega_{0}=M\setminus J\left(K^{\prime}\right)$
of $M$ ($J_{\pm}\left(\Omega\right)$ are open subsets of $M$ as
it is shown in \cite[Lem. A.8, p. 48]{FV11}, while $J_{\pm}\left(K\right)$
are closed subsets of $M$ as it is shown in \cite[Lem. A.5.1, p. 173]{BGP07})
and we consider the open covering $\left\{ \Omega_{+},\Omega_{-},\Omega_{0}\right\} $
of $M$. Associated to such open covering, we can choose a partition
of unity $\left\{ \chi_{+},\chi_{-},\chi_{0}\right\} $. We set $v_{\pm}=\chi_{\pm}u\in\mathrm{C}^{\infty}\left(M,E\right)$
and $v_{0}=\chi_{0}u\in\mathrm{C}^{\infty}\left(M,E\right)$. From
Theorem \ref{thmCauchyProblem} we deduce that $\mathrm{supp}\left(u\right)\subseteq J\left(K^{\prime}\right)$
because $K^{\prime}$ includes the supports of $u_{0}$ and $u_{1}$.
Hence $v_{0}=0$ and therefore $Pv_{+}=-Pv_{-}$. In particular this
implies that $Pv_{+}$ is supported in
\[
\mathrm{supp}\left(\chi_{+}\right)\cap\mathrm{supp}\left(\chi_{-}\right)\subseteq J_{+}\left(\Omega\right)\cap J_{-}\left(\Omega\right)\subseteq J_{+}\left(\overline{\Omega}\right)\cap J_{-}\left(\overline{\Omega}\right)\mbox{.}
\]
Since $\Omega$ is relatively compact in $M$, $J_{+}\left(\overline{\Omega}\right)\cap J_{-}\left(\overline{\Omega}\right)$
is compact (cfr. Proposition \ref{propUsefulSubsetsOfGloballyHyperbolicSpacetimes})
and hence $Pv_{+}\in\mathscr{D}\left(M,E\right)$. Consider now $ePv_{+}$:
\[
ePv_{+}=e^{a}Pv_{+}-e^{r}Pv_{+}=e^{a}Pv_{+}+e^{r}Pv_{-}=v_{+}+v_{-}=u\mbox{,}
\]
where we applied Lemma \ref{lemExtensionOf ea(Pu)=00003Du} taking
into account that $v_{+}$ has past compact support and $v_{-}$ has
future compact support as a consequence of their definitions. This
completes the proof.
\end{proof}
The next proposition provides a characterization of the kernel of
the causal propagator. The proof is based on the defining properties
of Green operators.
\begin{prop}
\label{propCausalPropagatorKernel}Consider a globally hyperbolic
spacetime $\mathscr{M}=\left(M,g,\mathfrak{o},\mathfrak{t}\right)$,
a vector bundle $E$ over $M$ and a linear differential operator
$P:\mathrm{C}^{\infty}\left(M,E\right)\rightarrow\mathrm{C}^{\infty}\left(M,E\right)$
admitting advanced and retarded Green operators $e^{a}$ and $e^{r}$.
Then we have that the kernel of the causal propagator $e$ coincides
with the image through $P$ of $\mathscr{D}\left(M,E\right)$:
\[
\ker e=P\left(\mathscr{D}\left(M,E\right)\right)\mbox{.}
\]
\end{prop}
\begin{proof}
The inclusion $P\left(\mathscr{D}\left(M,E\right)\right)\subseteq\ker e$
is a trivial consequence of the second defining property of Green
operators (cfr. Definition \ref{defGreenOperators}). To prove the
converse inclusion, consider $f\in\mathscr{D}\left(M,E\right)$ such
that $ef=0$. We have to find $h\in\mathscr{D}\left(M,E\right)$ such
that $Ph=f$ to prove that $f$ falls in $P\left(\mathscr{D}\left(M,E\right)\right)$.
We do this in the following way: First we notice that $ef=0$ implies
$e^{a}f=e^{r}f$. From this it follows that
\[
\mathrm{supp}\left(e^{a}f\right)\subseteq J_{+}\left(\mathrm{supp}\left(f\right)\right)\cap J_{-}\left(\mathrm{supp}\left(f\right)\right)\mbox{.}
\]
The set on the RHS of the last inclusion is compact owing to Proposition
\ref{propUsefulSubsetsOfGloballyHyperbolicSpacetimes}, hence $e^{a}f$
has compact support. Moreover $Pe^{a}f=f$ because of the first property
in Definition \ref{defGreenOperators}. Hence we have found a section
of the type required: $h=e^{a}f$.
\end{proof}
The last proposition of this subsection establishes a relationship
that holds between the Green operators for a normally hyperbolic operator
and the Green operators for its formal adjoint (that is automatically
normally hyperbolic).
\begin{prop}
\label{prope*rIsFormallyAdjointToea}Consider a globally hyperbolic
spacetime $\mathscr{M}=\left(M,g,\mathfrak{o},\mathfrak{t}\right)$,
a vector bundle $E$ over $M$ and a normally hyperbolic operator
$P$ on $E$ over $\mathscr{M}$. Let $e^{a}$ and $e^{r}$ be the
advanced and retarded Green operators for $P$ and $e^{*a}$ and $e^{*r}$
be the advanced and retarded Green operators for $P^{*}$, the formal
adjoint of $P$ (cfr. Remark \ref{remLinearDifferentialOperatorInDistributionalSense}),
which is automatically normally hyperbolic. Then we have that $e^{*a/r}$
is formally adjoint to $e^{r/a}$, which is to say
\[
\int\limits _{M}\left(e^{*a/r}f\right)\left(h\right)\mathrm{d}\mu_{g}=\int\limits _{M}f\left(e^{r/a}h\right)\mathrm{d}\mu_{g}
\]
\textup{for each $f\in\mathscr{D}\left(M,E^{*}\right)$ and each $h\in\mathscr{D}\left(M,E\right)$,
}where the dual pairing between $E^{*}$ and $E$ is taken into account
and $\mathrm{d}\mu_{g}$ is the volume form on $\mathscr{M}$.\end{prop}
\begin{proof}
For each $f\in\mathscr{D}\left(M,E^{*}\right)$ and each $h\in\mathscr{D}\left(M,E\right)$
we have
\begin{alignat*}{1}
\int\limits _{M}\left(e^{*a/r}f\right)\left(h\right)\mathrm{d}\mu_{g} & =\int\limits _{M}\left(\mathrm{e}^{*a/r}f\right)\left(P\mathrm{e}^{r/a}h\right)\mathrm{d}\mu_{g}=\int\limits _{M}\left(P^{*}e^{*a/r}f\right)\left(e^{r/a}h\right)\mathrm{d}\mu_{g}\\
 & =\int\limits _{M}f\left(e^{r/a}h\right)\mathrm{d}\mu_{g}\mbox{.}
\end{alignat*}
In the last calculation we have used the first defining property of
Green operators and we have exploited the relation of formal adjointness
between $P^{*}$and $P$ noting that
\[
\mathrm{supp}\left(e^{*a/r}f\right)\cap\mathrm{supp}\left(e^{r/a}h\right)\subseteq J_{\pm}\left(\mathrm{supp}\left(f\right)\right)\cap J_{\mp}\left(\mathrm{supp}\left(h\right)\right)
\]
is compact due to Proposition \ref{propUsefulSubsetsOfGloballyHyperbolicSpacetimes}.
\end{proof}

\section{\label{secAlgebrasAndStates}Algebras and states}

To define a quantum field theory in a proper way, we need essentially
two different types of ingredients. On the one hand there are algebras,
whose elements play the role of abstract {}``quantum observables''.
On the other hand there are states, which contain all the information
pertaining to the physical system that they are expected to describe.
The quantum field theory of a physical system concretely arises only
from the interaction of such building blocks. By this we mean that
a physical prediction is obtained taking the expectation value of
an observables on a given state. This section is devoted to a brief
presentation of both these ingredients with particular attention to
the algebras that are needed for the quantization of bosonic fields.

\subsection{\label{subC*-algebrasWeylSystemsCCRRepresentations}C{*}-algebras,
Weyl systems and CCR representations}

In this subsection we recollect the essential algebraic equipment
that we will use in the next chapters to build quantum field theories.
As for Section \ref{secWaveEquations}, most of the theorems are stated
without proofs, however these can be found in \cite[Chap. 4, Sects. 1-2]{BGP07}.

We begin giving the definition of an algebra. We take the chance to
specify some particular types of algebras which are enriched with
some additional structures such as {*}-algebras and C{*}-algebras
\begin{defn}
\index{algebra}\index{multiplication}\index{involution}An \textsl{associative
$\mathbb{C}$-algebra} (or simply an \textsl{algebra}) $\mathcal{A}$
is a $\mathbb{C}$-vector space $V$ endowed with a map $V\times V\rightarrow V$,
called \textsl{multiplication}, that maps $\left(a,b\right)\in V\times V$
to an element of $V$ denoted by $ab$ and that fulfils the following
properties:
\begin{itemize}
\item \textsl{$\mathbb{C}$-bilinearity}: for each $a$, $b$, $c\in V$
and each $\eta$, $\xi\in\mathbb{C}$ it holds that
\begin{eqnarray*}
\left(\eta a+\xi b\right)c & = & \eta ac+\xi bc\mbox{,}\\
a\left(\eta b+\xi c\right) & = & \eta ab+\xi ac\mbox{;}
\end{eqnarray*}

\item \textsl{associativity}: $\left(ab\right)c=a\left(bc\right)$ for each
$a$, $b$, $c\in V$.
\end{itemize}
\index{{*}-algebra}A {*}-algebra $\mathcal{A}$ is an algebra endowed
with a map $*:V\rightarrow V$, called \textsl{involution}, that maps
$a\in V$ to an element of $V$ denoted by $a^{*}$ and that fulfils
the following properties:
\begin{itemize}
\item \textsl{involutive property}: $a^{**}=a$ for each $a\in V$;
\item \textsl{$\mathbb{C}$-antilinearity}: $\left(\eta a+\xi b\right)^{*}=\overline{\eta}a^{*}+\overline{\xi}b^{*}$
for each $a$, $b\in V$ and each $\eta$, $\xi\in\mathbb{C}$;
\item \textsl{relation between multiplication and involution}: $\left(ab\right)^{*}=b^{*}a^{*}$
for each $a$, $b\in V$.
\end{itemize}
\index{C{*}-algebra}A C{*}-algebra $\mathcal{A}$ is a {*}-algebra
endowed with a norm $\left\Vert \cdot\right\Vert $ defined on the
underlying vector space such that it becomes a Banach space and the
following properties hold:
\begin{itemize}
\item \textsl{submultiplicativity}: $\left\Vert ab\right\Vert \leq\left\Vert a\right\Vert \left\Vert b\right\Vert $
for each $a$, $b\in V$;
\item \textsl{the involution is an isometry}: $\left\Vert a^{*}\right\Vert =\left\Vert a\right\Vert $
for each $a\in V$;
\item \textsl{C{*}-property}: $\left\Vert a^{*}a\right\Vert =\left\Vert a\right\Vert ^{2}$
for each $a\in V$.
\end{itemize}
\end{defn}
In the following we will always write $a\in\mathcal{A}$ when we consider
an element of the algebra $\mathcal{A}$. This means that we are considering
the element $a$ of the underlying $\mathbb{C}$-vector space $V$,
that in turn is the element $a$ of the set on which the $\mathbb{C}$-vector
structure is defined giving rise to $V$.

A very important example (at least in the context of quantum field
theory) of C{*}-algebra is provided by the space $\mathcal{B}\left(\mathscr{H}\right)$
of linear and continuous operators on a Hilbert space $\mathscr{H}$
with the composition of the operators as multiplication and the assignment
of the adjoint as involution.

We find it useful to define subalgebras of given algebras.
\begin{defn}
\label{defSubalgebra}\index{subalgebra}Consider an algebra $\mathcal{A}$.
A \textsl{subalgebra} $\mathcal{S}$ of $\mathcal{A}$ is a subspace
$W$ of the vector space $V$ underlying $\mathcal{A}$ that is closed
with respect to the multiplication of $\mathcal{A}$ so that the multiplication
of $\mathcal{A}$ restricted to $W$ becomes an associative $\mathbb{C}$-bilinear
internal operation on $W$ giving rise to the algebra $\mathcal{S}$.

\index{sub-{*}-algebra}If $\mathcal{A}$ is also a {*}-algebra, we
say that $\mathcal{S}$ is a \textsl{sub-{*}-algebra} of $\mathcal{A}$
if it is a subalgebra of $\mathcal{A}$ and its underlying vector
space $W$ is closed with respect to the involution of $\mathcal{A}$
so that it can be endowed with the involution of $\mathcal{A}$ restricted
to $W$ thus becoming a {*}-algebra itself.

\index{sub-C{*}-algebra}Finally if $\mathcal{A}$ is a C{*}-algebra,
we say that $S$ is a \textsl{sub-C{*}-algebra} of $\mathcal{A}$
if it is a sub-{*}-algebra of $\mathcal{A}$ and its underlying vector
space is a closed subspace of the Banach space underlying $\mathcal{A}$
so that $\mathcal{S}$ becomes a C{*}-algebra in its own right when
endowed with the norm of $\mathcal{A}$.
\end{defn}
Notice that in each of the cases seen above a subalgebra of an algebra
$\mathcal{A}$ is itself an algebra constituted by a subspace of the
vector space underlying $\mathcal{A}$ which is closed with respect
to all the operations that can be performed in $\mathcal{A}$ and
which is endowed with the restrictions of all the structures defined
on $\mathcal{A}$.
\begin{rem}
\label{remGeneratedSubalgebra}\index{set of generators}\index{generated subalgebra}We
can obtain the smallest subalgebra $\mathcal{S}$ (of a desired type)
including a subset $S$ of an algebra $\mathcal{A}$ (of that type)
simply taking the intersection of all the subalgebras of $\mathcal{A}$
(of that type) that include $S$. In such situation we call \textsl{set
of generators} the chosen subset $S$ and \textsl{generated subalgebra}
the subalgebra $\mathcal{S}$ that we have obtained. Indeed it can
happen that $S$ is such that $\mathcal{S}=\mathcal{A}$.
\end{rem}
It will be important for us to consider C{*}-algebras that contain
particular elements called unities.
\begin{defn}
\index{unit}\index{unital algebra}An element $1$ of an algebra
$\mathcal{A}$ is called a \textsl{unit} of $\mathcal{A}$ if $1a=a=a1$
for each $a\in\mathcal{A}$. Each algebra possessing a unit is said
to be \textsl{unital}.\end{defn}
\begin{rem}
Notice that each algebra $\mathcal{A}$ has at most one unit. This
is immediately seen assuming that both $1$ and $1^{\prime}$ are
units of $\mathcal{A}$ because from this assumption it follows that
$1=11^{\prime}=1^{\prime}$.

Moreover in each {*}-algebra $1^{*}=1$ since for each $a\in\mathcal{A}$
it holds that
\[
1^{*}a=\left(1^{*}a\right)^{**}=\left(a^{*}1^{**}\right)^{*}=\left(a^{*}1\right)^{*}=a^{**}=a
\]
and similarly $a1^{*}=a$. Then $1^{*}$ is a unit of $\mathcal{A}$
and uniqueness of units implies $1^{*}=1$.

The last observation concerning units that we make is related to their
norm: if $1$ denotes the unique unit of a C{*}-algebra $\mathcal{A}$
whose underlying vector space is not trivial, we have $\left\Vert 1\right\Vert =1$.
To see how this works we consider the C{*}-property and we remember
that the involution is an isometry, hence $\left\Vert 1\right\Vert ^{2}=\left\Vert 1^{*}1\right\Vert =\left\Vert 1^{*}\right\Vert =\left\Vert 1\right\Vert $.
The last equation implies that $\left\Vert 1\right\Vert $ is either
0 or 1. In the first case we have $1=0$. 0 must be the only element
of $\mathcal{A}$ in order to be a unit of $\mathcal{A}$. This contradicts
the hypothesis, therefore it must be $\left\Vert 1\right\Vert =1$.
\end{rem}
Now we define maps between algebras that are compatible with the structures
defined on such algebras.
\begin{defn}
\index{algebraic homomorphism}\index{algebraic isomorphism}\index{algebraic automorphism}Let
$\mathcal{A}$ and $\mathcal{B}$ be two algebras. A map $H:\mathcal{A}\rightarrow\mathcal{B}$
is an \textsl{algebraic homomorphism} if it is compatible with the
vector structures and multiplications of $\mathcal{A}$ and $\mathcal{B}$,
i.e. for each $a$, $b\in\mathcal{A}$ and each $\eta$, $\xi\in\mathbb{C}$
the following conditions hold:
\begin{eqnarray*}
H\left(\eta a+\xi b\right) & = & \eta Ha+\xi Hb\mbox{,}\\
H\left(ab\right) & = & \left(Ha\right)\left(Hb\right)\mbox{,}
\end{eqnarray*}
where the first equation involves the $\mathcal{A}$-vector structure
on the LHS and the $\mathcal{B}$-vector structure on the RHS, while
the second equation involves the $\mathcal{A}$-multiplication on
the LHS and the $\mathcal{B}$-multiplication on the RHS. A map $I:\mathcal{A}\rightarrow\mathcal{B}$
is an \textsl{algebraic isomorphism} if it is a bijective algebraic
homomorphism (its inverse is automatically an algebraic homomorphism
and hence an algebraic isomorphism). A map $I:\mathcal{A}\rightarrow\mathcal{A}$
is an \textsl{algebraic automorphism} if it is an algebraic isomorphism.

\index{{*}-homomorphism}\index{{*}-isomorphism}\index{{*}-automorphism}If
$\mathcal{A}$ and $\mathcal{B}$ are also {*}-algebras, a map $H:\mathcal{A}\rightarrow\mathcal{B}$
is a \textsl{{*}-homomorphism} if it is an algebraic homomorphism
compatible with the involutions of both $\mathcal{A}$ and $\mathcal{B}$,
i.e. $H\left(a^{*}\right)=\left(Ha\right)^{*}$ for each $a\in\mathcal{A}$,
where the LHS involves the $\mathcal{A}$-involution and the RHS involves
the $\mathcal{B}$-involution. A map $I:\mathcal{A}\rightarrow\mathcal{B}$
is a \textsl{{*}-isomorphism} if it is a bijective {*}-homomorphism
(its inverse is automatically a {*}-homomorphism and hence a {*}-isomorphism).
A map $I:\mathcal{A}\rightarrow\mathcal{A}$ is a \textsl{{*}-automorphism}
if it is a {*}-isomorphism.
\end{defn}
The upcoming proposition provides a condition that ensures continuity
for {*}-homomorphisms between unital C{*}-algebras.
\begin{prop}
\label{propContinuityOfUnitPreserving*-HomomorphismsOnUnitalC*-Algebras}Let
$\mathcal{A}$ and $\mathcal{B}$ be unital C{*}-algebras and consider
a {*}-homomor\-phism $H:\mathcal{A}\rightarrow\mathcal{B}$. Then
if $H$ is unit preserving, for each $a\in\mathcal{A}$ we have $\left\Vert H\left(a\right)\right\Vert \leq\left\Vert a\right\Vert $,
in particular $H$ can be seen as a linear and continuous operator
between the Banach spaces $\mathcal{A}$ and $\mathcal{B}$ with operator
norm $\left\Vert H\right\Vert \leq1$. If $H$ is also injective,
for each $a\in\mathcal{A}$ we have that $\left\Vert H\left(a\right)\right\Vert =\left\Vert a\right\Vert $,
in particular $H$ can be seen as an isometry between the Banach spaces
$\mathcal{A}$ and $\mathcal{B}$.\end{prop}
\begin{rem}
\label{remSurjective*-HomomorphismFromUnitalC*-AlgebrasToC*-Algebras}As
a particular case of this proposition, we consider a surjective {*}-homomorphism
$H$ from a unital C{*}-algebras $\mathcal{A}$ to a C{*}-algebra
$\mathcal{B}$. We notice that $H1_{\mathcal{A}}\in\mathcal{B}$ and
that for each $b\in\mathcal{B}$ it holds that
\begin{alignat*}{5}
\left(H1_{\mathcal{A}}\right)b & = & \left(H1_{\mathcal{A}}\right)\left(Ha\vphantom{H1_{\mathcal{A}}}\right) & = & H\left(1_{\mathcal{A}}a\right) & = & Ha & = & b\mbox{,}\\
b\left(H1_{\mathcal{A}}\right) & = & \left(Ha\vphantom{H1_{\mathcal{A}}}\right)\left(H1_{\mathcal{A}}\right) & = & H\left(a1_{\mathcal{A}}\right) & = & Ha & = & b\mbox{,}
\end{alignat*}
where $a\in\mathcal{A}$ is such that $Ha=b$ ($a$ exists as a consequence
of the surjectivity of $H$). Then we recognize $H1_{\mathcal{A}}$
to be the unique unit of $\mathcal{B}$. Hence as a matter of fact
$\mathcal{B}$ is a unital C{*}-algebra and $H$ is unit preserving
so that we can apply the last proposition. We conclude that $H$ can
be seen as a continuous linear operator between the Banach spaces
$\mathcal{A}$ and $\mathcal{B}$ with operator norm $\left\Vert H\right\Vert \leq1$.
If $H$ is also a {*}-isomorphism, due to the additional hypothesis
of injectivity, $H$ becomes an isometric isomorphism between the
Banach spaces $\mathcal{A}$ and $\mathcal{B}$.
\end{rem}
The next step in our path towards the construction of a quantum field
theory for a bosonic field is the introduction of Weyl systems. Before
we do that, we need to define symplectic spaces and symplectic maps.
\begin{defn}
\index{symplectic form}Let $V$ be a real vector space. We call \textsl{(non
degenerate) symplectic form} each map $\sigma:V\times V\rightarrow\mathbb{R}$
that satisfies the following conditions:
\begin{itemize}
\item bilinearity: for each $u$, $v$, $w\in V$ and each $\eta$, $\xi\in\mathbb{R}$
it holds that 
\begin{eqnarray*}
\sigma\left(\eta u+\xi v,w\right) & = & \eta\sigma\left(u,w\right)+\xi\sigma\left(v,w\right)\mbox{,}\\
\sigma\left(u,\eta v+\xi w\right) & = & \eta\sigma\left(u,w\right)+\xi\sigma\left(v,w\right)\mbox{;}
\end{eqnarray*}

\item antisymmetry: $\sigma\left(u,v\right)=-\sigma\left(v,u\right)$ for
each $u$, $v\in V$;
\item non degeneracy: if $u\in V$ is such that $\sigma\left(u,v\right)=0$
for each $v\in V$ then $u=0$.
\end{itemize}
\index{symplectic space}A \textsl{symplectic space} is a pair $\left(V,\sigma\right)$,
where $V$ is a real vector space and $\sigma$ is a symplectic form
on $V$.

\index{symplectic map}Given two symplectic spaces $\left(V,\sigma\right)$
and $\left(W,\omega\right)$, we say that $s:V\rightarrow W$ is a
\textsl{symplectic map} if it is linear and it is compatible with
the symplectic forms $\sigma$ and $\rho$, i.e. $\omega\left(su,sv\right)=\sigma\left(u,v\right)$
for each $u$, $v\in V$.\end{defn}
\begin{rem}
\label{remSymplecticMapsAreInjective}Note that each symplectic map
$s$ between to arbitrary symplectic spaces $\left(V,\sigma\right)$
and $\left(W,\omega\right)$ is injective. We can see this taking
$u\in V$ such that $su=0$ and showing that $u=0$. Indeed this is
true because $\sigma\left(u,v\right)=\omega\left(su,sv\right)=0$
for each $v\in V$ and $\sigma$ is non degenerate.
\end{rem}
Now that we know what a symplectic space is, we are ready to define
Weyl systems.
\begin{defn}
\label{defWeylSystem}\index{Weyl system}\index{Weyl map}Let $\left(V,\sigma\right)$
be a symplectic space. A \textsl{Weyl system associated to $\left(V,\sigma\right)$}
is a pair $\left(\mathcal{W},\mathrm{W}\right)$ where $\mathcal{W}$
is a unital C{*}-algebra and $\mathrm{W}:V\rightarrow\mathcal{W}$
is a \textsl{Weyl map}, i.e. a map that fulfils the following requirements
for each $u$, $v\in V$:
\begin{eqnarray*}
\mathrm{W}\left(0\right) & = & 1\mbox{,}\\
\mathrm{W}\left(-u\right) & = & \mathrm{W}\left(u\right)^{*}\mbox{,}\\
\mathrm{W}\left(u\right)\mathrm{W}\left(v\right) & = & \mathrm{e}^{-\frac{\imath}{2}\sigma\left(u,v\right)}\mathrm{W}\left(u+v\right)\mbox{.}
\end{eqnarray*}
\end{defn}
\begin{rem}
\label{remWeylMapProvidesUnitaries}The three requirements that each
Weyl map $\mathrm{W}$ must satisfy entail that $\mathrm{W}\left(u\right)^{*}\mathrm{W}\left(u\right)=1=\mathrm{W}\left(u\right)\mathrm{W}\left(u\right)^{*}$.
We show for example the first equality, the proof of the other being
almost identical. We proceed in the following way. In the first step
we exploit the second requirement, in the second step we exploit the
third requirement keeping in mind that each symplectic form is antisymmetric
and in the third and last step we apply the last requirement:
\[
\mathrm{W}\left(u\right)^{*}\mathrm{W}\left(u\right)=\mathrm{W}\left(-u\right)\mathrm{W}\left(u\right)=\mathrm{W}\left(0\right)=1\mbox{.}
\]

\end{rem}
For a concrete example of Weyl system associated to an arbitrary symplectic
space we refer the reader to \cite[Ex. 4.2.2, p. 116]{BGP07}. Such
example shows that there exists at least one Weyl system for each
symplectic space.

The requirements that define the Weyl map are such that they reproduce
the canonical commutation relations of bosonic quantum fields in an
exponentiated form, thus eliminating all the potential mathematical
complications that can arise when we try to construct a quantum field
theory starting from an algebra of bosonic fields satisfying the canonical
commutation relations in their original form. This is the reason why
we are interested in Weyl systems. To be more precise we are interested
in a particular class of Weyl systems that we are going to define.
\begin{defn}
\label{defCCRRepresentation}\index{CCR representation}\index{CCR algebra}Let
$\left(V,\sigma\right)$ be a symplectic space. A \textsl{CCR representation
of $\left(V,\sigma\right)$} is a Weyl system $\left(\mathcal{W},\mathrm{W}\right)$
associated to $\left(V,\sigma\right)$ such that $\mathrm{W}\left(V\right)$
is a set of generators for the unital C{*}-algebra $\mathcal{W}$.
In such situation $\mathcal{W}$ is called \textsl{CCR algebra}.
\end{defn}
Once we are given a Weyl system $\left(\mathcal{W},\mathrm{W}\right)$
associated to a symplectic space $\left(V,\sigma\right)$, we can
easily find a CCR representation of $\left(V,\sigma\right)$ considering
the Weyl system associated to $\left(V,\sigma\right)$ consisting
of the sub-C{*}-algebra generated by $\mathrm{W}\left(V\right)$ and
the Weyl map $\mathrm{W}$.

This construction ensures that the existence of a CCR representation
for each symplectic space is a consequence of the existence of a Weyl
system for that symplectic space. The next proposition states uniqueness
for CCR representations of symplectic spaces in an appropriate sense.
\begin{prop}
\label{propUniquenessOfCCRRepresentations}Let $\left(V,\sigma\right)$
be a symplectic space and consider two CCR representation $\left(\mathcal{W}_{1},\mathrm{W}_{1}\right)$
and $\left(\mathcal{W}_{2},\mathrm{W}_{2}\right)$ of $\left(V,\sigma\right)$.
Then there exists a unique {*}-isomor\-phism $I:\mathcal{W}_{1}\rightarrow\mathcal{W}_{2}$
such that $I\circ\mathrm{W}_{1}=\mathrm{W}_{2}$.
\end{prop}
Since $\mathcal{W}_{1}$ is a unital C{*}-algebra, we can apply Remark
\ref{remSurjective*-HomomorphismFromUnitalC*-AlgebrasToC*-Algebras}
and conclude that $I$ is actually unit preserving and can be interpreted
as an isometric isomorphism between the Banach spaces $\mathcal{W}_{1}$
and $\mathcal{W}_{2}$. This proposition implies that a CCR representation
associated to some symplectic space is unique up to {*}-isomorphisms.

We conclude this section with two propositions that will be essential
when we will try to build quantum field theories in the next chapters.
\begin{prop}
\label{propUnitPreserving*HomomorphismsFromACCRAlgebraAreInjective}Let
$\mathcal{W}$ be a CCR algebra. Then each unit preserving {*}-homomorphism
from $\mathcal{W}$ to a unital C{*}-algebra $\mathcal{A}$ is injective.
\end{prop}
We obtain a particular case of this proposition applying Proposition
\ref{propContinuityOfUnitPreserving*-HomomorphismsOnUnitalC*-Algebras}
to $\mathcal{W}$: Each unit preserving {*}-homomorphism from $\mathcal{W}$
to a unital C{*}-algebra $\mathcal{A}$ is injective and can be seen
as an isometry between the Banach spaces $\mathcal{W}$ and $\mathcal{A}$.
\begin{prop}
\label{propInjective*HomomorphismsFromSymplecticMaps}Let $\left(V,\sigma\right)$
and $\left(W,\rho\right)$ be two symplectic spaces and let $s:V\rightarrow W$
be a symplectic map. Denoting with $\left(\mathcal{V},\mathrm{V}\right)$
and $\left(\mathcal{W},\mathrm{W}\right)$ two CCR representations
of $\left(V,\sigma\right)$ and respectively of $\left(W,\rho\right)$,
we have that there exists a unique injective {*}-homomorphism $H:\mathcal{V}\rightarrow\mathcal{W}$
such that $H\circ\mathrm{V}=\mathrm{W}\circ s$.
\end{prop}
We want to stress that the {*}-homomorphism $H$ provided by the theorem
is automatically unit preserving because
\[
H\left(1_{\mathcal{V}}\right)=H\left(\mathrm{V}\left(0\right)\right)=\mathrm{W}\left(s\left(0\right)\right)=\mathrm{W}\left(0\right)=1_{\mathcal{W}}\mbox{.}
\]
Since $\mathcal{V}$ and $\mathcal{W}$ are both unital C{*}-algebras,
applying Proposition \ref{propContinuityOfUnitPreserving*-HomomorphismsOnUnitalC*-Algebras},
we deduce that $H$ is also an isometry between the Banach spaces
$\mathcal{V}$ and $\mathcal{W}$.

\subsection{States and representations}

In this subsection we focus on states and representations for a given
C{*}-algebra. In particular we show that a given state on each C{*}-algebra
induces a representation of the C{*}-algebra itself on some Hilbert
space. A more detailed discussion on this topic can be found in \cite[Sect. 1.4]{BB09}.

We start defining states on a C{*}-algebra.
\begin{defn}
\label{defState}\index{linear functional}\index{positive linear functional}\index{state}Let
$\mathcal{A}$ be a C{*}-algebra. We call \textsl{linear functional
on $\mathcal{A}$} each $\tau:\mathcal{A}\rightarrow\mathbb{C}$ that
is linear and continuous. The norm of a linear functional $\tau$
on $\mathcal{A}$ is defined by the following formula:
\[
\left\Vert \tau\right\Vert =\sup_{a\in\mathcal{A}\setminus\left\{ 0\right\} }\frac{\tau\left(a\right)}{\left\Vert a\right\Vert }\mbox{.}
\]
We say that $\tau$ is \textsl{positive} if $\tau\left(a^{*}a\right)\geq0$
for each $a\in\mathcal{A}$.

A \textsl{state $\tau$ on $\mathcal{A}$} is a positive linear functional
with norm 1, i.e. $\left\Vert \tau\right\Vert =1\mbox{.}$ We denote
the \textsl{set of states on $\mathcal{A}$} with $\mathfrak{sts}\mathcal{A}$.
\end{defn}
One of the most common examples of state is the following. Consider
the C{*}-algebra $\mathcal{B}\left(\mathscr{H}\right)$ of linear
and continuous operators on a Hilbert space $\mathscr{H}$ and let
$\Omega$ denote a norm 1 element of $\mathscr{H}$. Then for each
$L\in\mathcal{B}\left(\mathscr{H}\right)$ a state is provided by
the following map
\begin{eqnarray*}
\tau_{\Omega}:\mathcal{B}\left(\mathscr{H}\right) & \rightarrow & \mathbb{C}\\
L & \mapsto & \left(\Omega,L\Omega\right)_{\mathscr{H}}
\end{eqnarray*}
where $\left(\cdot,\cdot\right)_{\mathscr{H}}$ denotes the scalar
product of $\mathscr{H}$.

Positive linear functionals on a C{*}-algebra enjoy several properties
(especially if the C{*}-algebra is unital). We present some of these
properties in the following proposition.
\begin{prop}
\label{propLinearPositiveFunctional}Let $\mathcal{A}$ be a C{*}-algebra
and let $\tau$ be a positive linear functional on $\mathcal{A}$.
Then the following conditions hold:
\begin{itemize}
\item the map
\begin{eqnarray*}
\mathcal{A}\times\mathcal{A} & \rightarrow & \mathcal{A}\\
\left(a,b\right) & \mapsto & \tau\left(a^{*}b\right)
\end{eqnarray*}
is a positive semidefinite Hermitian sesquilinear form on $\mathcal{A}$;
\item the Cauchy-Schwarz inequality holds for this form, i.e. for each $a$,
$b\in\mathcal{A}$ we have
\[
\left|\tau\left(a^{*}b\right)\right|^{2}\leq\tau\left(a^{*}a\right)\tau\left(b^{*}b\right)\mbox{;}
\]

\item for each $a\in\mathcal{A}$, $\tau\left(a^{*}a\right)=0$ if and only
if $\tau\left(ba\right)=0$ for each $b\in\mathcal{A}$.
\end{itemize}
If $\mathcal{A}$ possesses a unit $1$ then we have some other properties:
\begin{itemize}
\item $\tau\left(a^{*}\right)=\overline{\tau\left(a\right)}$ for each $a\in\mathcal{A}$;
\item $\tau\left(1\right)=\left\Vert \tau\right\Vert $.
\end{itemize}
\end{prop}
\begin{proof}
We immediately realize that the form in the first condition of the
statement is sesquilinear (antilinear in the first argument and linear
in the second) and positive semidefinite. The only complication comes
when we want to check that it is also Hermitian. To prove this fact,
fix $a$, $b\in\mathcal{A}$ and $\eta\in\mathbb{C}$ and define $c=\eta a+b$.
Then we find that
\[
0\leq\tau\left(c^{*}c\right)=\left|\eta\right|^{2}\tau\left(a^{*}a\right)+\overline{\eta}\tau\left(a^{*}b\right)+\eta\tau\left(b^{*}a\right)+\tau\left(b^{*}b\right)
\]
and we deduce that $\overline{\eta}\tau\left(a^{*}b\right)+\eta\tau\left(b^{*}a\right)$
must be real for each $\eta\in\mathbb{C}$. This condition for $\eta=1$
and $\eta=\imath$ implies that the form is actually Hermitian.

The Cauchy-Schwarz inequality is satisfied by each positive semidefinite
Hermitian sesquilinear form. Anyway we show how to proceed in this
case since part of the proof has already been done. In fact the last
equation implies also the Cauchy-Schwarz inequality: If $\tau\left(a^{*}a\right)=0$,
then $2\Re\left(\overline{\eta}\tau\left(a^{*}b\right)\right)+\tau\left(b^{*}b\right)$
must be non negative for each $\eta\in\mathbb{C}$ and hence $\tau\left(a^{*}b\right)$
must be zero, otherwise we can make the choice
\[
\eta=-\frac{\tau\left(a^{*}b\right)}{\tau\left(a^{*}a\right)}\mbox{.}
\]
In both cases we conclude that the Cauchy-Schwarz inequality holds.

If we take $a\in\mathcal{A}$ such that $\tau\left(a^{*}a\right)=0$,
from the Cauchy-Schwarz inequality it follows that
\[
\left|\tau\left(ba\right)\right|^{2}\leq\tau\left(bb^{*}\right)\tau\left(a^{*}a\right)=0
\]
for each $b\in\mathcal{A}$. Then $\tau\left(ba\right)=0$. The converse
implication is trivial.

Now we suppose that $\mathcal{A}$ has a unit. The first property
easily follows from hermiticity:
\[
\tau\left(a^{*}\right)=\tau\left(a^{*}1\right)=\overline{\tau\left(1^{*}a\right)}=\overline{\tau\left(a\right)}\mbox{.}
\]

For the second property we proceed in the following way: For each
$a\in\mathcal{A}$ we find
\[
\left|\tau\left(a\right)\right|^{2}=\left|\tau\left(1^{*}a\right)\right|^{2}\leq\tau\left(1^{*}1\right)\tau\left(a^{*}a\right)\leq\tau\left(1\right)\left\Vert \tau\right\Vert \left\Vert a^{*}a\right\Vert =\tau\left(1\right)\left\Vert \tau\right\Vert \left\Vert a\right\Vert ^{2}\mbox{;}
\]
if $\tau=0$ then $\tau\left(1\right)=0=\left\Vert \tau\right\Vert $,
otherwise we deduce $\left\Vert \tau\right\Vert \leq\tau\left(1\right)$
and then the thesis follows bearing in mind that $\tau\left(1\right)\leq\left\Vert \tau\right\Vert \left\Vert 1\right\Vert =\left\Vert \tau\right\Vert $.
\end{proof}
If we are dealing with states, we have some other properties that
will be very helpful when we will try to find a representation for
each C{*}-algebra with the assignment of a state.
\begin{prop}
\label{propState}Let $\mathcal{A}$ be a unital C{*}-algebra and
let $\tau$ be a state on $\mathcal{A}$. Then the following properties
hold:
\begin{itemize}
\item $\left|\tau\left(a\right)\right|^{2}\leq\tau\left(a^{*}a\right)$
for each $a\in\mathcal{A}$;
\item for each $a$, $b\in\mathcal{A}$ we have $\tau\left(b^{*}a^{*}ab\right)\leq\left\Vert a\right\Vert ^{2}\tau\left(b^{*}b\right)$.
\end{itemize}
\end{prop}
\begin{proof}
We start from the first point. For each $a\in\mathcal{A}$, using
Proposition \ref{propLinearPositiveFunctional}, we obtain 
\[
\left|\tau\left(a\right)\right|^{2}=\left|\tau\left(1^{*}a\right)\right|^{2}\leq\tau\left(1^{*}1\right)\tau\left(a^{*}a\right)=\tau\left(1\right)\tau\left(a^{*}a\right)=\tau\left(a^{*}a\right)\mbox{.}
\]

For the second point we fix $a$, $b\in\mathcal{A}$. Consider the
case $\tau\left(b^{*}b\right)=0$. From the first statement it follows
that $\tau\left(cb\right)=0$ for each $c\in\mathcal{A}$ and, choosing
$c=b^{*}a^{*}a$, we obtain $\tau\left(b^{*}a^{*}ab\right)=0$ so
that the thesis holds in such case. Secondly we consider the case
$\tau\left(b^{*}b\right)>0$ and we define the map $\rho:\mathcal{A}\rightarrow\mathbb{C}$
by setting
\[
\rho\left(a\right)=\frac{\tau\left(b^{*}ab\right)}{\tau\left(b^{*}b\right)}\mbox{.}
\]
$\rho$ is immediately recognized as a positive linear functional
on $\mathcal{A}$ and, applying Proposition \ref{propLinearPositiveFunctional},
we deduce that $\left\Vert \rho\right\Vert =\rho\left(1\right)=1$,
hence $\rho$ is also a state. Then we conclude that $\rho\left(a^{*}a\right)\leq\left\Vert a^{*}a\right\Vert =\left\Vert a\right\Vert ^{2}$,
that is exactly our thesis.
\end{proof}
We have discussed states in sufficient detail for our scope. It is
time to turn our attention to representations of C{*}-algebras on
Hilbert spaces.
\begin{defn}
\index{representation}\index{faithful representation}Let $\mathcal{A}$
be a C{*}-algebra and let $\mathscr{H}$ be a Hilbert space. A \textsl{representation
of $\mathcal{A}$ on $\mathscr{H}$} is a {*}-homomorphism $\pi$
from $\mathcal{A}$ to the C{*}-algebra $\mathcal{B}\left(\mathscr{H}\right)$
of linear and continuous operators on $\mathscr{H}$. Such a representation
is said to be \textsl{faithful} if $\pi$ is injective.

\index{invariant subset}\index{irreducible representation}Let $\pi$
be a representation of the C{*}-algebra $\mathcal{A}$ on the Hilbert
space $\mathscr{H}$. We say that a subset $S$ of $\mathscr{H}$
is \textsl{invariant under $\mathcal{A}$} if the following condition
holds:
\[
\pi\left(\mathcal{A}\right)S=\left\{ \pi\left(a\right)v:\, a\in\mathcal{A},\, v\in S\right\} \subseteq S\mbox{.}
\]
We say that $\pi$ is \textsl{irreducible} if the only invariant closed
subspaces of $\mathscr{H}$ are $\left\{ 0\right\} $ and $\mathscr{H}$
itself.

\index{unitarily equivalent representations}Moreover two representations
$\pi_{1}$ and $\pi_{2}$ of $\mathcal{A}$ on the Hilbert spaces
$\mathscr{H}_{1}$ and respectively $\mathscr{H}_{2}$ are said to
be \textsl{unitarily equivalent} if there exists a unitary operator
$U:\mathscr{H}_{1}\rightarrow\mathscr{H}_{2}$ such that $U\circ\pi_{1}\left(a\right)=\pi_{2}\left(a\right)\circ U$
for each $a\in\mathcal{A}$.
\end{defn}
Before we state the main theorem of this subsection, we still need
to define another ingredient.
\begin{defn}
Let $\mathcal{A}$ be a C{*}-algebra, let $\mathscr{H}$ be a vector
space and let $\pi:\mathcal{A}\rightarrow\mathcal{B}\left(\mathscr{H}\right)$
be a representation. A vector $\Omega\in\mathscr{H}$ is said to be
cyclic for the representation $\pi$ if $\pi\left(\mathcal{A}\right)\Omega=\left\{ \pi\left(a\right)\Omega:a\in\mathcal{A}\right\} $
is a dense subspace of $\mathscr{H}$.
\end{defn}
We are ready to state the main theorem of this subsection.
\begin{thm}
\label{thmGNSRepresentation}\index{GNS representation}\index{GNS triple}Let
$\mathcal{A}$ be a unital C{*}-algebra and let $\tau$ be a state
on $\mathcal{A}$. Then there exists a triple $\left(\mathscr{H},\pi,\Omega\right)$,
where $\mathscr{H}$ is a Hilbert space with scalar product denoted
by $\left(\cdot,\cdot\right)$, $\pi:\mathcal{A}\rightarrow\mathcal{B}\left(\mathscr{H}\right)$
is a unit preserving continuous representation of $\mathcal{A}$ on
$\mathscr{H}$ and $\Omega\in\mathscr{H}$ is a unit cyclic vector
for the representation $\pi$ such that for each $a\in\mathcal{A}$
it holds that
\[
\left(\Omega,\pi\left(a\right)\Omega\right)=\tau\left(a\right)\mbox{.}
\]

This triple $\left(\mathscr{H},\pi,\Omega\right)$ is unique (up to
unitary equivalence) and it is called the \textsl{GNS triple for $\mathcal{A}$
induced by $\tau$}.\end{thm}
\begin{proof}
In Proposition \ref{propLinearPositiveFunctional} we have seen that
$\tau$ defines a positive semidefinite Hermitian sesquilinear product
on $\mathcal{A}$. If we consider the set
\[
\mathcal{N}=\left\{ a\in\mathcal{A}:\tau\left(a^{*}a\right)=0\right\} \mbox{,}
\]
we easily realize that this is a closed vector subspace of $\mathcal{A}$
applying the first part of Proposition \ref{propLinearPositiveFunctional}.
Then we can consider the quotient
\[
\mathcal{A}_{\bullet}=\frac{\mathcal{A}}{\mathcal{N}}\mbox{,}
\]
which becomes a Banach space when endowed with the quotient norm $\left\Vert \cdot\right\Vert _{\bullet}$
defined by the formula
\[
\left\Vert a_{\bullet}\right\Vert _{\bullet}=\inf_{a\in a_{\bullet}}\left\Vert a\right\Vert \mbox{,}\quad a_{\bullet}\in\mathcal{A}_{\bullet}\mbox{.}
\]

Consider now two equivalence classes $a_{\bullet}$, $b_{\bullet}\in\mathcal{A}_{\bullet}$
and choose $a$, $a^{\prime}\in a_{\bullet}$ and $b$, $b^{\prime}\in b_{\bullet}$.
With this choice of representatives we evaluate $\tau\left(a^{\prime*}b^{\prime}\right)$.
For convenience we define $n_{a}=a^{\prime}-a$ and $n_{b}=b^{\prime}-b$
and we immediately realize that $n_{a}$, $n_{b}\in\mathcal{N}$.
We find that
\[
\tau\left(a^{\prime*}b^{\prime}\right)=\tau\left(a^{*}b\right)+\tau\left(a^{*}n_{b}\right)+\tau\left(n_{a}^{*}b\right)+\tau\left(n_{a}^{*}n_{b}\right)
\]
and, applying again Proposition \ref{propLinearPositiveFunctional},
we deduce
\[
\tau\left(a^{\prime*}b^{\prime}\right)=\tau\left(a^{*}b\right)
\]
because $\tau\left(a^{*}n_{b}\right)=0$, $\tau\left(n_{a}^{*}b\right)=\overline{\tau\left(b^{*}n_{a}\right)}=0$
and $\tau\left(n_{a}^{*}n_{b}\right)=0$. This shows that the map
\begin{eqnarray*}
\left(\cdot,\cdot\right)_{\bullet}:\mathcal{A}_{\bullet}\times\mathcal{A}_{\bullet} & \rightarrow & \mathbb{C}\\
\left(a_{\bullet},b_{\bullet}\right) & \mapsto & \tau\left(a^{*}b\right),\: a\in a_{\bullet},b\in b_{\bullet}
\end{eqnarray*}
is well defined. It is immediate to check that it is a positive semidefinite
Hermitian sesquilinear form. Now we show that it is also positive
definite. Consider $a_{\bullet}\in\mathcal{A}_{\bullet}$ such that
$\left(a_{\bullet},a_{\bullet}\right)_{\bullet}=0$. By definition
of $\left(\cdot,\cdot\right)_{\bullet}$, this means that we have
$\tau\left(a^{*}a\right)=0$ for each $a\in a_{\bullet}$. Then $a_{\bullet}$
coincides with $\mathcal{N}$, that is the zero element of $\mathcal{A}_{\bullet}$.
We conclude that $\left(\cdot,\cdot\right)_{\bullet}$ is a scalar
product on $\mathcal{A}_{\bullet}$, so that $\mathcal{A}_{\bullet}$
becomes a pre-Hilbert space when endowed with $\left(\cdot,\cdot\right)_{\bullet}$.
This can be completed and we obtain an Hilbert space $\mathscr{H}$.
We denote its scalar product with $\left(\cdot,\cdot\right)$ and
we remind the reader that the pre-Hilbert space $\mathcal{A}_{\bullet}$
is isometrically isomorphic to a certain subspace $\mathscr{S}$ of
$\mathscr{H}$, therefore the composition of the inclusion map of
$\mathscr{S}$ in $\mathscr{H}$ with the isometrical isomorphism
$J$ from $\mathcal{A}_{\bullet}$ to $\mathscr{S}$ is an isometry.
We denote this isometry with $I$.

Consider now $a\in\mathcal{A}$ and $b_{\bullet}\in\mathcal{A}_{\bullet}$
and choose two representatives $b$, $b^{\prime}\in b_{\bullet}$.
For convenience we define $n=b^{\prime}-b$ and we notice that $n\in\mathcal{N}$.
From Proposition \ref{propState} we deduce that $\tau\left(n^{*}a^{*}an\right)=0$.
Then $an$ falls in $\mathcal{N}$ and we conclude that $\left[ab\right]_{\bullet}=\left[ab^{\prime}\right]_{\bullet}$.
This shows that for each $a\in\mathcal{A}$, the map
\begin{eqnarray*}
L_{a}:\mathcal{A}_{\bullet} & \rightarrow & \mathcal{A}_{\bullet}\\
b_{\bullet} & \mapsto & \left[ab\right]_{\bullet},\: b\in b_{\bullet}
\end{eqnarray*}
is well defined. $L_{a}$ is also linear, as one immediately recognizes.
For each $a\in\mathcal{A}$ we show that $L_{a}$ is also continuous
on $\mathcal{A}_{\bullet}$ endowed with the norm induced by $\left(\cdot,\cdot\right)_{\bullet}$.
We fix $a\in\mathcal{A}$ for each $b_{\bullet}\in\mathcal{A}_{\bullet}$
and we apply again the last part of Proposition \ref{propState}.
Then we find
\[
\left(L_{a}b_{\bullet},L_{a}b_{\bullet}\right)_{\bullet}=\left(\left[ab\right]_{\bullet},\left[ab\right]_{\bullet}\right)_{\bullet}=\tau\left(b^{*}a^{*}ab\right)\leq\left\Vert a\right\Vert ^{2}\tau\left(b^{*}b\right)=\left\Vert a\right\Vert ^{2}\left(b_{\bullet},b_{\bullet}\right)_{\bullet}\mbox{.}
\]
This means exactly the continuity of $L_{a}$ with respect to the
norm induced by $\left(\cdot,\cdot\right)_{\bullet}$. Moreover its
norm as a linear and continuous operator on $\mathcal{A}_{\bullet}$
is controlled from above by $\left\Vert a\right\Vert $: we write
that $\left\Vert L_{a}\right\Vert \leq\left\Vert a\right\Vert $.

Recalling the isometry $I:\mathcal{A}_{\bullet}\rightarrow\mathscr{H}$
and the isometrical isomorphism $J:\mathcal{A}_{\bullet}\rightarrow\mathscr{S}$,
for each $a\in\mathcal{A}$ we define $L_{a}^{\prime}=I\circ L_{a}\circ J^{-1}$.
We have that $L_{a}^{\prime}$ is a linear and continuous operator
from $\mathscr{S}$ to $\mathscr{H}$ with norm $\left\Vert L_{a}^{\prime}\right\Vert \leq\left\Vert a\right\Vert $.
Since $\mathscr{H}$ is complete, we can find a unique linear and
continuous extension of $L_{a}^{\prime}$ defined on the closure of
$\mathscr{S}$, i.e. $\mathscr{H}$. We denote such linear and continuous
operator on $\mathscr{H}$ with $\pi\left(a\right)$ and we find that
$\left\Vert \pi\left(a\right)\right\Vert =\left\Vert L_{a}^{\prime}\right\Vert \leq\left\Vert a\right\Vert $.
In this way we the map $\pi:\mathcal{A}\rightarrow\mathcal{B}\left(\mathscr{H}\right)$
is automatically defined. $\pi$ is linear as the reader can directly
check from its definition. Moreover $\left\Vert \pi\left(a\right)\right\Vert \leq\left\Vert a\right\Vert $
shows that $\pi$ is continuous. We must only check that for each
$a$, $b\in\mathcal{A}$ the following equations hold:
\begin{alignat*}{2}
\pi\left(ab\right) & =\pi\left(a\right)\pi\left(b\right)\mbox{,} & \quad\forall a,b\in\mathcal{A}\mbox{;}\\
\pi\left(a^{*}\right) & =\pi\left(a\right)^{*}\mbox{,} & \forall a\in\mathcal{A}\mbox{.}
\end{alignat*}
and then we have a continuous representation of $\mathcal{A}$ on
$\mathscr{H}$ . Fix $a\in\mathcal{A}$ and $v\in\mathscr{H}$. To
simplify the inspection of these equations we give an expression of
$\pi\left(a\right)v$. From the definition of $\pi\left(a\right)$
as the unique linear and continuous extension of $L_{a}^{\prime}$,
we find a Cauchy sequence $\left\{ v_{n}\right\} \subseteq\mathscr{S}$
that converges to $v\in\mathscr{H}$ such that $\left\{ L_{a}^{\prime}v_{n}\right\} $
converges to $\pi\left(a\right)v$ in $\mathscr{H}$ and for each
$n$ we choose a representative $v_{n}^{\prime}$ of the equivalence
class $J^{-1}v_{n}\in\mathcal{A}_{\bullet}$. Therefore we have
\begin{equation}
\pi\left(a\right)v=\lim_{n\rightarrow\infty}\left(L_{a}^{\prime}v_{n}\right)=\lim_{n\rightarrow\infty}\left(\left(I\circ L_{a}\circ J^{-1}\right)v_{n}\right)=\lim_{n\rightarrow\infty}\left(I\left[av_{n}^{\prime}\right]_{\bullet}\right)\mbox{.}\label{eqExpressionForpi(a)v}
\end{equation}
This formula allows us to easily check the first equation above. For
the second equation we must also keep in mind that the involution
of $\mathcal{B}\left(\mathscr{H}\right)$ is the map $\mathcal{B}\left(\mathscr{H}\right)\rightarrow\mathcal{B}\left(\mathscr{H}\right)$,
$L\mapsto L^{\dagger}$, where $L^{\dagger}$ is the adjoint of $L$
with respect to the scalar product $\left(\cdot,\cdot\right)$ of
$\mathscr{H}$, that is
\[
\left(L^{\dagger}v,w\right)=\left(v,Lw\right)\quad\forall v,w\in\mathscr{H}\mbox{.}
\]
Therefore the condition that we must actually check is the following:
\[
\left(v,\pi\left(a\right)w\right)=\left(\pi\left(a^{*}\right)v,w\right)\quad\forall v,w\in\mathscr{H}\;\forall a\in\mathcal{A}\mbox{.}
\]
This is easily seen to hold using eq. \eqref{eqExpressionForpi(a)v}.

Now we have to find a cyclic unit vector in $\mathscr{H}$ that satisfies
the condition of the statement. Using the unit of $\mathcal{A}$,
we define $\Omega=I\left[1\right]_{\bullet}$. This is indeed an element
of $\mathscr{H}$. We fix $a\in\mathcal{A}$ and we try to evaluate
$\left(\Omega,\pi\left(a\right)\Omega\right)$. In first place we
use eq. \eqref{eqExpressionForpi(a)v} with $v=\Omega$. As a consequence
of the definition of $\Omega$ the formula becomes simpler:
\[
\pi\left(a\right)\Omega=I\left[a1\right]_{\bullet}=I\left[a\right]_{\bullet}\mbox{.}
\]
Hence we have that $\pi\left(\mathcal{A}\right)\Omega=I\left(\mathcal{A}_{\bullet}\right)=\mathscr{S}$
and, since $\mathscr{S}$ is dense in $\mathscr{H}$, $\Omega$ is
indeed cyclic. Moreover we get
\[
\left(\Omega,\pi\left(a\right)\Omega\right)=\left(I\left[1\right]_{\bullet},I\left[a\right]_{\bullet}\right)=\left(\left[1\right]_{\bullet},\left[a\right]_{\bullet}\right)_{\bullet}=\tau\left(1^{*}a\right)=\tau\left(a\right)\mbox{.}
\]
Since $\pi\left(1\right)=\mathrm{id}_{\mathscr{H}}$, as it can be
checked via direct inspection, the last equation for $a=1$ implies
that $\left(\Omega,\Omega\right)=1$.

We have built a triple $\left(\pi,\mathscr{H},\Omega\right)$ with
the properties required in the statement. To complete the proof we
must show that such triple is unique up to a unitary transformation.
To this end suppose that we have another triple $\left(\pi^{\prime},\mathscr{H}^{\prime},\Omega^{\prime}\right)$
of the same type and for convenience we denote with $\mathscr{S}$
the dense subspace $\pi\left(\mathcal{A}\right)\Omega$ of $\mathscr{H}$
and with $\mathscr{S}^{\prime}$ the dense subspace $\pi^{\prime}\left(\mathcal{A}\right)\Omega^{\prime}$
of $\mathscr{H}^{\prime}$. If we have $a$, $b\in\mathcal{A}$ such
that $\pi\left(a\right)\Omega=\pi\left(b\right)\Omega$, then it holds
also that $\pi^{\prime}\left(a\right)\Omega=\pi\left(b\right)\Omega^{\prime}$.
To check this fact fix an arbitrary $v^{\prime}\in\mathscr{H}$ and,
using the density of $\mathscr{S}^{\prime}$ in $\mathscr{H}^{\prime}$,
choose a sequence $\left\{ v_{n}^{\prime}\right\} \subseteq\mathscr{S}^{\prime}$
that converges to $v^{\prime}$ in $\mathscr{H}^{\prime}$. By definition
of $\mathscr{S}^{\prime}$, for each $n$ we also find $a_{n}\in\mathcal{A}$
such that $\pi^{\prime}\left(a_{n}\right)\Omega^{\prime}=v_{n}^{\prime}$.
Using $\left(\cdot,\cdot\right)$ and $\left(\cdot,\cdot\right)^{\prime}$
to denote the scalar products of $\mathscr{H}$ and respectively $\mathscr{H}^{\prime}$
and bearing in mind the properties fulfilled by each of the triples,
we deduce that for each $c$, $d\in\mathcal{A}$
\begin{gather*}
\left(\pi\left(c\right)\Omega,\pi\left(d\right)\Omega\right)=\left(\Omega,\pi\left(c\right)^{\dagger}\pi\left(d\right)\Omega\right)=\left(\Omega,\pi\left(c^{*}d\right)\Omega\right)=\tau\left(c^{*}d\right)\mbox{,}\\
\left(\pi^{\prime}\left(c\right)\Omega^{\prime},\pi^{\prime}\left(d\right)\Omega^{\prime}\right)=\left(\Omega^{\prime},\pi^{\prime}\left(c\right)^{\dagger}\pi^{\prime}\left(d\right)\Omega^{\prime}\right)=\left(\Omega^{\prime},\pi^{\prime}\left(c^{*}d\right)\Omega^{\prime}\right)=\tau\left(c^{*}d\right)\mbox{.}
\end{gather*}
Hence we find that
\begin{eqnarray*}
\left(v^{\prime},\pi^{\prime}\left(a\right)\Omega^{\prime}\right)^{\prime} & = & \lim_{n\rightarrow\infty}\left(\pi^{\prime}\left(a_{n}\right)\Omega^{\prime},\pi^{\prime}\left(a\right)\Omega^{\prime}\right)\\
 & = & \lim_{n\rightarrow\infty}\left(\pi\left(a_{n}\right)\Omega,\pi\left(a\right)\Omega\right)\\
 & = & \lim_{n\rightarrow\infty}\left(\pi\left(a_{n}\right)\Omega,\pi\left(b\right)\Omega\right)\\
 & = & \lim_{n\rightarrow\infty}\left(\pi^{\prime}\left(a_{n}\right)\Omega^{\prime},\pi^{\prime}\left(b\right)\Omega^{\prime}\right)\\
 & = & \left(v^{\prime},\pi^{\prime}\left(b\right)\Omega^{\prime}\right)^{\prime}\mbox{.}
\end{eqnarray*}
This holds for each $v^{\prime}\in\mathscr{H}$. Therefore the map
\begin{alignat*}{2}
V & : & \mathscr{S} & \rightarrow\mathscr{H}^{\prime}\\
 &  & \pi\left(a\right)\Omega & \mapsto\pi^{\prime}\left(a\right)\Omega^{\prime}
\end{alignat*}
is well defined. Moreover $V$ is trivially linear and, from the considerations
made above, we deduce that for each $a$, $b\in\mathcal{A}$ the following
equation holds:
\[
\left(V\left(\pi\left(a\right)\Omega\right),V\left(\pi\left(b\right)\Omega\right)\right)^{\prime}=\left(\pi\left(a\right)\Omega,\pi\left(b\right)\Omega\right)\mbox{.}
\]
In particular this implies that $V$ is a linear and continuous operator
from the dense subspace $\mathscr{S}$ of the Hilbert space $\mathscr{H}$
to the other Hilbert space $\mathscr{H}^{\prime}$. Then there exists
a unique linear and continuous extension $U$ of $V$ defined on the
closure of $\mathscr{S}$, i.e. $U:\mathscr{H}\rightarrow\mathscr{H}^{\prime}$.
With the help of our last equation we show that $U$ is unitary. Fix
$v$, $w\in\mathscr{H}$. We find Cauchy sequences $\left\{ v_{n}\right\} $,
$\left\{ w_{n}\right\} \subseteq\mathscr{S}$ that converge to $v$
and respectively $w$ in $\mathscr{H}$ such that $\left\{ Vv_{n}\right\} $
and $\left\{ Vw_{n}\right\} $ converge to $Uv$ and, respectively,
$Uw$ in $\mathscr{H}$. Then for each $n$ we find $a_{n}$, $b_{n}\in\mathcal{A}$
such that $\pi\left(a_{n}\right)\Omega=v_{n}$ and $\pi\left(b_{n}\right)\Omega=w_{n}$.
Recalling that the scalar product is always continuous in both its
arguments, we obtain:
\begin{alignat*}{1}
\left(Uv,Uw\right)^{\prime} & =\lim_{n\rightarrow\infty}\left(V\left(\pi\left(a_{n}\right)\Omega\right),V\left(\pi\left(b_{n}\right)\Omega\right)\right)^{\prime}\\
 & =\lim_{n\rightarrow\infty}\left(\pi\left(a_{n}\right)\Omega,\pi\left(b_{n}\right)\Omega\right)\\
 & =\lim_{n\rightarrow\infty}\left(v_{n},w_{n}\right)\\
 & =\left(v,w\right)\mbox{.}
\end{alignat*}
Since the last equation holds for each $v$, $w\in\mathscr{H}$, we
deduce that $U$ is unitary as required. The only property that must
still be checked is the following:
\[
U\circ\pi\left(a\right)=\pi^{\prime}\left(a\right)\circ U\quad\forall a\in\mathcal{A}\mbox{.}
\]
By construction $U$ coincides with $V$ on $\mathscr{S}$ and so
$U\left(\pi\left(a\right)\Omega\right)=\pi^{\prime}\left(a\right)\Omega^{\prime}$
for each $a\in\mathcal{A}$. From this it follows that for each $a\in\mathcal{A}$
we have
\[
\left(\Omega^{\prime},\pi^{\prime}\left(a\right)\Omega^{\prime}\right)^{\prime}=\tau\left(a\right)=\left(\Omega,\pi\left(a\right)\Omega\right)=\left(U\Omega,U\left(\pi\left(a\right)\Omega\right)\right)^{\prime}=\left(U\Omega,\pi^{\prime}\left(a\right)\Omega^{\prime}\right)^{\prime}
\]
and, since $\pi^{\prime}\left(\mathcal{A}\right)\Omega=\mathscr{S}^{\prime}$
is dense in $\mathscr{H}^{\prime}$ and $\left(\cdot,\cdot\right)^{\prime}$
is continuous in its second argument, it follows that $\left(\Omega^{\prime},v^{\prime}\right)^{\prime}=\left(U\Omega,v^{\prime}\right)^{\prime}$
for each $v^{\prime}\in\mathscr{H}^{\prime}$, which is to say $\Omega^{\prime}=U\Omega$.
This fact provides us the equation that leads to the conclusion of
the proof:
\[
U\left(\pi\left(a\right)\Omega\right)=\pi^{\prime}\left(a\right)\left(U\Omega\right)\quad\forall a\in\mathcal{A}\mbox{.}
\]
As a preliminary step, we observe that for each $a$, $b\in\mathcal{A}$
it holds
\begin{alignat*}{1}
U\left(\pi\left(a\right)\left(\pi\left(b\right)\Omega\right)\right) & =U\left(\pi\left(ab\right)\Omega\right)=\pi^{\prime}\left(ab\right)\left(U\Omega\right)=\pi^{\prime}\left(a\right)\left(\pi^{\prime}\left(b\right)\left(U\Omega\right)\right)\\
 & =\pi^{\prime}\left(a\right)\left(U\left(\pi^{\prime}\left(b\right)\Omega\right)\right)\mbox{.}
\end{alignat*}
Consider $a\in\mathcal{A}$ and $v\in\mathscr{H}$. As usual we find
a Cauchy sequence $\left\{ v_{n}\right\} \subseteq\mathscr{S}$ that
converges to $v$ in $\mathscr{H}$ and for each $n$ we find $a_{n}\in\mathcal{A}$
such that $\pi\left(a_{n}\right)\Omega=v_{n}$. Then, reminding of
the continuity of $U$, $\pi\left(a\right)$ and $\pi^{\prime}\left(a\right)$,
we have
\[
U\left(\pi\left(a\right)v\right)=\lim_{n\rightarrow\infty}U\left(\pi\left(a\right)\left(\pi\left(a_{n}\right)\Omega\right)\right)=\lim_{n\rightarrow\infty}\pi^{\prime}\left(a\right)\left(U\left(\pi\left(a_{n}\right)\Omega\right)\right)=\pi^{\prime}\left(a\right)\left(Uv\right)\mbox{.}
\]
Since this holds for each $a\in\mathcal{A}$ and each $v\in\mathscr{H}$,
the proof is complete.
\end{proof}

\section{\label{secCategoryTheory}Category theory}

This section concludes the preliminary part of the thesis. We devote
it to the presentation of some notions from category theory that will
be extensively used in the next chapters. This is essentially due
to the fact that it is possible to construct a quantum field theory
as a covariant functor between to appropriate categories. As a matter
of fact we only need very few notions of category theory so that,
despite of its brevity, the current section, unlike the previous ones,
is totally self contained and sufficient for our scopes. Anyway as
general reference about this topic we suggest \cite{ML98}.

We start defining what it is meant for a category.
\begin{defn}
\label{defCategory}\index{category}\index{object}\index{morphism}\index{composition law}\index{category axioms}A
\textsl{category $\mathfrak{C}$} consists of a set of \textsl{objects
$\mathsf{Obj}_{\mathfrak{C}}$}, a set of \textsl{morphisms $\mathsf{Mor}_{\mathfrak{C}}\left(A,B\right)$
from $A$ to $B$} for each pair of objects $\left(A,B\right)$ and
a map, called \textsl{composition law},
\begin{eqnarray*}
\circ:\mathsf{Mor}_{\mathfrak{C}}\left(B,C\right)\times\mathsf{Mor}_{\mathfrak{C}}\left(A,B\right) & \rightarrow & \mathsf{Mor}_{\mathfrak{C}}\left(A,C\right)\\
\left(g,f\right) & \mapsto & g\circ f
\end{eqnarray*}
for each triple of objects $\left(A,B,C\right)$. The following axioms
(we call them \textsl{category axioms}) are assumed to hold:
\begin{itemize}
\item \textsl{identity law}: for each $A\in\mathsf{Obj}_{\mathfrak{C}}$
the set $\mathsf{Mor}_{\mathfrak{C}}\left(A,A\right)$ must contain
at least an element $\mathrm{id}_{A}$ such that, for each $B\in\mathsf{Obj}_{\mathfrak{C}}$,
each $f\in\mathsf{Mor}_{\mathfrak{C}}\left(A,B\right)$ and each $g\in\mathsf{Mor}_{\mathfrak{C}}\left(B,A\right)$,
it holds that
\begin{eqnarray*}
f\circ\mathrm{id}_{A} & = & f\mbox{,}\\
\mathrm{id}_{A}\circ g & = & g\mbox{;}
\end{eqnarray*}

\item \textsl{associative law}: for each $A$, $B$, $C$, $D\in\mathsf{Obj}_{\mathfrak{C}}$,
each $f\in\mathsf{Mor}_{\mathfrak{C}}\left(A,B\right)$, each $g\in\mathsf{Mor}_{\mathfrak{C}}\left(B,C\right)$
and each $h\in\mathsf{Mor}_{\mathfrak{C}}\left(C,D\right)$ it holds
that
\[
h\circ\left(g\circ f\right)=\left(h\circ g\right)\circ f\mbox{.}
\]

\end{itemize}
\index{subcategory}Let $\mathfrak{C}$ be a category. A \textsl{subcategory
$\mathfrak{S}$ of $\mathfrak{C}$} is a category such that $\mathsf{Obj}_{\mathfrak{S}}\subseteq\mathsf{Obj}_{\mathfrak{C}}$,
$\mathsf{Mor}_{\mathfrak{S}}\left(A,B\right)\subseteq\mathsf{Mor}_{\mathfrak{C}}\left(A,B\right)$
for each $A$, $B\in\mathsf{Obj}_{\mathfrak{S}}$. Moreover we require
that:
\begin{itemize}
\item for each object $A$ of $\mathfrak{S}$ the identity morphism of $\mathsf{Mor}_{\mathfrak{S}}\left(A,A\right)$
coincides with the identity morphism of $\mathsf{Mor}_{\mathfrak{C}}\left(A,A\right)$;
\item for each $A$, $B$, $C\in\mathsf{Obj}_{\mathfrak{S}}$, each $f\in\mathsf{Mor}_{\mathfrak{S}}\left(A,B\right)$
and each $g\in\mathsf{Mor}_{\mathfrak{S}}\left(B,C\right)$ the composition
$g\circ f$ in $\mathfrak{S}$ coincides with the composition $g\circ f$
in $\mathfrak{C}$.
\end{itemize}
\index{full subcategory}We say that $\mathfrak{S}$ is a \textsl{full
subcategory of $\mathfrak{C}$} if it is a subcategory of $\mathfrak{C}$
and $\mathsf{Mor}_{\mathfrak{S}}\left(A,B\right)=\mathsf{Mor}_{\mathfrak{C}}\left(A,B\right)$
for each $A$, $B\in\mathsf{Obj}_{\mathfrak{S}}$.\end{defn}
\begin{example}
Examples of categories are:

the category whose objects are sets, whose morphisms are functions
between pairs of sets and whose composition law is provided by the
composition of functions;

the category whose objects are topological spaces, whose morphisms
are continuous functions between pairs of topological spaces and whose
composition law is provided by the composition of functions;

the category whose objects are groups, whose morphisms are homomorphisms
between pairs of groups and whose composition law is provided by the
composition of functions.

One easily checks the validity of the category axioms in these cases.
One may even note that the second category and the third category
are (non full) subcategories of the first one.
\end{example}
Now we define covariant and contravariant functors.
\begin{defn}
\label{defFunctor}\index{covariant functor}\index{covariant axioms}Let
$\mathfrak{A}$ and $\mathfrak{B}$ be two categories. A \textsl{covariant
functor $\mathscr{F}$ from $\mathfrak{A}$ to $\mathfrak{B}$} is
a map
\[
\mathscr{F}:\mathsf{Obj}_{\mathfrak{A}}\rightarrow\mathsf{Obj}_{\mathfrak{B}}
\]
together with a collection of maps
\[
\left\{ \mathscr{F}:\mathsf{Mor}_{\mathfrak{A}}\left(A,B\right)\rightarrow\mathsf{Mor}_{\mathfrak{B}}\left(\mathscr{F}\left(A\right),\mathscr{F}\left(B\right)\right)\mbox{ for }A,B\in\mathsf{Obj}_{\mathfrak{A}}\right\} 
\]
such that the following requirements, called \textsl{covariant axioms},
are fulfilled:
\begin{itemize}
\item the composition of morphisms is preserved, i.e. for each $A$, $B$,
$C\in\mathsf{Obj}_{\mathfrak{A}}$, each $f\in\mathsf{Mor}_{\mathfrak{A}}\left(A,B\right)$
and each $g\in\mathsf{Mor}_{\mathfrak{A}}\left(B,C\right)$ we have
\[
\mathscr{F}\left(g\circ f\right)=\mathscr{F}\left(g\right)\circ\mathscr{F}\left(f\right)\mbox{,}
\]
where on the LHS we have the $\mathfrak{A}$-composition law, while
ton the RHS we have the $\mathfrak{B}$-composition law;
\item the identity map of an object $A$ of $\mathfrak{A}$ is mapped to
the identity map of the corresponding object $\mathscr{F}\left(A\right)$
of $\mathfrak{B}$, i.e. for each $A\in\mathsf{Obj}_{\mathfrak{A}}$
we have
\[
\mathscr{F}\left(\mathrm{id}_{A}\right)=\mathrm{id}_{\mathscr{F}\left(A\right)}\mbox{.}
\]

\end{itemize}
\index{contravariant functor}\index{contravariant axioms}A \textsl{contravariant
functor $\mathscr{G}$ from $\mathfrak{A}$ to $\mathfrak{B}$} is
a map
\[
\mathscr{G}:\mathsf{Obj}_{\mathfrak{A}}\rightarrow\mathsf{Obj}_{\mathfrak{B}}
\]
together with a collection of maps
\[
\left\{ \mathscr{G}:\mathsf{Mor}_{\mathfrak{A}}\left(A,B\right)\rightarrow\mathsf{Mor}_{\mathfrak{B}}\left(\mathscr{G}\left(B\right),\mathscr{G}\left(A\right)\right)\mbox{ for }A,B\in\mathsf{Obj}_{\mathfrak{A}}\right\} 
\]
such that the following requirements, called \textsl{contravariant
axioms}, are fulfilled:
\begin{itemize}
\item the composition of morphisms is reversed, i.e. for each $A$, $B$,
$C\in\mathsf{Obj}_{\mathfrak{A}}$, each $f\in\mathsf{Mor}_{\mathfrak{B}}\left(A,B\right)$
and each $g\in\mathsf{Mor}_{\mathfrak{A}}\left(B,C\right)$ we have
\[
\mathscr{G}\left(g\circ f\right)=\mathscr{G}\left(f\right)\circ\mathscr{G}\left(g\right)\mbox{,}
\]
where on the LHS we have the $\mathfrak{A}$-composition law, while
ton the RHS we have the $\mathfrak{B}$-composition law;
\item the identity map of an object $A$ of $\mathfrak{A}$ is mapped to
the identity map of the corresponding object $\mathscr{G}\left(A\right)$
of $\mathfrak{B}$, i.e. for each $A\in\mathsf{Obj}_{\mathfrak{A}}$
we have
\[
\mathscr{G}\left(\mathrm{id}_{A}\right)=\mathrm{id}_{\mathscr{G}\left(A\right)}\mbox{.}
\]

\end{itemize}
\end{defn}
We sometimes denote a covariant functor $\mathscr{F}$ from a category
$\mathfrak{A}$ to a category $\mathfrak{B}$ with $\mathscr{F}:\mathfrak{A}\overset{\rightarrow}{\rightarrow}\mathfrak{B}$
(the direction of the upper arrow denotes that the composition is
preserved). On the contrary, for a contravariant functor $\mathscr{G}$
from $\mathfrak{A}$ to $\mathfrak{B}$ we write $\mathscr{F}:\mathfrak{A}\overset{\leftarrow}{\rightarrow}\mathfrak{B}$
(here the direction of the upper arrow denotes that the composition
is reversed).
\begin{example}
\index{forgetful functors}We show an example of a covariant functor.
Consider the category $\mathfrak{tsp}$ of topological spaces and
the category $\mathfrak{set}$ of sets. We define $\mathscr{F}$ imposing
$\mathscr{F}\left(X\right)=S$ for each $X\in\mathsf{Obj}_{\mathfrak{tsp}}$,
where $S$ is the underlying set of $X$ and $\tau$ is its topology,
and imposing $\mathscr{F}\left(f\right)=f$ for each $X_{1}$, $X_{2}\in\mathsf{Obj}_{\mathfrak{tsp}}$
and each $f\in\mathsf{Mor}_{\mathfrak{tsp}}\left(X_{1},X_{2}\right)$.
It is immediate to check that $\mathscr{F}$ satisfies the covariant
axioms. Notice that covariant functors like $\mathscr{F}$ are called
\textsl{forgetful functors}, since they {}``forget'' of some structure
or property possessed by the objects and morphisms of the starting
category.\end{example}
\begin{defn}
\label{defCompositionOfFunctors}\index{composition of functors}Let
$\mathscr{F}$ be a covariant functor from a category $\mathfrak{A}$
to a category $\mathfrak{B}$ and let $\mathscr{G}$ be a covariant
functor from $\mathfrak{B}$ to a category $\mathfrak{C}$. The composition
of $\mathscr{F}$ and $\mathscr{G}$ is the covariant functor whose
map between the objects $\mathscr{G}\circ\mathscr{F}:\mathsf{Obj}_{\mathfrak{A}}\rightarrow\mathsf{Obj}_{\mathfrak{C}}$
is the composition of the maps $\mathscr{F}:\mathsf{Obj}_{\mathfrak{A}}\rightarrow\mathsf{Obj}_{\mathfrak{B}}$
and $\mathscr{G}:\mathsf{Obj}_{\mathfrak{B}}\rightarrow\mathsf{Obj}_{\mathfrak{C}}$
and whose maps between the morphisms are defined in the following
way: for each $A$, \textbf{$B\in\mathsf{Obj}_{\mathfrak{A}}$}, we
obtain
\[
\mathscr{G}\circ\mathscr{F}:\mathsf{Mor}_{\mathfrak{A}}\left(A,B\right)\rightarrow\mathsf{Mor}_{\mathfrak{C}}\left(\left(\mathscr{G}\circ\mathscr{F}\right)\left(A\right),\left(\mathscr{G}\circ\mathscr{F}\right)\left(B\right)\right)
\]
composing the maps 
\begin{eqnarray*}
\mathscr{F}:\mathsf{Mor}_{\mathfrak{A}}\left(A,B\right) & \rightarrow & \mathsf{Mor}_{\mathfrak{B}}\left(\mathscr{F}\left(A\right),\mathscr{F}\left(B\right)\right)\mbox{,}\\
\mathscr{G}:\mathsf{Mor}_{\mathfrak{B}}\left(\mathscr{F}\left(A\right),\mathscr{F}\left(B\right)\right) & \rightarrow & \mathsf{Mor}_{\mathfrak{C}}\left(\left(\mathscr{G}\circ\mathscr{F}\right)\left(A\right),\left(\mathscr{G}\circ\mathscr{F}\right)\left(B\right)\right)\mbox{.}
\end{eqnarray*}

The composition of contravariant functors is a covariant functor defined
similarly, the only difference being that we must compose the maps
\begin{eqnarray*}
\mathscr{F}:\mathsf{Mor}_{\mathfrak{A}}\left(A,B\right) & \rightarrow & \mathsf{Mor}_{\mathfrak{B}}\left(\mathscr{F}\left(B\right),\mathscr{F}\left(A\right)\right)\mbox{,}\\
\mathscr{G}:\mathsf{Mor}_{\mathfrak{B}}\left(\mathscr{F}\left(B\right),\mathscr{F}\left(A\right)\right) & \rightarrow & \mathsf{Mor}_{\mathfrak{C}}\left(\left(\mathscr{G}\circ\mathscr{F}\right)\left(A\right),\left(\mathscr{G}\circ\mathscr{F}\right)\left(B\right)\right)
\end{eqnarray*}
to obtain
\[
\mathscr{G}\circ\mathscr{F}:\mathsf{Mor}_{\mathfrak{A}}\left(A,B\right)\rightarrow\mathsf{Mor}_{\mathfrak{C}}\left(\left(\mathscr{G}\circ\mathscr{F}\right)\left(A\right),\left(\mathscr{G}\circ\mathscr{F}\right)\left(B\right)\right)\mbox{.}
\]

Finally the composition of a covariant functor with a contravariant
functor (or vice versa) is the contravariant functor defined as above
paying attention to the reversal in the direction of the morphisms
caused by a contravariant functor.
\end{defn}
One can easily check that the definition above is well posed and that
the composed functors are actually covariant in the first two cases
and contravariant in last case. The composition of functors gives
us the opportunity to present a new example of category, the {}``category
of categories'', whose objects are categories, whose morphisms are
covariant and contravariant functors and whose composition law is
the composition of functors.

To conclude this section we want to introduce another notion from
category theory, specifically that of natural transformation.
\begin{defn}
\index{natural transformation}\index{naturality axiom}\index{component of a natural transformation}Let
$\mathfrak{A}$ and $\mathfrak{B}$ be categories and let $\mathscr{F}$
and $\mathscr{G}$ be covariant functors from $\mathfrak{A}$ to $\mathfrak{B}$.
A \textsl{covariant natural transformation $\mathtt{n}$ from $\mathscr{F}$
to $\mathscr{G}$} is a collection of morphisms of the category $\mathfrak{B}$
\[
\left\{ \mathtt{n}_{A}\in\mathsf{Mor}_{\mathfrak{B}}\left(\mathscr{F}\left(A\right),\mathscr{G}\left(A\right)\right)\mbox{ for }A\in\mathsf{Obj}_{\mathfrak{A}}\right\} 
\]
such that the following condition, called \textsl{covariant naturality
axiom}, is verified:
\begin{quote}
for each $A$, $B\in\mathsf{Obj}_{\mathfrak{A}}$ and each $f\in\mathsf{Mor}_{\mathfrak{A}}\left(A,B\right)$
we have that
\[
\mathtt{n}_{B}\circ\mathscr{F}\left(f\right)=\mathscr{G}\left(f\right)\circ\mathtt{n}_{A}\mbox{.}
\]

\end{quote}
Otherwise let $\mathscr{F}$ and $\mathscr{G}$ be contravariant functors
from $\mathfrak{A}$ to $\mathfrak{B}$. A \textsl{contravariant natural
transformation $\mathtt{n}$ from $\mathscr{F}$ to $\mathscr{G}$}
is again a collection of morphisms of the category $\mathfrak{B}$
\[
\left\{ \mathtt{n}_{A}\in\mathsf{Mor}_{\mathfrak{B}}\left(\mathscr{F}\left(A\right),\mathscr{G}\left(A\right)\right)\mbox{ for }A\in\mathsf{Obj}_{\mathfrak{A}}\right\} 
\]
such that the following condition, called \textsl{contravariant naturality
axiom}, is verified:
\begin{quote}
for each $A$, $B\in\mathsf{Obj}_{\mathfrak{A}}$ and each $f\in\mathsf{Mor}_{\mathfrak{A}}\left(A,B\right)$
we have that
\[
\mathtt{n}_{A}\circ\mathscr{F}\left(f\right)=\mathscr{G}\left(f\right)\circ\mathtt{n}_{B}\mbox{.}
\]

\end{quote}
For each $A\in\mathsf{Obj}_{\mathfrak{A}}$ we say that $\mathtt{n}_{A}$
is the \textsl{$A$-component of the natural transformation $\mathtt{n}$}
(whether $\mathtt{n}$ is covariant or contravariant).

\index{natural isomorphism}A \textsl{covariant (contravariant) natural
isomorphism $\mathtt{i}$} is a covariant (respectively contravariant)
natural transformation such that each of its components is an isomorphism
between the appropriate objects (i.e. a bijective morphism whose inverse
is a morphism).
\end{defn}
For natural transformations we introduce a notation (similar to the
one introduced for functors) that allows us to easily distinguish
the covariant case from the contravariant one: a covariant natural
transformation $\mathtt{n}$ from $\mathscr{F}:\mathfrak{A}\overset{\rightarrow}{\rightarrow}\mathfrak{B}$
to $\mathscr{G}:\mathfrak{A}\overset{\rightarrow}{\rightarrow}\mathfrak{B}$
will be denoted by $\mathtt{n}:\mathscr{F}\overset{\rightarrow}{\rightarrow}\mathscr{G}$,
whereas a contravariant natural transformation $\mathtt{m}$ from
$\mathscr{F}:\mathfrak{A}\overset{\leftarrow}{\rightarrow}\mathfrak{B}$
to $\mathscr{G}:\mathfrak{A}\overset{\leftarrow}{\rightarrow}\mathfrak{B}$
will be denoted by $\mathtt{m}:\mathscr{F}\overset{\leftarrow}{\rightarrow}\mathscr{G}$.

%% file: 7_chapter2_GCLP.tex
\chapter{\label{chapGCLP}The generally covariant locality principle}

This chapter is divided in three sections. In the first one, following
\cite{BFV03}, we present an approach to quantum field theory on curved
spacetimes known as \textsl{generally covariant locality principle}
(abbreviated by the acronym \textsl{GCLP}) and we study the properties
of \textsl{locally covariant quantum field theories} (or \textsl{LCQFT}),
that are quantum field theories formulated following the scheme provided
by the GCLP. Our main goal is to show that this family of quantum
field theories automatically satisfies the Haag-Kastler axioms, originally
stated in \cite{HK64}. Hence on the one hand the GCLP recovers exactly
the algebraic approach to quantum field theory suggested by Haag and
Kastler, while on the other hand it has the advantage of emphasizing
the common features of the quantization procedures on different spacetimes
and elegantly accounts for the covariance property required by general
relativity for any theory to be physical.

In the second section we show how a LCQFT can be constructed starting
from the Cauchy problem for a classical field over a globally hyperbolic
spacetime. Here we follow an approach similar to that in \cite[Sect. 4.3]{BFV03}.

We conclude this chapter showing some examples of concrete locally
covariant quantum field theories. Specifically we study the cases
of the Klein-Gordon field, of the Proca field and of the electromagnetic
field.

\section{\label{secLCQFT}Locally covariant quantum field theory}

Locally covariant quantum field theories are defined in terms of covariant
functors between appropriate categories. The first part of this section
is devoted to a detailed presentation of such categories.

\subsection{The categories $\mathfrak{ghs}$ and $\mathfrak{alg}$}

We start defining both $\mathfrak{ghs}$ and $\mathfrak{alg}$. In
the subsequent remarks we study in detail some properties of their
morphisms and then we check that they actually satisfy the category
axioms stated in Definition \ref{defCategory}.
\begin{defn}
\label{defghsalg}The category $\mathfrak{ghs}$ is defined in the
following way:
\begin{itemize}
\item Objects are $d$-dimensional globally hyperbolic spacetimes $\mathscr{M}=\left(M,g,\mathfrak{o},\mathfrak{t}\right)$;
\item The set of morphism $\mathsf{Mor}_{\mathfrak{ghs}}\left(\mathscr{M},\mathscr{N}\right)$
between the objects $\mathscr{M}=\left(M,g,\mathfrak{o},\mathfrak{t}\right)$
and $\mathscr{N}=\left(N,h,\mathfrak{p},\mathfrak{u}\right)$ encompasses
all the orientation ($\psi_{*}^{\prime}\mathfrak{o}=\left.\mathfrak{p}\right|_{\psi\left(M\right)}$)
and time orientation ($\psi_{*}^{\prime}\mathfrak{t}=\left.\mathfrak{u}\right|_{\psi\left(M\right)}$)
preserving isometric embeddings $\psi:\mathscr{M}\rightarrow\mathscr{N}$
whose images $\psi\left(M\right)$ are $\mathscr{N}$-causally convex
open subsets of $N$;
\item The composition law is provided by the usual composition of functions.
\end{itemize}
$\mathfrak{alg}$ is the category whose objects are unital C{*}-algebras,
whose set of morphisms $\mathsf{Mor}_{\mathfrak{alg}}\left(\mathcal{A},\mathcal{B}\right)$
between the objects $\mathcal{A}$ and $\mathcal{B}$ comprises all
the injective unit preserving {*}-homomorphisms $H:\mathcal{A}\rightarrow\mathcal{B}$
and whose composition law is again the usual composition of functions.
\end{defn}
Before the check of the category axioms for $\mathfrak{ghs}$ and
$\mathfrak{alg}$, we devote few lines to some comments on their morphisms.
\begin{rem}
\label{remghs}Dealing with $\mathfrak{ghs}$, consider $\mathscr{M}=\left(M,g,\mathfrak{o},\mathfrak{t}\right)$,
$\mathscr{N}=\left(N,h,\mathfrak{p},\mathfrak{u}\right)\in\mathsf{Obj}_{\mathfrak{ghs}}$
and $\psi\in\mathsf{Mor}_{\mathfrak{ghs}}\left(\mathscr{M},\mathscr{N}\right)$.
We have that $\psi\left(M\right)$ is a $\mathscr{N}$-causally convex
open subset of $N$. It is also connected because it is the image
through $\psi$ of $M$, which is connected being a manifold. Then,
recalling Remark \ref{remRestrictionOfOrientedAndTimeOrientedLorentzianManifolds},
we can consider the oriented and time oriented Lorentzian manifold
$\left.\mathscr{N}\right|_{\psi\left(M\right)}=\left(\psi\left(M\right),\left.h\right|_{\psi\left(M\right)},\left.\mathfrak{p}\right|_{\psi\left(M\right)},\left.\mathfrak{u}\right|_{\psi\left(M\right)}\right)$.
If we consider the diffeomorphism $\psi^{\prime}:M\rightarrow\psi\left(M\right)$
(see the end of Remark \ref{remSubmanifold}) and we recall that $\psi$
is isometric and preserves orientation and time orientation, we can
introduce on $\psi\left(M\right)$ the (fiberwise) symmetric and (fiberwise)
non degenerate section of $\mathrm{T}^{\left(0,2\right)}\psi\left(M\right)$
$\psi_{*}^{\prime}g=\left.h\right|_{\psi\left(M\right)}$, the set
of $d$-forms $\psi_{*}^{\prime}\mathfrak{o}=\left.\mathfrak{p}\right|_{\psi\left(M\right)}$
and the vector field $\psi_{*}^{\prime}\mathfrak{t}=\left.\mathfrak{u}\right|_{\psi\left(M\right)}$.
Hence we recognize that $\psi_{*}^{\prime}g$ is a Lorentzian metric
on $\psi\left(M\right)$, that $\psi\left(M\right)$ is orientable
and $\psi_{*}^{\prime}\mathfrak{o}$ is a choice of an orientation
and that $\left(\psi\left(M\right),\psi_{*}^{\prime}g\right)$ is
a time orientable Lorentzian manifold and $\psi_{*}^{\prime}\mathfrak{t}$
is a choice of a time orientation. Therefore we can define the oriented
and time oriented Lorentzian manifold $\left(\psi\left(M\right),\psi_{*}^{\prime}g,\psi_{*}^{\prime}\mathfrak{o},\psi_{*}^{\prime}\mathfrak{t}\right)$
that we denote with $\psi\left(\mathscr{M}\right)$ and it immediately
turns out that $\psi\left(\mathscr{M}\right)=\left.\mathscr{N}\right|_{\psi\left(M\right)}$.
So we will usually write $\psi\left(\mathscr{M}\right)$ in place
of $\left.\mathscr{N}\right|_{\psi\left(M\right)}$. There is even
more: applying Proposition \ref{propCausalConvexityImpliesGlobalHyperbolicity},
we realize that $\psi\left(M\right)$ is an $\mathscr{N}$-globally
hyperbolic connected open subset of $N$ and then, applying Remark
\ref{remRestrictionToGloballyHyperbolicSubsets}, we deduce that $\psi\left(\mathscr{M}\right)$
is itself a $d$-dimensional globally hyperbolic spacetime, i.e. an
object of $\mathfrak{ghs}$ in its own right, and we can easily recognize
that the following two maps are actually morphisms of $\mathfrak{ghs}$:
\begin{itemize}
\item $\psi^{\prime}$ becomes a bijective morphism from $\mathscr{M}$
to $\psi\left(\mathscr{M}\right)$ whose inverse $\psi^{\prime-1}$
is a morphism from $\psi\left(\mathscr{M}\right)$ to $\mathscr{M}$;
\item the inclusion map $\iota_{\psi\left(M\right)}^{N}$ of $\psi\left(M\right)$
into $N$ becomes a morphism from $\psi\left(\mathscr{M}\right)$
to $\mathscr{N}$: This is a consequence of a more general fact that
holds for each object $\mathscr{O}=\left(O,i,\mathfrak{q},\mathfrak{v}\right)$
of $\mathfrak{ghs}$ and each $\mathscr{O}$-causally convex connected
open subset $\Omega$ of $O$, specifically that the inclusion map
$\iota_{\Omega}^{O}$ of $\Omega$ in $O$ is actually a morphism
from $\left.\mathscr{O}\right|_{\Omega}$ to $\mathscr{O}$ (to check
this fact note that Remark \ref{remSubmanifold} implies that $\Omega$
is a submanifold of $O$ and that the inclusion map $\iota_{\Omega}^{O}$
is an embedding and apply Proposition \ref{propCausalConvexityImpliesGlobalHyperbolicity}
and Remark \ref{remRestrictionToGloballyHyperbolicSubsets} to obtain
the globally hyperbolic spacetime $\left.\mathscr{O}\right|_{\Omega}=\left(\Omega,\left.i\right|_{\Omega},\left.\mathfrak{q}\right|_{\Omega},\left.\mathfrak{v}\right|_{\Omega}\right)$).
\end{itemize}
Using these two facts we can decompose each $\psi\in\mathsf{Mor}_{\mathfrak{ghs}}\left(\mathscr{M},\mathscr{N}\right)$
in two morphisms $\iota_{\psi\left(M\right)}^{N}\in\mathsf{Mor}_{\mathfrak{ghs}}\left(\psi\left(\mathscr{M}\right),\mathscr{N}\right)$
and $\psi^{\prime}\in\mathsf{Mor}_{\mathfrak{ghs}}\left(\mathscr{M},\psi\left(\mathscr{M}\right)\right)$
(which is bijective and whose inverse is a morphism from $\psi\left(\mathscr{M}\right)$
to $\mathscr{M}$) according to the formula $\psi=\iota_{\psi\left(M\right)}^{N}\circ\psi^{\prime}$.
\end{rem}

\begin{rem}
\label{remalg}As anticipated, we make some observations also on the
morphisms of $\mathfrak{alg}$. Recalling Proposition \ref{propContinuityOfUnitPreserving*-HomomorphismsOnUnitalC*-Algebras}
and bearing in mind that all the objects of $\mathfrak{alg}$ are
unital C{*}-algebras, we see that each morphism of this category can
also be seen as an isometry between the Banach spaces underlying its
domain and its codomain. We can use this fact to obtain results similar
to that found for the morphisms of $\mathfrak{ghs}$. Specifically
consider two objects $\mathcal{A}$ and $\mathcal{B}$ and a morphism
$H:\mathcal{A}\rightarrow\mathcal{B}$ of $\mathfrak{alg}$. We consider
the vector spaces $A$ and $B$ that underlie $\mathcal{A}$ and respectively
$\mathcal{B}$ and we focus on the image $H\left(A\right)$ of $A$,
which is trivially a vector space because $H$ is linear. On a side
we consider the sub-C{*}-algebra $\mathcal{B}_{H\left(A\right)}$
of $\mathcal{B}$ generated by $H\left(A\right)$ (cfr. Remark \ref{remGeneratedSubalgebra}).
Since $H$ is compatible with the multiplications and the involutions
of $\mathcal{A}$ and $\mathcal{B}$, it follows that $H\left(A\right)$
endowed with the restriction of the product and of the involution
of $\mathcal{B}$ is a {*}-algebra with unit $H1_{\mathcal{A}}=1_{\mathcal{B}}$
and the map $H^{\prime}:A\rightarrow H\left(A\right)$, defined by
$H^{\prime}a=Ha$, is a {*}-isomorphism from $\mathcal{A}$ to the
{*}-algebra $H\left(A\right)$. We have seen that $H$ is an injective
isometry between the Banach spaces $\mathcal{A}$ and $\mathcal{B}$.
This allows us to recognize that $H\left(A\right)$ is a closed subspace
of $\mathcal{B}$. Consider in fact a sequence $\left\{ b_{n}\right\} $
of elements of the vector space $H\left(A\right)$ that converges
to $b\in\mathcal{B}$ with respect to the norm of $\mathcal{B}$ and
take the sequence $\left\{ a_{n}=H^{-1}b_{n}\right\} $ in $\mathcal{A}$:
since $\left\{ b_{n}\right\} $ is a Cauchy sequence in $\mathcal{B}$
(as a consequence of being convergent) and $H$ is an isometry, it
follows that $\left\{ a_{n}\right\} $ is a Cauchy sequence in $\mathcal{A}$:
\[
\left\Vert a_{n}-a_{m}\right\Vert =\left\Vert Ha_{n}-Ha_{m}\right\Vert =\left\Vert b_{n}-b_{m}\right\Vert \mbox{.}
\]
But $\mathcal{A}$ is a Banach space and hence we find the limit $a\in\mathcal{A}$
of the sequence $\left\{ a_{n}\right\} $ with respect to the norm
of $\mathcal{A}$. Hence, bearing in mind that $H$ is in particular
continuous between the Banach spaces $\mathcal{A}$ and $\mathcal{B}$,
we have the following situation:
\[
Ha\overset{\infty\leftarrow n}{\longleftarrow}Ha_{n}=b_{n}\overset{n\rightarrow\infty}{\longrightarrow}b\mbox{.}
\]
The uniqueness of the limit in $\mathcal{B}$ implies that $Ha=b$,
hence in particular $b\in H\left(A\right)$. This proves that $H\left(A\right)$
is actually a closed subspace of $\mathcal{B}$. Then the unital {*}-algebra
$H\left(A\right)$ endowed with the restriction of the norm of $\mathcal{B}$
defines a unital sub-C{*}-algebra of $\mathcal{B}$ (cfr. Definition
\ref{defSubalgebra}) that we denote with $H\left(\mathcal{A}\right)$.
Since $\mathcal{B}_{H\left(A\right)}$ is by definition the smallest
sub-C{*}-algebra of $\mathcal{B}$ including $H\left(A\right)$ and
the vector space underlying $H\left(\mathcal{A}\right)$ coincides
exactly with $H\left(A\right)$, we conclude that $H\left(\mathcal{A}\right)=\mathcal{B}_{H\left(A\right)}$.
It also turns out that we have at our disposal two new morphisms of
$\mathfrak{alg}$:
\begin{itemize}
\item $H^{\prime}:\mathcal{A}\rightarrow H\left(\mathcal{A}\right)$, which
is in particular a unit preserving {*}-isomorphism between unital
C{*}-algebras and hence, from Remark \ref{remSurjective*-HomomorphismFromUnitalC*-AlgebrasToC*-Algebras},
an isometric isomorphism between the Banach spaces $\mathcal{A}$
and $H\left(\mathcal{A}\right)$ too;
\item the inclusion map $\iota_{H\left(A\right)}^{B}$ of $H\left(A\right)$
in $B$, which is recognized to be an injective unit preserving {*}-homomorphism
between the unital C{*}-algebras $H\left(\mathcal{A}\right)$ and
$\mathcal{B}$: This is a consequence of a more general fact that
holds for each C{*}-algebra $\mathcal{C}$ and each sub-C{*}-algebra
$\mathcal{S}$ of $\mathcal{C}$, specifically that the inclusion
map $\iota_{S}^{C}$ of the vector space $S$ underlying $\mathcal{S}$
in the vector space $C$ underlying $\mathcal{C}$ is recognized to
be an injective unit preserving {*}-homomorphism between the C{*}-algebras
$\mathcal{S}$ and $\mathcal{C}$.
\end{itemize}
Using the construction above, we can decompose each morphism $H:\mathcal{A}\rightarrow\mathcal{B}$
of $\mathfrak{alg}$ in the morphisms $\iota_{H\left(A\right)}^{B}\in\mathsf{Mor}_{\mathfrak{alg}}\left(H\left(\mathcal{A}\right),\mathcal{B}\right)$
and $H^{\prime}\in\mathsf{Mor}_{\mathfrak{alg}}\left(\mathcal{A},H\left(\mathcal{A}\right)\right)$
(which is also a {*}-isomorphism) according to the formula $H=\iota_{H\left(A\right)}^{B}\circ H^{\prime}$.
\end{rem}
Now we are ready to check that $\mathfrak{ghs}$ and $\mathfrak{alg}$
are actually categories.
\begin{rem}
We begin from $\mathfrak{ghs}$. If we take $\mathscr{M}=\left(M,g,\mathfrak{o},\mathfrak{t}\right)$,
$\mathscr{N}=\left(N,h,\mathfrak{p},\mathfrak{u}\right)$, $\mathscr{O}=\left(O,i,\mathfrak{q},\mathfrak{v}\right)$
in $\mathsf{Obj}_{\mathfrak{ghs}}$ and $\phi\in\mathsf{Mor}_{\mathfrak{ghs}}\left(\mathscr{M},\mathscr{N}\right)$,
$\psi\in\mathsf{Mor}_{\mathfrak{ghs}}\left(\mathscr{N},\mathscr{O}\right)$,
we immediately realize that $\psi\circ\phi:M\rightarrow O$ is a smooth
map and an immersion as a consequence of the same properties for $\psi$
$\phi$ and $\psi$. To prove that it is also an embedding with open
image, in first place we must show that $\left(\psi\circ\phi\right)\left(M\right)=\psi\left(\phi\left(M\right)\right)$
is an open subset of $O$. This is true because $\phi\left(M\right)$
is an open subset of $N$ and $\psi$ is an open map from $N$ to
$O$ (see the end of Remark \ref{remSubmanifold}). After that one
applies Remark \ref{remSubmanifold} to $\left(\psi\circ\phi\right)\left(M\right)$,
obtains a $d$-dimensional submanifold of $O$ and realizes that $\psi\circ\phi$
is an embedding because $\left(\psi\circ\phi\right)^{\prime}$ can
be written as the composition of $\left.\psi^{\prime}\right|_{\phi\left(M\right)}:\phi\left(M\right)\rightarrow\psi\left(\phi\left(M\right)\right)$
and $\phi^{\prime}:M\rightarrow\phi\left(M\right)$, which are both
diffeomorphisms. Then we must check $\psi\circ\phi$ is isometric
and preserves orientation and time orientation. This can be directly
checked exploiting the same properties that are assumed to hold for
both $\phi$ and $\psi$. Now the question is whether the image of
$M$ through $\psi\circ\phi$ is a causally convex subset of $\mathscr{O}$
or not. We try to give an answer fixing $p$, $q\in\left(\psi\circ\phi\right)\left(M\right)$.
We take a causal curve $\gamma$ in $\mathscr{O}$ connecting $p$
and $q$ and we check that it is entirely contained in $\left(\psi\circ\phi\right)\left(M\right)$.
Since $p$ and $q$ are obviously in $\psi\left(N\right)$, that is
$\mathscr{O}$-causally convex by hypothesis, it follows that $\gamma$
is contained in $\psi\left(N\right)$. Then we can use the isometric
diffeomorphism $\psi^{\prime}$ to construct $\gamma^{\prime}=\psi^{\prime-1}\circ\gamma$.
This is an $h$-causal curve in $N$ due to the fact that $\psi^{\prime-1}:\psi\left(\mathscr{N}\right)\rightarrow\mathscr{N}$
is an isometric diffeomorphism and it connects the points $p^{\prime}=h^{\prime-1}\left(p\right)$
and $q^{\prime}=h^{\prime-1}\left(q\right)$ of $N$. But $p^{\prime}$
and $q^{\prime}$ are also points of $\phi\left(M\right)$ since $p$,
$q\in\left(\psi\circ\phi\right)\left(M\right)$. Then by the same
argument applied to $\phi$ in place of $\psi$, we obtain that $\gamma^{\prime}$
is entirely contained in $\phi\left(M\right)$. From this we conclude
that $\gamma$ is contained in $\left(\psi\circ\phi\right)\left(M\right)$
and hence this subset of $O$ is indeed $\mathscr{O}$-causally convex.
This proves that $\psi\circ\phi$ is actually an element of $\mathsf{Mor}_{\mathfrak{ghs}}\left(\mathscr{M},\mathscr{O}\right)$
and so the law of composition is well defined. We must still check
that the category axioms hold. For each $\mathscr{M}\in\mathsf{Obj}_{\mathfrak{ghs}}$
it is easy to check that the identity morphism is provided by the
function $M\rightarrow M$, $p\mapsto p$ and so also the identity
law is verified. As for the associativity of the composition law,
it holds because the ordinary composition of functions is always associative.

Now we focus on $\mathfrak{alg}$. Here the situation is even simpler.
Taking $\mathcal{A}$, $\mathcal{B}$, $\mathcal{C}\in\mathsf{Obj}_{\mathfrak{alg}}$
and $H\in\mathsf{Mor}_{\mathfrak{alg}}\left(\mathcal{A},\mathcal{B}\right)$,
$K\in\mathsf{Mor}_{\mathfrak{alg}}\left(\mathcal{B},\mathcal{C}\right)$,
we immediately realize that $K\circ H:\mathcal{A}\rightarrow\mathcal{C}$
makes sense and gives an injective unit preserving {*}-homomorphism.
In order to show the strategy of proof for the last statement, we
explicitly prove that $H\circ K$ is actually compatible with the
involutions of $\mathcal{A}$ and of $\mathcal{C}$. Fix $a\in\mathcal{A}$.
Since both $H$ and $K$ are {*}-homomorphisms between the appropriate
algebras by hypothesis, it follows that
\[
\left(H\circ K\right)\left(a^{*}\right)=HK\left(a^{*}\right)=H\left(\left(Ka\right)^{*}\right)=\left(HKa\right)^{*}=\left(\left(H\circ K\right)a\right)^{*}\mbox{.}
\]
For each $\mathcal{A}\in\mathsf{Obj}_{\mathfrak{alg}}$, we recognize
the map $\mathcal{A}\rightarrow\mathcal{A}$, $a\mapsto a$ to be
the identity morphism of $\mathcal{A}$. As before, the associativity
of the composition law is trivial.
\end{rem}
At this point we have at hand all the material needed to state the
generally covariant locality principle.

\subsection{\index{generally covariant locality principle}\index{GCLP}Formulation
of the generally covariant locality principle}
\begin{quote}
The \textsl{generally covariant locality principle} (briefly \textsl{GCLP})
imposes that each quantum field theory on each globally hyperbolic
spacetime must be formulated as a \textsl{locally covariant quantum
field theory} (\textsl{LCQFT}).
\end{quote}
Since we have not yet defined what it is meant for a LCQFT, the statement
of the GCLP is still an empty box. We fill this box with the next
definition and we take the chance to state two additional properties
that can be required to a LCQFT. Later we will see that the fulfilment
of these additional properties allows us to completely recover the
Haag-Kastler axioms starting from the GCLP.
\begin{defn}
\label{defLCQFT}\index{locally covariant quantum field theory}\index{LCQFT}\index{causality condition}\index{time slice axiom}We
call \textsl{locally covariant quantum field theory} (or \textsl{LCQFT})
any covariant functor $\mathscr{A}$ from the category $\mathfrak{ghs}$
to the category $\mathfrak{alg}$.

A locally covariant quantum field theory $\mathscr{A}$ is said to
be \textsl{causal} if the following condition (called \textsl{causality
condition}) holds for each $\mathscr{M}_{1}=\left(M_{1},g_{1},\mathfrak{o}_{1},\mathfrak{t}_{1}\right)$,
$\mathscr{M}_{2}=\left(M_{2},g_{2},\mathfrak{o}_{2},\mathfrak{t}_{2}\right)$,
$\mathscr{M}=\left(M,g,\mathfrak{o},\mathfrak{t}\right)\in\mathsf{Obj}_{\mathfrak{ghs}}$,
each $\psi_{1}\in\mathsf{Mor}_{\mathfrak{ghs}}\left(\mathscr{M}_{1},\mathscr{M}\right)$
and each $\psi_{2}\in\mathsf{Mor}_{\mathfrak{ghs}}\left(\mathscr{M}_{2},\mathscr{M}\right)$
such that $\psi_{1}\left(M_{1}\right)$ and $\psi_{2}\left(M_{2}\right)$
are $\mathscr{M}$-causally separated subsets of $M$:
\begin{quote}
the elements of the image through the morphism $\mathscr{A}\left(\psi_{1}\right)$
of the object $\mathscr{A}\left(\mathscr{M}_{1}\right)$ commute with
the elements of the image through the morphism $\mathscr{A}\left(\psi_{2}\right)$
of the object $\mathscr{A}\left(\mathscr{M}_{2}\right)$, i.e.
\[
\left[\mathscr{A}\left(\psi_{1}\right)\left(\mathscr{A}\left(\mathscr{M}_{1}\right)\right),\mathscr{A}\left(\psi_{2}\right)\left(\mathscr{A}\left(\mathscr{M}_{2}\right)\right)\right]=\left\{ 0\right\} \mbox{,}
\]
where $0$ is the zero element of the C{*}-algebra $\mathscr{A}\left(\mathscr{M}\right)$.
\end{quote}
Moreover $\mathscr{A}$ is said to fulfil the \textsl{time slice axiom}
if the following condition holds for each $\mathscr{M}=\left(M,g,\mathfrak{o},\mathfrak{t}\right)$,
$\mathscr{N}\in\mathfrak{ghs}$ and each $\psi\in\mathsf{Mor}_{\mathfrak{ghs}}\left(\mathscr{M},\mathscr{N}\right)$
such that $\psi\left(M\right)$ contains a smooth spacelike Cauchy
surface for $\mathscr{N}$:
\begin{quote}
the morphism $\mathscr{A}\left(\psi\right)$ is surjective, i.e.
\[
\mathscr{A}\left(\psi\right)\left(\mathscr{A}\left(\mathscr{M}\right)\right)=\mathscr{A}\left(\mathscr{N}\right)\mbox{.}
\]

\end{quote}
\end{defn}
\begin{rem}
\label{remInterpretation}Even if a precise discussion on the physical
meaning of the generally covariant locality principle could be conducted
after the recovering of the algebraic quantum field theory framework
proposed by Haag and Kastler (cfr. \cite{HK64}) simply borrowing
the interpretation of the Haag-Kastler axioms, we want to make some
considerations on the last definition (as a matter of fact on the
GCLP) from now.

The first thing that we notice is that the functorial structure of
any locally covariant quantum field theory implements a sort of geometrical
locality in quantum field theory. We realize this fact considering
a LCQFT $\mathscr{A}:\mathfrak{ghs}\overset{\rightarrow}{\rightarrow}\mathfrak{alg}$,
an arbitrary globally hyperbolic spacetime $\mathscr{M}=\left(M,g,\mathfrak{o},\mathfrak{t}\right)$
and a $\mathscr{M}$-causally convex connected open subset $\Omega$
of $M$. From the last part of Remark \ref{remghs} we deduce that
$\iota_{\Omega}^{M}\in\mathsf{Mor}_{\mathfrak{ghs}}\left(\left.\mathscr{M}\right|_{\Omega},\mathscr{M}\right)$,
hence we consider $\mathscr{A}\left(\iota_{\Omega}^{M}\right)$, which
is a morphism of $\mathfrak{alg}$ from $\mathscr{A}\left(\left.\mathscr{M}\right|_{\Omega}\right)$
to $\mathscr{A}\left(\mathscr{M}\right)$, and we focus on its image
$\mathscr{A}\left(\iota_{\Omega}^{M}\right)$$\left(\mathscr{A}\left(\left.\mathscr{M}\right|_{\Omega}\right)\right)$.
Recalling Remark \ref{remalg}, we realize that $\mathscr{A}\left(\iota_{\Omega}^{M}\right)$$\left(\mathscr{A}\left(\left.\mathscr{M}\right|_{\Omega}\right)\right)$
is a unital sub-C{*}-algebra of the unital C{*}-algebra $\mathscr{A}\left(\mathscr{M}\right)$.
This is exactly what we mean by geometrical locality: A causally convex
connected open subset of a globally hyperbolic spacetime, when intended
as a globally hyperbolic spacetime in its own right, is associated
by a LCQFT $\mathscr{A}$ to a unital C{*}-algebra whose image (through
the morphism of $\mathfrak{ghs}$ obtained via $\mathscr{A}$ from
the inclusion map of $\Omega$ in $M$) is a unital sub-C{*}-algebras
of the unital C{*}-algebra associated via $\mathscr{A}$ to the entire
globally hyperbolic spacetime.

This geometrical locality allows us to introduce a physical interpretation.
We assume that, given a LCQFT $\mathscr{A}$ and a globally hyperbolic
spacetime $\mathscr{M}=\left(M,g,\mathfrak{o},\mathfrak{t}\right)$,
for each $\mathscr{M}$-causally convex relatively compact connected
open subset $\Omega$ of $M$, the unital sub-C{*}-algebra $\mathscr{A}\left(\iota_{\Omega}^{M}\right)\mathscr{A}\left(\left.\mathscr{M}\right|_{\Omega}\right)$
of the full unital C{*}-algebra $\mathscr{A}\left(\mathscr{M}\right)$
is the mathematical representation of the quantum observables that
could be measured on $\Omega$. Notice that this interpretation cannot
be applied to the full algebra $\mathscr{A}\left(\mathscr{M}\right)$
because $M$ cannot be compact (if it were, it would violate the causality
condition, cfr. \cite[Chap. 14, Lem. 13, p. 407]{O'N83}). By this
assumption we mean that we consider as physical observables only those
that can be measured on {}``small'' regions of the spacetime (precisely
$\mathscr{M}$-causally convex relatively compact connected open subsets
of $M$). Such choice is done because it doesn't appear physically
sensible to deal with an observable on a too large region since we
are not able to realize an experimental apparatus that makes measurements
for an observable {}``everywhere in space and time'', or anyway
on a region to much extended {}``in space'' or {}``in time'' (or
both). The entire algebra of quantum observables on a given globally
hyperbolic spacetime is obtained as the unital sub-C{*}-algebra of
$\mathscr{A}\left(\mathscr{M}\right)$ generated by all the observables
that we classified as physical. We use this interpretation to explore
the physical meaning of some properties of a locally covariant quantum
field theory.

Returning to the definition of a LCQFT, we notice that it is nothing
but a covariant functor from $\mathfrak{ghs}$ to $\mathfrak{alg}$,
which is to say that the GCLP simply states that each quantum field
theory must be formulated as a covariant functor that assigns a unital
C{*}-algebra to each globally hyperbolic spacetime and an injective
unit preserving {*}-homomorphism between the appropriate unital C{*}-algebras
to each orientation and time orientation preserving isometric embedding
between globally hyperbolic spacetimes whose image is a causally convex
open subset of the target spacetime. The physical sense that we obtain
in light of our interpretation is the following: For each globally
hyperbolic spacetime and each {}``sufficiently small'' region, we
have a unital sub-C{*}-algebra that represents the quantum observables
on that region and all these unital sub-C{*}-algebras generate the
entire algebra of observables on the given globally hyperbolic spacetime.
The power of the GCLP resides in this fact, that is the possibility
of discussing a quantum field theory on all the globally hyperbolic
spacetimes at once.

This functorial structure automatically incorporates in quantum field
theory the notion of general covariance under the transformations
induced by a group of isometric diffeomorphisms of the globally hyperbolic
spacetime. We will see this in detail when the Haag-Kastler axioms
will be recovered. In our interpretation this means that we expect
to find a representation of the group of isometric diffeomorphisms
in terms of a group of automorphisms on the algebra of observables
and that we require that such representation satisfies covariance
(as intended in the language of category theory).

To give a physical interpretation of the property of geometrical locality
encountered before, we proceed in the following way. Let $\Omega$
and $\Theta$ be $\mathscr{M}$-causally convex relatively compact
connected open subsets of $M$ such that $\Omega\subseteq\Theta$.
We can consider the globally hyperbolic spacetime $\left.\mathscr{M}\right|_{\Theta}=\left(\Theta,\left.g\right|_{\Theta},\left.\mathfrak{o}\right|_{\Theta},\left.\mathfrak{t}\right|_{\Theta}\right)$
and we immediately recognize that $\Omega$ is a $\left.\mathscr{M}\right|_{\Theta}$-causally
convex connected open subsets of $\Theta$, so that we can also consider
the globally hyperbolic spacetime $\left.\left.\mathscr{M}\right|_{\Theta}\right|_{\Omega}$,
which coincides with $\left.\mathscr{M}\right|_{\Omega}$ as it is
easily seen. Hence we can consider the inclusion map $\iota_{\Omega}^{\Theta}$
and we realize that this is a morphism of $\mathfrak{ghs}$ from $\left.\left.\mathscr{M}\right|_{\Theta}\right|_{\Omega}=\left.\mathscr{M}\right|_{\Omega}$
to $\left.\mathscr{M}\right|_{\Theta}$. This leads us to the conclusion
that $\mathscr{A}\left(\iota_{\Omega}^{\Theta}\right)\left(\mathscr{A}\left(\left.\mathscr{M}\right|_{\Omega}\right)\right)$
is a unital sub-C{*}-algebra of the unital C{*}-algebra $\mathscr{A}\left(\left.\mathscr{M}\right|_{\Theta}\right)$.
This suggests that a sort of isotony holds for the algebras of observables
associated to proper regions of a globally hyperbolic spacetime: If
$\Omega$ is smaller than $\Theta$, then we expect that the algebra
of observables on $\Omega$ is a subalgebra of the algebra of observables
of $\Theta$ (and both are trivially subalgebras of the complete algebra
of observables associated to the given globally hyperbolic spacetime).

Now we turn our attention to the causality condition. We begin observing
that the causality condition makes sense because of the functorial
structure of each LCQFT $\mathscr{A}$: Taking three objects $\mathscr{M}$,
$\mathscr{M}_{1}=\left(M_{1},g_{1},\mathfrak{o}_{1},\mathfrak{t}_{1}\right)$
and $\mathscr{M}_{2}=\left(M_{2},g_{2},\mathfrak{o}_{2},\mathfrak{t}_{2}\right)$
and two morphisms $\psi_{1}:\mathscr{M}_{1}\rightarrow\mathscr{M}$
and $\psi_{2}:\mathscr{M}_{2}\rightarrow\mathscr{M}$ such that $\psi_{1}\left(M_{1}\right)$
and $\psi_{2}\left(M_{2}\right)$ are $\mathscr{M}$-causally separated,
we can evaluate the commutator of an element of $\mathscr{A}\left(\psi_{1}\right)\left(\mathscr{A}\left(\mathscr{M}_{1}\right)\right)$
with an element of $\mathscr{A}\left(\psi_{2}\right)\left(\mathscr{A}\left(\mathscr{M}_{2}\right)\right)$
because, owing to the functorial structure, both $\mathscr{A}\left(\psi_{1}\right)\left(\mathscr{A}\left(\mathscr{M}_{1}\right)\right)$,
$\mathscr{A}\left(\psi_{2}\right)\left(\mathscr{A}\left(\mathscr{M}_{2}\right)\right)$
are unital sub-C{*}-algebras of $\mathscr{A}\left(\mathscr{M}\right)$.

From a physical point of view the causality condition imposes some
restrictions to the causal structure of a LCQFT $\mathscr{A}$. We
can sketch the typology of such restrictions considering the globally
hyperbolic spacetime $\mathscr{M}=\left(M,g,\mathfrak{o},\mathfrak{t}\right)$
and two $\mathscr{M}$-causally convex relatively compact connected
open subsets $\Omega$ and $\Theta$ of $M$ that are $\mathscr{M}$-causally
separated. As usual we interpret $\Omega$ and $\Theta$ as been globally
hyperbolic spacetimes in their own right (denoted respectively by
$\left.\mathscr{M}\right|_{\Omega}$ and $\left.\mathscr{M}\right|_{\Theta}$)
and we take into account the inclusion maps $\iota_{\Omega}^{M}$
and $\iota_{\Theta}^{M}$ (which are actually morphisms of $\mathfrak{ghs}$
respectively from $\left.\mathscr{M}\right|_{\Omega}$ and from $\left.\mathscr{M}\right|_{\Theta}$
to $\mathscr{M}$). The causality condition imposes that
\[
\left[\mathscr{A}\left(\iota_{\Omega}^{M}\right)\left(\mathscr{A}\left(\left.\mathscr{M}\right|_{\Omega}\right)\right),\mathscr{A}\left(\iota_{\Theta}^{M}\right)\left(\mathscr{A}\left(\left.\mathscr{M}\right|_{\Theta}\right)\right)\right]=\left\{ 0\right\} \mbox{.}
\]
In light of our interpretation of the unital sub-C{*}-algebras associated
to proper regions as the algebras of the quantum observables on these
regions, the last equation means that the observables associated to
(causally convex relatively compact connected open) subsets which
are causally separated should be measurable independently. From physical
considerations this property is expected to hold for each quantum
field theory: we hardly admit a physical theory in which there are
observables associated to causally separated regions that cannot be
measured independently. Hence we may see the causality condition as
a restriction on the possible correlations between observables localized
in proper domains which are causally separated.

The time slice axiom seems to be a condition on the causal structure
of a LCQFT too. Consider a LCQFT $\mathscr{A}$ and a globally hyperbolic
spacetime $\mathscr{M}=\left(M,g,\mathfrak{o},\mathfrak{t}\right)$.
From Theorem \ref{thmGlobalHyperbolicity} we deduce that there exists
a smooth spacelike Cauchy surface $\Sigma$ for $\mathscr{M}$. If
we choose a causally convex connected open subset $\Omega$ of $M$
including $\Sigma$, taking into account the globally hyperbolic spacetime
$\left.\mathscr{M}\right|_{\Omega}$ and the morphism $\iota_{\Omega}^{M}:\left.\mathscr{M}\right|_{\Omega}\rightarrow\mathscr{M}$
of $\mathfrak{ghs}$, we see that the time slice axiom imposes that
\[
\mathscr{A}\left(\iota_{\Omega}^{M}\right)\left(\mathscr{A}\left(\mathscr{M}\right)\right)=\mathscr{A}\left(\mathscr{M}\right)\mbox{.}
\]

To give an interpretation of the time slice axiom in terms of quantum
observables, we must consider a $\mathscr{M}$-causally convex relatively
compact connected open subset $\Theta$ of $M$ and we think to it
as being itself a globally hyperbolic spacetime denoted by $\left.\mathscr{M}\right|_{\Theta}$.
Applying Remark \ref{remUsefulSubsetsOfGloballyHyperbolicSpacetimes}
to $\left.\mathscr{M}\right|_{\Theta}$, we obtain for $\varepsilon>0$
an $\left.\mathscr{M}\right|_{\Theta}$-causally convex connected
open subset $\Omega_{\varepsilon}$ of $\Theta$ that includes a Cauchy
surface of $\left.\mathscr{M}\right|_{\Theta}$. The closure of $\Omega_{\varepsilon}$
in $M$ is included in the closure of $\Theta$ in $M$, which is
compact in $M$ by hypothesis. Therefore $\Omega_{\varepsilon}$ is
relatively compact in $M$. This proves the existence of $\left.\mathscr{M}\right|_{\Theta}$-causally
convex relatively compact connected open subsets of $\Theta$ that
include Cauchy surfaces of $\left.\mathscr{M}\right|_{\Theta}$. We
choose a subset with these properties and we denote it with $\Omega$.
We recognize that $\Omega$ is also $\mathscr{M}$-causally convex
and that the globally hyperbolic spacetimes $\left.\mathscr{M}\right|_{\Omega}$
and $\left.\left.\mathscr{M}\right|_{\Theta}\right|_{\Omega}$ coincide
so that we can consider the inclusion map $\iota_{\Omega}^{\Theta}$
as a morphism of $\mathfrak{ghs}$ from $\left.\mathscr{M}\right|_{\Omega}$
to $\left.\mathscr{M}\right|_{\Theta}$. In the present situation
the time slice axiom imposes that
\[
\mathscr{A}\left(\iota_{\Omega}^{\Theta}\right)\left(\mathscr{A}\left(\left.\mathscr{M}\right|_{\Omega}\right)\right)=\mathscr{A}\left(\left.\mathscr{M}\right|_{\Theta}\right)\mbox{.}
\]
This relation means that, when $\Theta$ is a proper subset of some
globally hyperbolic spacetime $\mathscr{M}$ and $\Omega$ is a proper
subset of $\Theta$ including a Cauchy surface of $\left.\mathscr{M}\right|_{\Theta}$,
the quantum observables over $\Omega$ exhaust all the quantum observables
that are admitted by the physics on $\Theta$, even if $\Theta$ is
larger. Then the time slice axiom forces the physics over a proper
subset $\Theta$ of a globally hyperbolic spacetime to be completely
determined by the physics over a proper neighborhood $\Omega$ of
a Cauchy surface for $\left.\mathscr{M}\right|_{\Theta}$.
\end{rem}
The functorial approach of the GCLP allows us to introduce a notion
of equivalence between LCQFTs.
\begin{defn}
\index{equivalence of LCQFTs}Let $\mathscr{A}$ and $\mathscr{B}$
be two LCQFTs. We say that $\mathscr{A}$ and $\mathscr{B}$ are equivalent
if there exists a covariant natural isomorphism $\mathtt{i}:\mathscr{A}\overset{\rightarrow}{\rightarrow}\mathscr{B}$.
\end{defn}
The reader can easily check that this is an equivalence relation on
the set of LCQFTs. Such equivalence can be interpreted as physical
indistinguishability. Suppose that $\mathscr{A}$ and $\mathscr{B}$
are LCQFTs and that $\mathtt{i}$ is covariant natural isomorphism
from $\mathscr{A}$ to $\mathscr{B}$ and fix two globally hyperbolic
spacetimes $\mathscr{M}$, $\mathscr{N}$ and a morphism $\psi:\mathscr{M}\rightarrow\mathscr{N}$
of $\mathfrak{ghs}$. We have that $\mathscr{A}\left(\mathscr{M}\right)$
and $\mathscr{B}\left(\mathscr{M}\right)$ may be identified through
the unit preserving {*}-isomorphism $\mathtt{i}_{\mathscr{M}}:\mathscr{A}\left(\mathscr{M}\right)\rightarrow\mathscr{B}\left(\mathscr{M}\right)$
(similarly we can identify $\mathscr{A}\left(\mathscr{N}\right)$
and $\mathscr{B}\left(\mathscr{N}\right)$ through the unit preserving
{*}-isomorphism $\mathtt{i}_{\mathscr{N}}:\mathscr{A}\left(\mathscr{N}\right)\rightarrow\mathscr{B}\left(\mathscr{N}\right)$)
and that the injective unit preserving {*}-homomorphisms $\mathscr{A}\left(\psi\right)$
and $\mathscr{B}\left(\psi\right)$ satisfy the following relation:
\[
\mathtt{i}_{\mathscr{N}}\circ\mathscr{A}\left(\psi\right)=\mathscr{B}\left(\psi\right)\circ\mathtt{i}_{\mathscr{M}}\mbox{.}
\]
Then, with the above identifications, also $\mathscr{A}\left(\psi\right)$
and $\mathscr{B}\left(\psi\right)$ are identified. This identification
in our interpretation means that the quantum observables admitted
by the physics described by the theory $\mathscr{A}$ on some globally
hyperbolic spacetime are exactly the same as those admitted by the
physics described by the theory $\mathscr{B}$ on the same globally
hyperbolic spacetime, that is to say that the physics described by
$\mathscr{A}$ is exactly the same as the physics described by $\mathscr{B}$
on each globally hyperbolic spacetime.

\subsection{Recovering the Haag-Kastler framework}

\index{algebraic quantum field theory}\index{Haag-Kastler axioms}In
this subsection we check that our approach to quantum field theory
through the generally covariant locality principle leads us to the
complete recovery of the \textsl{Haag-Kastler axioms} for each globally
hyperbolic spacetime. By this we mean that each locally covariant
quantum field theory applied to an arbitrary globally hyperbolic spacetime
gives rise to a quantum field theory for that spacetime in the formulation
suggested by Haag-Kastler in their seminal paper \cite{HK64}. We
underline that, this formulation of quantum field theory, known as
\textsl{algebraic quantum field theory}, although being equivalent
to the traditional formulation, has the advantage of being stated
in a rigorous mathematical framework, specifically that of C{*}-algebras.

A relevant part of the problem of recovering the algebraic approach
to quantum field theory has already been discussed in Remark \ref{remInterpretation}
even if we did not stress this fact there. In the next theorem we
will complete this discussion so that it will become evident by comparison
with \cite{HK64} that the Haag-Kastler axioms are recovered on each
globally hyperbolic spacetime once that a LCQFT is given.

We begin with a definition.
\begin{defn}
\label{defLocalAlgebras}\index{local algebra}Let $\mathscr{A}$
be a LCQFT and let $\mathscr{M}=\left(M,g,\mathfrak{o},\mathfrak{t}\right)$
be a globally hyperbolic spacetime. We define the set $\mathcal{K}_{\mathscr{M}}$
of all $\mathscr{M}$-causally convex non empty relatively compact
connected open subsets of $M$ and the family $\left\{ \mathcal{A}_{\mathscr{M}}\left(\Omega\right)\right\} $
consisting of the unital sub-C{*}-algebras $\mathcal{A}_{\mathscr{M}}\left(\Omega\right)=\mathscr{A}\left(\iota_{\Omega}^{M}\right)\left(\mathscr{A}\left(\left.\mathscr{M}\right|_{\Omega}\right)\right)$,
called \textsl{local algebras}, of the unital C{*}-algebra $\mathscr{A}\left(\mathscr{M}\right)$
for $\Omega\in\mathcal{K}_{\mathscr{M}}$. Moreover we define $\mathcal{A}_{\mathscr{M}}$
as the unital sub-C{*}-algebra of $\mathscr{A}\left(\mathscr{M}\right)$
generated by the family $\left\{ \mathcal{A}_{\mathscr{M}}\left(\Omega\right)\right\} $.
\end{defn}
Notice that the elements of $\mathcal{K}\left(\mathscr{M}\right)$
are exactly those subsets of $\mathscr{M}$ that we used in our interpretation
of the GCLP (cfr. Remark \ref{remInterpretation}) to pick out the
physically acceptable observables on the globally hyperbolic spacetime
$\mathscr{M}$. There we did not specified the exclusion of the empty
set, however it appears obvious from a physical point of view that
it does not make sense to speak of the physics on a region with no
events.

In that context we already noticed that, for each $\Omega\in\mathcal{K}_{\mathscr{M}}$,
$\left.\mathscr{M}\right|_{\Omega}$ is actually a globally hyperbolic
spacetime, so that we can consider the unital C{*}-algebra $\mathscr{A}\left(\left.\mathscr{M}\right|_{\Omega}\right)$
and the morphism $\iota_{\Omega}^{M}$ of the category $\mathfrak{ghs}$.
Then we can actually define $\mathcal{A}_{\mathscr{M}}\left(\Omega\right)$
as above and we recognize that it is a unital sub-C{*}-algebra of
the larger unital C{*}-algebra $\mathscr{A}\left(\mathscr{M}\right)$.
This shows that the family $\left\{ \mathcal{A}_{\mathscr{M}}\left(\Omega\right)\right\} $
is well defined. In our interpretation we also specified that we cannot
consider $\mathscr{A}\left(\mathscr{M}\right)$ as an algebra of observables
because $M$ cannot be compact otherwise $\mathscr{M}$ would violate
the causality condition (cfr. \cite[Chap. 14, Lem. 13, p. 407]{O'N83}).
For the same reason $\mathscr{A}\left(\mathscr{M}\right)$ is not
included in the family $\left\{ \mathcal{A}_{\mathscr{M}}\left(\Omega\right)\right\} $.

When we define $\mathcal{A}_{\mathscr{M}}$ as the sub-C{*}-algebra
of $\mathscr{A}\left(\mathscr{M}\right)$ generated by the family
$\left\{ \mathcal{A}_{\mathscr{M}}\left(\Omega\right)\right\} $,
we intend that $\mathcal{A}_{\mathscr{M}}$ is the sub-C{*}-algebra
of $\mathscr{A}\left(\mathscr{M}\right)$ generated by the subset
\[
S=\bigcup_{\Omega\in\mathcal{K}_{\mathscr{M}}}\mathcal{A}_{\mathscr{M}}\left(\Omega\right)
\]
of $\mathscr{A}\left(\mathscr{M}\right)$ (refer to \ref{remGeneratedSubalgebra}
for the notion of generated sub-C{*}-algebra). That this definition
actually makes sense is assured by the fact that all elements of $\left\{ \mathcal{A}_{\mathscr{M}}\left(\Omega\right)\right\} $
are sub-C{*}-algebras of $\mathscr{A}\left(\mathscr{M}\right)$.

With the last definition we are ready to formulate the theorem that
recovers the Haag-Kastler axioms starting from a LCQFT applied to
an arbitrary globally hyperbolic spacetime.
\begin{thm}
\label{thmRecoveringHaagKastlerAxioms}\index{isotony}\index{algebra of observables}\index{covariance}\index{local commutativity}\index{time slice axiom}Let
$\mathscr{A}$ be a LCQFT and let $\mathscr{M}=\left(M,g,\mathfrak{o},\mathfrak{t}\right)$
be a globally hyperbolic spacetime. Consider $\mathcal{K}_{\mathscr{M}}$,
$\left\{ \mathcal{A}_{\mathscr{M}}\left(\Omega\right)\right\} $ and
$\mathcal{A}_{\mathscr{M}}$ as defined above. Then the Haag-Kastler
axioms (cfr. \cite{HK64}) are fully recovered. Specifically the following
properties hold:
\begin{itemize}
\item \textsl{isotony}: for each $\Omega$, $\Theta\in\mathcal{K}_{\mathscr{M}}$
such that $\Omega\subseteq\Theta$, $\mathcal{A}_{\mathscr{M}}\left(\Omega\right)$
is a sub-C{*}-algebra of $\mathcal{A}_{\mathscr{M}}\left(\Theta\right)$;
\item \textsl{common unit}: all the elements of $\left\{ \mathcal{A}_{\mathscr{M}}\left(\Omega\right)\right\} $
have a common unit;
\item \textsl{algebra of observables}: $\mathcal{A}_{\mathscr{M}}$ is the
closure in $\mathscr{A}\left(\mathscr{M}\right)$ of the union of
the family $\left\{ \mathcal{A}_{\mathscr{M}}\left(\Omega\right)\right\} $;
\item \textsl{covariance}: if $G$ is a group of orientation and time orientation
preserving isometric diffeomorphisms of $\mathscr{M}$, then there
exists a representation of $G$ in terms of {*}-automorphisms on $\mathcal{A}_{\mathscr{M}}$
such that, for each $f\in G$ and each $\Omega\in\mathcal{K}_{\mathscr{M}}$,
the {*}-automorphism $\alpha_{f}$ associated to $f$ satisfies the
condition
\[
\alpha_{f}\left(\mathcal{A}_{\mathscr{M}}\left(\Omega\right)\right)=\mathcal{A}_{\mathscr{M}}\left(f\left(\Omega\right)\right)\mbox{;}
\]

\item \textsl{local commutativity}: if $\mathscr{A}$ is causal then, for
each $\Omega$, $\Theta\in\mathcal{K}_{\mathscr{M}}$ such that $\Omega$
and $\Theta$ are $\mathscr{M}$-causally separated, we have that
\[
\left[\mathcal{A}_{\mathscr{M}}\left(\Omega\right),\mathcal{A}_{\mathscr{M}}\left(\Theta\right)\right]=\left\{ 0\right\} \mbox{;}
\]

\item \textsl{time slice axiom}: if $\mathscr{A}$ fulfils the time slice
axiom, $\Sigma$ is a smooth spacelike Cauchy surface for $\mathscr{M}$
and $S$ is a connected open subset of $\Sigma$ such that its Cauchy
development $D_{\mathscr{M}}\left(S\right)$ is relatively compact,
then for each $\Omega\in\mathcal{K}_{\mathscr{M}}$ such that $S\subseteq\Omega$
we have
\[
\mathcal{A}_{\mathscr{M}}\left(\Omega\right)\supseteq\mathcal{A}_{\mathscr{M}}\left(D_{\mathscr{M}}\left(S\right)\right)\mbox{.}
\]

\end{itemize}
\end{thm}
\begin{proof}
We start from isotony. Suppose that $\Omega$ and $\Theta$ are elements
of $\mathcal{K}_{\mathscr{M}}$ such that $\Omega\subseteq\Theta$.
In Remark \ref{remInterpretation} we showed that $\mathscr{A}\left(\iota_{\Omega}^{\Theta}\right)\left(\mathscr{A}\left(\left.\mathscr{M}\right|_{\Omega}\right)\right)$
is a unital sub-C{*}-algebra of the unital C{*}-algebra $\mathscr{A}\left(\left.\mathscr{M}\right|_{\Theta}\right)$.
If we consider the morphisms $\iota_{\Omega}^{M}$ and $\iota_{\Theta}^{M}$
of the category $\mathfrak{ghs}$, we immediately recognize that $\iota_{\Omega}^{M}=\iota_{\Theta}^{M}\circ\mbox{\ensuremath{\iota_{\Omega}^{\Theta}}}$.
Since $\mathscr{A}$ is a covariant functor, we have that $\mathscr{A}\left(\iota_{\Omega}^{M}\right)=\mathscr{A}\left(\iota_{\Theta}^{M}\right)\circ\mathscr{A}\left(\mbox{\ensuremath{\iota_{\Omega}^{\Theta}}}\right)$.
We deduce that
\begin{eqnarray*}
\mathcal{A}_{\mathscr{M}}\left(\Omega\right) & = & \mathscr{A}\left(\iota_{\Omega}^{M}\right)\left(\mathscr{A}\left(\left.\mathscr{M}\right|_{\Omega}\right)\right)\\
 & = & \left(\mathscr{A}\left(\iota_{\Theta}^{M}\right)\circ\mathscr{A}\left(\mbox{\ensuremath{\iota_{\Omega}^{\Theta}}}\right)\right)\left(\mathscr{A}\left(\left.\mathscr{M}\right|_{\Omega}\right)\right)\\
 & \subseteq & \mathscr{A}\left(\iota_{\Theta}^{M}\right)\left(\mathscr{A}\left(\left.\mathscr{M}\right|_{\Theta}\right)\right)\\
 & = & \mathcal{A}_{\mathscr{M}}\left(\Theta\right)\mbox{.}
\end{eqnarray*}
Since both $\mathcal{A}_{\mathscr{M}}\left(\Omega\right)$ and $\mathcal{A}_{\mathscr{M}}\left(\Theta\right)$
are unital sub-C{*}-algebras of $\mathscr{A}\left(\mathscr{M}\right)$,
the inclusion $\mathcal{A}_{\mathscr{M}}\left(\Omega\right)\subseteq\mathcal{A}_{\mathscr{M}}\left(\Theta\right)$
implies that $\mathcal{A}_{\mathscr{M}}\left(\Omega\right)$ is a
unital sub-C{*}-algebra of $\mathcal{A}_{\mathscr{M}}\left(\Theta\right)$.

Now we turn our attention to the units of the elements of the family
$\left\{ \mathcal{A}_{\mathscr{M}}\left(\Omega\right)\right\} $.
Let $\Omega$ and $\Theta$ be two arbitrary elements of $\mathcal{K}_{\mathscr{M}}$.
Applying Remark \ref{remghs}, we can consider the globally hyperbolic
spacetimes $\left.\mathscr{M}\right|_{\Omega}$ and $\left.\mathscr{M}\right|_{\Theta}$
and the morphisms $\iota_{\Omega}^{M}$ and $\iota_{\Theta}^{M}$
of $\mathfrak{ghs}$. Using $\mathscr{A}$, we obtain the corresponding
morphisms $\mathscr{A}\left(\iota_{\Omega}^{M}\right)$ and $\mathscr{A}\left(\iota_{\Theta}^{M}\right)$
of $\mathfrak{alg}$ that map each element of the unital C{*}-algebra
$\mathscr{A}\left(\left.\mathscr{M}\right|_{\Omega}\right)$ and respectively
$\mathscr{A}\left(\left.\mathscr{M}\right|_{\Theta}\right)$ into
an element of the unital C{*}-algebra $\mathscr{A}\left(\mathscr{M}\right)$.
Denoting with $1_{\Omega}$ the unit of $\mathscr{A}\left(\left.\mathscr{M}\right|_{\Omega}\right)$,
with $1_{\Theta}$ the unit of $\mathscr{A}\left(\left.\mathscr{M}\right|_{\Theta}\right)$
and with $1_{M}$ the unit of $\mathscr{A}\left(\mathscr{M}\right)$
and keeping in mind that all morphisms of $\mathfrak{alg}$ are unit
preserving, i.e. they map the unit of their domain algebra to the
unit of their codomain algebra, we conclude that
\[
\mathscr{A}\left(\iota_{\Omega}^{M}\right)1_{\Omega}=1_{M}=\mathscr{A}\left(\iota_{\Theta}^{M}\right)1_{\Theta}\mbox{.}
\]
From Remark \ref{remalg} we notice that $\mathscr{A}\left(\iota_{\Omega}^{M}\right)1_{\Omega}$
and $\mathscr{A}\left(\iota_{\Theta}^{M}\right)1_{\Theta}$ are respectively
the units of $\mathcal{A}_{\mathscr{M}}\left(\Omega\right)$ and $\mathcal{A}_{\mathscr{M}}\left(\Theta\right)$,
so that the last equation means that the unit of $\mathcal{A}_{\mathscr{M}}\left(\Omega\right)$
coincides with the unit of $\mathcal{A}_{\mathscr{M}}\left(\Theta\right)$.

$\mathcal{A}_{\mathscr{M}}$ is defined as the sub-C{*}-algebra of
$\mathscr{A}\left(\mathscr{M}\right)$ that is generated by the set
\[
S=\bigcup_{\Omega\in\mathcal{K}_{\mathscr{M}}}\mathcal{A}_{\mathscr{M}}\left(\Omega\right)\mbox{.}
\]
Consider $a$ and $b$ in $S$. Then $a$ is in $\mathcal{A}_{\mathscr{M}}\left(\Omega\right)$
for some $\Omega\in\mathcal{K}_{\mathscr{M}}$ and $b$ is in $\mathcal{A}_{\mathscr{M}}\left(\Theta\right)$
for some $\Theta\in\mathcal{K}_{\mathscr{M}}$. Since both $\Omega$
and $\Theta$ are relatively compact, we have that $K=\overline{\Omega}\cup\overline{\Theta}$
is compact and so we can apply the fourth point of Proposition \ref{propUsefulSubsetsOfGloballyHyperbolicSpacetimes}
to $K$ so that we find $\Delta\in\mathcal{K}_{\mathscr{M}}$ including
$K$. In particular both $\Omega$ and $\Theta$ are included in $\Delta$
and hence isotony implies that $a$ and $b$ are also elements of
$\mathcal{A}_{\mathscr{M}}\left(\Delta\right)$. Then we can take
linear combinations, products and involutions with them and we will
always get elements of $\mathcal{A}_{\mathscr{M}}\left(\Delta\right)$
since it is a C{*}-algebra. But $\mathcal{A}_{\mathscr{M}}\left(\Delta\right)$
is included in $S$ too, so linear combination, product and involution
are internal operations on $S$. Therefore $S$ is a vector space
endowed with two internal operations that are our candidates for being
a multiplication and an involution. They are actually such because
they fulfil the properties that qualify them as a multiplication and
an involution on the larger vector space $\mathscr{A}\left(\mathscr{M}\right)$.
Hence we can think of $S$ as a unital {*}-algebra (its unit being
$1_{M}$ as a consequence of what we have seen above). When we endow
$S$ with the norm of $\mathscr{A}\left(\mathscr{M}\right)$, we realize
that it lacks only of closure in $\mathscr{A}\left(\mathscr{M}\right)$
to become a unital C{*}-algebra itself. So we close $S$ in $\mathscr{A}\left(\mathscr{M}\right)$
and we denote with $\mathcal{A}_{\mathscr{M}}^{\prime}$ the unital
C{*}-algebra that we obtain. By construction $S\subseteq\mathcal{A}_{\mathscr{M}}^{\prime}$,
hence $\mathcal{A}_{\mathscr{M}}\subseteq\mathcal{A}_{\mathscr{M}}^{\prime}$
by definition of $\mathcal{A}_{\mathscr{M}}$ as the sub-C{*}-algebra
of $\mathscr{A}\left(\mathscr{M}\right)$ generated by $S$. We want
to prove that $\mathcal{A}_{\mathscr{M}}\supseteq\mathcal{A}_{\mathscr{M}}^{\prime}$.
To this end pick $a\in\mathcal{A}_{\mathscr{M}}^{\prime}$. By construction
$a$ is the limit in the norm of $\mathscr{A}\left(\mathscr{M}\right)$
of a sequence $\left\{ a_{n}\right\} $ of elements of $S$ that is
Cauchy with respect to the norm of $\mathscr{A}\left(\mathscr{M}\right)$.
Yet $S\subseteq\mathcal{A}_{\mathscr{M}}$ and the norm of $\mathcal{A}_{\mathscr{M}}$
is exactly the restriction of the norm of $\mathscr{A}\left(\mathscr{M}\right)$
because $\mathcal{A}_{\mathscr{M}}$ is a sub-C{*}-algebra of $\mathscr{A}\left(\mathscr{M}\right)$.
We deduce that $\left\{ a_{n}\right\} $ is also a Cauchy sequence
in $\mathcal{A}_{\mathscr{M}}$. But, being a C{*}-algebra, $\mathcal{A}_{\mathscr{M}}$
is also a Banach space and so we find a limit $b$. Then $\left\{ a_{n}\right\} $
converges to both $a$ and $b$ in $\mathscr{A}\left(\mathscr{M}\right)$
and hence $a=b$. We conclude that $a\in\mathcal{A}_{\mathscr{M}}$,
therefore $\mathcal{A}_{\mathscr{M}}^{\prime}\subseteq\mathcal{A}_{\mathscr{M}}$.

As for covariance, we proceed in the following way. First of all we
notice that the group $G$ consists of bijective morphisms of $\mathfrak{ghs}$
from $\mathscr{M}$ to $\mathscr{M}$ whose inverses are morphisms
too: In order to recognize that $f\in G$ is a morphism of $\mathfrak{ghs}$
we must only check that its image is $\mathscr{M}$-causally convex,
but this is trivial because $f\left(M\right)=M$; bijectivity of $f\in G$
is assumed by hypothesis and its inverse $f^{-1}$ is automatically
a morphism of $\mathfrak{ghs}$. At this point we can use the LCQFT
$\mathscr{A}$ to map each $f\in G$ to a morphism of $\mathfrak{alg}$.
From $f^{-1}$ we obtain its inverse morphism so that $\mathscr{A}\left(f\right)$
is a bijective morphism of $\mathfrak{alg}$ from $\mathscr{A}\left(\mathscr{M}\right)$
to $\mathscr{A}\left(\mathscr{M}\right)$ whose inverse is a morphism
too:
\begin{alignat*}{4}
\mathscr{A}\left(f\right)\circ\mathscr{A}\left(f^{-1}\right) & = & \mathscr{A}\left(f\circ f^{-1}\right) & = & \mathscr{A}\left(\mathrm{id}_{\mathscr{M}}\right) & = & \mathrm{id}_{\mathscr{A}\left(\mathscr{M}\right)}\mbox{;}\\
\mathscr{A}\left(f^{-1}\right)\circ\mathscr{A}\left(f\right) & = & \mathscr{A}\left(f^{-1}\circ f\right) & = & \mathscr{A}\left(\mathrm{id}_{\mathscr{M}}\right) & = & \mathrm{id}_{\mathscr{A}\left(\mathscr{M}\right)}\mbox{.}
\end{alignat*}
Fix $f\in G$ and $\Omega\in\mathcal{K}_{\mathscr{M}}$. Observe that
$f\left(\Omega\right)\in\mathcal{K}_{\mathscr{M}}$: It is a relatively
compact open subset of $M$ because it is the preimage of the relatively
compact open subset $\Omega$ of $M$ through the continuous map $f^{-1}$,
it is connected because $f$ is continuous and $\Omega$ is connected
and finally it is $\mathscr{M}$-causally convex because $f^{-1}$
is smooth and isometric and $\Omega$ is $\mathscr{M}$-causally convex.
As usual, we can consider the globally hyperbolic spacetimes $\left.\mathscr{M}\right|_{\Omega}=\left(\Omega,\left.g\right|_{\Omega},\left.\mathfrak{o}\right|_{\Omega},\left.\mathfrak{t}\right|_{\Omega}\right)$
and $\left.\mathscr{M}\right|_{f\left(\Omega\right)}=\left(\Omega,\left.g\right|_{f\left(\Omega\right)},\left.\mathfrak{o}\right|_{f\left(\Omega\right)},\left.\mathfrak{t}\right|_{f\left(\Omega\right)}\right)$
and the morphisms $\iota_{\Omega}^{M}$ and $\iota_{f\left(\Omega\right)}^{M}$
of $\mathfrak{ghs}$. If we define the map $f_{\Omega}:\Omega\rightarrow f\left(\Omega\right)$,
$p\mapsto f\left(p\right)$, as a consequence of the properties of
$f$, we recognize that $f_{\Omega}$ is an orientation and time orientation
preserving isometric diffeomorphism from $\left.\mathscr{M}\right|_{\Omega}$
to $\left.\mathscr{M}\right|_{f\left(\Omega\right)}$:
\begin{eqnarray*}
f_{\Omega} & \in & \mathsf{Mor}_{\mathfrak{ghs}}\left(\left.\mathscr{M}\right|_{\Omega},\left.\mathscr{M}\right|_{f\left(\Omega\right)}\right)\mbox{,}\\
f_{\Omega}^{-1} & \in & \mathsf{Mor}_{\mathfrak{ghs}}\left(\left.\mathscr{M}\right|_{f\left(\Omega\right)},\left.\mathscr{M}\right|_{\Omega}\right)\mbox{.}
\end{eqnarray*}
Then it follows that
\begin{eqnarray*}
\mathscr{A}\left(f_{\Omega}\right) & \in & \mathsf{Mor}_{\mathfrak{alg}}\left(\mathscr{A}\left(\left.\mathscr{M}\right|_{\Omega}\right),\mathscr{A}\left(\left.\mathscr{M}\right|_{f\left(\Omega\right)}\right)\right)\mbox{,}\\
\mathscr{A}\left(f_{\Omega}^{-1}\right) & \in & \mathsf{Mor}_{\mathfrak{alg}}\left(\mathscr{A}\left(\left.\mathscr{M}\right|_{f\left(\Omega\right)}\right),\mathscr{A}\left(\left.\mathscr{M}\right|_{\Omega}\right)\right)
\end{eqnarray*}
are inverses one of the other. In particular we have that $\mathscr{A}\left(f_{\Omega}\right)$
is surjective:
\[
\mathscr{A}\left(f_{\Omega}\right)\left(\mathscr{A}\left(\left.\mathscr{M}\right|_{\Omega}\right)\right)=\mathscr{A}\left(\left.\mathscr{M}\right|_{f\left(\Omega\right)}\right)\mbox{.}
\]
It is easy to check that $\iota_{f\left(\Omega\right)}^{M}\circ f_{\Omega}=f\circ\iota_{\Omega}^{M}$
and hence we have 
\[
\mathscr{A}\left(\iota_{f\left(\Omega\right)}^{M}\right)\circ\mathscr{A}\left(f_{\Omega}\right)=\mathscr{A}\left(f\right)\circ\mathscr{A}\left(\iota_{\Omega}^{M}\right)\mbox{.}
\]
Therefore we find
\begin{eqnarray*}
\mathcal{A}_{\mathscr{M}}\left(f\left(\Omega\right)\right) & = & \mathscr{A}\left(\iota_{f\left(\Omega\right)}^{M}\right)\left(\mathscr{A}\left(\left.\mathscr{M}\right|_{f\left(\Omega\right)}\right)\right)\\
 & = & \left(\mathscr{A}\left(\iota_{f\left(\Omega\right)}^{M}\right)\circ\mathscr{A}\left(f_{\Omega}\right)\right)\left(\mathscr{A}\left(\left.\mathscr{M}\right|_{\Omega}\right)\right)\\
 & = & \left(\mathscr{A}\left(f\right)\circ\mathscr{A}\left(\iota_{\Omega}^{M}\right)\right)\left(\mathscr{A}\left(\left.\mathscr{M}\right|_{\Omega}\right)\right)\\
 & = & \mathscr{A}\left(f\right)\left(\mathcal{A}_{\mathscr{M}}\left(\Omega\right)\right)\mbox{.}
\end{eqnarray*}
Above we observed that $f\left(\Omega\right)\in\mathcal{K}_{\mathscr{M}}$
for each $\Omega\in\mathcal{K}_{\mathscr{M}}$. A similar argument
applied to $f^{-1}$ tells us also that $f^{-1}\left(\Omega\right)\in\mathcal{K}_{\mathscr{M}}$
for each $\Omega\in\mathcal{K}_{\mathscr{M}}$. This observation,
together with the last formula, implies that
\[
\bigcup_{\Omega\in\mathcal{K}_{\mathscr{M}}}\mathscr{A}\left(f\right)\left(\mathcal{A}_{\mathscr{M}}\left(\Omega\right)\right)=\bigcup_{\Omega\in\mathcal{K}_{\mathscr{M}}}\mathcal{A}_{\mathscr{M}}\left(f\left(\Omega\right)\right)=\bigcup_{\Omega^{\prime}\in\mathcal{K}_{\mathscr{M}}}\mathcal{A}_{\mathscr{M}}\left(\Omega^{\prime}\right)\mbox{.}
\]
Applying the third point of this theorem and bearing in mind that
$\mathscr{A}\left(f\right)$ is continuous with respect to the norm
of $\mathscr{A}\left(\mathscr{M}\right)$, we draw the following conclusion:
\begin{alignat*}{1}
\mathscr{A}\left(f\right)\left(\mathcal{A}_{\mathscr{M}}\right) & =\mathscr{A}\left(f\right)\left(\overline{\bigcup_{\Omega\in\mathcal{K}_{\mathscr{M}}}\mathcal{A}_{\mathscr{M}}\left(\Omega\right)}\right)=\overline{\bigcup_{\Omega\in\mathcal{K}_{\mathscr{M}}}\mathscr{A}\left(f\right)\left(\mathcal{A}_{\mathscr{M}}\left(\Omega\right)\right)}=\overline{\bigcup_{\Omega^{\prime}\in\mathcal{K}_{\mathscr{M}}}\mathcal{A}_{\mathscr{M}}\left(\Omega^{\prime}\right)}\\
 & =\mathcal{A}_{\mathscr{M}}\mbox{.}
\end{alignat*}
The last equation implies that for each $f\in G$ we can define the
map
\begin{eqnarray*}
\alpha_{f}:\mathcal{A}_{\mathscr{M}} & \rightarrow & \mathcal{A}_{\mathscr{M}}\\
a & \mapsto & \mathscr{A}\left(f\right)a
\end{eqnarray*}
and realize that it is a {*}-automorphism on the unital C{*}-algebra
$\mathcal{A}_{\mathscr{M}}$ satisfying $\alpha_{f}\left(\mathcal{A}_{\mathscr{M}}\right)=\mathcal{A}_{\mathscr{M}}$.
This defines a map $f\mapsto\alpha_{f}$ from the group $G$ to the
group of the {*}-automorphisms on the unital C{*}-algebra $\mathcal{A}_{\mathscr{M}}$
(the algebra of observables). In order to recognize this map as a
representation of the group $G$, we must still check that $\alpha_{f_{1}\circ f_{2}}=\alpha_{f_{1}}\circ\alpha_{f_{2}}$
for each $f_{1}$, $f_{2}\in G$. Fix $f_{1}$ and $f_{2}$ in $G$.
From covariant functoriality we deduce $\mathscr{A}\left(f_{1}\circ f_{2}\right)=\mathscr{A}\left(f_{1}\right)\circ\mathscr{A}\left(f_{2}\right)$.
For an arbitrary $a\in\mathcal{A}_{\mathscr{M}}$ we obtain
\[
\alpha_{f_{1}\circ f_{2}}a=\mathscr{A}\left(f_{1}\circ f_{2}\right)a=\mathscr{A}\left(f_{1}\right)\left(\mathscr{A}\left(f_{2}\right)a\right)=\alpha_{f_{1}}\left(\alpha_{f_{2}}a\right)
\]
because $\mathscr{A}\left(f_{2}\right)a\in\mathcal{A}_{\mathscr{M}}$
and therefore $\alpha_{f_{1}\circ f_{2}}=\alpha_{f_{1}}\circ\alpha_{f_{2}}$
actually holds for each $f_{1}$, $f_{2}\in G$.

We have already faced the problem of local commutativity when we gave
an interpretation of the causality condition in terms of local observables.
Anyway we briefly recollect the proof here for completeness. For this
scope assume that $\mathscr{A}$ is causal and fix $\Omega$ and $\Theta$
in $\mathcal{K}_{\mathscr{M}}$ such that they are causally separated
in $\mathscr{M}$. In the category $\mathfrak{ghs}$ we can consider
the objects $\left.\mathscr{M}\right|_{\Omega}$ and $\left.\mathscr{M}\right|_{\Theta}$
and the morphisms $\iota_{\Omega}^{M}$ and $\iota_{\Theta}^{M}$.
In the present situation we apply the causality condition (cfr. Definition
\ref{defLCQFT}) and we obtain
\[
\left[\mathscr{A}\left(\iota_{\Omega}^{M}\right)\left(\mathscr{A}\left(\left.\mathscr{M}\right|_{\Omega}\right)\right),\mathscr{A}\left(\iota_{\Theta}^{M}\right)\left(\mathscr{A}\left(\left.\mathscr{M}\right|_{\Theta}\right)\right)\right]=\left\{ 0\right\} \mbox{,}
\]
which is exactly our thesis because of Definition \ref{defLocalAlgebras}.

To prove the last part of the theorem we assume that $\mathscr{A}$
fulfils the time slice axiom. Let $\Sigma$ be a spacelike (hence
acausal due to \cite[Chap. 14, Lem. 42, p. 425]{O'N83}) Cauchy surface
for $\mathscr{M}$ and let $S$ be a connected open subset of $\Sigma$
such that $D_{\mathscr{M}}\left(S\right)$ is relatively compact in
$M$. For convenience we write $D$ in place of $D_{\mathscr{M}}\left(S\right)$.
In first place we must check that $D$ is in $\mathcal{K}_{\mathscr{M}}$,
otherwise our thesis doesn't make sense. From \cite[Lem. A.9, p. 48]{FV11}
we deduce that $D$ is an open subset of $M$. Now we show that $D$
is $\mathscr{M}$-causally convex. Take a $\mathfrak{t}$-future directed
$g$-causal curve $\gamma$ in $M$ starting from $p\in D$ and ending
in $q\in D$ and assume by contradiction that $\gamma$ is not entirely
contained in $D$. Then we find a point $r$ along $\gamma$ such
that there exists an inextensible $\mathfrak{t}$-future directed
$g$-timelike curve $\gamma^{\prime}$ in $M$ passing through $r$
which doesn't meet $S$. Hence we can use proper pieces of $\gamma$
and $\gamma^{\prime}$ to easily build an inextensible $\mathfrak{t}$-future
directed $g$-causal curve in $M$ passing through $p$ (or otherwise
$q$) which doesn't meet $S$. This undoubtedly violates the hypothesis
that both $p$ and $q$ are in $D$. Therefore $D$ is actually $\mathscr{M}$-causally
convex. We still need to show that $D$ is connected. Suppose that
$p$ and $q$ are points in $D$. Because of the definition of $D$,
it is not hard to find two $g$-causal curves $\gamma_{1}$ and $\gamma_{3}$
in $M$ connecting respectively $p$ to some point $r$ and $q$ to
some point $s$, with $r$ and $s$ in $S$. Since trivially $S\subseteq D$,
we deduce from $\mathscr{M}$-causally convexity that both $\gamma_{1}$
and $\gamma_{3}$ are included in $D$. $S$ is connected by hypothesis
and so we find a curve $\gamma_{2}$ connecting $r$ and $s$. If
we paste $\gamma_{1}$, $\gamma_{2}$ and $\gamma_{3}$ we obtain
a curve connecting $p$ to $q$ and this proves that $D$ is actually
connected. With this preparatory results and the hypothesis that $D$
is relatively compact, we can conclude that $D$ is an element of
$\mathcal{K}_{\mathscr{M}}$ and hence the thesis makes sense. Now
we take also $\Omega$ in $\mathcal{K}_{\mathscr{M}}$ such that $S\subseteq\Omega$
and we start the real proof. As usual we can consider the globally
hyperbolic spacetimes $\left.\mathscr{M}\right|_{\Omega}$ and $\left.\mathscr{M}\right|_{D}$
and the morphisms $\iota_{\Omega}^{M}$ and $\iota_{D}^{M}$ that
immerse these spacetimes in $\mathscr{M}$. We make a useful observation:
$S$ is a Cauchy surface for $\left.\mathscr{M}\right|_{D}$. This
is seen in the following way: $S$ is a subset of a Cauchy surface
$\Sigma$ for $\mathscr{M}$, hence each inextensible $\mathfrak{t}$-future
directed $g$-timelike curve in $M$ meets $S$ at most once; take
now an arbitrary inextensible $\left.\mathfrak{t}\right|_{D}$-future
directed $\left.g\right|_{D}$-timelike curve $\gamma$ in $D$; in
$M$ we can extend $\gamma$ to an inextensible $\mathfrak{t}$-future
directed $g$-timelike curve $\gamma^{\prime}$ in $M$; undoubtedly
$\gamma^{\prime}$ passes through some point in $D$, hence we deduce
that it meets $S$ (remember that $D$ is the Cauchy development of
$S$ in $\mathscr{M}$), therefore it meets $S$ exactly once; now
we restrict $\gamma^{\prime}$ to $D$ and we realize that such restriction
$\gamma^{\prime\prime}$ is a $\left.\mathfrak{t}\right|_{D}$-future
directed $\left.g\right|_{D}$-timelike curve in $D$ that meets $S$
exactly once and extends $\gamma$; but $\gamma$ was inextensible
by our assumption, hence $\gamma$ and $\gamma^{\prime\prime}$ coincide
so that $\gamma$ meets $S$ exactly once, proving that $S$ is a
Cauchy surface for $\left.\mathscr{M}\right|_{D}$. To proceed we
introduce the subset $\Theta=\Omega\cap D$. We realize at once that
$\Theta$ is an open subset of $M$. Furthermore we see that $\overline{\Theta}\subseteq\overline{\Omega}$,
hence $\Theta$ is relatively compact in $M$ since both $\Omega$
is such. If we take a $\mathfrak{t}$-future directed $g$-causal
curve $\gamma$ in $M$ starting at $p\in\Theta$ and ending at $q\in\Theta$,
we recognize that $\gamma$ must be included in both $\Omega$ and
$D$ because they are $\mathscr{M}$-causally convex. This implies
that $\Theta$ is $\mathscr{M}$-causally convex too. Now pick too
arbitrary points $p$ and $q$ of $\Theta$. Since $p$ and $q$ fall
in $D$, it is easy to find two $\mathfrak{t}$-future directed $g$-causal
curves $\gamma_{1}$ and $\gamma_{3}$ in $M$ connecting respectively
the point $p$ to some point $r\in S$ and the point $q$ to some
point $s\in S$. By hypothesis $S\subseteq\Omega$, hence also $S\subseteq\Theta$.
Then both $\gamma_{1}$ and $\gamma_{3}$ are contained in $\Theta$
as a consequence of $\mathscr{M}$-causal convexity. $S$ is connected
by hypothesis and so we find $\gamma_{2}$ (automatically included
in $\Theta$) connecting $r$ and $s$. Then pasting $\gamma_{1}$,
$\gamma_{2}$ and $\gamma_{3}$, we connect $p$ and $q$, therefore
$\Theta$ is also connected. With this we have shown that $\Theta\in\mathcal{K}_{\mathscr{M}}$.
Essentially this is the situation: We have a globally hyperbolic spacetime
$\left.\mathscr{M}\right|_{D}$ with a Cauchy surface $S$ included
in $\Theta\in\mathcal{K}_{\mathscr{M}}$, with $\Theta\subseteq D$.
Being $\mathscr{M}$-causally convex, $\Theta$ is also $\left.\mathscr{M}\right|_{D}$-causally
convex and so we can consider both the globally hyperbolic spacetimes
$\left.\mathscr{M}\right|_{\Theta}$ and $\left.\left.\mathscr{M}\right|_{D}\right|_{\Theta}$.
We realize immediately that $\left.\left.\mathscr{M}\right|_{D}\right|_{\Theta}=\left.\mathscr{M}\right|_{\Theta}$
and so the morphism $\iota_{\Theta}^{D}$ immerses $\left.\mathscr{M}\right|_{\Theta}$
in $\left.\mathscr{M}\right|_{D}$. As we said above, the image $\iota_{\Theta}^{D}\left(\Theta\right)=\Theta$
includes the Cauchy surface $S$ for $\left.\mathscr{M}\right|_{D}$.
Then it is possible to apply the time slice axiom obtaining
\[
\mathscr{A}\left(\iota_{\Theta}^{D}\right)\left(\mathscr{A}\left(\left.\mathscr{M}\right|_{\Theta}\right)\right)=\mathscr{A}\left(\left.\mathscr{M}\right|_{D}\right)\mbox{.}
\]
There is still another morphism of $\mathfrak{ghs}$ at our disposal:
$\iota_{\Theta}^{M}\in\mathsf{Mor}_{\mathfrak{ghs}}\left(\left.\mathscr{M}\right|_{\Theta},\mathscr{M}\right)$.
It is easy to check that $\iota_{\Theta}^{M}=\iota_{D}^{M}\circ\iota_{\Theta}^{D}$
and hence, via covariant functoriality, we deduce $\mathscr{A}\left(\iota_{\Theta}^{M}\right)=\mathscr{A}\left(\iota_{D}^{M}\right)\circ\mathscr{A}\left(\iota_{\Theta}^{D}\right)$.
Applying $\mathscr{A}\left(\iota_{D}^{M}\right)$ to both sides of
our last equation, we get
\[
\mathcal{A}_{\mathscr{M}}\left(\Theta\right)=\mathscr{A}\left(\iota_{\Theta}^{M}\right)\left(\mathscr{A}\left(\left.\mathscr{M}\right|_{\Theta}\right)\right)=\mathcal{A}_{\mathscr{M}}\left(D\right)\mbox{.}
\]
Remembering the inclusion $\Theta\subseteq\Omega$ and applying isotony,
we conclude the proof:
\[
\mathcal{A}_{\mathscr{M}}\left(\Omega\right)\supseteq\mathcal{A}_{\mathscr{M}}\left(\Theta\right)=\mathcal{A}_{\mathscr{M}}\left(D\right)\mbox{.}
\]
\end{proof}
\begin{rem}
\label{remPrimitivity}We warn the reader that one of the properties
required by the Haag-Kastler axioms is not included in our theorem,
specifically we did not show that the unital C{*}-algebra $\mathcal{A}_{\mathscr{M}}$
is primitive, i.e. there exists a faithful irreducible representation
of $\mathcal{A}_{\mathscr{M}}$ on a Hilbert space. Hence the conclusion
that the Haag-Kastler framework is completely recovered via the last
theorem is not correct at all. Anyway, we will see later that the
concrete locally covariant quantum field theories that we construct
satisfy also this property (see the upcoming Remark \ref{remPrimitityHoldsDueToTheQuantizationFunctor}).
\end{rem}

\section[Construction of a locally covariant quantum field theory]{\label{secBuildingALCQFT}Construction of a locally covariant quantum
field theory\sectionmark{Construction of a LCQFT}}

\sectionmark{Costruction of a LCQFT}

In this section we deal with the problem of building concrete locally
covariant quantum field theories for situations of physical interest.
In the first part we will show a procedure that leads to the construction
of a causal LCQFT fulfilling the time slice axiom starting from the
wave equation of a classical field represented by a section in an
arbitrary vector bundle over some globally hyperbolic spacetime. To
do this we will need to specialize some more the category $\mathfrak{ghs}$
of globally hyperbolic spacetimes. As a matter of fact Definition
\ref{defghsalg} contains all the knowledge that is required to state
the generally covariant locality principle without specifying anything
about the physical problem to which we want to apply such principle,
except the fact that it takes place over a globally hyperbolic spacetime.
This is the approach followed by Brunetti, Fredenhagen and Verch in
\cite{BFV03} when they proposed the GCLP, as well as by other authors
even in the more recent papers on this topic (e.g. \cite{FV11,Dap11}).
This choice is done to underline that the consequences of the GCLP
(specifically Theorem \ref{thmRecoveringHaagKastlerAxioms} in the
present case) do not depend upon any of the properties of the specific
quantum field that one may consider, except for the fact that it is
set over a globally hyperbolic spacetime. This is one of the strong
points of the GCLP.

Following Fewster and Verch \cite{FV11}, we could even enlarge the
category $\mathfrak{alg}$ in order to take into account a very wide
range of physical situations (not only quantum fields, but also classical
dynamical systems too, such as classical fields or mechanical systems).
There is not a precise way to define the new target category: which
is the more convenient setting for a theory actually depends upon
the type of physical problem the theory deals with (e.g. C{*}-algebras
for quantum fields and symplectic spaces for classical fields). The
key point is that there exists a common mathematical framework in
which it is possible to formulate all those theories: they are recognized
to be covariant functors from the category $\mathfrak{ghs}$ (eventually
with some more data concerning the specific problem under consideration)
to a convenient category that fit the physical problem in the best
way. It is the functorial approach that unifies all these physical
theories and for all of them it is possible to speak of causality
and time slice axiom, although this must be done in a way that is
adapted to the mathematical framework chosen to describe the physical
system we are interested in.

During the construction of a LCQFT we will encounter a relevant example
of what we are saying. Specifically, we will see that a classical
field is comfortably described by a covariant functor from $\mathfrak{ghs}$
(with some structure that pertains to the field itself, essentially
the wave equation governing its dynamics) to the category having symplectic
spaces as objects and symplectic maps as morphisms.

\subsection{\label{subClassicalFieldTheory}From classical field theory...}

We want to describe a classical field over some $d$-dimensional globally
hyperbolic spacetime $\mathscr{M}=\left(M,g,\mathfrak{o},\mathfrak{t}\right)$
that is modeled by a smooth section $u$ in a vector bundle $E$ of
rank $n$ over $M$ satisfying the normally hyperbolic equation $Au=0$
on each point of $M$, where $E$ is endowed with an inner product
denoted by $\overset{E}{\cdot}$ and $A$ is a formally selfadjoint
normally hyperbolic operator on $E$ over $\mathscr{M}$.

As we anticipated above, we are going to build a functor from a slightly
modified version of $\mathfrak{ghs}$ to a proper category that we
define right now.

\subsubsection{Categories}
\begin{defn}
\label{defghsfssp}The category $\mathfrak{ghs}^{f}$ is defined in
the following way:
\begin{itemize}
\item objects are triples $\left(\mathscr{M},E,A\right)$, where $\mathscr{M}=\left(M,g,\mathfrak{o},\mathfrak{t}\right)$
is a $d$-dimensional globally hyperbolic spacetime, $E$ is a vector
bundle of rank $n$ over $M$ endowed with an inner product denoted
by $\overset{E}{\cdot}$ and $A$ is a formally selfadjoint normally
hyperbolic operator on $E$ over $\mathscr{M}$;
\item morphisms between two arbitrary objects $\left(\mathscr{M},E,A\right)$
and $\left(\mathscr{N},F,B\right)$ are vector bundle homomorphisms
$\left(\psi,\Psi\right)$ compatible with the inner products $\overset{E}{\cdot}$
and $\overset{F}{\cdot}$ and the formally selfadjoint normally hyperbolic
operators $A$ and $B$ (we will explain the meaning of these conditions
immediately after this definition), where $\psi$ is a morphism of
$\mathfrak{ghs}$ from $\mathscr{M}$ to $\mathscr{N}$;
\item the composition law is the ordinary composition of vector bundle homomorphisms,
i.e. the composition of function for both members of the pairs.
\end{itemize}
We also define $\mathfrak{ssp}$ as the category whose objects are
symplectic spaces $\left(V,\omega\right)$, whose morphisms between
two arbitrary objects $\left(V,\sigma\right)$ and $\left(W,\omega\right)$
are symplectic maps $\xi$ and whose composition law is the usual
composition of functions.
\end{defn}
Before proceeding, we want to specify the meaning of the condition
of compatibility with the inner products and with the normally hyperbolic
operators that is required to the morphisms of $\mathfrak{ghs}^{f}$.
This completes the definition of $\mathfrak{ghs}^{f}$. We also take
the chance to underline some particular properties of the vector bundle
homomorphisms we are going to deal with.

The condition of compatibility with inner products means that each
vector bundle homomorphism that we take into account must be fiberwise
an isometry with respect to the vector space non degenerate inner
products induced on each fiber by the inner products on the vector
bundles. To be precise, we require that each vector bundle homomorphisms
$\left(\psi,\Psi\right)$ from $\left(\mathscr{M},E,A\right)$ to
$\left(\mathscr{N},F,B\right)$ satisfies the following condition:
\[
\left(\Psi_{p}\mu\right)\overset{F}{\cdot}_{\psi\left(p\right)}\left(\Psi_{p}\nu\right)=\mu\overset{E}{\cdot}_{p}\nu
\]
for each $p\in M$ and each $\mu$, $\nu\in E_{p}$, where $M$ is
the manifold underlying $\mathscr{M}$. This ensures that $\Psi$
is fiberwise isometric, hence, in particular, $\Psi_{p}$ is an injective
vector space homomorphism for each $p\in M$ because of the non degeneracy
of inner products. This observation has a relevant consequence: for
each $p\in M$
\[
n=\dim E_{p}=\dim\left(\Psi_{p}\left(E_{p}\right)\right)\leq\dim F_{\psi\left(p\right)}=n\mbox{.}
\]
This fact implies that $\Psi_{p}$ is a vector space isomorphism for
each $p\in M$ (however $\Psi$ is not a vector bundle isomorphism
unless $\psi$ is bijective).

The condition of compatibility with the formally selfadjoint normally
hyperbolic operators is slightly more involved. First of all notice
that we are in position to apply Remark \ref{remRestrictionOfVectorBundleHomomorphisms}:
$\psi$ is an embedding whose image is an open subset of its codomain
and $\Psi$ is fiberwise a vector space isomorphism. Then we obtain
the new vector bundle $\Psi\left(E\right)$ of rank $n$ over the
$d$-dimensional manifold $\psi\left(M\right)$ and the vector bundle
isomorphism $\left(\psi^{\prime},\Psi^{\prime}\right)$ obtained from
the restriction of $\left(\psi,\Psi\right)$ to $\Psi\left(E\right)$.
Now we take $u\in\mathscr{D}\left(M,E\right)$ and, using the vector
bundle isomorphism $\left(\psi^{\prime},\Psi^{\prime}\right)$ and
Remark \ref{remPushforwardPullbackSections}, we introduce the section
\[
u^{\prime}=\Psi^{\prime}\circ u\circ\psi^{\prime-1}:\psi\left(M\right)\rightarrow\Psi\left(E\right)\mbox{.}
\]
Since $\psi^{\prime}$ is a homeomorphism, it holds that
\[
\mathrm{supp}\left(u^{\prime}\right)=\psi^{\prime}\left(\mathrm{supp}\left(u\right)\right)
\]
and so it turns out that $u^{\prime}$ is a compactly supported section
because $\mathrm{supp}\left(u\right)$ is compact in $M$. Using the
fact that $u^{\prime}$ is null outside a compact subset of $\psi\left(M\right)$,
we can define the smooth compactly supported section $u^{\prime\prime}:N\rightarrow F$
via the formula
\[
u^{\prime\prime}\left(q\right)=\begin{cases}
u^{\prime}\left(q\right) & \mbox{if }q\in\psi\left(M\right)\mbox{,}\\
0 & \mbox{if }q\in N\setminus\psi\left(M\right)\mbox{.}
\end{cases}
\]
This defines a map between the vector spaces $\mathscr{D}\left(M,E\right)$
and $\mathscr{D}\left(N,F\right)$:
\begin{eqnarray}
\mathrm{ext}_{\Psi}:\mathscr{D}\left(M,E\right) & \rightarrow & \mathscr{D}\left(N,F\right)\label{eqDefinitionOfextPsi}\\
u & \mapsto & u^{\prime\prime}\mbox{.}\nonumber 
\end{eqnarray}
Notice that such map is trivially linear and that it transforms the
support through $\psi$:
\[
\mathrm{supp}\left(\mathrm{ext}_{\Psi}u\right)=\psi\left(\mathrm{supp}\left(u\right)\right)\mbox{.}
\]
This construction was made to be in a position that allows us to correctly
state the condition of compatibility with the normally hyperbolic
operators $A$ and $B$: for each $u\in\mathscr{D}\left(M,E\right)$
it holds that
\[
\mathrm{ext}_{\Psi}\left(Au\right)=B\left(\mathrm{ext}_{\Psi}u\right)\mbox{.}
\]

Till now, we have used $\left(\psi^{\prime},\Psi^{\prime}\right)$
simply to define the extension map $\mathrm{ext}_{\Psi}$. However
such map could be also defined directly using simply $\psi^{\prime}$
and $\Psi$. The real reason that prompted us to the construction
of $\left(\psi^{\prime},\Psi^{\prime}\right)$ is that it gives us
the opportunity to build a new object of $\mathfrak{ghs}^{f}$. Consider
the globally hyperbolic spacetime $\psi\left(\mathscr{M}\right)$
(cfr. Remark \ref{remghs}) and the vector bundle $\Psi\left(E\right)$.
On $\Psi\left(E\right)$ we put the restriction of the inner product
of $F$ as inner product and automatically we find that $\Psi^{\prime}$
is fiberwise an isometry. We define $A_{\Psi}$ on in a way that $\Psi^{\prime}$
automatically satisfies the condition of compatibility with $A_{\Psi}$
and $B$: $A_{\Psi}$ is the linear operator from $\mathrm{C}^{\infty}\left(\psi\left(M\right),\Psi\left(E\right)\right)$
to itself defined by the formula
\begin{equation}
A_{\Psi}u=\Psi^{\prime}\circ\left(A\left(\Psi^{\prime-1}\circ u\circ\psi^{\prime}\right)\right)\circ\psi^{\prime-1}\quad\forall u\in\mathrm{C}^{\infty}\left(\psi\left(M\right),\Psi\left(E\right)\right)\mbox{.}\label{eqDefinitionOfAPsi}
\end{equation}
In this way $A_{\Psi}$ is well defined because $\left(\psi^{\prime},\Psi^{\prime}\right)$
is a vector bundle isomorphism (see Remark \ref{remPushforwardPullbackSections})
and one can easily check that $A_{\Psi}$ is a formally selfadjoint
normally hyperbolic operator on $\Psi\left(E\right)$ over $\psi\left(\mathscr{M}\right)$
(such properties are directly inherited from the same properties that
are known to hold for $A$). As we anticipated, the definition of
$A_{\Psi}$ is given in such a way that automatically $\left(\psi^{\prime},\Psi^{\prime}\right)$
becomes compatible with $A$ and $A_{\Psi}$: noting that
\begin{equation}
\left(\mathrm{ext}_{\Psi^{\prime}}u\right)\left(q\right)=\begin{cases}
\left(\Psi^{\prime}\circ u\circ\psi^{\prime-1}\right)\left(q\right) & \mbox{if }q\in\psi\left(M\right)\mbox{,}\\
0 & \mbox{if }q\in\psi\left(M\right)\setminus\psi\left(M\right)
\end{cases}=\left(\Psi^{\prime}\circ u\circ\psi^{\prime-1}\right)\left(q\right)\mbox{,}\label{eqextPsi'}
\end{equation}
for each $u\in\mathscr{D}\left(M,E\right)$ and each $q\in\psi\left(M\right)$
and choosing $u=\mathrm{ext}_{\Psi^{\prime}}v$ for any $v\in\mathscr{D}\left(M,E\right)$
in eq. \eqref{eqDefinitionOfAPsi}, we read
\[
A_{\Psi}\left(\mathrm{ext}_{\Psi^{\prime}}v\right)=\mathrm{ext}_{\Psi^{\prime}}\left(Av\right)
\]
for each $v\in\mathscr{D}\left(M,E\right)$. Then we have built the
object $\left(\psi\left(\mathscr{M}\right),\Psi\left(E\right),A_{\Psi}\right)$
of $\mathfrak{ghs}^{f}$. Since $\psi^{\prime}$ is a morphism of
$\mathfrak{ghs}$ from $\mathscr{M}$ to $\psi\left(\mathscr{M}\right)$
whose inverse is also a morphism, we recognize $\left(\psi^{\prime},\Psi^{\prime}\right)$
to be a morphism of $\mathfrak{ghs}^{f}$ from $\left(\mathscr{M},E,A\right)$
to $\left(\psi\left(\mathscr{M}\right),\Psi\left(E\right),A_{\Psi}\right)$
whose inverse $\left(\psi^{\prime-1},\Psi^{\prime-1}\right)$ is a
morphism too. The situation of eq. \eqref{eqextPsi'} holds whenever
we deal with a vector bundle isomorphism, in particular we can define
similarly $\mathrm{ext}_{\Psi^{\prime-1}}$.

In Remark \ref{remRestrictionOfVectorBundles} we showed that there
is also a vector bundle homomorphism $\left(\iota_{\psi\left(M\right)}^{N},\iota_{\Psi\left(E\right)}^{F}\right)$
from $\Psi\left(E\right)$ to $F$. We already know from our discussion
on the category $\mathfrak{ghs}$ that $\iota_{\psi\left(M\right)}^{N}$
is a morphism of $\mathfrak{ghs}$ from $\psi\left(\mathscr{M}\right)$
to $\mathscr{N}$. If we show that $\left(\iota_{\psi\left(M\right)}^{N},\iota_{\Psi\left(E\right)}^{F}\right)$
is compatible with the inner products of $\Psi\left(E\right)$ and
$F$ and with the normally hyperbolic operators $A_{\Psi}$ and $B$,
we can conclude that $\left(\iota_{\psi\left(M\right)}^{N},\iota_{\Psi\left(E\right)}^{F}\right)$
is a morphism of $\mathfrak{ghs}^{f}$. Both requirements actually
hold because of the definitions of $\left(\iota_{\psi\left(M\right)}^{N},\iota_{\Psi\left(E\right)}^{F}\right)$,
of the inner product on $\Psi\left(E\right)$ as the restriction of
the inner product of $F$ and of the normally hyperbolic operator
$A_{\Psi}$. We explicitly check the compatibility with $A_{\Psi}$
and $B$. Notice that for each $v\in\mathscr{D}\left(\psi\left(M\right),\Psi\left(E\right)\right)$
and each $q\in\psi\left(M\right)$
\begin{eqnarray*}
\left(\mathrm{ext}_{\iota_{\Psi\left(E\right)}^{F}}v\right)\left(q\right) & = & \begin{cases}
\left(\left(\iota_{\Psi\left(E\right)}^{F}\right)^{\prime}\circ v\circ\left(\iota_{\psi\left(M\right)}^{N}\right)^{\prime-1}\right)\left(q\right) & \mbox{if }q\in\psi\left(M\right)\mbox{,}\\
0 & \mbox{if }q\in N\setminus\psi\left(M\right)
\end{cases}\\
 & = & \begin{cases}
v\left(q\right) & \mbox{if }q\in\psi\left(M\right)\mbox{,}\\
0 & \mbox{if }q\in N\setminus\psi\left(M\right)\mbox{.}
\end{cases}
\end{eqnarray*}
For each $v\in\mathscr{D}\left(\psi\left(M\right),\Psi\left(E\right)\right)$,
exploiting eq. \eqref{eqextPsi'}, we find
\[
\mathrm{ext}_{\iota_{\Psi\left(E\right)}^{F}}\left(A_{\Psi}v\right)=\mathrm{ext}_{\iota_{\Psi\left(E\right)}^{F}}\left(A_{\Psi}\left(\mathrm{ext}_{\Psi^{\prime}}\left(\mathrm{ext}_{\Psi^{\prime-1}}v\right)\right)\right)=\mathrm{ext}_{\iota_{\Psi\left(E\right)}^{F}}\left(\mathrm{ext}_{\Psi^{\prime}}\left(A\left(\mathrm{ext}_{\Psi^{\prime-1}}v\right)\right)\right)\mbox{.}
\]
For each $u\in\mathscr{D}\left(M,E\right)$ and each $q\in\psi\left(M\right)$
we also have
\begin{eqnarray*}
\left(\mathrm{ext}_{\iota_{\Psi\left(E\right)}^{F}}\left(\mathrm{ext}_{\Psi^{\prime}}u\right)\right)\left(q\right) & = & \begin{cases}
\left(\Psi^{\prime}\circ u\circ\psi^{\prime-1}\right)\left(q\right) & \mbox{if }q\in\psi\left(M\right)\mbox{,}\\
0 & \mbox{if }q\in N\setminus\psi\left(M\right)
\end{cases}\\
 & = & \left(\mathrm{ext}_{\Psi}u\right)\left(q\right)\mbox{.}
\end{eqnarray*}
We insert our last equation in the previous one, we exploit the compatibility
property of $\left(\psi,\Psi\right)$ with $A$ and $B$ and we recall
the definitions of $\mathrm{ext}_{\Psi}$ and $\mathrm{ext}_{\iota_{\Psi\left(E\right)}^{F}}$.
In this way we obtain
\[
\mathrm{ext}_{\iota_{\Psi\left(E\right)}^{F}}\left(A_{\Psi}v\right)=\mathrm{ext}_{\Psi}\left(A\left(\mathrm{ext}_{\Psi^{\prime-1}}v\right)\right)=B\left(\mathrm{ext}_{\Psi}\left(\mathrm{ext}_{\Psi^{\prime-1}}v\right)\right)=B\left(\mathrm{ext}_{\iota_{\Psi\left(E\right)}^{F}}v\right)
\]
for each $v\in\mathscr{D}\left(\psi\left(M\right),\Psi\left(E\right)\right)$.
This shows that $\left(\iota_{\psi\left(M\right)}^{N},\iota_{\Psi\left(E\right)}^{F}\right)$
is compatible with $A_{\Psi}$ and $B$ and hence it is actually a
morphism of $\mathfrak{ghs}^{f}$. Using $\left(\iota_{\psi\left(M\right)}^{N},\iota_{\Psi\left(E\right)}^{F}\right)$
and $\left(\psi^{\prime},\Psi^{\prime}\right)$, we can decompose
the original morphism $\left(\psi,\Psi\right)$ according the formula
\[
\left(\psi,\Psi\right)=\left(\iota_{\psi\left(M\right)}^{N},\iota_{\Psi\left(E\right)}^{F}\right)\circ\left(\psi^{\prime},\Psi^{\prime}\right)\mbox{.}
\]

Having explicated the meaning of all the requirements in Definition
\ref{defghsfssp}, we can ask whether $\mathfrak{ghs}^{f}$ and $\mathfrak{ssp}$
are actually categories. This question is faced in the subsequent
remark.
\begin{rem}
We check that $\mathfrak{ghs}^{f}$ is actually a category. The first
thing to be done is to verify that the composition law is well defined.
To this end consider three objects $\left(\mathscr{M},E,A\right)$,
$\left(\mathscr{N},F,B\right)$ and $\left(\mathscr{O},G,C\right)$,
a morphism $\left(\phi,\Phi\right)$ from $\left(\mathscr{M},E,A\right)$
to $\left(\mathscr{N},F,B\right)$ and a morphism $\left(\psi,\Psi\right)$
from $\left(\mathscr{N},F,B\right)$ to $\left(\mathscr{O},G,C\right)$.
As we have seen in Remark \ref{remghs}, $\psi\circ\phi$ is still
a morphism of $\mathfrak{ghs}$. Since $\Phi:E\rightarrow F$ and
$\Psi:F\rightarrow G$ are smooth maps, using coordinate charts of
the manifolds $E$, $F$, $G$ we immediately realize that also $\Psi\circ\Phi:E\rightarrow G$
is a smooth map. Let $\pi_{E}$, $\pi_{F}$ and $\pi_{G}$ be the
projection maps respectively of $E$, $F$ and $G$. We know that
$\pi_{F}\circ\Phi=\phi\circ\pi_{E}$ and that $\pi_{G}\circ\Psi=\psi\circ\pi_{F}$.
Then, applying the associativity of the composition of functions,
we find
\[
\pi_{G}\circ\left(\Psi\circ\Phi\right)=\psi\circ\pi_{F}\circ\Phi=\left(\psi\circ\phi\right)\circ\pi_{E}\mbox{.}
\]
As for fiberwise linearity, consider a point $p\in M$. Taking into
account the map
\begin{eqnarray*}
\left(\Psi\circ\Phi\right)_{p}:E_{p} & \rightarrow & G_{\psi\left(\phi\left(p\right)\right)}\\
\mu & \mapsto & \left(\Psi\circ\Phi\right)\mu\mbox{,}
\end{eqnarray*}
we can easily check that
\[
\left(\Psi\circ\Phi\right)_{p}=\Psi_{\phi\left(p\right)}\circ\Phi_{p}
\]
and hence $\left(\Psi\circ\Phi\right)_{p}$ is linear being the composition
of linear maps. This shows that $\left(\psi,\Psi\right)\circ\left(\phi,\Phi\right)=\left(\psi\circ\phi,\Psi\circ\Phi\right)$
is a vector bundle homomorphism from $E$ to $G$. We check its compatibility
with the inner products of the vector bundles involved exploiting
the same property that is assumed to hold for both $\left(\phi,\Phi\right)$
and $\left(\psi,\Psi\right)$: for each $p\in M$ and each $\mu$,
$\nu\in E_{p}$ we have
\[
\left(\left(\Psi\circ\Phi\right)_{p}\mu\right)\overset{G}{\cdot}_{\psi\left(\phi\left(p\right)\right)}\left(\left(\Psi\circ\Phi\right)_{p}\nu\right)=\left(\Phi_{p}\mu\right)\overset{F}{\cdot}_{\phi\left(p\right)}\left(\Phi_{p}\nu\right)=\mu\overset{E}{\cdot}_{p}\nu\mbox{.}
\]
As for the compatibility with the normally hyperbolic operators, for
each $u\in\mathscr{D}\left(M,E\right)$ it holds that
\begin{equation}
\mathrm{ext}_{\Psi}\left(\mathrm{ext}_{\Phi}\left(Au\right)\right)=\mathrm{ext}_{\Psi}\left(B\left(\mathrm{ext}_{\Phi}u\right)\right)=C\left(\mathrm{ext}_{\Psi}\left(\mathrm{ext}_{\Phi}u\right)\right)\mbox{,}\label{eqextPsiextPhiAu=00003DCextPsiextPhiu}
\end{equation}
where $\mathrm{ext}_{\Phi}:\mathscr{D}\left(M,E\right)\rightarrow\mathscr{D}\left(N,F\right)$
and $\mathrm{ext}_{\Psi}:\mathscr{D}\left(N,F\right)\rightarrow\mathscr{D}\left(O,G\right)$
are the extension maps obtained applying the discussion that led to
eq. \eqref{eqDefinitionOfextPsi} to $\left(\phi,\Phi\right)$ with
$\phi\left(M\right)$ and respectively to $\left(\psi,\Psi\right)$
with $\psi\left(N\right)$. In the same way from $\left(\psi,\Psi\right)\circ\left(\phi,\Phi\right)$
with $\left(\psi\circ\phi\right)\left(M\right)$, we obtain the extension
map $\mathrm{ext}_{\Psi\circ\Phi}:\mathscr{D}\left(M,E\right)\rightarrow\mathscr{D}\left(O,E\right)$.
Our scope now is to show that $\mathrm{ext}_{\Psi\circ\Phi}=\mathrm{ext}_{\Psi}\circ\mathrm{ext}_{\Phi}$:
for each $u\in\mathscr{D}\left(M,E\right)$ and each $r\in O$
\begin{eqnarray*}
\left(\mathrm{ext}_{\Psi\circ\Phi}u\right)\left(r\right) & = & \begin{cases}
\left(\left(\Psi\circ\Phi\right)^{\prime}\circ u\circ\left(\psi\circ\phi\right)^{\prime-1}\right)\left(r\right) & \mbox{if }r\in\left(\psi\circ\phi\right)\left(M\right)\mbox{,}\\
0 & \mbox{if }r\in O\setminus\left(\psi\circ\phi\right)\left(M\right)
\end{cases}\\
 & = & \begin{cases}
\left(\Psi\circ\Phi\right)^{\prime}\left(u\left(\left(\psi\circ\phi\right)^{\prime-1}\left(r\right)\right)\right) & \mbox{if }r\in\left(\psi\circ\phi\right)\left(M\right)\mbox{,}\\
0 & \mbox{if }r\in O\setminus\left(\psi\circ\phi\right)\left(M\right)
\end{cases}\\
 & = & \begin{cases}
\Psi^{\prime}\left(\Phi^{\prime}\left(u\left(\psi^{\prime-1}\left(\phi^{\prime-1}\left(r\right)\right)\right)\right)\right) & \mbox{if }r\in\left(\psi\circ\phi\right)\left(M\right)\mbox{,}\\
0 & \mbox{if }r\in O\setminus\left(\psi\circ\phi\right)\left(M\right)\mbox{,}
\end{cases}
\end{eqnarray*}
while
\begin{eqnarray*}
\left(\mathrm{ext}_{\Psi}\left(\mathrm{ext}_{\Phi}u\right)\right)\left(r\right) & = & \begin{cases}
\left(\Psi^{\prime}\circ\left(\mathrm{ext}_{\Phi}u\right)\circ\psi^{\prime-1}\right)\left(r\right) & \mbox{if }r\in\psi\left(N\right)\mbox{,}\\
0 & \mbox{if }r\in O\setminus\psi\left(N\right)
\end{cases}\\
 & = & \begin{cases}
\Psi^{\prime}\left(\Phi^{\prime}\left(u\left(\phi^{\prime-1}\left(\psi^{\prime-1}\left(r\right)\right)\right)\right)\right) & \mbox{if }r\in\psi\left(\phi\left(M\right)\right)\mbox{,}\\
0 & \mbox{if }r\in\psi\left(N\right)\setminus\psi\left(\phi\left(N\right)\right)\mbox{,}\\
0 & \mbox{if }r\in O\setminus\psi\left(N\right)\mbox{,}
\end{cases}
\end{eqnarray*}
hence the equation
\begin{equation}
\mathrm{ext}_{\Psi\circ\Phi}=\mathrm{ext}_{\Psi}\circ\mathrm{ext}_{\Phi}\label{eqextPsiPhi=00003DextPsiextPhi}
\end{equation}
holds as expected. Inserting such equation in eq. \eqref{eqextPsiextPhiAu=00003DCextPsiextPhiu},
we conclude that for each $u\in\mathscr{D}\left(M,E\right)$
\[
\mathrm{ext}_{\Psi\circ\Phi}\left(Au\right)=B\left(\mathrm{ext}_{\Psi\circ\Phi}u\right)\mbox{.}
\]
Then $\left(\psi,\Psi\right)\circ\left(\phi,\Phi\right)$ is actually
a morphism of $\mathfrak{ghs}^{f}$. This proves that the composition
law is well defined. To conclude, we have to check the category axioms.
If we take an arbitrary object $\left(\mathscr{M},E,A\right)$, the
identity law is satisfied by the vector bundle homomorphism $\mathrm{id}_{\left(\mathscr{M},E,A\right)}=\left(\mathrm{id}_{M},\mathrm{id}_{E}\right)$,
where $\mathrm{id}_{M}:M\rightarrow M$, $p\mapsto p$ and $\mathrm{id}_{E}:E\rightarrow E$,
$\mu\mapsto\mu$. The associativity of the composition law is trivial
because this property is inherited from the associativity of the ordinary
composition of functions.

Turning our attention to $\mathfrak{ssp}$, we take three objects
$\left(U,\rho\right)$, $\left(V,\sigma\right)$ and $\left(W,\omega\right)$,
a morphism $\xi$ from $\left(U,\rho\right)$ to $\left(V,\sigma\right)$
and a morphism $\eta$ from $\left(V,\sigma\right)$ to $\left(W,\omega\right)$
and we consider the function $\eta\circ\xi:U\rightarrow W$. We obtain
a linear map between the vector spaces $U$ and $W$ that preserves
the symplectic forms because both $\xi$ and $\eta$ are symplectic
maps:
\[
\omega\left(\eta\left(\xi v\right),\eta\left(\xi w\right)\right)=\sigma\left(\xi v,\xi w\right)=\rho\left(v,w\right)\quad\forall v,w\in U\mbox{.}
\]
Hence we realize that $\eta\circ\xi$ is a morphism from $\left(U,\rho\right)$
to $\left(W,\omega\right)$ and this proves that the composition law
is well defined. For each object $\left(V,\sigma\right)$, the identity
law is satisfied by the morphism $\mathrm{id}_{\left(V,\sigma\right)}$
from $\left(V,\sigma\right)$ to itself defined by $\mathrm{id}_{\left(V,\sigma\right)}v=v$
for each $v\in V$. Also in this case the associativity of the composition
law follows from same property of the composition of functions.
\end{rem}

\subsubsection{Functor}

We begin now the construction of a covariant functor that maps each
object of $\mathfrak{ghs}^{f}$ to an object of $\mathfrak{ssp}$.
In our intention the object of $\mathfrak{ghs}^{f}$ should describe
the physical problem that we deal with (in this case a wave equation
describing a field over a globally hyperbolic spacetime) and the corresponding
object of $\mathfrak{ssp}$ should be the solution for such problem
(all the dynamical configurations of the field, i.e. the solutions
of all the Cauchy problems with compactly supported initial data related
to the wave equation mentioned above). We may say that the object
of $\mathfrak{ghs}^{f}$ describes the system we want to study, while
the corresponding object of $\mathfrak{ssp}$ contains all the knowledge
about the dynamics of that system. In the following we will make extensive
use of the results in Subsection \ref{subGreenOperators}.

Assume that we are given an object $\left(\mathscr{M},E,A\right)$
of $\mathfrak{ghs}^{f}$. Applying Corollary \ref{corGreenOperators},
we obtain the advanced and retarded Green operators $e_{A}^{a}$ and
$e_{A}^{r}$ for $A$. With them we can introduce the causal propagator
$e_{A}=e_{A}^{a}-e_{A}^{r}$ for $A$ and Corollary \ref{corSpaceOfSolutions}
tells us that the space of the solutions for all the homogeneous Cauchy
problems with compactly supported initial data associated to $A$
is given by the vector space $V=e_{A}\left(\mathscr{D}\left(M,E\right)\right)$.
Moreover Proposition \ref{propCausalPropagatorKernel} provides an
important information on the structure of the kernel of the causal
propagator $e_{A}$, precisely $\ker e_{A}=A\left(\mathscr{D}\left(M,E\right)\right)$.
We also notice that in the current situation $A$ is supposed to be
formally selfadjoint, i.e. $A^{*}=A$ (cfr. Remark \ref{remFormalSelfadjointness}).
It follows in particular that $A^{*}$ is a normally hyperbolic operator
too and that its advanced and retarded Green operators are exactly
$e_{A}^{a}$ and $e_{A}^{r}$. Hence Proposition \ref{prope*rIsFormallyAdjointToea}
in the present situation means that $e_{A}^{a}$ and $e_{A}^{r}$
are the formally adjoints of each other. As a consequence of this
fact, we see that $e_{A}$ is formally antiselfadjoint. All these
observations will be exploited soon.
\begin{lem}
\label{lemObjghsf->Objssp}Consider the situation presented above
and bear in mind that $\mathscr{M}=\left(M,g,\mathfrak{o},\mathfrak{t}\right)$.
Taking into account the vector space $V=e_{A}\left(\mathscr{D}\left(M,E\right)\right)$,
the map
\begin{eqnarray*}
\sigma:V\times V & \rightarrow & \mathbb{R}\\
\left(u,v\right) & \mapsto & \int\limits _{M}\left(\left(e_{A}f\right)\overset{E}{\cdot}h\right)\mathrm{d}\mu_{g}\mbox{,}
\end{eqnarray*}
where $f$ and $h$ in $\mathscr{D}\left(M,E\right)$ are such that
$e_{A}f=u$ and $e_{A}h=v$ and $\mathrm{d}\mu_{g}$ is the standard
volume form on $\mathscr{M}$, is well defined, bilinear, antisymmetric
and non degenerate, i.e. $\sigma$ is a symplectic form on $V$ and
hence $\left(V,\sigma\right)$ is a symplectic space, as a matter
of fact an object of $\mathfrak{ssp}$.\end{lem}
\begin{proof}
We check that $\sigma$ is well defined. Fix $u$ and $v$ in $V=e_{A}\left(\mathscr{D}\left(M,E\right)\right)$
and take $f_{1}$, $f_{2}$, $h_{1}$, $h_{2}$ in $\mathscr{D}\left(M,E\right)$
such that $e_{A}f_{1}=u=e_{A}f_{2}$ and $e_{A}h_{1}=v=e_{A}h_{2}$.
Exploiting twice the antiselfadjointness of $e_{A}$, we deduce that
\begin{alignat*}{3}
\int\limits _{M}\left(\left(e_{A}f_{1}\right)\overset{E}{\cdot}h_{1}\right)\mathrm{d}\mu_{g} & = & \int\limits _{M}\left(\left(e_{A}f_{2}\right)\overset{E}{\cdot}h_{1}\right)\mathrm{d}\mu_{g} & = & -\int\limits _{M}\left(f_{2}\overset{E}{\cdot}\left(e_{A}h_{1}\right)\right)\mathrm{d}\mu_{g}\\
 & = & -\int\limits _{M}\left(f_{2}\overset{E}{\cdot}\left(e_{A}h_{2}\right)\right)\mathrm{d}\mu_{g} & = & \int\limits _{M}\left(\left(e_{A}f_{2}\right)\overset{E}{\cdot}h_{2}\right)\mathrm{d}\mu_{g}\mbox{.}
\end{alignat*}

Till this point we have shown that $\sigma$ is well defined. Bilinearity
follows directly from the linearity of the causal propagator, fiberwise
bilinearity of the inner product in $E$ and the linearity of the
integral. As for antisymmetry, we take $u$ and $v$ in $V$. Then
we find $f$ and $h$ in $\mathscr{D}\left(M,E\right)$ such that
$e_{A}f=u$ and $e_{A}h=v$. With this ingredients we evaluate $\sigma\left(u,v\right)$
bearing in mind that $e_{A}$ is antiselfadjoint and that the inner
product of $E$ is fiberwise symmetric:
\begin{alignat*}{1}
\sigma\left(u,v\right) & =\int\limits _{M}\left(\left(e_{A}f\right)\overset{E}{\cdot}h\right)\mathrm{d}\mu_{g}=-\int\limits _{M}\left(f\overset{E}{\cdot}\left(e_{A}h\right)\right)\mathrm{d}\mu_{g}=-\int\limits _{M}\left(\left(e_{A}h\right)\overset{E}{\cdot}f\right)\mathrm{d}\mu_{g}\\
 & =-\sigma\left(v,u\right)\mbox{.}
\end{alignat*}

We are left with the proof of non degeneracy. Suppose that we have
$u\in V$ such that $\sigma\left(u,v\right)=0$ for each $v\in V$.
This means that
\[
\int\limits _{M}\left(u\overset{E}{\cdot}f\right)\mathrm{d}\mu_{g}=0
\]
for each $f\in\mathscr{D}\left(M,E\right)$. But this implies that
$u=0$. Hence $\sigma$ is actually non degenerate.
\end{proof}
The last theorem provides the first part of our candidate covariant
functor, specifically the map
\begin{alignat*}{2}
\mathscr{B} & : & \mathsf{Obj}_{\mathfrak{ghs}^{f}} & \rightarrow\mathsf{Obj}_{\mathfrak{ghs}^{f}}\\
 &  & \left(\mathscr{M},E,A\right) & \mapsto\left(V,\sigma\right)\mbox{.}
\end{alignat*}
The second part should consist of a map
\[
\mathsf{Mor}_{\mathfrak{ghs}^{f}}\left(\left(\mathscr{M},E,A\right),\left(\mathscr{N},F,B\right)\right)\rightarrow\mathsf{Mor}_{\mathfrak{ssp}}\left(\mathscr{B}\left(\mathscr{M},E,A\right),\mathscr{B}\left(\mathscr{N},F,B\right)\right)
\]
for each pair of objects $\left(\mathscr{M},E,A\right)$ and $\left(\mathscr{N},F,B\right)$
of $\mathfrak{ghs}^{f}$. To build such map we need a preliminary
result.
\begin{lem}
\label{lemresPsieBextPsi=00003DeA}Let $\left(\mathscr{M}=\left(M,g,\mathfrak{o},\mathfrak{t}\right),E,A\right)$
and $\left(\mathscr{N}=\left(N,h,\mathfrak{p},\mathfrak{u}\right),F,B\right)$
be two objects of $\mathfrak{ghs}^{f}$ and let $\left(\psi,\Psi\right)$
be a morphism between these objects. Denote the advanced/retarded
Green operators for $A$ and $B$ respectively with $e_{A}^{a/r}$
and $e_{B}^{a/r}$ and consider the maps $\mathrm{ext}_{\Psi}$ defined
in eq. \eqref{eqDefinitionOfextPsi} and the map
\begin{eqnarray*}
\mathrm{res}_{\Psi}:\mathrm{C^{\infty}}\left(N,F\right) & \rightarrow & \mathrm{C}^{\infty}\left(M,E\right)\\
v & \mapsto & \Psi^{\prime-1}\circ\left(\left.v\right|_{\psi\left(M\right)}\right)\circ\psi^{\prime}.
\end{eqnarray*}
Then we have $\mathrm{res}_{\Psi}\circ e_{B}^{a/r}\circ\mathrm{ext}_{\Psi}=e_{A}^{a/r}$.\end{lem}
\begin{proof}
We start showing that the map $\mathrm{res}_{\Psi}$ is well defined.
Consider a section $v\in\mathrm{C}^{\infty}\left(N,F\right)$. By
$\left.v\right|_{\psi\left(M\right)}$ we mean the function from $\psi\left(M\right)$
to $\Psi\left(E\right)$ defined by $\left.v\right|_{\psi\left(M\right)}\left(q\right)=v\left(q\right)$
for each $q\in\psi\left(M\right)$. The domain and the codomain of
$\left.v\right|_{\psi\left(M\right)}$ are manifolds, hence we can
ask whether this function is continuous and, in this case, whether
it is also smooth. Both questions have positive answer because of
the topologies and the atlases of $\psi\left(M\right)$ and $\Psi\left(E\right)$,
which are open subsets of $N$ and respectively $\Psi\left(E\right)$,
are induced via restriction from those of $N$ and $F$ (cfr. Remark
\ref{remSubmanifold} for $\psi\left(M\right)$ and Remark \ref{remRestrictionOfVectorBundleHomomorphisms}
for $\Psi\left(E\right)$). We have recognized $\left.v\right|_{\psi\left(M\right)}$
to be a smooth map from $\psi\left(M\right)$ to $\Psi\left(E\right)$.
But the remark cited above tells us also that $\Psi\left(E\right)$
is a vector bundle over the manifold $\psi\left(M\right)$. Hence
we can also ask whether $\left.v\right|_{\psi\left(M\right)}$ is
a section in $\Psi\left(E\right)$ and again we get a positive answer
because $\pi_{\Psi\left(E\right)}$ is defined as the restriction
of $\pi_{F}$:
\[
\pi_{\Psi\left(E\right)}\left(\left.v\right|_{\psi\left(M\right)}\left(q\right)\right)=\pi_{F}\left(v\left(q\right)\right)=q\quad\forall q\in\psi\left(M\right)\mbox{.}
\]
From Remark \ref{remPushforwardPullbackSections} applied to the section
$\left.v\right|_{\psi\left(M\right)}$ and the vector bundle isomorphism
$\left(\psi^{\prime},\Psi^{\prime}\right)$, we finally obtain the
section in $E$ we were looking for:
\[
\Psi^{\prime-1}\circ\left(\left.v\right|_{\psi\left(M\right)}\right)\circ\psi^{\prime}\mbox{.}
\]
This proves that the definition of $\mathrm{res}_{\Psi}$ makes sense.

Our strategy to prove the thesis of this lemma consists in showing
that $\mathrm{res}_{\Psi}\circ e_{B}^{a/r}\circ\mathrm{ext}_{\Psi}$
is an advanced/retarded Green operator for $A$ (cfr. Definition \ref{defGreenOperators}),
so that it must coincide with $e_{A}^{a/r}$ because Corollary \ref{corGreenOperators}
assures uniqueness. We consider only the case of the advanced Green
operator, the other case being similar.

To show that $\mathrm{res}_{\Psi}\circ e_{B}^{a}\circ\mathrm{ext}_{\Psi}$
is the advanced Green operator for $A$ we do not check that it fulfils
the requirements of Definition \ref{defGreenOperators}, but we prefer
to show that it generates exactly one fundamental solution $U_{A}^{r}\left(p\right)$
for $A^{*}$ (in this case $A^{*}=A$, but we will not use such property)
with $\mathscr{M}$-future compact support for each point $p$ in
$M$ according to the formula
\[
U_{A}^{r}\left(p\right)\left[u\right]=\left(\left(\mathrm{res}_{\Psi}\circ e_{B}^{a}\circ\mathrm{ext}_{\Psi}\right)u\right)\left(p\right)\quad\forall u\in\mathscr{D}\left(M,E\right)\mbox{.}
\]
If we succeed in our scope, via Corollary \ref{corGreenOperators}
we obtain an advanced Green operator for $A$ from the collection
of fundamental solutions with future compact support $\left\{ U_{A}^{r}\left(p\right):\, p\in M\right\} $
for $A^{*}$. This operator is defined through a formula identical
to the one written above, but intended in the opposite sense, hence
we find that $\mathrm{res}_{\Psi}\circ e_{B}^{a}\circ\mathrm{ext}_{\Psi}$
is exactly this operator. In particular it follows that $\mathrm{res}_{\Psi}\circ e_{B}^{a}\circ\mathrm{ext}_{\Psi}$
is an advanced Green operator for $A$.

Now we fix $p\in M$. Together with the normally hyperbolic operator
$A^{*}$ (normal hyperbolicity of $A^{*}$ follows from the hypothesis
that $A$ is normally hyperbolic even if $A^{*}\neq A$), we consider
also its distributional version (still denoted by $A^{*}$) as it
is defined following the procedure shown in Remark \ref{remLinearDifferentialOperatorInDistributionalSense}
using $E_{p}$ as target vector space for the space of distributions
(see the discussion before Definition \ref{defFundamentalSolution}):
\[
A^{*}:\mathscr{D}^{\prime}\left(M,E^{*},E_{p}\right)\rightarrow\mathscr{D}^{\prime}\left(M,E^{*},E_{p}\right)
\]
Indeed the hypothesis of formal selfadjointness implies that $A=A^{*}$
also for the operators in distributional sense (cfr. Remark \ref{remFormalSelfadjointness}),
but this fact is not necessary for our conclusions and hence in this
proof we will distinguish between $A^{*}$ and $A$ as it would have
been without the hypothesis of formal selfadjointness. We note that
for each $u\in\mathscr{D}\left(M,E\right)$ it holds
\[
\left(A^{*}U_{A}^{r}\left(p\right)\right)\left[u\right]=U_{A}^{r}\left(p\right)\left[Au\right]=\left(\left(\mathrm{res}_{\Psi}\circ e_{B}^{a}\circ\mathrm{ext}_{\Psi}\right)Au\right)\left(p\right)=u\left(p\right)=\delta_{p}\left[u\right]
\]
due to the compatibility of $\left(\psi,\Psi\right)$ with $A$ and
$B$ and the fact that $e_{B}^{a}$ is the advanced Green operator
for $B$. This means that $U_{A}^{r}\left(p\right)$ is a fundamental
solution for $A^{*}$ at $p$ (cfr. Definition \ref{defFundamentalSolution}).
The support of the distribution $U_{A}^{r}\left(p\right)$ is given
by
\[
\mathrm{supp}\left(U_{A}^{r}\left(p\right)\right)=\left\{ \begin{array}{rl}
q\in M: & \mbox{for each neighborhood }U\mbox{ of }q\mbox{ in }M\mbox{ there}\\
 & \mbox{exists a section }u\in\mathscr{D}\left(M,E\right)\mbox{ with support }\\
 & \mbox{included in }U\mbox{ such that }U_{A}^{r}\left(p\right)\left[u\right]\neq0
\end{array}\right\} \mbox{.}
\]
But $U_{A}^{r}\left(p\right)\left[u\right]\neq0$ means that $\left(e_{B}^{a}\left(\mathrm{ext}_{\Psi}u\right)\right)\left(\psi\left(p\right)\right)\neq0$,
that is $U_{B}^{r}\left(\psi\left(p\right)\right)\left[\mathrm{ext}_{\Psi}u\right]\neq0$,
where $U_{B}^{r}\left(\psi\left(p\right)\right)$ is the unique fundamental
solution for $B^{*}$ at $\psi\left(p\right)$ with future compact
support generated by the advanced Green operator $e_{B}^{a}$ for
$B$ according to Corollary \ref{corGreenOperators}. On the one hand
we deduce that
\[
\mathrm{supp}\left(U_{A}^{r}\left(p\right)\right)\subseteq\psi^{-1}\left(\left\{ \begin{array}{rl}
q\in\psi\left(M\right): & \mbox{for each neighborhood }V\mbox{ of }q\mbox{ in }N\\
 & \mbox{there exists a section }v\in\mathscr{D}\left(N,F\right)\\
 & \mbox{with support included in }V\mbox{ such}\\
 & \mbox{such that }U_{B}^{r}\left(\psi\left(p\right)\right)\left[v\right]\neq0
\end{array}\right\} \right)\mbox{.}
\]
On the other hand the support of $U_{B}^{r}\left(\psi\left(p\right)\right)$
is of the form
\[
\mathrm{supp}\left(U_{B}^{r}\left(\psi\left(p\right)\right)\right)=\left\{ \begin{array}{rl}
q\in N: & \mbox{for each neighborhood }V\mbox{ of }q\mbox{ in }N\mbox{ there}\\
 & \mbox{exists a section }v\in\mathscr{D}\left(N,F\right)\mbox{ with support}\\
 & \mbox{included in }V\mbox{ such that }U_{B}^{r}\left(\psi\left(p\right)\right)\left[u\right]\neq0
\end{array}\right\} \mbox{.}
\]
From the comparison of the last two equations we conclude that
\[
\mathrm{supp}\left(U_{A}^{r}\left(p\right)\right)\subseteq\psi^{-1}\left(\mathrm{supp}\left(U_{B}^{r}\left(\psi\left(p\right)\right)\right)\right)\mbox{.}
\]
Theorem \ref{thmFundamentalSolutions} gives us an important information
about the support of $U_{B}^{r}\left(\psi\left(p\right)\right)$,
namely the inclusion
\[
\mathrm{supp}\left(U_{B}^{r}\left(\psi\left(p\right)\right)\right)\subseteq J_{-}^{\mathscr{N}}\left(\psi\left(p\right)\right)\mbox{,}
\]
therefore we obtain
\[
\mathrm{supp}\left(U_{A}^{r}\left(p\right)\right)\subseteq\psi^{-1}\left(J_{-}^{\mathscr{N}}\left(\psi\left(p\right)\right)\right)\mbox{.}
\]
Consider now a point $q\in\psi^{-1}\left(J_{-}^{\mathscr{N}}\left(\psi\left(p\right)\right)\right)$.
We recognize that $\psi\left(p\right)$ and $\psi\left(q\right)$
are both in $\psi\left(M\right)$ and moreover $\psi\left(q\right)$
falls in $J_{-}^{\mathscr{N}}\left(\psi\left(p\right)\right)$. Hence
we find a $\mathfrak{u}$-past directed $h$-causal curve $\gamma$
in $N$ starting at $\psi\left(p\right)$ and ending at $\psi\left(q\right)$.
Since we assumed that $\psi$ is a morphism of $\mathfrak{ghs}$,
$\psi\left(M\right)$ is $\mathscr{N}$-causally convex and so $\gamma$
is entirely included in $\psi\left(M\right)$. Then we can consider
the curve $\psi^{\prime-1}\circ\gamma$. Since $\psi$ is isometric
and preserves time orientation, we deduce that $\psi^{\prime-1}\circ\gamma$
is a $\mathfrak{t}$-past directed $g$-causal curve in $M$ starting
from $p$ and ending in $q$. This implies that $q\in J_{-}^{\mathscr{M}}\left(p\right)$.
Then we have the inclusion
\[
\mathrm{supp}\left(U_{A}^{r}\left(p\right)\right)\subseteq J_{-}^{\mathscr{M}}\left(p\right)\mbox{.}
\]
By assumption, $\mathscr{M}$ is a globally hyperbolic spacetime and
so $J_{-}^{\mathscr{M}}\left(p\right)\cap J_{+}^{\mathscr{M}}\left(q\right)$
is a compact subset of $M$ for each $q\in M$. Hence $J_{-}^{\mathscr{M}}\left(p\right)$
is $\mathscr{M}$-future compact and then $\mathrm{supp}\left(U_{A}^{r}\left(p\right)\right)$
is $\mathscr{M}$-future compact too being a closed subset of $J_{-}^{\mathscr{M}}\left(p\right)$.
This shows that $U_{A}^{r}\left(p\right)$ is a fundamental solution
for $A^{*}$ at $p$ with $\mathscr{M}$-future compact support for
each $p\in M$. Uniqueness follows from Lemma \ref{lem PU=00003D0 implies U=00003D0}.
Hence our strategy of proof can be carried out without difficulties
and the thesis is proved.
\end{proof}
Now we are ready to face the main problem, that is. the determination
of a function that maps each morphism between two objects of $\mathfrak{ghs}^{f}$
to a morphism between the corresponding two objects of $\mathfrak{ssp}$.
\begin{lem}
\label{lemMorghsf->Morssp}Let $\left(\mathscr{M}=\left(M,g,\mathfrak{o},\mathfrak{t}\right),E,A\right)$
and $\left(\mathscr{N}=\left(M,h,\mathfrak{p},\mathfrak{u}\right),F,B\right)$
be two objects of $\mathfrak{ghs}^{f}$ and denote with $\left(V,\sigma\right)$
and $\left(W,\omega\right)$ the corresponding objects of $\mathfrak{ssp}$
provided by Lemma \ref{lemObjghsf->Objssp}. Consider a morphism $\left(\psi,\Psi\right)$
of $\mathfrak{ghs}^{f}$ from $\left(\mathscr{M},E,A\right)$ to $\left(\mathscr{N},F,B\right)$.
Denote with $e_{A}$ and $e_{B}$ the causal propagators for $A$
and $B$ respectively. Then the map
\begin{eqnarray*}
\xi:V & \rightarrow & W\\
u & \mapsto & e_{B}\left(\mathrm{ext}_{\Psi}f\right)\mbox{,}
\end{eqnarray*}
where $f\in\mathscr{D}\left(M,E\right)$ is such that $e_{A}f=u$,
is well defined, linear and compatible with $\sigma$ and $\omega$,
i.e. $\xi$ is a symplectic map from $\left(V,\sigma\right)$ to $\left(W,\omega\right)$,
which is to say that $\xi$ is a morphism of $\mathfrak{ssp}$ between
the objects $\left(V,\sigma\right)$ and $\left(W,\omega\right)$.\end{lem}
\begin{proof}
Fix $u\in V=e_{A}\left(\mathscr{D}\left(M,E\right)\right)$ and consider
$f_{1}$ and $f_{2}$ in $\mathscr{D}\left(M,E\right)$ such that
$e_{A}f_{1}=u=e_{A}f_{2}$. In order to have $\xi$ well defined we
must show that $e_{B}\left(\mathrm{ext}_{\Psi}f_{1}\right)=e_{B}\left(\mathrm{ext}_{\Psi}f_{2}\right)$.
Because of the linearity of the causal propagator and of the extension
map (see how $\mathrm{ext}_{\Psi}$ was defined in eq. \eqref{eqDefinitionOfextPsi}),
this is equivalent to prove that $f^{\prime}=\mathrm{ext}_{\Psi}f$
falls in the kernel of $e_{B}$, where $f$ denotes $f_{1}-f_{2}$.
We know that $\ker e_{A}=A\left(\mathscr{D}\left(N,F\right)\right)$
and that $f\in\ker e_{A}$. Hence we find $h\in\mathscr{D}\left(M,E\right)$
such that $Ah=f$. Then, exploiting the compatibility of $\left(\psi,\Psi\right)$
with $A$ and $B$, we obtain
\[
\mathrm{ext}_{\Psi}f=\mathrm{ext}_{\Psi}\left(Ah\right)=B\left(\mathrm{ext}_{\Psi}h\right)\mbox{.}
\]
We have just found $h^{\prime}=\mathrm{ext}_{\Psi}h$ in $\mathscr{D}\left(N,F\right)$
such that $Bh^{\prime}=f^{\prime}$. This implies that $f^{\prime}$
falls in $B\left(\mathscr{D}\left(N,F\right)\right)=\ker e_{B}$.

Linearity of the causal propagators and of the extension map assures
that $\xi$ is linear too.

Consider now $u_{1}$ and $u_{2}$ in $V$ and evaluate $\omega\left(\xi u_{1},\xi u_{2}\right)$.
We find $f_{1}$ and $f_{2}$ in $\mathscr{D}\left(M,E\right)$ such
that $e_{A}f_{1}=u_{1}$ and $e_{A}f_{2}=u_{2}$. Then, exploiting
the definition of $\xi$, we get
\[
\omega\left(\xi u_{1},\xi u_{2}\right)=\int\limits _{N}\left(\left(e_{B}\left(\mathrm{ext}_{\Psi}f_{1}\right)\right)\overset{F}{\cdot}\left(\mathrm{ext}_{\Psi}f_{2}\right)\right)\mathrm{d}\mu_{h}\mbox{,}
\]
where $\mathrm{d}\mu_{h}$ is the standard volume element of $\mathscr{N}$.
Notice that the argument of the last integral is null at least outside
$\psi\left(M\right)$ because of the definition of $\mathrm{ext}_{\Psi}$.
Moreover the restriction to $\psi\left(M\right)$ of the vector bundle
$F$ is the vector bundle $\Psi\left(E\right)$ (cfr. Remark \ref{remRestrictionOfVectorBundleHomomorphisms}),
whose inner product is the restriction of the inner product of $F$
(see few lines before eq. \eqref{eqDefinitionOfAPsi}). This gives
us the opportunity to write
\[
\omega\left(\xi u_{1},\xi u_{2}\right)=\int\limits _{\psi\left(M\right)}\left(\left.\left(e_{B}\left(\mathrm{ext}_{\Psi}f_{1}\right)\right)\right|_{\psi\left(M\right)}\overset{\Psi\left(E\right)}{\cdot}\left.\left(\mathrm{ext}_{\Psi}f_{2}\right)\right|_{\psi\left(M\right)}\right)\mathrm{d}\mu_{\left.h\right|_{\psi\left(M\right)}}\mbox{.}
\]
Exploiting the definition of $\mathrm{ext}_{\Psi}$, we find that
\[
\left.\left(\mathrm{ext}_{\Psi}f_{2}\right)\right|_{\psi\left(M\right)}=\Psi^{\prime}\circ f_{2}\circ\psi^{\prime-1}\in\mathscr{D}\left(\psi\left(M\right),\Psi\left(E\right)\right)\mbox{.}
\]
Recalling that $\psi$ is an isometric embedding, we also have $\psi_{*}^{\prime}g=\left.h\right|_{\psi\left(M\right)}$
and, as a consequence of the fact that $\left(\psi^{\prime-1},\Psi^{\prime-1}\right)$
is a morphism of $\mathfrak{ghs}^{f}$ from $\left(\psi\left(\mathscr{M}\right),\Psi\left(E\right),A_{\Psi}\right)$
to $\left(\mathscr{M},E,A\right)$ (we noted this fact few lines after
eq. \eqref{eqextPsi'}), we deduce that
\begin{eqnarray*}
\omega\left(\xi u_{1},\xi u_{2}\right) & = & \int\limits _{\psi\left(M\right)}\left(\left.\left(e_{B}\left(\mathrm{ext}_{\Psi}f_{1}\right)\right)\right|_{\psi\left(M\right)}\overset{\Psi\left(E\right)}{\cdot}\left(\Psi^{\prime}\circ f_{2}\circ\psi^{\prime-1}\right)\right)\mathrm{d}\mu_{\psi_{*}^{\prime}g}\\
 & = & \int\limits _{M}\left(\left(\Psi^{\prime-1}\circ\left.\left(e_{B}\left(\mathrm{ext}_{\Psi}f_{1}\right)\right)\right|_{\psi\left(M\right)}\circ\psi^{\prime}\right)\overset{E}{\cdot}f_{2}\right)\mathrm{d}\mu_{g}\\
 & = & \int\limits _{M}\left(\left(\left(\mathrm{res}_{\Psi}\circ e_{B}\circ\mathrm{ext}_{\Psi}\right)f_{1}\right)\overset{E}{\cdot}f_{2}\right)\mathrm{d}\mu_{g}\mbox{.}
\end{eqnarray*}
We apply Lemma \ref{lemresPsieBextPsi=00003DeA} and, recalling the
definition of $\sigma$ given in Lemma \ref{lemObjghsf->Objssp},
we conclude the proof:
\begin{alignat*}{1}
\omega\left(\xi u_{1},\xi u_{2}\right) & =\int\limits _{M}\left(\left(\left(\mathrm{res}_{\Psi}\circ e_{B}\circ\mathrm{ext}_{\Psi}\right)f_{1}\right)\overset{E}{\cdot}f_{2}\right)\mathrm{d}\mu_{g}=\int\limits _{M}\left(\left(e_{A}f_{1}\right)\overset{E}{\cdot}f_{2}\right)\mathrm{d}\mu_{g}\\
 & =\sigma\left(u_{1},u_{2}\right)\mbox{.}
\end{alignat*}

\end{proof}
Now we have the second part of our candidate covariant functor: for
each pair of objects $\left(\mathscr{M},E,A\right)$ and $\left(\mathscr{N},F,B\right)$
of $\mathfrak{ghs}^{f}$ there exists a map
\begin{eqnarray*}
\mathscr{B}:\mathsf{Mor}_{\mathfrak{ghs}^{f}}\left(\left(\mathscr{M},E,A\right),\left(\mathscr{N},F,B\right)\right) & \rightarrow & \mathsf{Mor}_{\mathfrak{ssp}}\left(\mathscr{B}\left(\mathscr{M},E,A\right),\mathscr{B}\left(\mathscr{N},F,B\right)\right)\\
\left(\psi,\Psi\right) & \mapsto & \xi
\end{eqnarray*}
 defined in accordance with Lemma \ref{lemMorghsf->Morssp}. To complete
the theory of the classical field under consideration, it remains
only to check that $\mathscr{B}$ is actually a covariant functor.
The next theorem answers to this question.
\begin{thm}
\label{thmClassicalFieldFunctor}The map $\mathscr{B}:\mathsf{Obj}_{\mathfrak{ghs}^{f}}\rightarrow\mathsf{Obj}_{\mathfrak{ghs}^{f}}$
defined in accordance with Lemma \ref{lemObjghsf->Objssp}, together
with the collection of maps
\[
\left\{ \begin{array}{rr}
\mathscr{B}: & \mathsf{Mor}_{\mathfrak{ghs}^{f}}\left(\left(\mathscr{M},E,A\right),\left(\mathscr{N},F,B\right)\right)\rightarrow\mathsf{Mor}_{\mathfrak{ssp}}\left(\mathscr{B}\left(\mathscr{M},E,A\right),\mathscr{B}\left(\mathscr{N},F,B\right)\right)\\
 & \mbox{for }\left(\mathscr{M},E,A\right),\left(\mathscr{N},F,B\right)\in\mathsf{Obj}_{\mathfrak{ghs}^{f}}
\end{array}\right\} 
\]
defined few lines above, gives rise to a covariant functor $\mathscr{B}$
from the category $\mathfrak{ghs}^{f}$ to the category $\mathfrak{ssp}$.
Moreover $\mathscr{B}$ possesses the following properties:
\begin{itemize}
\item \textsl{causality}: for each $\left(\mathscr{M}=\left(M,g,\mathfrak{o},\mathfrak{t}\right),E,A\right)$,
$\left(\mathscr{M}_{1}=\left(M_{1},g_{1},\mathfrak{o}_{1},\mathfrak{t}_{1}\right),E_{1},A_{1}\right)$,
$\left(\mathscr{M}_{2}=\left(M_{2},g_{2},\mathfrak{o}_{2},\mathfrak{t}_{2}\right),E_{2},A_{2}\right)$
in $\mathsf{Obj}_{\mathfrak{ghs}^{f}}$, each morphism $\left(\psi_{1},\Psi_{1}\right)$
of $\mathfrak{ghs}^{f}$ from $\left(\mathscr{M}_{1},E_{1},A_{1}\right)$
to $\left(\mathscr{M},E,A\right)$ and each morphism $\left(\psi_{2},\Psi_{2}\right)$
from $\left(\mathscr{M}_{2},E_{2},A_{2}\right)$ to $\left(\mathscr{M},E,A\right)$
such that $\psi_{1}\left(M_{1}\right)$ and $\psi_{2}\left(M_{2}\right)$
are $\mathscr{M}$-causally separated subsets of $M$, it holds that
\[
\sigma\left(\xi_{1}u_{1},\xi_{2}u_{2}\right)=0
\]
for each $u_{1}\in V_{1}$ and each $u_{2}\in V_{2}$, where $\left(V,\sigma\right)$,
$\left(V_{1},\sigma_{1}\right)$, $\left(V_{2},\sigma_{2}\right)$
are the symplectic spaces obtained applying $\mathscr{B}$ respectively
to $\left(\mathscr{M},E,A\right)$, $\left(\mathscr{M}_{1},E_{1},A_{1}\right)$,
$\left(\mathscr{M}_{2},E_{2},A_{2}\right)$ and $\xi_{1}$, $\xi_{2}$
are the symplectic maps obtained applying $\mathscr{B}$ respectively
to $\left(\psi_{1},\Psi_{1}\right)$, $\left(\psi_{2},\Psi_{2}\right)$;
\item \textsl{time slice axiom}: for each $\left(\mathscr{M}=\left(M,g,\mathfrak{o},\mathfrak{t}\right),E,A\right)$,
$\left(\mathscr{N}=\left(N,h,\mathfrak{p},u\right),F,B\right)$ in
$\mathsf{Obj}_{\mathfrak{ghs}^{f}}$ and each morphism $\left(\psi,\Psi\right)$
of $\mathfrak{ghs}^{f}$ from $\left(\mathscr{M},E,A\right)$ to $\left(\mathscr{N},F,B\right)$
such that $\psi\left(M\right)$ contains a smooth spacelike Cauchy
surface $\Sigma$ for $\mathscr{N}$, it holds that
\[
\xi\left(V\right)=W\mbox{,}
\]
where $\left(V,\sigma\right)$, $\left(W,\omega\right)$ are the symplectic
spaces obtained applying $\mathscr{B}$ respectively to $\left(\mathscr{M},E,A\right)$,
$\left(\mathscr{N},F,B\right)$ and $\xi$ is the symplectic map obtained
applying $\mathscr{B}$ to $\left(\psi,\Psi\right)$. In particular
$\xi$ is bijective and its inverse $\xi^{-1}$ is a morphism of $\mathfrak{ssp}$
from $\left(W,\omega\right)$ to $\left(V,\sigma\right)$.
\end{itemize}
\end{thm}
\begin{proof}
We must check that $\mathscr{B}$ satisfies the covariant axioms of
Definition \ref{defFunctor}. Consider three objects $\left(\mathscr{M},E,A\right)$,
$\left(\mathscr{N},F,B\right)$ and $\left(\mathscr{O},G,C\right)$
of $\mathfrak{ghs}^{f}$, a morphism $\left(\phi,\Phi\right)$ from
$\left(\mathscr{M},E,A\right)$ to $\left(\mathscr{N},F,B\right)$
and a morphism $\left(\psi,\Psi\right)$ from $\left(\mathscr{N},F,B\right)$
to $\left(\mathscr{O},G,C\right)$. Our aim is to show that the composition
is preserved by $\mathscr{B}$, i.e.
\[
\mathscr{B}\left(\left(\psi,\Psi\right)\circ\left(\phi,\Phi\right)\right)=\mathscr{B}\left(\psi,\Psi\right)\circ\mathscr{B}\left(\phi,\Phi\right)\mbox{.}
\]
For each $u\in V$, where $V=\mathscr{B}\left(\mathscr{M},E,A\right)$,
we find $f\in\mathscr{D}\left(M,E\right)$ such that $e_{A}f=u$.
This allows us to evaluate the LHS of our last equation:
\[
\mathscr{B}\left(\left(\psi,\Psi\right)\circ\left(\phi,\Phi\right)\right)u=e_{C}\left(\mathrm{ext}_{\Psi\circ\Phi}f\right)\mbox{.}
\]
For the RHS we have
\[
\left(\mathscr{B}\left(\psi,\Psi\right)\circ\mathscr{B}\left(\phi,\Phi\right)\right)u=\mathscr{B}\left(\psi,\Psi\right)\left(e_{B}\left(\mathrm{ext}_{\Phi}f\right)\right)=e_{C}\left(\mathrm{ext}_{\Psi}\left(\mathrm{ext}_{\Phi}f\right)\right)\mbox{.}
\]
Recalling eq. \eqref{eqextPsiPhi=00003DextPsiextPhi} and comparing
our last two equations, we deduce that
\[
\mathscr{B}\left(\left(\psi,\Psi\right)\circ\left(\phi,\Phi\right)\right)u=\left(\mathscr{B}\left(\psi,\Psi\right)\circ\mathscr{B}\left(\phi,\Phi\right)\right)u
\]
for each $u\in V$, that is exactly what we wanted to prove. Now we
consider the identity morphism $\mathrm{id}_{\left(\mathscr{M},E,A\right)}$
of $\mathsf{Mor}_{\mathfrak{ghs}^{f}}\left(\left(\mathscr{M},E,A\right),\left(\mathscr{M},E,A\right)\right)$.
We immediately realize that such morphism is provided by the identity
maps of the sets $M$ and $E$:
\[
\mathrm{id}_{\left(\mathscr{M},E,A\right)}=\left(\mathrm{id}_{M},\mathrm{id}_{E}\right)\mbox{.}
\]
The identity morphism $\mathrm{id}_{\left(V,\sigma\right)}\in\mathsf{Mor}_{\mathfrak{ssp}}\left(\left(V,\sigma\right),\left(V,\sigma\right)\right)$,
for $\left(V,\sigma\right)=\mathscr{B}\left(\mathscr{M},E,A\right)$,
is provided by the identity map of the set $V$ too:
\[
\mathrm{id}_{\left(V,\sigma\right)}=\mathrm{id}_{V}\mbox{.}
\]
We want to show that
\[
\mathscr{B}\left(\mathrm{id}_{\left(\mathscr{M},E,A\right)}\right)=\mathrm{id}_{\left(V,\sigma\right)}\mbox{.}
\]
We consider $u\in V$ and, taking $f\in\mathscr{D}\left(M,E\right)$
such that $e_{A}f=u$, we obtain
\[
\mathscr{B}\left(\mathrm{id}_{\left(\mathscr{M},E,A\right)}\right)u=e_{A}\left(\mathrm{ext}_{\mathrm{id}_{E}}f\right)=e_{A}\left(\mathrm{id}_{E}\circ f\circ\mathrm{id}_{M}^{-1}\right)=e_{A}f=u=\mathrm{id}_{\left(V,\sigma\right)}u\mbox{.}
\]
This equation holds for each $u\in V$. We deduce that $\mathscr{B}$
maps the identity morphisms of $\mathfrak{ghs}^{f}$ to the identity
morphisms of $\mathfrak{ssp}$. We have shown that $\mathscr{B}$
is actually a covariant functor from $\mathfrak{ghs}^{f}$ to $\mathfrak{ssp}$.

We turn our attention to the causality property. Let $\left(\mathscr{M}=\left(M,g,\mathfrak{o},\mathfrak{t}\right),E,A\right)$,
$\left(\mathscr{M}_{1}=\left(M_{1},g_{1},\mathfrak{o}_{1},\mathfrak{t}_{1}\right),E_{1},A_{1}\right)$
and $\left(\mathscr{M}_{2}=\left(M_{2},g_{2},\mathfrak{o}_{2},\mathfrak{t}_{2}\right),E_{2},A_{2}\right)$
be objects of $\mathfrak{ghs}^{f}$ and suppose that $\left(\psi_{1},\Psi_{1}\right)$
is a morphism from $\left(\mathscr{M}_{1},E_{1},A_{1}\right)$ to
$\left(\mathscr{M},E,A\right)$ and that $\left(\psi_{2},\Psi_{2}\right)$
is a morphism from $\left(\mathscr{M}_{2},E_{2},A_{2}\right)$ to
$\left(\mathscr{M},E,A\right)$. Moreover assume that $\psi_{1}\left(M_{1}\right)$
and $\psi_{2}\left(M_{2}\right)$ are $\mathscr{M}$-causally separated
subsets of $M$. We denote with $\left(V,\sigma\right)$, $\left(V_{1},\sigma_{1}\right)$
and $\left(V_{2},\sigma_{2}\right)$ the symplectic spaces associated
respectively to $\left(\mathscr{M},E,A\right)$, $\left(\mathscr{M}_{1},E_{1},A_{1}\right)$
and $\left(\mathscr{M}_{2},E_{2},A_{2}\right)$ via $\mathscr{B}$
and with $\xi_{1}$ and $\xi_{2}$ the symplectic maps associated
respectively to $\left(\psi_{1},\Psi_{1}\right)$ and $\left(\psi_{2},\Psi_{2}\right)$.
Consider two elements $u_{1}\in V_{1}$ and $u_{2}\in V_{2}$. We
surely find $f_{1}\in\mathscr{D}\left(M_{1},E_{1}\right)$ such that
$e_{A_{1}}f_{1}=u_{1}$ and $f_{2}\in\mathscr{D}\left(M_{2},E_{2}\right)$
such that $e_{A_{2}}f_{2}=u_{2}$. This allows us to evaluate $\xi_{1}u_{1}$
and $\xi_{2}u_{2}$:
\begin{eqnarray*}
\xi_{1}u_{1} & = & e_{A}\left(\mathrm{ext}_{\Psi_{1}}f_{1}\right)\mbox{,}\\
\xi_{2}u_{2} & = & e_{A}\left(\mathrm{ext}_{\Psi_{2}}f_{2}\right)\mbox{.}
\end{eqnarray*}
Notice that $\mathrm{supp}\left(f_{1}\right)$ is a compact subset
of $M_{1}$ and therefore $\mathrm{supp}\left(\mathrm{ext}_{\Psi_{1}}f_{1}\right)$
is a compact subset of $M$ included in $\psi_{1}\left(M_{1}\right)$.
Similarly $\mathrm{supp}\left(\mathrm{ext}_{\Psi_{2}}f_{2}\right)$
is a compact subset of $M$ included in $\psi_{2}\left(M_{2}\right)$.
Due to the support property of the advanced and retarded Green operators
for $A$ (see Definition \ref{defGreenOperators}), we have that
\[
\mathrm{supp}\left(\xi_{1}u_{1}\right)\subseteq J^{\mathscr{M}}\left(\mathrm{supp}\left(\mathrm{ext}_{\Psi_{1}}f_{1}\right)\right)\subseteq J^{\mathscr{M}}\left(\psi_{1}\left(M_{1}\right)\right)\mbox{.}
\]
We assumed that $\psi_{1}\left(M_{1}\right)$ and $\psi_{2}\left(M_{2}\right)$
are $\mathscr{M}$-causally separated subsets of $M$, therefore,
via Remark \ref{remCausalSeparation}, we obtain
\[
\mathrm{supp}\left(\xi_{1}u_{1}\right)\cap\mathrm{supp}\left(\mathrm{ext}_{\Psi_{2}}f_{2}\right)\subseteq J^{\mathscr{M}}\left(\psi_{1}\left(M_{1}\right)\right)\cap\psi_{2}\left(M_{2}\right)=\emptyset\mbox{.}
\]
These observations give us the opportunity to evaluate $\sigma\left(\xi_{1}u_{1},\xi_{2}u_{2}\right)$:
\[
\sigma\left(\xi_{1}u_{1},\xi_{2}u_{2}\right)=\int\limits _{M}\left(\left(\xi_{1}u_{1}\right)\overset{E}{\cdot}\left(\mathrm{ext}_{\Psi_{2}}f_{2}\right)\right)\mathrm{d}\mu_{g}=0
\]
because the support of the integrand is empty. This shows that causality
holds.

We are left only with the check of the time slice axiom. Consider
two objects $\left(\mathscr{M}=\left(M,g,\mathfrak{o},\mathfrak{t}\right),E,A\right)$
and $\left(\mathscr{N}=\left(N,h,\mathfrak{p},u\right),F,B\right)$
of $\mathfrak{ghs}^{f}$ and suppose that $\left(\psi,\Psi\right)$
is a morphism from $\left(\mathscr{M},E,A\right)$ to $\left(\mathscr{N},F,B\right)$
whose image $\psi\left(M\right)$ includes a smooth spacelike Cauchy
surface $\Sigma$ for $\mathscr{N}$. We denote with $\left(V,\sigma\right)$
and $\left(W,\omega\right)$ the symplectic spaces obtained through
$\mathscr{B}$ from $\left(\mathscr{M},E,A\right)$ and respectively
$\left(\mathscr{N},F,B\right)$ and we impose $\xi=\mathscr{B}\left(\psi,\Psi\right)$.
$W$ is codomain of $\xi$, hence the inclusion $\xi\left(V\right)\subseteq W$
is trivial and we must prove the converse inclusion to complete the
proof. To this end consider $u\in W$. We look for a section $f\in\mathscr{D}\left(M,E\right)$
such that $e_{B}\left(\mathrm{ext}_{\Psi}f\right)=u$. We observe
that $u$ is obtained from a compactly supported section in $F$ through
the causal propagator $e_{B}$. Hence, exploiting the support properties
of the Green operators, we find a compact subset $K$ of $N$ such
that
\[
\mathrm{supp}\left(u\right)\subseteq J^{\mathscr{N}}\left(K\right)\mbox{.}
\]
The problem is that, in general, $K$ is not included in $\psi\left(M\right)$.
Anyway we can go around this obstacle with the following procedure.
We note that $\mathrm{supp}\left(u\right)\cap\Sigma$ is a compact
subset of $\Sigma$ (intended as a topological space in its own right
with the topology induced by the topology of $N$) because of Proposition
\ref{propUsefulSubsetsOfGloballyHyperbolicSpacetimes}. Take a $\mathfrak{u}$-future
directed $h$-timelike unit vector field $\mathfrak{n}$ over $\Sigma$
normal to $\Sigma$ (such vector field actually exists because $\Sigma$
is spacelike). Considering $\Sigma$ as a $\left(d-1\right)$-dimensional
submanifold of $N$ and introducing the vector bundle $\left.F\right|_{\Sigma}=\pi_{F}^{-1}\left(\Sigma\right)$,
we can define two compactly supported sections over $\Sigma$:
\begin{alignat*}{3}
u_{0}:\Sigma & \rightarrow\left.F\right|_{\Sigma} & \quad\mbox{and} & \quad & u_{1}:\Sigma & \rightarrow\left.F\right|_{\Sigma}\\
q & \mapsto u\left(q\right) &  &  & q & \mapsto\left(\nabla_{\mathfrak{n}}u\right)\left(q\right)\mbox{,}
\end{alignat*}
where $\nabla$ is the $B$-compatible connection in $F$. Since $u_{0}$
and $u_{1}$ fall in $\mathscr{D}\left(\Sigma,\left.E\right|_{\Sigma}\right)$,
we can use them to formulate a well posed Cauchy problem for the normally
hyperbolic operator $B$:
\[
\left\{ \begin{array}{rcl}
Bv & = & 0\mbox{,}\\
\left.v\right|_{\Sigma} & = & u_{0}\mbox{,}\\
\left.\nabla_{\mathfrak{n}}v\right|_{\Sigma} & = & u_{1}\mbox{.}
\end{array}\right.
\]
Theorem \ref{thmCauchyProblem} tells us that the Cauchy problem stated
above admits exactly one solution $v\in\mathrm{C}^{\infty}\left(N,F\right)$
whose support is contained in
\[
J^{\mathscr{N}}\left(\mathrm{supp}\left(u_{0}\right)\cup\mathrm{supp}\left(u_{1}\right)\right)\subseteq J^{\mathscr{N}}\left(\mathrm{supp}\left(u\right)\cap\Sigma\right)\mbox{.}
\]
By construction $u$ satisfies that Cauchy problem and therefore $u=v$
for uniqueness, in particular $u$ and $v$ have the same support.
Since $\mathrm{supp}\left(u\right)\cap\Sigma$ is a compact subset
of $\Sigma\subseteq\psi\left(M\right)$, it is also a compact subset
of $N$ included in $\psi\left(M\right)$. This fact gives us the
chance to find a compact subset $K$ of $N$ that contains $\mathrm{supp}\left(u\right)\cap\Sigma$
and that is included in $\psi\left(M\right)$. Since $K$ is compact,
we can also find a relatively compact open subset $\Omega$ of $N$
such that $K\subseteq\Omega\subseteq\psi\left(M\right)$. Using $\Omega$,
we can introduce a covering of $N$:
\[
\left\{ J_{+}^{\mathscr{N}}\left(\Omega\right),J_{-}^{\mathscr{N}}\left(\Omega\right),N\setminus J^{\mathscr{N}}\left(K\right)\right\} \mbox{.}
\]
This is an open covering because $J_{\pm}^{\mathscr{N}}\left(\Omega\right)$
are open subsets of $N$ (see \cite[Lem. A.8, p. 48]{FV11}) and $J_{\pm}^{\mathscr{N}}\left(K\right)$
are closed subsets of $N$ (see \cite[Lem. A.5.1, p. 173]{BGP07}).
Then we can introduce a partition of unity subordinate to such covering:
\[
\left\{ \chi^{+},\chi^{-},\chi^{0}\right\} \mbox{.}
\]
Then we define $u^{\pm}=\chi^{\pm}u$ and $u^{0}=\chi^{0}u$ and we
have $u=u^{+}+u^{-}+u^{0}$. As a consequence of our construction
\begin{alignat*}{2}
\mathrm{supp}\left(u^{0}\right) & =\mathrm{supp}\left(\chi^{0}\right)\cap\mathrm{supp}\left(u\right) & \subseteq & \left(N\setminus J^{\mathscr{N}}\left(K\right)\right)\cap J^{\mathscr{N}}\left(\mathrm{supp}\left(u\right)\cap\Sigma\right)\\
 & \subseteq\left(N\setminus J^{\mathscr{N}}\left(K\right)\right)\cap J^{\mathscr{N}}\left(K\right) & = & \emptyset
\end{alignat*}
and therefore $u^{0}$ is everywhere null. This implies that $u=u^{+}+u^{-}$.
Since we know that $Bu=0$, we deduce that $Bu^{+}=-Bu^{-}$. In particular
this relation implies that
\[
\mathrm{supp}\left(Bu^{+}\right)\subseteq\mathrm{supp}\left(\chi^{+}\right)\cap\mathrm{supp}\left(\chi^{-}\right)\subseteq J_{+}^{\mathscr{N}}\left(\Omega\right)\cap J_{-}^{\mathscr{N}}\left(\Omega\right)\subseteq J_{+}^{\mathscr{N}}\left(\overline{\Omega}\right)\cap J_{-}^{\mathscr{N}}\left(\overline{\Omega}\right)\mbox{,}
\]
where $\overline{\Omega}$ denotes the closure of $\Omega$ in $N$.
Since $\Omega$ is relatively compact in $N$, $\overline{\Omega}$
is compact in $N$. Applying Proposition \ref{propUsefulSubsetsOfGloballyHyperbolicSpacetimes},
we deduce that $J_{+}^{\mathscr{N}}\left(\overline{\Omega}\right)\cap J_{-}^{\mathscr{N}}\left(\overline{\Omega}\right)$
is a compact subset of $N$ and therefore $Bu^{+}=-Bu^{-}$ is a section
in $F$ with compact support. We are able to find more information
about its support:
\[
\mathrm{supp}\left(Bu^{+}\right)\subseteq J_{+}^{\mathscr{N}}\left(\Omega\right)\cap J_{-}^{\mathscr{N}}\left(\Omega\right)\subseteq\psi\left(M\right)\mbox{.}
\]
We prove this inclusion: Consider $p\in J_{+}^{\mathscr{N}}\left(\Omega\right)\cap J_{-}^{\mathscr{N}}\left(\Omega\right)$;
we find a $\mathfrak{u}$-future directed $h$-causal curve $\gamma_{1}$
in $N$ from $q\in\Omega$ to $p$ and a $\mathfrak{u}$-past directed
$h$-causal curve $\gamma_{2}$ in $N$ from $r\in\Omega$ to $p$;
reversing the direction of $\gamma_{2}$ and pasting the result with
$\gamma_{1}$, we obtain a $\mathfrak{u}$-future directed $h$-causal
curve $\gamma$ in $N$ from $q$ to $r$; both $q$ and $r$ fall
in $\psi\left(M\right)$ because $\Omega\subseteq\psi\left(M\right)$
by construction; since $\psi\left(M\right)$ is $\mathscr{N}$-causally
convex by hypothesis, $\gamma$ must be entirely contained in $\psi\left(M\right)$,
in particular $p\in\psi\left(M\right)$. At this point we have a section
$Bu^{+}\in\mathscr{D}\left(N,F\right)$ with support included in $\psi\left(M\right)$.
We use it to define a compactly supported section in $E$ via restriction:
\[
f=\mathrm{res}_{\Psi}\left(Bu^{+}\right)=-\mathrm{res}_{\Psi}\left(Bu^{-}\right)\in\mathscr{D}\left(M,E\right)\mbox{.}
\]
Now we check that $f$ is exactly the one we were looking for. First
of all $f$ has compact support so that $\mathrm{ext}_{\Psi}f$ has
compact support too, hence we can apply $e_{B}^{a/r}$ to it and we
obtain
\begin{eqnarray*}
e_{B}^{a}\left(\mathrm{ext}_{\Psi}f\right) & = & e_{B}^{a}\left(Bu^{+}\right)\mbox{,}\\
e_{B}^{r}\left(\mathrm{ext}_{\Psi}f\right) & = & -e_{B}^{r}\left(Bu^{-}\right)\mbox{.}
\end{eqnarray*}
Now we observe that
\begin{eqnarray*}
\mathrm{supp}\left(u^{+}\right) & \subseteq & J_{+}^{\mathscr{N}}\left(\overline{\Omega}\right)\mbox{,}\\
\mathrm{supp}\left(u^{-}\right) & \subseteq & J_{-}^{\mathscr{N}}\left(\overline{\Omega}\right)
\end{eqnarray*}
and Proposition \ref{propUsefulSubsetsOfGloballyHyperbolicSpacetimes}
implies that $u^{+}$ has $\mathscr{N}$-past compact support, while
$u^{-}$has $\mathscr{N}$-future compact support. This fact allows
us to apply Lemma \ref{lemExtensionOf ea(Pu)=00003Du} to obtain
\begin{eqnarray*}
e_{B}^{a}\left(\mathrm{ext}_{\Psi}f\right) & = & u^{+}\mbox{,}\\
e_{B}^{r}\left(\mathrm{ext}_{\Psi}f\right) & = & -u^{-}
\end{eqnarray*}
and therefore
\[
e_{B}\left(\mathrm{ext}_{\Psi}f\right)=u^{+}-\left(-u^{-}\right)=u\mbox{.}
\]
This completes our proof because, setting $w=e_{A}f\in V$, we have
$\xi w=e_{B}\left(\mathrm{ext}_{\Psi}f\right)=u$, hence in particular
$u\in\xi\left(V\right)$ and this fact, for the freedom in the choice
of $u\in W$, implies the inclusion $W\subseteq\xi\left(V\right)$.
The last part of the statement of the time slice axiom follows directly
because each symplectic map is automatically injective (cfr. Remark
\ref{remSymplecticMapsAreInjective}) and the time slice axiom assures
that $\xi$ is also surjective, hence the inverse $\xi^{-1}$ exists
and it is trivial to check that it is a symplectic map too.
\end{proof}

\subsection{\label{subQuantumFieldTheory}...to quantum field theory}

In the last subsection we built the theory of a classical field over
some $d$-dimensional globally hyperbolic spacetime $\mathscr{M}=\left(M,g,\mathfrak{o},\mathfrak{t}\right)$
modeled by a smooth section $u$ in a vector bundle $E$ of rank $n$
over $M$ satisfying the normally hyperbolic equation $Au=0$ on each
point of $M$, where $E$ is endowed with an inner product denoted
by $\overset{E}{\cdot}$ and $A$ is a formally selfadjoint normally
hyperbolic operator on $E$ over $\mathscr{M}$. Now we want to use
this result to build the quantum field theory that corresponds to
this situation. Most of the work has already been done in the previous
subsection and in Subsection \ref{subC*-algebrasWeylSystemsCCRRepresentations}.
Here we simply put the pieces of the puzzle together. We start building
a new covariant functor from the category $\mathfrak{ssp}$ to the
category $\mathfrak{alg}$ and then we compose it with $\mathscr{B}$.
As we will see, this will give us a covariant functor $\mathscr{A}$
that is actually a locally covariant quantum field theory fulfilling
both the causality condition and the time slice axiom. As a consequence
of our Theorem \ref{thmRecoveringHaagKastlerAxioms}, on each globally
hyperbolic spacetime $\mathscr{M}$ the LCQFT $\mathscr{A}$ provides
the quantum field theory (in the sense of the Haag-Kastler approach)
of the field under consideration.
\begin{lem}
\label{lemQuantizationFunctor}Consider a map $\mathscr{C}$ that
associates to each symplectic space $\left(V,\sigma\right)$ its unique
(up to {*}-isomorphisms) CCR representation $\left(\mathcal{V},\mathrm{V}\right)$
in accordance with Definition \ref{defCCRRepresentation} and for
each pair of symplectic spaces $\left(V,\sigma\right)$ and $\left(W,\omega\right)$,
whose corresponding CCR representations are respectively $\left(\mathcal{V},\mathrm{V}\right)=\mathscr{C}\left(V,\sigma\right)$
and $\left(\mathcal{W},\mathrm{W}\right)=\mathscr{C}\left(W,\omega\right)$,
consider a map $\mathscr{C}$ that associates to each symplectic map
$\xi$ from $\left(V,\sigma\right)$ to $\left(W,\omega\right)$ the
unique injective unit preserving {*}-homomorphism $H$ from $\left(\mathcal{V},\mathrm{V}\right)$
to $\left(\mathcal{W},\mathrm{W}\right)$ in accordance with Proposition
\ref{propInjective*HomomorphismsFromSymplecticMaps} and the subsequent
observation. Then $\mathscr{C}$ is a covariant functor from the category
$\mathfrak{ssp}$ to the category $\mathfrak{alg}$.\end{lem}
\begin{proof}
Consider a symplectic space $\left(V,\sigma\right)$. Shortly after
Definition \ref{defWeylSystem}, we observed that there exists at
least one Weyl system associated to each symplectic space. We consider
the unital sub-C{*}-algebra $\mathcal{V}$ of the Weyl algebra under
consideration generated by the image of the Weyl map. This gives rise
to a CCR representation $\left(\mathcal{V},\mathrm{V}\right)$ of
$\left(V,\sigma\right)$ as it can be directly checked via Definition
\ref{defCCRRepresentation}. $\left(\mathcal{V},\mathrm{V}\right)$
is unique up to {*}-isomorphism as a consequence of Proposition\ref{propUniquenessOfCCRRepresentations}.
Then a map of the type required in the statement is obtained imposing
$\mathscr{C}\left(V,\sigma\right)=\left(\mathcal{V},\mathrm{V}\right)$.
Note that we have just defined $\mathscr{C}$ as a map from $\mathsf{Obj}_{\mathfrak{ssp}}$
to $\mathsf{Obj}_{\mathfrak{alg}}$.

Consider now a pair of symplectic spaces $\left(V,\sigma\right)$
and $\left(W,\omega\right)$, let $\left(\mathcal{V},\mathrm{V}\right)$
and $\left(\mathcal{W},\mathrm{W}\right)$ denote respectively $\mathscr{C}\left(V,\sigma\right)$
and $\mathscr{C}\left(W,\omega\right)$ and suppose that $\xi$ is
a symplectic map from $\left(V,\sigma\right)$ to $\left(W,\omega\right)$.
Applying Proposition \ref{propInjective*HomomorphismsFromSymplecticMaps}
and the subsequent observation, we obtain a unique injective unit
preserving {*}-homomorphism $H:\mathcal{V}\rightarrow\mathcal{W}$
satisfying $H\circ\mathrm{V}=\mathrm{W}\circ\xi$. Then we define
a map $\mathscr{C}$ as required by the statement setting $\mathscr{C}\left(\xi\right)=H$.
Note that we have just defined $\mathscr{C}$ as a map from $\mathsf{Mor}_{\mathfrak{ssp}}\left(\left(V,\sigma\right),\left(W,\omega\right)\right)$
to $\mathsf{Mor}_{\mathfrak{alg}}\left(\left(\mathcal{V},\mathrm{V}\right),\left(\mathcal{W},\mathrm{W}\right)\right)$.

At this point $\mathscr{C}$ is a good candidate to become a covariant
functor from $\mathfrak{ssp}$ to $\mathfrak{alg}$, but we have still
to check the covariant axioms. We begin checking that $\mathscr{C}$
preserves the composition. To this end we consider three objects $\left(U,\rho\right)$,
$\left(V,\sigma\right)$ and $\left(W,\omega\right)$ of $\mathfrak{ssp}$
and we denote the images of these objects through $\mathscr{C}$ respectively
with $\left(\mathcal{U},\mathrm{U}\right)$, $\left(\mathcal{V},\mathrm{V}\right)$
and $\left(\mathcal{W},\mathrm{W}\right)$. Moreover we take a morphism
$\xi$ of $\mathfrak{ssp}$ from $\left(U,\rho\right)$ to $\left(V,\sigma\right)$
and a morphism $\eta$ of $\mathfrak{ssp}$ from $\left(V,\sigma\right)$
to $\left(W,\omega\right)$. $\eta\circ\xi$ is undoubtedly a morphism
of $\mathfrak{ssp}$ from $\left(U,\rho\right)$ to $\left(W,\omega\right)$.
Then we can consider the following morphisms of $\mathfrak{alg}$:
\begin{eqnarray*}
\mathscr{C}\left(\xi\right) & \in & \mathsf{Mor}_{\mathfrak{alg}}\left(\left(\mathcal{U},\mathrm{U}\right),\left(\mathcal{V},\mathrm{V}\right)\right)\mbox{,}\\
\mathscr{C}\left(\eta\right) & \in & \mathsf{Mor}_{\mathfrak{alg}}\left(\left(\mathcal{V},\mathrm{V}\right),\left(\mathcal{W},\mathrm{W}\right)\right)\mbox{,}\\
\mathscr{C}\left(\eta\circ\xi\right) & \in & \mathsf{Mor}_{\mathfrak{alg}}\left(\left(\mathcal{U},\mathrm{U}\right),\left(\mathcal{W},\mathrm{W}\right)\right)\mbox{.}
\end{eqnarray*}
We also have that these morphisms satisfy the following relations:
\begin{eqnarray*}
\mathscr{C}\left(\xi\right)\circ\mathrm{U} & = & \mathrm{V}\circ\xi\mbox{,}\\
\mathscr{C}\left(\eta\right)\circ\mathrm{V} & = & \mathrm{W}\circ\eta\mbox{,}\\
\mathscr{C}\left(\eta\circ\xi\right)\circ\mathrm{U} & = & \mathrm{W}\circ\left(\eta\circ\xi\right)\mbox{.}
\end{eqnarray*}
Surely $\mathscr{C}\left(\eta\right)\circ\mathscr{C}\left(\xi\right)$
is a morphism of $\mathfrak{alg}$ from $\left(\mathcal{U},\mathrm{U}\right)$
to $\left(\mathcal{W},\mathrm{W}\right)$ such as $\mathscr{C}\left(\eta\circ\xi\right)$
and, exploiting the first two equations, we get
\[
\mathscr{C}\left(\eta\right)\circ\mathscr{C}\left(\xi\right)\circ\mathrm{U}=\mathscr{C}\left(\eta\right)\circ\mathrm{V}\circ\xi=\mathrm{W}\circ\eta\circ\xi\mbox{.}
\]
Proposition \ref{propInjective*HomomorphismsFromSymplecticMaps} tells
us that there exists a unique injective {*}-homomorphism $H$ from
$\left(\mathcal{U},\mathrm{U}\right)$ to $\left(\mathcal{W},\mathrm{W}\right)$
such that $H\circ\mathrm{W}=\mathrm{U}\circ\left(\eta\circ\xi\right)$,
hence $\mathscr{C}\left(\eta\circ\xi\right)=\mathscr{C}\left(\eta\right)\circ\mathscr{C}\left(\xi\right)$
and then $\mathscr{C}$ preserves the composition of morphisms. To
conclude we check that $\mathscr{C}$ maps the identity morphisms
to the identity morphisms. To this end we consider the object $\left(V,\sigma\right)$
of $\mathfrak{ssp}$ and its image $\left(\mathcal{V},\mathrm{V}\right)$
through $\mathscr{C}$. It is easy to check that the identity morphism
$\mathrm{id}_{\left(V,\sigma\right)}$ of $\left(V,\sigma\right)$
is provided by the identity map $\mathrm{id}_{V}$ of the set $V$
and that the identity morphism $\mathrm{id}_{\left(\mathcal{V},\mathrm{V}\right)}$
of $\left(\mathcal{V},\mathrm{V}\right)$ is provided by the identity
map $\mathrm{id}_{\mathcal{V}}$ of the set $\mathcal{V}$. Together
with $\mathrm{id}_{\left(\mathcal{V},\mathrm{V}\right)}$, we can
consider another morphism of $\mathfrak{alg}$ from $\left(\mathcal{V},\mathrm{V}\right)$
to itself, specifically $\mathscr{C}\left(\mathrm{id}_{\left(V,\sigma\right)}\right)$.
On the one hand we have
\[
\mathrm{id}_{\left(\mathcal{V},\mathrm{V}\right)}\left(\mathrm{V}\left(v\right)\right)=\mathrm{V}\left(v\right)=\mathrm{V}\left(\mathrm{id}_{\left(V,\sigma\right)}v\right)\quad\forall v\in V\mbox{,}
\]
which means exactly
\[
\mathrm{id}_{\left(\mathcal{V},\mathrm{V}\right)}\circ\mathrm{V}=\mathrm{V}\circ\mathrm{id}_{\left(V,\sigma\right)}\mbox{,}
\]
while on the other side, exploiting the definition of $\mathscr{C}$,
we obtain
\[
\mathscr{C}\left(\mathrm{id}_{\left(V,\sigma\right)}\right)\circ\mathrm{V}=\mathrm{V}\circ\mathrm{id}_{\left(V,\sigma\right)}\mbox{.}
\]
Applying Proposition \ref{propInjective*HomomorphismsFromSymplecticMaps}
as we did above, we find that
\[
\mathscr{C}\left(\mathrm{id}_{\left(V,\sigma\right)}\right)=\mathrm{id}_{\left(\mathcal{V},\mathrm{V}\right)}\mbox{,}
\]
which is to say that $\mathscr{C}$ maps identity morphisms to identity
morphisms. This completes the proof.
\end{proof}
At this point we have the covariant functors $\mathscr{B}:\mathfrak{ghs}^{f}\overset{\rightarrow}{\rightarrow}\mathfrak{ssp}$
and $\mathscr{C}:\mathfrak{ssp}\overset{\rightarrow}{\rightarrow}\mathfrak{alg}$
and we can compose them in accordance with Definition \ref{defCompositionOfFunctors}
to obtain a new covariant functor. We present the result in the next
theorem.
\begin{thm}
\label{thmLCQFTForANormallyHyperbolicField}Consider the covariant
functor $\mathscr{B}:\mathfrak{ghs}^{f}\overset{\rightarrow}{\rightarrow}\mathfrak{ssp}$
defined in Theorem \ref{thmClassicalFieldFunctor} and the covariant
functor $\mathscr{C}:\mathfrak{ssp}\overset{\rightarrow}{\rightarrow}\mathfrak{alg}$
defined in Lemma \ref{lemQuantizationFunctor}. Then $\mathscr{A}=\mathscr{C}\circ\mathscr{B}$
is a locally covariant quantum field theory that fulfils the causality
condition and the time slice axiom.\end{thm}
\begin{proof}
The composition of covariant functors yields a covariant functor (see
Definition \ref{defCompositionOfFunctors}), hence $\mathscr{A}=\mathscr{C}\circ\mathscr{B}$
is a covariant functor from the category $\mathfrak{ghs}^{f}$ to
the category $\mathfrak{alg}$. Besides the richer content of the
category $\mathfrak{ghs}^{f}$ compared to $\mathfrak{ghs}$ (recall
the discussion at the beginning of this section), nonetheless we recognize
$\mathscr{A}$ to be a LCQFT (cfr. Definition \ref{defLCQFT}) in
light of the discussion at the beginning of this section at page \pageref{secBuildingALCQFT}.

Now we check that $\mathscr{A}$ fulfils the causality condition of
Definition \ref{defLCQFT}. To this end consider three objects $\left(\mathscr{M}=\left(M,g,\mathfrak{o},\mathfrak{t}\right),E,A\right)$,
$\left(\mathscr{M}_{1}=\left(M_{1},g_{1},\mathfrak{o}_{1},\mathfrak{t}_{1}\right),E_{1},A_{1}\right)$
and $\left(\mathscr{M}_{2}=\left(M_{2},g_{2},\mathfrak{o}_{2},\mathfrak{t}_{2}\right),E_{2},A_{2}\right)$
in $\mathfrak{ghs}^{f}$, a morphism $\left(\psi_{1},\Psi_{1}\right)$
from $\left(\mathscr{M}_{1},E_{1},A_{1}\right)$ to $\left(\mathscr{M},E,A\right)$
and a morphism $\left(\psi_{2},\Psi_{2}\right)$ from $\left(\mathscr{M}_{2},E_{2},A_{2}\right)$
to $\left(\mathscr{M},E,A\right)$ and suppose that $\psi_{1}\left(M_{1}\right)$
and $\psi_{2}\left(M_{2}\right)$ are $\mathscr{M}$-causally separated
subsets of $M$. Denote the symplectic spaces $\mathscr{B}\left(\mathscr{M},E,A\right)$,
$\mathscr{B}\left(\mathscr{M}_{1},E_{1},A_{1}\right)$ and $\mathscr{B}\left(\mathscr{M}_{2},E_{2},A_{2}\right)$
respectively with $\left(V,\sigma\right)$, $\left(V_{1},\sigma_{1}\right)$
and $\left(V_{2},\sigma_{2}\right)$ and the symplectic maps $\mathscr{B}\left(\psi_{1},\Psi_{1}\right)$
and $\mathscr{B}\left(\psi_{2},\Psi_{2}\right)$ respectively with
$\xi_{1}$ and $\xi_{2}$. Moreover denote the CCR representations
$\mathscr{A}\left(\mathscr{M},E,A\right)=\mathscr{C}\left(V,\sigma\right)$,
$\mathscr{A}\left(\mathscr{M}_{1},E_{1},A_{1}\right)=\mathscr{C}\left(V_{1},\sigma_{1}\right)$
and $\mathscr{A}\left(\mathscr{M}_{2},E_{2},A_{2}\right)=\mathscr{C}\left(V_{2},\sigma_{2}\right)$
respectively with $\left(\mathcal{V},\mathrm{V}\right)$, $\left(\mathcal{V}_{1},\mathrm{V}_{1}\right)$
and $\left(\mathcal{V}_{2},\mathrm{V}_{2}\right)$ and the injective
unit preserving {*}-homomorphisms $\mathscr{A}\left(\psi_{1},\Psi_{1}\right)=\mathscr{C}\left(\xi_{1}\right)$
and $\mathscr{A}\left(\psi_{2},\Psi_{2}\right)=\mathscr{C}\left(\xi_{2}\right)$
respectively with $H_{1}$ and $H_{2}$. Theorem \ref{thmClassicalFieldFunctor}
tells us that $\mathscr{B}$ satisfies the causality property, i.e.
\begin{equation}
\sigma\left(\xi_{1}u_{1},\xi_{2}u_{2}\right)=0\label{eqCausalityOfTheClassicalFieldFunctor}
\end{equation}
for each $u_{1}\in V_{1}$ and each $u_{2}\in V_{2}$. We want to
show that
\[
\left[H_{1}\left(\mathrm{V}_{1}\left(u_{1}\right)\right),H_{2}\left(\mathrm{V}_{2}\left(u_{2}\right)\right)\right]=0\mbox{.}
\]
Exploiting the definitions of $H_{1}$ and $H_{2}$ (cfr. Lemma \ref{lemQuantizationFunctor}),
we find
\begin{eqnarray*}
H_{1}\left(\mathrm{V}_{1}\left(u_{1}\right)\right) & = & \mathrm{V}\left(\xi_{1}u_{1}\right)\mbox{,}\\
H_{2}\left(\mathrm{V}_{2}\left(u_{2}\right)\right) & = & \mathrm{V}\left(\xi_{2}u_{2}\right)\mbox{.}
\end{eqnarray*}
This fact, together with the properties of the Weyl map $\mathrm{V}$
(cfr. Definition \ref{defWeylSystem}) and eq. \eqref{eqCausalityOfTheClassicalFieldFunctor},
allows us to evaluate the commutator above:
\begin{eqnarray*}
\left[H_{1}\left(\mathrm{V}_{1}\left(u_{1}\right)\right),H_{2}\left(\mathrm{V}_{2}\left(u_{2}\right)\right)\right] & = & \left[\mathrm{V}\left(\xi_{1}\left(u_{1}\right)\right),\mathrm{V}\left(\xi_{2}\left(u_{2}\right)\right)\right]\\
 & = & \mathrm{V}\left(\xi_{1}u_{1}\right)\mathrm{V}\left(\xi_{2}u_{2}\right)-\mathrm{V}\left(\xi_{2}u_{2}\right)\mathrm{V}\left(\xi_{1}u_{1}\right)\\
 & = & \left(\mathrm{e}^{-\frac{\imath}{2}\sigma\left(\xi_{1}u_{1},\xi_{2}u_{2}\right)}-\mathrm{e}^{-\frac{\imath}{2}\sigma\left(\xi_{2}u_{2},\xi_{1}u_{1}\right)}\right)\mathrm{V}\left(\xi_{1}u_{1}+\xi_{2}u_{2}\right)\\
 & = & 0\mbox{.}
\end{eqnarray*}
The last relation implies that
\[
\left[H_{1}\left(a_{1}\right),H_{2}\left(a_{2}\right)\right]=0
\]
for each $a_{1}\in\mathcal{V}_{1}$ and each $a_{2}\in\mathcal{V}_{2}$
because $\mathrm{V}_{1}\left(V_{1}\right)$ is the set of generators
of $\mathcal{V}_{1}$, $\mathrm{V}_{2}\left(V_{2}\right)$ is the
set of generators of $\mathcal{V}_{2}$ (by Definition of CCR representation),
both $H_{1}$ and $H_{2}$ are continuous (cfr. Proposition \ref{propContinuityOfUnitPreserving*-HomomorphismsOnUnitalC*-Algebras})
and also the sum and the multiplication of $\mathcal{V}$ are continuous.
This means that $\mathscr{A}$ fulfils the causality condition as
we stated it in Definition \ref{defLCQFT}.

As for the time slice axiom, consider two objects $\left(\mathscr{M}=\left(M,g,\mathfrak{o},\mathfrak{t}\right),E,A\right)$
and $\left(\mathscr{N}=\left(N,h,\mathfrak{p},\mathfrak{u}\right),F,B\right)$
of $\mathfrak{ghs}^{f}$ and a morphism $\left(\psi,\Psi\right)$
between $\left(\mathscr{M},E,A\right)$ and $\left(\mathscr{N},F,B\right)$
such that $\psi\left(M\right)$ contains a smooth spacelike Cauchy
surface $\Sigma$ for $\mathscr{N}$. We denote the symplectic spaces
$\mathscr{B}\left(\mathscr{M},E,A\right)$ and $\mathscr{B}\left(\mathscr{N},F,B\right)$
respectively with $\left(V,\sigma\right)$ and $\left(W,\omega\right)$
and the symplectic map $\mathscr{B}\left(\psi,\Psi\right)$ with $\xi$.
Moreover we denote the CCR representations $\mathscr{A}\left(\mathscr{M},E,A\right)=\mathscr{C}\left(V,\sigma\right)$
and $\mathscr{A}\left(\mathscr{N},F,B\right)=\mathscr{C}\left(W,\omega\right)$
respectively with $\left(\mathcal{V},\mathrm{V}\right)$ and $\left(\mathcal{W},\mathrm{W}\right)$
and the injective unit preserving {*}-homomorphism $\mathscr{A}\left(\psi,\Psi\right)=\mathscr{C}\left(\xi\right)$
with $H$. From Theorem \ref{thmClassicalFieldFunctor} we know that
$\mathscr{B}$ satisfies the version of the time slice axiom for covariant
functors from $\mathfrak{ghs}^{f}$ to $\mathfrak{alg}$, i.e. $\xi\left(V\right)=W$,
and our aim is to show that $H$ is surjective, which is to say that
$\mathscr{A}$ satisfies the time slice axiom as a LCQFT. As we noted
in Theorem \ref{thmClassicalFieldFunctor}, in the present situation
$\xi$ is bijective and its inverse $\xi^{-1}$ is a symplectic map
from $\left(W,\omega\right)$ to $\left(V,\sigma\right)$. Via the
functor $\mathscr{C}$ we obtain the injective unit preserving {*}-homomorphism
$\mathscr{C}\left(\xi^{-1}\right)$ and then the covariant axioms
imply that
\begin{alignat*}{2}
H\circ\mathscr{C}\left(\xi^{-1}\right) & =\mathscr{C}\left(\xi\right)\circ\mathscr{C}\left(\xi^{-1}\right) & = & \mathrm{id}_{\left(\mathcal{W},\mathrm{W}\right)}\mbox{,}\\
\mathscr{C}\left(\xi^{-1}\right)\circ H & =\mathscr{C}\left(\xi^{-1}\right)\circ\mathscr{C}\left(\xi\right) & = & \mathrm{id}_{\left(\mathcal{V},\mathrm{V}\right)}\mbox{.}
\end{alignat*}
This means that $H$ is bijective and its inverse is $H^{-1}=\mathscr{C}\left(\xi^{-1}\right)$.
In particular $H$ is surjective, as we wanted to show.\end{proof}
\begin{rem}
\label{remPrimitityHoldsDueToTheQuantizationFunctor}Since we recognized
$\mathscr{A}$ to be a LCQFT, we are allowed to apply Theorem \ref{thmRecoveringHaagKastlerAxioms}.
This gives us the opportunity to recover the Haag-Kastler framework
for the description of the quantum theory of the field we are dealing
with. Therefore the functor $\mathscr{A}$ gives actually a quantum
field theory (in its axiomatic definition by Haag and Kastler) for
the field under consideration on each globally hyperbolic spacetime.
At this point however this conclusion is not true at all because,
as we noted in Remark \ref{remPrimitivity}, we have not yet shown
that on each globally hyperbolic spacetime the unital C{*}-algebra
obtained through a LCQFT is primitive, i.e. it admits a faithful irreducible
representation on a Hilbert space. Anyway we can see that this property
holds for the LCQFT $\mathscr{A}$ that we have built right now. Actually
this is a property of our functor $\mathscr{C}$ because it maps objects
of $\mathfrak{ssp}$ to CCR representations and each CCR algebra is
primitive: Each unital C{*}-algebra admits an irreducible representation
$\pi$ on a Hilbert space $\mathscr{H}$ (cfr. \cite[Lem. 2.3.23, p. 59]{BR02})
and $\pi$ is indeed a unit preserving {*}-homomorphism from the unital
C{*}-algebra to the unital C{*}-algebra of bounded operators on $\mathscr{H}$;
in our case the unital C{*}-algebra is also a CCR representation,
therefore, applying Proposition \ref{propUnitPreserving*HomomorphismsFromACCRAlgebraAreInjective},
we see that $\pi$ must be injective too, i.e. faithful, and hence
we have just found a faithful irreducible representation $\pi$ on
a Hilbert space for each CCR representation.
\end{rem}

\section{\label{secExamples}Examples}

In the last section we have shown how to build a locally covariant
quantum field theory that describes a field over an arbitrary globally
hyperbolic spacetime which is ruled at a classical level by a wave
equation on that spacetime written in terms of a normally hyperbolic
operator acting on sections in a proper vector bundle. More precisely,
Subsection \ref{subClassicalFieldTheory} was devoted to the construction
of the field theory at a classical level consisting of a covariant
functor that provides the solutions to all homogeneous Cauchy problems
with compactly supported initial data on a given globally hyperbolic
spacetime. We also established a causality property and a form of
time slice axiom for such functor (cfr. Theorem \ref{thmClassicalFieldFunctor}).
After that, in Subsection \ref{subQuantumFieldTheory} we introduced
a covariant functor that maps each symplectic space to a unital C{*}-algebras,
actually the unique (up to {*}-isomorphisms) CCR representation of
that symplectic space. We may regard this covariant functor as a {}``quantization''
functor because, when composed with the covariant functor describing
the classical theory, it gives rise to a LCQFT which is causal and
fulfils the time slice axiom (in the sense of Definition \ref{defLCQFT}).
Theorem \ref{thmRecoveringHaagKastlerAxioms} and Remark \ref{remPrimitityHoldsDueToTheQuantizationFunctor}
recover the Haag-Kastler axioms and in this way they assure that this
LCQFT actually provides the quantum field theory (in its axiomatic
definition made by Haag and Kastler) for the field under consideration
on each globally hyperbolic spacetime.

In this section we want to show some realizations of LCQFTs in situations
of physical interest, specifically we discuss the real Klein-Gordon
field, the real Proca field and the electromagnetic field. We will
discuss these fields in terms of $k$-forms. This is the typical approach
for the Maxwell equations, but it is quite unusual to treat in this
way the Klein-Gordon equation and the Proca equation. Anyway we will
see that the more familiar equations are equivalent to those written
in terms of $k$-forms.

We want to show from now that the d'Alembertian operator
\[
\mathrm{\Box}_{k}=\mathrm{d\delta}+\mathrm{\delta d}:\mathrm{\Omega}^{k}M\rightarrow\mathrm{\Omega}^{k}M\mbox{,}
\]
defined in terms of the exterior derivative $\mathrm{d}$ (see Proposition
\ref{propExteriorDerivative}) and the codifferential $\mathrm{\delta}$
(see Definition \ref{defCodifferential}), is a formally selfadjoint
normally hyperbolic operator for each $k$.
\begin{prop}
\label{propBox}Let $\left(M,g\right)$ be an orientable Lorentzian
$d$-dimensional manifold and let $\mathfrak{o}$ be a choice of the
orientation. Then for each $k$ the d'Alembertian operator $\mathrm{\Box}_{k}$
defined above is a formally selfadjoint normally hyperbolic operator
on $\mathrm{\Lambda}^{k}M$ over $M$. Moreover the following identities
hold on $\mathrm{\Omega}^{k}M$:
\begin{eqnarray*}
\mathrm{\Box}_{k+1}\mathrm{d} & = & \mathrm{d}\Box_{k}\mbox{,}\\
\mathrm{\Box}_{k-1}\mathrm{\delta} & = & \mathrm{\delta}\Box_{k}\mbox{.}
\end{eqnarray*}
\end{prop}
\begin{proof}
$\mathrm{\Lambda}^{k}M$ reduces to $M\times\left\{ 0\right\} $ for
$k>d$ (see the comments immediately after Definition \ref{defAlternatingTensor})
so that the statement of the proposition becomes trivial. Then, without
loss of generality, we can fix $k\in\left\{ 0,\dots,d\right\} $.

In first place we show that $\mathrm{\Box}_{k}$ is a linear differential
operator from $\mathrm{\Lambda}^{k}M$ to itself. To this end we fix
a section $\mu\in\mathrm{\Omega}^{k}M$ and a point $p\in M$ and
we choose a coordinate neighborhood $\left(U,V,\phi\right)$ for $p$
in $M$. On $V$ we put the orientation $\phi_{*}\left(\left.\mathfrak{o}\right|_{U}\right)$
and on $\mathrm{T}V=V\times\mathbb{R}^{d}$ we set the inner product
$\phi_{*}\left(\left.g\right|_{U}\right)$. Then we choose an oriented
orthonormal basis $\left\{ e_{1},\dots,e_{d}\right\} $ of $\mathrm{\Lambda}_{\phi\left(p\right)}^{1}V$
and we define the local 1-forms $\mathrm{d}x^{i}\in\mathrm{\Lambda}^{1}V$
through the formula $\mathrm{d}x^{i}=e_{i}$ on each point of $V$:
$\left\{ \mathrm{d}x^{1},\dots,\mathrm{d}x^{d}\right\} $ is a basis
of $\mathrm{\Lambda}^{1}V$. Now we are ready to express $\mathrm{\Box}_{k}\mu$
in local coordinates applying Proposition \ref{propExteriorDerivative}
and the comments just after Definition \ref{defCodifferential}:
\[
\phi_{*}\left(\left.\mathrm{\Box}_{k}\mu\right|_{U}\right)=\left(\mathrm{d\delta}+\mathrm{\delta d}\right)\left(\phi_{*}\left(\left.\mu\right|_{U}\right)\right)\mbox{.}
\]
We rewrite $\phi_{*}\left(\left.\mu\right|_{U}\right)$ using the
base of $\mathrm{\Lambda}^{k}V$:
\[
\left(\phi_{*}\left(\left.\mu\right|_{U}\right)\right)\left(x\right)=\frac{1}{k!}f_{i_{1}\dots i_{k}}\left(x\right)\mathrm{d}x^{i_{1}}\wedge\cdots\wedge\mathrm{d}x^{i_{k}}\mbox{.}
\]
The result of the calculation gives a $k$-form over $V$ whose coefficients
in the basis $\left\{ \mathrm{d}x^{i_{1}}\wedge\cdots\wedge\mathrm{d}x^{i_{k}}\right\} $
consist of (very long) linear combinations of partial derivatives
of $f_{i_{1}\dots i_{k}}$ up to the second order with coefficients
involving the metric (and its first order partial derivatives) and
the Levi-Civita symbol. This is sufficient to understand that $\mathrm{\Box}_{k}$
is actually a linear differential operator of second order. To show
that it is also normally hyperbolic we report the final expression
of the term involving second order partial derivatives of $f_{i_{1}\dots i_{k}}$:
\[
-g^{lm}\left(x\right)\frac{\partial^{2}f_{i_{1}\dots i_{k}}}{\partial x^{l}\partial x^{m}}\left(x\right)\mathrm{d}x^{i_{1}}\wedge\cdots\wedge\mathrm{d}x^{i_{k}}\mbox{,}
\]
where $\left(g^{ij}\left(x\right)\right)$ is the inverse of the matrix
$\left(g_{ij}\left(x\right)\right)$ whose coefficients are defined
by
\[
g_{ij}\left(x\right)=g_{\phi^{-1}\left(x\right)}\left(\phi^{*}\left(\mathrm{d}x^{i}\right),\phi^{*}\left(\mathrm{d}x^{j}\right)\right)\mbox{.}
\]
From this we deduce that the principal symbol $\sigma_{\mathrm{\Box_{k}}}$
of $\mathrm{\Box}_{k}$ is the map
\begin{eqnarray*}
\mathrm{T}^{*}M & \rightarrow & \mathrm{Hom}\left(\mathrm{\Lambda}^{k}M,\mathrm{\Lambda}^{k}M\right)\\
\left(p,\omega\right) & \mapsto & -g^{lm}\left(\phi\left(p\right)\right)\left(\phi_{*}\omega\right)_{l}\left(\phi_{*}\omega\right)_{m}\mathrm{id}_{\mathrm{\Lambda}^{k}M}\mbox{,}
\end{eqnarray*}
where $\phi_{*}\omega=\left(\phi_{*}\omega\right)_{i}\mathrm{d}x^{i}\in\mathrm{T}_{\phi\left(p\right)}^{*}V$.
Noting that
\[
g^{lm}\left(\phi\left(p\right)\right)\left(\phi_{*}\omega\right)_{l}\left(\phi_{*}\omega\right)_{m}=g_{p}\left(\omega^{\sharp},\omega^{\sharp}\right)\mbox{,}
\]
we conclude that $\mathrm{\Box}_{k}$ is normally hyperbolic.

Formal selfadjointness is deduced from Proposition \ref{propCodifferentialIsFormallyAdjointToExteriorDerivative}
using the non degenerate inner product on $\mathrm{\Omega}^{k}M$
defined in Proposition \ref{propInnerProductOnkForms}. For each $\mu$,
$\nu\in\mathrm{\Omega}_{0}^{k}M$, we have
\begin{eqnarray*}
\left(\mathrm{\Box}_{k}\mu,\nu\right)_{g,k} & = & \left(\mathrm{d\delta}\mu,\nu\right)_{g,k}+\left(\mathrm{\delta d}\mu,\nu\right)_{g,k}\\
 & = & \left(\mathrm{\delta}\mu,\mathrm{\delta}\nu\right)_{g,k-1}+\left(\mathrm{d}\mu,\mathrm{d}\nu\right)_{g,k+1}\\
 & = & \left(\mathrm{\delta d}\mu,\nu\right)_{g,k}+\left(\mu,\mathrm{d\delta}\nu\right)_{g,k}\\
 & = & \left(\mu,\mathrm{\Box}_{k}\nu\right)_{g,k}
\end{eqnarray*}
and this means exactly that $\mathrm{\Box}_{k}$ is formally selfadjoint.

The stated identities follow from $\mathrm{d}^{2}=0$ and $\mathrm{\delta}^{2}=0$:
\begin{alignat*}{4}
\mathrm{\Box}_{k+1}\mathrm{d} & =\left(\mathrm{d\delta}+\mathrm{\delta d}\right)\mathrm{d} & \:=\: & \mathrm{d\delta d} & \:=\: & \mathrm{d}\left(\mathrm{d\delta}+\mathrm{\delta d}\right) & \,=\: & \mathrm{d}\mathrm{\Box}_{k}\mbox{,}\\
\mathrm{\Box}_{k-1}\mathrm{\delta} & =\left(\mathrm{d\delta}+\mathrm{\delta d}\right)\mathrm{\delta} & \:=\: & \mathrm{\delta d\delta} & \:=\: & \mathrm{\delta}\left(\mathrm{d\delta}+\mathrm{\delta d}\right) & \,=\: & \mathrm{\delta}\mathrm{\Box}_{k}\mbox{.}
\end{alignat*}

\end{proof}

\subsection{\label{subKleinGordonField}\index{Klein-Gordon Field}The Klein-Gordon
field}

This is the easiest of our examples because, as we will see, we can
apply completely the procedure of Section \ref{secBuildingALCQFT}.

We fix a value of the mass $m\geq0$. The Klein-Gordon field of mass
$m$ on a 4-dimensional globally hyperbolic spacetime $\mathscr{M}=\left(M,g,\mathfrak{o},\mathfrak{t}\right)$
is described by a section $\varphi$ in the trivial tensor bundle
$\mathrm{\Lambda}^{0}M=M\times\mathbb{R}$ (sometimes this bundle
is called line bundle) that satisfies the equation
\begin{equation}
A\varphi=\mathrm{\Box}_{0}\varphi+m^{2}\varphi=0\mbox{.}\label{eqKleinGordonEquation}
\end{equation}
The inner product of $\mathrm{\Lambda}^{0}M$ is provided by multiplication
of real numbers on each fiber. The operator $m^{2}\mathrm{id}_{\mathrm{\Omega}^{0}M}$
is trivially formally selfadjoint, hence Proposition \ref{propBox}
implies that $A$ is a formally selfadjoint normally hyperbolic operator.
These considerations allow us to recognize $\left(\mathscr{M},\mathrm{\Omega}^{0}M,A\right)$
as an object of $\mathfrak{ghs}^{f}$. Consider a morphism $\left(\psi,\Psi\right)$
of $\mathfrak{ghs}^{f}$ from $\left(\mathscr{M},\mathrm{\Omega}^{0}M,A\right)$
to another object $\left(\mathscr{N},\mathrm{\Omega}^{0}N,B\right)$.
In this case the situation is considerably simplified if compared
to the general case of a morphism of $\mathfrak{ghs}^{f}$ because
now each fiber is nothing but the real line and hence the fact that
$\Psi$ must be fiberwise an isometric isomorphism of $\mathbb{R}$
to itself, together with the condition of compatibility with $A$
and $B$ (that are always of the form $\mathrm{\Box}_{0}+m^{2}\mathrm{id}_{\mathrm{\Omega}^{0}M}$),
implies that $\Psi=\mathrm{id}_{\mathbb{R}}$.

Then, when we want to describe the Klein-Gordon field, we restrict
our category $\mathfrak{ghs}^{f}$ to a category $\mathfrak{ghs}^{KG}$
whose objects are 4-dimensional globally hyperbolic spacetimes $\mathscr{M}=\left(M,g,\mathfrak{o},\mathfrak{t}\right)$
with the line bundle $\mathrm{\Lambda}^{0}M$ as vector bundle on
which we set the inner product induced by fiberwise multiplication
of real numbers and the formally selfadjoint normally hyperbolic operator
$\mathrm{\Box}_{0}+m^{2}\mathrm{id}_{\mathrm{\Omega}^{0}M}$, which
is completely determined by the metric and the orientation of $\mathscr{M}$.
The morphisms that we consider are the morphisms of $\mathfrak{ghs}^{f}$
between the objects of $\mathfrak{ghs}^{KG}$. This entails that $\mathfrak{ghs}^{KG}$
is a full subcategory of $\mathfrak{ghs}^{f}$.

Due to the conditions of compatibility with the inner products and
the normally hyperbolic operators, a morphism of $\mathfrak{ghs}^{KG}$
reduces to a map from $\mathrm{\Lambda}^{0}M=M\times\mathbb{R}$ to
$\mathrm{\Lambda}^{0}N=N\times\mathbb{R}$ that acts on each fiber
as the identity: $\Psi_{p}\left(p,\mu\right)=\left(\psi\left(p\right),\mu\right)$
for each $p\in M$ and each $\mu\in\mathbb{R}$. Hence for each $u\in\mathrm{\Omega}^{0}M$
the extension of $u$ through $\left(\psi,\Psi\right)$ is nothing
but an {}``extended'' push-forward through $\psi^{\prime}$ of $u$
(refer to eq. \eqref{eqDefinitionOfextPsi}):
\[
\mathrm{ext}_{\Psi}u=\begin{cases}
\left(u\circ\psi^{\prime-1}\right)\left(q\right) & \mbox{if }q\in\psi\left(M\right)\mbox{,}\\
0 & \mbox{if }q\in N\setminus\psi\left(M\right)
\end{cases}=\mathrm{ext}_{\iota_{\mathrm{\Lambda}^{0}\psi\left(M\right)}^{\mathrm{\Lambda}^{0}N}}\left(\psi_{*}^{\prime}u\right)\mbox{.}
\]

With these considerations we realize that the Klein-Gordon field is
simply a special case of our general discussion so that we can apply
the procedure of Section \ref{secBuildingALCQFT} obtaining first
the covariant functor describing the classical theory and then the
quantization functor. By composition of these covariant functors we
obtain a locally covariant quantum field theory for the Klein-Gordon
field and Theorem \ref{thmRecoveringHaagKastlerAxioms} (see Remark
\ref{remPrimitityHoldsDueToTheQuantizationFunctor} for primitivity)
assures that this LCQFT provides on each globally hyperbolic spacetime
$\mathscr{M}$ a unital C{*}-algebra satisfying the Haag-Kastler axioms,
hence it is actually the quantum field theory of the Klein-Gordon
field on $\mathscr{M}$.
\begin{rem}
Our conclusion rely upon the assumption that the description we gave
of the Klein-Gordon field in terms of a 0-form $\varphi$ over a globally
hyperbolic spacetime $\mathscr{M}$ satisfying the equation $\mathrm{\Box}_{0}\varphi+m^{2}\varphi=0$
is equivalent to the usual formulation consisting of a real valued
smooth function $\varphi$ over $\mathscr{M}$ ruled by the equation
\begin{equation}
-\nabla^{i}\nabla_{i}\varphi+m^{2}\varphi=0\mbox{,}\label{eqKleinGordonEquationInIndexNotation}
\end{equation}
where $\nabla$ is the Levi-Civita connection, $\left(g^{ij}\left(p\right)\right)$
is the inverse of the matrix $\left(g_{ij}\left(p\right)\right)$
whose coefficients are defined by $g_{ij}\left(p\right)=g_{p}\left(e_{i},e_{j}\right)$
using a base $\left\{ e_{1},\dots,e_{4}\right\} $ of $\mathrm{T}_{p}M$,
$\nabla_{i}=\nabla_{e_{i}}$ and $\nabla^{i}=g^{ij}\nabla_{j}$.

Since $\mathrm{\Omega}^{0}M=\mathrm{C}^{\infty}\left(M,\mathrm{\Lambda}^{0}M\right)=\mathrm{C}^{\infty}\left(M\right)$,
the Klein-Gordon field in our description is actually a real valued
smooth function, as it is in the usual approach. The equivalence of
the equations is a special case of a more general formula by Lichnerowicz
(cfr. \cite[eq. (3.4), p. 17]{Lic64}) that we report here:
\begin{eqnarray}
\left(\mathrm{\Box}_{k}\omega\right)_{i_{1}\dots i_{k}} & = & -\nabla^{l}\nabla_{l}\omega_{i_{1}\dots i_{k}}+\sum_{n=1}^{k}R_{i_{n}l}g^{ll^{\prime}}\omega_{i_{1}\dots i_{n-1}l^{\prime}i_{n+1}\dots i_{k}}\nonumber \\
 &  & -\sum_{n=1}^{k}\sum_{n^{\prime}\neq n}C_{i_{n}li_{n^{\prime}}}^{\hphantom{i_{n}li_{n^{\prime}}}m}g^{ll^{\prime}}\omega_{i_{1}\dots i_{n-1}l^{\prime}i_{n+1}\dots i_{n^{\prime}-1}mi_{n^{\prime}+1}\dots i_{k}},\label{eqLichnerowiczFormula}
\end{eqnarray}
where $\omega\in\mathrm{\Lambda}^{k}M$, $R_{ij}$ denotes the Ricci
tensor and $C_{ijk}^{\hphantom{ijk}l}$ denotes the curvature of the
Levi-Civita connection $\nabla$ (see Subsection \ref{subVectorBundlesConnectionsInnerProducts}
for their definitions). This formula for $k=0$ shows the exact coincidence
of eq. \eqref{eqKleinGordonEquation} and eq. \eqref{eqKleinGordonEquationInIndexNotation}.
\end{rem}

\begin{rem}
Notice that it is possible to consider also a non minimally coupled
version of the Klein-Gordon equation on a globally hyperbolic spacetime
$\mathscr{M}=\left(M,g\mathfrak{o},\mathfrak{t}\right)$, namely we
can introduce a linear term that introduces a coupling between the
field and the scalar curvature:
\[
A_{R}\varphi=\mathrm{\Box}_{0}\varphi+\left(m^{2}+kR\right)\varphi=0\mbox{,}
\]
where $k$ is a constant and $R$ is the scalar curvature of the Levi-Civita
connection on $\mathscr{M}$ (see the end of Subsection \ref{subVectorBundlesConnectionsInnerProducts}).
Indeed $A_{R}$ is still a formally selfadjoint normally hyperbolic
operator since we added a linear term of 0-th order in the derivatives,
hence we can again apply the general construction of Section \ref{secBuildingALCQFT}.
\end{rem}

\subsection{\label{subProcaField}\index{Proca field}The Proca field}

At a classical level we describe the Proca field of mass $m>0$ on
a 4-dimensional globally hyperbolic spacetime $\mathscr{M}=\left(M,g,\mathfrak{o},\mathfrak{t}\right)$
as a 1-form $\Theta\in\mathrm{\Omega}^{1}M$ satisfying the equation
\begin{equation}
\mathrm{\delta d}\Theta+m^{2}\Theta=0\mbox{.}\label{eqProcaEquation}
\end{equation}

\begin{rem}
The standard expression in index notation for the equation of the
minimally coupled Proca field on $\mathscr{M}$ is the following:
\begin{equation}
-\nabla^{i}\nabla_{i}\Theta_{j}+\nabla^{i}\nabla_{j}\Theta_{i}+m^{2}\Theta_{j}=0\mbox{,}\label{eqProcaEquationInIndexNotation}
\end{equation}
where $\nabla$ is the Levi-Civita connection. To check that our formulation
(eq. \eqref{eqProcaEquation}) is equivalent to the standard one (eq.
\eqref{eqProcaEquationInIndexNotation}) we need to rewrite the standard
equation in a convenient form. The first step consists in the observation
that
\[
\nabla^{i}\nabla_{j}\Theta_{i}-\nabla_{j}\nabla^{i}\Theta_{i}=g^{ik}R_{ij}\Theta_{k}\mbox{.}
\]
This result is obtained through the direct computation of $\nabla_{i}\nabla_{j}\Theta^{k}$
and using the expression of the Ricci tensor $R_{ij}$ for the Levi-Civita
connection in terms of the Christoffel symbols (eq. \eqref{eqRicciTensorForTheLeviCivitaConnection}).
The substitution of the last equation in eq. \eqref{eqProcaEquationInIndexNotation}
gives
\[
-\nabla^{i}\nabla_{i}\Theta_{j}+\nabla_{j}\nabla^{i}\Theta_{i}+g^{ik}R_{ij}\Theta_{k}+m^{2}\Theta_{j}=0
\]
and, recalling the Lichnerowicz formula, eq. \eqref{eqLichnerowiczFormula},
for $k=1$, we deduce that
\[
\left(\mathrm{\Box}_{1}\Theta\right)_{j}+\nabla_{j}\nabla^{i}\Theta_{i}+m^{2}\Theta_{j}=0\mbox{.}
\]
Moreover one can check that $\mathrm{\delta}\Theta=-\nabla^{i}\Theta_{i}$
and hence
\[
\nabla_{j}\nabla^{i}\Theta_{i}=\partial_{j}\nabla^{i}\Theta_{i}=-\left(\mathrm{d\delta}\Theta\right)_{j}\mbox{.}
\]
With this we conclude that eq. \eqref{eqProcaEquation} and \eqref{eqProcaEquationInIndexNotation}
are actually equivalent.
\end{rem}
The case of the Proca field is more involved if compared to the case
of the Klein-Gordon field. The difficulty arises at a classical level
because, although being a formally selfadjoint linear differential
operator of second order on $\mathrm{\Lambda}^{1}M$ (as one might
easily check from eq. \eqref{eqProcaEquationInIndexNotation} and
exploiting the fact that $\mathrm{d}$ and $\mathrm{\delta}$ are
formal adjoints of each other), $\mathrm{\delta}d$ is not normally
hyperbolic (another glance at eq. \eqref{eqProcaEquationInIndexNotation}
shows that the term $\nabla^{i}\nabla_{j}\Theta_{i}$ breaks normal
hyperbolicity). This fact makes the results of Subsection \ref{subClassicalFieldTheory}
inapplicable to the current problem. Anyway one might observe that,
since the Proca field $\Theta$ must satisfy eq. \eqref{eqProcaEquation},
then it follows that it must also be coclosed, i.e. $\mathrm{\delta}\Theta=0$,
because $m>0$ and
\[
m^{2}\mathrm{\delta}\Theta=\mathrm{\delta}\left(\mathrm{\delta d}\Theta+m^{2}\Theta\right)=0\mbox{,}
\]
where we used the property $\mathrm{\delta}^{2}=0$. But then $\mathrm{d\delta}\Theta=0$
too and so the Proca field $\Theta$ satisfies also the equation
\[
\mathrm{\Box}_{1}\Theta+m^{2}\Theta=0\mbox{.}
\]
There is no doubt that $\mathrm{\Box}_{1}+m^{2}\mathrm{id}_{\mathrm{\Omega}_{1}M}$
is a formally selfadjoint normally hyperbolic operator and that we
could apply the procedure of Subsection \ref{subClassicalFieldTheory}
if we consider this operator. The problem is that, proceeding in this
way, we do not describe the Proca field because equation $\mathrm{\Box}_{1}\Theta+m^{2}\Theta=0$
does not imply $\mathrm{\delta}\Theta=0$. However the system
\begin{equation}
\left\{ \begin{array}{rcl}
\mathrm{\Box}_{1}\Theta+m^{2}\Theta & = & 0\\
\mathrm{\delta}\Theta & = & 0
\end{array}\right.\label{eqProcaSystem}
\end{equation}
is absolutely equivalent to eq. \eqref{eqProcaEquation} as one immediately
realizes. We already know how to obtain the solutions of all the Cauchy
problems with compactly supported initial data formulated using the
first equation of the system above. The trick that allows us to select
only those solutions that satisfy also the second equation can be
found in \cite[p. 9]{Dap11}. Before we present it, a lemma is required.
\begin{lem}
\label{lemGreenOperatorsCommuteWithExteriorDerivativeAndCodifferential}Let
$\mathscr{M}=\left(M,g,\mathfrak{o},\mathfrak{t}\right)$ be a $d$-dimensional
globally hyperbolic spacetime, let $m\geq0$ and let $k\in\left\{ 0,\dots,d\right\} $.
Consider the advanced/retarded Green operator $e_{k}^{a/r}$ for the
formally selfadjoint normally hyperbolic operator
\[
P_{k}=\mathrm{\Box}_{k}+m^{2}\mathrm{id}_{\mathrm{\Omega}^{k}M}:\mathrm{\Omega}^{k}M\rightarrow\mathrm{\Omega}^{k}M\mbox{.}
\]
We have that for each $\theta\in\mathrm{\Omega}_{0}^{k}M$ the following
identities hold:
\begin{eqnarray*}
e_{k+1}^{a/r}\left(\mathrm{d}\theta\right) & = & \mathrm{d}\left(e_{k}^{a/r}\theta\right)\quad\mbox{for }k\in\left\{ 0,\dots,d-1\right\} \mbox{;}\\
e_{k-1}^{a/r}\left(\mathrm{\delta}\theta\right) & = & \mathrm{\delta}\left(e_{k}^{a/r}\theta\right)\quad\mbox{for }k\in\left\{ 1,\dots,d\right\} \mbox{.}
\end{eqnarray*}
\end{lem}
\begin{proof}
Fix $k\in\left\{ 0,\dots,d-1\right\} $ and $\theta\in\mathrm{\Omega}_{0}^{k}M$
and consider $e_{k+1}^{a}\mathrm{d}\theta$. From the properties of
$e_{k}^{a}$ we know that $P_{k}e_{k}^{a}\theta=\theta$ so that
\[
e_{k+1}^{a}\left(\mathrm{d}\theta\right)=e_{k+1}^{a}\left(\mathrm{d}P_{k}\left(e_{k}^{a}\theta\right)\right)\mbox{.}
\]
From Proposition \ref{propBox} we deduce that $\mathrm{d}\circ P_{k+1}=P_{k}\circ\mathrm{d}$
and hence we find
\[
e_{k+1}^{a}\left(\mathrm{d}\theta\right)=e_{k+1}^{a}\left(P_{k+1}\mathrm{d}\left(e_{k}^{a}\theta\right)\right)\mbox{.}
\]
Note that, exploiting the support properties of $e_{k}^{a}$, we obtain
\[
\mathrm{supp}\left(\mathrm{d}\left(e_{k}^{a}\theta\right)\right)\subseteq\mathrm{supp}\left(e_{k}^{a}\theta\right)\subseteq J_{+}^{\mathscr{M}}\left(\mathrm{supp}\left(\theta\right)\right)\mbox{.}
\]
This inclusion implies that the support of $\mathrm{d}\left(e_{k}^{a}\theta\right)$
is $\mathscr{M}$-past compact because also $J_{+}^{\mathscr{M}}\left(\mathrm{supp}\left(\theta\right)\right)$
is $\mathscr{M}$-past compact (cfr. Proposition \ref{propUsefulSubsetsOfGloballyHyperbolicSpacetimes}).
Then we can apply Lemma \ref{lemExtensionOf ea(Pu)=00003Du} to $e_{k+1}^{a}$
and conclude that
\[
e_{k+1}^{a}\left(\mathrm{d}\theta\right)=\mathrm{d}\left(e_{k}^{a}\theta\right)\mbox{.}
\]
The proof for $\mathrm{\delta}$ in place of $\mathrm{d}$ is identical
and we can proceed similarly also if we consider the retarded Green
operators in place of the advanced ones.
\end{proof}
Now we are ready to show the trick. The first step consists in the
determination of the advanced and retarded Green operators for the
operator $\mathrm{\delta d}+m^{2}\mathrm{id}_{\mathrm{\Omega}^{k}M}$.
Although we cannot apply Corollary \ref{corGreenOperators} because
the operator is not normally hyperbolic, we can exploit the advanced
and retarded Green operators for $\mathrm{\Box}_{k}+m^{2}\mathrm{id}_{\mathrm{\Omega}^{k}M}$
to find advanced and retarded Green operators for $\mathrm{\delta d}+m^{2}\mathrm{id}_{\mathrm{\Omega}^{k}M}$.
\begin{lem}
\label{lemGreenOperatorsForTheProcaField}Let $\mathscr{M}=\left(M,g,\mathfrak{o},\mathfrak{t}\right)$
be a $d$-dimensional globally hyperbolic spacetime and let $k\in\left\{ 1,\dots,d-1\right\} $.
Consider the formally selfadjoint linear differential operator of
second order
\[
A_{k}=\mathrm{\delta d}+m^{2}\mathrm{id}_{\mathrm{\Omega}^{k}M}:\mathrm{\Omega}^{k}M\rightarrow\mathrm{\Omega}^{k}M\mbox{.}
\]
Then we have that
\[
f_{k}^{a/r}=e_{k}^{a/r}\circ\left(\mathrm{id}_{\mathrm{\Omega}_{0}^{k}M}+\frac{1}{m^{2}}\mathrm{d\delta}\right):\mathrm{\Omega}_{0}^{k}M\rightarrow\mathrm{\Omega}^{k}M
\]
is an advanced/retarded Green operator for $A_{k}$, where $e_{k}^{a/r}$
is the advanced/retarded Green operator for the formally selfadjoint
normally hyperbolic operator
\[
P_{k}=\mathrm{\Box}_{k}+m^{2}\mathrm{id}_{\mathrm{\Omega}^{k}M}:\mathrm{\Omega}^{k}M\rightarrow\mathrm{\Omega}^{k}M\mbox{.}
\]
Moreover $f_{k}^{r/a}$ is formally adjoint to $f_{k}^{a/r}$.\end{lem}
\begin{proof}
Fix $k\in\left\{ 1,\dots,d-1\right\} $. We consider only the case
of the advanced Green operator (the other case being similar). First
of all we notice that $f_{k}^{a}$ is linear and that for each $\theta\in\mathrm{\Omega}_{0}^{1}M$
we find
\[
\mathrm{supp}\left(f_{k}^{a}\theta\right)\subseteq J_{+}^{\mathscr{M}}\left(\mathrm{supp}\left(\theta+\frac{1}{m^{2}}\mathrm{d\delta}\theta\right)\right)\subseteq J_{+}^{\mathscr{M}}\left(\mathrm{supp}\left(\theta\right)\right)
\]
exploiting the support property of the advanced Green operator $e_{k}^{a}$.
Now fix an arbitrary $\theta\in\mathrm{\Omega}_{0}^{k}M$ and evaluate
$A_{k}\left(f_{k}^{a}\theta\right)$ bearing in mind Lemma \ref{lemGreenOperatorsCommuteWithExteriorDerivativeAndCodifferential}
and the properties of the Green operators:
\begin{eqnarray*}
A_{k}\left(f_{k}^{a}\theta\right) & = & \left(\mathrm{\delta d}+m^{2}\mathrm{id}_{\mathrm{\Omega}^{k}M}\right)e_{k}^{a/r}\left(\mathrm{id}_{\mathrm{\Omega}_{0}^{k}M}+\frac{1}{m^{2}}\mathrm{d\delta}\right)\theta\\
 & = & \left(\mathrm{\delta d}+m^{2}\mathrm{id}_{\mathrm{\Omega}^{k}M}\right)\left(\mathrm{id}_{\mathrm{\Omega}^{k}M}+\frac{1}{m^{2}}\mathrm{d\delta}\right)e_{k}^{a/r}\theta\\
 & = & \left(\mathrm{\delta d}+m^{2}\mathrm{id}_{\mathrm{\Omega}^{k}M}+\mathrm{d\delta}\right)e_{k}^{a/r}\theta\\
 & = & P_{k}\left(e_{k}^{a/r}\theta\right)\\
 & = & \theta\mbox{.}
\end{eqnarray*}
The calculation is even simpler for $f_{k}^{a}\left(A_{k}\theta\right)$:
\begin{eqnarray*}
f_{k}^{a}A_{k}\theta & = & e_{k}^{a/r}\left(\mathrm{id}_{\mathrm{\Omega}_{0}^{k}M}+\frac{1}{m^{2}}\mathrm{d\delta}\right)\left(\mathrm{\delta d}+m^{2}\mathrm{id}_{\mathrm{\Omega}^{k}M}\right)\theta\\
 & = & e_{k}^{a/r}\left(\mathrm{\Box_{k}}+m^{2}\mathrm{id}_{\mathrm{\Omega}^{k}M}\right)\theta\\
 & = & \theta\mbox{.}
\end{eqnarray*}
Then we recognize that $f_{k}^{a}$ is an advanced Green operator
for $A_{k}$ (cfr. Definition \ref{defGreenOperators}).

To conclude the proof we must show that $f_{k}^{r/a}$ is formally
adjoint to $f_{k}^{a/r}$, which is to say that
\[
\left(f_{k}^{r/a}\theta,\zeta\right)_{g,k}=\left(\theta,f_{k}^{a/r}\zeta\right)_{g,k}
\]
for each $\theta$, $\zeta\in\mathrm{\Omega}_{0}^{k}M$, where $\left(\cdot,\cdot\right)_{g,k}$
is the map defined in Proposition \ref{propInnerProductOnkForms}.
Therefore fix $\theta$ and $\zeta$ in $\mathrm{\Omega}_{0}^{k}M$
and evaluate $\left(f_{k}^{r}\theta,\zeta\right)_{g,k}$. Recall that
$e_{k}^{r/a}$ is formally adjoint to $e_{k}^{a/r}$ because $P_{k}$
is formally selfadjoint (cfr. Proposition \ref{prope*rIsFormallyAdjointToea}),
hence we find
\begin{eqnarray*}
\left(f_{k}^{r/a}\theta,\zeta\right)_{g,k} & = & \left(e_{k}^{r/a}\left(\mathrm{id}_{\mathrm{\Omega}_{0}^{k}M}+\frac{1}{m^{2}}\mathrm{d\delta}\right)\theta,\zeta\right)_{g,k}\\
 & = & \left(\left(\mathrm{id}_{\mathrm{\Omega}_{0}^{k}M}+\frac{1}{m^{2}}\mathrm{d\delta}\right)\theta,e^{a/r}\zeta\right)_{g,k}\mbox{.}
\end{eqnarray*}
We know also that $\mathrm{d}$ and $\mathrm{\delta}$ are formal
adjoints of each other (see Proposition \ref{propCodifferentialIsFormallyAdjointToExteriorDerivative})
and then we can proceed in our calculation:
\[
\left(f_{k}^{r/a}\theta,\zeta\right)_{g,k}=\left(\theta,\left(\mathrm{id}_{\mathrm{\Omega}^{k}M}+\frac{1}{m^{2}}\mathrm{d\delta}\right)e^{a/r}\zeta\right)_{g,k}\mbox{.}
\]
In the last step we exploit Lemma \ref{lemGreenOperatorsCommuteWithExteriorDerivativeAndCodifferential}:
\[
\left(f_{k}^{r/a}\theta,\zeta\right)_{g,k}=\left(\theta,e^{a/r}\left(\mathrm{id}_{\mathrm{\Omega}_{0}^{k}M}+\frac{1}{m^{2}}\mathrm{d\delta}\right)\zeta\right)_{g,k}=\left(\theta,f_{k}^{a/r}\zeta\right)_{g,k}\mbox{.}
\]

\end{proof}
At this point we have the advanced and retarded Green operators for
the operator $A_{k}=\mathrm{\delta d}+m^{2}\mathrm{id}_{\mathrm{\Omega}^{k}M}$
and we know that they are formal adjoints of each other. We want to
use them to determine the space $V$ of the solutions to all homogeneous
Cauchy problems for the operator $A_{k}$ with compactly supported
initial data. Then we want to exploit the reciprocal formal adjointness
of the Green operators for $A_{k}$ to define a symplectic form on
$V$.
\begin{prop}
\label{propSymplecticSpaceForTheProcaField}Let $\mathscr{M}=\left(M,g,\mathfrak{o},\mathfrak{t}\right)$
be a $d$-dimensional globally hyperbolic spacetime and let $k\in\left\{ 1,\dots,d-1\right\} $.
Consider the operator $A_{k}=\mathrm{\delta d}+m^{2}\mathrm{id}_{\mathrm{\Omega}^{k}M}$
and its advanced/retarded Green operators $f_{k}^{a/r}$ provided
by Proposition \ref{lemGreenOperatorsForTheProcaField}. Denote with
$f_{k}=f_{k}^{a}-f_{k}^{r}$ the corresponding causal propagator.
Then the space $V$ of the solutions to all homogeneous Cauchy problems
for the operator $A_{k}$ with compactly supported initial data coincides
with the image through $f_{k}$ of $\mathrm{\Omega}_{0}^{k}M$, while
the kernel of the causal propagator $f_{k}$ coincides with the image
through $A_{k}$ of $\mathrm{\Omega}_{0}^{k}M$:
\[
V=f_{k}\left(\mathrm{\Omega}_{0}^{k}M\right)\quad\mbox{and}\quad\ker f_{k}=A_{k}\left(\mathrm{\Omega}_{0}^{k}M\right)\mbox{.}
\]

Moreover the map
\begin{alignat*}{2}
\sigma & : & V\times V & \rightarrow\mathbb{R}\\
 &  & \left(\Theta,\Pi\right) & \mapsto\left(f_{k}\theta,\pi\right)_{g,k}\mbox{,}
\end{alignat*}
where $\left(\cdot,\cdot\right)_{g,k}$ is the map defined in Proposition
\ref{propInnerProductOnkForms} and $\theta$, $\pi\in\mathrm{\Omega}_{0}^{k}M$
are such that $f_{k}\theta=\Theta$ and $f_{k}\pi=\Pi$, is well defined,
bilinear, non degenerate and antisymmetric, i.e. it is a symplectic
form on $V$, hence $\left(V,\sigma\right)$ is a symplectic space. \end{prop}
\begin{proof}
We start from the inclusion $f_{k}\left(\mathrm{\Omega}_{0}^{k}M\right)\subseteq V$.
Take $\Theta\in f_{k}\left(\mathrm{\Omega}_{0}^{k}M\right)$ and consider
$\theta\in\mathrm{\Omega}_{0}^{k}M$ such that $f_{k}\theta=\Theta$.
As a consequence of Lemma \ref{lemGreenOperatorsForTheProcaField}
we have that $\Theta$ is also an element of $e_{k}\left(\mathrm{\Omega}_{0}^{k}M\right)$,
where $e_{k}$ denotes the causal propagator for $P_{k}=\mathrm{\Box}_{k}+m^{2}\mathrm{id}_{\mathrm{\Omega}^{k}M}$.
Therefore from Corollary \ref{corSpaceOfSolutions} we deduce that
$\Theta$ is the solution of a homogeneous Cauchy problem for the
normally hyperbolic operator $P_{k}$ with compactly supported initial
data. Since we have shown that eq. \eqref{eqProcaEquation} is equivalent
to eq. \eqref{eqProcaSystem}, it is sufficient to prove that $\mathrm{\delta}\Theta=0$
to conclude that the expected inclusion holds. We try to evaluate
$\mathrm{\delta}\Theta$ exploiting Lemma \ref{lemGreenOperatorsForTheProcaField}
and Lemma \ref{lemGreenOperatorsCommuteWithExteriorDerivativeAndCodifferential}:
\begin{alignat*}{2}
\mathrm{\delta}\Theta & =\mathrm{\delta}\left(e_{k}\left(\mathrm{id}_{\mathrm{\Omega}_{0}^{k}M}+\frac{1}{m^{2}}\mathrm{d\delta}\right)\theta\right) & = & e_{k}\left(\mathrm{id}_{\mathrm{\Omega}_{0}^{k}M}+\frac{1}{m^{2}}\mathrm{\delta d}\right)\mathrm{\delta}\theta\\
 & =\frac{1}{m^{2}}e_{k}\left(m^{2}+\mathrm{\delta d}+\mathrm{d\delta}\right)\mathrm{\delta}\theta & = & \frac{1}{m^{2}}e_{k}P_{k}\left(\mathrm{\delta}\theta\right)\\
 & =0\mbox{.}
\end{alignat*}

We turn our attention to the converse inclusion $V\subseteq f_{k}\left(\mathrm{\Omega}_{0}^{k}M\right)$.
To this end take $\Theta\in V$. Since eq. \eqref{eqProcaEquation}
is equivalent to eq. \eqref{eqProcaSystem}, $\Theta$ is also a coclosed
solution of a homogeneous Cauchy problem for the normally hyperbolic
operator $P_{k}$ with compactly supported initial data. Applying
Corollary \ref{corSpaceOfSolutions}, we find $\theta\in\mathrm{\Omega}_{0}^{k}M$
such that $e_{k}\theta=\Theta$. Consider now $f_{k}\theta$ bearing
in mind that $\mathrm{\delta}\Theta=0$ and exploiting Lemma \ref{lemGreenOperatorsForTheProcaField}
and Lemma \ref{lemGreenOperatorsCommuteWithExteriorDerivativeAndCodifferential}:
\[
f_{k}\theta=e_{k}\left(\mathrm{id}_{\mathrm{\Omega}_{0}^{k}M}+\frac{1}{m^{2}}\mathrm{d\delta}\right)\theta=\left(\mathrm{id}_{\mathrm{\Omega}^{k}M}+\frac{1}{m^{2}}\mathrm{d\delta}\right)e_{k}\theta=\Theta+\frac{1}{m^{2}}\mathrm{d\delta}\Theta=\Theta\mbox{.}
\]
This implies that $\Theta\in f_{k}\left(\mathrm{\Omega}_{0}^{k}M\right)$.

Now we show that $\ker f_{k}=A_{k}\left(\mathrm{\Omega}_{0}^{k}M\right)$.
The inclusion $A_{k}\left(\mathrm{\Omega}_{0}^{k}M\right)\subseteq\ker f_{k}$
is a trivial consequence of the properties of the Green operators
$f_{k}^{a}$ and $f_{k}^{r}$. To prove the other inclusion take $\theta$
in $\ker f_{k}$. This implies that $f_{k}^{a}\theta=f_{k}^{r}\theta$
and that $\theta$ is an element of $\mathrm{\Omega}_{0}^{k}M$, hence
in particular
\[
\mathrm{supp}\left(f_{k}^{a}\theta\right)\subseteq J_{+}^{\mathscr{M}}\left(\mathrm{supp}\left(\theta\right)\right)\cap J_{-}^{\mathscr{M}}\left(\mathrm{supp}\left(\theta\right)\right)\mbox{.}
\]
Exploiting Proposition \ref{propUsefulSubsetsOfGloballyHyperbolicSpacetimes},
we realize that $\mathrm{supp}\left(f_{k}^{a}\theta\right)$ is a
closed subset of $M$ included in a compact subset of $M$, hence
it is compact too. This shows that $\theta^{\prime}=f_{k}^{a}\theta\in\mathrm{\Omega}_{0}^{k}M$
and, evaluating $A_{k}\theta^{\prime}$ we see that $A_{k}\theta^{\prime}=A_{k}\left(f_{k}^{a}\theta\right)=\theta$.
Therefore we conclude that $\theta\in A_{k}\left(\mathrm{\Omega}_{0}^{k}M\right)$.

The proof of the last part of this proposition is identical to the
proof of Lemma \ref{lemObjghsf->Objssp}.
\end{proof}
We have associated a symplectic space $\left(V,\sigma\right)$ to
each triple $\left(\mathscr{M},\mathrm{\Lambda}^{k}M,A_{k}\right)$,
where $\mathscr{M}=\left(M,g,\mathfrak{o},\mathfrak{t}\right)$ is
a globally hyperbolic spacetime, $\mathrm{\Lambda}^{k}M$ is endowed
with the inner product $\left\langle \cdot,\cdot\right\rangle _{g,k}$
induced by $g$ (cfr. Proposition \ref{propHodgeDualProperties})
and $A_{k}=\mathrm{\delta d}+m^{2}\mathrm{id}_{\mathrm{\Omega}^{k}M}$.

Before we proceed with the ingredients needed for the construction
of a covariant functor describing the classical theory of the Proca
field, we want to introduce the category that we use as domain.
\begin{defn}
\label{defghsP}For $k\in\left\{ 1,\dots,d-1\right\} $, $\mathfrak{ghs}^{P}$
is the category whose objects are triples $\left(\mathscr{M},\mathrm{\Lambda}^{k}M,A_{k}\right)$,
where $\mathscr{M}=\left(M,g,\mathfrak{o},\mathfrak{t}\right)$ is
a $d$-dimensional globally hyperbolic spacetime, $\mathrm{\Lambda}^{k}M$
is endowed with the inner product induced by $g$ and $A_{k}=\mathrm{\delta d}+m^{2}\mathrm{id}_{\mathrm{\Omega}^{k}M}$,
whose morphisms are vector bundle homomorphisms $\left(\psi,\Psi\right)$
from $\mathrm{\Lambda}^{k}M$ to $\mathrm{\Lambda}^{k}N$ over some
morphism $\psi$ from $\mathscr{M}$ to $\mathscr{N}$ of $\mathfrak{ghs}$
that are compatible with the inner products and the linear differential
operators $\mathrm{\delta d}$ and $\mathrm{d\delta}$ of both the
domain and the codomain (for the meaning of this condition see the
comments after Definition \ref{defghsfssp} and bear in mind that
now normal hyperbolicity does not hold). As for the composition law,
it is the usual composition of functions.
\end{defn}
Except for the fact that the operators considered are not normally
hyperbolic, $\mathfrak{ghs}^{P}$ can be considered as a (possibly
non full) subcategory of $\mathfrak{ghs}^{f}$. This claim becomes
precise if we replace in all the objects of this category the operator
$A_{k}$ with the formally selfadjoint normally hyperbolic operator
$\mathrm{\Box}_{k}+m^{2}\mathrm{id}_{\mathrm{\Omega}^{k}M}$ because
the condition of compatibility with both $\mathrm{\delta d}$ and
$\mathrm{d\delta}$ of domain and codomain entails also compatibility
with $\mathrm{\Box}_{k}+m^{2}\mathrm{id}$ of domain and codomain.
As a consequence of this fact all the conclusion that we have drawn
for the morphism of $\mathfrak{ghs}^{f}$ hold also for the morphisms
of $\mathfrak{ghs}^{P}$ (and maybe these morphisms have even richer
properties since the compatibility condition seems to be more stringent).

Note that compatibility with $\mathrm{\delta d}$ trivially implies
also compatibility with the operators $\mathrm{\delta d}+m^{2}\mathrm{id}$
of both the domain and the codomain.

Now that we have specified the (restricted) class of morphisms that
we are going to take into account we can proceed further.
\begin{prop}
\label{propMorghsP->Morssp}Let $\mathscr{M}=\left(M,g,\mathfrak{o},\mathfrak{t}\right)$
and $\mathscr{N}=\left(N,h,\mathfrak{p},\mathfrak{u}\right)$ be $d$-dimensional
globally hyperbolic spacetimes, let $\left(\psi,\Psi\right)$ be a
morphism of $\mathfrak{ghs}^{P}$ from $\left(\mathscr{M},\mathrm{\Lambda}^{k}M,A_{k}\right)$
to $\left(\mathscr{N},\mathrm{\Lambda}^{k}N,B_{k}\right)$ and let
$k\in\left\{ 1,\dots,d-1\right\} $. Consider the advanced/retarded
Green operators $f_{k,M}^{a/r}$ and $f_{k,N}^{a/r}$ for $A_{k}$
and respectively $B_{k}$ provided by Lemma \ref{lemGreenOperatorsForTheProcaField}.
Denote with $\left(V,\sigma\right)$ and $\left(W,\omega\right)$
the symplectic spaces associated respectively to the triples \textup{$\left(\mathscr{M},\mathrm{\Lambda}^{k}M,A_{k}\right)$
and $\left(\mathscr{N},\mathrm{\Lambda}^{k}N,B_{k}\right)$.} Then
\[
\mathrm{res}_{\Psi}\circ f_{k,N}^{a/r}\circ\mathrm{ext}_{\Psi}=f_{k,M}^{a/r}\mbox{,}
\]
where the extension map is defined in eq. \eqref{eqDefinitionOfextPsi}
and the restriction map is defined in Lemma \ref{lemresPsieBextPsi=00003DeA},
and the map
\begin{alignat*}{2}
\xi & : & V & \rightarrow W\\
 &  & \Theta & \mapsto f_{k,N}\left(\mathrm{ext}_{\Psi}\theta\right)\mbox{,}
\end{alignat*}
where $\theta\in\mathrm{\Omega}_{0}^{k}M$ is such that $f_{k,M}\theta=\Theta$,
is well defined, linear and compatible with the symplectic forms $\sigma$
and $\omega$, i.e. it is a symplectic map from $\left(V,\sigma\right)$
to $\left(W,\omega\right)$.\end{prop}
\begin{proof}
The advanced/retarded Green operators $f_{k,M}^{a/r}$ and $f_{k,N}^{a/r}$
for $A_{k}$ and respectively $B_{k}$ provided by Lemma \ref{lemGreenOperatorsForTheProcaField}
have the following expressions:
\begin{eqnarray*}
f_{k,M}^{a/r} & = & e_{k,M}^{a/r}\circ\left(\mathrm{id}_{\mathrm{\Omega}_{0}^{k}M}+\frac{1}{m^{2}}\mathrm{d\delta}\right)\mbox{,}\\
f_{k,N}^{a/r} & = & e_{k,N}^{a/r}\circ\left(\mathrm{id}_{\mathrm{\Omega}_{0}^{k}N}+\frac{1}{m^{2}}\mathrm{d\delta}\right)\mbox{.}
\end{eqnarray*}
Since we know that $\left(\psi,\Psi\right)$ is compatible with the
operators $\mathrm{\Box}_{k}+m^{2}\mathrm{id}_{\mathrm{\Omega}^{k}M}$
and $\mathrm{\Box}_{k}+m^{2}\mathrm{id}_{\mathrm{\Omega}^{k}N}$,
we can apply Lemma \ref{lemresPsieBextPsi=00003DeA} to deduce that
\[
\mathrm{res}_{\Psi}\circ e_{k,N}^{a/r}\circ\mathrm{ext}_{\Psi}=e_{k,M}^{a/r}\mbox{,}
\]
while the condition of compatibility with $\mathrm{d\delta}:\mathrm{\Omega}^{k}M\rightarrow\mathrm{\Omega}^{k}M$
and $\mathrm{d\delta}:\mathrm{\Omega}^{k}N\rightarrow\mathrm{\Omega}^{k}N$
trivially implies that
\[
\mathrm{ext}_{\Psi}\left(\mathrm{id}_{\mathrm{\Omega}_{0}^{k}M}+\frac{1}{m^{2}}\mathrm{d\delta}\right)\theta=\left(\mathrm{id}_{\mathrm{\Omega}_{0}^{k}N}+\frac{1}{m^{2}}\mathrm{d\delta}\right)\left(\mathrm{ext}_{\Psi}\theta\right)
\]
for each $\theta\in\mathrm{\Omega}_{0}^{k}M$. From these facts we
deduce that
\begin{eqnarray*}
\left(\mathrm{res}_{\Psi}\circ f_{k,N}^{a/r}\circ\mathrm{ext}_{\Psi}\right)\theta & = & \left(\mathrm{res}_{\Psi}\circ e_{k,N}^{a/r}\circ\mathrm{ext}_{\Psi}\right)\left(\mathrm{id}_{\mathrm{\Omega}_{0}^{k}M}+\frac{1}{m^{2}}\mathrm{d\delta}\right)\theta\\
 & = & e_{k,M}^{a/r}\left(\mathrm{id}_{\mathrm{\Omega}_{0}^{k}M}+\frac{1}{m^{2}}\mathrm{d\delta}\right)\theta\\
 & = & f_{k,M}^{a/r}\theta
\end{eqnarray*}
for each $\theta\in\mathrm{\Omega}_{0}^{k}M$. This shows the first
part of the thesis.

The proof of the second part is identical to the proof of Lemma \ref{lemMorghsf->Morssp}:
we can proceed in the same way simply considering the Green operators
for $A_{k}$ and $B_{k}$ provided by Lemma \ref{lemGreenOperatorsForTheProcaField}
thanks to the result of the first part of our proof.
\end{proof}
Exploiting Proposition \ref{propSymplecticSpaceForTheProcaField}
and Proposition \ref{propMorghsP->Morssp}, we can introduce the covariant
functor describing the classical theory of the Proca field as shown
by the next theorem. In the following we choose the dimension of the
spacetimes $d=4$ and we drop the subscript $k$ since we fix $k=1$.
However note that the theorem holds also for each $d\in\mathbb{N}$
and each $k\in\left(1,\dots,d-1\right)$. The choices $d=4$ and $k=1$
are made for compatibility with the physical problem.
\begin{thm}
\label{thmClassicalFieldFunctorProcaField}Consider the map
\begin{alignat*}{2}
\mathscr{B} & : & \mathsf{Obj}_{\mathfrak{ghs}^{P}} & \rightarrow\mathsf{Obj}_{\mathfrak{ssp}}\\
 &  & \left(\mathscr{M},\mathrm{\Lambda}^{1}M,A\right) & \mapsto\left(V,\sigma\right)
\end{alignat*}
defined following Proposition \ref{propSymplecticSpaceForTheProcaField}
and for each pair $\left(\mathscr{M},\mathrm{\Lambda}^{1}M,A\right)$,
$\left(\mathscr{N},\mathrm{\Lambda}^{1}N,B\right)$ in $\mathsf{Obj}_{\mathfrak{ghs}^{P}}$
consider the map
\begin{eqnarray*}
\mathscr{B}:\mathsf{Mor}_{\mathfrak{ghs}^{P}}\left(\left(\mathscr{M},\mathrm{\Lambda}^{1}M,A\right),\left(\mathscr{N},\mathrm{\Lambda}^{1}N,B\right)\right) & \rightarrow & \mathsf{Mor}_{\mathfrak{ssp}}\left(\left(V,\sigma\right),\left(W,\omega\right)\right)\\
\left(\psi,\Psi\right) & \mapsto & \xi
\end{eqnarray*}
defined in accordance with Proposition \ref{propMorghsP->Morssp},
where $\left(V,\sigma\right)$ and $\left(W,\omega\right)$ respectively
denote the symplectic spaces $\mathscr{B}\left(\mathscr{M},\mathrm{\Lambda}^{1}M,A\right)$
and $\mathscr{B}\left(\mathscr{N},\mathrm{\Lambda}^{1}N,B\right)$.
These maps give rise to a covariant functor $\mathscr{B}$ from $\mathfrak{ghs}^{P}$
to $\mathfrak{ssp}$ that fulfils the following properties:
\begin{itemize}
\item causality: for each $\left(\mathscr{M}_{1},\mathrm{\Lambda}^{1}M_{1},A_{1}\right)$,
$\left(\mathscr{M}_{2},\mathrm{\Lambda}^{1}M_{2},A_{2}\right)$, $\left(\mathscr{M},\mathrm{\Lambda}^{1}M,A\right)$
in $\mathsf{Obj}_{\mathfrak{ghs}^{P}}$, each morphism $\left(\psi_{1},\Psi_{1}\right)$
from $\left(\mathscr{M}_{1},\mathrm{\Lambda}^{1}M_{1},A_{1}\right)$
to $\left(\mathscr{M},\mathrm{\Lambda}^{1}M,A\right)$ and each morphism
$\left(\psi_{2},\Psi_{2}\right)$ from $\left(\mathscr{M}_{2},\mathrm{\Lambda}^{1}M_{2},A_{2}\right)$
to $\left(\mathscr{M},\mathrm{\Lambda}^{1}M,A\right)$ such that $\psi_{1}\left(M_{1}\right)$
and $\psi_{2}\left(M_{2}\right)$ are $\mathscr{M}$-causally separated,
we have that
\[
\sigma\left(\xi_{1}\Theta_{1},\xi_{2}\Theta_{2}\right)=0
\]
for each $\Theta_{1}\in V_{1}$ and each $\Theta_{2}\in V_{2}$, where
$\left(V_{1},\sigma_{1}\right)$, $\left(V_{2},\sigma_{2}\right)$
and $\left(V,\sigma\right)$ denote the symplectic spaces corresponding
respectively to $\left(\mathscr{M}_{1},\mathrm{\Lambda}^{1}M_{1},A_{1}\right)$,
$\left(\mathscr{M}_{2},\mathrm{\Lambda}^{1}M_{2},A_{2}\right)$ and
$\left(\mathscr{M},\mathrm{\Lambda}^{1}M,A\right)$, while $\xi_{1}$
and $\xi_{2}$ denote the symplectic maps corresponding respectively
to $\left(\psi_{1},\Psi_{1}\right)$ and $\left(\psi_{2},\Psi_{2}\right)$;
\item time slice axiom: for each $\left(\mathscr{M},\mathrm{\Lambda}^{1}M,A\right)$,
$\left(\mathscr{N},\mathrm{\Lambda}^{1}N,B\right)$ in $\mathsf{Obj}_{\mathfrak{ghs}^{P}}$
and each morphism $\left(\psi,\Psi\right)$ from $\left(\mathscr{M},\mathrm{\Lambda}^{1}M,A\right)$
to $\left(\mathscr{N},\mathrm{\Lambda}^{1}N,B\right)$such that $\psi\left(M\right)$
includes a smooth spacelike Cauchy surface $\Sigma$ for $\mathscr{N}$,
we have that
\[
\xi\left(V\right)=W\mbox{,}
\]
where $\left(V,\sigma\right)$ and $\left(W,\omega\right)$ denote
the symplectic spaces corresponding respectively to $\left(\mathscr{M},\mathrm{\Lambda}^{1}M,A\right)$
and $\left(\mathscr{N},\mathrm{\Lambda}^{1}N,B\right)$, while $\xi$
denotes the symplectic map corresponding to $\left(\psi,\Psi\right)$.
\end{itemize}
\end{thm}
\begin{proof}
The check of the covariant axioms, as well as the proof of the causality
property, are identical to those in the proof of Theorem \ref{thmClassicalFieldFunctor}.
For the proof of the time slice axiom again we can largely imitate
the proof of the above mentioned theorem. We must only remember to
write the Cauchy problem used to pick out the compact subset $K$
for the normally hyperbolic operator $\mathrm{\Box}+m^{2}\mathrm{id}_{\mathrm{\Omega}^{1}M}$
in place of the operator $B$ (which is not normally hyperbolic),
otherwise we cannot apply Theorem \ref{thmCauchyProblem} to deduce
uniqueness of the solution. The use of $\mathrm{\Box}+m^{2}\mathrm{id}_{\mathrm{\Omega}^{1}M}$
in place of $B$ does not give rise to problems because sections $\Theta\in\mathrm{\Omega}^{1}N$
satisfying $B\Theta=0$ also satisfy $\mathrm{\Box}\Theta+m^{2}\Theta=0$
(bear in mind the equivalence between eq. \eqref{eqProcaEquation}
and eq. \eqref{eqProcaSystem}). At a certain point of the proof we
should find an identity of the type $B\Theta^{+}=-B\Theta^{-}$. This
entails $\mathrm{\delta}\Theta^{+}=-\mathrm{\delta}\Theta^{-}$ and
hence also
\[
\left(\mathrm{\Box}+m^{2}\mathrm{id}_{\mathrm{\Omega}^{1}M}\right)\Theta^{+}=-\left(\mathrm{\Box}+m^{2}\mathrm{id}_{\mathrm{\Omega}^{1}M}\right)\Theta^{-}\mbox{.}
\]
As we deduce from $B\Theta^{+}=-B\Theta^{-}$ that $B\Theta^{+}$
is a section of $\mathrm{\Omega}_{0}^{1}N$ with support included
in $\psi\left(M\right)$, in a similar manner we deduce from the equation
above that $\left(\mathrm{\Box}+m^{2}\mathrm{id}_{\mathrm{\Omega}^{1}M}\right)\Theta^{+}$
is a section of $\mathrm{\Omega}_{0}^{1}N$ with support included
in $\psi\left(M\right)$. Towards the end we resorted to Lemma \ref{lemExtensionOf ea(Pu)=00003Du}.
This is not directly applicable in the present situation because $B$
is not normally hyperbolic, however it holds that
\[
f_{N}^{a}\left(B\Theta^{+}\right)=e_{N}^{a}\left(\mathrm{\Box}+m^{2}\mathrm{id}_{\mathrm{\Omega}^{1}M}\right)\Theta^{+}
\]
where $e_{N}^{a}$ denotes the advanced/retarded Green operator for
the normally hyperbolic operator $\mathrm{\Box}+m^{2}\mathrm{id}_{\mathrm{\Omega}^{1}M}$.
Since we have just shown that $\left(\mathrm{\Box}+m^{2}\mathrm{id}_{\mathrm{\Omega}^{1}M}\right)\Theta^{+}$
has compact support, we are again in position to apply Lemma \ref{lemExtensionOf ea(Pu)=00003Du}.
A similar procedure applies to $f_{N}^{r}\left(B\Theta^{-}\right)$
and this leads us to the end of the proof.
\end{proof}
With the last theorem we have completed the classical theory of the
Proca field. Now we must proceed with the quantization of the classical
theory that can be done composing our functor $\mathscr{B}:\mathfrak{ghs}^{P}\overset{\rightarrow}{\rightarrow}\mathfrak{ssp}$
with the functor $\mathscr{C}:\mathfrak{ssp}\overset{\rightarrow}{\rightarrow}\mathfrak{alg}$
built in Lemma \ref{lemQuantizationFunctor}. As we proved in Theorem
\ref{thmLCQFTForANormallyHyperbolicField}, the result is a locally
covariant quantum field theory $\mathscr{A}=\mathscr{C}\circ\mathscr{B}$
that satisfies both the causality condition and the time slice axiom.
In turn this implies that we can apply Theorem \ref{thmRecoveringHaagKastlerAxioms}
and Remark \ref{remPrimitityHoldsDueToTheQuantizationFunctor}. Therefore
on each globally hyperbolic spacetime $\mathscr{A}$ provides the
quantum field theory of the Proca field according to the algebraic
approach suggested by Haag and Kastler.

Before we pass to the last example, we want to make some remarks about
the morphisms that are usually taken into account when dealing with
the realization of a LCQFT for a field of physical interest, such
as the Klein-Gordon field or the Proca field.
\begin{rem}
\label{remMorphismsOfConcreteFields}In the discussion of the classical
theory of a concrete field, for example the Klein-Gordon field or
the Proca field, it is usual to consider only push-forwards and pull-backs
as vector bundle homomorphisms. For the case of the Klein-Gordon field
we noted that these two approaches are equivalent. We show now that
push-forwards and pull-backs are morphisms of $\mathfrak{ghs}^{P}$
so that our approach surely includes the usual one.

Note that if $\psi$ is an orientation and time orientation preserving
isometric diffeomorphism from $\mathscr{M}=\left(M,g,\mathfrak{o},\mathfrak{t}\right)$
to $\mathscr{N}=\left(N,h,\mathfrak{p},\mathfrak{u}\right)$, we realize
immediately that $\left(\psi,\psi_{*}\right)$ is a bijective morphism
of $\mathfrak{ghs}^{P}$ between the objects $\left(\mathscr{M},\mathrm{\Lambda}^{k}M,A_{k}\right)$
and $\left(\mathscr{N},\mathrm{\Lambda}^{k}N,B_{k}\right)$ whose
inverse $\left(\psi^{-1},\psi^{*}\right)$ is a morphism of $\mathfrak{ghs}^{f}$
too: $\psi_{*}:\mathrm{\Lambda}^{k}M\rightarrow\mathrm{\Lambda}^{k}N$
is defined as the pull-back through $\psi^{-1}$ (see. Remark \ref{remPushforwardPullback})
and $\left(\psi,\psi_{*}\right)$ is indeed a vector bundle isomorphism
from $\mathrm{\Lambda}^{k}M$ to $\mathrm{\Lambda}^{k}N$ (cfr. Remark
\ref{remPushforwardPullbackSections}) which is compatible with the
inner products induced by the metrics and with the operators $A_{k}$
and $B_{k}$ because $\psi$ is isometric and the following identities
hold (see Proposition \ref{propExteriorDerivative} and the comments
after Definition \ref{defCodifferential}):
\begin{eqnarray*}
\psi_{*}\circ\mathrm{d} & = & \mathrm{d}\circ\psi_{*}\mbox{,}\\
\psi_{*}\circ\mathrm{\delta} & = & \mathrm{\delta}\circ\psi_{*}\mbox{.}
\end{eqnarray*}

When $\psi$ is only an orientation and time orientation preserving
isometric embedding from $\mathscr{M}=\left(M,g,\mathfrak{o},\mathfrak{t}\right)$
to $\mathscr{N}=\left(N,h,\mathfrak{p},\mathfrak{u}\right)$ whose
image $\psi\left(M\right)$ is an open subset of $N$ (i.e. a morphism
of our category $\mathfrak{ghs}$), we can apply the conclusions above
to the diffeomorphism $\psi^{\prime}:M\rightarrow\psi\left(M\right)$,
$p\mapsto\psi\left(p\right)$ and obtain a morphism of $\mathfrak{ghs}^{P}$:
\[
\left(\psi^{\prime},\psi_{*}^{\prime}\right):\left(\mathscr{M},\mathrm{\Lambda}^{k}M,A_{k}\right)\rightarrow\left(\psi\left(\mathscr{M}\right)=\left.\mathscr{N}\right|_{\psi\left(M\right)},\mathrm{\Lambda}^{k}\psi\left(M\right)=\left.\mathrm{\Lambda}^{k}N\right|_{\psi\left(M\right)},B_{k}\right)\mbox{.}
\]
Then we find a new morphism of $\mathfrak{ghs}^{P}$ from $\left(\mathscr{M},\mathrm{\Lambda}^{1}M,A\right)$
to $\left(\mathscr{N},\mathrm{\Lambda}^{1}N,B\right)$ defining the
vector bundle homomorphism
\begin{eqnarray*}
\left(\psi,\psi_{*}\right):\mathrm{\Lambda}^{1}M & \rightarrow & \mathrm{\Lambda}^{1}N\\
\left(p,\omega\right) & \mapsto & \left(\psi^{\prime}\left(p\right),\psi_{*}^{\prime}\omega\right)\mbox{.}
\end{eqnarray*}
$\left(\psi,\psi_{*}\right)$ inherits all the properties of $\left(\psi^{\prime},\psi_{*}^{\prime}\right)$
with the only exception that it is not surjective and hence it is
actually a morphism of $\mathfrak{ghs}^{P}$. As a matter of fact
we have simply defined $\left(\psi,\psi_{*}\right)$ as the composition
of $\left(\iota_{\psi\left(M\right)}^{N},\iota_{\mathrm{\Lambda}^{1}\psi\left(M\right)}^{\mathrm{\Lambda}^{1}N}\right)$
with $\left(\psi^{\prime},\psi_{*}^{\prime}\right)$, which are indeed
morphisms of $\mathfrak{ghs}^{P}$.

On the contrary one may find morphisms $\left(\psi,\Psi\right)$ of
$\mathfrak{ghs}^{P}$ which are not of the form $\left(\psi,\psi_{*}\right)$:
For example consider the Minkowski spacetime as globally hyperbolic
spacetime $\mathscr{M}$; the vector bundle isomorphism $\left(\mathrm{id}_{\mathbb{R}^{4}},\Psi\right):\mathrm{\Lambda}^{k}M\rightarrow\mathrm{\Lambda}^{k}M$,
where $\Psi$ acts on each fiber as a fixed Lorentz transformation
$L$ for tensors of type $\left(0,k\right)$, is a bijective morphism
of $\mathfrak{ghs}^{P}$ form $\left(\mathscr{M},\mathrm{\Lambda}^{k}M,A_{k}\right)$
whose inverse is a morphism too, but it is not of the form $\left(\mathrm{id}_{\mathbb{R}^{4}},\mathrm{id}_{\mathbb{R}^{4}*}\right)$
because $\mathrm{id}_{\mathbb{R}^{4}*}=\mathrm{id}_{\mathrm{\Lambda}^{k}M}$,
where $\mathrm{\Lambda}^{k}M=\mathbb{R}^{4}\times\mathbb{R}^{n}$,
$n=\binom{4}{k}$, in the present situation. This means that we are
dealing with a potential enlargement of the family of morphisms usually
considered (that is comprised by pull-backs and push-forwards through
morphisms of $\mathfrak{ghs}$).

We take the chance to anticipate that for the upcoming example, the
electromagnetic field, we will be forced to reduce to usual approach,
that is our morphisms will be only pull-backs and push-forwards through
morphisms of $\mathfrak{ghs}$.
\end{rem}

\subsection{\label{subElectromagneticField}\index{electromagnetic field}The
electromagnetic field}

Consider a globally hyperbolic spacetime $\mathscr{M}=\left(M,g,\mathfrak{o},\mathfrak{t}\right)$.
The electromagnetic field is usually described by a section $\mathtt{F}$
(known as field strength) in the vector bundle $\mathrm{\Lambda}^{2}M$,
i.e. a 2-form over $M$, satisfying Maxwell equations
\[
\left\{ \begin{array}{rcl}
\mathrm{d}\mathtt{F} & = & 0\mbox{,}\\
\mathrm{\delta}\mathtt{F} & = & 0\mbox{.}
\end{array}\right.
\]
If the second de Rham cohomology group is trivial, that is $H^{2}\left(M\right)=\left\{ 0\right\} $,
then all closed 2-forms over $M$ are also exact (cfr. Definition
\ref{defdeRhamCohomologyGroup}). This means that we can find a 1-form
$\mathtt{A}$ over $M$ (called vector potential) such that $\mathrm{d}\mathtt{A}=\mathtt{F}$
and the Maxwell equations reduce to 
\begin{equation}
\mathrm{\delta d}\mathtt{A}=0\mbox{,}\label{eqEMEquation}
\end{equation}
that is a version of the Proca equation with $m=0$ (cfr. eq. \eqref{eqProcaEquation}).
But when $M$ is such that $H^{2}\left(M\right)$ is not trivial it
happens that there are closed 2-forms $\mathtt{F}$ such that the
equation $\mathrm{d}\mathtt{A}=\mathtt{F}$ cannot be verified by
any 1-form $\mathtt{A}$, hence we cannot deduce eq. \eqref{eqEMEquation}
from Maxwell equations. This means that there exist field strengths
which are indeed solutions of the Maxwell equations, but are not generated
by a vector potential satisfying eq. \eqref{eqEMEquation}.

The problem in dealing directly with the Maxwell equations is the
absence of a normally hyperbolic operator that allows us to apply
the theory about wave equations we presented in Section \ref{secWaveEquations}.
Then we are induced to the choice of an approach based on the vector
potential $\mathtt{A}$ and eq. \eqref{eqEMEquation} in place of
the field strength and the Maxwell equations, although the essential
physical observable in our description is still the field strength
$\mathtt{F}$ (not the vector potential $\mathtt{A}$), as it was
in the approach based on the Maxwell equations. Indeed we recover
the Maxwell equations simply defining $\mathtt{F}=\mathrm{d}\mathtt{A}$,
but we automatically exclude from our description all those field
strengths that are not closed. In conclusion we renounce to the description
of all the field strengths admitted by the Maxwell equations to obtain
an equation which seems to be more convenient. However, exactly as
in the case of the Proca field, $\mathrm{\delta d}$ is formally selfadjoint
linear differential operator of second order, but it fails to be normally
hyperbolic and hence we cannot automatically obtain advanced and retarded
Green operators on each globally hyperbolic spacetime. Moreover now
eq. \eqref{eqEMEquation} does not imply that $\mathrm{\delta}\mathtt{A}=0$
because of the absence of the mass term and hence the system
\begin{equation}
\left\{ \begin{array}{rcl}
\mathrm{\Box}_{1}\mathtt{A} & = & 0\mbox{,}\\
\mathrm{\delta}\mathtt{A} & = & 0
\end{array}\right.\label{eqEMSystem}
\end{equation}
is not equivalent to eq. \eqref{eqEMEquation}, although solutions
$\mathtt{A}$ of the system are solutions of eq. \eqref{eqEMEquation}
too. Then we cannot attempt a procedure similar to that followed for
the Proca field to show that the Green operators for $A$ are related
to those for $\mathrm{\Box}_{1}$.

\index{gauge equivalence}Luckily there is \textsl{gauge equivalence}
that comes to our aid. We said that the physical observable is the
field strength $\mathtt{F}=\mathrm{d}\mathtt{A}$. It may happen that
different vector potentials $\mathtt{A}$ and $\mathtt{A}^{\prime}$
satisfying eq. \eqref{eqEMEquation} generate the same field strength
$\mathtt{F}$, in which case they are said to be gauge equivalent.
Then from a physical point of view $\mathtt{A}$ and $\mathtt{A}^{\prime}$
are indistinguishable since they generate the same observable. Hence
we do not want to have in our classical theory of the electromagnetic
field both $\mathtt{A}$ and $\mathtt{A}^{\prime}$ as distinguished
dynamical configurations of the vector potential. The next lemma puts
together these facts showing that eq. \eqref{eqEMEquation} and eq.
\eqref{eqEMSystem} become equivalent when we identify gauge equivalent
configurations.
\begin{lem}
\label{lemGaugeEquivalenceAndSolutionsOfdDeltaAndBox}\index{Lorentz gauge}Let
$\mathscr{M}=\left(M,g,\mathfrak{o},\mathfrak{t}\right)$ be a globally
hyperbolic spacetime and consider $\mathtt{A}\in\mathrm{\Omega}^{1}M$.
Then the following conditions are equivalent:
\begin{itemize}
\item $\mathtt{A}$ satisfies the equation $\mathrm{\delta d}\mathtt{A}=0$;
\item there exists $\mathtt{A}^{\prime}\in\mathrm{\Omega}^{1}M$, which
is gauge equivalent to $\mathtt{A}$, i.e. $\mathrm{d}\left(\mathtt{A}^{\prime}-\mathtt{A}\right)=0$,
that satisfies the equation $\mathrm{\Box}_{1}\mathtt{A}^{\prime}=0$
and the \textsl{Lorentz gauge} condition $\mathrm{\delta}\mathtt{A}^{\prime}=0$.
\end{itemize}
Moreover consider the space $S_{1}$ of gauge inequivalent classes
of $1$-forms satisfying $\mathrm{\delta d}\mathtt{A}=0$,
\[
S_{1}=\frac{\left\{ \mathtt{A}\in\mathrm{\Omega}^{1}M:\,\mathrm{\delta d}\mathtt{A}=0\right\} }{\left\{ \mathtt{A}\in\mathrm{\Omega}^{1}M:\,\mathrm{d}\mathtt{A}=0\right\} }\mbox{,}
\]
and the space $S_{2}$ of gauge inequivalent classes of $1$-forms
satisfying $\mathrm{\Box}_{1}\mathtt{A}^{\prime}=0$ and the Lorentz
gauge condition,
\[
S_{2}=\frac{\left\{ \mathtt{A}^{\prime}\in\mathrm{\Omega}^{1}M:\,\mathrm{\Box}_{1}\mathtt{A}^{\prime}=0,\,\mathrm{\delta}\mathtt{A}^{\prime}=0\right\} }{\left\{ \mathtt{A}^{\prime}\in\mathrm{\Omega}^{1}M:\,\mathrm{d}\mathtt{A}^{\prime}=0,\,\mathrm{\delta}\mathtt{A}^{\prime}=0\right\} }\mbox{.}
\]
Then the map $I:S_{1}\rightarrow S_{2}$ defined by $I\left[\mathtt{A}\right]_{1}=\left[\mathtt{A}^{\prime}\right]_{2}$,
where $\mathtt{A}$ is a representative of the class $\left[\mathtt{A}\right]_{1}$,
$\mathtt{A}^{\prime}$, which is gauge equivalent to $\mathtt{A}$,
satisfies $\mathrm{\Box}_{1}\mathtt{A}^{\prime}=0$ and the Lorentz
gauge condition and $\left[\mathtt{A}^{\prime}\right]_{2}$ denotes
the class that has $\mathtt{A}^{\prime}$ as representative, is a
vector space isomorphism.\end{lem}
\begin{proof}
Fix $\mathtt{A}\in\mathrm{\Omega}^{1}M$. If we suppose that there
exists $\mathtt{A}^{\prime}\in\mathrm{\Omega}^{1}M$ such that
\[
\left\{ \begin{array}{rcl}
\mathrm{d}\left(\mathtt{A}^{\prime}-\mathtt{A}\right) & = & 0\mbox{,}\\
\mathrm{\Box}_{1}\mathtt{A}^{\prime} & = & 0\mbox{,}\\
\mathrm{\delta}\mathtt{A}^{\prime} & = & 0\mbox{,}
\end{array}\right.
\]
then we immediately deduce that
\[
\mathrm{\delta d}\mathtt{A}=\mathrm{\delta d}\mathtt{A}^{\prime}=\mathrm{\delta d}\mathtt{A}^{\prime}+\mathrm{d\delta}\mathtt{A}^{\prime}=\mathrm{\Box}_{1}\mathtt{A}^{\prime}=0\mbox{.}
\]
Conversely suppose that $\mathrm{\delta d}\mathtt{A}=0$. Consider
the equation $\mathrm{\Box}_{0}f=-\mathrm{\delta}\mathtt{A}$. In
\cite[Cor. 5, p. 78]{Gin09} we find a procedure that extends the
result of Theorem \ref{thmCauchyProblem} stating the existence and
uniqueness of the solution of a Cauchy problem for a normally hyperbolic
operator even when the initial data are not compactly supported. We
deduce that there exists $f\in\mathrm{\Omega}^{0}M$ satisfying $\mathrm{\Box}_{0}f=-\mathrm{\delta}\mathtt{A}$.
We set $\mathtt{A}^{\prime}=\mathtt{A}+\mathrm{d}f$ and we check
that $\mathtt{A}^{\prime}$ fulfils the requirements of the second
condition in the statement of the proposition. Indeed $\mathrm{d}\left(\mathtt{A}^{\prime}-\mathtt{A}\right)=\mathrm{d}\left(\mathrm{d}f\right)=0$.
Moreover, applying Proposition \ref{propBox}, we find 
\[
\mathrm{\Box}_{1}\mathtt{A}^{\prime}=\mathrm{\Box}_{1}\mathtt{A}+\mathrm{\Box}_{1}\mathrm{d}f=\mathrm{d\delta}\mathtt{A}+\mathrm{d}\mathrm{\Box}_{0}f=0\mbox{.}
\]
It remains to check only the Lorentz gauge condition:
\[
\mathrm{\delta}\mathtt{A}^{\prime}=\mathrm{\delta}\mathtt{A}+\mathrm{\delta d}f=\mathrm{\delta}\mathtt{A}+\mathrm{\Box}_{0}f=0\mbox{.}
\]

Now we turn our attention to the definition of $I$. Take $\left[\mathtt{A}\right]_{1}\in S_{1}$
and consider two representatives $\mathtt{A}$ and $\mathtt{B}$ of
$\left[\mathtt{A}\right]_{1}$. Then $\mathrm{\delta d}\mathtt{A}=0=\mathrm{\delta d}\mathtt{B}$
and, applying the first part of this lemma, we find $\mathtt{A}^{\prime}$
and $\mathtt{B}^{\prime}$ in $\mathrm{\Omega}^{1}M$ such that
\[
\left\{ \begin{array}{rcccl}
\mathrm{\Box}_{1}\mathtt{A}^{\prime} & = & 0 & = & \mathrm{\Box}_{1}\mathtt{B}^{\prime}\mbox{,}\\
\mathrm{\delta}\mathtt{A}^{\prime} & = & 0 & = & \mathrm{\delta}\mathtt{B}^{\prime}\mbox{,}\\
\mathrm{d}\left(\mathtt{A}^{\prime}-\mathtt{A}\right) & = & 0 & = & \mathrm{d}\left(\mathtt{B}^{\prime}-\mathtt{B}\right)\mbox{.}
\end{array}\right.
\]
In particular we deduce that
\[
\mathrm{d}\left(\mathtt{A}^{\prime}-\mathtt{B}^{\prime}\right)=\mathrm{d}\left(\mathtt{A}-\mathtt{B}\right)=0
\]
because $\mathtt{A}$ and $\mathtt{B}$ are gauge equivalent being
representatives of the same equivalence class of $S_{1}$. Moreover
trivially $\mathrm{\delta}\left(\mathtt{A}^{\prime}-\mathtt{B}^{\prime}\right)=0$.
This proves that $I$ is well defined. Linearity can be directly checked
from the definition of $I$. Consider now $\left[\mathtt{A}\right]_{1}$
such that $I\left[\mathtt{A}\right]_{1}=\left[0\right]_{2}$, where
$\left[0\right]_{2}$ denotes the class of $S_{2}$ that has the null
section as representative (this is actually the zero element of the
vector space $S_{2}$). This means that each representative $\mathtt{A}$
of the class $\left[\mathtt{A}\right]_{1}$ is gauge equivalent to
each representative $\mathtt{A}^{\prime}$ of the class $\left[0\right]_{2}$.
In particular we choose the null section 0 as representative of $\left[0\right]_{2}$
and we deduce that each representative $\mathtt{A}$ of the class
$\left[\mathtt{A}\right]_{1}$ is such that $\mathrm{d}\mathtt{A}=0$,
i.e. $\left[\mathtt{A}\right]_{1}$ is the zero element of the vector
space $S_{1}$ (we may write $\left[\mathtt{A}\right]_{1}=\left[0\right]_{1}$).
Then we conclude that $I$ is injective. To conclude the proof take
$\left[\mathtt{A}^{\prime}\right]_{2}\in S_{2}$. We look for $\left[\mathtt{A}\right]_{1}\in S_{1}$
such that $I\left[\mathtt{A}\right]_{1}=\left[\mathtt{A}^{\prime}\right]_{2}$.
Take a representative $\mathtt{A}\in\left[\mathtt{A}^{\prime}\right]_{2}$.
This in particular verifies $\mathrm{\delta d}\mathtt{A}=0$ and therefore
we can consider the class $\left[\mathtt{A}\right]_{1}$ that has
$\mathtt{A}$ as representative. Applying the definition of $I$,
we see that $I\left[\mathtt{A}\right]_{1}=\left[\mathtt{A}\right]_{2}=\left[\mathtt{A}^{\prime}\right]_{2}$.
This shows that $I$ is also surjective and hence it is a vector space
isomorphism as expected.
\end{proof}
The last theorem gives us the opportunity to identify $S_{1}$ with
$S_{2}$. This means that we can equivalently consider gauge inequivalent
classes of 1-forms satisfying $\mathrm{\delta d\mathtt{A}}=0$ or
gauge inequivalent classes of vector potentials satisfying both $\mathrm{\Box}_{1}\mathtt{A}=0$
and $\mathrm{\delta}\mathtt{A}=0$ as classical observables of the
electromagnetic field.

For the construction of a covariant functor describing the classical
theory of the electromagnetic field, we need to determine a symplectic
space comprised by all the gauge inequivalent classes of solutions
for homogeneous Cauchy problems with compactly supported initial data
associated to the operator $\mathrm{\delta d}$. This must be done
for each globally hyperbolic spacetime. Unfortunately the lack of
normal hyperbolicity and the presence gauge invariance significantly
alter the situation of Subsection \ref{subClassicalFieldTheory} so
that we are forced to start the construction of the classical theory
from the beginning.

\index{Lorentz 1-form}In first place we try to determine the vector
space on which we will define a symplectic form. The solution of this
problem is suggested by \cite[Prop. 4, p. 228]{Dim92}. Note that
from now on we say that $\mathtt{A}$ is a \textsl{Lorentz 1-form}
if it is a 1-form satisfying the Lorentz gauge condition, i.e. $\mathrm{\delta}\mathtt{A}=0$.
\begin{lem}
\label{lemSpaceOfGaugeInequivalentDynamicalConfigurationsOfTheEMField}Let
$\mathscr{M}=\left(M,g,\mathfrak{o},\mathfrak{t}\right)$ be a globally
hyperbolic spacetime and define the space of compactly supported coclosed
$1$-forms over $M$:
\[
\mathrm{\Omega}_{0,\mathrm{\delta}}^{1}M=\left\{ \theta\in\mathrm{\Omega}_{0}^{1}M:\,\mathrm{\delta}\theta=0\right\} \mbox{.}
\]
The Lorentz solution of a homogeneous Cauchy problem for the normally
hyperbolic operator $\mathrm{\Box}_{1}$ with compactly supported
initial data is gauge equivalent to $e\theta$, where $e$ is the
causal propagator for the normally hyperbolic operator $\mathrm{\Box}_{1}$
and $\theta$ is some element of $\mathrm{\Omega}_{0,\mathrm{\delta}}^{1}M$.
Conversely, for each $\theta\in\mathrm{\Omega}_{0,\mathrm{\delta}}^{1}M$,
$e\theta$ is a Lorentz solution of a homogeneous Cauchy problem for
the normally hyperbolic operator $\mathrm{\Box}_{1}$ with compactly
supported initial data.

It follows immediately that the space of gauge inequivalent classes
of Lorentz solutions of homogeneous Cauchy problems for the normally
hyperbolic operator $\mathrm{\Box}_{1}$ with compactly supported
initial data coincides with the following subset of $S_{2}$ (for
the definition of $S_{2}$ refer to Lemma \ref{lemGaugeEquivalenceAndSolutionsOfdDeltaAndBox}):
\[
V=\left\{ \left[e\theta\right]_{2}:\,\theta\in\mathrm{\Omega}_{0,\mathrm{\delta}}^{1}M\right\} \subseteq S_{2}\mbox{,}
\]
where $\left[e\theta\right]_{2}$ denotes the class of $S_{2}$ that
has $e\theta$ among its representatives.\end{lem}
\begin{proof}
Consider a Lorentz solution $\mathtt{A}\in\mathrm{\Omega}^{1}M$ of
a homogeneous Cauchy problem for the normally hyperbolic operator
$\mathrm{\Box}_{1}$ with compactly supported initial data on a given
Cauchy surface $\Sigma$ for $M$. Then we find a compact subset $K$
of $M$ including the support of the initial data for the Cauchy problem
and we take a relatively compact open subset $O$ of $M$ including
$K$. We deduce that
\[
\left\{ J_{+}^{\mathscr{M}}\left(O\right),J_{-}^{\mathscr{M}}\left(O\right),M\setminus J^{\mathscr{M}}\left(K\right)\right\} 
\]
is an open covering of $M$ because $J_{\pm}^{\mathscr{M}}\left(O\right)$
are open subsets of $M$ (see \cite[Lem. A.8, p. 48]{FV11}) and $J_{\pm}^{\mathscr{M}}\left(K\right)$
are closed subsets of $M$ (see \cite[Lem. A.5.1, p. 173]{BGP07})
and we can introduce a partition of unity subordinate to such covering:
\[
\left\{ \chi^{+},\chi^{-},\chi^{0}\right\} \mbox{.}
\]
Defining $\mathtt{A}^{\pm}=\chi^{\pm}\mathtt{A}$ and $\mathtt{A}^{0}=\chi^{0}\mathtt{A}$,
we see that $\mathtt{A}=\mathtt{A}^{+}+\mathtt{A}^{-}+\mathtt{A}^{0}$.
But $K$ includes the support of the initial data for the solution
$\mathtt{A}$ so that $\mathrm{supp}\left(\mathtt{A}\right)\subseteq J^{\mathscr{M}}\left(K\right)$
(this is a consequence of Theorem \ref{thmCauchyProblem}, which can
be applied because $\mathrm{\Box}_{1}$ is normally hyperbolic) and
hence
\[
\mathrm{supp}\left(\mathtt{A}^{0}\right)=\mathrm{supp}\left(\chi^{0}\right)\cap\mathrm{supp}\left(\mathtt{A}\right)\subseteq\left(M\setminus J^{\mathscr{M}}\left(K\right)\right)\cap J^{\mathscr{M}}\left(K\right)=\emptyset\mbox{.}
\]
This means that $\mathtt{A}^{0}=0$ and so $\mathtt{A}=\mathtt{A}^{+}+\mathtt{A}^{-}$.
From $\mathrm{\Box}_{1}\mathtt{A}=0$ and $\mathrm{\delta}\mathtt{A}=0$
we deduce that $\mathrm{\Box}_{1}\mathtt{A}^{+}=-\mathrm{\Box}_{1}\mathtt{A}^{-}$
and $\mathrm{\delta}\mathtt{A}^{+}=-\mathrm{\delta}\mathtt{A}^{-}$.
The first of these identities implies that $\mathrm{\Box}_{1}\mathtt{A}^{+}$
has compact support because we can apply Proposition \ref{propUsefulSubsetsOfGloballyHyperbolicSpacetimes}
to
\[
\mathrm{supp}\left(\mathrm{\delta d}\mathtt{A}^{+}\right)\subseteq\mathrm{supp}\left(\chi^{+}\right)\cap\mathrm{supp}\left(\chi^{-}\right)\subseteq J_{+}^{\mathscr{M}}\left(\overline{O}\right)\cap J_{-}^{\mathscr{M}}\left(\overline{O}\right)
\]
noting that $\overline{O}$ is a compact subset of $M$ since by construction
$O$ is a relatively compact subset of $M$. A similar procedure shows
also that $\mathrm{\delta}\mathtt{A}^{+}$ has compact support. Then,
considering $\theta=\mathrm{\delta d}\mathtt{A}^{+}$, we have a compactly
supported 1-form that trivially satisfies $\mathrm{\delta}\theta=0$.
We must only check that $e\theta$ is gauge equivalent to $\mathtt{A}$.
Applying Lemma \ref{lemGreenOperatorsCommuteWithExteriorDerivativeAndCodifferential},
we see that $\mathrm{d}\left(e^{a}\theta\right)=e^{a}\left(\mathrm{d}\theta\right)$.
Evaluating $\mathrm{d}\theta$ and keeping in mind that $\mathrm{d}^{2}=0$,
we obtain
\[
\mathrm{d}\theta=\mathrm{d\delta d}\mathtt{A}^{+}=\mathrm{\Box}_{1}\mathrm{d}\mathtt{A}^{+}\mbox{.}
\]
Proposition \ref{propUsefulSubsetsOfGloballyHyperbolicSpacetimes}
implies that $\mathtt{A}^{+}$ has past compact support so that we
can exploit Lemma \ref{lemExtensionOf ea(Pu)=00003Du} to obtain
\[
\mathrm{d}\left(e^{a}\theta\right)=e^{a}\mathrm{\Box}_{1}\mathrm{d}\mathtt{A}^{+}=\mathrm{d}\mathtt{A}^{+}\mbox{.}
\]
A similar procedure shows that $\mathrm{d}\left(e^{r}\theta\right)=-\mathrm{d}\mathtt{A}^{-}$
and therefore we conclude $\mathrm{d}\left(e\theta\right)=\mathrm{d}\mathtt{A}$,
which means exactly that $e\theta$ is gauge equivalent to $\mathtt{A}$.

Now take $\theta\in\mathrm{\Omega}_{0,\mathrm{\delta}}^{1}M$ and
consider $e\theta$. Trivially $\mathrm{\Box}_{1}\left(e\theta\right)=0$
and by Lemma \ref{lemGreenOperatorsCommuteWithExteriorDerivativeAndCodifferential}
we see that $\mathrm{\delta}\left(e\theta\right)=e\left(\mathrm{\delta}\theta\right)=0$
($e\theta$ is a Lorentz 1-form). To see that $e\theta$ is also a
solution of a homogeneous Cauchy problem for the normally hyperbolic
operator $\mathrm{\Box}_{1}$ with compactly supported initial data,
we take a spacelike smooth Cauchy surface $\Sigma$ for $\mathscr{M}$
(the existence is assured by Theorem \ref{thmGlobalHyperbolicity})
and we define on it a $\mathfrak{t}$-future directed $g$-timelike
unit vector field $\mathfrak{n}$ over $\Sigma$ normal to $\Sigma$.
Then we take $\alpha_{0}$ as the restriction of $e\theta$ to $\Sigma$
and $\alpha_{1}$ as the restriction of $\nabla_{n}\left(e\theta\right)$
to $\Sigma$, where $\nabla$denotes the Levi-Civita connection. $\alpha_{0}$
and $\alpha_{1}$ are indeed sections in the restriction of $\mathrm{\Lambda}^{1}M$
to $\Sigma$ and their supports are compact because we know that $\theta$
has compact support and we can apply Proposition \ref{propUsefulSubsetsOfGloballyHyperbolicSpacetimes}
to 
\[
\mathrm{supp}\left(e\theta\right)\cap\Sigma\subseteq J^{\mathscr{M}}\left(\mathrm{supp}\left(\theta\right)\right)\cap\Sigma\mbox{.}
\]
Then we can consider the following Cauchy problem:
\[
\left\{ \begin{array}{rcl}
\mathrm{\Box}_{1}\mathtt{A} & = & 0\mbox{,}\\
\left.\mathtt{A}\right|_{\Sigma} & = & \alpha_{0}\mbox{,}\\
\left.\nabla_{\mathfrak{n}}\mathtt{A}\right|_{\Sigma} & = & \alpha_{1}\mbox{.}
\end{array}\right.
\]
By construction $e\theta$ is a solution (actually the unique solution
due to Theorem \ref{thmCauchyProblem}, which holds because $\mathrm{\Box}_{1}$
is normally hyperbolic). Since we have shown at the beginning of the
proof that $\mathrm{\delta}\left(e\theta\right)=0$, we conclude that
$e\theta$ is a Lorentz solution of a homogeneous Cauchy problem for
the normally hyperbolic operator $\mathrm{\Box}_{1}$ with compactly
supported initial data. This completes the proof.
\end{proof}
We have a vector space $V$. Now we need a symplectic form on it.
A new difficulty associated to the first de Rham cohomology group
of $M$ arises in this situation as we will see in the proof of the
next Lemma. To go around this obstacle we assume that $H^{1}\left(M\right)=\left\{ 0\right\} $
following the approach of \cite{Dap11}.
\begin{lem}
\label{lemSymplecticFormForEMField}Let $\mathscr{M}=\left(M,g,\mathfrak{o},\mathfrak{t}\right)$
be a globally hyperbolic spacetime such that $H^{1}\left(M\right)=\left\{ 0\right\} $
and consider the vector space $V$ defined in Lemma \ref{lemSpaceOfGaugeInequivalentDynamicalConfigurationsOfTheEMField}.
The map
\begin{alignat*}{2}
\sigma & : & V\times V & \rightarrow\mathbb{R}\\
 &  & \left(\left[\mathtt{A}\right]_{2},\left[\mathtt{B}\right]_{2}\right) & \mapsto\left(e\theta,\zeta\right)_{g,1}\mbox{,}
\end{alignat*}
where $e$ is the causal propagator for the formally selfadjoint normally
hyperbolic operator $\mathrm{\Box}_{1}$, $\theta$ and $\zeta$ are
elements of $\mathrm{\Omega}_{0,\mathrm{\delta}}^{1}M$ such that
$e\theta$ and $e\zeta$ are representatives of $\left[\mathtt{A}\right]_{2}$
and respectively $\left[\mathtt{B}\right]_{2}$ and $\left(\cdot,\cdot\right)_{g,1}$
is defined in Proposition \ref{propInnerProductOnkForms}, is well
defined, bilinear, antisymmetric and non degenerate, i.e. it is a
symplectic map on $V$. Hence $\left(V,\sigma\right)$ is a symplectic
space.\end{lem}
\begin{proof}
To show that $\sigma$ is well defined, take $\left[\mathtt{A}\right]_{2}$
and $\left[\mathtt{B}\right]_{2}$ in $V$. Because of the definition
of $V$ we find $\theta$ and $\zeta$ in $\mathrm{\Omega}_{0,\mathrm{\delta}}^{1}M$
such that $e\theta\in\left[\mathtt{A}\right]_{2}$ and $e\zeta\in\left[\mathtt{B}\right]_{2}$.
Since $\zeta$ is compactly supported, we can evaluate $\left(e\theta,\zeta\right)_{g,1}$
and indeed we get a real number. If we consider also $\theta^{\prime}$
and $\zeta^{\prime}$ in $\mathrm{\Omega}_{0,\mathrm{\delta}}^{1}M$
such that $e\theta^{\prime}\in\left[\mathtt{A}\right]_{2}$ and $e\zeta^{\prime}\in\left[\mathtt{B}\right]_{2}$,
we have $\left(e\theta^{\prime},\zeta^{\prime}\right)_{g,1}$ and
we must check that it coincides with $\left(e\theta,\zeta\right)_{g,1}$
in order to have $\sigma$ well defined. Since $e\theta$ and $e\theta^{\prime}$
are both representatives of $\left[\mathtt{A}\right]_{2}$, they are
gauge equivalent, i.e. $\mathrm{d}\left(e\theta-e\theta^{\prime}\right)=0$.
Now the hypothesis $H^{1}\left(M\right)=\left\{ 0\right\} $ comes
into play because it implies that we can find $\alpha\in\mathrm{\Omega}^{0}M$
such that $e\theta-e\theta^{\prime}=\mathrm{d}\alpha$. Similarly
we find $\beta\in\mathrm{\Omega}^{0}M$ such that $e\zeta-e\zeta^{\prime}=\mathrm{d}\beta$.
bearing in mind that $e$ is formally antiselfadjoint (because $\mathrm{\Box}_{1}$
is formally selfadjoint, cfr. Proposition \ref{prope*rIsFormallyAdjointToea})
and that $\mathrm{d}$ and $\mathrm{\delta}$ are formal adjoints
of each other, we can tackle the evaluation of $\left(e\theta^{\prime},\zeta^{\prime}\right)_{g,1}$:
\begin{alignat*}{2}
\left(e\theta^{\prime},\zeta^{\prime}\right)_{g,1} & = & \left(e\theta,\zeta^{\prime}\right)_{g,1} & -\left(\mathrm{d}\alpha,\zeta^{\prime}\right)_{g,1}\\
 & = & -\left(\theta,e\zeta^{\prime}\right)_{g,1} & -\left(\alpha,\delta\zeta^{\prime}\right)_{0,g}\\
 & = & -\left(\theta,e\zeta\right)_{g,1} & +\left(\theta,\mathrm{d}\beta\right)_{g,1}\\
 & = & \left(e\theta,\zeta\right)_{g,1} & +\left(\mathrm{\delta}\theta,\beta\right)_{g,0}\\
 & = & \left(e\theta,\zeta\right)_{g,1} & \mbox{,}
\end{alignat*}
where we exploited $\mathrm{\delta}\zeta^{\prime}=0$ and $\mathrm{\delta}\beta=0$.
This shows that $\sigma$ is well defined. Notice that without the
hypothesis $H^{1}\left(M\right)=\left\{ 0\right\} $ this proof does
not work.

Bilinearity of $\sigma$ easily follows from bilinearity of $\left(\cdot,\cdot\right)_{g,1}$
and linearity of $e$. As for antisymmetry, consider $\left[\mathtt{A}\right]_{2}$
and $\left[\mathtt{B}\right]_{2}$ in $V$. By definition of $V$
we find $\theta$ and $\zeta$ in $\mathrm{\Omega}_{0,\mathrm{\delta}}^{1}M$
such that $e\theta\in\left[\mathtt{A}\right]_{2}$ and $e\zeta\in\left[\mathtt{B}\right]_{2}$.
Exploiting the definition of $\sigma$, the antiselfadjointness of
$e$ and the symmetry of $\left(\cdot,\cdot\right)_{g,1}$, we obtain
\[
\sigma\left(\left[\mathtt{A}\right]_{2},\left[\mathtt{B}\right]_{2}\right)=\left(e\theta,\zeta\right)_{g,1}=-\left(\theta,e\zeta\right)_{g,1}=-\left(e\zeta,\theta\right)_{g,1}=-\sigma\left(\left[\mathtt{B}\right]_{2},\left[\mathtt{A}\right]_{2}\right)\mbox{.}
\]
It remains only to check that $\sigma$ is non degenerate. To this
end consider $\left[\mathtt{A}\right]_{2}\in V$ such that $\sigma\left(\left[\mathtt{A}\right]_{2},\left[\mathtt{B}\right]_{2}\right)=0$
for each $\left[\mathtt{B}\right]_{2}\in V$. Taking $\theta\in\mathrm{\Omega}_{0,\mathrm{\delta}}^{1}M$
such that $e\theta\in\left[\mathtt{A}\right]_{2}$, we deduce that
$\left(e\theta,\zeta\right)_{g,1}=0$ for each $\zeta\in\mathrm{\Omega}_{0,\mathrm{\delta}}^{1}M$.
In particular we have $\left(e\theta,\mathrm{\delta}\alpha\right)_{g,1}=0$
for each $\alpha\in\mathrm{\Omega}_{0}^{2}M$ and hence $\left(\mathrm{d}\left(e\theta\right),\alpha\right)_{g,2}=0$
for each $\alpha\in\mathrm{\Omega}_{0}^{2}M$. Since $\left(\cdot,\cdot\right)_{g,2}$
is non degenerate, we conclude that $\mathrm{d}\left(e\theta\right)=0$.
This fact means that $e\theta$ is a representative of the zero class
of $V$, i.e. $\left[A\right]_{2}=\left[0\right]_{2}$.
\end{proof}
At this point we are able to associate a symplectic space $\left(V,\sigma\right)$
comprised by all the gauge inequivalent classes of dynamical configuration
for the electromagnetic field on each globally hyperbolic spacetime
$\mathscr{M}=\left(M,g,\mathfrak{o},\mathfrak{t}\right)$ such that
the first de Rham cohomology group of $M$ is trivial.

This fact induces us to introduce of a special category $\mathfrak{ghs}^{EM}$
for the electromagnetic field.
\begin{defn}
\label{defghsEM}We define the category $\mathfrak{ghs}^{EM}$ in
the following way:
\begin{itemize}
\item objects are triples $\left(\mathscr{M},\mathrm{\Lambda}^{1}M,\mathrm{\delta d}\right)$,
where $\mathscr{M}=\left(M,g,\mathfrak{o},\mathfrak{t}\right)$ is
a globally hyperbolic spacetime with $H^{1}\left(M\right)=\left\{ 0\right\} $,
$\mathrm{\Lambda}^{1}M$ is the vector bundle over $M$ that we consider,
$\left\langle \cdot,\cdot\right\rangle _{g,1}$ is the inner product
on $\mathrm{\Lambda}^{1}M$ induced by the metric $g$ (refer to Proposition
\ref{propHodgeDualProperties} for a characterization of this inner
product) and $\mathrm{\delta d}$ is the linear differential operator
on $\mathrm{\Lambda}^{1}M$ over $\mathscr{M}$ governing the dynamics
of the electromagnetic field;
\item morphisms from $\left(\mathscr{M},\mathrm{\Lambda}^{1}M,\mathrm{\delta d}\right)$
to $\left(\mathscr{N},\mathrm{\Lambda}^{1}N,\mathrm{\delta d}\right)$
are vector bundle homomorphisms of the form $\left(\psi,\psi_{*}\right)$
from $\mathrm{\Lambda}^{1}M$ to $\mathrm{\Lambda}^{1}N$ such that
$\psi$ is a morphism of $\mathfrak{ghs}$ (note that $H^{1}\left(M\right)=\left\{ 0\right\} $
entails $H^{1}\left(\psi\left(M\right)\right)=\left\{ 0\right\} $);%
\footnote{\label{fnMorghsEM}For the electromagnetic field we are forced to
restrict our class of morphisms to that usually considered, i.e. only
vector bundle homomorphisms that are push-forwards of morphisms of
$\mathfrak{ghs}$. For the precise definition of these morphisms and
some comments refer to Remark \ref{remMorphismsOfConcreteFields}.
This choice is done because push-forwards (and similarly pull-backs)
have the property of being defined on $k$-forms for arbitrary $k$
and moreover they intertwine with both $\mathrm{d}$ and $\mathrm{\delta}$
(see Proposition \ref{propExteriorDerivative} and comments after
Definition \ref{defCodifferential}).%
}
\item the composition law is simply the composition of functions.
\end{itemize}
\end{defn}
This is a specialization of the category $\mathfrak{ghs}^{f}$, actually
a subcategory (but not a full subcategory because we consider only
push-forwards). This statement is not at all correct because here
$\mathrm{\delta d}$ is not normally hyperbolic, but it becomes rigorous
if we replace $\mathrm{\delta d}$ with the normally hyperbolic operator
$\mathrm{\Box}_{1}$. Then all the observations referred to $\mathfrak{ghs}^{f}$
hold also for $\mathfrak{ghs}^{EM}$.

Applying Lemma \ref{lemSpaceOfGaugeInequivalentDynamicalConfigurationsOfTheEMField}
and Lemma \ref{lemSymplecticFormForEMField}, we can define the map
\begin{alignat*}{2}
\mathscr{B} & : & \mathsf{Obj}_{\mathfrak{ghs}^{EM}} & \rightarrow\mathsf{Obj}_{\mathfrak{ssp}}\\
 &  & \left(\mathscr{M},\mathrm{\Lambda}^{1}M,\mathrm{\delta d}\right) & \mapsto\left(V,\sigma\right)\mbox{.}
\end{alignat*}
This is the first part of our candidate functor describing the classical
theory of the electromagnetic field. The second part comes from the
next lemma.

Before presenting the statement, we introduce some notation. From
now on the vector space that was denoted by $S_{2}$ in Lemma \ref{lemGaugeEquivalenceAndSolutionsOfdDeltaAndBox}
will be denoted by $S_{M}$ to keep trace of the manifold we are working
on. Similarly the equivalence class previously indicated with $\left[\cdot\right]_{2}$
will be denoted by $\left[\cdot\right]_{M}$.
\begin{lem}
\label{lemMorghsEM->Morssp}Let $\left(\psi,\psi_{*}\right)$ be a
morphism of $\mathfrak{ghs}^{EM}$ from the object $\left(\mathscr{M},\mathrm{\Lambda}^{1}M,\mathrm{\delta d}\right)$
to the object $\left(\mathscr{N},\mathrm{\Lambda}^{1}N,\mathrm{\delta d}\right)$.
Denote with $\left(V,\sigma\right)$ and $\left(W,\omega\right)$
the symplectic spaces associated to $\left(\mathscr{M},\mathrm{\Lambda}^{1}M,\mathrm{\delta d}\right)$
and respectively $\left(\mathscr{N},\mathrm{\Lambda}^{1}N,\mathrm{\delta d}\right)$
by the map $\mathscr{B}$ defined few lines above. Then the map
\begin{alignat*}{2}
\xi & : & V & \rightarrow W\\
 &  & \left[\mathtt{A}\right]_{M} & \mapsto\left[e_{N}\left(\mathrm{ext}_{\psi_{*}}\theta\right)\right]_{N}\mbox{,}
\end{alignat*}
where $e_{M}$ and $e_{N}$ are the causal propagators for the formally
selfadjoint normally hyperbolic operator $\mathrm{\Box}_{1}$ on $\mathrm{\Lambda}^{1}M$
over $\mathscr{M}$ and respectively on $\mathrm{\Lambda}^{1}N$ over
$\mathscr{N}$ and $\theta\in\mathrm{\Omega}_{0,\mathrm{\delta}}^{1}M$
is such that $e_{M}\theta\in\left[\mathtt{A}\right]_{M}$, is well
defined, linear and compatible with the symplectic forms $\sigma$
and $\omega$, that is to say that $\xi$ is a symplectic map from
$\left(V,\sigma\right)$ to $\left(W,\omega\right)$.\end{lem}
\begin{proof}
The first step of this proof is devoted to show that $\xi$ is well
defined. To this end take $\left[\mathtt{A}\right]_{M}\in V$. By
definition we find $\theta\in\mathrm{\Omega}_{0,\mathrm{\delta}}^{1}M$
such that $e_{M}\theta\in\left[\mathtt{A}\right]_{M}$. It follows
that $\mathrm{\delta}\left(\mathrm{ext}_{\psi_{*}}\theta\right)=0$
because 
\[
\left(\mathrm{ext}_{\psi_{*}}\theta\right)\left(q\right)=\begin{cases}
\left(\psi_{*}^{\prime}\theta\right)\left(q\right) & \mbox{if }q\in\psi\left(M\right)\mbox{,}\\
0 & \mbox{if }q\in N\setminus\psi\left(M\right)
\end{cases}
\]
and $\delta\circ\psi_{*}^{\prime}=\psi_{*}^{\prime}\circ\delta$.
This implies that $\left[e_{N}\left(\mathrm{ext}_{\psi_{*}}\theta\right)\right]_{N}$
is indeed an element of $W$. Suppose now that also $\zeta\in\mathrm{\Omega}_{0,\mathrm{\delta}}^{1}M$
is such that $e_{M}\zeta$ is a representative of $\left[\mathtt{A}\right]_{M}$.
Then we also have $\left[e_{N}\left(\mathrm{ext}_{\psi_{*}}\zeta\right)\right]_{N}$,
and we must prove that this is equal to $\left[e_{N}\left(\mathrm{ext}_{\psi_{*}}\theta\right)\right]_{N}$
for $\xi$ to be well defined. We know that $e_{M}\theta$ and $e_{M}\zeta$
are gauge equivalent, i.e. $\mathrm{d}\left(e_{M}\theta-e_{M}\zeta\right)=0$.
Exploiting Lemma \ref{lemGreenOperatorsCommuteWithExteriorDerivativeAndCodifferential},
we deduce that $\mathrm{d}\left(\theta-\zeta\right)$ falls in the
kernel of $e_{M}:\mathrm{\Omega}_{0}^{2}M\rightarrow\mathrm{\Omega}^{2}M$,
which is the causal propagator for the normally hyperbolic operator
$\mathrm{\Box}_{2}$. Applying Proposition \ref{propCausalPropagatorKernel}
to $e_{M}:\mathrm{\Omega}_{0}^{2}M\rightarrow\mathrm{\Omega}^{2}M$,
we find $\eta\in\mathrm{\Omega}_{0}^{2}M$ such that $\mathrm{\Box}_{2}\eta=\mathrm{d}\left(\theta-\zeta\right)$.
We have already seen one of the advantages of dealing with push-forwards
of isometric embeddings, that is $\delta\circ\psi_{*}^{\prime}=\psi_{*}^{\prime}\circ\delta$.
Besides this there are also the identity $\mathrm{d}\circ\psi_{*}^{\prime}=\psi_{*}^{\prime}\circ\mathrm{d}$
and, above all, the possibility to give sense to $\mathrm{ext}_{\psi_{*}}$
also for $k$-forms with $k\neq1$. From these observations it follows
that
\[
\mathrm{d}\left(\mathrm{ext}_{\psi_{*}}\left(\theta-\zeta\right)\right)=\mathrm{ext}_{\psi_{*}}\left(\mathrm{d}\left(\theta-\zeta\right)\right)=\mathrm{ext}_{\psi_{*}}\left(\mathrm{\Box}_{2}\eta\right)=\mathrm{\Box}_{2}\left(\mathrm{ext}_{\psi_{*}}\eta\right)\mbox{.}
\]
Exploiting Lemma \ref{lemGreenOperatorsCommuteWithExteriorDerivativeAndCodifferential},
we deduce that
\[
\mathrm{d}\left(e_{N}\left(\mathrm{ext}_{\psi_{*}}\left(\theta-\zeta\right)\right)\right)=e_{N}\left(\mathrm{d}\left(\mathrm{ext}_{\psi_{*}}\left(\theta-\zeta\right)\right)\right)=e_{N}\left(\mathrm{\Box}_{2}\left(\mathrm{ext}_{\psi_{*}}\eta\right)\right)=0
\]
because $\mathrm{ext}_{\psi_{*}}\eta$ has compact support as $\eta$.

Linearity is a direct consequence of the definition of $\xi$. We
focus on the compatibility with the symplectic forms $\sigma$ and
$\omega$. To this end we consider $\left[\mathtt{A}\right]_{M}$
and $\left[\mathtt{B}\right]_{M}$ in $V$. Then we find $\theta$
and $\zeta$ in $\mathrm{\Omega}_{0,\mathrm{\delta}}^{1}M$ such that
$e_{M}\theta\in\left[\mathtt{A}\right]_{M}$ and $e_{M}\zeta\in\left[\mathtt{B}\right]_{M}$.
We are ready to evaluate $\omega\left(\xi\left[\mathtt{A}\right]_{M},\xi\left[\mathtt{B}\right]_{M}\right)$:
\begin{eqnarray*}
\omega\left(\xi\left[\mathtt{A}\right]_{M},\xi\left[\mathtt{B}\right]_{M}\right) & = & \omega\left(\left[e_{N}\left(\mathrm{ext}_{\psi_{*}}\theta\right)\right]_{N},\left[e_{N}\left(\mathrm{ext}_{\psi_{*}}\zeta\right)\right]_{N}\right)\\
 & = & \left(e_{N}\left(\mathrm{ext}_{\psi_{*}}\theta\right),\mathrm{ext}_{\psi_{*}}\zeta\right)_{h,1}\\
 & = & \left(\left(\mathrm{res}_{\psi_{*}}\circ e_{N}\circ\mathrm{ext}_{\psi_{*}}\right)\theta,\zeta\right)_{1,g}\\
 & = & \left(e_{M}\theta,\zeta\right)_{1,g}\\
 & = & \sigma\left(\left[\mathtt{A}\right]_{M},\left[\mathtt{B}\right]_{M}\right)\mbox{,}
\end{eqnarray*}
where we used also the identity $\mathrm{res}_{\psi_{*}}\circ e_{N}\circ\mathrm{ext}_{\psi_{*}}=e_{M}$
(cfr. Lemma \ref{lemresPsieBextPsi=00003DeA}).
\end{proof}
Now we have the second part of our candidate covariant functor. For
each pair of objects $\left(\mathscr{M},\mathrm{\Lambda}^{1}M,\mathrm{\delta d}\right)$
and $\left(\mathscr{N},\mathrm{\Lambda}^{1}N,\mathrm{\delta d}\right)$
of $\mathfrak{ghs}^{EM}$ there exists a map
\begin{eqnarray*}
\mathscr{B}:\mathsf{Mor}_{\mathfrak{ghs}^{EM}}\left(\left(\mathscr{M},\mathrm{\Lambda}^{1}M,\mathrm{\delta d}\right),\left(\mathscr{N},\mathrm{\Lambda}^{1}N,\mathrm{\delta d}\right)\right) & \rightarrow & \mathsf{Mor}_{\mathfrak{ssp}}\left(\left(V,\sigma\right),\left(W,\omega\right)\right)\\
\left(\psi,\psi_{*}\right) & \mapsto & \xi
\end{eqnarray*}
defined in accordance with our last lemma, where $\left(V,\sigma\right)$
and $\left(W,\omega\right)$ respectively denote the symplectic spaces
$\mathscr{B}\left(\mathscr{M},\mathrm{\Lambda}^{1}M,\mathrm{\delta d}\right)$
and $\mathscr{B}\left(\mathscr{N},\mathrm{\Lambda}^{1}N,\mathrm{\delta d}\right)$.
To complete the classical theory of the electromagnetic field, it
remains only to check that $\mathscr{B}$ is actually a covariant
functor. The next theorem answers to this question and provides also
the causality property and the time slice axiom for $\mathscr{B}$.
\begin{thm}
\label{thmClassicalFieldFunctorEMField}Consider the map
\begin{alignat*}{2}
\mathscr{B} & : & \mathsf{Obj}_{\mathfrak{ghs}^{EM}} & \rightarrow\mathsf{Obj}_{\mathfrak{ghs}^{EM}}\\
 &  & \left(\mathscr{M},\mathrm{\Lambda}^{1}M,\mathrm{\delta d}\right) & \mapsto\left(V,\sigma\right)
\end{alignat*}
defined in accordance with Lemma \ref{lemSpaceOfGaugeInequivalentDynamicalConfigurationsOfTheEMField}
and Lemma \ref{lemSymplecticFormForEMField} and for each pair of
objects $\left(\mathscr{M},\mathrm{\Lambda}^{1}M,\mathrm{\delta d}\right)$,
\textup{$\left(\mathscr{N},\mathrm{\Lambda}^{1}N,\mathrm{\delta d}\right)$
of $\mathfrak{ghs}^{EM}$} consider the map
\begin{eqnarray*}
\mathscr{B}:\mathsf{Mor}_{\mathfrak{ghs}^{EM}}\left(\left(\mathscr{M},\mathrm{\Lambda}^{1}M,\mathrm{\delta d}\right),\left(\mathscr{N},\mathrm{\Lambda}^{1}N,\mathrm{\delta d}\right)\right) & \rightarrow & \mathsf{Mor}_{\mathfrak{ssp}}\left(\left(V,\sigma\right),\left(W,\omega\right)\right)\\
\left(\psi,\psi_{*}\right) & \mapsto & \xi
\end{eqnarray*}
defined in accordance with Lemma \ref{lemMorghsEM->Morssp}, where
$\left(V,\sigma\right)$ and $\left(W,\omega\right)$ respectively
denote the symplectic spaces $\mathscr{B}\left(\mathscr{M},\mathrm{\Lambda}^{1}M,\mathrm{\delta d}\right)$
and $\mathscr{B}\left(\mathscr{N},\mathrm{\Lambda}^{1}N,\mathrm{\delta d}\right)$.
These maps give rise to a covariant functor $\mathscr{B}$ from the
category $\mathfrak{ghs}^{EM}$ to the category $\mathfrak{ssp}$.
Moreover $\mathscr{B}$ possesses the following properties:
\begin{itemize}
\item \textsl{causality}: for each $\left(\mathscr{M}_{1},\mathrm{\Lambda}^{1}M_{1},\mathrm{\delta d}\right)$,
$\left(\mathscr{M}_{2},\mathrm{\Lambda}^{1}M_{2},\mathrm{\delta d}\right)$,
$\left(\mathscr{M},\mathrm{\Lambda}^{1}M,\mathrm{\delta d}\right)$
in $\mathfrak{ghs}^{EM}$, each morphism $\left(\psi_{1},\psi_{1*}\right)$
from $\left(\mathscr{M}_{1},\mathrm{\Lambda}^{1}M_{1},\mathrm{\delta d}\right)$
to $\left(\mathscr{M},\mathrm{\Lambda}^{1}M,\mathrm{\delta d}\right)$
and each morphism $\left(\psi_{2},\psi_{2*}\right)$ from $\left(\mathscr{M}_{2},\mathrm{\Lambda}^{1}M_{2},\mathrm{\delta d}\right)$
to $\left(\mathscr{M},\mathrm{\Lambda}^{1}M,\mathrm{\delta d}\right)$
such that $\psi_{1}\left(M_{1}\right)$ and $\psi_{2}\left(M_{2}\right)$
are $\mathscr{M}$-causally separated subsets of $M$, it holds that
\[
\sigma\left(\xi_{1}\left[\mathtt{A}_{1}\right]_{M_{1}},\xi_{2}\left[\mathtt{A}_{2}\right]_{M_{2}}\right)=0
\]
for each $\left[\mathtt{A}_{1}\right]_{M_{1}}\in V_{1}$ and each
$\left[\mathtt{A}_{2}\right]_{M_{2}}\in V_{2}$, where $\left(V_{1},\sigma_{1}\right)$,
$\left(V_{2},\sigma_{2}\right)$ and $\left(V,\sigma\right)$ are
the symplectic spaces obtained with the application of $\mathscr{B}$
respectively to $\left(\mathscr{M}_{1},\mathrm{\Lambda}^{1}M_{1},\mathrm{\delta d}\right)$,
$\left(\mathscr{M}_{2},\mathrm{\Lambda}^{1}M_{2},\mathrm{\delta d}\right)$
and $\left(\mathscr{M},\mathrm{\Lambda}^{1}M,\mathrm{\delta d}\right)$,
while $\xi_{1}=\mathscr{B}\left(\psi_{1},\psi_{*1}\right)$ and $\xi_{2}=\mathscr{B}\left(\psi_{2},\psi_{*2}\right)$;
\item \textsl{time slice axiom}: for each $\left(\mathscr{M},\mathrm{\Lambda}^{1}M,\mathrm{\delta d}\right)$
and $\left(\mathscr{N},\mathrm{\Lambda}^{1}N,\mathrm{\delta d}\right)$
in $\mathsf{Obj}_{\mathfrak{ghs}^{EM}}$ and each morphism $\left(\psi,\psi_{*}\right)$
from $\left(\mathscr{M},\mathrm{\Lambda}^{1}M,\mathrm{\delta d}\right)$
to $\left(\mathscr{N},\mathrm{\Lambda}^{1}N,\mathrm{\delta d}\right)$
such that $\psi\left(M\right)$ includes a smooth spacelike Cauchy
surface $\Sigma$ for $\mathscr{N}$, it holds that
\[
\xi\left(V\right)=W\mbox{,}
\]
where $\left(V,\sigma\right)$ and $\left(W,\omega\right)$ are the
symplectic spaces obtained with the application of $\mathscr{B}$
respectively to $\left(\mathscr{M},\mathrm{\Lambda}^{1}M,\mathrm{\delta d}\right)$
and $\left(\mathscr{N},\mathrm{\Lambda}^{1}N,\mathrm{\delta d}\right)$,
while $\xi=\mathscr{B}\left(\psi,\psi_{*}\right)$. In particular
$\xi$ is bijective and its inverse $\xi^{-1}$ is a morphism of $\mathfrak{ssp}$
from $\left(W,\omega\right)$ to $\left(V,\sigma\right)$.
\end{itemize}
\end{thm}
\begin{proof}
Whenever it is possible, this proof imitates that of Theorem \ref{thmClassicalFieldFunctor},
the main difference being due to the presence of the equivalence classes.
Once that this fact is kept in mind, the verification of the covariant
axioms and of the causality property is identical.

We must still check the time slice axiom. Since $W$ is codomain of
$\xi$, the inclusion $\xi\left(V\right)\subseteq W$ is trivial and
we must prove the converse inclusion to complete the proof. To this
end consider $\left[\mathtt{A}\right]_{N}\in W$. We look for a section
$\theta\in\mathrm{\Omega}_{0,\mathrm{\delta}}^{1}M$ such that $e_{N}\left(\mathrm{ext}_{\psi_{*}}\theta\right)\in\left[\mathtt{A}\right]_{N}$.
We observe that $\left[\mathtt{A}\right]_{N}$ has a representative
of the form $e_{N}\zeta$ for $\zeta\in\mathrm{\Omega}_{0,\mathrm{\delta}}^{1}N$
that we denote with $\mathtt{A}$, hence, exploiting the support properties
of the Green operators and Proposition \ref{propUsefulSubsetsOfGloballyHyperbolicSpacetimes},
we deduce that $\mathrm{supp}\left(\mathtt{A}\right)\cap\Sigma$ is
a compact subset of $\Sigma$. Then we start with the usual procedure
(refer to the proof of Theorem \ref{thmClassicalFieldFunctor}) applied
to the normally hyperbolic operator $\mathrm{\Box}_{1}$. Remember
that now we have $\mathrm{\Box}_{1}\mathtt{A}=0$, but also $\mathrm{\delta}\mathtt{A}=0$
because we can exploit Lemma \ref{lemGreenOperatorsCommuteWithExteriorDerivativeAndCodifferential}.

This entails that we find a decomposition $\mathtt{A}=\mathtt{A}^{+}+\mathtt{A}^{-}$,
where $\mathtt{A}^{\pm}$ has $\mathscr{M}$-past/future compact support.
Moreover we have that $\mathrm{\Box}_{1}\mathtt{A}^{+}=-\mathrm{\Box}_{1}\mathtt{A}^{-}$
and $\mathrm{\delta}\mathtt{A}^{+}=-\mathrm{\delta}\mathtt{A}^{-}$
are elements of $\mathrm{\Omega}_{0}^{1}N$ with support included
in $\psi\left(M\right)$. We use them to define an element of $\mathrm{\Omega}_{0}^{1}M$
via restriction:
\[
\theta=\mathrm{res}_{\psi_{*}}\left(\mathrm{\delta d}\mathtt{A}^{+}\right)=\mathrm{res}_{\psi_{*}}\left(\mathrm{\Box}_{1}\mathtt{A}^{+}-\mathrm{d\delta}\mathtt{A}^{+}\right)\mbox{.}
\]
Trivially $\mathrm{\delta}\theta=0$ because $\delta\circ\psi_{*}^{\prime}=\psi_{*}^{\prime}\circ\mathrm{\delta}$
so that $\theta\in\mathrm{\Omega}_{0,\mathrm{\delta}}^{1}M$. Now
we check that $\theta$ is exactly the one we were looking for. First
of all $\theta$ has compact support so that $\mathrm{ext}_{\psi_{*}}\theta$
has compact support too and hence we can apply $e_{N}^{a/r}$ to it
obtaining
\begin{eqnarray*}
e_{N}^{a}\left(\mathrm{ext}_{\psi_{*}}\theta\right) & = & +e_{N}^{a}\left(\mathrm{\Box}_{1}\mathtt{A}^{+}-\mathrm{d\delta}\mathtt{A}^{+}\right)\mbox{,}\\
e_{N}^{r}\left(\mathrm{ext}_{\psi_{*}}\theta\right) & = & -e_{N}^{r}\left(\mathrm{\Box}_{1}\mathtt{A}^{-}-\mathrm{d\delta}\mathtt{A}^{-}\right)\mbox{.}
\end{eqnarray*}
Now we exploit the $\mathscr{M}$-past/future compact support of $\mathtt{A}^{\pm}$
to apply Lemma \ref{lemExtensionOf ea(Pu)=00003Du}. Furthermore we
bear in mind that $\mathrm{\delta}\mathtt{A}^{+}=-\mathrm{\delta}\mathtt{A}^{+}$
has compact support so that we can apply also Lemma \ref{lemGreenOperatorsCommuteWithExteriorDerivativeAndCodifferential}.
In this way we find
\begin{eqnarray*}
e_{N}^{a}\left(\mathrm{ext}_{\psi_{*}}\theta\right) & = & +\mathtt{A}^{+}-\mathrm{d}\left(e_{N}^{a}\left(\mathrm{\delta}\mathtt{A}^{+}\right)\right)\mbox{,}\\
e_{N}^{r}\left(\mathrm{ext}_{\psi_{*}}\theta\right) & = & -\mathtt{A}^{-}+\mathrm{d}\left(e_{N}^{r}\left(\mathrm{\delta}\mathtt{A}^{-}\right)\right)\mbox{.}
\end{eqnarray*}
The last two equations together give 
\[
e_{N}\left(\mathrm{ext}_{\psi_{*}}\theta\right)=\mathtt{A}-\mathrm{d}\left(e_{N}^{a}\left(\mathrm{\delta}\mathtt{A}^{+}\right)+e_{N}^{r}\left(\mathrm{\delta}\mathtt{A}^{-}\right)\right)=e_{N}\zeta-\mathrm{d}\left(e_{N}\left(\mathrm{\delta}\mathtt{A}^{+}\right)\right)\mbox{.}
\]
This completes the proof because
\[
\mathrm{d}\left(e_{N}\left(\mathrm{ext}_{\psi_{*}}\theta\right)-e_{N}\zeta\right)=-\mathrm{d}\left(\mathrm{d}\left(e_{N}\left(\mathrm{\delta}\mathtt{A}^{+}\right)\right)\right)=0\mbox{,}
\]
hence
\[
\xi\left[e_{M}\theta\right]_{M}=\left[e_{N}\left(\mathrm{ext}_{\psi_{*}}\theta\right)\right]_{N}=\left[e_{N}\zeta\right]_{N}=\left[\mathtt{A}\right]_{N}
\]
and in particular we deduce that $\left[\mathtt{A}\right]_{N}\in\xi\left(V\right)$.
For the freedom in the choice of $\left[\mathtt{A}\right]_{N}\in W$,
this fact implies the inclusion $W\subseteq\xi\left(V\right)$. The
last part of the statement of the time slice axiom follows directly
because each symplectic map is automatically injective (cfr. Remark
\ref{remSymplecticMapsAreInjective}) and the time slice axiom assures
that $\xi$ is also surjective, hence the inverse $\xi^{-1}$ exists
and it is trivial to check that it is a symplectic map too.
\end{proof}
Now that we have the covariant functor $\mathscr{B}$ describing the
classical theory of the electromagnetic field and we know that it
satisfies both the causality condition and the slice axiom. We can
proceed with the quantization procedure composing $\mathscr{B}$ with
the covariant functor $\mathscr{C}:\mathfrak{ssp}\overset{\rightarrow}{\rightarrow}\mathfrak{alg}$
defined in Lemma \ref{lemQuantizationFunctor}. In this way we obtain
a locally covariant quantum field theory $\mathscr{A}=\mathscr{C}\circ\mathscr{B}:\mathfrak{ghs}^{EM}\overset{\rightarrow}{\rightarrow}\mathfrak{alg}$
for the electromagnetic field which is causal and fulfils the time
slice axiom (cfr. Theorem \ref{thmLCQFTForANormallyHyperbolicField}).
Then Theorem \ref{thmRecoveringHaagKastlerAxioms}, together with
Remark \ref{remPrimitityHoldsDueToTheQuantizationFunctor}, entails
that, on each globally hyperbolic spacetime $\mathscr{M}$, $\mathscr{A}$
provides the quantum field theory of the electromagnetic field $\mathscr{A}\left(\mathscr{M}\right)$
in accordance with the Haag-Kastler axioms.

%% file: 8_chapter3_RCE.tex
\chapter{\label{chapRCE}\index{relative Cauchy evolution}Relative Cauchy
evolution}

The current chapter is devoted to the presentation of the \textsl{relative
Cauchy evolution} (in the following often indicated by the acronym
\textsl{RCE}) as it has been recently defined in \cite{FV11}. We
give a sketch of the idea: Suppose that a locally covariant quantum
field theory $\mathscr{A}$ fulfilling the time slice axiom is given
(if the time slice axiom does not hold, we cannot define the RCE at
all). The assignment of a globally hyperbolic spacetime $\mathscr{M}$
induces via $\mathscr{A}$ the assignment of a unital C{*}-algebra
$\mathscr{A}\left(\mathscr{M}\right)$ (cfr. Definition \ref{defLCQFT}).
Consider now another globally hyperbolic spacetime $\mathscr{M}^{\prime}$
with the same underlying manifold that coincides with $\mathscr{M}$
outside a compact subset in which the metric of $\mathscr{M}^{\prime}$
is a perturbation (in a proper sense) of the metric of $\mathscr{M}$.
Then also on $\mathscr{M}^{\prime}$ we have a unital C{*}-algebra
$\mathscr{A}\left(\mathscr{M}^{\prime}\right)$. The RCE establishes
the relation between the perturbed unital C{*}-algebra $\mathscr{A}\left(\mathscr{M}^{\prime}\right)$
and the original unital C{*}-algebra $\mathscr{A}\left(\mathscr{M}\right)$.

If the LCQFT $\mathscr{A}$ we are dealing with satisfies also the
causality condition, as it was shown in Theorem \ref{thmRecoveringHaagKastlerAxioms},
via $\mathscr{A}$ we can obtain on each globally hyperbolic spacetime
a quantum field theory according to the axiomatic approach proposed
by Haag and Kastler in \cite{HK64}. Therefore, when $\mathscr{A}$
is also causal, we may interpret the RCE as a relation between the
perturbed quantum field theory $\mathscr{A}\left(\mathscr{M}^{\prime}\right)$
and the original quantum field theory $\mathscr{A}\left(\mathscr{M}\right)$,
namely it tells us how the observables over $\mathscr{M}$ are transformed
when we change $\mathscr{M}$ into $\mathscr{M}^{\prime}$ and then
we go back to $\mathscr{M}$, i.e. when we perform a fluctuation of
the spacetime metric.

We conclude that we have at our disposal an instrument that makes
it possible to study the effects of fluctuations of the underlying
metric on the quantum theory of some field for which we are able to
construct a LCQFT fulfilling both the causality condition and the
time slice axiom (the causality condition being required only to give
sense to the interpretation in terms of observables, but not really
indispensable for the definition of the RCE). The importance of this
tool relies in the subsequent considerations. Till this point we dealt
with quantum field theories on fixed globally hyperbolic spacetimes.
However we know that the spacetime where we live is a solution of
the Einstein's equation, hence it depends on the energy-matter content
of the whole universe. Indeed if we have a quantum field, we also
expect to have its contribution to the stress-energy tensor appearing
on the RHS of the Einstein equation and we may try to account for
this contribution adding the expectation value of the stress-energy
tensor associated to the quantum field. In this way the so-called
semiclassical Einstein's equation arise (for a detailed discussion
on this topic refer to \cite[Sect. 4.6, p. 85]{Wal94}). What we expect
from such equation is a back-reaction effect: Quantum fields contribute
to the stress-energy tensor which affects the solution of the semiclassical
Einstein's equation, hence the spacetime metric, giving rise to a
sort of perturbation of the quantum field itself.

When we are looking for solutions of the semiclassical Einstein's
equation in the presence of a quantum field, we cannot forget of this
back-reaction effect. Our aim is to show that the RCE is the proper
tool to account for this effect when we deal with the Klein-Gordon
field, the Proca field or the electromagnetic field. This fact was
originally conjectured by Brunetti, Fredenhagen and Verch in \cite{BFV03}:
They supposed that the action of the functional derivative of the
RCE with respect to the spacetime metric agrees with the action of
the quantized stress-energy tensor and they showed that in any case
the functional derivative of the RCE is symmetric and divergence free
(both these properties are required to hold for any stress-energy
tensor to be consistent with the LHS of the Einstein's equation).
Moreover they verified their conjecture in the case of the Klein-Gordon
field.

In the first part of this chapter we define the RCE following \cite{FV11}.
Although it is equivalent to the definition originally proposed in
\cite{BFV03} (for the proof of the equivalence refer to \cite{FV11}),
this approach seems to be more practical in some respects. Then we
define the functional derivative of the RCE with respect to the spacetime
metric (with reference to \cite{BFV03}) and we show that this object
is symmetric and divergence free. In the second part we deal with
the Klein-Gordon field, the Proca field and the electromagnetic field.
In first place we present the relation between the functional derivative
of the RCE and the stress-energy tensor found by Brunetti, Fredenhagen
and Verch in the case of the Klein-Gordon field and in second place
we show that similar results hold also for the Proca field and for
the electromagnetic field. In this way it is proved that the action
of the functional derivative of the RCE agrees with the action of
the quantized stress-energy tensor not only for the Klein-Gordon field,
but also in the cases of the Proca field and of the electromagnetic
field, thus confirming the conjecture that the functional derivative
of the RCE behaves like the quantized stress-energy tensor associated
to the field.

\section{\label{secRCEDefinitionAndProperties}Definition and some properties}

\subsection{Procedure to define the relative Cauchy evolution}

Following \cite{FV11}, we assume that a globally hyperbolic spacetime
$\mathscr{M}=\left(M,g,\mathfrak{o},\mathfrak{t}\right)$ is given
and we consider a compactly supported section $h\in\mathscr{D}\left(M,\mathrm{T}^{*}M\otimes_{s}\mathrm{T}^{*}M\right)$
in the symmetric tensor product of $\mathrm{T}^{*}M$ with itself.
Then $g_{h}=g+h$ is indeed a section in $\mathrm{T}^{*}M\otimes_{s}\mathrm{T}^{*}M$
(cfr. Remark \ref{remInnerProductsAsSections}). If we assume that
$h$ is such that $g_{h}$ is a Lorentzian metric, then $\left(M,g_{h}\right)$
is a Lorentzian manifold. We can also require that $h$ is such that
$\left(M,g_{h}\right)$ is time orientable. Since $g_{h}$ coincides
with $g$ outside $\mathrm{supp}\left(h\right)$, there exists only
one connected component of the set of everywhere $g_{h}$-timelike
vector fields over $M$ which includes an element that coincides with
some element of $\mathfrak{t}$ outside $\mathrm{supp}\left(h\right)$,
i.e. there exists only one time orientation $\mathfrak{t}_{h}$ for
the time orientable Lorentzian manifold $\left(M,g_{h}\right)$ that
agrees with $\mathfrak{t}$ outside the support of $h$. In this way
we obtain the oriented an time oriented Lorentzian manifold $\left(M,g_{h},\mathfrak{o},\mathfrak{t}_{h}\right)$.
\begin{defn}
\index{globally hyperbolic perturbation}Let $\mathscr{M}=\left(M,g,\mathfrak{o},\mathfrak{t}\right)$
be a globally hyperbolic spacetime. $h\in\mathscr{D}\left(M,\mathrm{T}^{*}M\otimes_{s}\mathrm{T}^{*}M\right)$
is said to be an \textsl{$\mathscr{M}$-globally hyperbolic perturbation}
of the metric $g$ if the oriented and time oriented Lorentzian manifold
$\left(M,g_{h},\mathfrak{o},\mathfrak{t}\right)$ built above is actually
a globally hyperbolic spacetime. In this case we denote the globally
hyperbolic spacetime generated by the perturbation with $\mathscr{M}\left[h\right]=\left(M,g_{h},\mathfrak{o},\mathfrak{t}_{h}\right)$.

We denote the set of the $\mathscr{M}$-globally hyperbolic perturbations
of the metric $g$ with $GHP\left(\mathscr{M}\right)$ and we endow
such set with the topology induced by the usual topology of $\mathscr{D}\left(M,\mathrm{T}^{*}M\otimes_{s}\mathrm{T}^{*}M\right)$.
Moreover for each compact subset $K$ of $M$ we define the subset
$GHP\left(\mathscr{M},K\right)$ of the $\mathscr{M}$-globally hyperbolic
perturbations of the metric $g$ with support contained in $K$.
\end{defn}
Note that for each $\mathscr{M}\in\mathsf{Obj}_{\mathfrak{ghs}}$
the set $GHP\left(\mathscr{M}\right)$ is not empty because it contains
at least the null section in $\mathrm{T}^{*}M\otimes_{s}\mathrm{T}^{*}M$.
As a matter of fact one can show that for each $\mathscr{M}\in\mathsf{Obj}_{\mathfrak{ghs}}$
there exists a neighborhood of the null section in $\mathrm{C}^{\infty}\left(M,\mathrm{T}^{*}M\otimes_{s}\mathrm{T}^{*}M\right)$
which is included in $GHP\left(\mathscr{M}\right)$. A similar conclusion
holds also for $GHP\left(\mathscr{M},K\right)$ for each compact subset
$K$ of $M$.

Before we define the relative Cauchy evolution, a lemma showing that
the upcoming definition makes sense is required. The statement holds
choosing all the upper signs or, alternatively, all the lower signs
when $\pm$ and $\mp$ appear.
\begin{lem}
\label{lemMorphismsForRCE}Let $\mathscr{M}=\left(M,g,\mathfrak{o},\mathfrak{t}\right)$
be a globally hyperbolic spacetime and consider a compact subset $K$
of $M$. We set $M_{\pm}=M\setminus J_{\mp}^{\mathscr{M}}\left(K\right)$.
Then the following conclusions hold true:
\begin{itemize}
\item for each $h\in GHP\left(\mathscr{M},K\right)$, $M_{\pm}$ is an $\mathscr{M}$-causally
convex and $\mathscr{M}\left[h\right]$-causally convex connected
open subset of $M$ and the globally hyperbolic spacetimes $\left.\mathscr{M}\right|_{M_{\pm}}$
and $\left.\mathscr{M}\left[h\right]\right|_{M_{\pm}}$ coincide;
\item there exists a smooth spacelike Cauchy surface $\Sigma_{\pm}$ for
$\mathscr{M}$ contained in $M_{\pm}$ which is also a smooth spacelike
Cauchy surface for $\mathscr{M}\left[h\right]$ for each $h\in GHP\left(\mathscr{M},K\right)$;
\item the inclusion map $\iota_{M_{\pm}}^{M}:M_{\pm}\rightarrow M$ can
be seen as a morphism of $\mathfrak{ghs}$ from $\left.\mathscr{M}\right|_{M_{\pm}}=\left.\mathscr{M}\left[h\right]\right|_{M_{\pm}}$
to $\mathscr{M}$ and as a morphism of $\mathfrak{ghs}$ from $\left.\mathscr{M}\left[h\right]\right|_{M_{\pm}}=\left.\mathscr{M}\right|_{M_{\pm}}$
to $\mathscr{M}\left[h\right]$ and its image includes a smooth spacelike
Cauchy surface $\Sigma_{\pm}$ for both $\mathscr{M}$ and $\mathscr{M}\left[h\right]$.
\end{itemize}
\end{lem}
\begin{proof}
We focus on $M_{+}$ (the other case being similar). First of all
we note that $J_{-}^{\mathscr{M}}\left(K\right)$ is a closed subset
of $M$ (see \cite[Lem. A.5.1, p. 173]{BGP07}) and hence $M_{+}$
is open.

Now we show that $M_{+}$ is $\mathscr{M}$-causally convex. By contradiction
suppose that there exists a $\mathfrak{t}$-future directed $g$-causal
curve $\gamma$ starting at $p\in M_{+}$ and ending at $q\in M_{+}$
which is not entirely contained in $M_{+}$. Then we find a point
$r$ along $\gamma$ that falls in $J_{-}^{\mathscr{M}}\left(K\right)$.
It follows directly that $p$ is a point of $J_{-}^{\mathscr{M}}\left(K\right)$
in contrast with the hypothesis that $p\in M_{+}$.

We fix $h\in GHP\left(\mathscr{M},K\right)$ and we show that $M_{+}$
is also $\mathscr{M}\left[h\right]$-causally convex. Again by contradiction
suppose that there exists a $\mathfrak{t}_{h}$-future directed $g_{h}$-causal
curve $\gamma$ starting at $p\in M_{+}$ and ending at $q\in M_{+}$
which is not entirely contained in $M_{+}$. We deduce that $\gamma$
intersects $J_{-}^{\mathscr{M}}\left(K\right)$. We consider the piece
$\gamma^{\prime}$ of $\gamma$ starting from $p$ and ending in a
point $r$ of the boundary of $J_{-}^{\mathscr{M}}\left(K\right)$
so that $\gamma^{\prime}$ is outside  $J_{-}^{\mathscr{M}}\left(K\right)$,
except for the point $r$ (which falls in $J_{-}^{\mathscr{M}}\left(K\right)$
because it is closed). Since $\mathrm{supp}\left(h\right)\subseteq K$
and $\mathfrak{t}_{h}$ agrees with $\mathfrak{t}$ outside $\mathrm{supp}\left(h\right)$,
we conclude that $\gamma^{\prime}$ is also a $\mathfrak{t}$-future
directed $g$-causal curve from $p$ to $r\in J_{-}^{\mathscr{M}}\left(K\right)$.
Then we deduce that $p\in J_{-}^{\mathscr{M}}\left(K\right)$ in contrast
with the hypothesis $p\in M_{+}$.

To prove connectedness, we apply Theorem \ref{thmGlobalHyperbolicity}
to $\mathscr{M}$. In this way we find a smooth spacelike Cauchy surface
$\Sigma$ for $\mathscr{M}$ and a diffeomorphism $\psi:M\rightarrow\mathbb{R}\times\Sigma$.
Then we define $\tau=\mathrm{pr}_{1}\circ\psi:M\rightarrow\mathbb{R}$,
where $\mathrm{pr}_{1}$ denotes the projection upon the first argument
of the Cartesian product $\mathbb{R}\times\Sigma$. We realize immediately
that $\tau$ is continuous (actually smooth). Since $K$ is compact,
we find $t\in\mathbb{R}$ such that $t>\sup\left\{ \tau\left(p\right):\, p\in K\right\} $.
Then we consider $\Sigma_{t}=\psi^{-1}\left(\left\{ t\right\} \times\Sigma\right)$.
This is a smooth spacelike Cauchy surface for $\mathscr{M}$ due to
Theorem \ref{thmGlobalHyperbolicity}, in particular it is also connected.
Moreover by construction $\Sigma_{t}\subseteq M_{+}$. Take now two
arbitrary points $p$ and $q$ in $M_{+}$ and consider two inextensible
$\mathfrak{t}$-future directed $g$-timelike curves $\gamma_{p}$
and $\gamma_{q}$ such that $\gamma_{p}$ passes through $p$ and
$\gamma_{q}$ passes through $q$. These curves indeed meet $\Sigma_{t}$
because it is a Cauchy surface. We denote with $r$ and $s$ the intersections
of $\gamma_{p}$ and respectively $\gamma_{q}$ with $\Sigma_{t}$
and we consider the pieces $\gamma_{pr}$ and $\gamma_{qs}$ of $\gamma_{p}$
and respectively $\gamma_{q}$ connecting $p$ to $r$ and $q$ to
$s$. From $\mathscr{M}$-causal convexity it follows that $\gamma_{pr}$
and $\gamma_{qs}$ are entirely contained in $M_{+}$ because also
$r$ and $s$ fall in $M_{+}$. Exploiting connectedness of $\Sigma_{t}$,
we find $\gamma_{rs}$ connecting $r$ to $s$. Reversing $\gamma_{qs}$
and pasting the result with $\gamma_{pr}$ and $\gamma_{rs}$, we
obtain a curve which connects $p$ to $q$. This shows that $M_{+}$
is connected.

Up to now we have shown that $M_{+}$ is an $\mathscr{M}$-causally
convex and $\mathscr{M}\left[h\right]$-causally convex connected
open subset of $M$ for each $h\in GHP\left(\mathscr{M},K\right)$.
Applying Proposition \ref{propCausalConvexityImpliesGlobalHyperbolicity}
and Remark \ref{remCausalConvexityImpliesCausalCompatibility}, we
deduce that $\left.\mathscr{M}\right|_{M_{+}}=\left(M_{+},\left.g\right|_{M_{+}},\left.\mathfrak{o}\right|_{M_{+}},\left.\mathfrak{t}\right|_{M_{+}}\right)$
and $\left.\mathscr{M}\left[h\right]\right|_{M_{+}}=\left(M_{+},\left.g_{h}\right|_{M_{+}},\left.\mathfrak{o}\right|_{M_{+}},\left.\mathfrak{t}_{h}\right|_{M_{+}}\right)$
are globally hyperbolic spacetimes for each $h\in GHP\left(\mathscr{M},K\right)$.
Now fix an arbitrary $h\in GHP\left(\mathscr{M},K\right)$. Then $\mathrm{supp}\left(h\right)\subseteq K$
and $\mathfrak{t}_{h}$ agrees with $\mathfrak{t}$ outside $\mathrm{supp}\left(h\right)$.
These facts entail that $\left.g_{h}\right|_{M_{+}}=\left.g\right|_{M_{+}}$
and that $\left.\mathfrak{t}_{h}\right|_{M_{+}}$ and $\left.\mathfrak{t}\right|_{M_{+}}$
induce the same time orientations on the Lorentzian manifolds $\left(M_{+},\left.g\right|_{M_{+}}\right)$
and $\left(M_{+},\left.g_{h}\right|_{M_{+}}\right)$, therefore we
conclude $\left.\mathscr{M}\left[h\right]\right|_{M_{+}}=\left.\mathscr{M}\right|_{M_{+}}$
and the proof of the first point is complete.

As for the second point, we already determined a smooth spacelike
Cauchy surface $\Sigma_{t}\subseteq M_{+}$ for $\mathscr{M}$. Now
we show that this one is also a Cauchy surface for $\mathscr{M}\left[h\right]$
for each $h\in GHP\left(\mathscr{M},K\right)$. Fix $h\in GHP\left(\mathscr{M},K\right)$.
Since $g_{h}$ and $g$ coincide on $M_{+}$, $\Sigma_{t}$ is spacelike
also with respect to $g_{h}$. Consider an inextensible $\mathfrak{t}_{h}$-future
directed $g_{h}$-timelike curve $\gamma$ in $M$. There are two
possibilities: If $\gamma$ does not meet $\mathrm{supp}\left(h\right)$,
then it is also an inextensible $\mathfrak{t}$-future directed $g$-timelike
curve in $M$ and hence it must meet $\Sigma_{t}$ exactly once; conversely
if $\gamma$ meets $\mathrm{supp}\left(h\right)$, we can consider
the piece $\gamma^{\prime}$ of $\gamma$ that lies in $J_{+}^{\mathscr{M}}\left(K\right)\setminus K$.
$\gamma^{\prime}$ is a $\mathfrak{t}$-future directed $g$-timelike
curve in $M$ which is by construction inextensible in the future.
We can extend it in the past in such a way that the result is an inextensible
$\mathfrak{t}$-future directed $g$-timelike curve $\gamma^{\prime\prime}$
in $M$. Then $\gamma^{\prime\prime}$ meets $\Sigma_{t}$ exactly
once. The choice of $t>\sup\left\{ \tau\left(p\right):\, p\in K\right\} $
entails that $K$ lies in the $\mathscr{M}$-causal past of $\Sigma_{t}$
and that $K\cap\Sigma_{t}=0$. Then the only intersection of $\gamma^{\prime\prime}$
with $\Sigma_{t}$ must fall in the $\mathscr{M}$-causal future of
$K$. We deduce that $\gamma^{\prime}$ already met $\Sigma_{t}$,
and hence also $\gamma$. Note that the other piece of $\gamma$ (the
one not contained in $J_{+}^{\mathscr{M}}\left(K\right)\setminus K$)
cannot intersect $\Sigma_{t}$ because it is contained in $J_{-}^{\mathscr{M}}\left(K\right)$.
Hence also in the second case $\gamma$ meets $\Sigma_{t}$ exactly
once.

We turn our attention to the last point and we begin noting that $\iota_{M_{+}}^{M}$
is an embedding (cfr. Remark \ref{remSubmanifold}). Now fix $h\in GHP\left(\mathscr{M},K\right)$.
Undoubtedly $\iota_{M_{+}}^{M}$ is isometric and preserves both orientation
and time orientation whether we consider $\mathscr{M}$ or $\mathscr{M}\left[h\right]$
as target since $\iota_{M_{+}}^{M*}g=\left.g\right|_{M_{+}}=\left.g_{h}\right|_{M_{+}}$,
$\iota_{M_{+}}^{M*}\mathfrak{o}=\left.\mathfrak{o}\right|_{M_{+}}$
and $\iota_{M_{+}}^{M*}\mathfrak{t}_{h}=\left.\mathfrak{t}_{h}\right|_{M_{+}}$
and $\iota_{M_{+}}^{M*}\mathfrak{t}=\left.\mathfrak{t}\right|_{M_{+}}$%
\footnote{Note that pulling back $\mathfrak{t}$ and $\mathfrak{t}_{h}$ through
$\iota_{M_{+}}^{M}$ means that we are taking any representative (which
is a vector field) restricted to $M_{+}$ and we are pushing it forward
through the diffeomorphism $\iota_{M_{+}}^{M\prime}:M_{+}\rightarrow M_{+}$
induced by the embedding $\iota_{M_{+}}^{M}$.%
} induce the same time orientations on the Lorentzian manifolds $\left(M_{+},\left.g\right|_{M_{+}}\right)$
and $\left(M_{+},\left.g_{h}\right|_{M_{+}}\right)$ (which are the
same as a matter of fact). Hence $\iota_{M_{+}}^{M}$ is an isometric
embedding which preserves both orientation and time orientation whether
we consider $\mathscr{M}$ or $\mathscr{M}\left[h\right]$ as target
(we are considering $\left.\mathscr{M}\left[h\right]\right|_{M_{+}}=\left.\mathscr{M}\right|_{M_{+}}$
as source). The image of $\iota_{M_{+}}^{M}$ is trivially $M_{+}$,
which is causally convex with respect to both $\mathscr{M}$ and $\mathscr{M}\left[h\right]$.
Moreover we showed that $\Sigma_{t}$ is a smooth spacelike Cauchy
surface for both $\mathscr{M}$ and $\mathscr{M}\left[h\right]$ that
is contained in $M_{+}$. These observations concludes the proof.
\end{proof}
Consider a globally hyperbolic spacetime $\mathscr{M}$ and take $h\in GHP\left(\mathscr{M}\right)$.
Applying the last lemma with $K=\mathrm{supp}\left(h\right)$, we
have the following diagrams:
\[
\begin{array}{rcccl}
\mathscr{M} & \overset{\iota_{M_{-}}^{M}}{\longleftarrow} & \left.\mathscr{M}\right|_{M_{-}}=\left.\mathscr{M}\left[h\right]\right|_{M_{-}} & \overset{\iota_{M_{-}}^{M}}{\longrightarrow} & \mathscr{M}\left[h\right]\mbox{,}\\
\mathscr{M} & \overset{\iota_{M_{+}}^{M}}{\longleftarrow} & \left.\mathscr{M}\right|_{M_{+}}=\left.\mathscr{M}\left[h\right]\right|_{M_{+}} & \overset{\iota_{M_{+}}^{M}}{\longrightarrow} & \mathscr{M}\left[h\right]\mbox{.}
\end{array}
\]
Note that here the arrows represent morphisms of the category $\mathfrak{ghs}$
whose image includes a smooth spacelike Cauchy surface of the target
object (namely a globally hyperbolic spacetime). We introduce a convenient
notation rewriting the diagrams above (each element of the new diagrams
is defined by the element of the old diagram which occupies the same
position):
\[
\begin{array}{rcccl}
\mathscr{M} & \overset{\imath_{-}^{\mathscr{M}}\left[h\right]}{\longleftarrow} & \mathscr{M}_{-}\left[h\right] & \overset{\jmath_{-}^{\mathscr{M}}\left[h\right]}{\longrightarrow} & \mathscr{M}\left[h\right]\mbox{,}\\
\mathscr{M} & \overset{\imath_{+}^{\mathscr{M}}\left[h\right]}{\longleftarrow} & \mathscr{M}_{+}\left[h\right] & \overset{\jmath_{+}^{\mathscr{M}}\left[h\right]}{\longrightarrow} & \mathscr{M}\left[h\right]\mbox{.}
\end{array}
\]
The main advantage of the new notation relies in the fact that we
can recognize from the name if we are considering $\iota_{M_{\pm}}^{M}$
as a morphism from $\left.\mathscr{M}\right|_{M_{\pm}}$ to $\mathscr{M}$
(in which case we use the symbol $\imath$) or as a morphism from
$\left.\mathscr{M}\left[h\right]\right|_{M_{\pm}}$ to $\mathscr{M}\left[h\right]$
(in which case we use the symbol $\jmath$). Moreover this notation
emphasizes the dependence on $h$ of all the elements actually depend
in some way on the choice of $h$ in $GHP\left(\mathscr{M}\right)$.

If we consider a locally covariant quantum field theory $\mathscr{A}$,
the diagrams above are mapped to
\[
\begin{array}{rcccl}
\mathscr{A}\left(\mathscr{M}\right) & \overset{\mathscr{A}\left(\imath_{-}^{\mathscr{M}}\left[h\right]\right)}{\longleftarrow} & \mathscr{A}\left(\mathscr{M}_{-}\left[h\right]\right) & \overset{\mathscr{A}\left(\jmath_{-}^{\mathscr{M}}\left[h\right]\right)}{\longrightarrow} & \mathscr{A}\left(\mathscr{M}\left[h\right]\right)\mbox{,}\\
\mathscr{A}\left(\mathscr{M}\right) & \overset{\mathscr{A}\left(\imath_{+}^{\mathscr{M}}\left[h\right]\right)}{\longleftarrow} & \mathscr{A}\left(\mathscr{M}_{+}\left[h\right]\right) & \overset{\mathscr{A}\left(\jmath_{+}^{\mathscr{M}}\left[h\right]\right)}{\longrightarrow} & \mathscr{A}\left(\mathscr{M}\left[h\right]\right)\mbox{,}
\end{array}
\]
where all the arrows now are morphisms of the category $\mathfrak{alg}$.
If we suppose that $\mathscr{A}$ fulfils the time slice axiom, we
deduce that all the morphisms are actually unit preserving {*}-isomorphisms
between unital C{*}-algebras. This fact is a consequence of the time
slice axiom, together with Lemma \ref{lemMorphismsForRCE}. Reversing
the arrows on the left in the last two diagrams, we can define the
following {*}-isomorphisms between unital C{*}-algebras:
\begin{gather*}
\tau_{-}^{\mathscr{M}}\left[h\right]=\mathscr{A}\left(\jmath_{-}^{\mathscr{M}}\left[h\right]\right)\circ\mathscr{A}\left(\imath_{-}^{\mathscr{M}}\left[h\right]\right)^{-1}:\mathscr{A}\left(\mathscr{M}\right)\rightarrow\mathscr{A}\left(\mathscr{M}\left[h\right]\right)\mbox{,}\\
\tau_{+}^{\mathscr{M}}\left[h\right]=\mathscr{A}\left(\jmath_{+}^{\mathscr{M}}\left[h\right]\right)\circ\mathscr{A}\left(\imath_{+}^{\mathscr{M}}\left[h\right]\right)^{-1}:\mathscr{A}\left(\mathscr{M}\right)\rightarrow\mathscr{A}\left(\mathscr{M}\left[h\right]\right)\mbox{.}
\end{gather*}

We are ready to define the relative Cauchy evolution.
\begin{defn}
\label{defRCE}\index{relative Cauchy evolution}\index{RCE}Consider
a LCQFT $\mathscr{A}$ fulfilling the time slice axiom. For each globally
hyperbolic spacetime $\mathscr{M}$ and each $h\in GHP\left(\mathscr{M}\right)$,
we call \textsl{relative Cauchy evolution} (or briefly \textsl{RCE})
induced by $h$ on $\mathscr{M}$ the following {*}-automorphism of
the unital C{*}-algebra $\mathscr{A}\left(\mathscr{M}\right)$:
\[
R_{h}^{\mathscr{M}}=\left(\tau_{-}^{\mathscr{M}}\left[h\right]\right)^{-1}\circ\tau_{+}^{\mathscr{M}}\left[h\right]:\mathscr{A}\left(\mathscr{M}\right)\rightarrow\mathscr{A}\left(\mathscr{M}\right)\mbox{.}
\]

\end{defn}
Exploiting the expressions of $\tau_{\pm}^{\mathscr{M}}\left[h\right]$,
we may rewrite the RCE in the following way:
\begin{equation}
R_{h}^{\mathscr{M}}=\mathscr{A}\left(\imath_{-}^{\mathscr{M}}\left[h\right]\right)\circ\mathscr{A}\left(\jmath_{-}^{\mathscr{M}}\left[h\right]\right)^{-1}\circ\mathscr{A}\left(\jmath_{+}^{\mathscr{M}}\left[h\right]\right)\circ\mathscr{A}\left(\imath_{+}^{\mathscr{M}}\left[h\right]\right)^{-1}\mbox{.}\label{eqDefinitionOfRCE}
\end{equation}

As a consequence of the functorial properties of $\mathscr{A}$, we
expect that the RCE on a globally hyperbolic spacetime $\mathscr{M}=\left(M,g,\mathfrak{o},\mathfrak{t}\right)$
is insensitive to changes in the fluctuations $h$ of the spacetime
metric $g$ produced by an orientation preserving diffeomorphism from
the oriented manifold $\left(M,\mathfrak{o}\right)$ to itself that
acts trivially outside of a compact subset of $M$ including the support
of $h$.
\begin{prop}
\label{propRCEInsensitiveToDiffemorphisms}Let $\mathscr{A}$ be a
LCQFT fulfilling the time slice axiom, let $\mathscr{M}=\left(M,g,\mathfrak{o},\mathfrak{t}\right)$
be a globally hyperbolic spacetime and let $\psi$ be an orientation
preserving diffeomorphism from $\left(M,\mathfrak{o}\right)$ to itself
acting trivially outside of a compact subset $K$ of $M$. Consider
$h\in GHP\left(\mathscr{M},K\right)$ such that $h^{\prime}=\psi_{*}g_{h}-g\in GHP\left(\mathscr{M},K\right)$.
Then the diffeomorphism $\psi:M\rightarrow M$ may be seen as an orientation
and time orientation preserving isometric diffeomorphism from $\mathscr{M}\left[h\right]$
to $\mathscr{M}\left[h^{\prime}\right]$ and $R_{h^{\prime}}^{\mathscr{M}}=R_{h}^{\mathscr{M}}$.\end{prop}
\begin{proof}
Recall that $\mathscr{M}\left[h\right]=\left(M,g_{h},\mathfrak{o},\mathfrak{t}_{h}\right)$
and $\mathscr{M}\left[h^{\prime}\right]=\left(M,g_{h^{\prime}},\mathfrak{o},\mathfrak{t}_{h^{\prime}}\right)$.
Exploiting the hypothesis, we deduce
\[
\psi_{*}g_{h}=g+h^{\prime}=g_{h^{\prime}}\mbox{.}
\]
From this fact it follows that $\mathfrak{t}_{h^{\prime}}$ is one
of the connected components of the set of everywhere $\psi_{*}g_{h}$-timelike
vector fields over $M$. Furthermore $\mathfrak{t}_{h}$ is by definition
one of the connected components of the set of everywhere $g_{h}$-timelike
vector fields over $M$, hence $\psi_{*}\mathfrak{t}_{h}$ is one
of the connected components of the set of everywhere $\psi_{*}g_{h}$-timelike
vector fields over $M$. $\mathfrak{t}_{h}$ agrees with $\mathfrak{t}$
outside $\mathrm{supp}\left(h\right)$ by definition of $\mathscr{M}\left[h\right]$,
while $\mathfrak{t}_{h^{\prime}}$ agrees with $\mathfrak{t}$ outside
$\mathrm{supp}\left(h^{\prime}\right)$ by definition of $\mathscr{M}\left[h^{\prime}\right]$,
hence $\mathfrak{t}_{h}$ and $\mathfrak{t}_{h^{\prime}}$ agree outside
$K$. Moreover $\psi_{*}\mathfrak{t}_{h}=\mathfrak{t}_{h}$ outside
$K$ because by hypothesis $\psi$ acts trivially outside $K$. Then
we conclude that $\psi_{*}\mathfrak{t}_{h}$ and $\mathfrak{t}_{h^{\prime}}$
agree outside $K$, therefore they are the same connected component
of the set of everywhere $\psi_{*}g_{h}$-timelike vector fields over
$M$, i.e. $\psi_{*}\mathfrak{t}_{h}=\mathfrak{t}_{h^{\prime}}$.
This shows that actually $\psi$ may be interpreted as an orientation
and time orientation isometric diffeomorphism from $\mathscr{M}\left[h\right]$
to $\mathscr{M}\left[h^{\prime}\right]$.

Now we focus on the second part of the statement. We begin defining
$M_{\pm}=M\setminus J_{\mp}^{\mathscr{M}}\left(K\right)$. Since $h$
and $h^{\prime}$ are elements of $GHP\left(\mathscr{M},K\right)$,
we can apply Lemma \ref{lemMorphismsForRCE} to deduce that $M_{\pm}$
is an $\mathscr{M}$-causally convex, $\mathscr{M}\left[h\right]$-causally
convex and $\mathscr{M}\left[h^{\prime}\right]$-causally convex connected
open subset of $M$ containing a smooth spacelike Cauchy surface $\Sigma_{\pm}$
for $\mathscr{M}$ that is also a smooth spacelike Cauchy surface
for both $\mathscr{M}\left[h\right]$ and $\mathscr{M}\left[h^{\prime}\right]$.
Taking into account $\mathscr{M}_{\pm}\left[h\right]$ and $\mathscr{M}_{\pm}\left[h^{\prime}\right]$
(whose underlying manifolds are respectively $M\setminus J_{\mp}\left(\mathrm{supp}\left(h\right)\right)$
and $M\setminus J_{\mp}\left(\mathrm{supp}\left(h^{\prime}\right)\right)$),
we see that both include $M_{\pm}$ and then also $\Sigma_{\pm}$.
It turns out almost trivially that $\Sigma_{\pm}$ is a smooth spacelike
Cauchy surface for both $\mathscr{M}_{\pm}\left[h\right]$ and $\mathscr{M}_{\pm}\left[h^{\prime}\right]$.
Hence $M_{\pm}$ is also an $\mathscr{M}_{\pm}\left[h\right]$-causally
convex and $\mathscr{M}_{\pm}\left[h^{\prime}\right]$-causally convex
connected open subset containing a smooth spacelike Cauchy surface
$\Sigma_{\pm}$ for both $\mathscr{M}_{\pm}\left[h\right]$ and $\mathscr{M}_{\pm}\left[h^{\prime}\right]$.
This fact entails that we can consider the globally hyperbolic spacetimes
$\left.\mathscr{M}_{\pm}\left[h\right]\right|_{M_{\pm}}$ and $\left.\mathscr{M}_{\pm}\left[h^{\prime}\right]\right|_{M_{\pm}}$
and interpret the inclusion maps of $M_{\pm}$ in $M\setminus J_{\mp}^{\mathscr{M}}\left(\mathrm{supp}\left(h\right)\right)$
and in $M\setminus J_{\mp}^{\mathscr{M}}\left(\mathrm{supp}\left(h^{\prime}\right)\right)$
as morphisms of the category $\mathfrak{ghs}$ whose image includes
a smooth spacelike Cauchy surface of the target:
\begin{eqnarray*}
\alpha_{\pm} & \in & \mathsf{Mor}_{\mathfrak{ghs}}\left(\left.\mathscr{M}_{\pm}\left[h\right]\right|_{M_{\pm}},\mathscr{M}_{\pm}\left[h\right]\right)\mbox{,}\\
\beta_{\pm} & \in & \mathsf{Mor}_{\mathfrak{ghs}}\left(\left.\mathscr{M}_{\pm}\left[h^{\prime}\right]\right|_{M_{\pm}},\mathscr{M}_{\pm}\left[h^{\prime}\right]\right)\mbox{.}
\end{eqnarray*}
We may also consider the globally hyperbolic spacetime $\left.\mathscr{M}\right|_{M_{\pm}}$
and we realize that
\[
\left.\mathscr{M}_{\pm}\left[h\right]\right|_{M_{\pm}}=\left.\mathscr{M}\left[h\right]\right|_{M_{\pm}}=\left.\mathscr{M}\right|_{M_{\pm}}=\left.\mathscr{M}\left[h^{\prime}\right]\right|_{M_{\pm}}=\left.\mathscr{M}_{\pm}\left[h^{\prime}\right]\right|_{M_{\pm}}
\]
because $g_{h}=g=g_{h^{\prime}}$ outside $K$ and $\mathfrak{t}_{h}$,
$\mathfrak{t}$ and $\mathfrak{t}_{h^{\prime}}$ agree outside $K$.
Hence we can consider $\alpha_{\pm}$ and $\beta_{\pm}$ as morphisms
starting from $\left.\mathscr{M}\right|_{M_{\pm}}$. For convenience
we recollect here the morphisms generated by the globally hyperbolic
perturbations $h$ and $h^{\prime}$:
\begin{eqnarray*}
\imath_{\pm}^{\mathscr{M}}\left[h\right] & \in & \mathsf{Mor}_{\mathfrak{ghs}}\left(\mathscr{M}_{\pm}\left[h\right],\mathscr{M}\right)\mbox{,}\\
\jmath_{\pm}^{\mathscr{M}}\left[h\right] & \in & \mathsf{Mor}_{\mathfrak{ghs}}\left(\mathscr{M}_{\pm}\left[h\right],\mathscr{M}\left[h\right]\right)\mbox{,}\\
\imath_{\pm}^{\mathscr{M}}\left[h^{\prime}\right] & \in & \mathsf{Mor}_{\mathfrak{ghs}}\left(\mathscr{M}_{\pm}\left[h^{\prime}\right],\mathscr{M}\right)\mbox{,}\\
\jmath_{\pm}^{\mathscr{M}}\left[h^{\prime}\right] & \in & \mathsf{Mor}_{\mathfrak{ghs}}\left(\mathscr{M}_{\pm}\left[h^{\prime}\right],\mathscr{M}\left[h^{\prime}\right]\right)\mbox{.}
\end{eqnarray*}
We exploit $\alpha_{\pm}$ and $\beta_{\pm}$ and the fact that their
images include a smooth spacelike Cauchy surface of the target, together
with the hypothesis that the time slice axiom holds for $\mathscr{M}$,
to rewrite both $R_{h}^{\mathscr{M}}$ and $R_{h^{\prime}}^{\mathscr{M}}$:
\begin{eqnarray*}
R_{h}^{\mathscr{M}} & = & \mathscr{A}\left(\imath_{-}^{\mathscr{M}}\left[h\right]\right)\circ\mathscr{A}\left(\jmath_{-}^{\mathscr{M}}\left[h\right]\right)^{-1}\circ\mathscr{A}\left(\jmath_{+}^{\mathscr{M}}\left[h\right]\right)\circ\mathscr{A}\left(\imath_{+}^{\mathscr{M}}\left[h\right]\right)^{-1}\\
 & = & \mathscr{A}\left(\imath_{-}^{\mathscr{M}}\left[h\right]\right)\circ\mathscr{A}\left(\alpha_{-}\right)\circ\mathscr{A}\left(\alpha_{-}\right)^{-1}\circ\mathscr{A}\left(\jmath_{-}^{\mathscr{M}}\left[h\right]\right)^{-1}\\
 &  & \circ\mathscr{A}\left(\jmath_{+}^{\mathscr{M}}\left[h\right]\right)\circ\mathscr{A}\left(\alpha_{+}\right)\circ\mathscr{A}\left(\alpha_{+}\right)^{-1}\circ\mathscr{A}\left(\imath_{+}^{\mathscr{M}}\left[h\right]\right)^{-1}\\
 & = & \mathscr{A}\left(\imath_{-}^{\mathscr{M}}\left[h\right]\circ\alpha_{-}\right)\circ\mathscr{A}\left(\jmath_{-}^{\mathscr{M}}\left[h\right]\circ\alpha_{-}\right)^{-1}\\
 &  & \circ\mathscr{A}\left(\jmath_{+}^{\mathscr{M}}\left[h\right]\circ\alpha_{+}\right)\circ\mathscr{A}\left(\imath_{+}^{\mathscr{M}}\left[h\right]\circ\alpha_{+}\right)^{-1}\mbox{,}
\end{eqnarray*}
\begin{eqnarray*}
R_{h^{\prime}}^{\mathscr{M}} & = & \mathscr{A}\left(\imath_{-}^{\mathscr{M}}\left[h^{\prime}\right]\right)\circ\mathscr{A}\left(\jmath_{-}^{\mathscr{M}}\left[h^{\prime}\right]\right)^{-1}\circ\mathscr{A}\left(\jmath_{+}^{\mathscr{M}}\left[h^{\prime}\right]\right)\circ\mathscr{A}\left(\imath_{+}^{\mathscr{M}}\left[h^{\prime}\right]\right)^{-1}\\
 & = & \mathscr{A}\left(\imath_{-}^{\mathscr{M}}\left[h^{\prime}\right]\right)\circ\mathscr{A}\left(\beta_{-}\right)\circ\mathscr{A}\left(\beta_{-}\right)^{-1}\circ\mathscr{A}\left(\jmath_{-}^{\mathscr{M}}\left[h^{\prime}\right]\right)^{-1}\\
 &  & \circ\mathscr{A}\left(\jmath_{+}^{\mathscr{M}}\left[h^{\prime}\right]\right)\circ\mathscr{A}\left(\beta_{+}\right)\circ\mathscr{A}\left(\beta_{+}\right)^{-1}\circ\mathscr{A}\left(\imath_{+}^{\mathscr{M}}\left[h^{\prime}\right]\right)^{-1}\\
 & = & \mathscr{A}\left(\imath_{-}^{\mathscr{M}}\left[h^{\prime}\right]\circ\beta_{-}\right)\circ\mathscr{A}\left(\jmath_{-}^{\mathscr{M}}\left[h^{\prime}\right]\circ\beta_{-}\right)^{-1}\\
 &  & \circ\mathscr{A}\left(\jmath_{+}^{\mathscr{M}}\left[h^{\prime}\right]\circ\beta_{+}\right)\circ\mathscr{A}\left(\imath_{+}^{\mathscr{M}}\left[h^{\prime}\right]\circ\beta_{+}\right)^{-1}\mbox{.}
\end{eqnarray*}
We can observe that $\imath_{\pm}^{\mathscr{M}}\left[h\right]\circ\alpha_{\pm}$
and $\imath_{\pm}^{\mathscr{M}}\left[h^{\prime}\right]\circ\beta_{\pm}$
are both morphisms from $\left.\mathscr{M}\right|_{M_{\pm}}$ to $\mathscr{M}$
whose underlying map is nothing but the inclusion map of $M_{\pm}$
into $M$, hence these morphisms are exactly the same and we denote
both of them with $\psi_{\pm}$. Now we exploit the morphism $\psi$
from $\mathscr{M}\left[h\right]$ to $\mathscr{M}\left[h^{\prime}\right]$:
\begin{eqnarray*}
R_{h}^{\mathscr{M}} & = & \mathscr{A}\left(\psi_{-}\right)\circ\mathscr{A}\left(\jmath_{-}^{\mathscr{M}}\left[h\right]\circ\alpha_{-}\right)^{-1}\circ\mathscr{A}\left(\jmath_{+}^{\mathscr{M}}\left[h\right]\circ\alpha_{+}\right)\circ\mathscr{A}\left(\psi_{+}\right)^{-1}\\
 & = & \mathscr{A}\left(\psi_{-}\right)\circ\mathscr{A}\left(\jmath_{-}^{\mathscr{M}}\left[h\right]\circ\alpha_{-}\right)^{-1}\circ\mathscr{A}\left(\psi\right)^{-1}\\
 &  & \circ\mathscr{A}\left(\psi\right)\circ\mathscr{A}\left(\jmath_{+}^{\mathscr{M}}\left[h\right]\circ\alpha_{+}\right)\circ\mathscr{A}\left(\psi_{+}\right)^{-1}\\
 & = & \mathscr{A}\left(\psi_{-}\right)\circ\mathscr{A}\left(\psi\circ\jmath_{-}^{\mathscr{M}}\left[h\right]\circ\alpha_{-}\right)^{-1}\circ\mathscr{A}\left(\psi\circ\jmath_{+}^{\mathscr{M}}\left[h\right]\circ\alpha_{+}\right)\circ\mathscr{A}\left(\psi_{+}\right)^{-1}\mbox{.}
\end{eqnarray*}
We note that both $\psi\circ\jmath_{\pm}^{\mathscr{M}}\left[h\right]\circ\alpha_{\pm}$
and $\jmath_{\pm}^{\mathscr{M}}\left[h^{\prime}\right]\circ\beta_{\pm}$
are morphisms from $\left.\mathscr{M}\right|_{M_{\pm}}$ to $\mathscr{M}\left[h^{\prime}\right]$
and, since $\psi$ acts trivially outside $K\subseteq M\setminus M_{\pm}$,
we deduce that the underlying maps coincide. Hence $\psi\circ\jmath_{\pm}^{\mathscr{M}}\left[h\right]\circ\alpha_{\pm}$
and $\jmath_{\pm}^{\mathscr{M}}\left[h^{\prime}\right]\circ\beta_{\pm}$
are actually the same morphism and we denote them with $\psi^{\pm}$.
At this point we have
\[
R_{h^{\prime}}^{\mathscr{M}}=\mathscr{A}\left(\psi_{-}\right)\circ\mathscr{A}\left(\psi^{-}\right)^{-1}\circ\mathscr{A}\left(\psi^{+}\right)\circ\mathscr{A}\left(\psi_{+}\right)^{-1}=R_{h}^{\mathscr{M}}
\]
and this concludes the proof.
\end{proof}

\subsection{Functional derivative of the relative Cauchy evolution}

In this subsection we define the functional derivative of the relative
Cauchy evolution with respect to the spacetime metric following the
procedure presented in \cite{BFV03} (adapted to the current definition
of the RCE).

We consider a globally hyperbolic spacetime $\mathscr{M}=\left(M,g,\mathfrak{o},\mathfrak{t}\right)$.
For each $h\in GHP\left(\mathscr{M}\right)$ we know that $\mathscr{M}\left[h\right]=\left(M,g_{h},\mathfrak{o},\mathfrak{t}_{h}\right)$
is a globally hyperbolic spacetime in its own right. Moreover, for
each $h\in GHP\left(\mathscr{M}\right)$, taking $M_{\pm}=M\setminus J_{\mp}^{\mathscr{M}}\left(\mathrm{supp}\left(h\right)\right)$
and applying Lemma \ref{lemMorphismsForRCE} with $K=\mathrm{supp}\left(h\right)$,
we find that $\mathscr{M}_{\pm}\left[h\right]=\left.\mathscr{M}\right|_{M_{\pm}}$
is a globally hyperbolic spacetimes including a smooth spacelike Cauchy
surface for both $\mathscr{M}$ and $\mathscr{M}\left[h\right]$.

We consider a locally covariant quantum field theory $\mathscr{A}$
fulfilling the time slice axiom and we take into account the unital
C{*}-algebra $\mathscr{A}\left(\mathscr{M}\right)$.
\begin{assumption}
\label{assFunctionalDerivativeRCE}Suppose that $\pi$ is a representation
of $\mathscr{A}\left(\mathscr{M}\right)$ on a Hilbert space $\mathscr{H}$.
Assume that there exist a dense subspace $\mathscr{V}$ of $\mathscr{H}$
and a dense unital sub-{*}-algebra $\mathcal{B}$ of $\mathscr{A}\left(\mathscr{M}\right)$
such that for each $\Omega\in\mathscr{V}$ and each $b\in\mathcal{B}$
the following conditions are satisfied:
\begin{itemize}
\item for each compact subset $K$ of $M$ and each smooth $1$-parameter
family
\begin{eqnarray*}
\left(-1,1\right) & \rightarrow & GHP\left(\mathscr{M},K\right)\\
s & \mapsto & h^{s}
\end{eqnarray*}
such that $h^{0}=0$, the map
\begin{eqnarray*}
\left(-1,1\right) & \rightarrow & \mathbb{C}\\
s & \mapsto & \left\langle \Omega,\pi\left(R_{h^{s}}^{\mathscr{M}}b\right)\Omega\right\rangle \mbox{,}
\end{eqnarray*}
where $\left\langle \cdot,\cdot\right\rangle $ denotes the scalar
product of $\mathscr{H}$, is continuously differentiable;
\item there exists a section $\beta\in\mathrm{C}^{\infty}\left(M,\mathrm{T}M\otimes_{s}\mathrm{T}M\right)$
such that, for each compact subset $K$ of $M$ and each smooth $1$-parameter
family
\begin{eqnarray*}
\left(-1,1\right) & \rightarrow & GHP\left(\mathscr{M},K\right)\\
s & \mapsto & h^{s}
\end{eqnarray*}
verifying $h^{0}=0$, it holds that
\begin{equation}
\int\limits _{M}\left(\left.\frac{\mathrm{d}h^{s}}{\mathrm{d}s}\right|_{0}\right)\left(\beta\right)\mathrm{d}\mu_{g}=\left.\frac{\mathrm{d}}{\mathrm{d}s}\left\langle \Omega,\pi\left(R_{h^{s}}^{\mathscr{M}}b\right)\Omega\right\rangle \right|_{0}\mbox{,}\label{eqFunctionalDerivativeRCE}
\end{equation}
where the dual pairing between $\mathrm{T}^{*}M\otimes_{s}\mathrm{T}^{*}M$
and $\mathrm{T}M\otimes_{s}\mathrm{T}M$ is taken into account and
$\mathrm{d}\mu_{g}$ is the standard volume form on $\mathscr{M}$.
\end{itemize}
\end{assumption}
\begin{rem}
Some remarks about the last assumption are required. First of all
we explain the meaning of the integrand appearing on the LHS of eq.
\eqref{eqFunctionalDerivativeRCE}. Fix a compact subset $K$ of $M$
and consider a smooth 1-parameter family $s\mapsto h^{s}$ of the
type required above. Using local coordinates at a point $p\in M$,
we have the following expression for the components of $\left.\nicefrac{\mathrm{d}h^{s}}{\mathrm{d}s}\right|_{0}$
evaluated at $p$: 
\[
\left(\left.\frac{\mathrm{d}h^{s}}{\mathrm{d}s}\right|_{0}\left(p\right)\right)_{ij}=\left.\frac{\mathrm{d}}{\mathrm{d}s}h_{ij}^{s}\left(p\right)\right|_{0}\mbox{,}
\]
where $h_{ij}^{s}\left(p\right)$ are the components of $h^{s}$ evaluated
at $p$ for some $s\in\left(-1,1\right)$. From the assumption that
$s\mapsto h^{s}$ is smooth, it follows that $\left.\nicefrac{\mathrm{d}h^{s}}{\mathrm{d}s}\right|_{0}$
is a section in $\mathrm{T}^{*}M\otimes_{s}\mathrm{T}^{*}M$. Since
$\mathrm{supp}\left(h^{s}\right)$ is contained in $K$ for each $s\in\left(-1,1\right)$,
we deduce also that $\left.\nicefrac{\mathrm{d}h^{s}}{\mathrm{d}s}\right|_{0}$
has support included in $K$, hence compact. This fact assures that
the integral makes sense.

Secondly we consider the term that appears on the RHS. The derivative
appearing here is well defined as a direct consequence of the first
point in the assumption above.

Now that we have understood the meaning of both the LHS and the RHS
of eq. \eqref{eqFunctionalDerivativeRCE}, we can try to understand
the consequences of this equation on the section $\beta$ (which is
supposed to exist). Fix a compact subset $K$ of $M$. The freedom
in the choice of the family $s\mapsto h^{s}$, together with the fact
that $GHP\left(\mathscr{M},K\right)$ includes a neighborhood of the
null section in $\mathrm{C}^{\infty}\left(M,\mathrm{T}^{*}M\otimes_{s}\mathrm{T}^{*}M\right)$,
entails that $\beta$ is uniquely determined on $K$: If we suppose
that there exists another section $\beta^{\prime}$ of the same type
satisfying the same equation, we deduce that
\[
\int\limits _{M}f\left(\beta-\beta^{\prime}\right)\mathrm{d}\mu_{g}=0\quad\forall f\in\mathscr{D}\left(M,\mathrm{T}^{*}M\otimes_{s}\mathrm{T}^{*}M\right):\,\mathrm{supp}\left(f\right)\subseteq K
\]
and, working with sections $f$ with support contained in open subsets
of $M$ included in $K$ on which the vector bundle $\mathrm{T}^{*}M\otimes_{s}\mathrm{T}^{*}M$
is trivialized, we conclude that $\beta-\beta^{\prime}=0$ on $K$
due to the density of the vector space $\mathscr{D}\left(O,O\times\mathbb{R}^{n}\right)$
in the Banach space $L^{p}\left(O,O\times\mathbb{R}^{n}\right)$ for
each open subset $O$ of $\mathbb{R}^{d}$, each $d$, $n\in\mathbb{N}$
and each $p\in\left[1,\infty\right)$. Then the freedom in the choice
of $K$ entails that $\beta$ is uniquely determined everywhere on
$M$.

These observations entail that the assumption made above assures the
uniqueness of the functional derivative $\beta$ of $\left\langle \Omega,\pi\left(R_{h}^{\mathscr{M}}b\right)\Omega\right\rangle $
with respect to $\mathscr{M}$-globally hyperbolic perturbations of
the spacetime metric. For brevity we will simply say that $\beta$
is the functional derivative of $\left\langle \Omega,\pi\left(R_{h}^{\mathscr{M}}b\right)\Omega\right\rangle $
with respect to the spacetime metric and we will write $\frac{\mathrm{\delta}}{\delta h}\left\langle \Omega,\pi\left(R_{h}^{\mathscr{M}}b\right)\Omega\right\rangle $
in place of $\beta$.
\end{rem}
In the last remark we saw how Assumption \ref{assFunctionalDerivativeRCE}
implies that $\frac{\mathrm{\delta}}{\delta h}\left\langle \Omega,\pi\left(R_{h}^{\mathscr{M}}b\right)\Omega\right\rangle $
is uniquely defined for each $\Omega$ in a dense subspace $\mathscr{V}$
of a proper Hilbert space $\mathscr{H}$ and for each $b$ in a proper
dense sub-{*}-algebra of $\mathscr{A}\left(\mathscr{M}\right)$. We
are ready to define the functional derivative of the RCE with respect
to the spacetime metric.
\begin{defn}
\index{functional derivative of the RCE}Let $\mathscr{A}$ be a LCQFT
fulfilling the time slice axiom and let $\mathscr{M}$ be a globally
hyperbolic spacetime. Consider a representation $\pi$ of $\mathscr{A}\left(\mathscr{M}\right)$
on a Hilbert space $\mathscr{H}$. If Assumption \ref{assFunctionalDerivativeRCE}
holds, there exist a dense subspace $\mathscr{V}$ of $\mathscr{H}$
and a dense unital sub-{*}-algebra $\mathcal{B}$ of $\mathscr{A}\left(\mathscr{M}\right)$
such that we can uniquely define for each $b\in\mathcal{B}$ the \textsl{functional
derivative with respect to the spacetime metric of the relative Cauchy
evolution acting on $b$} (briefly \textsl{functional derivative of
the RCE}), denoted by $\frac{\mathrm{\delta}}{\delta h}\pi\left(R_{h}^{\mathscr{M}}b\right)$,
as a quadratic form on $\mathscr{V}$:
\[
\left\langle \Omega,\left(\frac{\mathrm{\delta}}{\delta h}\pi\left(R_{h}^{\mathscr{M}}b\right)\right)\Omega\right\rangle =\frac{\mathrm{\delta}}{\delta h}\left\langle \Omega,\pi\left(R_{h}^{\mathscr{M}}b\right)\Omega\right\rangle \quad\forall\Omega\in\mathscr{V}\mbox{.}
\]

\end{defn}
In \cite{BFV03} it was conjectured that the action of the functional
derivative of the RCE with respect to the spacetime metric agrees
with the action of the quantized-stress energy tensor. The first properties
to be checked in order to support such hypothesis are the symmetry
and the null divergence. Per definition $\frac{\mathrm{\delta}}{\delta h}\left\langle \Omega,\pi\left(R_{h}^{\mathscr{M}}b\right)\Omega\right\rangle $
is an element of $\mathrm{C}^{\infty}\left(M,\mathrm{T}M\otimes_{s}\mathrm{T}M\right)$
for each $b\in\mathcal{B}$ and each $\Omega\in\mathscr{V}$, hence
$\frac{\mathrm{\delta}}{\delta h}\pi\left(R_{h}^{\mathscr{M}}b\right)$
is symmetric for each $b\in\mathcal{B}$ (in the sense of the quadratic
forms on $\mathscr{V}$). The evaluation of the divergence is adressed
in the following proposition.

First we need to introduce some notation previously. Since $\frac{\mathrm{\delta}}{\delta h}\left\langle \Omega,\pi\left(R_{h}^{\mathscr{M}}b\right)\Omega\right\rangle $
is an element of $\mathrm{C}^{\infty}\left(M,\mathrm{T}M\otimes_{s}\mathrm{T}M\right)$,
at each point $p$ of $M$ we may write it in local coordinates. We
denote its components at $p$ with
\[
\frac{\mathrm{\delta}}{\delta h_{ij}\left(p\right)}\left\langle \Omega,\pi\left(R_{h}^{\mathscr{M}}b\right)\Omega\right\rangle \mbox{.}
\]
Note that the indices are {}``doubly'' covariant, hence contravariant,
in accordance with the fact that $\frac{\mathrm{\delta}}{\delta h}\left\langle \Omega,\pi\left(R_{h}^{\mathscr{M}}b\right)\Omega\right\rangle $
evaluated in a point $p\in M$ is an element of $\mathrm{T}^{\left(2,0\right)}M$.
Similarly we denote the components of $\frac{\mathrm{\delta}}{\delta h}\pi\left(R_{h}^{\mathscr{M}}b\right)$
at $p$ with
\[
\frac{\mathrm{\delta}}{\delta h_{ij}\left(p\right)}\pi\left(R_{h}^{\mathscr{M}}b\right)\mbox{.}
\]

\begin{prop}
\label{propFunctionalDerivativeOfTheRCEHasNullDivergence}Let $\mathscr{A}$
be a LCQFT fulfilling the time slice axiom and let $\mathscr{M}=\left(M,g,\mathfrak{o},\mathfrak{t}\right)$
be a globally hyperbolic spacetime. Consider a representation $\pi$
of $\mathscr{A}\left(\mathscr{M}\right)$ on a Hilbert space $\mathscr{H}$
such that Assumption \ref{assFunctionalDerivativeRCE} holds so that
we find a dense subspace $\mathscr{V}$ of $\mathscr{H}$ and a dense
unital sub-{*}-algebra $\mathcal{B}$ of $\mathscr{A}\left(\mathscr{M}\right)$
on which the functional derivative of the RCE is defined. Then for
each $b\in\mathcal{B}$ we have
\[
\nabla_{i}\left(\frac{\mathrm{\delta}}{\delta h_{ij}\left(p\right)}\pi\left(R_{h}^{\mathscr{M}}b\right)\right)=0\quad\forall p\in M
\]
in the sense of the quadratic forms on $\mathscr{V}$, where $\nabla$
denotes the Levi-Civita connection with respect to the metric $g$.\end{prop}
\begin{proof}
The thesis is a formal expression meaning that for each $b\in\mathcal{B}$,
each $\Omega\in\mathscr{V}$ and each $p\in M$
\[
\nabla_{i}\left(\frac{\mathrm{\delta}}{\delta h_{ij}\left(p\right)}\left\langle \Omega,\pi\left(R_{h}^{\mathscr{M}}b\right)\Omega\right\rangle \right)=0\mbox{.}
\]
We fix $b\in\mathcal{B}$ and $\Omega\in\mathscr{V}$ and, denoting
$\frac{\mathrm{\delta}}{\delta h}\left\langle \Omega,\pi\left(R_{h}^{\mathscr{M}}b\right)\Omega\right\rangle $
with $\beta$, we may rewrite the thesis in the following way: for
each oriented local coordinate neighborhood $\left(U,V,\phi\right)$
and for each vector field $X\in\mathscr{D}\left(M,\mathrm{T}M\right)$
with support included in $U$ it holds
\[
\int\limits _{V}\left(\nabla_{i}\beta^{ij}\right)g_{jk}X^{k}\sqrt{\left|\det g\right|}\mathrm{d}V=0\mbox{,}
\]
where $\mathrm{d}V$ denotes the standard volume form on $\mathbb{R}^{d}$
and all sections in the integrand are meant in local coordinates.
Via an integration by parts and since $X$ is null on the boundary
of $U$, we deduce that the last equation is equivalent to
\[
\int\limits _{V}\beta^{ij}\left(\nabla_{i}X_{j}\right)\sqrt{\left|\det g\right|}\mathrm{d}V=0\mbox{.}
\]
We know that $\beta$ is symmetric so that we can write
\[
\beta^{ij}\left(\nabla_{i}X_{j}\right)=\frac{1}{2}\left(\beta^{ij}+\beta^{ji}\right)\left(\nabla_{i}X_{j}\right)=\frac{1}{2}\beta^{ij}\left(\nabla_{i}X_{j}+\nabla_{j}X_{i}\right)\mbox{,}
\]
hence our thesis finally becomes
\begin{equation}
\int\limits _{V}\beta^{ij}\left(\nabla_{i}X_{j}+\nabla_{j}X_{i}\right)\sqrt{\left|\det g\right|}\mathrm{d}V=0\label{eqFunctionalDerivativeOfTheRCEHasNullDivergenceThesisRevised}
\end{equation}
for each local coordinate neighborhood $\left(U,V,\phi\right)$ and
for each vector field $X\in\mathscr{D}\left(M,\mathrm{T}M\right)$
with support included in $U$.

Fix now a local coordinate neighborhood $\left(U,V,\phi\right)$ and
a compactly supported vector field $X\in\mathscr{D}\left(M,\mathrm{T}M\right)$
with support $K$ included in $U$. We know that each compactly supported
vector field on $M$ generates a 1-parameter group of diffeomorphisms
of $M$ $s\in\mathbb{R}\mapsto\psi^{s}$ acting trivially outside
of $K$ with $\psi^{0}=\mathrm{id}_{M}$ (cfr. \cite[Thm. 1.9.2, p. 49]{Jos95}).
Note that $\psi^{s}$ is necessarily orientation preserving for each
$s\in\mathbb{R}$: Outside of $K$ it acts trivially (hence its Jacobian
determinant is positive); if it reverses some coordinate neighborhood
inside $K$ (i.e. its Jacobian determinant in that coordinate neighborhood
is negative), then there exists a point in some coordinate neighborhood
in which its Jacobian determinant is null, in contradiction with the
fact that it is a diffeomorphism. Consider now $GHP\left(\mathscr{M},K\right)$
and remember that it includes at least a neighborhood of the null
section in $\mathrm{C}^{\infty}\left(M,\mathrm{T}^{*}M\otimes_{s}\mathrm{T}^{*}M\right)$.
Since trivially $\psi_{*}^{0}g-g=0$, we may find $\varepsilon>0$
such that $\psi_{*}^{s}g-g$ falls in $GHP\left(\mathscr{M},K\right)$
for each $\left|s\right|<\epsilon$. Defining $h^{s}=\psi_{*}^{\varepsilon s}g-g$,
we obtain a $1$-parameter family $\left(-1,1\right)\rightarrow GHP\left(\mathscr{M},K\right)$,
$s\mapsto h^{s}$ such that $h^{0}=0$. By definition of $\beta$,
we have
\begin{equation}
\int\limits _{M}\left(\left.\frac{\mathrm{d}h^{s}}{\mathrm{d}s}\right|_{0}\right)\left(\beta\right)\mathrm{d}\mu_{g}=\left.\frac{\mathrm{d}}{\mathrm{d}s}\left\langle \Omega,\pi\left(R_{h^{s}}^{\mathscr{M}}b\right)\Omega\right\rangle \right|_{0}\mbox{.}\label{eqFunctionalDerivativeOfTheRCEHasNullDivergenceKeyPoint}
\end{equation}
On the one hand $h^{s}=\psi_{*}^{\varepsilon s}g-g=\psi_{*}^{\varepsilon s}g_{h^{0}}-g$
is an element of $GHP\left(\mathscr{M},K\right)$ for each $s\in\left(-1,1\right)$,
hence we can apply Proposition \ref{propRCEInsensitiveToDiffemorphisms}
(we choose $h=h^{0}=0$ as original perturbation and $h^{\prime}=h^{s}$
for each $s\in\left(-1,1\right)$) and we deduce that $R_{h^{s}}^{\mathscr{M}}=R_{h^{0}}^{\mathscr{M}}=\mathrm{id}_{\mathscr{A}\left(\mathscr{M}\right)}$
(the last equality follows from the fact that $h^{0}=0$) for each
$s\in\left(-1,1\right)$. This fact entails
\[
\left.\frac{\mathrm{d}}{\mathrm{d}s}\left\langle \Omega,\pi\left(R_{h^{s}}^{\mathscr{M}}b\right)\Omega\right\rangle \right|_{0}=0\mbox{.}
\]
On the other hand for each $p\in M$ we have
\[
\left.\frac{\mathrm{d}}{\mathrm{d}s}h_{ij}^{s}\left(p\right)\right|_{0}=\varepsilon\left.\frac{\mathrm{d}}{\mathrm{d}s}\left(\psi_{*}^{s}g\right)_{ij}\left(p\right)\right|_{0}=\varepsilon\left(\nabla_{i}X_{j}+\nabla_{j}X_{i}\right)\mbox{.}
\]
In fact $\left.\frac{\mathrm{d}}{\mathrm{d}s}\left(\psi_{*}^{s}g\right)_{ij}\left(p\right)\right|_{0}$
is exactly the definition of the Lie derivative of $g$ along the
vector field $X$ (cfr. \cite[eq. C.2.1, p. 439]{Wal84}) and the
last equivalence follows from \cite[eq. C.2.16, p. 441]{Wal84}. Inserting
the last two equations into eq. \eqref{eqFunctionalDerivativeOfTheRCEHasNullDivergenceKeyPoint},
we get the following result:
\[
\varepsilon\int\limits _{V}\beta^{ij}\left(\nabla_{i}X_{j}+\nabla_{j}X_{i}\right)\sqrt{\left|\det g\right|}\mathrm{d}V=0\mbox{,}
\]
where we used the fixed coordinate neighborhood $\left(U,V,\phi\right)$
to express the integral in local coordinates (this can actually be
done since the integrand is supported in $K\subseteq U$). With the
exception of $\varepsilon$, which can be thrown away being a positive
number, this is exactly our last reformulation of the thesis, eq.
\eqref{eqFunctionalDerivativeOfTheRCEHasNullDivergenceThesisRevised}.
\end{proof}

\section{\label{secRCEForConcreteFields}Relative Cauchy evolution for concrete
fields}

The functional derivative of the relative Cauchy evolution with respect
to the spacetime metric was defined as a section in $\mathrm{T}^{*}M\otimes_{s}\mathrm{T}^{*}M$,
hence, as we already observed, it is symmetric by construction. Moreover
in Proposition \ref{propFunctionalDerivativeOfTheRCEHasNullDivergence}
we proved that its divergence is null. Both these properties are good
hints to support the conjecture that the functional derivative of
the RCE has the meaning of a quantized stress-energy tensor. In this
section we settle this question once and for all for the cases of
the Klein-Gordon field (already discussed in \cite{BFV03}), the Proca
field and the electromagnetic field on a globally hyperbolic spacetime
$\mathscr{M}$.

\subsection{\label{subQuasiFreeHadamardStates}Quasi-free Hadamard states}

This subsection is devoted to introduce quasi-free Hadamard states,
an essential ingredient in our way to the proof of the theorems stating
the compatibility between the action of the quantized stress-energy
tensor of the Klein-Gordon, Proca or electromagnetic field and the
functional derivative of the corresponding relative Cauchy evolution
with respect to the spacetime metric.

To start, we consider the locally covariant quantum field theory $\mathscr{A}:\mathfrak{ghs}^{f}\overset{\rightarrow}{\rightarrow}\mathfrak{alg}$
built in Section \ref{secBuildingALCQFT} (in the next subsections
$\mathscr{A}$ will be one of the LCQFTs built for the concrete examples
discussed in Section \ref{secExamples}) and we choose an object $\left(\mathscr{M},E,A\right)$
of $\mathfrak{ghs}^{f}$ so that we have at our disposal the unital
C{*}-algebra $\mathscr{A}\left(\mathscr{M}\right)$. Now we take a
state $\tau$ (see Definition \ref{defState}) and we apply Theorem
\ref{thmGNSRepresentation}. In this way we obtain the GNS triple
$\left(\mathscr{H}_{\tau}^{\mathscr{M}},\pi_{\tau}^{\mathscr{M}},\Omega_{\tau}^{\mathscr{M}}\right)$
associated to the state $\tau$ on the unital C{*}-algebra $\mathscr{A}\left(\mathscr{M},E,A\right)$.

Recalling the procedure of Section \ref{secBuildingALCQFT}, we see
that $\mathscr{A}\left(\mathscr{M},E,A\right)$ is the (unique up
to {*}-isomorphisms) CCR representation $\left(\mathcal{V},\mathrm{V}\right)$
of the symplectic space $\left(V,\sigma\right)=\mathscr{B}\left(\mathscr{M},E,A\right)$
provided by the covariant functor $\mathscr{B}:\mathfrak{ghs}^{f}\overset{\rightarrow}{\rightarrow}\mathfrak{ssp}$
describing the theory of the field at a classical level (as a matter
of fact $\mathscr{A}$ was obtained as the composition of $\mathscr{B}$
with the covariant functor $\mathscr{C}:\mathfrak{ssp}\overset{\rightarrow}{\rightarrow}\mathfrak{alg}$
embodying the quantization procedure). We define the represented counterpart
of the Weyl map $\mathrm{V}$ setting $\mathrm{V}_{\tau}^{\mathscr{M}}=\pi_{\tau}^{\mathscr{M}}\circ\mathrm{V}:V\rightarrow\mathcal{B}\left(\mathscr{H}_{\tau}^{\mathscr{M}}\right)$,
where $\mathcal{B}\left(\mathscr{H}_{\tau}^{\mathscr{M}}\right)$
denotes the unital C{*}-algebra of the linear and continuous operators
on the Hilbert space $\mathscr{H}_{\tau}^{\mathscr{M}}$. Note that
$\mathrm{V}_{\tau}^{\mathscr{M}}$ maps each $u\in V$ to a unitary
operator $\mathrm{V}_{\tau}^{\mathscr{M}}\left(u\right)$ on the Hilbert
space $\mathscr{H}_{\tau}^{\mathscr{M}}$, as one easily deduces from
Remark \ref{remWeylMapProvidesUnitaries} and the fact that $\pi_{\tau}^{\mathscr{M}}$
is a unit preserving {*}-homomorphism from $\mathcal{V}$ to $\mathcal{B}\left(\mathscr{H}_{\tau}^{\mathscr{M}}\right)$.
With reference to \cite[Chap VI, Sect. 62, p. 16]{AG93} and \cite[Chap VI, Sect. 74, p. 74]{AG93},
we can find a selfadjoint operator $\varPhi_{\tau}^{\mathscr{M}}\left(u\right)\in\mathcal{B}\left(\mathscr{H}_{\tau}^{\mathscr{M}}\right)$
such that
\begin{equation}
\mathrm{e}^{\imath\varPhi_{\tau}^{\mathscr{M}}\left(u\right)}=\mathrm{V}_{\tau}^{\mathscr{M}}\left(u\right)\mbox{.}\label{eqWeylGeneratorsInExponentialForm}
\end{equation}
We may interpret the selfadjoint operator $\varPhi_{\tau}^{\mathscr{M}}\left(u\right)$
as the quantum field corresponding to the classical field $u\in V$.
It turns out that a map from $V$ to $\mathcal{B}\left(\mathscr{H}_{\tau}^{\mathscr{M}}\right)$
is automatically defined:
\begin{eqnarray*}
\varPhi_{\tau}^{\mathscr{M}}:V & \rightarrow & \mathcal{B}\left(\mathscr{H}_{\tau}^{\mathscr{M}}\right)\\
u & \mapsto & \varPhi_{\tau}^{\mathscr{M}}\left(u\right)\mbox{.}
\end{eqnarray*}

Using the map $\Phi_{\tau}^{\mathscr{M}}$ we can define the $n$-point
functions on the state $\tau$ and then characterize quasi-free states.
\begin{defn}
\index{n-point function}\index{quasi-free state}Denote with $\mathscr{A}:\mathfrak{ghs}^{f}\overset{\rightarrow}{\rightarrow}\mathfrak{alg}$
the locally covariant quantum field theory and with $\mathscr{B}:\mathfrak{ghs}^{f}\overset{\rightarrow}{\rightarrow}\mathfrak{ssp}$
the covariant functor describing the classical field theory (cfr.
Section \ref{secBuildingALCQFT}). For each object $\left(\mathscr{M},E,A\right)$
of $\mathfrak{ghs}^{f}$ we consider the CCR representation $\left(\mathcal{V},\mathrm{V}\right)=\mathscr{A}\left(\mathscr{M},E,A\right)$
and the symplectic space $\left(V,\sigma\right)=\mathscr{B}\left(\mathscr{M},E,A\right)$
and for each state $\tau$ on the unital C{*}-algebra $\mathcal{V}$
we take the (unique up to unitary transformations) GNS triple $\left(\mathscr{H}_{\tau}^{\mathscr{M}},\pi_{\tau}^{\mathscr{M}},\Omega_{\tau}^{\mathscr{M}}\right)$
provided by Theorem \ref{thmGNSRepresentation} applied to the state
$\tau$ on the unital C{*}-algebra $\mathcal{V}$. Following the procedure
shown above, we obtain a map
\begin{eqnarray*}
\varPhi_{\tau}^{\mathscr{M}}:V & \rightarrow & \mathcal{B}\left(\mathscr{H}_{\tau}^{\mathscr{M}}\right)\\
u & \mapsto & \varPhi_{\tau}^{\mathscr{M}}\left(u\right)
\end{eqnarray*}
for each $\left(\mathscr{M},E,A\right)\in\mathsf{Obj}_{\mathfrak{ghs}^{f}}$
and each state $\tau$ on the unital C{*}-algebra $\mathscr{A}\left(\mathscr{M},E,A\right)$.

We define the \textsl{$n$-point function} on $\theta\in\mathscr{H}_{\tau}^{\mathscr{M}}$
as the map 
\begin{eqnarray*}
w_{\tau,n}^{\mathscr{M},\theta}:\overset{n\mbox{ times}}{\overbrace{\mathscr{D}\left(M,E\right)\times\dots\times\mathscr{D}\left(M,E\right)}} & \rightarrow & \mathbb{R}\\
\left(f_{1},\dots,f_{n}\right) & \mapsto & \left\langle \theta,\varPhi_{\tau}^{\mathscr{M}}\left(e_{A}f_{1}\right)\cdots\varPhi_{\tau}^{\mathscr{M}}\left(e_{A}f_{n}\right)\theta\right\rangle _{\tau}^{\mathscr{M}}\mbox{,}
\end{eqnarray*}
where $\left\langle \cdot,\cdot\right\rangle _{\tau}^{\mathscr{M}}$
denotes the scalar product of the Hilbert space $\mathscr{H}_{\tau}^{\mathscr{M}}$.
The $n$-point function $w_{\tau,n}^{\mathscr{M},\Omega_{\tau}^{\mathscr{M}}}$
on the vector $\Omega_{\tau}^{\mathscr{M}}$ of the GNS triple is
simply denoted by $w_{\tau,n}^{\mathscr{M}}$.

We say that the state $\tau$ is \textsl{quasi-free} if its $\left(2n+1\right)$-point
function $w_{\tau,2n+1}^{\mathscr{M}}$ vanishes for each $n\in\mathbb{N}$,
while its $2n$-point function satisfies the following identity for
each $n\in\mathbb{N}$:
\[
w_{\tau,2n}^{\mathscr{M}}\left(f_{1},\dots,f_{2n}\right)=\sum_{s}w_{\tau,2}^{\mathscr{M}}\left(f_{s\left(1\right)},f_{s\left(2\right)}\right)\cdots w_{\tau,2}^{\mathscr{M}}\left(f_{s\left(2n-1\right)},f_{s\left(2n\right)}\right)
\]
for each $f_{1}$, $\dots$, $f_{2n}\in\mathscr{D}\left(M,E\right)$,
where the sum is taken over all the permutations $s$ of $\left\{ 1,\dots,2n\right\} $
such that $s\left(1\right)<s\left(3\right)<\dots<s\left(2n-1\right)$
and $s\left(2\right)<s\left(4\right)<\dots<s\left(2n\right)$.
\end{defn}
Note that for each quasi-free state all the $n$-point functions $w_{\tau,n}^{\mathscr{M}}$
are completely determined by the 2-point function $w_{\tau,2}^{\mathscr{M}}$.

\index{Hadamard state}Now we want to spend few words about \textsl{Hadamard
states}. These states are widely accepted as the physically meaningful
states for quantum field theories on curved spacetimes. This is due
to the fact that the short distance behavior of their 2-point functions
mimics the short distance behavior of vacuum states for quantum field
theories on Minkowski spacetime. Although singularities are present,
they are controlled in such a way that the expectation values of physical
observables (e.g. the stress-energy tensor) on Hadamard states are
prevented from taking unbounded fluctuations.

To give an idea of what it is meant for a Hadamard state we give the
following definition according Kay and Wald, \cite{KW91}. Indeed
this is specific for the Klein-Gordon field, yet it already gives
a sketch of the constraints on the singularities of a Hadamard state.
A precise extension of the notion of Hadamard state to fields in arbitrary
vector bundles can be found in \cite[Sect. 5.1, p. 20]{SV01}.
\begin{defn}
Let $\mathscr{A}:\mathfrak{ghs}^{KG}\rightarrow\mathfrak{alg}$ be
the LCQFT for the Klein-Gordon field (cfr. Subsection \ref{subKleinGordonField})
and let $\left(\mathscr{M},\mathrm{\Lambda}^{0}M,A\right)$ be an
object of the category $\mathfrak{ghs}^{KG}$. Consider a diffeomorphism
$\psi:M\rightarrow\mathbb{R}\times S$ provided by Theorem \ref{thmGlobalHyperbolicity}
($S$ is some smooth spacelike Cauchy surface for $\mathscr{M}$)
and define the smooth function $T=\mathrm{pr}_{1}\circ\psi:M\rightarrow\mathbb{R}$,
where $\mathrm{pr}_{1}$ denotes the projection on the first argument
of the Cartesian product. We define the squared geodesic distance
$d$ on an open neighborhood $O$ in $M\times M$ of the set of causally
related points $\left(p,q\right)$ such that $J_{+}^{\mathscr{M}}\left(p\right)\cap J_{-}^{\mathscr{M}}\left(q\right)$
and $J_{-}^{\mathscr{M}}\left(p\right)\cap J_{+}^{\mathscr{M}}\left(q\right)$
are included in a convex normal neighborhood. It turns out that $d$
is well defined and smooth. Then for each $n\in\left\{ 0,1,2,\dots\right\} $
and each $\varepsilon>0$ we define the function $G_{n,\varepsilon}:O\rightarrow\mathbb{C}$
according to the formula
\[
G_{n,\varepsilon}\left(p,q\right)=\frac{1}{\left(2\pi\right)^{2}}\left(\frac{\Delta^{\nicefrac{1}{2}}}{\gamma\left(p,q\right)}+v^{\left(n\right)}\left(p,q\right)\ln\gamma\left(p,q\right)\right)\mbox{,}
\]
where the branch-cut for the logarithm is taken on the negative half
of the real line, $\Delta$ is the van Vleck-Morette determinant (refer
to \cite{DB60}),
\[
v^{\left(n\right)}\left(p,q\right)=\sum_{m=1}^{n}v_{n}\left(p,q\right)\left(\sigma\left(p,q\right)\right)^{m}\mbox{,}
\]
the functions $v_{n}$ are uniquely determined via the Hadamard recursion
relations (refer to \cite{DB60,Gar98}) and 
\[
\gamma\left(p,q\right)=d\left(p,q\right)+2\imath\varepsilon\left(T\left(p\right)-T\left(q\right)\right)+\varepsilon^{2}\mbox{.}
\]

Now let $\Sigma$ be a smooth spacelike Cauchy surface for $M$ and
take a causal normal neighborhood of $\Sigma$ in $M$ (its existence
is proved in \cite[Lem. 2.2, p. 62]{KW91}). Consider an open neighborhood
$O^{\prime}$ in $N\times N$ of the set of pairs of causally related
points such that the closure of $O^{\prime}$ in $N\times N$ is contained
in $O$. Let $\chi$ be a smooth real valued function on $N\times N$
which is null outside $O$ and equal to 1 inside $O^{\prime}$. Then
we say that a state $\tau$ on the unital C{*}-algebra $\mathscr{A}\left(\mathscr{M},E,A\right)$
is a Hadamard state if its 2-point function $w_{\tau,2}^{\mathscr{M}}$
is such that for each $n\in\left\{ 0,1,2,\dots\right\} $ there exists
a function $H_{n}\in\mathrm{C}^{n}\left(N\times N\right)$ which satisfies
the following condition:
\[
w_{\tau,2}^{\mathscr{M}}\left(f_{1},f_{2}\right)=\lim_{\varepsilon\rightarrow0}\iint\limits _{N}\Lambda_{n,\varepsilon}\left(p,q\right)f_{1}\left(p\right)f_{2}\left(q\right)\mathrm{d}\mu_{g}\left(p\right)\mathrm{d}\mu_{g}\left(q\right)
\]
for each $f_{1}$, $f_{2}\in\mathscr{D}\left(N\right)$, where
\[
\Lambda_{n,\varepsilon}\left(p,q\right)=\chi\left(p,q\right)G_{n,\varepsilon}\left(p,q\right)+H_{n}\left(p,q\right)\mbox{.}
\]

\end{defn}
The problem of the determination of Hadamard states on curved spacetimes
for the various quantum fields one may consider is not discussed here,
neither we analyze the properties of Hadamard states in detail because
this would require the introduction of several notions from microlocal
analysis. Anyway we provide some references:
\begin{itemize}
\item \cite{Hor90} for the necessary tools of microlocal analysis;
\item \cite{Rad96,SV01,SVW02,FV03,San10b} are only some of the publications
discussing conditions (in the context of microlocal analysis) for
a state on some C{*}-algebra that are equivalent to the requirement
of being Hadamard (both for the case of a specific fields or for more
general situations) and showing the existence of Hadamard states for
specific fields.
\end{itemize}
In the present context we are mainly interested in the existence of
Hadamard states for spin 1 fields. Such result was established by
Fewster and Pfenning in \cite{FP03}.

\index{smeared field}Anyway few remarks about some of the properties
of the GNS representation induced by a Hadamard state are required.
Let $\mathscr{A}:\mathfrak{ghs}^{f}\rightarrow\mathfrak{alg}$ be
the LCQFT built in Section \ref{secBuildingALCQFT} (remember that
it is causal and, above all, it fulfils the time slice axiom) and
let $\left(\mathscr{M},E,A\right)$ be an object of $\mathfrak{ghs}^{f}$.
Consider a Hadamard state $\tau$ on the CCR representation $\left(\mathcal{V},\mathrm{V}\right)=\mathscr{A}\left(\mathscr{M},E,A\right)$
(recall that $\left(V,\sigma\right)=\mathscr{B}\left(\mathscr{M},E,A\right)$
denotes the symplectic space from which $\left(\mathcal{V},\mathrm{V}\right)$
arises via the quantization functor $\mathscr{C}$) and denote with
$\left(\mathscr{H}_{\tau}^{\mathscr{M}},\pi_{\tau}^{\mathscr{M}},\Omega_{\tau}^{\mathscr{M}}\right)$
its associated GNS triple. Then the state $\tau$ is sufficiently
regular to allow us to regard the function
\[
t\in\mathbb{R}\mapsto\mathrm{V}_{\tau}^{\mathscr{M}}\left(tu\right)
\]
as a differentiable function whatever choice of $u\in V$ we make.
This gives us the opportunity to define the map:
\begin{eqnarray*}
\varPsi_{\tau}^{\mathscr{M}}:\mathscr{D}\left(M,E\right) & \rightarrow & \mathcal{B}\left(\mathscr{H}_{\tau}^{\mathscr{M}}\right)\\
f & \mapsto & -\imath\left.\frac{\mathrm{d}}{\mathrm{d}t}\mathrm{V}_{\tau}^{\mathscr{M}}\left(te_{A}f\right)\right|_{0}\mbox{,}
\end{eqnarray*}
where $e_{A}$ denotes the causal propagator for $A$. For each $f\in\mathscr{D}\left(M,E\right)$
we call $\varPsi_{\tau}^{\mathscr{M}}\left(f\right)$ \textsl{smeared
field}. We deduce from its definition that the map $\varPsi_{\tau}^{\mathscr{M}}$
is linear and that the corresponding smeared fields allow us to write
$\mathrm{V}_{\tau}^{\mathscr{M}}\left(e_{A}f\right)$ in exponential
form (cfr. eq. \eqref{eqWeylGeneratorsInExponentialForm}) for each
$f\in\mathscr{D}\left(M,E\right)$: one easily checks that
\[
\imath\varPsi_{\tau}^{\mathscr{M}}\left(f\right)=\left.\frac{\mathrm{d}}{\mathrm{d}t}\mathrm{V}_{\tau}^{\mathscr{M}}\left(te_{A}f\right)\right|_{0}
\]
agrees with
\[
\mathrm{e}^{\imath\varPsi_{\tau}^{\mathscr{M}}\left(f\right)}=\mathrm{V}_{\tau}^{\mathscr{M}}\left(e_{A}f\right)\mbox{.}
\]
In this way we can also see that $\varPsi_{\tau}^{\mathscr{M}}\left(f\right)=\varPhi_{\tau}^{\mathscr{M}}\left(e_{A}f\right)$
for each $f\in\mathscr{D}\left(M,E\right)$. Moreover we can deduce
the commutation relation between smeared fields from the Weyl relations
(cfr. Definition \ref{defWeylSystem}) and also the commutation relation
between a smeared field and a represented Weyl generator. We find
\begin{eqnarray*}
\left[\varPsi_{\tau}^{\mathscr{M}}\left(f\right),\varPsi_{\tau}^{\mathscr{M}}\left(f^{\prime}\right)\right] & = & \imath\sigma\left(e_{A}f,e_{A}f^{\prime}\right)\mbox{,}\\
\left[\varPsi_{\tau}^{\mathscr{M}}\left(f\right),\mathrm{V}_{\tau}^{\mathscr{M}}\left(e_{A}f^{\prime}\right)\right] & = & -\sigma\left(e_{A}f,e_{A}f^{\prime}\right)\mathrm{V}_{\tau}^{\mathscr{M}}\left(e_{A}f^{\prime}\right)
\end{eqnarray*}
for each $f$, $f^{\prime}\in\mathscr{D}\left(M,E\right)$. These
relations will be useful in the proof of the theorems stating the
agreement between the action of the functional derivative of the relative
Cauchy evolution and the quantized stress-energy tensor. In particular
it is interesting for this purpose to consider the commutator of the
product of two smeared fields with some represented Weyl generator.
Exploiting the second commutation relation given above, we find
\begin{eqnarray}
\left[\varPsi_{\tau}^{\mathscr{M}}\left(f\right)\varPsi_{\tau}^{\mathscr{M}}\left(f^{\prime}\right),\mathrm{V}_{\tau}^{\mathscr{M}}\left(u\right)\right] & = & -\sigma\left(e_{A}f,u\right)\mathrm{V}_{\tau}^{\mathscr{M}}\left(u\right)\varPsi_{\tau}^{\mathscr{M}}\left(f^{\prime}\right)\nonumber \\
 &  & -\sigma\left(e_{A}f^{\prime},u\right)\varPsi_{\tau}^{\mathscr{M}}\left(f\right)\mathrm{V}_{\tau}^{\mathscr{M}}\left(u\right)\label{eqCommutatorBetween2SmearedFieldsAndAWeylGenerator}
\end{eqnarray}
for each $f$, $f^{\prime}\in\mathscr{D}\left(M,E\right)$ and each
$u\in V$.

There is still another very important consequence of the choice of
a quasi-free Hadamard state $\tau$: We find a dense subspace $\mathscr{V}_{\tau}^{\mathscr{M}}$
of the Hilbert space $\mathscr{H}_{\tau}^{\mathscr{M}}$, namely the
one constituted by all the vectors obtained applying an arbitrary
polynomial in $\varPsi_{\tau}^{\mathscr{M}}\left(f\right)$ and $\mathrm{V}_{\tau}^{\mathscr{M}}\left(u\right)$
(for any choice of $f\in\mathscr{D}\left(M,E\right)$ and $u\in V$)
to the GNS vector $\Omega_{\tau}^{\mathscr{M}}$, and a dense unital
sub-{*}-algebra $\mathcal{B}_{\tau}^{\mathscr{M}}$ of $\mathscr{A}\left(\mathscr{M},E,A\right)$
such that Assumption \ref{assFunctionalDerivativeRCE} holds. This
fact entails that we can actually give sense to the functional derivative
of the RCE.

Moreover it is possible to establish a relation that will be the key
for the proof of our theorems from now on. First of all we have to
define a classical counterpart of the relative Cauchy evolution which
is obtained simply replacing the covariant functor $\mathscr{A}$
with the covariant functor $\mathscr{B}$ in eq. \eqref{eqDefinitionOfRCE}:
for each object $\left(\mathscr{M},E,A\right)$ of $\mathfrak{ghs}^{f}$
and each $h\in GHP\left(\mathscr{M}\right)$ we set
\[
r_{h}^{\mathscr{M}}=\mathscr{B}\left(\imath_{-}^{\mathscr{M}}\left[h\right]\right)\circ\mathscr{B}\left(\jmath_{-}^{\mathscr{M}}\left[h\right]\right)^{-1}\circ\mathscr{B}\left(\jmath_{+}^{\mathscr{M}}\left[h\right]\right)\circ\mathscr{B}\left(\imath_{+}^{\mathscr{M}}\left[h\right]\right)^{-1}\mbox{.}
\]
Note that the definition is well posed because a proper version of
the time slice axiom holds also for the covariant functor $\mathscr{B}:\mathfrak{ghs}^{f}\overset{\rightarrow}{\rightarrow}\mathfrak{ssp}$
describing the classical field theory (see Subsection \ref{subClassicalFieldTheory})
and that $R_{h}^{\mathscr{M}}=\mathscr{C}\left(r_{h}^{\mathscr{M}}\right)$,
where $\mathscr{C}:\mathfrak{ssp}\overset{\rightarrow}{\rightarrow}\mathfrak{alg}$
is the covariant functor that realizes the quantization procedure
(cfr. Subsection \ref{subQuantumFieldTheory}).%
\footnote{In the following we will study in some detail the classical RCE for
the specific fields we will consider.%
} With this definition we are ready to state the key relation which
can be found in \cite[Prop. A.8, p. 363]{FP03}:
\begin{equation}
\left.\frac{\mathrm{d}}{\mathrm{d}s}\left\langle \theta,\mathrm{V}_{\tau}^{\mathscr{M}}\left(r_{h^{s}}^{\mathscr{M}}u\right)\theta\right\rangle _{\tau}^{\mathscr{M}}\right|_{0}=\frac{\imath}{2}\left\langle \theta,\left\{ \varPhi_{\tau}^{\mathscr{M}}\left(\left.\frac{\mathrm{d}}{\mathrm{d}s}\left(r_{h^{s}}^{\mathscr{M}}u\right)\right|_{0}\right),\mathrm{V}_{\tau}^{\mathscr{M}}\left(u\right)\right\} \theta\right\rangle _{\tau}^{\mathscr{M}}\label{eqKeyPointForTheMainTheoremAboutRCE}
\end{equation}
for each compact subset $K$ of $M$, each smooth 1-parameter family
of globally hyperbolic perturbations $\left(-1,1\right)\rightarrow GHP\left(\mathscr{M},K\right)$,
$s\mapsto h^{s}$ such that $h^{0}=0$, each $\theta\in\mathscr{V}_{\tau}^{\mathscr{M}}$
and each $u\in V$. In \cite{FV03} the proof is performed in the
context of the Klein-Gordon field, however it holds in general since
it relies only on the properties of the CCR representation of some
symplectic space and on the choice of a Hadamard state which gives
rise to a Hilbert space representation with the {}``good'' properties
mentioned above.

In the upcoming subsections, in which we deal with concrete fields,
we will always fix a quasi-free Hadamard state on the unital C{*}-algebra
provided by the LCQFT for such field on some globally hyperbolic spacetime
(note that all the LCQFTs we built in Section \ref{secExamples} fulfil
the time slice axiom) and we will perform calculations exploiting
all the properties that we presented here.

\subsection{\label{subKGRCE}Relative Cauchy evolution for the Klein-Gordon field}

In this subsection we follow the calculations in \cite{BFV03} to
show a relation between the functional derivative of the relative
Cauchy evolution for the Klein-Gordon field and its quantized stress-energy
tensor. This relation will be proved in the theorem concluding this
subsection. First of all we need to introduce all the building blocks.

\subsubsection{Relative Cauchy evolution for the classical Klein-Gordon field}

As a starting point we consider Subsection \ref{subKleinGordonField},
where we discussed the construction of a locally covariant quantum
field theory for the Klein-Gordon field applying a specialization
of the general procedure (Section \ref{secBuildingALCQFT}). Here
we use the notation introduced in Subsection \ref{subKleinGordonField}
and in Section \ref{secBuildingALCQFT}.

The first ingredient that we need to consider pertains to the classical
theory of the Klein-Gordon field. Denote with $\mathscr{B}:\mathfrak{ghs}^{KG}\overset{\rightarrow}{\rightarrow}\mathfrak{ssp}$
the covariant functor describing the classical theory of the Klein-Gordon
field built following the procedure of Subsection \ref{subClassicalFieldTheory}
(specialized to the case of the Klein-Gordon field along the line
sketched in Subsection \ref{subKleinGordonField}). In the upcoming
proposition it appears an almost self-explanatory notation, namely
we write $\left.A\right|_{O}$, where $A$ is the formally selfadjoint
normally hyperbolic operator $\mathrm{\Box}_{0}+m^{2}\mathrm{id}_{\mathrm{\Omega}^{0}M}$
governing the Klein-Gordon field. In any case the beginning of the
proof clarifies precisely what $\left.A\right|_{O}$ stands for.
\begin{prop}
\label{propKGExpressionForTheInverseOfAProperSymplecticMap}Let $\mathscr{B}:\mathfrak{ghs}^{KG}\overset{\rightarrow}{\rightarrow}\mathfrak{ssp}$
be the covariant functor describing the classical theory of the Klein-Gordon
field, let $\left(\mathscr{M}=\left(M,g,\mathfrak{o},\mathfrak{t}\right),\mathrm{\Lambda}^{0}M,A\right)$
be an object of $\mathfrak{ghs}^{KG}$ and let $O$ be an $\mathscr{M}$-causally
convex connected open subset of $M$ including a smooth spacelike
Cauchy surface $\Sigma$ for $\mathscr{M}$. Consider $\left(\left.\mathscr{M}\right|_{O},\mathrm{\Lambda}^{0}O,\left.A\right|_{O}\right)\in\mathsf{Obj}_{\mathfrak{ghs}^{KG}}$
and the morphism $\left(\iota_{O}^{M},\iota_{\mathrm{\Lambda}^{0}O}^{\mathrm{\Lambda}^{0}M}\right)$
of $\mathfrak{ghs}^{KG}$ from $\left(\left.\mathscr{M}\right|_{O},\mathrm{\Lambda}^{0}O,\left.A\right|_{O}\right)$
to $\left(\mathscr{M},\mathrm{\Lambda}^{0}M,A\right)$ induced by
the inclusion maps $\iota_{O}^{M}:O\rightarrow M$ and $\iota_{\mathrm{\Lambda}^{0}O}^{\mathrm{\Lambda}^{0}M}:\mathrm{\Lambda}^{0}O\rightarrow\mathrm{\Lambda}^{0}M$.
Then there exists a partition of unity $\left\{ \chi^{a},\chi^{r}\right\} $
on $M$ such that the inverse $\mathscr{B}\left(\iota_{O}^{M},\iota_{\mathrm{\Lambda}^{0}O}^{\mathrm{\Lambda}^{0}M}\right)^{-1}$
of the bijective morphism $\mathscr{B}\left(\iota_{O}^{M},\iota_{\mathrm{\Lambda}^{0}O}^{\mathrm{\Lambda}^{0}M}\right)$
of $\mathfrak{ssp}$ from $\left(V,\sigma\right)=\mathscr{B}\left(\left.\mathscr{M}\right|_{O},\mathrm{\Lambda}^{0}O,\left.A\right|_{O}\right)$
to $\left(W,\omega\right)=\mathscr{B}\left(\mathscr{M},\mathrm{\Lambda}^{0}M,A\right)$
satisfies the following equation:
\[
\mathscr{B}\left(\iota_{O}^{M},\iota_{\mathrm{\Lambda}^{0}O}^{\mathrm{\Lambda}^{0}M}\right)^{-1}\varphi=\pm e_{\left.A\right|_{O}}\left(\mathrm{res}_{\iota_{\mathrm{\Lambda}^{0}O}^{\mathrm{\Lambda}^{0}M}}\left(A\left(\chi^{a/r}\varphi\right)\right)\right)\quad\forall\varphi\in W\mbox{,}
\]
where $e_{\left.A\right|_{O}}$ is the causal propagator for $\left.A\right|_{O}$
and the restriction map is defined in Lemma \ref{lemresPsieBextPsi=00003DeA}.\end{prop}
\begin{proof}
We fix a globally hyperbolic spacetime $\mathscr{M}=\left(M,g,\mathfrak{o},\mathfrak{t}\right)$
and we consider an $\mathscr{M}$-causally convex connected open subset
$O$ of $M$ including a smooth spacelike Cauchy surface $\Sigma$
for $\mathscr{M}$. In Remark \ref{remghs} we saw that we can consider
the globally hyperbolic spacetime $\left.\mathscr{M}\right|_{O}$
and that the inclusion map $\iota_{O}^{M}:O\rightarrow M$ can be
interpreted as a morphism of $\mathfrak{ghs}$ from $\left.\mathscr{M}\right|_{O}$
to $\mathscr{M}$. Exploiting \ref{remRestrictionOfVectorBundles},
we realize that $\left.\mathrm{\Lambda}^{0}M\right|_{O}=\mathrm{\Lambda}^{0}O$
is a vector bundle and that $\left(\iota_{O}^{M},\iota_{\mathrm{\Lambda}^{0}O}^{\mathrm{\Lambda}^{0}M}\right)$
is a vector bundle homomorphism. It follows from the comments made
after Definition \ref{defghsfssp} that $\mathrm{\Lambda}^{0}O$ can
be endowed with the restriction of the inner product on $\mathrm{\Lambda}^{0}M$
and that we can consider the formally selfadjoint normally hyperbolic
operator $A_{\iota_{\left.\mathrm{\Lambda}^{0}M\right|_{O}}^{\mathrm{\Lambda}^{0}M}}$
(for convenience we denote it with $\left.A\right|_{O}$). Hence we
have the object $\left(\left.\mathscr{M}\right|_{O},\mathrm{\Lambda}^{0}O,\left.A\right|_{O}\right)$
of $\mathfrak{ghs}^{KG}$ and, exploiting again the comments made
after Definition \ref{defghsfssp}, we immediately see that $\left(\iota_{O}^{M},\iota_{\mathrm{\Lambda}^{0}O}^{\mathrm{\Lambda}^{0}M}\right)$
is a morphism of $\mathfrak{ghs}^{KG}$ from $\left(\left.\mathscr{M}\right|_{O},\mathrm{\Lambda}^{0}O,\left.A\right|_{O}\right)$
to $\left(\mathscr{M},\mathrm{\Lambda}^{0}M,A\right)$.

Now the main part of the proof begins. We exploit \cite[Thm. 1.2]{BS06}
that provides us (among other things) a diffeomorphism $\psi:M\rightarrow\mathbb{R}\times\Sigma$
such that $\psi^{-1}\left(\left\{ 0\right\} \times\Sigma\right)=\Sigma$
and $\Sigma_{t}=\psi\left(\left\{ t\right\} \times\Sigma\right)$
is a smooth spacelike Cauchy surface for $\mathscr{M}$ for each $t\in\mathbb{R}$.
Since $\Sigma$ is included in $O$ by hypothesis and $O$ is open,
we deduce that $O$ is a neighborhood of $\Sigma$. $\psi^{-1}$ is
continuous, therefore we find $\varepsilon>0$ such that $\psi^{-1}\left(\left[-\varepsilon,\varepsilon\right]\times\Sigma\right)\subseteq O$.
This entails that $\Sigma_{-\varepsilon}$ and $\Sigma_{\varepsilon}$
are smooth spacelike Cauchy surfaces for $\mathscr{M}$ that are contained
in $O$. We consider the open covering $\left\{ I_{+}^{\mathscr{M}}\left(\Sigma_{-\varepsilon}\right),I_{-}^{\mathscr{M}}\left(\Sigma_{\varepsilon}\right)\right\} $
of $M$ and its subordinate partition of unity $\left\{ \chi^{a},\chi^{r}\right\} $.

Take $\varphi\in W$ and denote the causal propagator for the formally
selfadjoint normally hyperbolic operator $A=\mathrm{\Box}_{0}+m^{2}\mathrm{id}_{\mathrm{\Omega}^{0}M}$
with $e_{A}$. As a consequence of the construction of the functor
$\mathscr{B}$, $W=e_{A}\left(\mathrm{\Omega}_{0}^{0}M\right)$. Hence
there exists a compact subset $K$ of $M$ such that $\mathrm{supp}\left(\varphi\right)\subseteq J^{\mathscr{M}}\left(K\right)$.
We define $\varphi^{a/r}=\chi^{a/r}\varphi$:
\[
\mathrm{supp}\left(\varphi^{a/r}\right)\subseteq J_{\pm}^{\mathscr{M}}\left(\Sigma_{\mp\varepsilon}\right)\mbox{.}
\]
We deduce that $\varphi^{a/r}$ is an element of $\mathrm{\Omega}^{0}M$
with $\mathscr{M}$-past/future compact support. Another consequence
of $W=e_{A}\left(\mathrm{\Omega}_{0}^{0}M\right)$ is $A\varphi=0$.
From this fact, together with $\chi^{a}+\chi^{r}=1$, we deduce $A\varphi^{a}=-A\varphi^{r}$,
hence
\[
\mathrm{supp}\left(A\varphi^{a}\right)\subseteq J^{\mathscr{M}}\left(K\right)\cap J_{+}^{\mathscr{M}}\left(\Sigma_{-\varepsilon}\right)\cap J_{-}^{\mathscr{M}}\left(\Sigma_{\varepsilon}\right)\subseteq O\mbox{.}
\]
Exploiting Proposition \ref{propUsefulSubsetsOfGloballyHyperbolicSpacetimes},
we realize that $A\varphi^{a/r}\in\mathrm{\Omega}_{0}^{0}M$ with
support contained in $O$. At this point we can apply the restriction
map (its definition in the general context of arbitrary vector bundles
can be found in Lemma \ref{lemresPsieBextPsi=00003DeA}) in order
to obtain
\[
\mathrm{res}_{\iota_{\mathrm{\Lambda}^{0}O}^{\mathrm{\Lambda}^{0}M}}\left(A\left(\chi^{a/r}\varphi\right)\right)\in\mathrm{\Omega}_{0}^{0}O\mbox{.}
\]
Therefore it makes sense to consider
\[
\pm e_{\left.A\right|_{O}}\left(\mathrm{res}_{\iota_{\mathrm{\Lambda}^{0}O}^{\mathrm{\Lambda}^{0}M}}\left(A\left(\chi^{a/r}\varphi\right)\right)\right)\mbox{.}
\]
This shows that the map
\begin{eqnarray*}
\alpha:W & \rightarrow & V\\
\varphi & \mapsto & \pm e_{\left.A\right|_{O}}\left(\mathrm{res}_{\iota_{\mathrm{\Lambda}^{0}O}^{\mathrm{\Lambda}^{0}M}}\left(A\left(\chi^{a/r}\varphi\right)\right)\right)
\end{eqnarray*}
is well defined.

Note that, from the hypothesis made, we know that the image $\iota_{O}^{M}\left(O\right)=O$
includes a smooth spacelike Cauchy surface for $\mathscr{M}$. Hence
$\mathscr{B}\left(\iota_{O}^{M},\iota_{\mathrm{\Lambda}^{0}O}^{\mathrm{\Lambda}^{0}M}\right)^{-1}$
is a morphism of $\mathfrak{ssp}$ from $\left(W,\omega\right)$ to
$\left(V,\sigma\right)$ because the time slice axiom holds for $\mathscr{B}$
(cfr. Theorem \ref{thmClassicalFieldFunctor}). We must check that
$\alpha=\mathscr{B}\left(\iota_{O}^{M},\iota_{\mathrm{\Lambda}^{0}O}^{\mathrm{\Lambda}^{0}M}\right)^{-1}$.
Take $\varphi\in W$, recall Lemma \ref{lemMorghsf->Morssp} and observe
that the restriction followed by the corresponding extension leaves
the argument of the restriction unchanged:
\begin{eqnarray*}
\mathscr{B}\left(\iota_{O}^{M},\iota_{\mathrm{\Lambda}^{0}O}^{\mathrm{\Lambda}^{0}M}\right)\left(\alpha\varphi\right) & = & \mathscr{B}\left(\iota_{O}^{M},\iota_{\mathrm{\Lambda}^{0}O}^{\mathrm{\Lambda}^{0}M}\right)\left(\pm e_{\left.A\right|_{O}}\left(\mathrm{res}_{\iota_{\mathrm{\Lambda}^{0}O}^{\mathrm{\Lambda}^{0}M}}\left(A\varphi^{a/r}\right)\right)\right)\\
 & = & \pm e_{A}\left(\mathrm{ext}_{\iota_{\mathrm{\Lambda}^{0}O}^{\mathrm{\Lambda}^{0}M}}\left(\mathrm{res}_{\iota_{\mathrm{\Lambda}^{0}O}^{\mathrm{\Lambda}^{0}M}}\left(A\varphi^{a/r}\right)\right)\right)\\
 & = & \pm e_{A}A\varphi^{a/r}\mbox{.}
\end{eqnarray*}
The support properties of $\varphi^{a/r}$ and $A\varphi^{a/r}$ allow
us to apply Lemma \ref{lemExtensionOf ea(Pu)=00003Du}:
\[
\pm e_{A}A\varphi^{a/r}=\pm\left(e_{A}^{a}A\varphi^{a/r}-e_{A}^{r}A\varphi^{a/r}\right)=\varphi^{a}+\varphi^{r}=\varphi\mbox{.}
\]
With this we conclude that
\[
\mathscr{B}\left(\iota_{O}^{M},\iota_{\mathrm{\Lambda}^{0}O}^{\mathrm{\Lambda}^{0}M}\right)\left(\alpha\varphi\right)=\varphi=\mathscr{B}\left(\iota_{O}^{M},\iota_{\mathrm{\Lambda}^{0}O}^{\mathrm{\Lambda}^{0}M}\right)\left(\mathscr{B}\left(\iota_{O}^{M},\iota_{\mathrm{\Lambda}^{0}O}^{\mathrm{\Lambda}^{0}M}\right)^{-1}\varphi\right)\quad\forall\varphi\in W\mbox{.}
\]
Since $\mathscr{B}\left(\iota_{O}^{M},\iota_{\mathrm{\Lambda}^{0}O}^{\mathrm{\Lambda}^{0}M}\right)$
is injective, the last equation entails
\[
\alpha\varphi=\mathscr{B}\left(\iota_{O}^{M},\iota_{\mathrm{\Lambda}^{0}O}^{\mathrm{\Lambda}^{0}M}\right)^{-1}\varphi\quad\forall\varphi\in W\mbox{,}
\]
therefore we realize that the thesis actually holds.
\end{proof}
Now we specialize the definition of the RCE to the case of the Klein-Gordon
field. Consider an object $\left(\mathscr{M},\mathrm{\Lambda}^{0}M,A\right)$
of $\mathfrak{ghs}^{KG}$, take $h\in GHP\left(\mathscr{M}\right)$
and recall the definitions of the morphisms $\imath_{\pm}^{\mathscr{M}}\left[h\right]$
and $\jmath_{\pm}^{\mathscr{M}}\left[h\right]$ introduced before
Definition \ref{defRCE}. Together with the perturbed spacetime $\mathscr{M}\left[h\right]$,
we must also consider the effect of the perturbation of the spacetime
metric on the vector bundle (especially the inner product defined
on it) and on the differential operator $A=\mathrm{\Box}_{0}+m^{2}\mathrm{id}_{\mathrm{\Omega}^{0}M}$.
In this case $\mathrm{\Lambda}^{0}M$ and its inner product (being
simply the fiberwise multiplication of real numbers) remain unchanged,
while we define $A\left[h\right]=\mathrm{\Box}_{0}\left[h\right]+m^{2}\mathrm{id}_{\mathrm{\Omega}^{0}M}$,
where $\mathrm{\Box}_{0}\left[h\right]$ is the d'Alembert operator
defined on $\mathscr{M}\left[h\right]$ for 0-forms, specifically
the metric involved here is $g_{h}=g+h$ in place of $g$. We may
consider the inclusion map $\iota_{\mathrm{\Lambda}^{0}M_{\pm}}^{\mathrm{\Lambda}^{0}M}$,
where $M_{\pm}=M\setminus J_{\mp}^{\mathscr{M}}\left(\mathrm{supp}\left(h\right)\right)$
in accordance with the definitions of $\imath_{\pm}^{\mathscr{M}}\left[h\right]$
and $\jmath_{\pm}^{\mathscr{M}}\left[h\right]$. $\left.A\right|_{M_{\pm}}$
is compatible with $A$ via $\left(\iota_{M_{\pm}}^{M},\iota_{\mathrm{\Lambda}^{0}M_{\pm}}^{\mathrm{\Lambda}^{0}M}\right)$
(see the comments made after Definition \ref{defghsfssp}):
\[
\mathrm{ext}_{\iota_{\mathrm{\Lambda}^{0}M_{\pm}}^{\mathrm{\Lambda}^{0}M}}\left(\left.A\right|_{M_{\pm}}f\right)=A\left(\mathrm{ext}_{\iota_{\mathrm{\Lambda}^{0}M_{\pm}}^{\mathrm{\Lambda}^{0}M}}f\right)\quad\forall f\in\mathrm{\Omega}_{0}^{0}M_{\pm}\mbox{.}
\]
Since the effects of the perturbation $h$ are relevant only inside
$\mathrm{supp}\left(h\right)$, we realize that $A\left[h\right]$
and $A$ act exactly in the same way on sections supported outside
$\mathrm{supp}\left(h\right)$. Together with $\left.A\right|_{M_{\pm}}$,
we may consider $\left.A\left[h\right]\right|_{M_{\pm}}$ and we immediately
recognize that they coincide (we denote both of them with $A_{\pm}\left[h\right]$
in a fashion similar to that used when we introduced $\mathscr{M}_{\pm}\left[h\right]$
to denote $\left.\mathscr{M}\right|_{M_{\pm}}=\left.\mathscr{M}\left[h\right]\right|_{M_{\pm}}$).
All these observations are made in order to introduce the objects
$\left(\mathscr{M}\left[h\right],\mathrm{\Lambda}^{0}M,A\left[h\right]\right)$
and $\left(\mathscr{M}_{\pm}\left[h\right],\mathrm{\Lambda}^{0}M_{\pm},A_{\pm}\left[h\right]\right)$
of $\mathfrak{ghs}^{KG}$ and to interpret the vector bundle homomorphism
$\left(\iota_{M_{\pm}}^{M},\iota_{\mathrm{\Lambda}^{0}M_{\pm}}^{\mathrm{\Lambda}^{0}M}\right):\mathrm{\Lambda}^{0}M_{\pm}\rightarrow\mathrm{\Lambda}^{0}M$
in the following (generally inequivalent) ways (note the analogy with
the definitions of $\imath_{\pm}^{\mathscr{M}}\left[h\right]$ and
$\jmath_{\pm}^{\mathscr{M}}\left[h\right]$ as different morphisms
obtained from the same inclusion map $\iota_{M_{\pm}}^{M}$): 
\begin{eqnarray*}
\left(\imath_{\pm}^{\mathscr{M}}\left[h\right],\imath_{\pm}^{\mathscr{M},\mathrm{\Lambda}^{0}}\left[h\right]\right) & \in & \mathsf{Mor}_{\mathfrak{ghs}^{KG}}\left(\left(\mathscr{M}_{\pm}\left[h\right],\mathrm{\Lambda}^{0}M_{\pm},A_{\pm}\left[h\right]\right),\left(\mathscr{M},\mathrm{\Lambda}^{0}M,A\right)\right)\mbox{,}\\
\left(\jmath_{\pm}^{\mathscr{M}}\left[h\right],\jmath_{\pm}^{\mathscr{M},\mathrm{\Lambda}^{0}}\left[h\right]\right) & \in & \mathsf{Mor}_{\mathfrak{ghs}^{KG}}\left(\left(\mathscr{M}_{\pm}\left[h\right],\mathrm{\Lambda}^{0}M_{\pm},A_{\pm}\left[h\right]\right),\left(\mathscr{M}\left[h\right],\mathrm{\Lambda}^{0}M,A\left[h\right]\right)\right)\mbox{.}
\end{eqnarray*}
Denote with $\mathscr{A}$ the LCQFT (fulfilling both the causality
condition and the time slice axiom) built following the procedure
of Section \ref{secBuildingALCQFT} specialized according to Subsection
\ref{subKleinGordonField}. For $\left(\mathscr{M},\mathrm{\Lambda}^{0}M,A\right)\in\mathsf{Obj}_{\mathfrak{ghs}^{KG}}$
and $h\in GHP\left(\mathscr{M}\right)$ we define the RCE for the
Klein-Gordon field as:
\begin{eqnarray*}
R_{h}^{\mathscr{M}} & = & \mathscr{A}\left(\imath_{-}^{\mathscr{M}}\left[h\right],\imath_{-}^{\mathscr{M},\mathrm{\Lambda}^{0}}\left[h\right]\right)\circ\mathscr{A}\left(\jmath_{-}^{\mathscr{M}}\left[h\right],\jmath_{-}^{\mathscr{M},\mathrm{\Lambda}^{0}}\left[h\right]\right)^{-1}\\
 &  & \circ\mathscr{A}\left(\jmath_{+}^{\mathscr{M}}\left[h\right],\jmath_{+}^{\mathscr{M},\mathrm{\Lambda}^{0}}\left[h\right]\right)\circ\mathscr{A}\left(\imath_{+}^{\mathscr{M}}\left[h\right],\imath_{+}^{\mathscr{M},\mathrm{\Lambda}^{0}}\left[h\right]\right)^{-1}\mbox{.}
\end{eqnarray*}
In a similar way one can consider a classical version of the RCE based
on the covariant functor $\mathscr{B}$ describing the classical theory
of the Klein-Gordon field (this is actually possible due to version
of the time slice axiom satisfied by $\mathscr{B}$, cfr. Theorem
\ref{thmClassicalFieldFunctor}):
\begin{eqnarray*}
r_{h}^{\mathscr{M}} & = & \mathscr{B}\left(\imath_{-}^{\mathscr{M}}\left[h\right],\imath_{-}^{\mathscr{M},\mathrm{\Lambda}^{0}}\left[h\right]\right)\circ\mathscr{B}\left(\jmath_{-}^{\mathscr{M}}\left[h\right],\jmath_{-}^{\mathscr{M},\mathrm{\Lambda}^{0}}\left[h\right]\right)^{-1}\\
 &  & \circ\mathscr{B}\left(\jmath_{+}^{\mathscr{M}}\left[h\right],\jmath_{+}^{\mathscr{M},\mathrm{\Lambda}^{0}}\left[h\right]\right)\circ\mathscr{B}\left(\imath_{+}^{\mathscr{M}}\left[h\right],\imath_{+}^{\mathscr{M},\mathrm{\Lambda}^{0}}\left[h\right]\right)^{-1}\mbox{.}
\end{eqnarray*}
Since the LCQFT $\mathscr{A}$ is obtained via composition of $\mathscr{B}$
with the quantization functor $\mathscr{C}$ presented in Subsection
\ref{subQuantumFieldTheory}, we almost immediately realize that
\begin{equation}
R_{h}^{\mathscr{M}}=\mathscr{C}\left(r_{h}^{\mathscr{M}}\right)\label{eqKGQuantumVsClassicalRCE}
\end{equation}
(this is simply a consequence of the covariant axioms fulfilled by
any covariant functor). We can determine the action of $r_{h}^{\mathscr{M}}$
applying Proposition \ref{propKGExpressionForTheInverseOfAProperSymplecticMap}
and Lemma \ref{lemMorghsf->Morssp}. We find proper partitions of
unity $\left\{ \chi_{+}^{a},\chi_{+}^{r}\right\} $ and $\left\{ \chi_{-}^{a},\chi_{-}^{r}\right\} $
on $M$ such that we can express $\mathscr{B}\left(\imath_{+}^{\mathscr{M}}\left[h\right],\imath_{+}^{\mathscr{M},\mathrm{\Lambda}^{0}}\left[h\right]\right)^{-1}$
and respectively $\mathscr{B}\left(\jmath_{-}^{\mathscr{M}}\left[h\right],\jmath_{-}^{\mathscr{M},\mathrm{\Lambda}^{0}}\left[h\right]\right)^{-1}$
according to Proposition \ref{propKGExpressionForTheInverseOfAProperSymplecticMap}.
If we take $\varphi\in\mathscr{B}\left(\mathscr{M},\mathrm{\Lambda}^{0}M,A\right)$
and evaluate $r_{h}^{\mathscr{M}}\varphi$, we easily obtain the following
result: 
\begin{equation}
r_{h}^{\mathscr{M}}\varphi=e_{A}A\left[h\right]\left(\chi_{-}^{a/r}e_{A\left[h\right]}A\left(\chi_{+}^{a/r}\varphi\right)\right)\mbox{.}\label{eqKGExpressionForTheClassicalRCE}
\end{equation}

In the following we will need the expression of the derivative $\left.\frac{\mathrm{d}}{\mathrm{d}s}r_{h^{s}}^{\mathscr{M}}\varphi\right|_{0}$
for an arbitrary smooth 1-parameter family of perturbations of the
metric. For convenience in the upcoming calculation we write $\mathrm{\delta}_{s}$
in place of $\left.\frac{\mathrm{d}}{\mathrm{d}s}\left(\cdot\right)\right|_{0}$.
Fix now $\varphi\in\mathscr{B}\left(\mathscr{M},\mathrm{\Lambda}^{0}M,A\right)$,
a compact subset $K$ of $M$ and a smooth 1-parameter family of globally
hyperbolic perturbations $\left(-1,1\right)\rightarrow GHP\left(\mathscr{M},K\right)$,
$s\mapsto h^{s}$ and evaluate $\mathrm{\delta}_{s}r_{h^{s}}^{\mathscr{M}}\varphi$.
Our starting point is eq. \eqref{eqKGExpressionForTheClassicalRCE}
with the choice of the superscript $r$ (if we choose $a$, we face
a very similar calculation and we indeed obtain the same result).
In the present situation apparently we would have to consider different
partitions of unity $\left\{ \chi_{+}^{a},\chi_{+}^{r}\right\} $
and $\left\{ \chi_{-}^{a},\chi_{-}^{r}\right\} $ for each of the
values assumed by $s$. Anyway this difficulty can be avoided making
an intelligent choice of the smooth spacelike Cauchy surfaces used
to define the partitions of unity: We use always the same foliation
of $\mathscr{M}$ (induced by some fixed smooth Cauchy surface $\Sigma$
for $\mathscr{M}$) and take the smooth spacelike Cauchy surfaces
inside $M_{\pm}=M\setminus J_{\mp}^{\mathscr{M}}\left(K\right)$ instead
of choosing, for each value of $s$, a pair of proper smooth spacelike
Cauchy surfaces inside $M_{\pm}=M\setminus J_{\mp}^{\mathscr{M}}\left(\mathrm{supp}\left(h^{s}\right)\right)$.
In this way a single choice of the smooth spacelike Cauchy surfaces
is satisfactory for each value of $s$. Such choice is possible because
the supports of all the elements $h^{s}$ in the family of perturbations
are controlled by the compact subset $K$ of $M$.

In the first step we apply the Leibniz rule%
\footnote{\label{fnSequentialContinuityOfCausalPropagators}note that causal
propagators are sequentially continuous with respect to a proper notion
of convergence, refer to \cite[Def. 3.4.6, p. 90 and Prop. 3.4.8, p. 91]{BGP07}%
}: 
\[
\mathrm{\delta}_{s}r_{h^{s}}^{\mathscr{M}}\varphi=e_{A}\left(\left(\mathrm{\delta}_{s}A\left[h^{s}\right]\right)\left(\chi_{-}^{r}e_{A}A\left(\chi_{+}^{r}\varphi\right)\right)+A\left(\chi_{-}^{r}\left(\mathrm{\delta}_{s}e_{A\left[h^{s}\right]}\right)A\left(\chi_{+}^{r}\varphi\right)\right)\right)\mbox{.}
\]
We focus on the first addend: On the one hand, following the proof
of Proposition \ref{propKGExpressionForTheInverseOfAProperSymplecticMap}
(we are considering $M_{-}=M\setminus J_{+}^{\mathscr{M}}\left(K\right)$
as $O$), we can easily see that $\mathrm{supp}\left(\chi_{-}^{r}\right)\subseteq J_{-}^{\mathscr{M}}\left(M_{-}\right)$,
while on the other hand $\mathrm{\delta}_{s}A\left[h^{s}\right]$
can have coefficients different from zero only inside $K$. This entails
that
\[
\left(\mathrm{\delta}_{s}A\left[h^{s}\right]\right)\left(\chi_{-}^{r}e_{A}A\left(\chi_{+}^{r}\varphi\right)\right)=0\mbox{,}
\]
therefore we obtain
\[
\mathrm{\delta}_{s}r_{h^{s}}^{\mathscr{M}}\varphi=e_{A}A\left(\chi_{-}^{r}\left(\mathrm{\delta}_{s}e_{A\left[h^{s}\right]}\right)A\left(\chi_{+}^{r}\varphi\right)\right)\mbox{.}
\]
Recalling again the proof of Proposition \ref{propKGExpressionForTheInverseOfAProperSymplecticMap},
we realize that $A\left(\chi_{+}^{r}\varphi\right)=-A\left(\chi_{+}^{a}\varphi\right)$
and deduces that its support is compact and lies in the causal future
of a smooth spacelike Cauchy surface for $\mathscr{M}$ included in
$M_{+}=M\setminus J_{-}^{\mathscr{M}}\left(K\right)$ (that by construction
lies outside $K$ and intersects its causal future). On the contrary
$\chi_{-}^{r}$ is supported in the causal past of a smooth spacelike
Cauchy surface for $\mathscr{M}$ included in $M_{-}$ (that by construction
lies outside $K$ and intersects its causal past). These observations
entail that $\chi_{-}^{r}e_{A\left[h^{s}\right]}^{a}A\left(\chi_{+}^{r}\varphi\right)$
has empty support, hence it is null. Therefore from the last equation
we obtain
\begin{equation}
\mathrm{\delta}_{s}r_{h^{s}}^{\mathscr{M}}\varphi=-e_{A}A\left(\chi_{-}^{r}\left(\mathrm{\delta}_{s}e_{A\left[h^{s}\right]}^{r}\right)A\left(\chi_{+}^{r}\varphi\right)\right)\mbox{.}\label{eqKGExpressionForTheFunctionalDerivativeOfTheClassicalRCEStep1}
\end{equation}

Now we take a closer look to the term $e_{A\left[h^{s}\right]}^{r}A\left[h^{s}\right]\left(\chi_{+}^{r}\varphi\right)$.
In order for this term to make sense it must be shown that $A\left[h^{s}\right]\left(\chi_{+}^{r}\varphi\right)$
has compact support. This follows from the the following facts:
\begin{itemize}
\item $A\left(\chi_{+}^{r}\varphi\right)=-A\left(\chi_{+}^{a}\varphi\right)$
implies that $A\left(\chi_{+}^{a/r}\varphi\right)$ has compact support
(note that $\chi_{+}^{a/r}$ is supported in the causal future/past
of a proper smooth spacelike Cauchy surface for $\mathscr{M}$ and
remember that $\mathrm{supp}\left(\varphi\right)\subseteq J^{\mathscr{M}}\left(K^{\prime}\right)$
for a proper compact subset $K^{\prime}$ of $M$);
\item $A\left[h^{s}\right]$ differs from $A$ only inside $K$, which is
compact.
\end{itemize}
This two facts imply that
\[
\mathrm{supp}\left(A\left[h^{s}\right]\left(\chi_{+}^{a/r}\varphi\right)\right)\subseteq\mathrm{supp}\left(A\left(\chi_{+}^{a/r}\varphi\right)\right)\cup K\mbox{,}
\]
hence $A\left[h^{s}\right]\left(\chi_{+}^{a/r}\varphi\right)$ is
compactly supported too. From the first point above it follows also
that $\chi_{+}^{a/r}\varphi$ has past/future compact support (we
are exploiting Proposition \ref{propUsefulSubsetsOfGloballyHyperbolicSpacetimes}).
Hence we can apply Lemma \ref{lemExtensionOf ea(Pu)=00003Du} to conclude
that for each $s$ we have
\[
e_{A\left[h^{s}\right]}^{a/r}A\left[h^{s}\right]\left(\chi_{+}^{a/r}\varphi\right)=\chi_{+}^{a/r}\varphi\mbox{.}
\]
Exploiting the Leibniz rule, we find
\[
0=\mathrm{\delta}_{s}\left(\chi_{+}^{r}\varphi\right)=\mathrm{\delta}_{s}\left(e_{A\left[h^{s}\right]}^{r}A\left[h^{s}\right]\left(\chi_{+}^{r}\varphi\right)\right)=\left(\mathrm{\delta}_{s}e_{A\left[h^{s}\right]}^{r}\right)A\left(\chi_{+}^{r}\varphi\right)+e_{A}^{r}\left(\mathrm{\delta}_{s}A\left[h^{s}\right]\right)\left(\chi_{+}^{r}\varphi\right)\mbox{.}
\]
With this identity we can rewrite eq. \eqref{eqKGExpressionForTheFunctionalDerivativeOfTheClassicalRCEStep1}:
\[
\mathrm{\delta}_{s}r_{h^{s}}^{\mathscr{M}}\varphi=e_{A}A\left(\chi_{-}^{r}e_{A}^{r}\left(\mathrm{\delta}_{s}A\left[h^{s}\right]\right)\left(\chi_{+}^{r}\varphi\right)\right)\mbox{.}
\]
Notice that $\left(\mathrm{\delta}_{s}A\left[h^{s}\right]\right)\left(\chi_{+}^{a}\varphi\right)=0$
because the coefficients of $\mathrm{\delta}_{s}A\left[h^{s}\right]$
are supported inside $K$ while $\chi_{+}^{a}$ is supported in the
causal future of $M_{+}$. Hence we can add such term without modifying
the result:
\[
\left(\mathrm{\delta}_{s}A\left[h^{s}\right]\right)\left(\chi_{+}^{r}\varphi\right)=\left(\mathrm{\delta}_{s}A\left[h^{s}\right]\right)\left(\chi_{+}^{r}\varphi\right)+\left(\mathrm{\delta}_{s}A\left[h^{s}\right]\right)\left(\chi_{+}^{a}\varphi\right)=\left(\mathrm{\delta}_{s}A\left[h^{s}\right]\right)\varphi\mbox{.}
\]
In this way we obtain
\[
\mathrm{\delta}_{s}r_{h}^{\mathscr{M}}\varphi=e_{A}A\left(\chi_{-}^{r}e_{A}^{r}\left(\mathrm{\delta}_{s}A\left[h^{s}\right]\right)\varphi\right)\mbox{.}
\]

Take into account the term $\chi_{-}^{r}e_{A}^{a}\left(\mathrm{\delta}_{s}A\left[h^{s}\right]\right)\varphi$:
the coefficients of $\mathrm{\delta}_{s}A\left[h^{s}\right]$ are
supported inside $K$, hence
\[
\mathrm{supp}\left(e_{A}^{a}\left(\mathrm{\delta}_{s}A\left[h^{s}\right]\right)\varphi\right)\subseteq J_{+}^{\mathscr{M}}\left(K\right)\mbox{,}
\]
while $\chi_{-}^{r}$ is supported inside $J_{-}^{\mathscr{M}}\left(M_{-}\right)$.
This entails that $\chi_{-}^{r}e_{A}^{a}\left(\mathrm{\delta}_{s}A\left[h^{s}\right]\right)\varphi=0$,
therefore we can modify again our last equation with the subtraction
of this term leaving the result unchanged:
\begin{eqnarray*}
\mathrm{\delta}_{s}r_{h^{s}}^{\mathscr{M}}\varphi & = & e_{A}A\left(\chi_{-}^{r}e_{A}^{r}\left(\mathrm{\delta}_{s}A\left[h^{s}\right]\right)\varphi\right)-e_{A}A\left(\chi_{-}^{r}e_{A}^{a}\left(\mathrm{\delta}_{s}A\left[h^{s}\right]\right)\varphi\right)\\
 & = & -e_{A}A\left(\chi_{-}^{r}e_{A}\left(\mathrm{\delta}_{s}A\left[h^{s}\right]\right)\varphi\right)\mbox{.}
\end{eqnarray*}

The observation about the support of the coefficients appearing in
the linear differential operator $\mathrm{\delta}_{s}A\left[h^{s}\right]$
entails that $f=\left(\mathrm{\delta}_{s}A\left[h^{s}\right]\right)\varphi$
is an element of $\mathrm{\Omega}_{0}^{0}M$ with support included
in $K$ and trivially we have $Ae_{A}f=0$, so that 
\[
A\left(\chi_{-}^{r}e_{A}f\right)=-A\left(\chi_{-}^{a}e_{A}f\right)\mbox{.}
\]
On account of the last identity, the inclusion $\mathrm{supp}\left(e_{A}f\right)\subseteq J^{\mathscr{M}}\left(K\right)$
and the identity $\mathrm{supp}\left(\chi_{-}^{a/r}\right)=J_{\pm}^{\mathscr{M}}\left(\Sigma_{-}^{a/r}\right)$
for proper smooth spacelike Cauchy surfaces $\Sigma_{-}^{a/r}$ and
applying Proposition \ref{propUsefulSubsetsOfGloballyHyperbolicSpacetimes}
and Lemma \ref{lemExtensionOf ea(Pu)=00003Du}, we obtain the following
result:
\[
-e_{A}A\left(\chi_{-}^{r}e_{A}f\right)=e_{A}^{a}A\left(\chi_{-}^{a}e_{A}f\right)+e_{A}^{r}A\left(\chi_{-}^{r}e_{A}f\right)=\chi_{-}^{a}e_{A}f+\chi_{-}^{r}e_{A}f=e_{A}f\mbox{.}
\]
With the last identity we conclude
\begin{equation}
\left.\frac{\mathrm{d}}{\mathrm{d}s}r_{h^{s}}^{\mathscr{M}}\varphi\right|_{0}=e_{A}\left(\left.\frac{\mathrm{d}}{\mathrm{d}s}A\left[h^{s}\right]\right|_{0}\right)\varphi\mbox{.}\label{eqKGExpressionForTheDerivativeOfTheClassicalRCE}
\end{equation}

We are left with the problem of the expression for $\mathrm{\delta}_{s}A\left[h^{s}\right]\varphi$.
We know that $A\left[h^{s}\right]=\mathrm{\Box}_{0}\left[h^{s}\right]+m^{2}\mathrm{id}_{\mathrm{\Omega}^{0}M}$,
where $\mathrm{\Box}_{0}\left[h^{s}\right]$ denotes the d'Alembert
operator built with the perturbed metric $g_{h^{s}}$. Indeed the
term $m^{2}\varphi$ gives null contribution to $\mathrm{\delta}_{s}A\left[h^{s}\right]\varphi$,
hence we are interested in the evaluation of $\mathrm{\delta}_{s}\mathrm{\Box}_{0}\left[h^{s}\right]\varphi$.
Using an arbitrary coordinate neighborhood, we see from eq. \eqref{eqKleinGordonEquationInIndexNotation}
that
\[
\mathrm{\Box}_{0}\left[h^{s}\right]\varphi=-g_{h^{s}}^{ij}\nabla\left[h^{s}\right]_{i}\partial_{j}\varphi=-g_{h^{s}}^{ij}\partial_{i}\partial_{j}\varphi+g_{h^{s}}^{ij}\Gamma\left[h^{s}\right]_{ij}^{k}\partial_{k}\varphi\mbox{,}
\]
where $\Gamma\left[h^{s}\right]_{ij}^{k}$ are the Christoffel symbols
of the Levi-Civita connection $\nabla\left[h^{s}\right]$ on $\mathscr{M}\left[h^{s}\right]$,
and therefore
\begin{equation}
\left.\frac{\mathrm{d}}{\mathrm{d}s}A\left[h^{s}\right]\varphi\right|_{0}=\left.\frac{\mathrm{d}}{\mathrm{d}s}\mathrm{\Box}_{0}\left[h^{s}\right]\varphi\right|_{0}=\left.\frac{\mathrm{d}}{\mathrm{d}s}h_{ij}^{s}\right|_{0}\nabla^{j}\nabla^{j}\varphi+\left.\frac{\mathrm{d}}{\mathrm{d}s}\Gamma\left[h^{s}\right]_{ij}^{k}\right|_{0}g^{ij}\nabla_{k}\varphi\mbox{,}\label{eqKGDifferentialOperatorVariation}
\end{equation}
where in the last step we exploited the relation
\begin{equation}
\left.\frac{\mathrm{d}}{\mathrm{d}s}h_{ij}^{s}\right|_{0}=\left.\frac{\mathrm{d}}{\mathrm{d}s}g_{h^{s}ij}\right|_{0}=-\left.\frac{\mathrm{d}}{\mathrm{d}s}g_{h^{s}}^{kl}\right|_{0}g_{ki}g_{lj}\label{eqLoweringTheIndicesOfTheMetricVariation}
\end{equation}
that follows from $g_{h^{s}}=g+h^{s}$ and $g_{h^{s}}^{kl}g_{h^{s}ki}g_{h^{s}lj}=g_{h^{s}ij}$.

\subsubsection{Properties of the GNS representation induced by a quasi-free Hadamard
state for the Klein-Gordon field}

The second preparatory step is the choice of a quasi-free Hadamard
state $\tau$ for the unital C{*}-algebra $\left(\mathcal{V},\mathrm{V}\right)=\mathscr{A}\left(\mathscr{M},\mathrm{\Lambda}^{0}M,A\right)$
(which is actually a CCR representation) describing the Klein-Gordon
field on the globally hyperbolic spacetime $\mathscr{M}$. With this
choice, we introduce the (unique up to unitary equivalence) GNS triple
$\left(\mathscr{H}_{\tau}^{\mathscr{M}},\pi_{\tau}^{\mathscr{M}},\Omega_{\tau}^{\mathscr{M}}\right)$
induced by $\tau$ and we follow the discussion made in Subsection
\ref{subQuasiFreeHadamardStates}. In this way we obtain the represented
version
\begin{equation}
\mathrm{V}_{\tau}^{\mathscr{M}}=\pi_{\tau}^{\mathscr{M}}\circ\mathrm{V}:V\rightarrow\mathcal{B}\left(\mathscr{H}_{\tau}^{\mathscr{M}}\right)\label{eqKGRepresentedWeylMap}
\end{equation}
of the Weyl map $\mathrm{V}$, where $\left(V,\sigma\right)=\mathscr{B}\left(\mathscr{M},\mathrm{\Lambda}^{0}M,A\right)$
is the symplectic space provided by the covariant functor $\mathscr{B}$
describing the classical theory of the Klein-Gordon field, together
with the map
\begin{eqnarray*}
\varPhi_{\tau}^{\mathscr{M}}:V & \rightarrow & \mathcal{B}\left(\mathscr{H}_{\tau}^{\mathscr{M}}\right)\\
\varphi & \mapsto & \varPhi_{\tau}^{\mathscr{M}}\left(\varphi\right)
\end{eqnarray*}
that allows us to express the unitary operator $\mathrm{V}_{\tau}^{\mathscr{M}}\left(\varphi\right)$
as the complex exponential of a selfadjoint operator for each $\varphi\in V$,
namely $\varPhi_{\tau}^{\mathscr{M}}\left(\varphi\right)$ is selfadjoint
and satisfies $\mathrm{e}^{\imath\varPhi_{\tau}^{\mathscr{M}}\left(\varphi\right)}=\mathrm{V}_{\tau}^{\mathscr{M}}\left(\varphi\right)$.
Together with this map, we have the smeared fields (by virtue of the
choice of a Hadamard state):
\begin{eqnarray*}
\varPsi_{\tau}^{\mathscr{M}}:\mathrm{\Omega}_{0}^{0}M & \rightarrow & \mathcal{B}\left(\mathscr{H}_{\tau}^{\mathscr{M}}\right)\\
f & \mapsto & -\imath\left.\frac{\mathrm{d}}{\mathrm{d}t}\mathrm{V}_{\tau}^{\mathscr{M}}\left(te_{A}f\right)\right|_{0}\mbox{.}
\end{eqnarray*}
As for the general case, it holds that
\begin{equation}
\varPsi_{\tau}^{\mathscr{M}}\left(f\right)=\varPhi_{\tau}^{\mathscr{M}}\left(e_{A}f\right)\label{eqKGRelationBetweenvarPsiAndvarPhi}
\end{equation}
for each $f\in\mathrm{\Omega}_{0}^{0}M$ and we recognize $\varPsi_{\tau}^{\mathscr{M}}$
to be linear.

We stated the most relevant consequence of the choice of a quasi-free
Hadamard state $\tau$ in Subsection \ref{subQuasiFreeHadamardStates}:
\begin{itemize}
\item Assumption \ref{assFunctionalDerivativeRCE} is verified, i.e. we
find a dense subspace $\mathscr{V}_{\tau}^{\mathscr{M}}$ of $\mathscr{H}_{\tau}^{\mathscr{M}}$
and a dense sub-{*}-algebra $\mathcal{B}_{\tau}^{\mathscr{M}}$ of
$\mathscr{A}\left(\mathscr{M},\mathrm{\Lambda}^{0}M,A\right)$ such
that the functional derivative of the RCE with respect to the spacetime
metric can be defined;
\item specializing eq. \eqref{eqKeyPointForTheMainTheoremAboutRCE}  to
the case of the Klein-Gordon field, we see that the following equation
holds for each $\theta\in\mathscr{V}_{\tau}^{\mathscr{M}}$, each
$\varphi\in V$, each compact subset $K$ of $M$ and each smooth
$1$-parameter family $\left(-1,1\right)\rightarrow GHP\left(\mathscr{M},K\right)$,
$s\mapsto h^{s}$ such that $h^{0}=0$:
\begin{equation}
\left.\frac{\mathrm{d}}{\mathrm{d}s}\left\langle \theta,\mathrm{V}_{\tau}^{\mathscr{M}}\left(r_{h^{s}}^{\mathscr{M}}\varphi\right)\theta\right\rangle _{\tau}^{\mathscr{M}}\right|_{0}=\frac{\imath}{2}\left\langle \theta,\left\{ \varPhi_{\tau}^{\mathscr{M}}\left(\left.\frac{\mathrm{d}}{\mathrm{d}s}\left(r_{h^{s}}^{\mathscr{M}}\varphi\right)\right|_{0}\right),\mathrm{V}_{\tau}^{\mathscr{M}}\left(\varphi\right)\right\} \theta\right\rangle _{\tau}^{\mathscr{M}}\mbox{.}\label{eqKGKeyPointForTheMainTheoremAboutRCE}
\end{equation}

\end{itemize}
Finally one can show that for each $\eta$, $\xi\in\mathscr{V}_{\tau}^{\mathscr{M}}$
there exists a smooth section, that we denote with
\begin{eqnarray}
M & \rightarrow & \mathbb{C}\label{eqKGUnsmearedField}\\
p & \mapsto & \left\langle \eta,\varPsi_{\tau}^{\mathscr{M}}\left(p\right)\xi\right\rangle _{\tau}^{\mathscr{M}}\mbox{,}\nonumber 
\end{eqnarray}
where $\left\langle \cdot,\cdot\right\rangle _{\tau}^{\mathscr{M}}$
denotes the scalar product of the Hilbert space $\mathscr{H}_{\tau}^{\mathscr{M}}$,
such that
\begin{equation}
\left\langle \eta,\varPsi_{\tau}^{\mathscr{M}}\left(f\right)\xi\right\rangle _{\tau}^{\mathscr{M}}=\int\limits _{M}\left\langle \eta,\varPsi_{\tau}^{\mathscr{M}}\left(p\right)\xi\right\rangle _{\tau}^{\mathscr{M}}f\left(p\right)\mathrm{d}\mu_{g}\label{eqKGRelationBetweenUnsmearedAndSmearedFields}
\end{equation}
for each $f\in\mathrm{\Omega}_{0}^{0}M$, where $\mathrm{d}\mu_{g}$
is the standard volume form on $\mathscr{M}$. We may regard this
section as (the matrix element of) the unsmeared field. Uniqueness
of the unsmeared field is a direct consequence of the last equation.

\subsubsection{Quantized stress-energy tensor for the Klein-Gordon field}

We still need to find the expression of the quantized stress-energy
tensor. This is obtained starting from the action of the Klein-Gordon
field on the globally hyperbolic spacetime $\mathscr{M}$, which in
turn comes from the differential operator $A=\mathrm{\Box}_{0}+m^{2}\mathrm{id}_{\mathrm{\Omega}^{0}M}$
governing the field:
\[
S_{\mathscr{M}}=\frac{1}{2}\left(\varphi,A\varphi\right)_{g,0}=\frac{1}{2}\left(\mathrm{d}\varphi,\mathrm{d}\varphi\right)_{g,1}+\frac{1}{2}m^{2}\left(\varphi,\varphi\right)_{g,0}=\frac{1}{2}\int\limits _{M}\left(\mathrm{d}\varphi\wedge*\mathrm{d}\varphi+m^{2}\varphi\wedge*\varphi\right)\mbox{.}
\]
From the expression of $S_{\mathscr{M}}$, we find the classical stress-energy
tensor (written in some coordinate neighborhood) via functional differentiation
with respect to the metric:
\begin{eqnarray*}
T_{ij}^{\mathscr{M}}\left(p\right) & = & \left.\frac{2}{\sqrt{\left|\det g_{h}\left(p\right)\right|}}\frac{\mathrm{\delta}S_{\mathscr{M}\left[h\right]}}{\mathrm{\delta}g_{h}^{ij}\left(p\right)}\right|_{0}\\
 & = & \nabla_{i}\varphi\left(p\right)\nabla_{j}\varphi\left(p\right)-\frac{1}{2}g_{ij}\left(p\right)g^{kl}\left(p\right)\nabla_{k}\varphi\left(p\right)\nabla_{l}\varphi\left(p\right)-\frac{1}{2}m^{2}g_{ij}\left(p\right)\varphi^{2}\left(p\right)\mbox{.}
\end{eqnarray*}

The choice of a quasi-free Hadamard state $\tau$ allows us to promote
$T_{ij}^{\mathscr{M}}$ to the renormalized quantum stress-energy
tensor $\mathcal{T}_{\tau\, ij}^{\mathscr{M}}$ simply with the formal
replacement of the classical field $\varphi\left(p\right)$ with (the
matrix elements of) the unsmeared field $p\in M\mapsto\left\langle \eta,\varPsi_{\tau}^{\mathscr{M}}\left(p\right)\xi\right\rangle _{\tau}^{\mathscr{M}}$
defined in eq. \eqref{eqKGUnsmearedField} for each $\eta$, $\xi\in\mathscr{V}_{\tau}^{\mathscr{M}}$.
This regularization procedure is known as point-splitting (refer to
\cite[eq. 4.6.5, p. 88]{Wal94}): For each $\eta$, $\xi\in\mathscr{V}_{\tau}^{\mathscr{M}}$,
we choose two {}``near'' points $p$ and $q$ in $M$ and a curve
$\gamma$ connecting them and, parallel transporting along the curve
$\gamma$, we write
\begin{eqnarray}
\left\langle \eta,\mathcal{T}_{\tau}^{\mathscr{M}\, ij}\left(p,q\right)\xi\right\rangle _{\tau}^{\mathscr{M}} & = & \left\langle \eta,\nabla^{i}\varPsi_{\tau}^{\mathscr{M}}\left(p\right)\nabla^{j}\varPsi_{\tau}^{\mathscr{M}}\left(q\right)\xi\right\rangle _{\tau}^{\mathscr{M}}\nonumber \\
 &  & -\frac{1}{2}g^{ia}\left(p\right)Y_{\gamma\, a}^{j}g^{kb}\left(p\right)Y_{\gamma\, b}^{l}\left\langle \eta,\nabla_{k}\varPsi_{\tau}^{\mathscr{M}}\left(p\right)\nabla_{l}\varPsi_{\tau}^{\mathscr{M}}\left(q\right)\xi\right\rangle _{\tau}^{\mathscr{M}}\nonumber \\
 &  & -\frac{1}{2}m^{2}g^{ia}\left(p\right)Y_{\gamma\, a}^{j}\left\langle \eta,\varPsi_{\tau}^{\mathscr{M}}\left(p\right)\varPsi_{\tau}^{\mathscr{M}}\left(q\right)\xi\right\rangle _{\tau}^{\mathscr{M}}\mbox{.}\label{eqKGStressEnergyTensorViaPointSplitting}
\end{eqnarray}
Finally we must take the limit $q\rightarrow p$ once that all the
divergences are removed. The point-splitting procedure involves the
parallel transport $Y_{\gamma}$ (see Definition \ref{defParallelTransport}),
which depends upon the choice of the curve $\gamma$ connecting the
point $p$ to the point $q$. It follows that the expression above
depends on the choice of such curve. Anyway this ambiguity is avoided
if we assume that $p$ and $q$ are sufficiently close to have a unique
geodesic connecting them and we choose such geodesic as $\gamma$.
This assumption can be done because in our calculation we will finally
take the limit $q\rightarrow p$ along the chosen curve. As a matter
of fact the expression of the stress-energy tensor renormalized with
respect to the state $\tau$ as reference differs from the expression
given above by a multiple of the identity operator. However such term
is irrelevant for our calculations since the stress-energy tensor
will appear only inside a commutator.

As we said, the stress-energy tensor appears in our subsequent calculations
only in a commutator, specifically a commutator with an arbitrary
Weyl generator (represented via $\pi_{\tau}^{\mathscr{M}}$). A cursory
glance to eq. \eqref{eqKGStressEnergyTensorViaPointSplitting} shows
that it is useful for us to evaluate the matrix elements of the commutator
of two unsmeared fields with an arbitrary represented Weyl generator.
To this end we evaluate separately the matrix elements arising from
the LHS and the RHS of eq. \eqref{eqCommutatorBetween2SmearedFieldsAndAWeylGenerator}.
We fix $\eta$, $\xi\in\mathscr{V}_{\tau}^{\mathscr{M}}$, $f$, $f^{\prime}\in\mathrm{\Omega}_{0}^{0}M$
and $\varphi\in V$, where $\left(V,\sigma\right)=\mathscr{B}\left(\mathscr{M},\mathrm{\Lambda}^{0}M,A\right)$,
and we use eq. \eqref{eqKGRelationBetweenUnsmearedAndSmearedFields}
twice: 
\begin{multline*}
\iint\limits _{M}\left\langle \eta,\left[\varPsi_{\tau}^{\mathscr{M}}\left(p\right)\varPsi_{\tau}^{\mathscr{M}}\left(q\right),\mathrm{V}_{\tau}^{\mathscr{M}}\left(\varphi\right)\right]\xi\right\rangle _{\tau}^{\mathscr{M}}f\left(p\right)f^{\prime}\left(q\right)\mathrm{d}\mu_{g}\left(p\right)\mathrm{d}\mu_{g}\left(q\right)\\
=\left\langle \eta,\left[\varPsi_{\tau}^{\mathscr{M}}\left(f\right)\varPsi_{\tau}^{\mathscr{M}}\left(f^{\prime}\right),\mathrm{V}_{\tau}^{\mathscr{M}}\left(\varphi\right)\right]\xi\right\rangle _{\tau}^{\mathscr{M}}\mbox{.}
\end{multline*}
Now we exploit also the definition of the symplectic form $\sigma$
(cfr. Lemma \ref{lemObjghsf->Objssp}):
\begin{multline*}
-\sigma\left(e_{A}f,\varphi\right)\left\langle \eta,\mathrm{V}_{\tau}^{\mathscr{M}}\left(\varphi\right)\varPsi_{\tau}^{\mathscr{M}}\left(f^{\prime}\right)\xi\right\rangle _{\tau}^{\mathscr{M}}\\
=\iint\limits _{M}\varphi\left(p\right)f\left(p\right)f^{\prime}\left(q\right)\left\langle \eta,\mathrm{V}_{\tau}^{\mathscr{M}}\left(\varphi\right)\varPsi_{\tau}^{\mathscr{M}}\left(q\right)\xi\right\rangle _{\tau}^{\mathscr{M}}\mathrm{d}\mu_{g}\left(p\right)\mathrm{d}\mu_{g}\left(q\right)\mbox{,}
\end{multline*}
\begin{multline*}
-\sigma\left(e_{A}f^{\prime},\varphi\right)\left\langle \eta,\varPsi_{\tau}^{\mathscr{M}}\left(f\right)\mathrm{V}_{\tau}^{\mathscr{M}}\left(\varphi\right)\xi\right\rangle _{\tau}^{\mathscr{M}}\\
=\iint\limits _{M}\varphi\left(q\right)f^{\prime}\left(q\right)f\left(p\right)\left\langle \eta,\varPsi_{\tau}^{\mathscr{M}}\left(p\right)\mathrm{V}_{\tau}^{\mathscr{M}}\left(\varphi\right)\xi\right\rangle _{\tau}^{\mathscr{M}}\mathrm{d}\mu_{g}\left(p\right)\mathrm{d}\mu_{g}\left(q\right)\mbox{.}
\end{multline*}
From eq. \eqref{eqCommutatorBetween2SmearedFieldsAndAWeylGenerator}
and the freedom in the choice of $f$ and $f^{\prime}$ we deduce
that
\begin{multline}
\left\langle \eta,\left[\varPsi_{\tau}^{\mathscr{M}}\left(p\right)\varPsi_{\tau}^{\mathscr{M}}\left(q\right),\mathrm{V}_{\tau}^{\mathscr{M}}\left(\varphi\right)\right]\xi\right\rangle _{\tau}^{\mathscr{M}}\\
=\varphi\left(p\right)\left\langle \eta,\mathrm{V}_{\tau}^{\mathscr{M}}\left(\varphi\right)\varPsi_{\tau}^{\mathscr{M}}\left(q\right)\xi\right\rangle _{\tau}^{\mathscr{M}}+\varphi\left(q\right)\left\langle \eta,\varPsi_{\tau}^{\mathscr{M}}\left(p\right)\mathrm{V}_{\tau}^{\mathscr{M}}\left(\varphi\right)\xi\right\rangle _{\tau}^{\mathscr{M}}\label{eqKGCommutatorBetween2UnsmearedFieldsAndAWeylGenerator}
\end{multline}
for each $\varphi\in V$ and each $\eta$, $\xi\in\mathscr{V}_{\tau}^{\mathscr{M}}$.

\subsubsection{Main theorem}

We are ready to prove that the action of the functional derivative
of the relative Cauchy evolution with respect to the spacetime metric
agrees with the action of the quantum stress-energy tensor in the
case of the Klein-Gordon field. As a matter of fact the main part
of the proof has already been discussed in the previous parts of the
current subsection. Here we simply state the theorem and put together
all the ingredients.
\begin{thm}
\label{thmKGFunctionalDerivativeOfRCEAgreesWithStressEnergyTensor}Let
$\mathscr{A}:\mathfrak{ghs}^{KG}\overset{\rightarrow}{\rightarrow}\mathfrak{alg}$
be the locally covariant quantum field theory for the Klein-Gordon
field obtained specializing the result of Section \ref{secBuildingALCQFT}
to the situation of Subsection \ref{subKleinGordonField} and let
$\left(\mathscr{M},\mathrm{\Lambda}^{0}M,A\right)$ be an object of
the category $\mathfrak{ghs}^{KG}$ defined there. Consider a quasi-free
Hadamard state $\tau$ on the CCR representation $\left(\mathcal{V},\mathrm{V}\right)=\mathscr{A}\left(\mathscr{M},\mathrm{\Lambda}^{0}M,A\right)$
and denote the GNS triple induced by $\tau$ with $\left(\mathscr{H}_{\tau}^{\mathscr{M}},\pi_{\tau}^{\mathscr{M}},\Omega_{\tau}^{\mathscr{M}}\right)$.
We denote with $\mathrm{V}_{\tau}^{\mathscr{M}}$ the represented
counterpart of the Weyl map $\mathrm{V}$ (cfr. eq. \eqref{eqKGRepresentedWeylMap})
and with $\mathcal{T}_{\tau}^{\mathscr{M}}$ the quantum stress-energy
tensor for the Klein-Gordon field on $\mathscr{M}$ obtained via point-splitting
in the representation induced by the state $\tau$ (cfr. eq. \eqref{eqKGStressEnergyTensorViaPointSplitting}).
Then there exists a dense subspace $\mathscr{V}_{\tau}^{\mathscr{M}}$
of $\mathscr{H}_{\tau}^{\mathscr{M}}$ such that
\[
\frac{\mathrm{\delta}}{\mathrm{\delta}h}\pi_{\tau}^{\mathscr{M}}\left(R_{h}^{\mathscr{M}}\left(\mathrm{V}\left(\varphi\right)\right)\right)=-\frac{\imath}{2}\left[\mathcal{T}_{\tau}^{\mathscr{M}},\mathrm{V}_{\tau}^{\mathscr{M}}\left(\varphi\right)\right]\quad\forall\varphi\in V
\]
in the sense of quadratic forms on $\mathscr{V}_{\tau}^{\mathscr{M}}$.\end{thm}
\begin{proof}
A dense subspace $\mathscr{V}_{\tau}^{\mathscr{M}}$ of $\mathscr{H}_{\tau}^{\mathscr{M}}$
exists by virtue of the choice of a quasi-free Hadamard state $\tau$
(see few lines before eq. \eqref{eqKGKeyPointForTheMainTheoremAboutRCE}).
The thesis means that
\[
\left\langle \theta,\frac{\mathrm{\delta}}{\mathrm{\delta}h}\pi_{\tau}^{\mathscr{M}}\left(R_{h}^{\mathscr{M}}\left(\mathrm{V}\left(\varphi\right)\right)\right)\theta\right\rangle _{\tau}^{\mathscr{M}}=-\frac{\imath}{2}\left\langle \theta,\left[\mathcal{T}_{\tau}^{\mathscr{M}},\mathrm{V}_{\tau}^{\mathscr{M}}\left(\varphi\right)\right]\theta\right\rangle _{\tau}^{\mathscr{M}}
\]
for each $\theta\in\mathscr{V}_{\tau}^{\mathscr{M}}$ and each $\varphi\in V$,
where $\left\langle \cdot,\cdot\right\rangle _{\tau}^{\mathscr{M}}$
denotes the scalar product on the Hilbert space $\mathscr{H}_{\tau}^{\mathscr{M}}$.

We fix a compact subset $K$ of $M$ and 1-parameter family of globally
hyperbolic perturbations $\left(-1,1\right)\rightarrow GHP\left(\mathscr{M},K\right)$,
$s\mapsto h^{s}$ such that $h^{0}=0$. Using the definition of $\frac{\mathrm{\delta}}{\mathrm{\delta}h}R_{h}^{\mathscr{M}}\left(\mathrm{V}\left(\varphi\right)\right)$,
we may find an equivalent form of our thesis (we still adopt the notation
$\mathrm{\delta}_{s}=\left.\nicefrac{\mathrm{d}}{\mathrm{d}s}\right|_{0}$):
\[
\mathrm{\delta}_{s}\left\langle \theta,\pi_{\tau}^{\mathscr{M}}\left(R_{h^{s}}^{\mathscr{M}}\left(\mathrm{V}\left(\varphi\right)\right)\right)\theta\right\rangle _{\tau}^{\mathscr{M}}=-\frac{\imath}{2}\int\limits _{M}\left(\mathrm{\delta}_{s}h^{s}\right)\left(\left\langle \theta,\left[\mathcal{T}_{\tau}^{\mathscr{M}},\mathrm{V}_{\tau}^{\mathscr{M}}\left(\varphi\right)\right]\theta\right\rangle _{\tau}^{\mathscr{M}}\right)\mathrm{d}\mu_{g}\mbox{,}
\]
where we are considering the dual pairing between $\mathrm{T}^{*}M\otimes_{s}\mathrm{T}^{*}M$
and $\mathrm{T}M\otimes_{s}\mathrm{T}M$ in the integrand appearing
on the RHS.

Recall that $R_{h}^{\mathscr{M}}=\mathscr{C}\left(r_{h}^{\mathscr{M}}\right)$
(cfr. eq. \eqref{eqKGQuantumVsClassicalRCE}) and the properties of
the quantization functor $\mathscr{C}$ defined in Subsection \ref{subQuantumFieldTheory}.
We deduce that $\pi_{\tau}^{\mathscr{M}}\circ R_{h^{s}}^{\mathscr{M}}\circ\mathrm{V}=\pi_{\tau}^{\mathscr{M}}\circ\mathrm{V}\circ r_{h^{s}}^{\mathscr{M}}=\mathrm{V}_{\tau}^{\mathscr{M}}\circ r_{h^{s}}^{\mathscr{M}}$.
This observation entails another slight modification of the thesis:
\[
\mathrm{\delta}_{s}\left\langle \theta,\mathrm{V}_{\tau}^{\mathscr{M}}\left(r_{h^{s}}^{\mathscr{M}}\varphi\right)\theta\right\rangle _{\tau}^{\mathscr{M}}=-\frac{\imath}{2}\int\limits _{M}\left(\mathrm{\delta}_{s}h^{s}\right)\left(\left\langle \theta,\left[\mathcal{T}_{\tau}^{\mathscr{M}},\mathrm{V}_{\tau}^{\mathscr{M}}\left(\varphi\right)\right]\theta\right\rangle _{\tau}^{\mathscr{M}}\right)\mathrm{d}\mu_{g}\mbox{.}
\]
Now we exploit eq. \eqref{eqKGKeyPointForTheMainTheoremAboutRCE}
and we eliminate the factor $\nicefrac{\imath}{2}$ on both sides
of the resulting equation:
\[
\left\langle \theta,\left\{ \varPhi_{\tau}^{\mathscr{M}}\left(\mathrm{\delta}_{s}r_{h^{s}}^{\mathscr{M}}\varphi\right),\mathrm{V}_{\tau}^{\mathscr{M}}\left(\varphi\right)\right\} \theta\right\rangle _{\tau}^{\mathscr{M}}=-\int\limits _{M}\left(\mathrm{\delta}_{s}h^{s}\right)\left(\left\langle \theta,\left[\mathcal{T}_{\tau}^{\mathscr{M}},\mathrm{V}_{\tau}^{\mathscr{M}}\left(\varphi\right)\right]\theta\right\rangle _{\tau}^{\mathscr{M}}\right)\mathrm{d}\mu_{g}\mbox{.}
\]
We still want to reformulate the thesis a little bit using eq. \eqref{eqKGExpressionForTheDerivativeOfTheClassicalRCE}
and eq. \eqref{eqKGRelationBetweenvarPsiAndvarPhi}:
\[
\underset{\mathsf{L}}{\underbrace{\left\langle \theta,\left\{ \varPsi_{\tau}^{\mathscr{M}}\left(\mathrm{\delta}_{s}A\left[h^{s}\right]\varphi\right),\mathrm{V}_{\tau}^{\mathscr{M}}\left(\varphi\right)\right\} \theta\right\rangle _{\tau}^{\mathscr{M}}}}=\underset{\mathsf{R}}{\underbrace{-\int\limits _{M}\left(\mathrm{\delta}_{s}h^{s}\right)\left(\left\langle \theta,\left[\mathcal{T}_{\tau}^{\mathscr{M}},\mathrm{V}_{\tau}^{\mathscr{M}}\left(\varphi\right)\right]\theta\right\rangle _{\tau}^{\mathscr{M}}\right)\mathrm{d}\mu_{g}}}\mbox{.}
\]

Now we work with the LHS of the last equation (denoted by $\mathsf{L}$)
and the RHS (denoted by $\mathsf{R}$) separately. Starting from $\mathsf{L}$,
we exploit the relation between smeared and unsmeared fields, eq.
\eqref{eqKGRelationBetweenUnsmearedAndSmearedFields}:
\[
\mathsf{L}=\int\limits _{M}\left\langle \theta,\left\{ \varPsi_{\tau}^{\mathscr{M}}\left(p\right),\mathrm{V}_{\tau}^{\mathscr{M}}\left(\varphi\right)\right\} \theta\right\rangle _{\tau}^{\mathscr{M}}\left(\mathrm{\delta}_{s}A\left[h^{s}\right]\varphi\right)\left(p\right)\mathrm{d}\mu_{g}\mbox{.}
\]

We want to express $\mathsf{L}$ using oriented coordinate neighborhoods.
Indeed we can find an open covering of $M$ constituted by coordinate
neighborhoods. In order to make calculations easier, we choose these
coordinate neighborhoods in such a way that on each of them $\left|\det g\right|=1$.
We can exploit the paracompactness of the manifold $M$ to pick out
a locally finite refinement and we introduce a partition of unity
subordinate to the refined covering. Since $\mathrm{supp}\left(h^{s}\right)\subseteq K$
for each $s\in\left(-1,1\right)$, the support of the coefficients
in $\mathrm{\delta}_{s}A\left[h^{s}\right]$ must be included in $K$
too. Exploiting the compactness of $K$, we can find that only a finite
number of the coordinate neighborhoods considered so far intersect
it. We denote them with $\left\{ \left(U_{\alpha},V_{\alpha},\phi_{\alpha}\right)\right\} $
and we consider only the corresponding members $\left\{ \chi_{\alpha}\right\} $
in the partition of unity (the other members indeed have null product
with the integrand). This entails that we can use this finite collection
of coordinate neighborhoods (together with the corresponding members
of the original partition of unity) to express $\mathsf{L}$ in local
coordinates:
\[
\mathsf{L}=\sum_{\alpha}\int\limits _{V_{\alpha}}\chi_{\alpha}\left\langle \theta,\left\{ \varPsi_{\tau}^{\mathscr{M}}\left(x\right),\mathrm{V}_{\tau}^{\mathscr{M}}\left(\varphi\right)\right\} \theta\right\rangle _{\tau}^{\mathscr{M}}\left(\mathrm{\delta}_{s}A\left[h^{s}\right]\varphi\right)\left(x\right)\mathrm{d}V\mbox{,}
\]
where $\mathrm{d}V$ denotes the standard volume form on $\mathbb{R}^{4}$
and all the sections that appear inside the integral are now written
in local coordinates%
\footnote{by this we mean that, inside the integral over $V_{\alpha}$, $\mathrm{\delta}_{s}A\left[h^{s}\right]\varphi$
now denotes the push-forward through $\phi_{\alpha}$ of the original
$\mathrm{\delta}_{s}A\left[h^{s}\right]\varphi$ restricted to $U_{\alpha}$
and similarly for the other sections inside the integral%
}. It is convenient to define
\begin{eqnarray*}
\zeta:M & \rightarrow & \mathbb{C}\\
p & \mapsto & \left\langle \theta,\left\{ \varPsi_{\tau}^{\mathscr{M}}\left(x\right),\mathrm{V}_{\tau}^{\mathscr{M}}\left(\varphi\right)\right\} \theta\right\rangle _{\tau}^{\mathscr{M}}
\end{eqnarray*}
in order to simplify our notation. Now we use eq. \eqref{eqKGDifferentialOperatorVariation}.
In this way we obtain
\[
\mathsf{L}=\underset{\mathsf{L}_{1}}{\underbrace{\sum_{\alpha}\int\limits _{V_{\alpha}}\chi_{\alpha}\zeta\left(\nabla^{i}\nabla^{j}\varphi\right)\mathrm{\delta}_{s}h_{ij}^{s}\mathrm{d}V}}+\underset{L_{2}}{\underbrace{\sum_{\alpha}\int\limits _{V_{\alpha}}\chi_{\alpha}\zeta\left(\nabla_{k}\varphi\right)\mathrm{\delta}_{s}\Gamma\left[h^{s}\right]_{ij}^{k}g^{ij}\mathrm{d}V}}\mbox{,}
\]
where the dependence of the integrand on the point $x\in V_{\alpha}$
now is understood. We denote the first addend appearing on the RHS
of the last equation with $\mathsf{L}_{1}$ and the second with $\mathsf{L}_{2}$.
We integrate $\mathsf{L}_{1}$ by parts noting that $\chi_{\alpha}$
is null on the boundary of $V_{\alpha}$, hence no surface term appears:
\begin{eqnarray*}
\mathsf{L}_{1} & = & \overset{\mathsf{X}}{\overbrace{-\sum_{\alpha}\int\limits _{V_{\alpha}}\chi_{\alpha}\left(\nabla^{i}\zeta\right)\left(\nabla^{j}\varphi\right)\mathrm{\delta}_{s}h_{ij}^{s}\mathrm{d}V}}\overset{\mathsf{L}_{3}}{\overbrace{-\sum_{\alpha}\int\limits _{V_{\alpha}}\chi_{\alpha}\zeta\left(\nabla^{j}\varphi\right)\nabla^{i}\mathrm{\delta}_{s}h_{ij}^{s}\mathrm{d}V}}\\
 &  & \underset{=0}{\underbrace{-\sum_{\alpha}\int\limits _{V_{\alpha}}\left(\nabla^{i}\chi_{\alpha}\right)\zeta\left(\nabla^{j}\varphi\right)\mathrm{\delta}_{s}h_{ij}^{s}\mathrm{d}V}}\mbox{.}
\end{eqnarray*}
The last term in the equation above gives null contribution. We can
check this fact observing that, on each point of the support of $\mathrm{\delta}_{s}h^{s}$,
the finite number of $\chi_{\alpha}$ sum up to 1, hence their derivatives
sum up to zero. We denote the first of the remaining terms with $\mathsf{X}$
and the second with $\mathsf{L}_{3}$. Up to now we have
\[
\mathsf{L}=\mathsf{X}+\mathsf{L}_{2}+\mathsf{L}_{3}\mbox{.}
\]

Now we investigate $\mathsf{R}$. As we did for $\mathsf{L}$, we
express it using the chosen local coordinates:
\[
\mathsf{R}=-\sum_{\alpha}\int\limits _{V_{\alpha}}\chi_{\alpha}\left(x\right)\left(\mathrm{\delta}_{s}h_{ij}^{s}\right)\left(x\right)\left\langle \theta,\left[\mathcal{T}_{\tau}^{\mathscr{M}\, ij}\left(x\right),\mathrm{V}_{\tau}^{\mathscr{M}}\left(\varphi\right)\right]\theta\right\rangle _{\tau}^{\mathscr{M}}\mathrm{d}V\mbox{.}
\]
Consider the integrand (dropping $\chi_{\alpha}$ for the moment).
Inside the commutator appears the quantized stress-energy tensor.
Indeed we have eq. \eqref{eqKGStressEnergyTensorViaPointSplitting}
that tells us about its form, but we must perform the coincidence
limit before we can insert such equation inside the integral in place
of $\mathcal{T}_{\tau}^{\mathscr{M}}$. As a matter of fact we previously
calculate the expectation value of the commutator recalling the commutation
relation found in eq. \eqref{eqKGCommutatorBetween2UnsmearedFieldsAndAWeylGenerator}
and only after that we take the coincidence limit as required by the
point-splitting procedure realizing that no divergences arise. Exploiting
also the symmetry of $\mathrm{\delta}_{s}h^{s}$ and $g$, anticommutators
appear. All these operations produce the following result (to shorten
the expression we replace $\left\langle \theta,\left\{ \varPsi_{\tau}^{\mathscr{M}}\left(x\right),\mathrm{V}_{\tau}^{\mathscr{M}}\left(\varphi\right)\right\} \theta\right\rangle _{\tau}^{\mathscr{M}}$
with $\zeta$ as above):
\begin{eqnarray*}
\mathsf{R} & = & \overset{=\mathsf{X}}{\overbrace{-\sum_{\alpha}\int\limits _{V_{\alpha}}\chi_{\alpha}\left(\nabla^{i}\varphi\right)\left(\nabla^{j}\zeta\right)\mathrm{\delta}_{s}h_{ij}^{s}\mathrm{d}V}}\\
 &  & \underset{\mathsf{R}_{1}}{+\underbrace{\frac{1}{2}\sum_{\alpha}\int\limits _{V_{\alpha}}\chi_{\alpha}g^{kl}\left(\nabla_{k}\varphi\right)\left(\nabla_{l}\zeta\right)g^{ij}\mathrm{\delta}_{s}h_{ij}^{s}\mathrm{d}V}}+\underset{\mathsf{R}_{2}}{\underbrace{\frac{1}{2}m^{2}\sum_{\alpha}\int\limits _{V_{\alpha}}\chi_{\alpha}\varphi\zeta g^{ij}\mathrm{\delta}_{s}h_{ij}^{s}\mathrm{d}V}}\mbox{.}
\end{eqnarray*}
The first term coincides with the term $\mathsf{X}$ in $\mathsf{L}_{1}$
once that the indices $i$ and $j$ are interchanged taking into account
the symmetry of $\mathrm{\delta}_{s}h^{s}$. As for the other two
terms, some more work is required. We denote the first one with $\mathsf{R}_{1}$
and the second one with $\mathsf{R}_{2}$ and we integrate $\mathsf{R}_{1}$
by parts (this time we directly omit the term containing derivatives
of $\chi_{\alpha}$ since it gives null contribution as noted above):
\[
\mathsf{R}_{1}=-\frac{1}{2}\sum_{\alpha}\int\limits _{V_{\alpha}}\chi_{\alpha}g^{kl}\left(\nabla_{l}\nabla_{k}\varphi\right)\zeta g^{ij}\mathrm{\delta}_{s}h_{ij}^{s}\mathrm{d}V-\frac{1}{2}\sum_{\alpha}\int\limits _{V_{\alpha}}\chi_{\alpha}g^{kl}\left(\nabla_{k}\varphi\right)\zeta g^{ij}\nabla_{l}\mathrm{\delta}_{s}h_{ij}^{s}\mathrm{d}V\mbox{.}
\]
If we put together $\mathsf{R}_{1}$ and $\mathsf{R}_{2}$ and we
remind that $A\varphi=0$ since $\varphi\in V$, we get
\begin{eqnarray*}
\mathsf{R}_{1}+\mathsf{R}_{2} & = & \frac{1}{2}\sum_{\alpha}\int\limits _{V_{\alpha}}\chi_{\alpha}\overset{=A\varphi=0}{\overbrace{\left(-g^{kl}\nabla_{l}\nabla_{k}\varphi+m^{2}\varphi\right)}}\zeta g^{ij}\mathrm{\delta}h_{ij}^{s}\mathrm{d}V\\
 &  & -\frac{1}{2}\sum_{\alpha}\int\limits _{V_{\alpha}}\chi_{\alpha}g^{kl}\left(\nabla_{k}\varphi\right)\zeta g^{ij}\nabla_{l}\mathrm{\delta}h_{ij}^{s}\mathrm{d}V\\
 & = & -\frac{1}{2}\sum_{\alpha}\int\limits _{V_{\alpha}}\chi_{\alpha}\zeta\left(\nabla_{k}\varphi\right)g^{ij}g^{kl}\nabla_{l}\mathrm{\delta}h_{ij}^{s}\mathrm{d}V\mbox{.}
\end{eqnarray*}
At this stage our thesis $\mathsf{L}=\mathsf{R}$ is reduced to the
following identity:
\begin{equation}
\mathsf{L}_{2}+\mathsf{L}_{3}=\mathsf{R}_{1}+\mathsf{R}_{2}\mbox{.}\label{eqKGRCEFinalIntegralIdentity}
\end{equation}

The next step consist in the proof of the identity 
\begin{equation}
\mathrm{\delta}_{s}\Gamma\left[h^{s}\right]_{ij}^{k}g^{ij}-g^{jk}\nabla^{i}\mathrm{\delta}_{s}h_{ij}^{s}=-\frac{1}{2}g^{ij}g^{kl}\nabla_{l}\mathrm{\delta}_{s}h_{ij}^{s}\label{eqKGRCEGeometricalIdentity}
\end{equation}
in each point of $M$. If this identity actually holds everywhere,
it follows that eq. \eqref{eqKGRCEFinalIntegralIdentity} holds too
and hence the proof is complete: In fact, as the reader might easily
check, eq. \eqref{eqKGRCEGeometricalIdentity} written using the coordinate
neighborhoods $\left(U_{\alpha},V_{\alpha},\phi_{\alpha}\right)$,
integrated on both sides on each $V_{\alpha}$ together with the factor
$\chi_{\alpha}\zeta\nabla_{k}\varphi$ and summed over the finite
number of indices $\alpha$ gives exactly eq. \eqref{eqKGRCEFinalIntegralIdentity}.

The first thing we do is to use the metric to lower the index on $\nabla$
in the second term on the LHS of eq. \eqref{eqKGRCEGeometricalIdentity}
and, after that, we rename some summation indices (bear in mind that
$g$ and $\mathrm{\delta}_{s}h^{s}$ are symmetric). In this way the
identity in eq. \eqref{eqKGRCEGeometricalIdentity} to be checked
becomes:
\begin{equation}
\mathrm{\delta}_{s}\Gamma\left[h^{s}\right]_{ij}^{k}g^{ij}-g^{ij}g^{lk}\nabla_{i}\mathrm{\delta}_{s}h_{lj}^{s}=-\frac{1}{2}g^{ij}g^{kl}\nabla_{l}\mathrm{\delta}_{s}h_{ij}^{s}\mbox{.}\label{eqKGRCEGeometricalIdentityRevised}
\end{equation}
Now we fix an arbitrary point $p$ in $M$ and we choose Riemannian
normal coordinates in a (sufficiently small) neighborhood of $p$
(cfr. e.g. \cite[Sect. 3.3, p. 42]{Wal84}). Doing so, we put ourselves
in a favorable situation from a computational point of view since
with this choice of coordinates the Christoffel symbols $\Gamma_{ij}^{k}$
are null at $p$ and hence we can freely replace $\nabla$ with $\partial$
(note that a similar result does not hold for the Christoffel symbols
of a {}``perturbed'' connection $\nabla\left[h^{s}\right]$). With
this considerations we evaluate $\mathrm{\delta}_{s}\Gamma\left[h^{s}\right]_{ij}^{k}$
(recall eq. \eqref{eqChristoffelSymbolsOfTheLeviCivitaConnection}
which provides the expression of the Christoffel symbols for the Levi-Civita
connection):
\begin{eqnarray}
\mathrm{\delta}_{s}\Gamma\left[h^{s}\right]_{ij}^{k} & = & \left(\mathrm{\delta}_{s}g_{h^{s}}^{kl}\right)\overset{=g_{lm}\Gamma_{ij}^{m}=0}{\overbrace{\frac{1}{2}\left(\partial_{i}g_{lj}+\partial_{j}g_{il}-\partial_{l}g_{ij}\right)}}+g^{kl}\frac{1}{2}\left(\partial_{i}\mathrm{\delta}_{s}h_{lj}^{s}+\partial_{j}\mathrm{\delta}_{s}h_{il}^{s}-\partial_{l}\mathrm{\delta}_{s}h_{ij}^{s}\right)\nonumber \\
 & = & g^{kl}\frac{1}{2}\left(\partial_{i}\mathrm{\delta}_{s}h_{lj}^{s}+\partial_{j}\mathrm{\delta}_{s}h_{il}^{s}-\partial_{l}\mathrm{\delta}_{s}h_{ij}^{s}\right)\mbox{.}\label{eqVariationOfTheChristoffelSymbol}
\end{eqnarray}
As a matter of fact we are interested in the contraction of $\mathrm{\delta}_{s}\Gamma\left[h^{s}\right]_{ij}^{k}$
with $g^{ij}$:
\begin{equation}
\mathrm{\delta}_{s}\Gamma\left[h^{s}\right]_{ij}^{k}g^{ij}=g^{ij}g^{kl}\partial_{i}\mathrm{\delta}_{s}h_{lj}^{s}-\frac{1}{2}g^{ij}g^{kl}\partial_{l}\mathrm{\delta}_{s}h_{ij}^{s}\mbox{,}\label{eqVariationOfTheChristoffelSymbolContractedWithTheInverseMetric}
\end{equation}
where we exploited the identity $\left(\partial_{j}\mathrm{\delta}_{s}h_{il}^{s}\right)g^{ij}=\left(\partial_{i}\mathrm{\delta}_{s}h_{jl}^{s}\right)g^{ij}$.
With this result we evaluate the LHS of eq. \eqref{eqKGRCEGeometricalIdentityRevised}:
\begin{eqnarray*}
\mathrm{\delta}_{s}\Gamma\left[h^{s}\right]_{ij}^{k}g^{ij}-g^{ij}g^{lk}\nabla_{i}\mathrm{\delta}_{s}h_{lj}^{s} & = & g^{ij}g^{kl}\partial_{i}\mathrm{\delta}_{s}h_{lj}^{s}-\frac{1}{2}g^{ij}g^{kl}\partial_{l}\mathrm{\delta}_{s}h_{ij}^{s}-g^{ij}g^{lk}\partial_{i}\mathrm{\delta}_{s}h_{lj}^{s}\\
 &  & =-\frac{1}{2}g^{ij}g^{kl}\partial_{l}\mathrm{\delta}_{s}h_{ij}^{s}\mbox{.}
\end{eqnarray*}
It is sufficient to restore $\nabla$ in place of $\partial$ on the
RHS of the last equation to realize that eq. \eqref{eqKGRCEGeometricalIdentityRevised}
actually is proved. We already showed that this one is equivalent
to eq. \eqref{eqKGRCEGeometricalIdentity}, which in turn entails
\eqref{eqKGRCEFinalIntegralIdentity}. This completes the proof.
\end{proof}

\subsection{\label{subPRCE}Relative Cauchy evolution for the Proca field}

Now we turn our attention to the Proca field. Our aim is to extend
the result obtained for the Klein-Gordon field also in the present
context, that is to prove the agreement of the action of the functional
derivative of the relative Cauchy evolution with the action of the
quantized stress-energy tensor for the Proca field.

We need some preparation also in this case. We will follow an approach
very similar to that of the previous subsection. The main difference
lies in the fact that now we are going to take into account the results
of Subsection \ref{subProcaField} in place of those from Subsection
\ref{subKleinGordonField}, specifically we consider the locally covariant
quantum field theory $\mathscr{A}:\mathfrak{ghs}^{P}\overset{\rightarrow}{\rightarrow}\mathfrak{alg}$
(cfr. Definition \ref{defghsP} for the definition of the category
) defined as the composition of the covariant functor $\mathscr{B}:\mathfrak{ghs}^{P}\overset{\rightarrow}{\rightarrow}\mathfrak{ssp}$
describing the classical theory of the Proca field (see Theorem \ref{thmClassicalFieldFunctorProcaField})
with the usual quantization functor $\mathscr{C}:\mathfrak{ssp}\overset{\rightarrow}{\rightarrow}\mathfrak{alg}$
(see Lemma \ref{lemQuantizationFunctor}). After the proof of Theorem
\ref{thmClassicalFieldFunctorProcaField} we argued that $\mathscr{A}$
fulfils the time slice axiom as a LCQFT (indeed the causality condition
holds too, but this fact is not relevant in this context). This ensures
that one can actually consider the RCE for the Proca field as presented
in Section \ref{secRCEDefinitionAndProperties} and all the results
found there still hold since now we are only considering a richer
structure on each spacetime, but the morphisms considered there are
easily recognized to induce morphisms also in this context.

From now on we use the notation of Subsection \ref{subProcaField}.
In particular we recall that the differential operators considered
here (which are formally selfadjoint, but fail to be normally hyperbolic)
are of the form
\[
A=\mathrm{\delta d}+m^{2}\mathrm{id}_{\mathrm{\Omega}^{1}M}:\mathrm{\Omega}^{1}M\rightarrow\mathrm{\Omega}^{1}M
\]
on each globally hyperbolic spacetime $\mathscr{M}=\left(M,g,\mathfrak{o},\mathfrak{t}\right)$.
At the same time we also consider a formally selfadjoint normally
hyperbolic operator 
\begin{eqnarray*}
P_{A} & = & \mathrm{\Box}_{1}+m^{2}\mathrm{id}_{\mathrm{\Omega}^{1}M}:\mathrm{\Omega}^{1}M\rightarrow\mathrm{\Omega}^{1}M\mbox{.}
\end{eqnarray*}
We denote with $e_{A}^{a/r}$ its associated advanced/retarded Green
operator and we use it to define the advanced/retarded Green operator
for $A$ (cfr. Lemma \ref{lemGreenOperatorsForTheProcaField}):
\[
f_{A}^{a/r}=e_{P}^{a/r}\circ\left(\mathrm{id}_{\mathrm{\Omega}_{0}^{1}M}+\frac{1}{m^{2}}\mathrm{d\delta}\right):\mathrm{\Omega}_{0}^{1}M\rightarrow\mathrm{\Omega}^{1}M\mbox{.}
\]

\subsubsection{Relative Cauchy evolution for the classical Proca field}

Our first purpose is to find a convenient expression for the relative
Cauchy evolution of the Proca field at a classical level and to relate
it to the original RCE. To do this we need a result similar to Proposition
\ref{propKGExpressionForTheInverseOfAProperSymplecticMap}. Note that
the object $\left(\left.\mathscr{M}\right|_{O},\mathrm{\Lambda}^{1}O,\left.A\right|_{O}\right)$
of $\mathfrak{ghs}^{P}$ that we are going to take into account is
defined exactly with the procedure followed for the corresponding
object of $\mathfrak{ghs}^{KG}$ with the only replacement of $\mathrm{\Lambda}^{0}$
with $\mathrm{\Lambda}^{1}$ (see the first part of the proof in Proposition
\ref{propKGExpressionForTheInverseOfAProperSymplecticMap}).
\begin{prop}
\label{propPExpressionForTheInverseOfAProperSymplecticMap}Let $\mathscr{B}:\mathfrak{ghs}^{P}\overset{\rightarrow}{\rightarrow}\mathfrak{ssp}$
be the covariant functor describing the classical theory of the Proca
field (cfr. Subsection \ref{subProcaField}), consider an object $\left(\mathscr{M}=\left(M,g,\mathfrak{o},\mathfrak{t}\right),\mathrm{\Lambda}^{1}M,A\right)$
of $\mathfrak{ghs}^{P}$ and let $O$ be an $\mathscr{M}$-causally
convex connected open subset of $M$ including a smooth spacelike
Cauchy surface $\Sigma$ for $\mathscr{M}$. Consider the object $\left(\left.\mathscr{M}\right|_{O},\mathrm{\Lambda}^{1}O,\left.A\right|_{O}\right)$
of $\mathfrak{ghs}^{P}$ and the morphism $\left(\iota_{O}^{M},\iota_{\mathrm{\Lambda}^{1}O}^{\mathrm{\Lambda}^{1}M}\right)$
of $\mathfrak{ghs}^{P}$ from $\left(\left.\mathscr{M}\right|_{O},\mathrm{\Lambda}^{1}O,\left.A\right|_{O}\right)$
to $\left(\mathscr{M},\mathrm{\Lambda}^{1}M,A\right)$ induced by
the inclusion maps $\iota_{O}^{M}:O\rightarrow M$ and $\iota_{\mathrm{\Lambda}^{1}O}^{\mathrm{\Lambda}^{1}M}:\mathrm{\Lambda}^{1}O\rightarrow\mathrm{\Lambda}^{1}M$.
Then there exists a partition of unity $\left\{ \chi^{a},\chi^{r}\right\} $
on $M$ such that the inverse $\mathscr{B}\left(\iota_{O}^{M},\iota_{\mathrm{\Lambda}^{1}O}^{\mathrm{\Lambda}^{1}M}\right)^{-1}$
of the bijective morphism $\mathscr{B}\left(\iota_{O}^{M},\iota_{\mathrm{\Lambda}^{1}O}^{\mathrm{\Lambda}^{1}M}\right)$
of $\mathfrak{ssp}$ from $\left(V,\sigma\right)=\mathscr{B}\left(\left.\mathscr{M}\right|_{O},\mathrm{\Lambda}^{1}O,\left.A\right|_{O}\right)$
to $\left(W,\omega\right)=\mathscr{B}\left(\mathscr{M},\mathrm{\Lambda}^{1}M,A\right)$
satisfies the following equation:
\[
\mathscr{B}\left(\iota_{O}^{M},\iota_{\mathrm{\Lambda}^{1}O}^{\mathrm{\Lambda}^{1}M}\right)^{-1}\Theta=\pm f_{\left.A\right|_{O}}\left(\mathrm{res}_{\iota_{\mathrm{\Lambda}^{1}O}^{\mathrm{\Lambda}^{1}M}}\left(A\left(\chi^{a/r}\Theta\right)\right)\right)\quad\forall\Theta\in W\mbox{,}
\]
where $f_{\left.A\right|_{O}}$ is the causal propagator for $\left.A\right|_{O}$
and the restriction map is defined in Lemma \ref{lemresPsieBextPsi=00003DeA}.\end{prop}
\begin{proof}
The most part of this proof is identical to the proof of Proposition
\ref{propKGExpressionForTheInverseOfAProperSymplecticMap}, provided
that you replace everywhere $\mathrm{\Lambda}^{0}$, $\mathrm{\Omega}^{0}$,
$\varphi$, $A=\mathrm{\Box}_{0}+m^{2}\mathrm{id}_{\mathrm{\Omega}^{0}M}$
and its causal propagator $e_{A}$ with $\mathrm{\Lambda}^{1}$, $\mathrm{\Omega}^{1}$,
$\Theta$, the current linear differential operator $A=\mathrm{\delta d}+m^{2}\mathrm{id}_{\mathrm{\Omega}^{1}M}$
and its causal propagator $f_{A}$ (whose existence follows from Lemma
\ref{lemGreenOperatorsForTheProcaField}). You should also remember
that in the present situation there is a (potentially) stricter condition
on the morphisms of $\mathfrak{ghs}^{P}$, that is compatibility with
both $\mathrm{\delta d}$ and $\mathrm{d\delta}$, but this does not
give rise to problems of any sort because the inclusion maps easily
satisfy this requirement. The time slice axiom holds also in this
situation as we proved in Theorem \ref{thmClassicalFieldFunctorProcaField},
hence $\mathscr{B}\left(\iota_{O}^{M},\iota_{\mathrm{\Lambda}^{1}O}^{\mathrm{\Lambda}^{1}M}\right)$
is bijective and its inverse $\mathscr{B}\left(\iota_{O}^{M},\iota_{\mathrm{\Lambda}^{1}O}^{\mathrm{\Lambda}^{1}M}\right)^{-1}$
is a morphism of $\mathfrak{ssp}$. Our aim is to find a convenient
expression for $\mathscr{B}\left(\iota_{O}^{M},\iota_{\mathrm{\Lambda}^{1}O}^{\mathrm{\Lambda}^{1}M}\right)^{-1}$.
The only slight difference arises when we check the identity in the
statement. To be precise, we obtain the next equation following exactly
the same reasoning:

\[
\mathscr{B}\left(\iota_{O}^{M},\iota_{\mathrm{\Lambda}^{1}O}^{\mathrm{\Lambda}^{1}M}\right)\left(\alpha\Theta\right)=\pm f_{A}A\left(\chi^{a/r}\Theta\right)\mbox{,}
\]
where $\alpha$ denotes the map from $W$ to $V$ defined by
\[
\alpha\Theta=\mathrm{res}_{\iota_{\mathrm{\Lambda}^{1}O}^{\mathrm{\Lambda}^{1}M}}\left(A\left(\chi^{a/r}\Theta\right)\right)
\]
for each $\Theta\in W$ and $\chi^{a/r}$ is the partition of unity
that we find imitating the first part of the proof of Proposition
\ref{propKGExpressionForTheInverseOfAProperSymplecticMap}. Now we
would like to apply Lemma \ref{lemExtensionOf ea(Pu)=00003Du}, but
this cannot be done directly since no normally hyperbolic operator
is immediately available. Anyway this problem is easily circumvented
recalling that
\[
f_{A}=e_{A}\circ\left(\mathrm{id}_{\Omega_{0}^{1}M}+\frac{1}{m^{2}}\mathrm{d\delta}\right)
\]
and that
\[
\left(\mathrm{id}_{\Omega^{1}M}+\frac{1}{m^{2}}\mathrm{d\delta}\right)\circ A=P_{A}\mbox{.}
\]
With these observations we find
\[
\mathscr{B}\left(\iota_{O}^{M},\iota_{\mathrm{\Lambda}^{1}O}^{\mathrm{\Lambda}^{1}M}\right)\left(\alpha\Theta\right)=\pm e_{A}P_{A}\Theta^{a/r}\mbox{,}
\]
where $\Theta^{a/r}=\chi^{a/r}\Theta$. Now a normally hyperbolic
operator $P_{A}$ is available, but we need to show that $P_{A}\Theta^{a/r}$
has compact support in order to exploit Lemma \ref{lemExtensionOf ea(Pu)=00003Du}.
This can be done easily because $A\Theta=0$ trivially entails $\mathrm{\delta}\Theta=0$;
therefore we have $\mathrm{\delta}\Theta^{a}=-\mathrm{\delta}\Theta^{r}$.
From this identity we deduce that $\mathrm{\delta}\Theta^{a}$ has
compact support (the proof is based on the support properties of the
causal propagator $f_{A}$ and of the partition of unity). Since $P_{A}\Theta^{a}=A\Theta^{a}+\mathrm{d\delta}\Theta^{a}=-P_{A}\Theta^{r}$,
we can conclude that $P_{A}\Theta^{a}$ actually has compact support
and hence we are allowed to apply Lemma \ref{lemExtensionOf ea(Pu)=00003Du}
obtaining
\[
\pm e_{A}P_{A}\Theta^{a/r}=\pm\left(e_{A}^{a}P_{A}\Theta^{a/r}-e_{A}^{r}P_{A}\Theta^{a/r}\right)=\Theta^{a}+\Theta^{r}=\Theta\mbox{.}
\]
With this we conclude
\[
\mathscr{B}\left(\iota_{O}^{M},\iota_{\mathrm{\Lambda}^{1}O}^{\mathrm{\Lambda}^{1}M}\right)\left(\alpha\Theta\right)=\Theta=\mathscr{B}\left(\iota_{O}^{M},\iota_{\mathrm{\Lambda}^{1}O}^{\mathrm{\Lambda}^{1}M}\right)\left(\mathscr{B}\left(\iota_{O}^{M},\iota_{\mathrm{\Lambda}^{1}O}^{\mathrm{\Lambda}^{1}M}\right)^{-1}\Theta\right)\quad\forall\Theta\in W\mbox{.}
\]
Since $\mathscr{B}\left(\iota_{O}^{M},\iota_{\mathrm{\Lambda}^{1}O}^{\mathrm{\Lambda}^{1}M}\right)$
is injective, the last equation entails
\[
\mathscr{B}\left(\iota_{O}^{M},\iota_{\mathrm{\Lambda}^{0}O}^{\mathrm{\Lambda}^{0}M}\right)^{-1}\Theta=\alpha\Theta\quad\forall\varphi\in W\mbox{,}
\]
which is exactly our thesis.
\end{proof}
As we did in the case of the Klein-Gordon field, we specialize the
definition of the RCE to the current situation. Consider an object
$\left(\mathscr{M},\mathrm{\Lambda}^{1}M,A\right)$ of $\mathfrak{ghs}^{P}$,
take $h\in GHP\left(\mathscr{M}\right)$ and recall the definitions
of the morphisms $\imath_{\pm}^{\mathscr{M}}\left[h\right]$ and $\jmath_{\pm}^{\mathscr{M}}\left[h\right]$
introduced before Definition \ref{defRCE}. Together with the perturbed
spacetime $\mathscr{M}\left[h\right]$, we must also consider the
effects of $h$ on the inner product defined on the vector bundle
$\mathrm{\Lambda}^{1}M$ and on the differential operator $A=\mathrm{\delta d}+m^{2}\mathrm{id}_{\mathrm{\Omega}^{1}M}$.
The inner product on $\mathrm{\Lambda}^{1}M$ is induced by the metric,
hence we should consider the inner product induced by the perturbed
metric $g_{h}$. As for the linear differential operator we define
$A\left[h\right]=\mathrm{\delta}\left[h\right]\mathrm{d}+m^{2}\mathrm{id}_{\mathrm{\Omega}^{1}M}$,
where $\mathrm{\delta}\left[h\right]$ is the codifferential over
$\mathscr{M}\left[h\right]$. Similarly we have to consider $P_{A}\left[h\right]=\mathrm{\Box}_{1}\left[h\right]+m^{2}\mathrm{id}_{\mathrm{\Omega}^{1}M}$,
where $\mathrm{\Box}_{1}\left[h\right]=\mathrm{d\delta}\left[h\right]+\mathrm{\delta}\left[h\right]\mathrm{d}$
is the d'Alembert operator defined over $\mathscr{M}\left[h\right]$
for 1-forms. As a matter of fact we are replacing the metric $g$
with $g_{h}=g+h$ whenever there is something related to the metric.
We may consider the inclusion map $\iota_{\mathrm{\Lambda}^{1}M_{\pm}}^{\mathrm{\Lambda}^{1}M}$,
where $M_{\pm}=M\setminus J_{\mp}^{\mathscr{M}}\left(\mathrm{supp}\left(h\right)\right)$
in accordance with the definitions of $\imath_{\pm}^{\mathscr{M}}\left[h\right]$
and $\jmath_{\pm}^{\mathscr{M}}\left[h\right]$. Compatibility with
both $\mathrm{\delta d}$ and $\mathrm{d\delta}$ via $\left(\iota_{M_{\pm}}^{M},\iota_{\mathrm{\Lambda}^{1}M_{\pm}}^{\mathrm{\Lambda}^{1}M}\right)$
holds (cfr. Definition \ref{defghsP}). Since the effects of the perturbation
$h$ are relevant only inside $\mathrm{supp}\left(h\right)$, we realize
that $\mathrm{\delta}\left[h\right]$ and $\mathrm{\delta}$ act exactly
in the same way on sections supported outside $\mathrm{supp}\left(h\right)$.
Together with $\left.A\right|_{M_{\pm}}$, we may consider $\left.A\left[h\right]\right|_{M_{\pm}}$
and we immediately recognize that they coincide (we denote both of
them with $A_{\pm}\left[h\right]$). Similarly $\left.P_{A}\right|_{M_{\pm}}=\left.P_{A}\left[h\right]\right|_{M_{\pm}}$
so that we denote both with $P_{A\pm}\left[h\right]$. Hence we can
introduce the objects $\left(\mathscr{M}\left[h\right],\mathrm{\Lambda}^{1}M,A\left[h\right]\right)$
and $\left(\mathscr{M}_{\pm}\left[h\right],\mathrm{\Lambda}^{1}M_{\pm},A_{\pm}\left[h\right]\right)$
of $\mathfrak{ghs}^{P}$ and interpret the vector bundle homomorphism
$\left(\iota_{M_{\pm}}^{M},\iota_{\mathrm{\Lambda}^{0}M_{\pm}}^{\mathrm{\Lambda}^{0}M}\right):\mathrm{\Lambda}^{1}M_{\pm}\rightarrow\mathrm{\Lambda}^{1}M$
in the following ways: 
\begin{eqnarray*}
\left(\imath_{\pm}^{\mathscr{M}}\left[h\right],\imath_{\pm}^{\mathscr{M},\mathrm{\Lambda}^{1}}\left[h\right]\right) & \in & \mathsf{Mor}_{\mathfrak{ghs}^{P}}\left(\left(\mathscr{M}_{\pm}\left[h\right],\mathrm{\Lambda}^{1}M_{\pm},A_{\pm}\left[h\right]\right),\left(\mathscr{M},\mathrm{\Lambda}^{1}M,A\right)\right)\mbox{,}\\
\left(\jmath_{\pm}^{\mathscr{M}}\left[h\right],\jmath_{\pm}^{\mathscr{M},\mathrm{\Lambda}^{1}}\left[h\right]\right) & \in & \mathsf{Mor}_{\mathfrak{ghs}^{P}}\left(\left(\mathscr{M}_{\pm}\left[h\right],\mathrm{\Lambda}^{1}M_{\pm},A_{\pm}\left[h\right]\right),\left(\mathscr{M}\left[h\right],\mathrm{\Lambda}^{1}M,A\left[h\right]\right)\right)\mbox{.}
\end{eqnarray*}
Denote with $\mathscr{A}$ the LCQFT (fulfilling both the causality
condition and the time slice axiom) built following the procedure
of Subsection \ref{subProcaField}. For $\left(\mathscr{M},\mathrm{\Lambda}^{1}M,A\right)\in\mathsf{Obj}_{\mathfrak{ghs}^{P}}$
and $h\in GHP\left(\mathscr{M}\right)$ we define the RCE for the
Proca field as
\begin{eqnarray*}
R_{h}^{\mathscr{M}} & = & \mathscr{A}\left(\imath_{-}^{\mathscr{M}}\left[h\right],\imath_{-}^{\mathscr{M},\mathrm{\Lambda}^{1}}\left[h\right]\right)\circ\mathscr{A}\left(\jmath_{-}^{\mathscr{M}}\left[h\right],\jmath_{-}^{\mathscr{M},\mathrm{\Lambda}^{1}}\left[h\right]\right)^{-1}\\
 &  & \circ\mathscr{A}\left(\jmath_{+}^{\mathscr{M}}\left[h\right],\jmath_{+}^{\mathscr{M},\mathrm{\Lambda}^{1}}\left[h\right]\right)\circ\mathscr{A}\left(\imath_{+}^{\mathscr{M}}\left[h\right],\imath_{+}^{\mathscr{M},\mathrm{\Lambda}^{1}}\left[h\right]\right)^{-1}\mbox{.}
\end{eqnarray*}
In a similar way one can consider a classical version of the RCE based
on the covariant functor $\mathscr{B}:\mathfrak{ghs}^{P}\overset{\rightarrow}{\rightarrow}\mathfrak{ssp}$
describing the classical theory of the Proca field (this is actually
possible due to version of the time slice axiom satisfied by $\mathscr{B}$,
cfr. Theorem \ref{thmClassicalFieldFunctorProcaField}):
\begin{eqnarray*}
r_{h}^{\mathscr{M}} & = & \mathscr{B}\left(\imath_{-}^{\mathscr{M}}\left[h\right],\imath_{-}^{\mathscr{M},\mathrm{\Lambda}^{1}}\left[h\right]\right)\circ\mathscr{B}\left(\jmath_{-}^{\mathscr{M}}\left[h\right],\jmath_{-}^{\mathscr{M},\mathrm{\Lambda}^{1}}\left[h\right]\right)^{-1}\\
 &  & \circ\mathscr{B}\left(\jmath_{+}^{\mathscr{M}}\left[h\right],\jmath_{+}^{\mathscr{M},\mathrm{\Lambda}^{1}}\left[h\right]\right)\circ\mathscr{B}\left(\imath_{+}^{\mathscr{M}}\left[h\right],\imath_{+}^{\mathscr{M},\mathrm{\Lambda}^{1}}\left[h\right]\right)^{-1}\mbox{.}
\end{eqnarray*}
Since the LCQFT $\mathscr{A}$ is obtained via composition of $\mathscr{B}$
with the quantization functor $\mathscr{C}$ presented in Subsection
\ref{subQuantumFieldTheory}, we realize that
\begin{equation}
R_{h}^{\mathscr{M}}=\mathscr{C}\left(r_{h}^{\mathscr{M}}\right)\mbox{.}\label{eqPQuantumVsClassicalRCE}
\end{equation}
We can determine the action of $r_{h}^{\mathscr{M}}$ applying Proposition
\ref{propPExpressionForTheInverseOfAProperSymplecticMap} and Proposition
\ref{propMorghsP->Morssp}. To be precise, we find proper partitions
of unity $\left\{ \chi_{+}^{a},\chi_{+}^{r}\right\} $ and $\left\{ \chi_{-}^{a},\chi_{-}^{r}\right\} $
on $M$ such that we can express the action of $\mathscr{B}\left(\imath_{+}^{\mathscr{M}}\left[h\right],\imath_{+}^{\mathscr{M},\mathrm{\Lambda}^{1}}\left[h\right]\right)^{-1}$
and respectively of $\mathscr{B}\left(\jmath_{-}^{\mathscr{M}}\left[h\right],\jmath_{-}^{\mathscr{M},\mathrm{\Lambda}^{1}}\left[h\right]\right)^{-1}$
according to Proposition \ref{propPExpressionForTheInverseOfAProperSymplecticMap}.
If we take $\Theta\in\mathscr{B}\left(\mathscr{M},\mathrm{\Lambda}^{1}M,A\right)$
and evaluate $r_{h}^{\mathscr{M}}\Theta$, we easily find the following
result: 
\[
r_{h}^{\mathscr{M}}\Theta=f_{A}A\left[h\right]\left(\chi_{-}^{a/r}f_{A\left[h\right]}A\left(\chi_{+}^{a/r}\Theta\right)\right)\mbox{.}
\]

In the following we will need the expression of $\left.\frac{\mathrm{d}}{\mathrm{d}s}r_{h^{s}}^{\mathscr{M}}\Theta\right|_{0}$
for an arbitrary smooth 1-parameter family of perturbations of the
metric $s\mapsto h^{s}$. For convenience in the upcoming calculation
we write $\mathrm{\delta}_{s}$ in place of $\left.\frac{\mathrm{d}}{\mathrm{d}s}\left(\cdot\right)\right|_{0}$.
Fix now $\Theta\in\mathscr{B}\left(\mathscr{M},\mathrm{\Lambda}^{1}M,A\right)$,
a compact subset $K$ of $M$ and a smooth 1-parameter family of globally
hyperbolic perturbations $\left(-1,1\right)\rightarrow GHP\left(\mathscr{M},K\right)$,
$s\mapsto h^{s}$ and evaluate $\mathrm{\delta}_{s}r_{h^{s}}^{\mathscr{M}}\Theta$.
We carry on such calculation with a procedure identical to the one
followed for the Klein-Gordon field. We must only pay attention to
the application of Lemma \ref{lemExtensionOf ea(Pu)=00003Du}, which
cannot be exploited directly. For example, if we are dealing with
a section $\Theta$ in $\mathrm{\Lambda}^{1}M$ with $\mathscr{M}$-past/future
compact support such that $A\Theta$ has compact support, we must
show that also $P_{A}\Theta$ has compact support and then we can
use Lemma \ref{lemExtensionOf ea(Pu)=00003Du} to conclude
\[
f_{A}^{a/r}A\Theta=e_{A}^{a/r}P_{A}\Theta=\Theta\mbox{.}
\]
In this way we obtain
\begin{equation}
\left.\frac{\mathrm{d}}{\mathrm{d}s}r_{h^{s}}^{\mathscr{M}}\Theta\right|_{0}=f_{A}\left(\left.\frac{\mathrm{d}}{\mathrm{d}s}A\left[h^{s}\right]\right|_{0}\right)\Theta\mbox{.}\label{eqPExpressionForTheDerivativeOfTheClassicalRCE}
\end{equation}

We are left with the problem of the expression for $\mathrm{\delta}_{s}A\left[h^{s}\right]\Theta$.
We know that $A\left[h^{s}\right]=\mathrm{\delta}\left[h^{s}\right]\mathrm{d}+m^{2}\mathrm{id}_{\mathrm{\Omega}^{1}M}$,
where $\mathrm{\delta}\left[h^{s}\right]$ denotes the codifferential
built with the perturbed metric $g_{h^{s}}$. Indeed the term $m^{2}\Theta$
gives null contribution to $\mathrm{\delta}_{s}A\left[h^{s}\right]\Theta$,
hence we are interested in the evaluation of $\mathrm{\delta}_{s}\mathrm{\delta}\left[h^{s}\right]\mathrm{d}\Theta$.
Using an arbitrary coordinate neighborhood, one can check that
\begin{eqnarray*}
\left(\mathrm{\delta}\left[h^{s}\right]\mathrm{d}\Theta\right)_{k} & = & g_{h^{s}}^{ij}\nabla\left[h^{s}\right]_{i}\left(-\nabla\left[h^{s}\right]_{j}\Theta_{k}+\nabla\left[h^{s}\right]_{k}\Theta_{j}\right)\\
 & = & -g_{h^{s}}^{ij}\partial_{i}\Pi_{jk}+g_{h^{s}}^{ij}\Gamma\left[h^{s}\right]_{ij}^{l}\Pi_{lk}+g_{h^{s}}^{ij}\Gamma\left[h^{s}\right]_{ik}^{l}\Pi_{jl}\mbox{,}
\end{eqnarray*}
where $\Gamma\left[h^{s}\right]_{ij}^{k}$ are the Christoffel symbols
of the Levi-Civita connection $\nabla\left[h^{s}\right]$ on $\mathscr{M}\left[h^{s}\right]$
and
\begin{equation}
\Pi_{ij}=\nabla\left[h^{s}\right]_{i}\Theta_{j}-\nabla\left[h^{s}\right]_{j}\Theta_{i}=\partial_{i}\Theta_{j}-\partial_{j}\Theta_{i}=\nabla_{i}\Theta_{j}-\nabla_{j}\Theta_{i}\mbox{.}\label{eqPClassicalFieldStrength}
\end{equation}
Therefore
\begin{eqnarray}
\left.\frac{\mathrm{d}}{\mathrm{d}s}\left(A\left[h^{s}\right]\Theta\right)_{k}\right|_{0} & = & \left.\frac{\mathrm{d}}{\mathrm{d}s}\left(\mathrm{\delta}\left[h^{s}\right]\mathrm{d}\Theta\right)_{k}\right|_{0}\label{eqPDifferentialOperatorVariation}\\
 & = & \left.\frac{\mathrm{d}}{\mathrm{d}s}h_{ij}^{s}\right|_{0}\nabla^{i}\Pi_{\phantom{j}k}^{j}+\left.\frac{\mathrm{d}}{\mathrm{d}s}\Gamma\left[h^{s}\right]_{ij}^{l}\right|_{0}g^{ij}\Pi_{lk}+\left.\frac{\mathrm{d}}{\mathrm{d}s}\Gamma\left[h^{s}\right]_{ik}^{l}\right|_{0}g^{ij}\Pi_{jl}\mbox{,}\nonumber 
\end{eqnarray}
where in the last step we exploited also eq. \eqref{eqLoweringTheIndicesOfTheMetricVariation}.

\subsubsection{Properties of the GNS representation induced by a quasi-free Hadamard
state for the Proca field}

We go on imitating what we have already done in the case of the Klein-Gordon
field. So we choose a quasi-free Hadamard state $\tau$ for the unital
C{*}-algebra $\left(\mathcal{V},\mathrm{V}\right)=\mathscr{A}\left(\mathscr{M},\mathrm{\Lambda}^{1}M,A\right)$
(which is actually a CCR representation) describing the quantum theory
of the Proca field on the globally hyperbolic spacetime $\mathscr{M}$.
With this choice, we introduce the (unique up to unitary equivalence)
GNS triple $\left(\mathscr{H}_{\tau}^{\mathscr{M}},\pi_{\tau}^{\mathscr{M}},\Omega_{\tau}^{\mathscr{M}}\right)$
induced by $\tau$ and we follow the discussion made in Subsection
\ref{subQuasiFreeHadamardStates}. In this way we obtain the represented
version of the Weyl map $\mathrm{V}$ associated to the CCR representation
$\left(\mathcal{V},\mathrm{V}\right)$:
\begin{equation}
\mathrm{V}_{\tau}^{\mathscr{M}}=\pi_{\tau}^{\mathscr{M}}\circ\mathrm{V}:V\rightarrow\mathcal{B}\left(\mathscr{H}_{\tau}^{\mathscr{M}}\right)\mbox{,}\label{eqPRepresentedWeylMap}
\end{equation}
where $\left(V,\sigma\right)=\mathscr{B}\left(\mathscr{M},\mathrm{\Lambda}^{1}M,A\right)$
is the symplectic space provided by the covariant functor $\mathscr{B}:\mathfrak{ghs}^{P}\overset{\rightarrow}{\rightarrow}\mathfrak{ssp}$
describing the classical theory of the Proca field. We find a map
\begin{eqnarray*}
\varPhi_{\tau}^{\mathscr{M}}:V & \rightarrow & \mathcal{B}\left(\mathscr{H}_{\tau}^{\mathscr{M}}\right)\\
\Theta & \mapsto & \varPhi_{\tau}^{\mathscr{M}}\left(\Theta\right)\mbox{.}
\end{eqnarray*}
satisfying $\mathrm{e}^{\imath\varPhi_{\tau}^{\mathscr{M}}\left(\Theta\right)}=\mathrm{V}_{\tau}^{\mathscr{M}}\left(\Theta\right)$
for each $\Theta\in V$, where $\varPhi_{\tau}^{\mathscr{M}}\left(\Theta\right)$
is selfadjoint. Together with this map, we have the smeared fields
(by virtue of the choice of a Hadamard state):
\begin{eqnarray*}
\varPsi_{\tau}^{\mathscr{M}}:\mathrm{\Omega}_{0}^{1}M & \rightarrow & \mathcal{B}\left(\mathscr{H}_{\tau}^{\mathscr{M}}\right)\\
\theta & \mapsto & -\imath\left.\frac{\mathrm{d}}{\mathrm{d}t}\mathrm{V}_{\tau}^{\mathscr{M}}\left(tf_{A}\theta\right)\right|_{0}\mbox{.}
\end{eqnarray*}
As for the general case, it holds that
\begin{equation}
\varPsi_{\tau}^{\mathscr{M}}\left(\theta\right)=\varPhi_{\tau}^{\mathscr{M}}\left(f_{A}\theta\right)\label{eqPRelationBetweenvarPsiAndvarPhi}
\end{equation}
for each $\theta\in\mathrm{\Omega}_{0}^{1}M$ and we recognize $\varPsi_{\tau}^{\mathscr{M}}$
to be linear. 

As we said in Subsection \ref{subQuasiFreeHadamardStates}, the choice
of a quasi-free Hadamard state $\tau$ assures that Assumption \ref{assFunctionalDerivativeRCE}
is satisfied, i.e. we are able to find a dense subspace $\mathscr{V}_{\tau}^{\mathscr{M}}$
of $\mathscr{H}_{\tau}^{\mathscr{M}}$ and a dense sub-{*}-algebra
$\mathcal{B}_{\tau}^{\mathscr{M}}$ of $\mathscr{A}\left(\mathscr{M},\mathrm{\Lambda}^{1}M,A\right)$
such that the functional derivative of the RCE with respect to the
spacetime metric can be defined. We also have a version of eq. \eqref{eqKeyPointForTheMainTheoremAboutRCE}
fitted to the Proca field: for each $\xi\in\mathscr{V}_{\tau}^{\mathscr{M}}$,
each $\Theta\in V$, each compact subset $K$ of $M$ and each smooth
$1$-parameter family $\left(-1,1\right)\rightarrow GHP\left(\mathscr{M},K\right)$,
$s\mapsto h^{s}$ such that $h^{0}=0$, it holds that
\begin{equation}
\left.\frac{\mathrm{d}}{\mathrm{d}s}\left\langle \xi,\mathrm{V}_{\tau}^{\mathscr{M}}\left(r_{h^{s}}^{\mathscr{M}}\Theta\right)\xi\right\rangle _{\tau}^{\mathscr{M}}\right|_{0}=\frac{\imath}{2}\left\langle \xi,\left\{ \varPhi_{\tau}^{\mathscr{M}}\left(\left.\frac{\mathrm{d}}{\mathrm{d}s}\left(r_{h^{s}}^{\mathscr{M}}\Theta\right)\right|_{0}\right),\mathrm{V}_{\tau}^{\mathscr{M}}\left(\Theta\right)\right\} \xi\right\rangle _{\tau}^{\mathscr{M}}\mbox{.}\label{eqPKeyPointForTheMainTheoremAboutRCE}
\end{equation}

Moreover one can show that for each $\eta$, $\xi\in\mathscr{V}_{\tau}^{\mathscr{M}}$
there exists a smooth section, denoted by
\begin{eqnarray*}
M & \rightarrow & \mathrm{T}_{\mathbb{C}}M\\
p & \mapsto & \left\langle \eta,\varPsi_{\tau}^{\mathscr{M}}\left(p\right)\xi\right\rangle _{\tau}^{\mathscr{M}}\mbox{,}
\end{eqnarray*}
where $\mathrm{T}_{\mathbb{C}}M$ stands for the complex vector bundle
obtained via the tensor product of each fiber of $\mathrm{T}M$ with
$\mathbb{C}$ and $\left\langle \cdot,\cdot\right\rangle _{\tau}^{\mathscr{M}}$
denotes the scalar product of the Hilbert space $\mathscr{H}_{\tau}^{\mathscr{M}}$,
such that
\begin{equation}
\left\langle \eta,\varPsi_{\tau}^{\mathscr{M}}\left(\theta\right)\xi\right\rangle _{\tau}^{\mathscr{M}}=\int\limits _{M}\left(\theta\left(p\right)\right)\left(\left\langle \eta,\varPsi_{\tau}^{\mathscr{M}}\left(p\right)\xi\right\rangle _{\tau}^{\mathscr{M}}\right)\mathrm{d}\mu_{g}\label{eqPRelationBetweenUnsmearedAndSmearedFields}
\end{equation}
for each $\theta\in\mathrm{\Omega}_{0}^{1}M$, where $\mathrm{d}\mu_{g}$
is the standard volume form on $\mathscr{M}$ and the dual pairing
between $\mathrm{T}^{*}M$ and $\mathrm{T}M$ has been taken into
account (note that one may indeed write the integrand in the abstract
index notation putting a contravariant index on the new smooth section
and a covariant index on the test function). We may regard this section
as the matrix element of the (unique) unsmeared Proca field induced
by the quasi-free Hadamard state $\tau$ on the globally hyperbolic
spacetime $\mathscr{M}$.

\subsubsection{Quantized stress-energy tensor for the Proca field}

We try to define the quantized stress-energy tensor associated to
the Proca field through the point-splitting procedure starting from
the expression of the classical stress-energy tensor. To obtain it,
we need the expression for the action of the Proca field on the globally
hyperbolic spacetime $\mathscr{M}$, which in turn comes from the
differential operator $A=\mathrm{\delta d}+m^{2}\mathrm{id}_{\mathrm{\Omega}^{1}M}$
governing the classical dynamics of the field:
\begin{alignat*}{1}
S_{\mathscr{M}} & =\frac{1}{2}\left(\Theta,A\Theta\right)_{g,1}=\frac{1}{2}\left(\mathrm{d}\Theta,\mathrm{d}\Theta\right)_{g,2}+\frac{1}{2}m^{2}\left(\Theta,\Theta\right)_{g,1}\\
 & =\frac{1}{2}\int\limits _{M}\left(\mathrm{d}\Theta\wedge*\mathrm{d}\Theta+m^{2}\Theta\wedge*\Theta\right)\mbox{.}
\end{alignat*}
Taking the functional derivative of $S_{\mathscr{M}}$ with respect
to the metric, we find the classical stress-energy tensor for the
Proca field (which we express in local coordinates):
\begin{eqnarray*}
T_{ij}^{\mathscr{M}}\left(p\right) & = & \left.\frac{2}{\sqrt{\left|\det g_{h}\left(p\right)\right|}}\frac{\mathrm{\delta}S_{\mathscr{M}\left[h\right]}}{\mathrm{\delta}g_{h}^{ij}\left(p\right)}\right|_{0}\\
 & = & g^{bd}\left(p\right)\Pi_{ib}\left(p\right)\Pi_{jd}\left(p\right)-\frac{1}{4}g_{ij}\left(p\right)g^{ac}\left(p\right)g^{bd}\left(p\right)\Pi_{ab}\left(p\right)\Pi_{cd}\left(p\right)\\
 &  & +m^{2}\Theta_{i}\left(p\right)\Theta_{j}\left(p\right)-\frac{1}{2}m^{2}g_{ij}\left(p\right)g^{ab}\left(p\right)\Theta_{a}\left(p\right)\Theta_{b}\left(p\right)\mbox{,}
\end{eqnarray*}
where $\Pi$ is defined in eq. \eqref{eqPClassicalFieldStrength}.

With the choice of a quasi-free Hadamard state $\tau$ we can promote
$T_{ij}^{\mathscr{M}}$ to the renormalized quantum stress-energy
tensor $\mathcal{T}_{\tau\, ij}^{\mathscr{M}}$ simply via point-splitting
(refer to \cite[eq. 4.6.5, p. 88]{Wal94}): For each $\eta$, $\xi\in\mathscr{V}_{\tau}^{\mathscr{M}}$,
we choose two {}``near'' points $p$ and $q$ in $M$ and a curve
$\gamma$ connecting them and, parallel transporting along the curve
$\gamma$, we write
\begin{multline}
\left\langle \eta,\mathcal{T}_{\tau}^{\mathscr{M}\, ij}\left(p,q\right)\xi\right\rangle _{\tau}^{\mathscr{M}}=g_{bf}\left(p\right)Y_{\gamma\, b}^{f}\left\langle \eta,\varPi_{\tau}^{\mathscr{M}\, ib}\left(p\right)\varPi_{\tau}^{\mathscr{M}\, jd}\left(q\right)\xi\right\rangle _{\tau}^{\mathscr{M}}\\
-\frac{1}{4}g^{ik}\left(p\right)Y_{\gamma\, k}^{j}g_{ae}\left(p\right)Y_{\gamma\, c}^{e}g_{bf}\left(p\right)Y_{\gamma\, d}^{f}\left\langle \eta,\varPi_{\tau}^{\mathscr{M}\, ab}\left(p\right)\varPi_{\tau}^{\mathscr{M}\, cd}\left(q\right)\xi\right\rangle _{\tau}^{\mathscr{M}}\\
-\frac{1}{2}m^{2}g^{ik}\left(p\right)Y_{\gamma\, k}^{j}g_{ad}\left(p\right)Y_{\gamma\, b}^{d}\left\langle \eta,\varPsi_{\tau}^{\mathscr{M}\, a}\left(p\right)\varPsi_{\tau}^{\mathscr{M}\, b}\left(q\right)\xi\right\rangle _{\tau}^{\mathscr{M}}\\
+m^{2}\left\langle \eta,\varPsi_{\tau}^{\mathscr{M}\, i}\left(p\right)\varPsi_{\tau}^{\mathscr{M}\, j}\left(q\right)\xi\right\rangle _{\tau}^{\mathscr{M}}\mbox{,}\label{eqPStressEnergyTensorViaPointSplitting}
\end{multline}
where we introduced
\[
\varPi_{\tau}^{\mathscr{M}\, ij}\left(p\right)=\nabla^{i}\varPsi_{\tau}^{\mathscr{M}\, j}\left(p\right)-\nabla^{j}\varPsi_{\tau}^{\mathscr{M}\, i}\left(p\right)
\]
to shorten the last formula.

All the remarks made for the Klein-Gordon quantized stress-energy
tensor hold also in this case. In particular the expression does not
depend upon the choice of the curve $\gamma$ provided that $p$ and
$q$ are in a sufficiently small neighborhood so that there exists
a unique geodesic connecting them and we choose $\gamma$ to be such
geodesic. Indeed such choice can be done since our scope is to take
the coincidence limit $q\rightarrow p$ along $\gamma$ (once that
we are sure that no divergence may arise). Indeed this is not the
standard regularization procedure with respect to $\tau$ as reference
state, but the result differs only by a multiple of the identity operator.
Since we are interested in the commutator of the stress-energy tensor
with some other operator, for our aims the point-splitting is equivalent
to the standard regularization procedure.

In our upcoming theorem the stress-energy tensor will appear only
in a commutator with some represented Weyl generator $\mathrm{V}_{\tau}^{\mathscr{M}}\left(\Theta\right)$,
$\Theta\in V$, where $\left(V,\sigma\right)=\mathscr{B}\left(\mathscr{M},\mathrm{\Lambda}^{1}M,A\right)$
for a fixed object $\left(\mathscr{M},\mathrm{\Lambda}^{1}M,A\right)$
in $\mathfrak{ghs}^{P}$. From eq. \eqref{eqPStressEnergyTensorViaPointSplitting}
we realize that it would be useful to evaluate the matrix elements
of the commutator of two unsmeared fields and of $\varPi_{\tau}^{\mathscr{M}}\left(p\right)\varPi_{\tau}^{\mathscr{M}}\left(q\right)$
with an arbitrary represented Weyl generator: we recall eq. \eqref{eqCommutatorBetween2SmearedFieldsAndAWeylGenerator}
and we evaluate its LHS and its RHS fixing $\eta$, $\xi\in\mathscr{V}_{\tau}^{\mathscr{M}}$,
$\theta$, $\theta^{\prime}\in\mathrm{\Omega}_{0}^{1}M$ and $\Theta\in V$
and exploiting eq. \eqref{eqPRelationBetweenUnsmearedAndSmearedFields}
twice (all the equations are written using the abstract index notation):
\begin{multline*}
\iint\limits _{M}\left\langle \eta,\left[\varPsi_{\tau}^{\mathscr{M}\, i}\left(p\right)\varPsi_{\tau}^{\mathscr{M}\, j}\left(q\right),\mathrm{V}_{\tau}^{\mathscr{M}}\left(\Theta\right)\right]\xi\right\rangle _{\tau}^{\mathscr{M}}\theta_{i}\left(p\right)\theta_{j}^{\prime}\left(q\right)\mathrm{d}\mu_{g}\left(p\right)\mathrm{d}\mu_{g}\left(q\right)\\
=\left\langle \eta,\left[\varPsi_{\tau}^{\mathscr{M}}\left(\theta\right)\varPsi_{\tau}^{\mathscr{M}}\left(\theta^{\prime}\right),\mathrm{V}_{\tau}^{\mathscr{M}}\left(\Theta\right)\right]\xi\right\rangle _{\tau}^{\mathscr{M}}\mbox{.}
\end{multline*}
Now we exploit also the definition of the symplectic form $\sigma$
(cfr. Proposition \ref{propSymplecticSpaceForTheProcaField}):
\begin{multline*}
-\sigma\left(f_{A}\theta,\Theta\right)\left\langle \eta,\mathrm{V}_{\tau}^{\mathscr{M}}\left(\Theta\right)\varPsi_{\tau}^{\mathscr{M}}\left(\theta^{\prime}\right)\xi\right\rangle _{\tau}^{\mathscr{M}}\\
=\iint\limits _{M}\Theta_{k}\left(p\right)g^{ki}\left(p\right)\theta_{i}\left(p\right)\theta_{j}^{\prime}\left(q\right)\left\langle \eta,\mathrm{V}_{\tau}^{\mathscr{M}}\left(\Theta\right)\varPsi_{\tau}^{\mathscr{M}\, j}\left(q\right)\xi\right\rangle _{\tau}^{\mathscr{M}}\mathrm{d}\mu_{g}\left(p\right)\mathrm{d}\mu_{g}\left(q\right)\mbox{,}
\end{multline*}
\begin{multline*}
-\sigma\left(f_{A}\theta^{\prime},\Theta\right)\left\langle \eta,\varPsi_{\tau}^{\mathscr{M}}\left(\theta\right)\mathrm{V}_{\tau}^{\mathscr{M}}\left(\Theta\right)\xi\right\rangle _{\tau}^{\mathscr{M}}\\
=\iint\limits _{M}\Theta_{k}\left(q\right)g^{kj}\left(q\right)\theta_{j}^{\prime}\left(q\right)\theta_{i}\left(p\right)\left\langle \eta,\varPsi_{\tau}^{\mathscr{M}\, i}\left(p\right)\mathrm{V}_{\tau}^{\mathscr{M}}\left(\Theta\right)\xi\right\rangle _{\tau}^{\mathscr{M}}\mathrm{d}\mu_{g}\left(p\right)\mathrm{d}\mu_{g}\left(q\right)\mbox{.}
\end{multline*}
From eq. \eqref{eqCommutatorBetween2SmearedFieldsAndAWeylGenerator}
and the freedom in the choice of $\theta$ and $\theta^{\prime}$
we deduce that
\begin{multline}
\left\langle \eta,\left[\varPsi_{\tau}^{\mathscr{M}\, i}\left(p\right)\varPsi_{\tau}^{\mathscr{M}\, j}\left(q\right),\mathrm{V}_{\tau}^{\mathscr{M}}\left(\Theta\right)\right]\xi\right\rangle _{\tau}^{\mathscr{M}}\\
=\Theta^{i}\left(p\right)\left\langle \eta,\mathrm{V}_{\tau}^{\mathscr{M}}\left(\Theta\right)\varPsi_{\tau}^{\mathscr{M}\, j}\left(q\right)\xi\right\rangle _{\tau}^{\mathscr{M}}+\Theta^{j}\left(q\right)\left\langle \eta,\varPsi_{\tau}^{\mathscr{M}\, i}\left(p\right)\mathrm{V}_{\tau}^{\mathscr{M}}\left(\Theta\right)\xi\right\rangle _{\tau}^{\mathscr{M}}\label{eqPCommutatorBetween2UnsmearedFieldsAndAWeylGenerator}
\end{multline}
for each $\Theta\in V$ and each $\eta$, $\xi\in\mathscr{V}_{\tau}^{\mathscr{M}}$.
From the last equation we deduce also that
\begin{multline}
\left\langle \eta,\left[\varPi_{\tau}^{\mathscr{M}\, ij}\left(p\right)\varPi_{\tau}^{\mathscr{M}\, kl}\left(q\right),\mathrm{V}_{\tau}^{\mathscr{M}}\left(\Theta\right)\right]\xi\right\rangle _{\tau}^{\mathscr{M}}\\
=\Pi^{ij}\left(p\right)\left\langle \eta,\mathrm{V}_{\tau}^{\mathscr{M}}\left(\Theta\right)\varPi_{\tau}^{\mathscr{M}\, kl}\left(q\right)\xi\right\rangle _{\tau}^{\mathscr{M}}+\Pi^{kl}\left(q\right)\left\langle \eta,\varPi_{\tau}^{\mathscr{M},ij}\left(p\right)\mathrm{V}_{\tau}^{\mathscr{M}}\left(\Theta\right)\xi\right\rangle _{\tau}^{\mathscr{M}}\mbox{.}\label{eqPCommutatorBetween2QuantizedFieldStrengthsAndAWeylGenerator}
\end{multline}

\subsubsection{Main theorem}

We devoted the discussion from the beginning of the current subsection
to prepare all the material needed to state and prove a theorem about
the compatibility between the action of the functional derivative
of the relative Cauchy evolution with respect of the spacetime metric
and the stress-energy tensor, namely a result similar to the one found
in Subsection \ref{subKGRCE} for the Klein-Gordon field.
\begin{thm}
\label{thmPFunctionalDerivativeOfRCEAgreesWithStressEnergyTensor}Let
$\mathscr{A}:\mathfrak{ghs}^{P}\overset{\rightarrow}{\rightarrow}\mathfrak{alg}$
be the locally covariant quantum field theory for the Proca field
built in Subsection \ref{subProcaField} and let $\left(\mathscr{M},\mathrm{\Lambda}^{1}M,A\right)$
be an object of the category $\mathfrak{ghs}^{P}$ (see Definition
\ref{defghsP}). Consider a quasi-free Hadamard state $\tau$ on the
CCR representation $\left(\mathcal{V},\mathrm{V}\right)=\mathscr{A}\left(\mathscr{M},\mathrm{\Lambda}^{1}M,A\right)$
and denote the GNS triple induced by $\tau$ with $\left(\mathscr{H}_{\tau}^{\mathscr{M}},\pi_{\tau}^{\mathscr{M}},\Omega_{\tau}^{\mathscr{M}}\right)$.
We denote with $\mathrm{V}_{\tau}^{\mathscr{M}}$ the represented
counterpart of the Weyl map $\mathrm{V}$ (cfr. eq. \eqref{eqPRepresentedWeylMap})
and with $\mathcal{T}_{\tau}^{\mathscr{M}}$ the quantum stress-energy
tensor for the Proca field on $\mathscr{M}$ obtained using the point-splitting
procedure in the representation induced by the state $\tau$ (cfr.
eq. \eqref{eqPStressEnergyTensorViaPointSplitting}). Then there exists
a dense subspace $\mathscr{V}_{\tau}^{\mathscr{M}}$ of $\mathscr{H}_{\tau}^{\mathscr{M}}$
such that
\[
\frac{\mathrm{\delta}}{\mathrm{\delta}h}\pi_{\tau}^{\mathscr{M}}\left(R_{h}^{\mathscr{M}}\left(\mathrm{V}\left(\Theta\right)\right)\right)=-\frac{\imath}{2}\left[\mathcal{T}_{\tau}^{\mathscr{M}},\mathrm{V}_{\tau}^{\mathscr{M}}\left(\Theta\right)\right]\quad\forall\Theta\in V
\]
in the sense of quadratic forms on $\mathscr{V}_{\tau}^{\mathscr{M}}$.\end{thm}
\begin{proof}
We consider the dense subspace $\mathscr{V}_{\tau}^{\mathscr{M}}$
of $\mathscr{H}_{\tau}^{\mathscr{M}}$ whose existence is assured
by the choice of a quasi-free Hadamard state $\tau$ on the CCR representation
$\left(\mathcal{V},\mathrm{V}\right)$ (see few lines before eq. \eqref{eqPKeyPointForTheMainTheoremAboutRCE}).

We fix $\xi\in\mathscr{V}_{\tau}^{\mathscr{M}}$, $\Theta\in V$,
a compact subset $K$ of $M$ and 1-parameter family of globally hyperbolic
perturbations $\left(-1,1\right)\rightarrow GHP\left(\mathscr{M},K\right)$,
$s\mapsto h^{s}$ such that $h^{0}=0$ and we adopt the notation $\mathrm{\delta}_{s}=\left.\nicefrac{\mathrm{d}}{\mathrm{d}s}\right|_{0}$.
We reformulate the thesis in a convenient manner imitating the first
part of the proof of the similar theorem for the Klein-Gordon field,
namely Theorem \ref{thmKGFunctionalDerivativeOfRCEAgreesWithStressEnergyTensor}.
The only difference is that here we consider the results for the Proca
field in place of the similar results for the Klein-Gordon field.
To be precise, we use eq. \eqref{eqPQuantumVsClassicalRCE} in place
of eq. \eqref{eqKGQuantumVsClassicalRCE}, eq. \eqref{eqPKeyPointForTheMainTheoremAboutRCE}
in place of eq. \eqref{eqKGKeyPointForTheMainTheoremAboutRCE}, eq.
\eqref{eqPExpressionForTheDerivativeOfTheClassicalRCE} and eq. \eqref{eqPRelationBetweenvarPsiAndvarPhi}
respectively in place of eq. \eqref{eqKGExpressionForTheDerivativeOfTheClassicalRCE}
and eq. \eqref{eqKGRelationBetweenvarPsiAndvarPhi}. In this way we
obtain the following equivalent formulation of our thesis:

\[
\underset{\mathsf{L}}{\underbrace{\left\langle \xi,\left\{ \varPsi_{\tau}^{\mathscr{M}}\left(\mathrm{\delta}_{s}A\left[h^{s}\right]\Theta\right),\mathrm{V}_{\tau}^{\mathscr{M}}\left(\Theta\right)\right\} \xi\right\rangle _{\tau}^{\mathscr{M}}}}=\underset{\mathsf{R}}{\underbrace{-\int\limits _{M}\left(\mathrm{\delta}_{s}h^{s}\right)\left(\left\langle \xi,\left[\mathcal{T}_{\tau}^{\mathscr{M}},\mathrm{V}_{\tau}^{\mathscr{M}}\left(\Theta\right)\right]\xi\right\rangle _{\tau}^{\mathscr{M}}\right)\mathrm{d}\mu_{g}}}\mbox{,}
\]
where the dual pairing between $\mathrm{T}^{*}M\otimes_{s}\mathrm{T}^{*}M$
and $\mathrm{T}M\otimes_{s}\mathrm{T}M$ is considered in the integrand
appearing on the RHS.

We begin with the analysis of the LHS of the last equation (denoted
by $\mathsf{L}$). The RHS (denoted by $\mathsf{R}$) will be discussed
later. We exploit the relation between smeared and unsmeared fields,
eq. \eqref{eqPRelationBetweenUnsmearedAndSmearedFields}:
\[
\mathsf{L}=\int\limits _{M}\left(\left(\mathrm{\delta}_{s}A\left[h^{s}\right]\Theta\right)\left(p\right)\right)\left(\left\langle \xi,\left\{ \varPsi_{\tau}^{\mathscr{M}}\left(p\right),\mathrm{V}_{\tau}^{\mathscr{M}}\left(\Theta\right)\right\} \xi\right\rangle _{\tau}^{\mathscr{M}}\right)\mathrm{d}\mu_{g}\mbox{,}
\]
where the dual pairing between $\mathrm{T}^{*}M$ and $\mathrm{T}M$
is considered. In order to find an expression for $\mathsf{L}$ in
local coordinates, we repeat the construction performed in the proof
of Theorem \ref{thmKGFunctionalDerivativeOfRCEAgreesWithStressEnergyTensor}
to obtain a convenient family of oriented coordinate neighborhoods.
In this way we find a finite family $\left\{ \left(U_{\alpha},V_{\alpha},\phi_{\alpha}\right)\right\} $
obtained choosing from a locally finite covering of $M$ constituted
by oriented coordinate neighborhoods all the elements that intersect
the fixed compact subset $K$ of $M$ (which includes the support
of the coefficients appearing in $\mathrm{\delta}_{s}A\left[h^{s}\right]$).
We stress that the choice of the oriented coordinate neighborhoods
is made in such a way that $\left|\det g\right|=1$. At the same time
we consider only the corresponding members $\left\{ \chi_{\alpha}\right\} $
in the partition of unity subordinate to the original locally finite
covering. Using this finite collection of coordinate neighborhoods,
together with the corresponding members of the partition of unity,
we can obtain the expression of $\mathsf{L}$ in local coordinates:
\[
\mathsf{L}=\sum_{\alpha}\int\limits _{V_{\alpha}}\chi_{\alpha}\left\langle \xi,\left\{ \varPsi_{\tau}^{\mathscr{M}\, i}\left(x\right),\mathrm{V}_{\tau}^{\mathscr{M}}\left(\Theta\right)\right\} \xi\right\rangle _{\tau}^{\mathscr{M}}\left(\mathrm{\delta}_{s}A\left[h^{s}\right]\Theta\right)_{i}\left(x\right)\mathrm{d}V\mbox{,}
\]
where $\mathrm{d}V$ denotes the standard volume form on $\mathbb{R}^{4}$
and all the sections that appear inside the integral are now written
in local coordinates%
\footnote{by this we mean that, inside the integral over $V_{\alpha}$, $\mathrm{\delta}_{s}A\left[h^{s}\right]\Theta$
now denotes the push-forward through $\phi_{\alpha}$ of the original
$\mathrm{\delta}_{s}A\left[h^{s}\right]\Theta$ restricted to $U_{\alpha}$
and similarly for the other sections inside the integral%
}. It is convenient to define the section
\begin{eqnarray*}
\zeta:M & \rightarrow & \mathrm{T}_{\mathbb{C}}M\\
p & \mapsto & \left\langle \xi,\left\{ \varPsi_{\tau}^{\mathscr{M}}\left(p\right),\mathrm{V}_{\tau}^{\mathscr{M}}\left(\Theta\right)\right\} \xi\right\rangle _{\tau}^{\mathscr{M}}\mbox{,}
\end{eqnarray*}
where $\mathrm{T}_{\mathbb{C}}M$ is the complex vector bundle obtained
taking the tensor product of each fiber in $\mathrm{T}M$ with $\mathbb{C}$.
Now we use eq. \eqref{eqPDifferentialOperatorVariation}. In this
way we obtain
\begin{eqnarray*}
\mathsf{L} & = & \overset{\mathsf{L}_{1}}{\overbrace{\sum_{\alpha}\int\limits _{V_{\alpha}}\chi_{\alpha}\zeta^{k}\left(\nabla^{i}\Pi_{\phantom{j}k}^{j}\right)\mathrm{\delta}_{s}h_{ij}^{s}\mathrm{d}V}}\\
 &  & \underset{\mathsf{L}_{2}}{+\underbrace{\sum_{\alpha}\int\limits _{V_{\alpha}}\chi_{\alpha}\zeta^{k}\Pi_{lk}\mathrm{\delta}_{s}\Gamma\left[h^{s}\right]_{ij}^{l}g^{ij}\mathrm{d}V}}+\underset{\mathsf{L}_{3}}{\underbrace{\sum_{\alpha}\int\limits _{V_{\alpha}}\chi_{\alpha}\zeta^{k}\Pi_{jl}g^{ij}\mathrm{\delta}_{s}\Gamma\left[h^{s}\right]_{ik}^{l}\mathrm{d}V}}\mbox{,}
\end{eqnarray*}
where $\Pi$ is defined as in eq. \eqref{eqPClassicalFieldStrength}
and the dependence of the integrand on the point $x\in V_{\alpha}$
is now understood. We denote the first addend appearing on the RHS
of the last equation with $\mathsf{L}_{1}$ and the others with $\mathsf{L}_{2}$
and $\mathsf{L}_{3}$. We integrate $\mathsf{L}_{1}$ by parts noting
that $\chi_{\alpha}$ is null on the boundary of $V_{\alpha}$, hence
no surface term appears:
\begin{eqnarray*}
\mathsf{L}_{1} & = & \overset{\mathsf{X}}{\overbrace{-\sum_{\alpha}\int\limits _{V_{\alpha}}\chi_{\alpha}\left(\nabla^{i}\zeta^{k}\right)\Pi_{\phantom{j}k}^{j}\mathrm{\delta}_{s}h_{ij}^{s}\mathrm{d}V}}\\
 &  & \underset{\mathsf{L}_{4}}{\underbrace{-\sum_{\alpha}\int\limits _{V_{\alpha}}\chi_{\alpha}\zeta_{k}\Pi^{jk}\nabla^{i}\mathrm{\delta}_{s}h_{ij}^{s}\mathrm{d}V}}-\underset{=0}{\underbrace{\sum_{\alpha}\int\limits _{V_{\alpha}}\left(\nabla^{i}\chi_{\alpha}\right)\zeta^{k}\Pi_{\phantom{j}k}^{j}\mathrm{\delta}_{s}h_{ij}^{s}\mathrm{d}V}}\mbox{.}
\end{eqnarray*}
The last term in the equation above vanishes because the sum of $\chi_{\alpha}$
gives 1 on each point of the support of $\mathrm{\delta}_{s}h^{s}$,
hence the sum of their derivatives is null on such region. We denote
the first of the remaining terms with $\mathsf{X}$ and the second
with $\mathsf{L}_{4}$. At the present moment we have
\[
\mathsf{L}=\mathsf{X}+\mathsf{L}_{2}+\mathsf{L}_{3}+\mathsf{L}_{4}\mbox{.}
\]

Now we investigate $\mathsf{R}$ expressing it in the local coordinates
$\left\{ \left(U_{\alpha},V_{\alpha},\phi_{\alpha}\right)\right\} $:
\[
\mathsf{R}=-\sum_{\alpha}\int\limits _{V_{\alpha}}\chi_{\alpha}\left\langle \xi,\left[\mathcal{T}_{\tau}^{\mathscr{M}\, ij}\left(x\right),\mathrm{V}_{\tau}^{\mathscr{M}}\left(\Theta\right)\right]\xi\right\rangle _{\tau}^{\mathscr{M}}\mathrm{\delta}_{s}h_{ij}^{s}\mathrm{d}V\mbox{.}
\]
Now recall the expression of the quantized stress-energy tensor, eq.
\eqref{eqPStressEnergyTensorViaPointSplitting}, and the commutation
relation found in eq. \eqref{eqPCommutatorBetween2UnsmearedFieldsAndAWeylGenerator}
and in eq. \eqref{eqPCommutatorBetween2QuantizedFieldStrengthsAndAWeylGenerator}
and use these data to evaluate $\left\langle \xi,\left[\mathcal{T}_{\tau}^{\mathscr{M}\, ij}\left(p,q\right),\mathrm{V}_{\tau}^{\mathscr{M}}\left(\Theta\right)\right]\xi\right\rangle _{\tau}^{\mathscr{M}}$.
That done, observe that no divergence arises in the limit $q\rightarrow p$.
Hence we can take the coincidence limit as required by the point-splitting
procedure and insert the result in the last equation. Exploiting the
symmetry of $\mathrm{\delta}_{s}h^{s}$ and $g$, we manage to simplify
the result (matrix elements of anticommutators should appear). We
obtain the following expression (as above we replace $\left\langle \xi,\left\{ \varPsi_{\tau}^{\mathscr{M}}\left(x\right),\mathrm{V}_{\tau}^{\mathscr{M}}\left(\Theta\right)\right\} \xi\right\rangle _{\tau}^{\mathscr{M}}$
with $\zeta$):
\begin{eqnarray*}
\mathsf{R} & = & \overset{\mathsf{R}_{1}}{\overbrace{-\sum_{\alpha}\int\limits _{V_{\alpha}}\chi_{\alpha}\left(\nabla^{i}\zeta^{b}-\nabla^{b}\zeta^{i}\right)\Pi_{\phantom{j}b}^{j}\mathrm{\delta}_{s}h_{ij}^{s}\mathrm{d}V-m^{2}\sum_{\alpha}\int\limits _{V_{\alpha}}\chi_{\alpha}\zeta^{i}\Theta^{j}\mathrm{\delta}_{s}h_{ij}^{s}\mathrm{d}V}}\\
 &  & \underset{\mathsf{R}_{2}}{+\underbrace{\frac{1}{4}\sum_{\alpha}\int\limits _{V_{\alpha}}\chi_{\alpha}\Pi_{ab}\left(\nabla^{a}\zeta^{b}-\nabla^{b}\zeta^{a}\right)\mathrm{\delta}_{s}h_{ij}^{s}g^{ij}\mathrm{d}V+\frac{1}{2}m^{2}\sum_{\alpha}\int\limits _{V_{\alpha}}\chi_{\alpha}\zeta^{a}\Theta_{a}\mathrm{\delta}_{s}h_{ij}^{s}g^{ij}\mathrm{d}V}}\mbox{.}
\end{eqnarray*}
We denote the term on the first line of the RHS in the last equation
with $\mathsf{R}_{1}$ and that on the second line with $\mathsf{R}_{2}$.
In first place we evaluate $\mathsf{R}_{1}$ performing a partial
integration on its first term (we directly omit the term containing
derivatives of the functions $\chi_{\alpha}$ its contribution being
null):
\begin{eqnarray*}
\mathsf{R}_{1} & = & \overset{=\mathsf{X}}{\overbrace{-\sum_{\alpha}\int\limits _{V_{\alpha}}\chi_{\alpha}\left(\nabla^{i}\zeta^{k}\right)\Pi_{\phantom{j}k}^{j}\mathrm{\delta}_{s}h_{ij}^{s}\mathrm{d}V}}\\
 &  & +\sum_{\alpha}\int\limits _{V_{\alpha}}\chi_{\alpha}\left(\nabla^{b}\zeta^{i}\right)\Pi_{\phantom{j}b}^{j}\mathrm{\delta}_{s}h_{ij}^{s}\mathrm{d}V-m^{2}\sum_{\alpha}\int\limits _{V_{\alpha}}\chi_{\alpha}\zeta^{i}\Theta^{j}\mathrm{\delta}_{s}h_{ij}^{s}\mathrm{d}V\\
 & = & \mathsf{X}-\sum_{\alpha}\int\limits _{V_{\alpha}}\chi_{\alpha}\zeta^{i}\Pi^{jb}\nabla_{b}\mathrm{\delta}_{s}h_{ij}^{s}\mathrm{d}V+\sum_{\alpha}\int\limits _{V_{\alpha}}\chi_{\alpha}\zeta^{i}g^{jk}\underset{=0}{\underbrace{\left(\nabla^{b}\Pi_{bk}-m^{2}\Theta_{k}\right)}}\mathrm{\delta}_{s}h_{ij}^{s}\mathrm{d}V\mbox{.}
\end{eqnarray*}
It appears the term $\mathsf{X}$ already found in $\mathsf{L}$ and
with trivial manipulations on summation indices we are able to show
a term involving the LHS of the Proca equation (cfr. eq. \eqref{eqProcaEquationInIndexNotation}
bearing in mind the definition of $\Pi$ given in eq. \eqref{eqPClassicalFieldStrength}).
In this way we get rid of another term since $\Theta\in V$, hence
it is a solution of the Proca equation. At the moment we have
\[
\mathsf{R}_{1}=\mathsf{X}\underset{\mathsf{R}_{3}}{\underbrace{-\sum_{\alpha}\int\limits _{V_{\alpha}}\chi_{\alpha}\zeta^{k}\Pi^{jb}\nabla_{b}\mathrm{\delta}_{s}h_{kj}^{s}\mathrm{d}V}}=\mathsf{X}+\mathsf{R}_{3}\mbox{.}
\]
In second place we evaluate $\mathsf{R}_{2}$ proceeding with the
same approach. First of all we notice that we can exploit the antisymmetry
of $\Pi$ to simplify a little bit the first integral. Then we partially
integrate such term with the intention of finding another integrand
that explicitly exhibits the structure of the Proca equation so that
we can get rid of it too (again we omit at all the null term containing
derivatives of $\chi_{\alpha}$): 
\begin{eqnarray*}
2\mathsf{R}_{2} & = & \sum_{\alpha}\int\limits _{V_{\alpha}}\chi_{\alpha}\Pi_{ab}\left(\nabla^{a}\zeta^{b}\right)\mathrm{\delta}_{s}h_{ij}^{s}g^{ij}\mathrm{d}V+m^{2}\sum_{\alpha}\int\limits _{V_{\alpha}}\chi_{\alpha}\zeta^{a}\Theta_{a}\mathrm{\delta}_{s}h_{ij}^{s}g^{ij}\mathrm{d}V\\
 & = & -\sum_{\alpha}\int\limits _{V_{\alpha}}\chi_{\alpha}\zeta^{b}\Pi_{ab}\nabla^{a}\mathrm{\delta}_{s}h_{ij}^{s}g^{ij}\mathrm{d}V-\sum_{\alpha}\int\limits _{V_{\alpha}}\chi_{\alpha}\zeta^{b}\underset{=0}{\underbrace{\left(\nabla^{a}\Pi_{ab}-m^{2}\Theta_{b}\right)}}\mathrm{\delta}_{s}h_{ij}^{s}g^{ij}\mathrm{d}V\mbox{.}
\end{eqnarray*}
Therefore, renaming some summation indices, we obtain the following
result:
\[
\mathsf{R}_{2}=-\frac{1}{2}\sum_{\alpha}\int\limits _{V_{\alpha}}\chi_{\alpha}\zeta^{k}\Pi_{lk}\nabla^{l}\mathrm{\delta}_{s}h_{ij}^{s}g^{ij}\mathrm{d}V\mbox{.}
\]

At this stage our thesis $\mathsf{L}=\mathsf{R}$ is reduced to the
following identity:
\begin{equation}
\mathsf{L}_{2}+\mathsf{L}_{3}+\mathsf{L}_{4}=\mathsf{R}_{2}+\mathsf{R}_{3}\mbox{.}\label{eqPRCEFinalIntegralIdentity}
\end{equation}

The remaining part of this proof is similar to the end of the proof
of Theorem \ref{thmKGFunctionalDerivativeOfRCEAgreesWithStressEnergyTensor}.
To be precise, this time we will prove two identities that hold everywhere
on $M$:
\begin{eqnarray}
\Pi_{jl}g^{ij}\mathrm{\delta}_{s}\Gamma\left[h^{s}\right]_{ik}^{l} & = & -\Pi^{jb}\nabla_{b}\mathrm{\delta}_{s}h_{kj}^{s}\mbox{,}\label{eqPGeometricalIdentity1}\\
\mathrm{\delta}_{s}\Gamma\left[h^{s}\right]_{ij}^{l}g^{ij}-g^{lj}\nabla^{i}\mathrm{\delta}_{s}h_{ij}^{s} & = & -\frac{1}{2}\nabla^{l}\mathrm{\delta}_{s}h_{ij}^{s}g^{ij}\mbox{.}\label{eqPGeometricalIdentity2}
\end{eqnarray}
A cursory glance to such identities shows that the first one entails
$\mathsf{L}_{3}=\mathsf{R}_{3}$ (it is sufficient to contract it
with $\chi_{\alpha}\zeta^{k}$ on each $V_{\alpha}$ and then integrate
over $V_{\alpha}$ and take the sum over $\alpha$), while the second
entails $\mathsf{L}_{2}+\mathsf{L}_{4}=\mathsf{R}_{2}$ (now you should
contract with $\chi_{\alpha}\zeta^{k}\Pi_{lk}$ and then proceed as
in the other case). Hence these identities together imply our thesis,
eq. \eqref{eqPRCEFinalIntegralIdentity}. We prove them fixing a point
$p$ of $M$ and choosing Riemannian normal coordinates in a (sufficiently
small) neighborhood of $p$ (cfr. e.g. \cite[Sect. 3.3, p. 42]{Wal84})
so that the Christoffel symbols of the connection $\nabla$ are null
at $p$ (note that nothing can be said about the Christoffel symbols
of a {}``perturbed'' connection $\nabla\left[h^{s}\right]$).

We begin evaluating the LHS of the first identity, eq. \eqref{eqPGeometricalIdentity1},
with the help of eq. \eqref{eqVariationOfTheChristoffelSymbol}:
\begin{eqnarray*}
\Pi_{jl}g^{ij}\mathrm{\delta}_{s}\Gamma\left[h^{s}\right]_{ik}^{l} & = & \Pi^{im}\frac{1}{2}\left(\partial_{i}\mathrm{\delta}_{s}h_{mk}^{s}+\partial_{k}\mathrm{\delta}_{s}h_{im}^{s}-\partial_{m}\mathrm{\delta}_{s}h_{ik}^{s}\right)\\
 & = & \Pi^{im}\partial_{i}\mathrm{\delta}_{s}h_{mk}^{s}\\
 & = & -\Pi^{jb}\nabla_{b}\mathrm{\delta}_{s}h_{kj}^{s}\mbox{,}
\end{eqnarray*}
where we exploited the symmetry of $\mathrm{\delta}_{s}h^{s}$, the
antisymmetry of $\Pi$ (note that in particular $\Pi^{im}\partial_{k}\mathrm{\delta}_{s}h_{im}^{s}=0$)
and we renamed some summation indices for convenience. This calculation
shows that eq. \eqref{eqPGeometricalIdentity1} actually holds.

Now we focus on the second identity, eq. \eqref{eqPGeometricalIdentity2}.
Specifically we evaluate the first term on its LHS using eq. \eqref{eqVariationOfTheChristoffelSymbolContractedWithTheInverseMetric},
exploiting the symmetry of $\mathrm{\delta}_{s}h_{ij}^{s}$, renaming
some summation indices and bearing in mind that our choice of coordinates
allows us to replace $\nabla$ with $\partial$ and vice versa at
the fixed point $p$:
\begin{eqnarray*}
\mathrm{\delta}_{s}\Gamma\left[h^{s}\right]_{ij}^{l}g^{ij} & = & g^{ij}g^{lk}\partial_{i}\mathrm{\delta}_{s}h_{kj}^{s}-\frac{1}{2}g^{ij}g^{lk}\partial_{k}\mathrm{\delta}_{s}h_{ij}^{s}\\
 & = & g^{lj}\nabla^{i}\mathrm{\delta}_{s}h_{ij}^{s}-\frac{1}{2}\nabla^{l}\mathrm{\delta}_{s}h_{ij}^{s}g^{ij}\mbox{.}
\end{eqnarray*}
This shows that eq. \eqref{eqPGeometricalIdentity2} holds too, hence
the proof is complete.
\end{proof}

\subsection{Relative Cauchy evolution for the electromagnetic\protect \\
field}

The last question we try to answer deals with the agreement between
the action of the functional derivative of the relative Cauchy evolution
for the electromagnetic field and its quantized stress-energy tensor.
To tackle such problem we resort to our discussion about the locally
covariant quantum field theory for the electromagnetic field (cfr.
Subsection \ref{subElectromagneticField}).

Our approach will be similar to the last two subsections, but now
we consider the electromagnetic field, hence we adopt the notation
introduced in Subsection \ref{subElectromagneticField} and we refer
to the results proved there. In particular here we consider the category
$\mathfrak{ghs}^{EM}$ (see Definition \ref{defghsEM}) and the covariant
functor $\mathscr{B}:\mathfrak{ghs}^{EM}\overset{\rightarrow}{\rightarrow}\mathfrak{ssp}$
describing the classical theory of the electromagnetic field which
fulfils both the causality condition and the time slice axiom in the
sense of functors describing classical field theories (see Theorem
\ref{thmClassicalFieldFunctorEMField}). Having $\mathscr{B}$ at
disposal, we follow the usual procedure (see Subsection \ref{subQuantumFieldTheory})
to obtain the locally covariant quantum field theory $\mathscr{A}:\mathfrak{ghs}^{EM}\overset{\rightarrow}{\rightarrow}\mathfrak{alg}$
for the electromagnetic field, i.e. we take the composition of $\mathscr{B}$
with the quantization functor $\mathscr{C}:\mathfrak{ssp}\overset{\rightarrow}{\rightarrow}\mathfrak{alg}$.
By virtue of the properties enjoyed by $\mathscr{B}$, we deduce that
$\mathscr{A}$ is causal and fulfils the time slice axiom in the sense
of LCQFTs (for the details refer to Subsection \ref{subElectromagneticField}).
In particular, the fulfilment of the time slice axiom is essential
for the upcoming discussion.

\subsubsection{Relative Cauchy evolution for the classical electromagnetic field}

The first building block for our final theorem is an expression for
the relative Cauchy evolution for the electromagnetic field at a classical
level. Such result will be achieved with the help of the next proposition.
We remind the reader that an object of $\mathfrak{ghs}^{EM}$ is a
triple $\left(\mathscr{M},\mathrm{\Lambda}^{1}M,A\right)$ where $\mathscr{M}=\left(M,g,\mathfrak{o},\mathfrak{t}\right)$
is a globally hyperbolic spacetime, $\mathrm{\Lambda}^{1}M$ denotes
the cotangent bundle over $M$ which we endow with the inner product
$\left\langle \cdot,\cdot\right\rangle _{g,1}$ induced by the metric
$g$ and $A$ is the linear differential operator $\mathrm{\delta d}$
acting on sections in $\mathrm{\Lambda}^{1}M$ (note that such operator
depends on the metric $g$ through the codifferential $\mathrm{\delta}$).

As we did for the Klein-Gordon field and the Proca field, we are going
to take into account an object of $\mathfrak{ghs}^{EM}$ denoted by
$\left(\left.\mathscr{M}\right|_{O},\mathrm{\Lambda}^{1}O,\left.A\right|_{O}\right)$
for some $\mathscr{M}$-causally convex connected open subset $O$
of $M$. To see how such object is defined and realize that it is
actually an object of $\mathfrak{ghs}^{EM}$ refer to the first part
of the proof of Proposition \ref{propKGExpressionForTheInverseOfAProperSymplecticMap}
replacing $\mathrm{\Lambda}^{0}$ with $\mathrm{\Lambda}^{1}$.

A slight difference appears at the level of morphisms since now we
consider only push-forwards of morphisms of $\mathfrak{ghs}$. This
has to be intended in a proper sense, namely that of Remark \ref{remMorphismsOfConcreteFields}:
We call push-forward of a morphism $\psi\in\mathsf{Mor}_{\mathfrak{ghs}}\left(\mathscr{M},\mathscr{N}\right)$
the composition of the inclusion map of the proper tensor bundle over
$\psi\left(M\right)$ into the tensor bundle over $N$ of the same
type and the push-forward $\psi_{*}^{\prime}=\left(\psi^{\prime-1}\right)^{*}:\mathrm{\Lambda}M\rightarrow\Lambda\psi\left(M\right)$
through the isometric diffeomorphism $\psi^{\prime}:M\rightarrow\psi\left(M\right)$
induced by $\psi$. For example a morphism $\left(\psi,\psi_{*}\right)$
of $\mathfrak{ghs}^{EM}$ from $\left(\mathscr{M},\mathrm{\Lambda}^{1}M,A\right)$
to $\left(\mathscr{N},\mathrm{\Lambda}^{1}N,B\right)$ acts on a element
of $\mathrm{\Lambda}^{k}M$ as $\iota_{\mathrm{\Lambda}^{k}M}^{\mathrm{\Lambda}^{k}N}\circ\left(\psi^{\prime-1}\right)^{*}$.

Here we are interested in morphisms of $\mathfrak{ghs}$ that are
generated by the inclusion maps of a causally convex connected open
subset of a globally hyperbolic spacetime into the whole spacetime.
In such cases the induced isometric diffeomorphism is nothing but
the identity map of the subset, hence the push-forward (in the sense
specified above) reduces to the inclusion map between the proper tensor
bundles.
\begin{prop}
\label{propEMExpressionForTheInverseOfAProperSymplecticMap}Let $\left(\mathscr{M},\mathrm{\Lambda}^{1}M,A\right)$
be an object of $\mathfrak{ghs}^{EM}$ and let $O$ be an $\mathscr{M}$-causally
convex connected open subset of $M$ including a smooth spacelike
Cauchy surface $\Sigma$ for $\mathscr{M}$. Consider the object $\left(\left.\mathscr{M}\right|_{O},\mathrm{\Lambda}^{1}O,\left.A\right|_{O}\right)$
of $\mathfrak{ghs}^{EM}$ and the morphism $\left(\iota_{O}^{M},\iota_{O*}^{M}\right)$
of $\mathfrak{ghs}^{EM}$ from $\left(\left.\mathscr{M}\right|_{O},\mathrm{\Lambda}^{1}O,\left.A\right|_{O}\right)$
to $\left(\mathscr{M},\mathrm{\Lambda}^{1}M,A\right)$ induced by
the inclusion map $\iota_{O}^{M}:O\rightarrow M$. Then there exists
a partition of unity $\left\{ \chi^{a},\chi^{r}\right\} $ on $M$
such that the inverse $\mathscr{B}\left(\iota_{O}^{M},\iota_{O*}^{M}\right)^{-1}$
of the bijective morphism $\mathscr{B}\left(\iota_{O}^{M},\iota_{O*}^{M}\right)$
of $\mathfrak{ssp}$ from $\left(V,\sigma\right)=\mathscr{B}\left(\left.\mathscr{M}\right|_{O},\mathrm{\Lambda}^{1}O,\left.A\right|_{O}\right)$
to $\left(W,\omega\right)=\mathscr{B}\left(\mathscr{M},\mathrm{\Lambda}^{1}M,A\right)$
satisfies the following equation:
\[
\mathscr{B}\left(\iota_{O}^{M},\iota_{O*}^{M}\right)^{-1}\left[\mathtt{A}\right]_{M}=\left[\pm e_{\left.A\right|_{O}}\left(\mathrm{res}_{\iota_{O*}^{M}}\left(A\left(\chi^{a/r}\mathtt{A}\right)\right)\right)\right]_{O}\quad\forall\left[\mathtt{A}\right]_{M}\in W\mbox{,}
\]
where $\mathtt{A}$ is a representative of the equivalence class $\left[\mathtt{A}\right]_{M}$,
$e_{\left.A\right|_{O}}$ is the causal propagator for the formally
selfadjoint normally hyperbolic operator $\left.\mathrm{\Box}_{1}\right|_{O}=\left.\left(A+\mathrm{d\delta}\right)\right|_{O}$
and the restriction map is defined in Lemma \ref{lemresPsieBextPsi=00003DeA}.\end{prop}
\begin{proof}
We apply the procedure presented in the first part of the proof of
Proposition \ref{propKGExpressionForTheInverseOfAProperSymplecticMap}
to choose two smooth spacelike Cauchy surfaces $\Sigma_{-\varepsilon}$
and $\Sigma_{\varepsilon}$ for $\mathscr{M}$ contained in $O$ among
the smooth spacelike Cauchy surfaces in the foliation of $\mathscr{M}$
induced by $\Sigma$. With this choice we consider the open covering
$\left\{ I_{+}^{\mathscr{M}}\left(\Sigma_{-\varepsilon}\right),I_{-}^{\mathscr{M}}\left(\Sigma_{\varepsilon}\right)\right\} $
of $M$ and its subordinate partition of unity $\left\{ \chi^{a},\chi^{r}\right\} $.

Take now $\left[\mathtt{A}\right]_{M}\in W$. As a consequence of
the construction of the functor $\mathscr{B}$, $W=\left\{ \left[e_{A}\theta\right]_{M}:\,\theta\in\mathrm{\Omega}_{0,\mathrm{\delta}}^{1}M\right\} $
(we remind the reader that $\mathrm{\Omega}_{0,\mathrm{\delta}}^{1}M$
denotes the space of compactly supported coclosed 1-forms). Hence,
choosing a representative $\mathtt{A}$ of the class $\left[\mathtt{A}\right]_{M}$,
we also find $\theta\in\mathrm{\Omega}_{0,\mathrm{\delta}}^{1}M$
and therefore we deduce $\mathrm{supp}\left(\mathtt{A}\right)\subseteq J^{\mathscr{M}}\left(K\right)$
for $K=\mathrm{supp}\left(\theta\right)$, which is a compact subset
of $M$. If we define $\mathtt{A}^{a/r}=\chi^{a/r}\mathtt{A}$, we
see that
\[
\mathrm{supp}\left(\mathtt{A}^{a/r}\right)\subseteq J_{\pm}^{\mathscr{M}}\left(\Sigma_{\mp\varepsilon}\right)\mbox{.}
\]
From the last inclusion it follows that $\mathtt{A}^{a/r}$ is an
element of $\mathrm{\Omega}^{1}M$ with $\mathscr{M}$-past/future
compact support. Moreover we know that $\mathrm{\Box}_{1}\mathtt{A}=0$
and $\mathrm{\delta}\mathtt{A}=0$ because $\mathtt{A}=e_{A}\theta$,
$\mathrm{\delta\theta}=0$ and $\mathrm{\delta}e_{A}\theta=e_{A}\mathrm{\delta}\theta$
(see Lemma \ref{lemGreenOperatorsCommuteWithExteriorDerivativeAndCodifferential}).
This entails that $A\mathtt{A}=0$. From this fact, together with
$\chi^{a}+\chi^{r}=1$, we deduce
\begin{eqnarray*}
A\mathtt{A}^{a} & = & -A\mathtt{A}^{r}\mbox{,}\\
\mathrm{\delta}\mathtt{A}^{a} & = & -\mathrm{\delta}\mathtt{A}^{r}\mbox{,}
\end{eqnarray*}
 hence
\begin{eqnarray*}
\mathrm{supp}\left(A\mathtt{A}^{a}\right) & \subseteq & J^{\mathscr{M}}\left(K\right)\cap J_{+}^{\mathscr{M}}\left(\Sigma_{-\varepsilon}\right)\cap J_{-}^{\mathscr{M}}\left(\Sigma_{\varepsilon}\right)\subseteq O\mbox{,}\\
\mathrm{supp}\left(\mathrm{\delta}\mathtt{A}^{a}\right) & \subseteq & J^{\mathscr{M}}\left(K\right)\cap J_{+}^{\mathscr{M}}\left(\Sigma_{-\varepsilon}\right)\cap J_{-}^{\mathscr{M}}\left(\Sigma_{\varepsilon}\right)\subseteq O\mbox{.}
\end{eqnarray*}
Exploiting Proposition \ref{propUsefulSubsetsOfGloballyHyperbolicSpacetimes},
we realize that both $A\mathtt{A}^{a/r}$ and $\mathrm{\delta}\mathtt{A}^{a/r}$
fall in $\mathrm{\Omega}_{0}^{1}M$ and their supports are contained
in $O$. At this point we know that we can apply the restriction map%
\footnote{the definition of the restriction map in the general context of vector
bundles was given in Lemma \ref{lemresPsieBextPsi=00003DeA}.%
} to $A\mathtt{A}^{a/r}$ (and indeed also to $\mathrm{\delta}\mathtt{A}^{a/r}$)
in order to obtain an element of $\mathrm{\Omega}_{0}^{1}O$:
\[
\theta^{\prime}=\mathrm{res}_{\iota_{O*}^{M}}\left(A\left(\chi^{a/r}\mathtt{A}\right)\right)\in\mathrm{\Omega}_{0}^{1}O\mbox{.}
\]
One can almost immediately recognize that $\mathrm{\delta}\theta^{\prime}=0$
because $\mathrm{\delta}\circ A=\mathrm{\delta\circ\delta\circ d}=0$
and
\[
\mathrm{\delta}\circ\mathrm{res}_{\psi_{*}}=\mathrm{res}_{\psi_{*}}\circ\delta
\]
for each morphism $\left(\psi,\psi_{*}\right)$ of $\mathfrak{ghs}^{EM}$
because the push-forward intertwines with both $\mathrm{d}$ and $\mathrm{\delta}$
(see the footnote at page \pageref{fnMorghsEM}). This proves that
it makes sense to consider
\[
\left[\pm e_{\left.A\right|_{O}}\left(\mathrm{res}_{\iota_{O*}^{M}}\left(A\left(\chi^{a/r}\mathtt{A}\right)\right)\right)\right]_{O}\in V=\left\{ \left[e_{\left.A\right|_{O}}\theta^{\prime}\right]_{M}:\,\theta^{\prime}\in\mathrm{\Omega}_{0,\mathrm{\delta}}^{1}O\right\} \mbox{.}
\]
Suppose that we choose a different representative $\mathtt{A}^{\prime}$
of $\left[\mathtt{A}\right]_{M}$. We obtain
\[
\left[\pm e_{\left.A\right|_{O}}\left(\mathrm{res}_{\iota_{O*}^{M}}\left(A\left(\chi^{a/r}\mathtt{A}^{\prime}\right)\right)\right)\right]_{O}\in V\mbox{.}
\]
Indeed we know that $\mathrm{d}\left(\mathtt{A}-\mathtt{A}^{\prime}\right)=0$
because $\mathtt{A}$ and $\mathtt{A}^{\prime}$ are in the same equivalence
class and we wonder if the new element of $V$ coincides with the
old one. To answer such question we have to take a representative
from each of the elements of $V$ considered and show that they differ
by a closed 1-form. Since all operators involved are linear, it is
sufficient to show that
\[
\mathrm{d}\left(e_{\left.A\right|_{O}}\left(\mathrm{res}_{\iota_{O*}^{M}}\left(A\left(\chi^{a/r}\tilde{\mathtt{A}}\right)\right)\right)\right)=0\mbox{,}
\]
where $\tilde{\mathtt{A}}=\mathtt{A}-\mathtt{A}^{\prime}$. We apply
again Lemma \ref{lemGreenOperatorsCommuteWithExteriorDerivativeAndCodifferential}
and, bearing in mind the footnote at page \pageref{fnMorghsEM}, we
obtain
\[
\mathrm{d}\left(e_{\left.A\right|_{O}}\left(\mathrm{res}_{\iota_{O*}^{M}}\left(A\left(\chi^{a/r}\tilde{\mathtt{A}}\right)\right)\right)\right)=e_{\left.A\right|_{O}}\left(\mathrm{res}_{\iota_{O*}^{M}}\left(\mathrm{d}A\left(\chi^{a/r}\tilde{\mathtt{A}}\right)\right)\right)\mbox{.}
\]
Since $\mathrm{d}\tilde{\mathtt{A}}=0$, we deduce that $\mathrm{d}\left(\chi^{a}\tilde{\mathtt{A}}\right)=-\mathrm{d}\left(\chi^{r}\tilde{\mathtt{A}}\right)$.
From this relation we deduce that $\mathrm{d}\left(\chi^{a}\tilde{\mathtt{A}}\right)$
has compact support contained in $O$ (the proof is identical to that
of the compactness and the inclusion in $O$ of the supports of $A\mathtt{A}^{a/r}$
and $\mathrm{\delta}\mathtt{A}^{a/r}$). Moreover $\mathrm{d}\circ A=\mathrm{\Box}_{1}\circ\mathrm{d}$
and
\[
\mathrm{\Box}_{1}\circ\mathrm{res}_{\iota_{O*}^{M}}=\mathrm{res}_{\iota_{O*}^{M}}\circ\mathrm{\Box}_{1}\mbox{.}
\]
From all these observations we conclude that
\[
\mathrm{d}\left(e_{\left.A\right|_{O}}\left(\mathrm{res}_{\iota_{O*}^{M}}\left(A\left(\chi^{a/r}\tilde{\mathtt{A}}\right)\right)\right)\right)=e_{\left.A\right|_{O}}\mathrm{\Box}_{1}\left(\mathrm{res}_{\iota_{O*}^{M}}\left(\mathrm{d}\left(\chi^{a/r}\tilde{\mathtt{A}}\right)\right)\right)=0\mbox{.}
\]
This proves that the map
\begin{eqnarray*}
\alpha:W & \rightarrow & V\\
\left[\mathtt{A}\right]_{M} & \mapsto & \left[\pm e_{\left.A\right|_{O}}\left(\mathrm{res}_{\iota_{O*}^{M}}\left(A\left(\chi^{a/r}\mathtt{A}\right)\right)\right)\right]_{O}\mbox{,}
\end{eqnarray*}
where $\mathtt{A}$ is a representative of $\left[\mathtt{A}\right]_{M}$,
is well defined.

Note that, from the hypothesis made, we know that the image $\iota_{O}^{M}\left(O\right)=O$
includes a smooth spacelike Cauchy surface for $\mathscr{M}$. Hence
$\mathscr{B}\left(\iota_{O}^{M},\iota_{O*}^{M}\right)^{-1}$ is a
morphism of $\mathfrak{ssp}$ from $\left(W,\omega\right)$ to $\left(V,\sigma\right)$
because the time slice axiom holds for $\mathscr{B}$ (cfr. Theorem
\ref{thmClassicalFieldFunctorEMField}). To conclude the proof we
must check that $\alpha=\mathscr{B}\left(\iota_{O}^{M},\iota_{O*}^{M}\right)^{-1}$.
Take $\left[\mathtt{A}\right]_{M}\in W$ and one of its representatives
$\mathtt{A}$, consider a partition of unity $\left\{ \chi^{a},\chi^{b}\right\} $
built following the prescriptions given above and define $\mathtt{A}^{a/r}=\chi^{a/r}\mathtt{A}$.
Recalling Lemma \ref{lemMorghsEM->Morssp} and observing that the
restriction map followed by the corresponding extension leaves the
argument of the restriction unchanged, we find
\begin{eqnarray*}
\mathscr{B}\left(\iota_{O}^{M},\iota_{O*}^{M}\right)\left(\alpha\left[\mathtt{A}\right]_{M}\right) & = & \mathscr{B}\left(\iota_{O}^{M},\iota_{O*}^{M}\right)\left[\pm e_{\left.A\right|_{O}}\left(\mathrm{res}_{\iota_{O*}^{M}}\left(A\mathtt{A}^{a/r}\right)\right)\right]_{O}\\
 & = & \left[\pm e_{A}\left(\mathrm{ext}_{\iota_{O*}^{M}}\circ\mathrm{res}_{\iota_{O*}^{M}}\right)A\mathtt{A}^{a/r}\right]_{M}\\
 & = & \left[\pm e_{A}A\mathtt{A}^{a/r}\right]_{M}\mbox{.}
\end{eqnarray*}
The support properties of $\mathtt{A}^{a/r}$, $A\mathtt{A}^{a/r}$
and $\mathrm{\delta}\mathtt{A}^{a/r}$ allow us to apply Lemma \ref{lemGreenOperatorsCommuteWithExteriorDerivativeAndCodifferential}
and Lemma \ref{lemExtensionOf ea(Pu)=00003Du}:
\begin{eqnarray*}
\pm e_{A}A\mathtt{A}^{a/r} & = & \pm\left(e_{A}\mathrm{\Box}_{1}\mathtt{A}^{a/r}-e_{A}\mathrm{d\delta}\mathtt{A}^{a/r}\right)\\
 & = & \pm\left[\left(e_{A}^{a}\mathrm{\Box}_{1}\mathtt{A}^{a/r}-e_{A}^{r}\mathrm{\Box}_{1}\mathtt{A}^{a/r}\right)-e_{A}\mathrm{d\delta}\mathtt{A}^{a/r}\right]\\
 & = & \mathtt{A}^{a}+\mathtt{A}^{r}\mp\mathrm{d}e_{A}\mathrm{\delta}\mathtt{A}^{a/r}\\
 & = & \mathtt{A}\mp\mathrm{d}e_{A}\mathrm{\delta}\mathtt{A}^{a/r}\mbox{.}
\end{eqnarray*}
The result of the last calculation entails that $\pm e_{A}A\mathtt{A}^{a/r}$
and $\mathtt{A}$ are gauge equivalent, hence 
\[
\left[\pm e_{A}A\mathtt{A}^{a/r}\right]_{M}=\left[\mathtt{A}\right]_{M}\mbox{.}
\]
With this we conclude
\[
\mathscr{B}\left(\iota_{O}^{M},\iota_{O*}^{M}\right)\left(\alpha\left[\mathtt{A}\right]_{M}\right)=\left[\mathtt{A}\right]_{M}=\mathscr{B}\left(\iota_{O}^{M},\iota_{O*}^{M}\right)\left(\mathscr{B}\left(\iota_{O}^{M},\iota_{O*}^{M}\right)^{-1}\left[\mathtt{A}\right]_{M}\right)\quad\forall\left[\mathtt{A}\right]_{M}\in W\mbox{.}
\]
Since $\mathscr{B}\left(\iota_{O}^{M},\iota_{O*}^{M}\right)$ is injective,
the last equation entails
\[
\alpha\left[\mathtt{A}\right]_{M}=\mathscr{B}\left(\iota_{O}^{M},\iota_{O*}^{M}\right)^{-1}\left[\mathtt{A}\right]_{M}\quad\forall\left[\mathtt{A}\right]_{M}\in W\mbox{,}
\]
therefore we realize that the thesis actually holds.
\end{proof}
We specialize the definition of the RCE to the case of the electromagnetic
field. Consider an object $\left(\mathscr{M},\mathrm{\Lambda}^{1}M,A\right)$
of $\mathfrak{ghs}^{EM}$, take $h\in GHP\left(\mathscr{M}\right)$
and recall the definitions of the morphisms $\imath_{\pm}^{\mathscr{M}}\left[h\right]$
and $\jmath_{\pm}^{\mathscr{M}}\left[h\right]$ of the category $\mathfrak{ghs}$
introduced before Definition \ref{defRCE}. Together with the perturbed
spacetime $\mathscr{M}\left[h\right]$, we must also consider the
effects of the perturbation $h$ on the vector bundle (specifically
on the inner product defined on it) and on the linear differential
operator $A=\mathrm{\delta d}$. The inner product on the cotangent
bundle over the perturbed spacetime is induced by the perturbed metric
$g_{h}=g+h$ and the perturbed linear differential operator is $A\left[h\right]=\mathrm{\delta}\left[h\right]\mathrm{d}$,
where $\mathrm{\delta}\left[h\right]$ is the codifferential defined
on $\mathscr{M}\left[h\right]$. It can be useful to consider also
the perturbed d'Alembert operator $\mathrm{\Box}_{1}\left[h\right]=\mathrm{\delta}\left[h\right]\mathrm{d}+\mathrm{d}\mathrm{\delta}\left[h\right]$
acting on 1-form over $\mathscr{M}\left[h\right]$. We take into account
the inclusion map $\iota_{M_{\pm}*}^{M}$, where $M_{\pm}=M\setminus J_{\mp}^{\mathscr{M}}\left(\mathrm{supp}\left(h\right)\right)$
in accordance with the definitions of $\imath_{\pm}^{\mathscr{M}}\left[h\right]$
and $\jmath_{\pm}^{\mathscr{M}}\left[h\right]$. Compatibility of
$\left(\iota_{M_{\pm}}^{M},\iota_{M_{\pm}*}^{M}\right)$ with both
$\mathrm{d}$ and $\delta$ holds (see the footnote at page \pageref{fnMorghsEM}):
\begin{eqnarray*}
\mathrm{ext}_{\iota_{M_{\pm}*}^{M}}\left(\mathrm{d}\theta\right) & = & \mathrm{d}\left(\mathrm{ext}_{\iota_{M_{\pm}*}^{M}}\theta\right)\quad\forall\theta\in\mathrm{\Omega}_{0}^{k}M_{\pm}\mbox{,}\\
\mathrm{ext}_{\iota_{M_{\pm}*}^{M}}\left(\mathrm{\delta}\theta\right) & = & \mathrm{\delta}\left(\mathrm{ext}_{\iota_{M_{\pm}*}^{M}}\theta\right)\quad\forall\theta\in\mathrm{\Omega}_{0}^{k}M_{\pm}\mbox{.}
\end{eqnarray*}
Since the effects of the perturbation $h$ are relevant only inside
$\mathrm{supp}\left(h\right)$, we realize that $\mathrm{\delta}\left[h\right]$
and $\mathrm{\delta}$ act exactly in the same way on sections supported
outside $\mathrm{supp}\left(h\right)$. Together with $\left.A\right|_{M_{\pm}}$,
we may consider $\left.A\left[h\right]\right|_{M_{\pm}}$ and we immediately
recognize that they are the same linear differential operator acting
on sections in $\mathrm{\Lambda}^{1}M_{\pm}$ (we denote both of them
with $A_{\pm}\left[h\right]$). All these observations are made in
order to introduce the objects $\left(\mathscr{M}\left[h\right],\mathrm{\Lambda}^{1}M,A\left[h\right]\right)$
and $\left(\mathscr{M}_{\pm}\left[h\right],\mathrm{\Lambda}^{1}M_{\pm},A_{\pm}\left[h\right]\right)$
of $\mathfrak{ghs}^{EM}$ and to interpret $\left(\iota_{M_{\pm}}^{M},\iota_{M_{\pm}*}^{M}\right)$
both as a morphism from $\left(\mathscr{M}_{\pm}\left[h\right],\mathrm{\Lambda}^{1}M_{\pm},A_{\pm}\left[h\right]\right)$
to $\left(\mathscr{M},\mathrm{\Lambda}^{1}M,A\right)$ and as a morphism
from $\left(\mathscr{M}_{\pm}\left[h\right],\mathrm{\Lambda}^{1}M_{\pm},A_{\pm}\left[h\right]\right)$
to $\left(\mathscr{M}\left[h\right],\mathrm{\Lambda}^{1}M,A\left[h\right]\right)$
(note the analogy with the definitions of $\imath_{\pm}^{\mathscr{M}}\left[h\right]$
and $\jmath_{\pm}^{\mathscr{M}}\left[h\right]$ as different morphisms
obtained from the inclusion map $\iota_{M_{\pm}}^{M}$). We denote
such morphisms in the following way: 
\begin{eqnarray*}
\left(\imath_{\pm}^{\mathscr{M}}\left[h\right],\imath_{\pm}^{\mathscr{M}}\left[h\right]_{*}\right) & \in & \mathsf{Mor}_{\mathfrak{ghs}^{EM}}\left(\left(\mathscr{M}_{\pm}\left[h\right],\mathrm{\Lambda}^{1}M_{\pm},A_{\pm}\left[h\right]\right),\left(\mathscr{M},\mathrm{\Lambda}^{1}M,A\right)\right)\mbox{,}\\
\left(\jmath_{\pm}^{\mathscr{M}}\left[h\right],\jmath_{\pm}^{\mathscr{M}}\left[h\right]_{*}\right) & \in & \mathsf{Mor}_{\mathfrak{ghs}^{EM}}\left(\left(\mathscr{M}_{\pm}\left[h\right],\mathrm{\Lambda}^{1}M_{\pm},A_{\pm}\left[h\right]\right),\left(\mathscr{M}\left[h\right],\mathrm{\Lambda}^{1}M,A\left[h\right]\right)\right)\mbox{.}
\end{eqnarray*}
Denote with $\mathscr{A}$ the LCQFT (fulfilling both the causality
condition and the time slice axiom) built in Subsection \ref{subElectromagneticField}.
For $\left(\mathscr{M},\mathrm{\Lambda}^{1}M,A\right)\in\mathsf{Obj}_{\mathfrak{ghs}^{EM}}$
and $h\in GHP\left(\mathscr{M}\right)$ we define the RCE for the
electromagnetic field as:
\begin{eqnarray*}
R_{h}^{\mathscr{M}} & = & \mathscr{A}\left(\imath_{-}^{\mathscr{M}}\left[h\right],\imath_{-}^{\mathscr{M}}\left[h\right]_{*}\right)\circ\mathscr{A}\left(\jmath_{-}^{\mathscr{M}}\left[h\right],\jmath_{-}^{\mathscr{M}}\left[h\right]_{*}\right)^{-1}\\
 &  & \circ\mathscr{A}\left(\jmath_{+}^{\mathscr{M}}\left[h\right],\jmath_{+}^{\mathscr{M}}\left[h\right]_{*}\right)\circ\mathscr{A}\left(\imath_{+}^{\mathscr{M}}\left[h\right],\imath_{+}^{\mathscr{M}}\left[h\right]_{*}\right)^{-1}\mbox{.}
\end{eqnarray*}
In a similar way one can consider a classical version of the RCE based
on the covariant functor $\mathscr{B}$ describing the classical theory
of the electromagnetic field (this is actually possible due to version
of the time slice axiom satisfied by $\mathscr{B}$, cfr. Theorem
\ref{thmClassicalFieldFunctorEMField}):
\begin{eqnarray*}
r_{h}^{\mathscr{M}} & = & \mathscr{B}\left(\imath_{-}^{\mathscr{M}}\left[h\right],\imath_{-}^{\mathscr{M}}\left[h\right]_{*}\right)\circ\mathscr{B}\left(\jmath_{-}^{\mathscr{M}}\left[h\right],\jmath_{-}^{\mathscr{M}}\left[h\right]_{*}\right)^{-1}\\
 &  & \circ\mathscr{B}\left(\jmath_{+}^{\mathscr{M}}\left[h\right],\jmath_{+}^{\mathscr{M}}\left[h\right]_{*}\right)\circ\mathscr{B}\left(\imath_{+}^{\mathscr{M}}\left[h\right],\imath_{+}^{\mathscr{M}}\left[h\right]_{*}\right)^{-1}\mbox{.}
\end{eqnarray*}
Since the LCQFT $\mathscr{A}$ is obtained via composition of $\mathscr{B}$
with the quantization functor $\mathscr{C}$ presented in Subsection
\ref{subQuantumFieldTheory}, we realize that%
\footnote{this is a direct consequence of the covariant axioms, which are required
to be verified by any covariant functor%
}
\begin{equation}
R_{h}^{\mathscr{M}}=\mathscr{C}\left(r_{h}^{\mathscr{M}}\right)\mbox{.}\label{eqEMQuantumVsClassicalRCE}
\end{equation}
We can determine the action of $r_{h}^{\mathscr{M}}$ applying Proposition
\ref{propEMExpressionForTheInverseOfAProperSymplecticMap} and Lemma
\ref{lemMorghsEM->Morssp}. We find proper partitions of unity $\left\{ \chi_{+}^{a},\chi_{+}^{r}\right\} $
and $\left\{ \chi_{-}^{a},\chi_{-}^{r}\right\} $ on $M$ such that
we can express the action of $\mathscr{B}\left(\imath_{+}^{\mathscr{M}}\left[h\right],\imath_{+}^{\mathscr{M}}\left[h\right]_{*}\right)^{-1}$
and respectively of $\mathscr{B}\left(\jmath_{-}^{\mathscr{M}}\left[h\right],\jmath_{-}^{\mathscr{M}}\left[h\right]_{*}\right)^{-1}$
according to Proposition \ref{propEMExpressionForTheInverseOfAProperSymplecticMap}.
If we take $\left[\mathtt{A}\right]_{M}\in\mathscr{B}\left(\mathscr{M},\mathrm{\Lambda}^{1}M,A\right)$
and evaluate $r_{h}^{\mathscr{M}}\left[\mathtt{A}\right]_{M}$, we
easily obtain the following result: 
\begin{equation}
r_{h}^{\mathscr{M}}\left[\mathtt{A}\right]_{M}=\left[e_{A}A\left[h\right]\left(\chi_{-}^{a/r}e_{A\left[h\right]}A\left(\chi_{+}^{a/r}\mathtt{A}\right)\right)\right]_{M}\mbox{,}\label{eqEMExpressionForTheClassicalRCE}
\end{equation}
whatever choice of the representative $\mathtt{A}$ of the equivalence
class $\left[\mathtt{A}\right]_{M}$ we make. The independence on
the choice of the representative follows from the fact that the same
property holds for all the morphisms that we composed to find the
expression above.

To prove our final theorem we will need the expression of $\left.\frac{\mathrm{d}}{\mathrm{d}s}r_{h^{s}}^{\mathscr{M}}\left[\mathtt{A}\right]_{M}\right|_{0}$
for an arbitrary smooth 1-parameter family of perturbations of the
metric. For convenience in the upcoming calculation we will denote
$\left.\frac{\mathrm{d}}{\mathrm{d}s}\left(\cdot\right)\right|_{0}$
with $\mathrm{\delta}_{s}$. Fix $\left[\mathtt{A}\right]_{M}\in\mathscr{B}\left(\mathscr{M},\mathrm{\Lambda}^{1}M,A\right)$,
a compact subset $K$ of $M$ and a smooth 1-parameter family of globally
hyperbolic perturbations $\left(-1,1\right)\rightarrow GHP\left(\mathscr{M},K\right)$,
$s\mapsto h^{s}$ such that $h^{0}=0$. To evaluate $\mathrm{\delta}_{s}r_{h^{s}}^{\mathscr{M}}\left[\mathtt{A}\right]_{M}$,
we start from eq. \eqref{eqEMExpressionForTheClassicalRCE} with the
choice of the superscript $r$ (indeed the choice of $a$ would produce
a similar calculation and the same result). In the present situation
apparently we would have to consider different partitions of unity
$\left\{ \chi_{+}^{a},\chi_{+}^{r}\right\} $ and $\left\{ \chi_{-}^{a},\chi_{-}^{r}\right\} $
for each of the values assumed by $s$. Anyway such complication can
be avoided making an intelligent choice of the smooth spacelike Cauchy
surfaces used to define the partitions of unity: We use always the
same foliation of $\mathscr{M}$ (induced by some fixed smooth spacelike
Cauchy surface $\Sigma$ for $\mathscr{M}$) and take the smooth spacelike
Cauchy surfaces inside $M_{\pm}=M\setminus J_{\mp}^{\mathscr{M}}\left(K\right)$
instead of choosing, for each value of $s$, a pair of proper smooth
spacelike Cauchy surfaces inside $M_{\pm}=M\setminus J_{\mp}^{\mathscr{M}}\left(\mathrm{supp}\left(h^{s}\right)\right)$.
In this way a single choice of the smooth spacelike Cauchy surfaces
is satisfactory for each value of $s$. Such choice is possible because
the supports of all the elements $h^{s}$ in the family of perturbations
are controlled by the compact subset $K$ of $M$.

We fix $\left[\mathtt{A}\right]_{M}$ and we take one of its representatives
$\mathtt{A}$. For convenience we define
\[
\mathtt{X}^{s}=e_{A}A\left[h^{s}\right]\left(\chi_{-}^{a/r}e_{A\left[h^{s}\right]}A\left(\chi_{+}^{a/r}\mathtt{A}\right)\right)
\]
so that eq. \eqref{eqEMExpressionForTheClassicalRCE} becomes $r_{h^{s}}^{\mathscr{M}}\left[\mathtt{A}\right]_{M}=\left[\mathtt{X}^{s}\right]_{M}$
and our problem reduces to the search of a convenient vector potential
that is gauge equivalent to $\mathrm{\delta}_{s}\mathtt{X}^{s}$.
We try to reproduce the calculations performed in the case of the
Klein-Gordon field. The important thing now is that we can add terms
to our representative vector potential without changing the equivalence
class in which it falls, provided that such terms are closed and coclosed
1-forms: actually this means that we can take Lorentz solutions (refer
to Lemma \ref{lemSpaceOfGaugeInequivalentDynamicalConfigurationsOfTheEMField})
that are gauge equivalent to our starting Lorentz solution $\mathrm{\delta}_{s}\mathtt{X}^{s}$.
In first place we apply the Leibniz rule (see the footnote at page
\pageref{fnSequentialContinuityOfCausalPropagators}): 
\[
\mathrm{\delta}_{s}\mathtt{X}^{s}=e_{A}\left(\left(\mathrm{\delta}_{s}A\left[h^{s}\right]\right)\left(\chi_{-}^{r}e_{A}A\left(\chi_{+}^{r}\mathtt{A}\right)\right)+A\left(\chi_{-}^{r}\left(\mathrm{\delta}_{s}e_{A\left[h^{s}\right]}\right)A\left(\chi_{+}^{r}\mathtt{A}\right)\right)\right)\mbox{.}
\]
We focus on the first addend: On the one hand, following the proof
of Proposition \ref{propEMExpressionForTheInverseOfAProperSymplecticMap}
(we are considering $M_{-}=M\setminus J_{+}^{\mathscr{M}}\left(K\right)$
as $O$), we can easily see that $\mathrm{supp}\left(\chi_{-}^{r}\right)\subseteq J_{-}^{\mathscr{M}}\left(M_{-}\right)$,
while on the other hand $\mathrm{\delta}_{s}A\left[h^{s}\right]$
can have non null coefficients only inside $K$. This entails that
\[
\left(\mathrm{\delta}_{s}A\left[h^{s}\right]\right)\left(\chi_{-}^{r}e_{A}A\left(\chi_{+}^{r}\mathtt{A}\right)\right)=0\mbox{,}
\]
so that we obtain
\[
\mathrm{\delta}_{s}\mathtt{X}^{s}=e_{A}A\left(\chi_{-}^{r}\left(\mathrm{\delta}_{s}e_{A\left[h^{s}\right]}\right)A\left(\chi_{+}^{r}\mathtt{A}\right)\right)\mbox{.}
\]
Recalling again the proof of Proposition \ref{propEMExpressionForTheInverseOfAProperSymplecticMap},
one sees that $A\left(\chi_{+}^{r}\mathtt{A}\right)=-A\left(\chi_{+}^{a}\mathtt{A}\right)$
and hence deduces that its support is compact and lies in the causal
future of a smooth spacelike Cauchy surface for $\mathscr{M}$ included
in $M_{+}=M\setminus J_{-}^{\mathscr{M}}\left(K\right)$ (that by
construction lies outside $K$ and intersects its causal future).
On the contrary $\chi_{-}^{r}$ is supported in the causal past of
a smooth spacelike Cauchy surface for $\mathscr{M}$ included in $M_{-}$
(that by construction lies outside $K$ and intersects its causal
past). These observations entail that $\chi_{-}^{r}e_{A\left[h^{s}\right]}^{a}A\left(\chi_{+}^{r}\mathtt{A}\right)$
has empty support for each $s$, hence it is null. Therefore from
the last equation we obtain
\begin{equation}
\mathrm{\delta}_{s}\mathtt{X}_{s}=-e_{A}A\left(\chi_{-}^{r}\left(\mathrm{\delta}_{s}e_{A\left[h^{s}\right]}^{r}\right)A\left(\chi_{+}^{r}\mathtt{A}\right)\right)\mbox{.}\label{eqEMExpressionForTheFunctionalDerivativeOfTheClassicalRCEStep1}
\end{equation}

Now we take a closer look to the term $e_{A\left[h^{s}\right]}^{r}A\left[h^{s}\right]\left(\chi_{+}^{r}\mathtt{A}\right)$
for an arbitrary but fixed value of $s$. In order for this term to
make sense it must be shown that $A\left[h^{s}\right]\left(\chi_{+}^{r}\mathtt{A}\right)$
has compact support. This follows from the the following facts:
\begin{itemize}
\item $A\left(\chi_{+}^{r}\mathtt{A}\right)=-A\left(\chi_{+}^{a}\mathtt{A}\right)$
implies that $A\left(\chi_{+}^{a/r}\mathtt{A}\right)$ has compact
support (note that $\chi_{+}^{a/r}$ is supported in the causal future/past
of a proper smooth spacelike Cauchy surface for $\mathscr{M}$ and
remember that $\mathrm{supp}\left(\mathtt{A}\right)\subseteq J^{\mathscr{M}}\left(K^{\prime}\right)$
for a proper compact subset $K^{\prime}$ of $M$);
\item $\mathrm{\delta}\mathtt{A}=0$, hence we also have $\mathrm{\delta}\left(\chi_{+}^{r}\mathtt{A}\right)=-\mathrm{\delta}\left(\chi_{+}^{a}\mathtt{A}\right)$,
which entails that $\mathrm{\delta}\left(\chi_{+}^{a/r}\mathtt{A}\right)$
has compact support by the argument exploited in the previous point;
\item $\mathrm{\delta}\left[h^{s}\right]$ acts as $\mathrm{\delta}$ on
sections whose support has empty intersection with $K$, which is
compact, hence the support of the codifferential of a section can
be enlarged at most by $K$ when we replace $\mathrm{\delta}$ with
the perturbed codifferential $\mathrm{\delta}\left[h^{s}\right]$
(obviously the same conclusion holds if we replace $A$ with $A\left[h^{s}\right]$).
\end{itemize}
These facts imply that
\begin{eqnarray*}
\mathrm{supp}\left(A\left[h^{s}\right]\left(\chi_{+}^{a/r}\mathtt{A}\right)\right) & \subseteq & \mathrm{supp}\left(A\left(\chi_{+}^{a/r}\mathtt{A}\right)\right)\cup K\mbox{,}\\
\mathrm{supp}\left(\mathrm{\delta}\left[h^{s}\right]\left(\chi_{+}^{a/r}\mathtt{A}\right)\right) & \subseteq & \mathrm{supp}\left(\mathrm{\delta}\left(\chi_{+}^{a/r}\mathtt{A}\right)\right)\cup K\mbox{,}\\
\mathrm{supp}\left(\mathrm{\Box}_{1}\left[h^{s}\right]\left(\chi_{+}^{a/r}\mathtt{A}\right)\right) & \subseteq & \mathrm{supp}\left(A\left(\chi_{+}^{a/r}\mathtt{A}\right)\right)\cup\mathrm{supp}\left(\mathrm{\delta}\left(\chi_{+}^{a/r}\mathtt{A}\right)\right)\cup K\mbox{,}
\end{eqnarray*}
hence the supports appearing on the LHS of the last inclusions are
compact subsets of $M$ since they are closed (by definition of support)
and contained in the union of a finite number of compact subsets of
$M$. From the first point above it follows also that $\chi_{+}^{a/r}\mathtt{A}$
has past/future compact support (we are exploiting Proposition \ref{propUsefulSubsetsOfGloballyHyperbolicSpacetimes}).
Hence we can apply Lemma \ref{lemExtensionOf ea(Pu)=00003Du} and
Lemma \ref{lemGreenOperatorsCommuteWithExteriorDerivativeAndCodifferential}
to conclude that
\begin{eqnarray}
e_{A\left[h^{s}\right]}^{a/r}A\left[h^{s}\right]\left(\chi_{+}^{a/r}\mathtt{A}\right) & = & e_{A\left[h^{s}\right]}^{a/r}\mathrm{\Box}_{1}\left[h^{s}\right]\left(\chi_{+}^{a/r}\mathtt{A}\right)-e_{A\left[h^{s}\right]}^{a/r}\mathrm{d}\mathrm{\delta}\left[h^{s}\right]\left(\chi_{+}^{a/r}\mathtt{A}\right)\nonumber \\
 & = & \chi_{+}^{a/r}\mathtt{A}-\mathrm{d}e_{A\left[h^{s}\right]}^{a/r}\mathrm{\delta}\left[h^{s}\right]\left(\chi_{+}^{a/r}\mathtt{A}\right)\mbox{.}\label{eqEMUsefulIdentityToFindTheConvenientFormOfTheClassicalRCE}
\end{eqnarray}
Applying $\mathrm{\delta}_{s}$ to both sides of the last equation
(with the superscript $r$) and exploiting the Leibniz rule, we find
\[
\left(\mathrm{\delta}_{s}e_{A\left[h^{s}\right]}^{r}\right)A\left(\chi_{+}^{r}\mathtt{A}\right)+e_{A}^{r}\left(\mathrm{\delta}_{s}A\left[h^{s}\right]\right)\left(\chi_{+}^{r}\mathtt{A}\right)=\mathrm{d}\left(\mathrm{\delta}_{s}\left(e_{A\left[h^{s}\right]}^{r}\mathrm{\delta}\left[h^{s}\right]\left(\chi_{+}^{r}\mathtt{A}\right)\right)\right)\mbox{,}
\]
which can be written as
\[
\left(\mathrm{\delta}_{s}e_{A\left[h^{s}\right]}^{r}\right)A\left(\chi_{+}^{r}\mathtt{A}\right)=-e_{A}^{r}\left(\mathrm{\delta}_{s}A\left[h^{s}\right]\right)\left(\chi_{+}^{r}\mathtt{A}\right)+\mathrm{d}\left(\mathrm{\delta}_{s}\left(e_{A\left[h^{s}\right]}^{r}\mathrm{\delta}\left[h^{s}\right]\left(\chi_{+}^{r}\mathtt{A}\right)\right)\right)\mbox{.}
\]
With this identity we can rewrite eq. \eqref{eqEMExpressionForTheFunctionalDerivativeOfTheClassicalRCEStep1}:
\[
\mathrm{\delta}_{s}\mathtt{X}^{s}=e_{A}A\left(\chi_{-}^{r}e_{A}^{r}\left(\mathrm{\delta}_{s}A\left[h^{s}\right]\right)\left(\chi_{+}^{r}\mathtt{A}\right)\right)-e_{A}A\left(\chi_{-}^{r}\mathrm{d}\left(\mathrm{\delta}_{s}\left(e_{A\left[h^{s}\right]}^{r}\mathrm{\delta}\left[h^{s}\right]\left(\chi_{+}^{r}\mathtt{A}\right)\right)\right)\right)\mbox{.}
\]
Now we show that the second term appearing on the RHS is both closed
and coclosed applying Lemma \ref{lemGreenOperatorsCommuteWithExteriorDerivativeAndCodifferential}:
\begin{alignat*}{2}
\mathrm{d}e_{A}A\left(\chi_{-}^{r}\mathrm{d}\left(\mathrm{\delta}_{s}\left(e_{A\left[h^{s}\right]}^{r}\mathrm{\delta}\left[h^{s}\right]\left(\chi_{+}^{r}\mathtt{A}\right)\right)\right)\right) & =e_{A}\mathrm{\Box}_{2}\mathrm{d}\left(\chi_{-}^{r}\mathrm{d}\left(\mathrm{\delta}_{s}\left(e_{A\left[h^{s}\right]}^{r}\mathrm{\delta}\left[h^{s}\right]\left(\chi_{+}^{r}\mathtt{A}\right)\right)\right)\right) & = & 0\mbox{,}\\
\mathrm{\delta}e_{A}A\left(\chi_{-}^{r}\mathrm{d}\left(\mathrm{\delta}_{s}\left(e_{A\left[h^{s}\right]}^{r}\mathrm{\delta}\left[h^{s}\right]\left(\chi_{+}^{r}\mathtt{A}\right)\right)\right)\right) & =e_{A}\mathrm{\delta}A\left(\chi_{-}^{r}\mathrm{d}\left(\mathrm{\delta}_{s}\left(e_{A\left[h^{s}\right]}^{r}\mathrm{\delta}\left[h^{s}\right]\left(\chi_{+}^{r}\mathtt{A}\right)\right)\right)\right) & = & 0\mbox{.}
\end{alignat*}
Therefore we are allowed to replace the representative $\mathrm{\delta}_{s}\mathtt{X}^{s}$
with the representative
\[
\mathtt{Y}=e_{A}A\left(\chi_{-}^{r}e_{A}^{r}\left(\mathrm{\delta}_{s}A\left[h^{s}\right]\right)\left(\chi_{+}^{r}\mathtt{A}\right)\right)
\]
without changing the equivalence class.

Observing that $\chi_{+}^{a}$ is supported inside $J_{+}^{\mathscr{M}}\left(M_{+}\right)$
(which does not intersect $K$ by definition of $M_{+}$) and recalling
that the coefficients of $\mathrm{\delta}_{s}A\left[h^{s}\right]$
are null outside $K$, we conclude that $\left(\mathrm{\delta}_{s}A\left[h^{s}\right]\right)\left(\chi_{+}^{a}\mathtt{A}\right)=0$,
hence we can add the term
\[
e_{A}A\left(\chi_{-}^{r}e_{A}^{r}\left(\mathrm{\delta}_{s}A\left[h^{s}\right]\right)\left(\chi_{+}^{a}\mathtt{A}\right)\right)
\]
to $\mathtt{Y}$ without any problem. In this way we obtain
\[
\mathtt{Y}=e_{A}A\left(\chi_{-}^{r}e_{A}^{r}\left(\mathrm{\delta}_{s}A\left[h^{s}\right]\right)\mathtt{A}\right)\mbox{.}
\]

Now take into account the term $\chi_{-}^{r}e_{A}^{a}\left(\mathrm{\delta}_{s}A\left[h^{s}\right]\right)\mathtt{A}$:
the coefficients of $\mathrm{\delta}_{s}A\left[h^{s}\right]$ are
supported inside $K$, hence
\[
\mathrm{supp}\left(e_{A}^{a}\left(\mathrm{\delta}_{s}A\left[h^{s}\right]\right)\mathtt{A}\right)\subseteq J_{+}^{\mathscr{M}}\left(K\right)\mbox{,}
\]
while $\chi_{-}^{r}$ is supported inside $J_{-}^{\mathscr{M}}\left(M_{-}\right)$.
This entails that $\chi_{-}^{r}e_{A}^{a}\left(\mathrm{\delta}_{s}A\left[h^{s}\right]\right)\mathtt{A}=0$,
therefore we can modify again our expression for $\mathtt{Y}$ subtracting
the term
\[
e_{A}A\left(\chi_{-}^{r}e_{A}^{a}\left(\mathrm{\delta}_{s}A\left[h^{s}\right]\right)\mathtt{A}\right)=0\mbox{.}
\]
The result is
\[
\mathtt{Y}=-e_{A}A\left(\chi_{-}^{r}e_{A}\left(\mathrm{\delta}_{s}A\left[h^{s}\right]\right)\mathtt{A}\right)\mbox{.}
\]

Now we focus our attention on the term $\theta=\left(\mathrm{\delta}_{s}A\left[h^{s}\right]\right)\mathtt{A}$.
First of all we notice that it is an element of $\mathrm{\Omega}_{0}^{1}M$
supported inside $K$ because of the support properties of the coefficients
of $\mathrm{\delta}_{s}A\left[h^{s}\right]$. Applying the Leibniz
rule in reverse we find the following chain of equalities:
\begin{equation}
\mathrm{\delta}\theta=\mathrm{\delta}_{s}\left(\mathrm{\delta}\left[h^{s}\right]A\left[h^{s}\right]\mathtt{A}\right)-\left(\delta_{s}\mathrm{\delta}\left[h^{s}\right]\right)A\mathtt{A}=0\mbox{,}\label{eqEMdeltasAhsAIsCoclosed}
\end{equation}
where we exploited $A\mathtt{A}=0$ and $\mathrm{\delta}\left[h^{s}\right]\circ A\left[h^{s}\right]=0$.
This proves that $\theta$ is also coclosed, hence $e_{A}\theta$
is a Lorentz solution. In particular we have $Ae_{A}\theta=0$ and
also $\mathrm{\delta}e_{A}\theta=0$\@. Then it follows that 
\begin{eqnarray*}
A\left(\chi_{-}^{r}e_{A}\theta\right) & = & -A\left(\chi_{-}^{a}e_{A}\theta\right)\mbox{,}\\
\mathrm{\delta}\left(\chi_{-}^{r}e_{A}\theta\right) & = & -\mathrm{\delta}\left(\chi_{-}^{a}e_{A}\theta\right)\mbox{.}
\end{eqnarray*}
From the last identities, exploiting $\mathrm{supp}\left(e_{A}\theta\right)\subseteq J^{\mathscr{M}}\left(K\right)$,
$\mathrm{supp}\left(\chi_{-}^{a/r}\right)=J_{\pm}^{\mathscr{M}}\left(\Sigma_{-}^{a/r}\right)$
for proper smooth spacelike Cauchy surfaces $\Sigma_{-}^{a/r}$ for
$\mathscr{M}$ and Proposition \ref{propUsefulSubsetsOfGloballyHyperbolicSpacetimes},
we deduce that the sections $A\left(\chi_{-}^{a/r}e_{A}\theta\right)$
and $\mathrm{\delta}\left(\chi_{-}^{a/r}e_{A}\theta\right)$ have
compact support. Therefore we can use eq. \eqref{eqEMUsefulIdentityToFindTheConvenientFormOfTheClassicalRCE}
for $s=0$ to obtain the following result: 
\begin{eqnarray*}
\mathtt{Y} & = & e_{A}^{a}A\left(\chi_{-}^{a}e_{A}\theta\right)+e_{A}^{r}A\left(\chi_{-}^{r}e_{A}\theta\right)\\
 & = & \chi_{-}^{a}e_{A}\theta-\mathrm{d}e_{A}^{a}\mathrm{\delta}\left(\chi_{+}^{a}e_{A}\theta\right)+\chi_{-}^{r}e_{A}\theta-\mathrm{d}e_{A}^{r}\mathrm{\delta}\left(\chi_{+}^{r}e_{a}\theta\right)\\
 & = & e_{A}\theta-\mathrm{d}e_{A}\mathrm{\delta}\left(\chi_{+}^{a}e_{A}\theta\right)\mbox{.}
\end{eqnarray*}
The last calculation proves that $\mathtt{Y}$ and $e_{A}\theta$
are gauge equivalent Lorentz solutions, hence we can consider
\[
\mathtt{Z}=e_{A}\theta=e_{A}\left(\mathrm{\delta}_{s}A\left[h^{s}\right]\right)\mathtt{A}
\]
as new representative of the same equivalence class, i.e.
\[
\mathtt{Z}\in\left[e_{A}A\left[h\right]\left(\chi_{-}^{a/r}e_{A\left[h\right]}A\left(\chi_{+}^{a/r}\mathtt{A}\right)\right)\right]_{M}\mbox{.}
\]
With this we conclude
\begin{equation}
\left.\frac{\mathrm{d}}{\mathrm{d}s}r_{h^{s}}^{\mathscr{M}}\left[\mathtt{A}\right]_{M}\right|_{0}=\left[e_{A}\left(\left.\frac{\mathrm{d}}{\mathrm{d}s}A\left[h^{s}\right]\right|_{0}\right)\mathtt{A}\right]_{M}\mbox{,}\label{eqEMExpressionForTheDerivativeOfTheClassicalRCE}
\end{equation}
where $\mathtt{A}$ is a representative of the class $\left[\mathtt{A}\right]_{M}$.
Note that also now the result does not depend on the choice of the
particular representative of $\left[\mathtt{A}\right]_{M}$ since,
if we consider two representatives in the same equivalence class,
they differ by a closed form $\tilde{\mathtt{A}}$, i.e. $\mathrm{d}\tilde{\mathtt{A}}=0$,
hence also $A\left[h^{s}\right]\tilde{\mathtt{A}}=0$ for each $s$.
This fact anyway is trivial since the original expression for $\mathrm{\delta}_{s}r_{h^{s}}^{\mathscr{M}}\left[\mathtt{A}\right]_{M}$
was independent of the choice of the representative and now we simply
looked for a convenient representative in the same equivalence class.

The evaluation of $\mathrm{\delta}_{s}A\left[h^{s}\right]\mathtt{A}$
can be carried on as for the case of the Proca field (as a matter
of fact the the only difference relies in the absence of the mass
term, which is irrelevant for this calculation since it does not depend
on $s$). For convenience we quote here the result:
\begin{equation}
\left.\frac{\mathrm{d}}{\mathrm{d}s}\left(A\left[h^{s}\right]\mathtt{A}\right)_{k}\right|_{0}=\left.\frac{\mathrm{d}}{\mathrm{d}s}h_{ij}^{s}\right|_{0}\nabla^{i}\mathtt{F}_{\phantom{j}k}^{j}+\left.\frac{\mathrm{d}}{\mathrm{d}s}\Gamma\left[h^{s}\right]_{ij}^{l}\right|_{0}g^{ij}\mathtt{F}_{lk}+\left.\frac{\mathrm{d}}{\mathrm{d}s}\Gamma\left[h^{s}\right]_{ik}^{l}\right|_{0}g^{ij}\mathtt{F}_{jl}\mbox{,}\label{eqEMDifferentialOperatorVariation}
\end{equation}
where $\mathtt{F}$ denotes the field strength associated to $\mathtt{A}$,
i.e. $\mathtt{F}=\mathrm{d}\mathtt{A}$ or, in local coordinates,
\begin{equation}
\mathtt{F}_{ij}=\partial_{i}\mathtt{A}_{j}-\partial_{j}\mathtt{A}_{i}\mbox{,}\label{eqEMClassicalFieldStrength}
\end{equation}
which can be expressed in a manifestly covariant manner on both $\mathscr{M}$
and $\mathscr{M}\left[h^{s}\right]$ (for each $s$) because the terms
involving the Christoffel symbols cancel out due to their symmetry
(cfr. eq. \eqref{eqChristoffelSymbolsOfTheLeviCivitaConnectionSymmetry}):
\[
\nabla_{i}\mathtt{A}_{j}-\nabla_{j}\mathtt{A}_{i}=\mathtt{F}_{ij}=\nabla_{i}\left[h^{s}\right]\mathtt{A}_{j}-\nabla_{j}\left[h^{s}\right]\mathtt{A}_{i}\mbox{.}
\]
Note that the RHS of eq. \eqref{eqEMDifferentialOperatorVariation}
is independent of the choice of the representative $\mathtt{A}$ since
only the field strength $\mathtt{F}=\mathrm{d}\mathtt{A}$ appears.

\subsubsection{Properties of the GNS representation induced by a quasi-free Hadamard
state for the electromagnetic field}

At this point we choose a quasi-free Hadamard state $\tau$ on the
CCR representation $\left(\mathcal{V},\mathrm{V}\right)=\mathscr{A}\left(\mathscr{M},\mathrm{\Lambda}^{1}M,A\right)$
describing the electromagnetic field on the globally hyperbolic spacetime
$\mathscr{M}$. With this choice, we introduce the (unique up to unitary
equivalence) GNS triple $\left(\mathscr{H}_{\tau}^{\mathscr{M}},\pi_{\tau}^{\mathscr{M}},\Omega_{\tau}^{\mathscr{M}}\right)$
induced by $\tau$ and we follow the discussion made in Subsection
\ref{subQuasiFreeHadamardStates}. In this way we obtain the represented
version
\begin{equation}
\mathrm{V}_{\tau}^{\mathscr{M}}=\pi_{\tau}^{\mathscr{M}}\circ\mathrm{V}:V\rightarrow\mathcal{B}\left(\mathscr{H}_{\tau}^{\mathscr{M}}\right)\label{eqEMRepresentedWeylMap}
\end{equation}
of the Weyl map $\mathrm{V}$, where $\left(V,\sigma\right)=\mathscr{B}\left(\mathscr{M},\mathrm{\Lambda}^{1}M,A\right)$
is the symplectic space provided by the covariant functor describing
the classical theory of the electromagnetic field, and the map
\begin{alignat*}{2}
\varPhi_{\tau}^{\mathscr{M}} & : & V & \rightarrow\mathcal{B}\left(\mathscr{H}_{\tau}^{\mathscr{M}}\right)\\
 &  & \left[\mathtt{A}\right]_{M} & \mapsto\varPhi_{\tau}^{\mathscr{M}}\left(\left[\mathtt{A}\right]_{M}\right)
\end{alignat*}
which maps each element $\left[\mathtt{A}\right]_{M}$ of $V$ to
a selfadjoint operator $\varPhi_{\tau}^{\mathscr{M}}\left(\left[\mathtt{A}\right]_{M}\right)$
on $\mathscr{H}_{\tau}^{\mathscr{M}}$. Moreover for each $\left[\mathtt{A}\right]_{M}\in V$
it holds that 
\[
\mathrm{e}^{\imath\varPhi_{\tau}^{\mathscr{M}}\left(\left[\mathtt{A}\right]_{M}\right)}=\mathrm{V}_{\tau}^{\mathscr{M}}\left(\left[\mathtt{A}\right]_{M}\right)\mbox{.}
\]
Together with the map $\varPhi_{\tau}^{\mathscr{M}}$, we have the
smeared fields (this is a consequence of the choice of a Hadamard
state):
\begin{eqnarray*}
\varPsi_{\tau}^{\mathscr{M}}:\mathrm{\Omega}_{0,\mathrm{\delta}}^{1}M & \rightarrow & \mathcal{B}\left(\mathscr{H}_{\tau}^{\mathscr{M}}\right)\\
\theta & \mapsto & -\imath\left.\frac{\mathrm{d}}{\mathrm{d}t}\mathrm{V}_{\tau}^{\mathscr{M}}\left(t\left[e_{A}\theta\right]_{M}\right)\right|_{0}\mbox{.}
\end{eqnarray*}
Note that here appears a slight difference with respect to the previous
cases, namely that the test section we consider are coclosed. As for
the general case, it holds that
\begin{equation}
\varPsi_{\tau}^{\mathscr{M}}\left(\theta\right)=\varPhi_{\tau}^{\mathscr{M}}\left(\left[e_{A}\theta\right]_{M}\right)\label{eqEMRelationBetweenvarPsiAndvarPhi}
\end{equation}
for each $\theta\in\mathrm{\Omega}_{0,\mathrm{\delta}}^{1}M$ and
we recognize $\varPsi_{\tau}^{\mathscr{M}}$ to be linear. 

Also in this case the choice of a quasi-free Hadamard state $\tau$
assures that Assumption \ref{assFunctionalDerivativeRCE} holds, i.e.
we are able to find a dense subspace $\mathscr{V}_{\tau}^{\mathscr{M}}$
of $\mathscr{H}_{\tau}^{\mathscr{M}}$ and a dense sub-{*}-algebra
$\mathcal{B}_{\tau}^{\mathscr{M}}$ of $\mathscr{A}\left(\mathscr{M},\mathrm{\Lambda}^{1}M,A\right)$
such that the functional derivative of the RCE with respect to the
spacetime metric can be defined. In particular $\mathscr{V}_{\tau}^{\mathscr{M}}$
is constituted by all the vectors of the form $L\Omega_{\tau}^{\mathscr{M}}$,
where $L$ is an arbitrary polynomial in $\mathrm{V}_{\tau}^{\mathscr{M}}\left(\left[\mathtt{A}\right]_{M}\right)$
and $\varPsi_{\tau}^{\mathscr{M}}\left(\theta\right)$ for arbitrary
$\left[\mathtt{A}\right]_{M}\in V$ and $\theta\in\mathrm{\Omega}_{0,\mathrm{\delta}}^{1}M$.

Again we have an equation similar to eq. \eqref{eqKeyPointForTheMainTheoremAboutRCE}:
for each $\xi\in\mathscr{V}_{\tau}^{\mathscr{M}}$, each $\left[\mathtt{A}\right]_{M}\in V$,
each compact subset $K$ of $M$ and each smooth $1$-parameter family
$\left(-1,1\right)\rightarrow GHP\left(\mathscr{M},K\right)$, $s\mapsto h^{s}$
such that $h^{0}=0$, it holds that
\begin{equation}
\left.\frac{\mathrm{d}}{\mathrm{d}s}\left\langle \xi,\mathrm{V}_{\tau}^{\mathscr{M}}\left(r_{h^{s}}^{\mathscr{M}}\left[\mathtt{A}\right]_{M}\right)\xi\right\rangle _{\tau}^{\mathscr{M}}\right|_{0}=\frac{\imath}{2}\left\langle \xi,\left\{ \varPhi_{\tau}^{\mathscr{M}}\left(\left.\frac{\mathrm{d}}{\mathrm{d}s}\left(r_{h^{s}}^{\mathscr{M}}\left[\mathtt{A}\right]_{M}\right)\right|_{0}\right),\mathrm{V}_{\tau}^{\mathscr{M}}\left(\left[\mathtt{A}\right]_{M}\right)\right\} \xi\right\rangle _{\tau}^{\mathscr{M}}\mbox{.}\label{eqEMKeyPointForTheMainTheoremAboutRCE}
\end{equation}

Moreover one can show that for each $\eta$, $\xi\in\mathscr{V}_{\tau}^{\mathscr{M}}$
there exists a smooth section, that we denote with
\begin{eqnarray}
M & \rightarrow & \mathrm{T}_{\mathbb{C}}M\label{eqEMUnsmearedField}\\
p & \mapsto & \left\langle \eta,\varPsi_{\tau}^{\mathscr{M}}\left(p\right)\xi\right\rangle _{\tau}^{\mathscr{M}}\mbox{,}\nonumber 
\end{eqnarray}
where $\mathrm{T}_{\mathbb{C}}M$ stands for the complex vector bundle
obtained via the tensor product of each fiber of $\mathrm{T}M$ with
$\mathbb{C}$ and $\left\langle \cdot,\cdot\right\rangle _{\tau}^{\mathscr{M}}$
denotes the scalar product of the Hilbert space $\mathscr{H}_{\tau}^{\mathscr{M}}$,
such that
\begin{equation}
\left\langle \eta,\varPsi_{\tau}^{\mathscr{M}}\left(\theta\right)\xi\right\rangle _{\tau}^{\mathscr{M}}=\int\limits _{M}\left(\theta\left(p\right)\right)\left(\left\langle \eta,\varPsi_{\tau}^{\mathscr{M}}\left(p\right)\xi\right\rangle _{\tau}^{\mathscr{M}}\right)\mathrm{d}\mu_{g}\label{eqEMRelationBetweenUnsmearedAndSmearedFields}
\end{equation}
for each $\theta\in\mathrm{\Omega}_{0,\mathrm{\delta}}^{1}M$, where
$\mathrm{d}\mu_{g}$ is the standard volume form on $\mathscr{M}$
and the dual pairing between $\mathrm{T}^{*}M$ and $\mathrm{T}M$
has been taken into account (note that one may indeed write the integrand
using the abstract index notation putting a contravariant index on
the new section and a covariant index on the test function).
\begin{rem}
We meet here the first consequence of the restriction of the set of
test sections to $\mathrm{\Omega}_{0,\mathrm{\delta}}^{1}M$, namely
that $\left\langle \eta,\varPsi_{\tau}^{\mathscr{M}}\left(p\right)\xi\right\rangle _{\tau}^{\mathscr{M}}$
fails to be unique: as a matter of fact each section that differs
from this one by an exact 1-form (with raised indices) will do the
work perfectly well. On the contrary, if there are two sections satisfying
eq. \eqref{eqEMRelationBetweenUnsmearedAndSmearedFields}, we deduce
that they differ by a closed one form (with raised indices).

It seems that we fail to have a characterization of $\left\langle \eta,\varPsi_{\tau}^{\mathscr{M}}\left(p\right)\xi\right\rangle _{\tau}^{\mathscr{M}}$
in terms of a class of gauge equivalent sections since we are not
sure that we obtain a section satisfying eq. \eqref{eqEMRelationBetweenUnsmearedAndSmearedFields}
if we add a closed form (with raised indices) to a section that satisfies
eq. \eqref{eqEMRelationBetweenUnsmearedAndSmearedFields}. However
we required that the first de Rham cohomology group of the manifolds
over which we discuss the electromagnetic field is trivial (see. Definition
\ref{defghsEM}), hence each closed 1-form is also exact. This hypothesis
restores the usual notion of gauge equivalence also for $\left\langle \eta,\varPsi_{\tau}^{\mathscr{M}}\left(p\right)\xi\right\rangle _{\tau}^{\mathscr{M}}$
because now exactness and closure of 1-forms coincide.

Indeed from a physical point of view we expected to find a counterpart
of the gauge equivalence for the electromagnetic field at the quantum
level.
\end{rem}
As we will see, the lack of uniqueness for \textbf{$\left\langle \eta,\varPsi_{\tau}^{\mathscr{M}}\left(p\right)\xi\right\rangle _{\tau}^{\mathscr{M}}$}
will not affect our calculation. For the moment we regard the section
in eq. \eqref{eqEMUnsmearedField} as (the matrix element of) one
of the gauge equivalent unsmeared electromagnetic fields induced by
the quasi-free Hadamard state $\tau$ on the globally hyperbolic spacetime
$\mathscr{M}$.

\subsubsection{Quantized stress-energy tensor for the electromagnetic field}

Our final theorem requires that we know how to express the quantized
stress-energy tensor for the electromagnetic field. As always, we
use as a starting point the equation governing the classical dynamics
of the field to obtain a natural expression for the action associated
to the field itself. After that, we determine the classical stress-energy
tensor for the electromagnetic field evaluating the functional derivative
of the action with respect to the spacetime metric and we try to determine
the corresponding quantum observable via the point-splitting procedure.
At the classical level the situation is identical to the Proca field
provided that we set $m=0$ (this is due to the fact that the linear
differential operator governing the classical dynamics of the electromagnetic
field, i.e. $\mathrm{\delta d}$, is nothing but the one for the Proca
field with $m=0$). For the action on the globally hyperbolic spacetime
$\mathscr{M}$ we obtain the following expression:

\[
S_{\mathscr{M}}=\frac{1}{2}\left(\mathtt{A},A\mathtt{A}\right)_{g,1}=\frac{1}{2}\left(\mathrm{d}\mathtt{A},\mathrm{d}\mathtt{A}\right)_{g,2}=\frac{1}{2}\int\limits _{M}\left(\mathrm{d}\mathtt{A}\wedge*\mathrm{d}\mathtt{A}\right)\mbox{.}
\]
Evaluating the functional derivative of $S_{\mathscr{M}}$ with respect
to the metric, we find the classical stress-energy tensor for the
electromagnetic field (we express it in local coordinates):
\begin{eqnarray*}
T_{ij}^{\mathscr{M}}\left(p\right) & = & \left.\frac{2}{\sqrt{\left|\det g_{h}\left(p\right)\right|}}\frac{\mathrm{\delta}S_{\mathscr{M}\left[h\right]}}{\mathrm{\delta}g_{h}^{ij}\left(p\right)}\right|_{0}\\
 & = & g^{bd}\left(p\right)\mathtt{F}_{ib}\left(p\right)\mathtt{F}_{jd}\left(p\right)-\frac{1}{4}g_{ij}\left(p\right)g^{ac}\left(p\right)g^{bd}\left(p\right)\mathtt{F}_{ab}\left(p\right)\mathtt{F}_{cd}\left(p\right)\mbox{,}
\end{eqnarray*}
where $\mathtt{F}$ is defined according to eq. \eqref{eqEMClassicalFieldStrength}.
The choice of a Hadamard state allows us to promote $T_{ij}^{\mathscr{M}}$
to the renormalized quantum stress-energy tensor $\mathcal{T}_{\tau\, ij}^{\mathscr{M}}$
simply via point-splitting (refer to \cite[eq. 4.6.5, p. 88]{Wal94}):
For each $\eta$, $\xi\in\mathscr{V}_{\tau}^{\mathscr{M}}$, we choose
two {}``near'' points $p$ and $q$ in $M$ and a curve $\gamma$
connecting them and, parallel transporting along the curve $\gamma$,
we write
\begin{multline}
\left\langle \eta,\mathcal{T}_{\tau}^{\mathscr{M}\, ij}\left(p,q\right)\xi\right\rangle _{\tau}^{\mathscr{M}}=g_{bf}\left(p\right)Y_{\gamma\, b}^{f}\left\langle \eta,\varPi_{\tau}^{\mathscr{M}\, ib}\left(p\right)\varPi_{\tau}^{\mathscr{M}\, jd}\left(q\right)\xi\right\rangle _{\tau}^{\mathscr{M}}\\
-\frac{1}{4}g^{ik}\left(p\right)Y_{\gamma\, k}^{j}g_{ae}\left(p\right)Y_{\gamma\, c}^{e}g_{bf}\left(p\right)Y_{\gamma\, d}^{f}\left\langle \eta,\varPi_{\tau}^{\mathscr{M}\, ab}\left(p\right)\varPi_{\tau}^{\mathscr{M}\, cd}\left(q\right)\xi\right\rangle _{\tau}^{\mathscr{M}}\mbox{,}\label{eqEMStressEnergyTensorViaPointSplitting}
\end{multline}
where we considered
\begin{equation}
\varPi_{\tau}^{\mathscr{M}\, ij}\left(p\right)=\nabla^{i}\varPsi_{\tau}^{\mathscr{M}\, j}\left(p\right)-\nabla^{j}\varPsi_{\tau}^{\mathscr{M}\, i}\left(p\right)\label{eqEMQuantizedFieldStrength}
\end{equation}
to shorten the expression.

If we lower all the indices in eq. \eqref{eqEMQuantizedFieldStrength},
we realize that $\varPi_{\tau}^{\mathscr{M}}$ is nothing but the
exterior derivative of the (non unique) unsmeared field. This fact
entails that $\varPi_{\tau}^{\mathscr{M}}$ does not depend on the
particular choice of the unsmeared field because, even if we add a
closed 1-form, then the exterior derivative set this contribution
to zero. A direct consequence of this fact is the independence of
eq. \eqref{eqEMStressEnergyTensorViaPointSplitting} on the choice
of an unsmeared field because only $\varPi_{\tau}^{\mathscr{M}\, ij}\left(p\right)$
appears on the RHS. Indeed all these observations must be intended
in the sense of matrix elements (which are the only ones that we defined
so far).

As we observed when we dealt with the Klein-Gordon field in $\mathcal{T}_{\tau\, ij}^{\mathscr{M}}$
there is no dependence upon the choice of the curve $\gamma$ along
which we parallel transport provided that the points $p$ and $q$
are sufficiently near so that there exists only one geodesic connecting
them and we consider such geodesic as $\gamma$. Indeed we can take
$p$ and $q$ in a sufficiently small neighborhood since our aim is
to take the limit $q\rightarrow p$ along $\gamma$ once that we have
found an expression that does not present divergences in this limit.
We also stress that the quantized stress-energy tensor obtained via
point-splitting differs by a multiple of the identity operator from
the quantized stress-energy tensor provided by the regularization
procedure with $\tau$ as reference state. Anyway we are only interested
in the commutator of the stress-energy tensor with some represented
Weyl generator, hence such difference is irrelevant in our computations.

As we said few lines above, our upcoming theorem will involve the
stress-energy tensor only in a commutator with some represented Weyl
generator $\mathrm{V}_{\tau}^{\mathscr{M}}\left(\left[\mathtt{A}\right]_{M}\right)$
for $\left[\mathtt{A}\right]_{M}\in V$, where $\left(V,\sigma\right)=\mathscr{B}\left(\mathscr{M},\mathrm{\Lambda}^{1}M,A\right)$
for a fixed object $\left(\mathscr{M},\mathrm{\Lambda}^{1}M,A\right)$
in $\mathfrak{ghs}^{EM}$. Reading eq. \eqref{eqEMStressEnergyTensorViaPointSplitting}
we realize that it would be useful to evaluate the matrix elements
of the commutator of $\varPi_{\tau}^{\mathscr{M}}\left(p\right)\varPi_{\tau}^{\mathscr{M}}\left(q\right)$
with an arbitrary represented Weyl generator. To this end we recall
eq. \eqref{eqCommutatorBetween2SmearedFieldsAndAWeylGenerator} and
we evaluate its LHS and its RHS fixing $\eta$, $\xi\in\mathscr{V}_{\tau}^{\mathscr{M}}$,
$\theta$, $\theta^{\prime}\in\mathrm{\Omega}_{0,\mathrm{\delta}}^{1}M$
and $\left[\mathtt{A}\right]_{M}\in V$ using eq. \eqref{eqEMRelationBetweenUnsmearedAndSmearedFields}
twice (all the equations are written using the abstract index notation):
\begin{multline*}
\iint\limits _{M}\left\langle \eta,\left[\varPsi_{\tau}^{\mathscr{M}\, i}\left(p\right)\varPsi_{\tau}^{\mathscr{M}\, j}\left(q\right),\mathrm{V}_{\tau}^{\mathscr{M}}\left(\left[\mathtt{A}\right]_{M}\right)\right]\xi\right\rangle _{\tau}^{\mathscr{M}}\theta_{i}\left(p\right)\theta_{j}^{\prime}\left(q\right)\mathrm{d}\mu_{g}\left(p\right)\mathrm{d}\mu_{g}\left(q\right)\\
=\left\langle \eta,\left[\varPsi_{\tau}^{\mathscr{M}}\left(\theta\right)\varPsi_{\tau}^{\mathscr{M}}\left(\theta^{\prime}\right),\mathrm{V}_{\tau}^{\mathscr{M}}\left(\left[\mathtt{A}\right]_{M}\right)\right]\xi\right\rangle _{\tau}^{\mathscr{M}}\mbox{.}
\end{multline*}
Indeed the matrix element inside the integral on the LHS of the last
equation is not unique because of the gauge invariance. Now we exploit
also the definition of the symplectic form $\sigma$ (cfr. Lemma \ref{lemSymplecticFormForEMField}):
\begin{multline*}
-\sigma\left(\left[e_{A}\theta\right]_{M},\left[\mathtt{A}\right]_{M}\right)\left\langle \eta,\mathrm{V}_{\tau}^{\mathscr{M}}\left(\left[\mathtt{A}\right]_{M}\right)\varPsi_{\tau}^{\mathscr{M}}\left(\theta^{\prime}\right)\xi\right\rangle _{\tau}^{\mathscr{M}}\\
=\iint\limits _{M}\mathtt{A}_{k}\left(p\right)g^{ki}\left(p\right)\theta_{i}\left(p\right)\theta_{j}^{\prime}\left(q\right)\left\langle \eta,\mathrm{V}_{\tau}^{\mathscr{M}}\left(\left[\mathtt{A}\right]_{M}\right)\varPsi_{\tau}^{\mathscr{M}\, j}\left(q\right)\xi\right\rangle _{\tau}^{\mathscr{M}}\mathrm{d}\mu_{g}\left(p\right)\mathrm{d}\mu_{g}\left(q\right)\mbox{,}
\end{multline*}
\begin{multline*}
-\sigma\left(\left[e_{A}\theta^{\prime}\right]_{M},\left[\mathtt{A}\right]_{M}\right)\left\langle \eta,\varPsi_{\tau}^{\mathscr{M}}\left(\theta\right)\mathrm{V}_{\tau}^{\mathscr{M}}\left(\left[\mathtt{A}\right]_{M}\right)\xi\right\rangle _{\tau}^{\mathscr{M}}\\
=\iint\limits _{M}\mathtt{A}_{k}\left(q\right)g^{kj}\left(q\right)\theta_{j}^{\prime}\left(q\right)\theta_{i}\left(p\right)\left\langle \eta,\varPsi_{\tau}^{\mathscr{M}\, i}\left(p\right)\mathrm{V}_{\tau}^{\mathscr{M}}\left(\left[\mathtt{A}\right]_{M}\right)\xi\right\rangle _{\tau}^{\mathscr{M}}\mathrm{d}\mu_{g}\left(p\right)\mathrm{d}\mu_{g}\left(q\right)\mbox{,}
\end{multline*}
where $\mathtt{A}$ is some representative of the class $\left[\mathtt{A}\right]_{M}$.
These integrals present the terms
\begin{gather*}
\mathtt{A}^{i}\left(p\right)\left\langle \eta,\mathrm{V}_{\tau}^{\mathscr{M}}\left(\left[\mathtt{A}\right]_{M}\right)\varPsi_{\tau}^{\mathscr{M}\, j}\left(q\right)\xi\right\rangle _{\tau}^{\mathscr{M}}\mbox{,}\\
\left\langle \eta,\mathrm{V}_{\tau}^{\mathscr{M}}\left(\left[\mathtt{A}\right]_{M}\right)\varPsi_{\tau}^{\mathscr{M}\, i}\left(qp\right)\xi\right\rangle _{\tau}^{\mathscr{M}}\mathtt{A}^{j}\left(q\right)
\end{gather*}
which are not uniquely defined exactly as seen above. From eq. \eqref{eqCommutatorBetween2SmearedFieldsAndAWeylGenerator}
and the freedom in the choice of $\theta$ and $\theta^{\prime}$
we deduce that
\begin{multline}
\left\langle \eta,\left[\varPsi_{\tau}^{\mathscr{M}\, i}\left(p\right)\varPsi_{\tau}^{\mathscr{M}\, j}\left(q\right),\mathrm{V}_{\tau}^{\mathscr{M}}\left(\left[\mathtt{A}\right]_{M}\right)\right]\xi\right\rangle _{\tau}^{\mathscr{M}}\\
\thicksim\mathtt{A}^{i}\left(p\right)\left\langle \eta,\mathrm{V}_{\tau}^{\mathscr{M}}\left(\left[\mathtt{A}\right]_{M}\right)\varPsi_{\tau}^{\mathscr{M}\, j}\left(q\right)\xi\right\rangle _{\tau}^{\mathscr{M}}+\mathtt{A}^{j}\left(q\right)\left\langle \eta,\varPsi_{\tau}^{\mathscr{M}\, i}\left(p\right)\mathrm{V}_{\tau}^{\mathscr{M}}\left(\left[\mathtt{A}\right]_{M}\right)\xi\right\rangle _{\tau}^{\mathscr{M}}\label{eqEMCommutatorBetween2UnsmearedFieldsAndAWeylGenerator}
\end{multline}
for each choice of $\mathtt{A}$ in the class $\left[\mathtt{A}\right]_{M}$,
where $\thicksim$ means gauge equivalence. All the terms that we
could add without affecting the relation $\thicksim$ cancel out once
that we evaluate
\[
\left\langle \eta,\left[\varPi_{\tau}^{\mathscr{M}\, ij}\left(p\right)\varPi_{\tau}^{\mathscr{M}\, kl}\left(q\right),\mathrm{V}_{\tau}^{\mathscr{M}}\left(\Theta\right)\right]\xi\right\rangle _{\tau}^{\mathscr{M}}\mbox{,}
\]
hence from eq. \eqref{eqEMCommutatorBetween2UnsmearedFieldsAndAWeylGenerator}
we deduce that
\begin{multline}
\left\langle \eta,\left[\varPi_{\tau}^{\mathscr{M}\, ij}\left(p\right)\varPi_{\tau}^{\mathscr{M}\, kl}\left(q\right),\mathrm{V}_{\tau}^{\mathscr{M}}\left(\Theta\right)\right]\xi\right\rangle _{\tau}^{\mathscr{M}}\\
=\mathtt{F}^{ij}\left(p\right)\left\langle \eta,\mathrm{V}_{\tau}^{\mathscr{M}}\left(\Theta\right)\varPi_{\tau}^{\mathscr{M}\, kl}\left(q\right)\xi\right\rangle _{\tau}^{\mathscr{M}}+\mathtt{F}^{kl}\left(q\right)\left\langle \eta,\varPi_{\tau}^{\mathscr{M},ij}\left(q\right)\mathrm{V}_{\tau}^{\mathscr{M}}\left(\Theta\right)\xi\right\rangle _{\tau}^{\mathscr{M}}\mbox{.}\label{eqEMCommutatorBetween2FieldStrengthsAndAWeylGenerator}
\end{multline}
This is the relation that we will use in the proof of the next theorem.

\subsubsection{Main theorem}

We are ready to state and prove the main theorem of this subsection.
Such theorem extends to the case of the electromagnetic field the
results of compatibility between the action of the functional derivative
of the relative Cauchy evolution and the stress-energy tensor, which
are already known to hold for the Klein-Gordon field and the Proca
field (refer to Subsection \ref{subKGRCE} and to Subsection \ref{subPRCE}).
\begin{thm}
Let $\mathscr{A}:\mathfrak{ghs}^{EM}\overset{\rightarrow}{\rightarrow}\mathfrak{alg}$
be the locally covariant quantum field theory for the electromagnetic
field built in Subsection \ref{subElectromagneticField} and let $\left(\mathscr{M},\mathrm{\Lambda}^{1}M,A\right)$
be an object of the category $\mathfrak{ghs}^{EM}$ (see Definition
\ref{defghsEM}). Consider a quasi-free Hadamard state $\tau$ on
the CCR representation $\left(\mathcal{V},\mathrm{V}\right)=\mathscr{A}\left(\mathscr{M},\mathrm{\Lambda}^{1}M,A\right)$
and denote the GNS triple induced by $\tau$ with $\left(\mathscr{H}_{\tau}^{\mathscr{M}},\pi_{\tau}^{\mathscr{M}},\Omega_{\tau}^{\mathscr{M}}\right)$.
We denote with $\mathrm{V}_{\tau}^{\mathscr{M}}$ the represented
counterpart of the Weyl map $\mathrm{V}$ (cfr. eq. \eqref{eqEMRepresentedWeylMap})
and with $\mathcal{T}_{\tau}^{\mathscr{M}}$ the quantum stress-energy
tensor for the electromagnetic field on $\mathscr{M}$ obtained via
point-splitting in the representation induced by the state $\tau$
(cfr. eq. \eqref{eqEMStressEnergyTensorViaPointSplitting}). Then
there exists a dense subspace $\mathscr{V}_{\tau}^{\mathscr{M}}$
of $\mathscr{H}_{\tau}^{\mathscr{M}}$ such that
\[
\frac{\mathrm{\delta}}{\mathrm{\delta}h}\pi_{\tau}^{\mathscr{M}}\left(R_{h}^{\mathscr{M}}\left(\mathrm{V}\left(\left[\mathtt{A}\right]_{M}\right)\right)\right)=-\frac{\imath}{2}\left[\mathcal{T}_{\tau}^{\mathscr{M}},\mathrm{V}_{\tau}^{\mathscr{M}}\left(\left[\mathtt{A}\right]_{M}\right)\right]\quad\forall\left[\mathtt{A}\right]_{M}\in V
\]
in the sense of quadratic forms on $\mathscr{V}_{\tau}^{\mathscr{M}}$.\end{thm}
\begin{proof}
A dense subspace $\mathscr{V}_{\tau}^{\mathscr{M}}$ of $\mathscr{H}_{\tau}^{\mathscr{M}}$
exists by virtue of the choice of a quasi-free Hadamard state $\tau$
(see few lines before eq. \eqref{eqEMKeyPointForTheMainTheoremAboutRCE}).

We fix $\xi\in\mathscr{V}_{\tau}^{\mathscr{M}}$, $\left[\mathtt{A}\right]_{M}\in V$,
a compact subset $K$ of $M$ and 1-parameter family $\left(-1,1\right)\rightarrow GHP\left(\mathscr{M},K\right)$,
$s\mapsto h^{s}$ such that $h^{0}=0$. We repeat the first part of
the proof of Theorem \ref{thmKGFunctionalDerivativeOfRCEAgreesWithStressEnergyTensor}
using eq. \eqref{eqEMQuantumVsClassicalRCE} and eq. \eqref{eqEMKeyPointForTheMainTheoremAboutRCE}
in place of eq. \eqref{eqKGQuantumVsClassicalRCE} and respectively
eq. \eqref{eqKGKeyPointForTheMainTheoremAboutRCE}. In this way we
find the following reformulation of the thesis:
\begin{multline*}
\left\langle \xi,\left\{ \varPhi_{\tau}^{\mathscr{M}}\left(\mathrm{\delta}_{s}r_{h^{s}}^{\mathscr{M}}\left[\mathtt{A}\right]_{M}\right),\mathrm{V}_{\tau}^{\mathscr{M}}\left(\left[\mathtt{A}\right]_{M}\right)\right\} \xi\right\rangle _{\tau}^{\mathscr{M}}\\
=-\int\limits _{M}\left(\mathrm{\delta}_{s}h^{s}\right)\left(\left\langle \xi,\left[\mathcal{T}_{\tau}^{\mathscr{M}},\mathrm{V}_{\tau}^{\mathscr{M}}\left(\left[\mathtt{A}\right]_{M}\right)\right]\xi\right\rangle _{\tau}^{\mathscr{M}}\right)\mathrm{d}\mu_{g}\mbox{,}
\end{multline*}
where the dual pairing between $\mathrm{T}^{*}M\otimes_{s}\mathrm{T}^{*}M$
and $\mathrm{T}M\otimes_{s}\mathrm{T}M$ is taken into account in
the integrand appearing on the RHS. Now we exploit eq. \eqref{eqEMExpressionForTheDerivativeOfTheClassicalRCE}
choosing a representative $\mathtt{A}$ of the fixed class $\left[\mathtt{A}\right]_{M}$:
\begin{multline*}
\left\langle \xi,\left\{ \varPhi_{\tau}^{\mathscr{M}}\left(\left[e_{A}\mathrm{\delta}_{s}A\left[h^{s}\right]\mathtt{A}\right]_{M}\right),\mathrm{V}_{\tau}^{\mathscr{M}}\left(\left[\mathtt{A}\right]_{M}\right)\right\} \xi\right\rangle _{\tau}^{\mathscr{M}}\\
=-\int\limits _{M}\left(\mathrm{\delta}_{s}h^{s}\right)\left(\left\langle \xi,\left[\mathcal{T}_{\tau}^{\mathscr{M}},\mathrm{V}_{\tau}^{\mathscr{M}}\left(\left[\mathtt{A}\right]_{M}\right)\right]\xi\right\rangle _{\tau}^{\mathscr{M}}\right)\mathrm{d}\mu_{g}\mbox{.}
\end{multline*}
Note that the result does not depend on the particular choice of $\mathtt{A}$
in the class $\left[\mathtt{A}\right]_{M}$ because the same is true
for eq. \eqref{eqEMExpressionForTheDerivativeOfTheClassicalRCE}.
We can apply also eq. \eqref{eqEMRelationBetweenvarPsiAndvarPhi}
since eq. \eqref{eqEMdeltasAhsAIsCoclosed} shows that $\mathrm{\delta}_{s}A\left[h^{s}\right]\mathtt{A}$
is coclosed whatever choice of $\mathtt{A}$ we make:
\begin{multline*}
\overset{\mathsf{L}}{\overbrace{\left\langle \xi,\left\{ \varPsi_{\tau}^{\mathscr{M}}\left(\mathrm{\delta}_{s}A\left[h^{s}\right]\mathtt{A}\right),\mathrm{V}_{\tau}^{\mathscr{M}}\left(\left[\mathtt{A}\right]_{M}\right)\right\} \xi\right\rangle _{\tau}^{\mathscr{M}}}}\\
=\underset{\mathsf{R}}{\underbrace{-\int\limits _{M}\left(\mathrm{\delta}_{s}h^{s}\right)\left(\left\langle \xi,\left[\mathcal{T}_{\tau}^{\mathscr{M}},\mathrm{V}_{\tau}^{\mathscr{M}}\left(\left[\mathtt{A}\right]_{M}\right)\right]\xi\right\rangle _{\tau}^{\mathscr{M}}\right)\mathrm{d}\mu_{g}}}\mbox{.}
\end{multline*}

Now we work with the LHS of the last equation (denoted by $\mathsf{L}$)
and the RHS (denoted by $\mathsf{R}$) separately. Starting from $\mathsf{L}$,
we exploit the relation between smeared and unsmeared fields, eq.
\eqref{eqEMRelationBetweenUnsmearedAndSmearedFields}:
\[
\mathsf{L}=\int\limits _{M}\left(\left(\mathrm{\delta}_{s}A\left[h^{s}\right]\mathtt{A}\right)\left(p\right)\right)\left(\left\langle \xi,\left\{ \varPsi_{\tau}^{\mathscr{M}}\left(p\right),\mathrm{V}_{\tau}^{\mathscr{M}}\left(\left[\mathtt{A}\right]_{M}\right)\right\} \xi\right\rangle _{\tau}^{\mathscr{M}}\right)\mathrm{d}\mu_{g}\mbox{,}
\]
where we consider the dual pairing between $\mathrm{T}^{*}M$ and
$\mathrm{T}M$. Indeed the integrand on the right is not uniquely
determined because of gauge equivalence. Anyway every admissible choice
of this section will give the same value for $\mathsf{L}$. In order
to find an expression for $\mathsf{L}$ in local coordinates, we perform
the usual construction which provides a finite family $\left\{ \left(U_{\alpha},V_{\alpha},\phi_{\alpha}\right)\right\} $
obtained choosing all the elements of a locally finite covering of
$M$ constituted by oriented coordinate neighborhoods that intersect
the fixed compact subset $K$ of $M$ (which includes the support
of the coefficients appearing in $\mathrm{\delta}_{s}A\left[h^{s}\right]$).
As usual the choice of the oriented coordinate neighborhoods is made
in such a way that $\left|\det g\right|=1$ so that $\mathrm{d}\mu_{g}$
reduces to the standard volume form $\mathrm{d}V$ on $\mathbb{R}^{4}$
on each coordinate neighborhood. At the same time we take only the
corresponding members $\left\{ \chi_{\alpha}\right\} $ in the partition
of unity subordinate to the original locally finite covering. In this
way we obtain the expression of $\mathsf{L}$ in local coordinates:
\[
\mathsf{L}=\sum_{\alpha}\int\limits _{V_{\alpha}}\chi_{\alpha}\left\langle \xi,\left\{ \varPsi_{\tau}^{\mathscr{M}\, i}\left(x\right),\mathrm{V}_{\tau}^{\mathscr{M}}\left(\left[\mathtt{A}\right]_{M}\right)\right\} \xi\right\rangle _{\tau}^{\mathscr{M}}\left(\mathrm{\delta}_{s}A\left[h^{s}\right]\mathtt{A}\right)_{i}\left(x\right)\mathrm{d}V\mbox{,}
\]
where all the sections that appear inside the integral are now written
in local coordinates, namely $\mathrm{\delta}_{s}A\left[h^{s}\right]\mathtt{A}$
inside the integral over $V_{\alpha}$ denotes the push-forward through
$\phi_{\alpha}$ of the original $\left(\mathrm{\delta}_{s}A\left[h^{s}\right]\mathtt{A}\right)$
restricted to $U_{\alpha}$ and similarly for the other sections inside
the integral. It is convenient to define
\begin{eqnarray*}
\zeta:M & \rightarrow & \mathrm{T}_{\mathbb{C}}M\\
p & \mapsto & \left\langle \xi,\left\{ \varPsi_{\tau}^{\mathscr{M}}\left(x\right),\mathrm{V}_{\tau}^{\mathscr{M}}\left(\left[\mathtt{A}\right]_{M}\right)\right\} \xi\right\rangle _{\tau}^{\mathscr{M}}
\end{eqnarray*}
in order to simplify our notation. Indeed $\zeta$ is not uniquely
determined so that we fix some proper $\zeta$ and we show that everything
works whatever choice of $\zeta$ we make. The next two steps are
identical to the corresponding ones in the proof of Theorem \ref{thmPFunctionalDerivativeOfRCEAgreesWithStressEnergyTensor},
provided that we use $\mathtt{F}$ defined in eq. \eqref{eqEMExpressionForTheDerivativeOfTheClassicalRCE}
in place of $\Pi$: In first place we use eq. \eqref{eqEMDifferentialOperatorVariation}
and in second place we partially integrate. We get the following result:
\begin{eqnarray*}
\mathsf{L} & = & \overset{\mathsf{X}}{\overbrace{-\sum_{\alpha}\int\limits _{V_{\alpha}}\chi_{\alpha}\left(\nabla^{i}\zeta^{k}\right)\mathtt{F}_{\phantom{j}k}^{j}\mathrm{\delta}_{s}h_{ij}^{s}\mathrm{d}V}}+\overset{\mathsf{L}_{2}}{\overbrace{\sum_{\alpha}\int\limits _{V_{\alpha}}\chi_{\alpha}\zeta^{k}\mathtt{F}_{lk}\mathrm{\delta}_{s}\Gamma\left[h^{s}\right]_{ij}^{l}g^{ij}\mathrm{d}V}}\\
 &  & +\underset{\mathsf{L}_{3}}{\underbrace{\sum_{\alpha}\int\limits _{V_{\alpha}}\chi_{\alpha}\zeta^{k}\mathtt{F}_{jl}g^{ij}\mathrm{\delta}_{s}\Gamma\left[h^{s}\right]_{ik}^{l}\mathrm{d}V}}\underset{\mathsf{L}_{4}}{\underbrace{-\sum_{\alpha}\int\limits _{V_{\alpha}}\chi_{\alpha}\zeta_{k}\mathtt{F}^{jk}\nabla^{i}\mathrm{\delta}_{s}h_{ij}^{s}\mathrm{d}V}}\mbox{.}
\end{eqnarray*}

Now we focus on $\mathsf{R}$ and we express it using the local coordinates
$\left\{ \left(U_{\alpha},V_{\alpha},\phi_{\alpha}\right)\right\} $:
\[
\mathsf{R}=-\sum_{\alpha}\int\limits _{V_{\alpha}}\chi_{\alpha}\left(x\right)\left(\mathrm{\delta}_{s}h_{ij}^{s}\right)\left(x\right)\left\langle \xi,\left[\mathcal{T}_{\tau}^{\mathscr{M}\, ij}\left(x\right),\mathrm{V}_{\tau}^{\mathscr{M}}\left(\left[\mathtt{A}\right]_{M}\right)\right]\xi\right\rangle _{\tau}^{\mathscr{M}}\mathrm{d}V\mbox{.}
\]
Now recall the expression of the quantized stress-energy tensor, eq.
\eqref{eqEMStressEnergyTensorViaPointSplitting}, and the commutation
relation found in eq. \eqref{eqEMCommutatorBetween2FieldStrengthsAndAWeylGenerator}.
Exploiting these results, evaluate $\left\langle \xi,\left[\mathcal{T}_{\tau}^{\mathscr{M}\, ij}\left(p,q\right),\mathrm{V}_{\tau}^{\mathscr{M}}\left(\left[\mathtt{A}\right]_{M}\right)\right]\xi\right\rangle _{\tau}^{\mathscr{M}}$.
That done, take the coincidence limit as required by the point-splitting
procedure (note that no divergence arises) and insert the result into
the last equation. After that, use the symmetry of $\mathrm{\delta}_{s}h^{s}$
and $g$ to simplify the expression (matrix elements of anticommutators
should appear). All these operations produce the following result
(to shorten the expression we denote $\left\langle \xi,\left\{ \varPi_{\tau}^{\mathscr{M}}\left(x\right),\mathrm{V}_{\tau}^{\mathscr{M}}\left(\left[\mathtt{A}\right]_{M}\right)\right\} \xi\right\rangle _{\tau}^{\mathscr{M}}$
with $\Xi$):
\[
\mathsf{R}=\underset{\mathsf{R}_{1}}{\underbrace{-\sum_{\alpha}\int\limits _{V_{\alpha}}\chi_{\alpha}\Xi^{ib}\mathtt{F}_{\phantom{j}b}^{j}\mathrm{\delta}_{s}h_{ij}^{s}\mathrm{d}V}}+\underset{\mathsf{R}_{2}}{\underbrace{\frac{1}{4}\sum_{\alpha}\int\limits _{V_{\alpha}}\chi_{\alpha}\mathtt{F}_{ab}\Xi^{ab}\mathrm{\delta}_{s}h_{ij}^{s}g^{ij}\mathrm{d}V}}\mbox{.}
\]
Eq. \eqref{eqEMQuantizedFieldStrength} and the subsequent remarks
entail that
\begin{eqnarray*}
\Xi^{ij} & = & \left\langle \xi,\left\{ \varPi_{\tau}^{\mathscr{M}\, ij},\mathrm{V}_{\tau}^{\mathscr{M}}\left(\left[\mathtt{A}\right]_{M}\right)\right\} \xi\right\rangle _{\tau}^{\mathscr{M}}\\
 & = & \nabla^{i}\left\langle \xi,\left\{ \varPsi_{\tau}^{\mathscr{M}\, j}\left(x\right),\mathrm{V}_{\tau}^{\mathscr{M}}\left(\left[\mathtt{A}\right]_{M}\right)\right\} \xi\right\rangle _{\tau}^{\mathscr{M}}-\nabla^{j}\left\langle \xi,\left\{ \varPsi_{\tau}^{\mathscr{M}\, i}\left(x\right),\mathrm{V}_{\tau}^{\mathscr{M}}\left(\left[\mathtt{A}\right]_{M}\right)\right\} \xi\right\rangle _{\tau}^{\mathscr{M}}
\end{eqnarray*}
does not depend on the particular choice of the non unique matrix
element of the unsmeared field. In particular we can use the section
$\zeta$ previously fixed:
\[
\Xi^{ij}=\nabla^{i}\zeta^{j}-\nabla^{j}\zeta^{i}\mbox{.}
\]
We denote the first part of $\mathsf{R}$ with $\mathsf{R}_{1}$ and
the second with $\mathsf{R}_{2}$. In first place we evaluate $\mathsf{R}_{1}$
by partial integration (we omit the term including derivatives of
$\chi_{\alpha}$ since as always they give null contribution):
\begin{eqnarray*}
\mathsf{R}_{1} & = & \overset{=\mathsf{X}}{\overbrace{-\sum_{\alpha}\int\limits _{V_{\alpha}}\chi_{\alpha}\left(\nabla^{i}\zeta^{k}\right)\mathtt{F}_{\phantom{j}k}^{j}\mathrm{\delta}_{s}h_{ij}^{s}\mathrm{d}V}}+\sum_{\alpha}\int\limits _{V_{\alpha}}\chi_{\alpha}\left(\nabla^{b}\zeta^{i}\right)\mathtt{F}_{\phantom{j}b}^{j}\mathrm{\delta}_{s}h_{ij}^{s}\mathrm{d}V\\
 & = & \mathsf{X}\underset{\mathsf{R}_{3}}{\underbrace{-\sum_{\alpha}\int\limits _{V_{\alpha}}\chi_{\alpha}\zeta^{k}\mathtt{F}^{jb}\nabla_{b}\mathrm{\delta}_{s}h_{kj}^{s}\mathrm{d}V}}+\sum_{\alpha}\int\limits _{V_{\alpha}}\chi_{\alpha}\zeta^{i}g^{jk}\underset{=0}{\underbrace{\left(\nabla^{b}\mathtt{F}_{bk}\right)}}\mathrm{\delta}_{s}h_{ij}^{s}\mathrm{d}V\mathrm{,}
\end{eqnarray*}
where we recognized the term $\mathsf{X}$ already present in $\mathsf{L}$,
we exploited the fact that
\[
\nabla^{i}\mathtt{F}_{ij}=\left(\mathrm{\delta d}\mathtt{A}\right)_{j}=0
\]
because every representative $\mathtt{A}$ of the class $\left[\mathtt{A}\right]_{M}$
satisfies $A\mathtt{A}=\mathrm{\delta d}\mathtt{A}=0$ and we denoted
with $\mathsf{R}_{3}$ the remaining term. We have the following result:
\[
\mathsf{R}_{1}=\mathsf{X}+\mathsf{R}_{3}\mbox{.}
\]
In second place we evaluate $\mathsf{R}_{2}$ proceeding with the
same approach. First of all we notice that we can exploit the antisymmetry
of $\mathtt{F}$ to simplify a little bit the first integral. Then
we partially integrate such term with the purpose of finding another
integrand that explicitly exhibits the structure of the field equation,
i.e. a term $\nabla^{i}\mathtt{F}_{ij}$, so that we can get rid of
it too (again we omit the null term containing derivatives of $\chi_{\alpha}$):
\begin{eqnarray*}
\mathsf{R}_{2} & = & \frac{1}{2}\sum_{\alpha}\int\limits _{V_{\alpha}}\chi_{\alpha}\mathtt{F}_{ab}\left(\nabla^{a}\zeta^{b}\right)\mathrm{\delta}_{s}h_{ij}^{s}g^{ij}\mathrm{d}V\\
 & = & -\frac{1}{2}\sum_{\alpha}\int\limits _{V_{\alpha}}\chi_{\alpha}\zeta^{b}\mathtt{F}_{ab}\nabla^{a}\mathrm{\delta}_{s}h_{ij}^{s}g^{ij}\mathrm{d}V-\frac{1}{2}\sum_{\alpha}\int\limits _{V_{\alpha}}\chi_{\alpha}\zeta^{b}\underset{=0}{\underbrace{\left(\nabla^{a}\mathtt{F}_{ab}\right)}}\mathrm{\delta}_{s}h_{ij}^{s}g^{ij}\mathrm{d}V\\
 & = & -\frac{1}{2}\sum_{\alpha}\int\limits _{V_{\alpha}}\chi_{\alpha}\zeta^{b}\mathtt{F}_{ab}\nabla^{a}\mathrm{\delta}_{s}h_{ij}^{s}g^{ij}\mathrm{d}V\mbox{.}
\end{eqnarray*}
Therefore, renaming some summation indices, we obtain the following
result:
\[
\mathsf{R}=\mathsf{X}+\mathsf{R}_{2}+\mathsf{R}_{3}\mbox{.}
\]

At this stage our thesis $\mathsf{L}=\mathsf{R}$ is reduced to the
following identity:
\[
\mathsf{L}_{2}+\mathsf{L}_{3}+\mathsf{L}_{4}=\mathsf{R}_{2}+\mathsf{R}_{3}\mbox{.}
\]
One immediately realizes that eq. \eqref{eqPGeometricalIdentity1}
(with $\mathtt{F}$ in place of $\Pi$) and eq. \eqref{eqPGeometricalIdentity2}
imply our last equation: to recognize this fact proceed as we did
after eq. \eqref{eqPGeometricalIdentity2} in the case of the Proca
field.

Eq. \eqref{eqPGeometricalIdentity2} is a purely geometrical identity,
hence holds also in this case without any further comment. On the
contrary eq. \eqref{eqPGeometricalIdentity1} involves an object strictly
connected with the dynamics of the Proca field (namely $\Pi$), however
the proof of this identity relies only on the antisymmetry of such
object, a property that indeed holds also for $\mathtt{F}$, hence
a similar identity holds for $\mathtt{F}$ in place of $\Pi$. These
observations entail that we have $\mathsf{L}=\mathsf{R}$ whatever
choice of $\zeta$ we make. This completes the proof.
\end{proof}

%% file: 9_conclusions.tex
\chapter*{Conclusions}

\addcontentsline{toc}{chapter}{Conclusions}

In Chapter \ref{chapMathematicalPreliminaries} we introduced almost
all the mathematical tools required for the entire thesis. Particular
attention was devoted to geometrical tools in the context of vector
bundles, which constitute the mathematical setting of the whole discussion,
together with globally hyperbolic spacetimes. We also recalled some
results about normally hyperbolic equations on globally hyperbolic
spacetimes. After that we turned our attention to the algebraic tools,
namely algebras and states, that are needed to discuss the algebraic
approach to quantum field theory. We focused mainly on particular
C{*}-algebras, namely Weyl systems and CCR representation, which are
well suited for the quantization of bosonic fields. To conclude some
definitions from category theory where presented, the language of
category theory being suitable for a number of notions presented in
the thesis.

After the required mathematical preliminaries, the main subject of
the thesis was tackled in Chapter \ref{chapGCLP} with the introduction
the \textsl{generally covariant locality principle} (\textsl{GCLP}),
originally formulated in \cite{BFV03}. To do this, in first place
we analyzed in detail the structure of the category of globally hyperbolic
spacetimes and the structure of the category of unital C{*}-algebras,
taking advantage of the remarks made in Chapter \ref{chapMathematicalPreliminaries}.
In second place we stated the GCLP giving the definition of \textsl{locally
covariant quantum field theory} (\textsl{LCQFT}). We devoted particular
attention to the physical interpretation of the GCLP, essentially
borrowing the interpretation of the Haag-Kastler axioms (refer to
\cite{HK64}). We also showed in full detail that it is possible to
completely recover the Haag-Kastler axioms starting from the assignment
of a locally covariant quantum field theory fulfilling both the causality
condition and the time slice axiom. In third place we showed how to
realize a LCQFT starting from a normally hyperbolic equation over
a globally hyperbolic spacetime. This was done in two steps. The first
one consisted in the realization of a covariant functor describing
the classical theory of the field whose dynamics is ruled by the assigned
normally hyperbolic equation, while the second was realized quantizing
such classical field theory via composition with another covariant
functor that embodies the quantization procedure. Great care was devoted
to study in full detail the properties of the starting category for
the classical field functor, which is a sort of enriched category
of globally hyperbolic spacetimes. We concluded Chapter \ref{chapGCLP}
with the construction of LCQFTs for the Klein-Gordon field, the Proca
field and the electromagnetic field. While the Klein-Gordon case is
nothing more than a specialization of the general procedure, the other
two cases required more attention as a consequence of the lack of
a normally hyperbolic equation governing their classical dynamics.
The case of the electromagnetic field proved to be the most involved.
To simplify the situation, we restricted to those field strengths
that could be described in terms of a vector potential. Therefore,
in place of the Maxwell equations, we considered the resulting equation
for the vector potential and we kept into account the effects of gauge
equivalence.

Chapter \ref{chapRCE} was devoted to the main argument of the thesis,
namely the \textsl{relative Cauchy evolution} (\textsl{RCE}). In fact
our original purpose was to show that a relation between the RCE and
the stress-energy tensor similar to the one proved in \cite{BFV03}
for the Klein-Gordon field holds also for the Proca field and the
electromagnetic field. In first place we defined in a general context
the RCE and its functional derivative with respect to the spacetime
metric. We proved that the functional derivative, which is symmetric
by construction, is also divergence free, thus finding a hint for
a possible strict relation with the stress-energy tensor. After that
we returned to the examples discussed at the end of Chapter \ref{chapGCLP}.
In first place we proved the relation between the functional derivative
of the RCE and the stress-energy tensor originally showed in \cite{BFV03}
for the case of the Klein-Gordon field. In second place we tried to
extend this result to the Proca field and the electromagnetic field.
While the case of the Proca field proved to be almost straightforward
(the main difference can be ascribed to the fact that the Proca field
is a 1-form, while the Klein-Gordon field is a 0-form), the electromagnetic
field presents some additional complications. Anyway we were able
to circumvent these obstructions exploiting the gauge equivalence.
In this way our purpose was achieved, namely we showed that the relation
between the functional derivative of the RCE and stress-energy tensor,
which was already known to hold for the Klein-Gordon field, holds
in an identical form in the cases of the Proca and the electromagnetic
fields too.

%% file: 12_bibliografy.tex
\bibliographystyle{bibliografy}
\addcontentsline{toc}{chapter}{\bibname}\bibliography{bibliografy}

%% file: 13_subject_index.tex
\addcontentsline{toc}{chapter}{Index}

\printindex{}